\documentclass[12pt]{report}
\pdfoutput=1
\usepackage{amsmath}
\usepackage{amssymb}
\usepackage{mathtools}
\usepackage{graphicx}
\usepackage{hyperref}
\usepackage{caption}
\usepackage{pdflscape}
\usepackage{pdfpages}
\usepackage{enumerate}
\usepackage{array}
\usepackage{makecell}
\usepackage{ctable}
\usepackage{morefloats}
\usepackage{setspace}
\usepackage[]{appendix}
\usepackage[superscript]{cite}
\usepackage{mathptmx}
\usepackage[parfill]{parskip}
\usepackage{afterpage}
\usepackage{attachfile}
\usepackage{geometry}
\usepackage{titlesec}
\usepackage{xcolor}

\hypersetup{colorlinks,linkcolor={red!50!black},citecolor={green!50!black},filecolor={blue!80!black}}
\titleformat*{\section}{\normalsize\bfseries}
\titleformat*{\subsection}{\normalsize\bfseries}
\titleformat*{\subsubsection}{\normalsize\bfseries}
\titleformat{\chapter}[display]
{\bfseries\Large}{Chapter \ \thechapter\Large}{5pt}{}[\vspace{-0.8cm}\rule{\textwidth}{2pt}]
\begin{document}
\begin{spacing}{1.5}


\begin{titlepage}
\thispagestyle{empty}
\begin{center}
\vspace*{0.2cm}
\LARGE
\textbf{Introduction to the Kalman Filter and Tuning its Statistics for Near Optimal Estimates and Cramer Rao Bound}

\vspace{1.5cm}
\large{by\\}
\vspace{0.5cm}
\normalsize{\textbf{Shyam Mohan M, Naren Naik, R.M.O. Gemson, M.R. Ananthasayanam}}\\

\vspace{2cm}
\large\textbf{Technical Report : TR/EE2015/401 } \\

\vspace{2cm}
\begin{figure}[!h]
\begin{center}
\scalebox{0.35}{\includegraphics{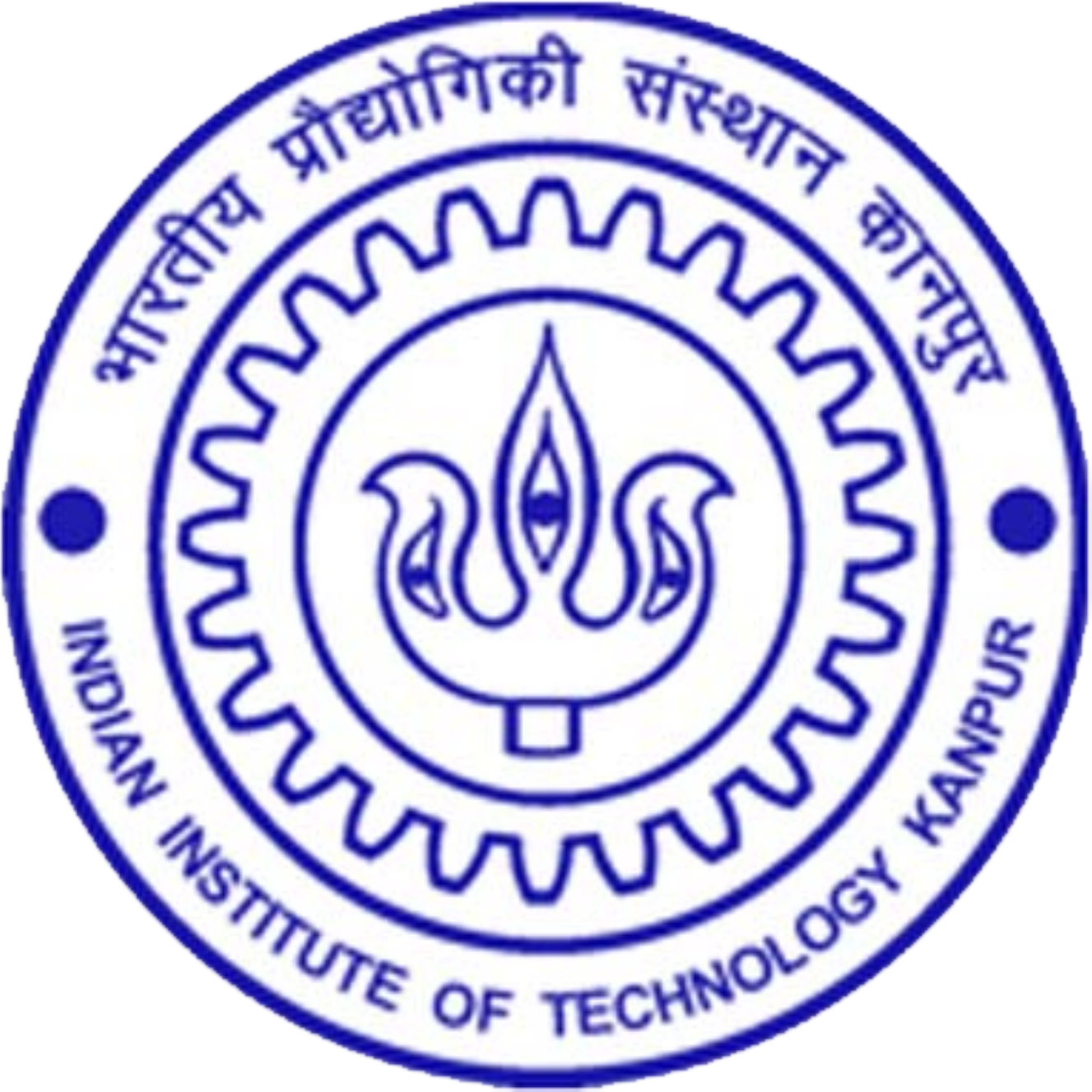}}
\end{center}
\end{figure}
\vspace{2cm}
\large{\textbf{DEPARTMENT OF ELECTRICAL ENGINEERING \\
        INDIAN INSTITUTE OF TECHNOLOGY KANPUR \\
                      FEBRUARY 2015}}
\end{center}
\end{titlepage}

%
%
%
%
%

\thispagestyle{empty}
\begin{center}
\vspace*{0.2cm}
\LARGE
\textbf{Introduction to the Kalman Filter and Tuning its Statistics for Near Optimal Estimates and Cramer Rao Bound}

\vspace{1cm}
\large{by}\\[12pt]
\normalsize{\textbf{Shyam Mohan M$^1$, Naren Naik$^2$, R.M.O. Gemson$^3$, M.R. Ananthasayanam$^4$}}\\[14pt]
{\normalsize $^1$Formerly Post Graduate Student, IIT, Kanpur, India}\\
{\normalsize $^2$Professor, Department of Electrical Engineering, IIT, Kanpur, India}\\
{\normalsize $^3$Formerly Additional General Manager, HAL, Bangalore, India}\\
{\normalsize $^4$Formerly Professor, Department of Aerospace Engineering, IISc, Banglore}\\

\vspace{1cm}
\large\textbf{Technical Report : TR/EE2015/401} \\
\vspace{2cm}
\begin{figure}[!h]
\begin{center}
\scalebox{0.35}{\includegraphics{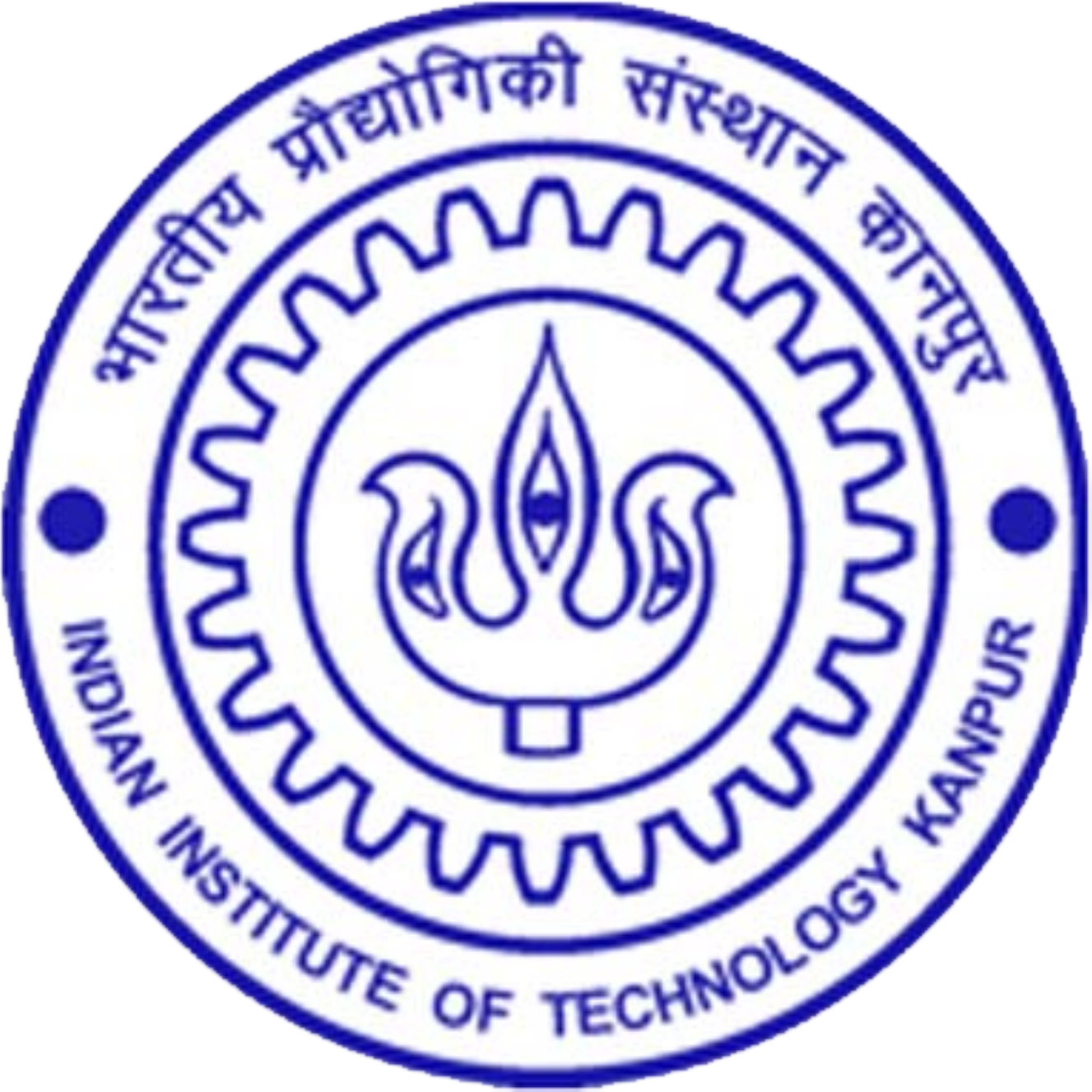}}
\end{center}
\end{figure}
\vspace{1.5cm}
\large{\textbf{DEPARTMENT OF ELECTRICAL ENGINEERING \\
INDIAN INSTITUTE OF TECHNOLOGY KANPUR \\
FEBRUARY 2015}}
\end{center}

\newpage
\pagenumbering{roman}
\text{}
\vspace{1cm}
\begin{center}
\textbf{\underline{ABSTRACT}}
\end{center}

\begin{spacing}{2}

This report provides a brief historical evolution of the concepts in the Kalman filtering theory since ancient times to the present. A brief description of the filter equations its aesthetics, beauty, truth, fascinating perspectives and competence are described. For a Kalman filter design to provide optimal estimates tuning of its statistics namely initial state and covariance, unknown parameters, and state and measurement noise covariances is important. The earlier tuning approaches are reviewed. The present approach is a reference recursive recipe based on multiple filter passes through the data without any optimization to reach a `statistical equilibrium' solution. It utilizes the a priori, a posteriori, and smoothed states, their corresponding predicted measurements and the actual measurements help to balance the measurement equation and similarly the state equation to help form a generalized likelihood cost function. The filter covariance at the end of each pass is heuristically scaled up by the number of data points is further trimmed to statistically match the exact estimates and Cramer Rao Bounds (CRBs) available with no process noise provided the initial covariance for subsequent passes. During simulation studies with process noise the matching of the input and estimated noise sequence over time and in real data the generalized cost functions helped to obtain confidence in the results. Simulation studies of a constant signal, a ramp, a spring, mass, damper system with a weak non linear spring constant, longitudinal and lateral motion of an airplane was followed by similar but more involved real airplane data was carried out in MATLAB\textsuperscript{\textregistered}. In all cases the present approach was shown to provide internally consistent and best possible estimates and their CRBs.
\end{spacing}

\begin{spacing}{1.8}

\newpage
\begin{center}
\textbf{\underline{ACKNOWLEDGEMENTS}}
\vspace{0.25cm}
\end{center}

It is a pleasure to thank many people with whom the authors interacted over a period of time in the topic of Kalman filtering and its Applications. Decades earlier this topic was started as a course in the Department of Aerospace Engineering and a Workshop was conducted along with Prof. S. M. Deshpande who started it and then moved over full time to Computational Fluid Dynamics. Subsequently MRA taught the course and spent many years carrying out research and development in this area during his tenure at the IISc, Bangalore. The problem of filter tuning has always been intriguing for MRA since most people in the area tweaked rather than tuned most of the time with the result that there is no one procedure that could be used routinely in applying the Kalman filter in its innumerable applications. The PhD thesis of RMOG has been the only effort for a proper tuning of the filter parameters but this was not too well known. In the recent past for a couple of years the interaction among the present authors helped to reach the present method which we believe is quite good for such routine applications. The report has been written in such a way to be useful as a teaching material. Our grateful thanks are due to Profs. R. M. Vasu (Dept. of Instrumentation and Applied Physics), D. Roy (Dept. of Civil Engineering), and M. R. Muralidharan (Supercomputer Education and Research Centre) for help in a number of observable and unobservable ways without which this work would just not have been possible at all and also for providing computational facilities at the IISc, Bangalore. It should be possible to further improve the present approach based on applications in a wide variety of problems in science and technology.
\begin{flushright}
\textbf{- AUTHORS.}
\end{flushright}
\noindent\line(1,0){470}\\
Shyam Mohan M (shyammoh.2014@iitkalumni.org)\\
Naren Naik (nnaik@iitk.ac.in) \\
R.M.O. Gemson (mogratnam@rediffmail.com)\\
M.R. Ananthasayanam (sayanam2005@yahoo.co.in)
\end{spacing}

\newpage
\tableofcontents

\newpage
\begin{table}[h]
\begin{center}
{\Large \underline{List of Symbols$^*$}}\\[14pt]
\begin{tabular}{ l  l  }
$x$ & State vector of size $n\times 1$\\[12pt]
$\Theta$ & Parameter vector of size $p\times 1$\\[12pt]
$X=[x,\Theta]^T$ & Augmented state vector of size $(n+p)\times 1$\\[12pt]
$Z_k$ & Measurement Vector of size $m\times 1$ at discrete time index `k'\\[12pt]
$\mathbf{X_0,P_{0}}$ & Initial state and its covariance\\[12pt]
\textbf{R, Q} & Measurement and Process noise covariance matrix\\[12pt]
\textbf{J0-J8} & Cost functions\\[12pt]
$X_{k|k-1}$ & Prior state estimate at time index k based on data upto k-1\\[12pt]
$X_{k|k}$ & Posterior state estimate at time index k based on data upto k \\[12pt]
$X_{k|N}$ & Smoothed state estimate at time index k based on data upto N\\[12pt]
$X{d_{k|N}}$ & Dynamical state estimate at time index k based on data upto N\\[12pt]
$P_{k|k-1}$ & Prior state covariance matrix at time index k given data upto k-1\\[12pt]
$P_{k|k}$ & Posterior state covariance matrix at time index k given data upto k\\[12pt]
$P_{k|N}$ & Smoothed state covariance matrix at time index k given data upto N\\[12pt]
$K_k$ & Kalman Gain based on data upto time index k\\[12pt]
$K_{k|N}$ & Smoothed Gain based on all data upto time index N\\[12pt]
$F_{k-1}=\left[\frac{\partial{f}}{\partial{X}}\right]_{X=X_{k-1|k-1}}$ & State Jacobian matrix using posterior state estimate\\[12pt]
$F_{k-1|N}=\left[\frac{\partial{f}}{\partial{X}}\right]_{X=X_{k-1|N}}$ & State Jacobian matrix using smoothed state estimate\\[12pt]
$F{d_{k-1|N}}=\left[\frac{\partial{f}}{\partial{X}}\right]_{X=X{d_{k-1|N}}}$ & State Jacobian matrix using dynamical state estimate\\[12pt]
$H_{k}= \left[\frac{\partial{h}}{\partial{X}}\right]_{X=X_{k|k-1}}$ & Measurement Jacobian matrix using prior state estimate\\[12pt]
$H_{k|k}= \left[\frac{\partial{h}}{\partial{X}}\right]_{X=X_{k|k}}$ & Measurement Jacobian matrix using posterior state estimate\\[12pt]
$H_{k|N}= \left[\frac{\partial{h}}{\partial{X}}\right]_{X=X_{k|N}}$ & Measurement Jacobian matrix using smoothed state estimate\\\\[14pt]
\end{tabular}
\begin{flushleft}
\textit{*Most other symbols are explained as and when they occur in the report. }
\end{flushleft}
\end{center}
\end{table}

\chapter{Historical Development of Concepts in Kalman Filter}
\pagenumbering{arabic}
\setcounter{page}{1}
\label{ch1}

\section{Randomness from Ancient Times to the Present}
It is useful to understand the concept of randomness before introducing the Kalman filter. Randomness occurs and is inevitable in all walks of life. The ontology (the true nature) is one thing and the epistemology (our understanding) is another thing. A computer generated sequence of random numbers is deterministic ontology but for the user who does not know how they are generated it is probabilistic epistemology. Randomness is patternless but not propertyless. Randomness could be our ignorance. Quantum Mechanics seems to possess true randomness. One feels that randomness is a nuisance, and should be avoided. However we have to live with it and compulsively need it in many situations. In a multiple choice question paper no examiner would dare to put all the correct answers at the same place! As another example the density, pressure, and temperature, or even many trace constituents in air can be measured with a confidence only due to the random mixing that invariably takes place over a suitable space and time scale. It is worthwhile to state the introduction of random process noise into the kinematic or dynamical equations of motion of aircraft, missiles, launch vehicles, and satellite system helps to inhibit the onset of Kalman filter instability and thus track these vehicles. Chance or randomness is not something to be worried about presently the most logical way to express our knowledge.

\begin{figure}[h]
\begin{center}
\includegraphics[width=5.0in,height=3.0in]{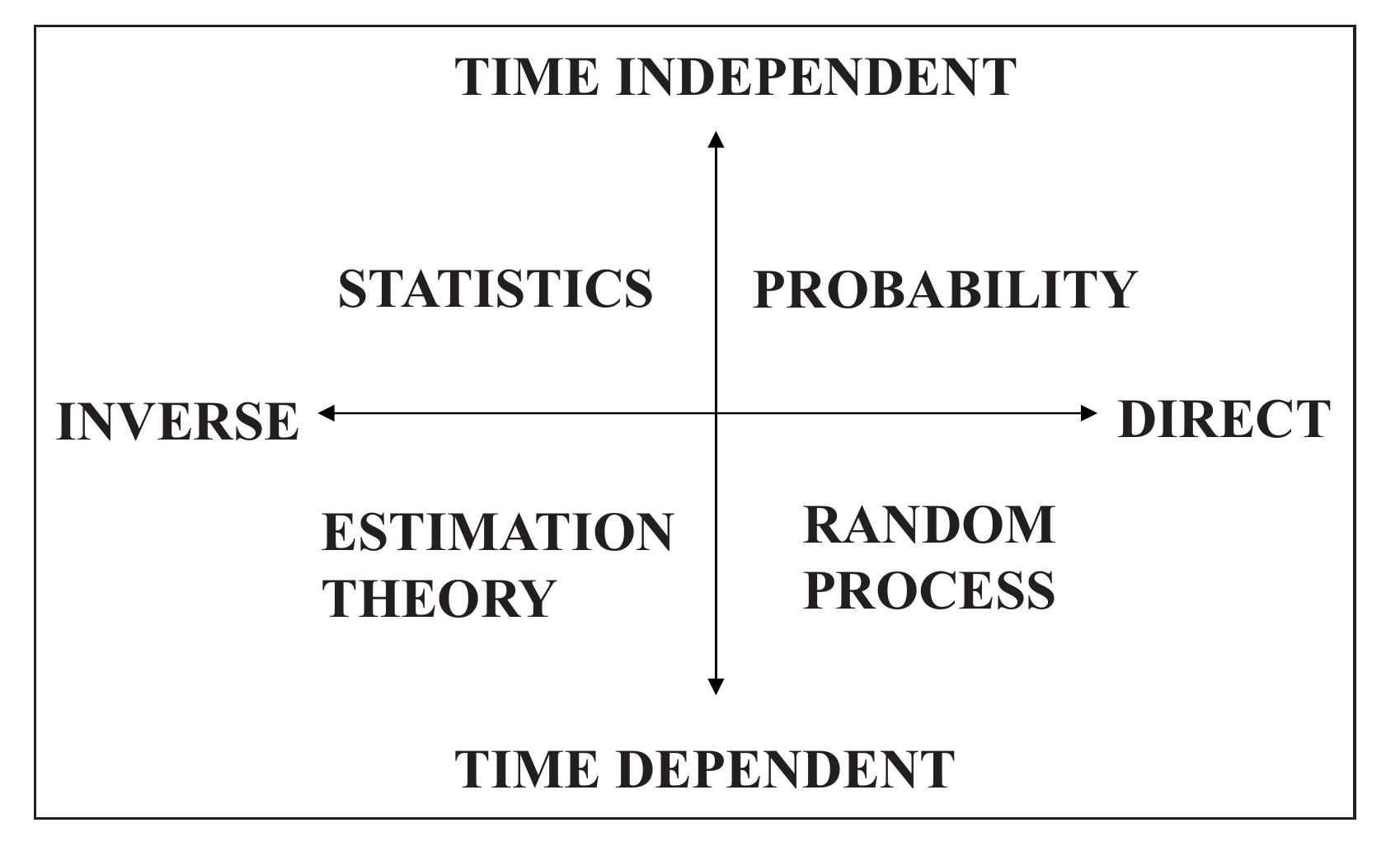}
\caption{Relationship among Probability, Statistics, Random process,}
\caption*{and Estimation Theory. Randomness is common to all.}
\label{rand}
\end{center}
\end{figure}

The well known statistician Rao \cite{Rao1987} (1987) states that statistics as a method of learning from experience and decision making under uncertainty must have been practiced from the beginning
of mankind. The inductive reasoning in these processes have never been codified due to the uncertain nature of the conclusions drawn from any data. The breakthrough occurred only at the beginning of the twentieth century with the realization that inductive reasoning can be made precise by specifying additionally just the amount of uncertainty in the conclusions. This helped to work out an optimum course of action involving minimum risk in uncertain situations in the deductive process. This is what Rao \cite{Rao1987} (1987) has stated as a new paradigm of thinking to handle uncertain situations,
\begin{center}
\textbf{Uncertain knowledge + Knowledge of the amount of uncertainty in it\\
=  Usable knowledge}
\end{center}
For our purpose randomness is common to probability, statistics, random process, and estimation theory as can be seen from Fig. \ref{rand}. The Kalman filter can be stated as a sequential statistical analysis of time dependent data. The way it does is to account for the change, capture the new knowledge or measurement, and assimilate it for change as explained in the next section.

\section{Conceptual Development of the Kalman Filter}
The triplets of change, capture and correct forms the conceptual basis for the Kalman filter and it can be stated as follows :
\begin{center}
\textbf{Change + Capture $\longrightarrow$ Correct}
\end{center}
The `+' sign above has a deep significance in the way the present and the new information are combined to have progress in the correct direction based on an appropriate criterion. The above can be stated among other possible ways in science and technology such as
\begin{align*}
\textbf{Present + New } &\longrightarrow \textbf{ Updated information or knowledge} \\
\textbf{Theory + Experiment } &\longrightarrow \textbf{ Progress} \\
\textbf{Model + Data } &\longrightarrow \textbf{ Estimate the unknowns} \\
\textbf{Intuition + Experiments } &\longrightarrow \textbf{ Rigorous Mathematical Framework}
\end{align*}
The origin of such an ubiquitous Kalman filter (KF) in Estimation Theory (ET) can be traced since ancient times. Its progress is similar to any physical theory moving randomly across intuition, experiments and mathematical framework.

\section{Estimation Theory in Ancient Indian Astronomy}
One of the simplest examples of estimation is to observe one quantity and infer another quantity connected to it by a certain relation which may be empirical or theoretical. Thus obtaining the `unobservables' from the `observables' is a very subtle concept that can be stated as `observe one and infer another'. For example by making kinematic measurements such as the position, velocity and accelerations on a system one can infer the internal dynamical properties of the system such its mass, moment of inertia, aerodynamic characteristics by suitable modeling.

The ancient Indian astronomers at least since AD 500 used the above concept to update the parameters for predicting the position of celestial objects for timing their Vedic rituals based on measurements carried out at various time intervals which can be stated as
\begin{center}
\textbf{Updated parameter =  Earlier parameter +}\\
\textbf{(Some quantity) $\times$ ( measured - predicted) Position of the celestial object}
\end{center}
The `some quantity' as we will see later on is the Kalman gain. Note the measured longitude of the celestial object is different from the state that is updated which is the number of revolutions in a yuga just as state and measurements are different in many Kalman filter applications!

The earliest astronomical manual Suryasiddhantha (Burgess \cite{Burgess1935} 1935) is dated before AD 500. When deviations were noted from its predictions the ancient Indian astronomers revised the parameters of the planetary positions based on observations at later times. Aryabhata made independent observations and gave corrections to it that provided accurate results around his time of AD 522. The constants or cannons of Aryabhata was revised by a group of astronomers in AD 683-684 known as Parahita system. After a long time lapse large deviations occurred from Parahita system and Parameswara of the Kerala School of astronomy revised it based on astronomical observations as promulgated in his work Drgganita in AD 1431.

The next revision was by Nilakantha as stated in his work Tantrasangraha (Ramasubramanian and Sriram \cite{Ramasubramanian2011} 2011) in AD 1443. Nilakantha had stated ``the eclipses cited in Siddhanthas can be computed and the details verified. Similarly other known eclipses as well as those currently observable are to be studied. In the light of such experience future ones can be computed and predicted (\textbf{extrapolation!}). Or eclipses occurring at other places can be studied taking  into account the latitude and longitude of the places and on this basis the methods providing the parameters for the Sun, Moon can be perfected (\textbf{data fusion!}). Based on these, the past and future eclipses of one's own place can be studied and verified with appropriate refinement of the technique." This is just the idea of \textbf{`smoothing'!} When Tantrasamgraha too was becoming inaccurate, observations were carried out by astronomers on the west coast for thirty years from AD 1577 to AD 1607 and a revision was promulgated in AD 1607 in Drkkarana which quotes about all the above earlier revisions (good literature survey!). It stated that henceforth also deviations would occur and these should be carefully observed.

Not all the above canons were based on the latest observations. The well known historian Billard \cite{Billard1971} (1971) noted that many canons were evolved and one such canon around AD 898 had a very high accuracy valid over a larger number of centuries. He inferred that these must have been based on \textbf{the astronomical elements of an earlier time and the new observations of the later time.} A more accurate canon valid over a larger time period is not possible without a weighted addition of the earlier canons (which are the ones that are available since no back dated observations are possible!) and present observations. They must have chosen such a weight or in other words the gain `K' subjectively to help in giving weights to earlier astronomical parameters and the present measurements. The present day Kalman filter is packed with many desirable and fashionable subjective assumptions regarding the state, measurement and noise characteristics to facilitate mathematical tractability (and thus obtain the Kalman gain) and the capability to handle far more involved applications. The Table-\ref{tbcape3} provides some of the well known updates of the planetary parameters by ancient Indian astronomers.

It is highly worthwhile to translate the works of Kerala astronomers and Billard, use the best available programs to calculate the earlier positions of the celestial objects to understand how the updates were done.

\section{Estimation Theory during the Medieval Period}

The asteroid Ceres sighted by Piazzi on the first day of 1801 was tracked subsequently for 41 days after which it vanished behind the Sun's rays. Piazzi's data covering only 9 degrees of arc of the celestial sphere consisted of the right ascension, declination and time was published in September of that year. Its orbital elements could not be determined by the then available methods. Newton had stated it as the most difficult nonlinear problem then in astronomy. Gauss made calculations using the observational data, estimated its orbit and sent it to Piazzi who found it again on the last day of 1801 (Tennenbaum and Director \cite{Tennenbaum2006} 2006) .

This was possible based on the Least Squares (LS) technique used by Gauss (Dover\cite{Dover1963} 1963, reprint of Gauss 1809) to make calculations regarding the trajectory and predict its location later in time. Independently the method of LS had been discovered and published by Legendre of France and Robert Adrian of the United States. In fact, even before Gauss was born, the physicist Lambert had used it! (Grewal and Andrews \cite{Grewal2008}  2008).

Gauss had provided almost all the foundations of estimation theory. He postulated that a system model should be available, minimum number of measurements for observability, the redundant data helping to reduce the influence of measurement errors, a cost function based on the difference between the measurement and that predicted by the model should be minimized. There should also be some a priori knowledge concerning the unknowns to be estimated. Further since the errors could be unknown or unknowable Gauss had given hints about probabilistic approach, normal distribution, method of maximum likelihood estimation, linearization and the Gaussian elimination procedure.

Gauss did not balance the governing differential equation, but tried to fit the measurement with the prediction. If he had tried the former, he would have been led to a biased solution! Fortune favours the brave! This is where a proper mathematical framework helps to understand if an algorithm converges to the correct value with more and more measurements. Begin with intuition, try out with some examples and finally cast in a mathematical framework. Mathematics itself grew like this, but present day students are taught the other way round!

\section{Estimation Theory (ET) During the Twentieth Century}
A very general characterization of ET when time independent data analysis is made it is statistics and when  sequential processing of the time dependent data is carried out it is the Kalman filter as mentioned earlier. The post Gaussian contributions in estimation theory consists of the method of moments, method of maximum likelihood estimation, the Kalman filter and its variants, frequency domain approach and the capability to handle time varying state and parameters. As far as experiments, the testing times are short and accurate, with optimal inputs to excite the system to obtain the best possible parameter estimates have been developed. Further the use of matrix theory, and real time processing by computers exist. We are dealing with more difficult situations, but the conceptual framework to solve these problems had been fully laid out by Gauss!

The Kalman filter equations are recursive and it is helpful to estimate the time varying state variables and  parameters. The sequential least squares was rediscovered by Plackett \cite{Plackett1950} (1950) and Kalman \cite{Kalman1960} (1960). Thus in fact Sprott \cite{Sprott1978} (1978) had questioned if the Kalman filter is really a significant contribution! The point is the frequency domain approach of the Wiener Filter \cite{Wiener1949} (1949) has been enhanced to the natural time domain approach by the Kalman filter. Further the shift from batch to the sequential approach is very convenient to handle continuous measurement data flow. It is to the credit of Kalman that apart from unifying earlier results that he introduced the concept of controllability and observability to make the estimation problem well posed. The only slight difference, but very momentous between the Recursive Least Squares (RLS) and the Kalman filter is the time propagation of the state and covariance estimates before their update by using the measurements.

During the mid-twentieth century the Kalman filter in discrete time (Kalman \cite{Kalman1960}, 1960), and in continuous time (Kalman and Bucy \cite{Kalman1961}, 1961) helped the Apollo project to the Moon. It is interesting to note that during the periods of its development whether in mythological, medieval, or modern times its connection with celestial objects is remarkable.

\section{Present day Applications of Estimation Theory}

Presently the scale and magnitude of many difficult and interesting problems that ET is handling could not have been comprehended by ancient Indian astronomers, Gauss or Kalman. In many present day applications one does not even know quite well the structure of the state and measurement equations as well as the parameters in them and the statistical characteristics of the state and measurement noises. It is possible to add the unknown initial conditions of the state as well. One can summarize that almost nothing is known but everything has to be determined or estimated from the measurements alone! This means the connecting relationship between the state and the measurement has to be found out. This has to be from previous knowledge or intuition such that it is meaningful, reasonable, acceptable, or useful. Even this has to be achieved only by the internal consistency of the various quantities and/or the variables that occur at different times during the filter operation through the data. Further the systems are large and the best possible optimal estimation has to be worked out in real time. However due to the enormous computing power that is presently available it has been possible to handle the above situations.

Typical present day applications of the Kalman filter include target tracking (Bar-Shalom et al. \cite{Shalom2001} 2001), evolution of the space debris scenario (Ananthasayanam et al. \cite{MRA2006} 2006), fusion of GPS and INS data (Grewal et al. \cite{Grewal2007} 2007), study of the tectonic plate movements (Kleusberg and Teunissen \cite{Kleusberg1996} 1996), high energy physics (Fruhwirth et al. \cite{Fruhwirth2000} 2000), agriculture, biology and medicine (Federer and Murthy \cite{Federer1998} 1998), dendroclimatology (Visser and Molenaar \cite{Visser1988} 1988), finance (Wells \cite{Wells1996} 1996), source separation problem in telecommunications, biomedicine, audio, speech, and in particular astrophysics (Costagli and Kuruoglu \cite{Costagli2007} 2007), and atmospheric data assimilation for weather prediction (Evensen \cite{Evensen2009} 2009).

\section{Three Main Types of Problems in Estimation Theory}

If we denote the system model as (S), measurement model as (M), state noise as (\textbf{Q}) and measurement noise as (\textbf{R}) three main types of problems emerge in estimation theory (Klein \cite{Klein1979} 1979). Generally the state equations are differential equations and the measurement equations related to the states are algebraic in nature. Both the above could have unknown or inaccurately known parameters and noise characteristics which will have to be estimated.

The first dealing with balancing the state differential equations without specifying the characteristics of the process noise is known as the Equation Error Method (EEM). The simplest EEM formulation is obtained by assuming that noise free state variables and noisy state derivatives measurements are available (Maine and Iliff  \cite{Maine1981} 1981) to balance the governing state differential equations. If the measurements of the states are noisy then EEM leads to biased parameter estimates. However if the noise in the state measurements are very small then good parameter estimates are still possible. The EEM is amenable for solution by simple least squares technique.

The second combination of state, measurement and measurement noise forms the Output Error method (OEM) thus with no process noise (Klein \cite{Klein1979} 1979). The OEM formulation matches the predicted measurement based on the states obtained from the state differential equations with the actual measurements. The matching based on the squared difference between the actual measurement and the predicted measurement summed over all the time points leads to a cost function \textbf{J} that can be minimized by any suitable batch processing optimization algorithm. The minimization of \textbf{J} leads to the estimates of the unknown parameters in both the states and measurements. It may be mentioned that the results from the above approach is used as an anchor in the present work for tuning the Kalman filter statistics when there is no process noise.

The most general third formulation is when the state and the measurement equations with both the state and measurement noises are present. Before an experiment is carried out we can talk of probabilistic outcomes. Once the data is available the unknown parameters and quantities can be treated as deterministic unknowns. It is useful to read the discussion in Rao \cite{Rao1987} (1987) wherein he describes a random trace of a signal as the sum of many deterministic functions.  Thus once again a matching of the measurements with those derivable by propagating the state differential equations with estimated parameters has to be carried out. Since the measurements are random they could correspond to random states. The propagation of the state being also random there is a difference between the predicted and updated states and this difference can be used to derive a cost function from the state differential equations which also can be minimized. Since the contributions from the various states and the measurements have different units it is desirable to normalize each contribution using their mean square values. Thus one can treat estimation operationally as a deterministic optimization problem for a given measurement data. One should note that even from deterministic approaches, based on the second derivatives of the cost function (based on the Hessian, the first derivative of \textbf{J} being zero at the minimum) a measure of the uncertainty for all the estimated unknown quantities can be obtained. In all the above formulations if a certain suitable probability distribution is not assumed for the \textbf{Q} and \textbf{R} then it is deterministic. Another simple example is a constant signal added with noise. It is trivial to get an estimate of the mean and standard deviation denoting the effect of dispersion. If only the noise is assumed to be Gaussian then quantitative values can be attached to the standard deviation of the estimate.

However if the state and measurement noises are characterized as random variables (in say the simplest form being zero mean, white and Gaussian) then all the above three types of estimation problems can be cast in a probabilistic framework and one of the most prolific is the Method of Maximum Likelihood Estimation (MMLE) helps to attach quantitative values for the uncertainties of the estimates. Due to the existence of the state noise the state propagation is not a deterministic trajectory but becomes a random process. This predicted state has to be statistically combined with the noisy measurements at every point based on a suitable criterion and updated at every measurement and then again propagated. This is the Kalman filter formalism as will be elaborated in the next Chapter.

When the parameters in the state and measurement equations are treated as augmented state it is called as the Extended Kalman Filter (EKF). Either the MMLE or EKF both of which are equivalent can be utilized for the estimation of unknown parameters and noise characteristics. For the sake of mathematical tractability and simplicity the various random variables in the Kalman filter are assumed to follow a Gaussian distribution. Thus the initial states are assumed to follow a multidimensional Gaussian distribution. If the governing state equations are linear then, only the parameters in the Gaussian distribution change. Then if the distribution of the measurement noise is Gaussian then after statistically combining the above once again leads to a Gaussian distribution for the updated state which is again propagated in time. The process is repeated till all the measurement data are utilised.

However when the state differential equations are nonlinear then between the measurements the time propagated Gaussian distribution becomes non Gaussian. It is then a suitable linearization of the state and measurement (if nonlinear) are carried out to keep the formalism similar to the linear case. This is known as the Extended Kalman Filter (EKF) formulation to handle nonlinear systems and measurements. In this report we consider the problem of estimating the unknown parameters and also \textbf{Q} and \textbf{R} if these are also unknown. This is also called the problem of tuning the Kalman filter statistics.

Generally, in the MMLE for dynamical systems, one deals with mainly where the measurement noise alone is present. However in many present day applications, modeling errors and random state noise input conditions occur. Hence it becomes compelling to deal with MMLE including the process noise as well. In such a situation one needs to estimate or account for the process noise while estimating the unknown parameters. This makes the problem difficult by an order of magnitude due to the requirement of more computer memory and time as well as the convergence difficulties of the various algorithms. Thus it is generally far more difficult to handle the estimation problem in the presence of unknown or uncertain process and measurement noises.

The MMLE is widely used, since it yields realistic results for practical problems and its estimates have many desirable statistical properties such as consistency, efficiency and sufficiency. The numerical effort in the EEM, OEM and the most general MMLE or EKF options are of order 1, 10 and 100 respectively.

\section{Summary of different aspects of Estimation Theory}
The basic framework of Estimation Theory (Eykhoff \cite{Eykhoff1981} 1981) consists of
\begin{enumerate}
\item Modeling the system, measurement and all the noise characteristics.
\item A criterion to match or mix the model output with the measurements.
\item A numerical algorithm for the above task and consequently obtain the estimates
and the uncertainty of the estimated quantities and
\item An internal consistency check to ensure that all the above steps are consistent and if not shows the need for modification.
\end{enumerate}

The first aspect requires some `a priori' knowledge about the system under investigation. The model can be true or best or adequate and generally one aims for the last one. The model in the form of mathematical equations is expected to characterize the essential or desired aspects of the state process.

The second needs the matching of model output with the measurements and it can be based either on deterministic or probabilistic criterion. In a comprehensive way the model is made to follow the output measurements under the given input conditions. In order to match in a quantitative sense between the postulated model and the measurements, one has to choose a criterion function or a cost function \textbf{J}. \par

The third step is the selection of a numerical algorithm to satisfy the above criterion. Generally the matching (or combining) of the model output with the measurements have to be carried out by a suitable optimization technique to minimize the above cost function.

The last one is the process of model validation. The task of model validation is crucial to understand the adequacy of the postulated state and measurement models and the convergence of the numerical algorithm. Statistical hypothesis testing procedure is the main tool to check the consistency of all the aspects of the estimation process.

The above four aspects help to evolve a suitable mathematical model of the system by properly estimating the unknown parameters and other noise characteristics based on the input and measurement data. Subsequently such a model can be utilized in further studies like prediction, control and system optimization. The present work concentrates in particular on the Step 3 of the ET framework.

\newpage
\begin{table}[h]
\caption{Corrections of Planetary Parameters in Ancient India.}
\caption*{Credit : Ananthasayanam and Bharadwaj \cite{MRA2011} 2011}
\label{tbcape3}
\begin{tiny}
\begin{tabular}{c }
\end{tabular}
\end{tiny}
\end{table}

\includegraphics[width=6in,height=8in]{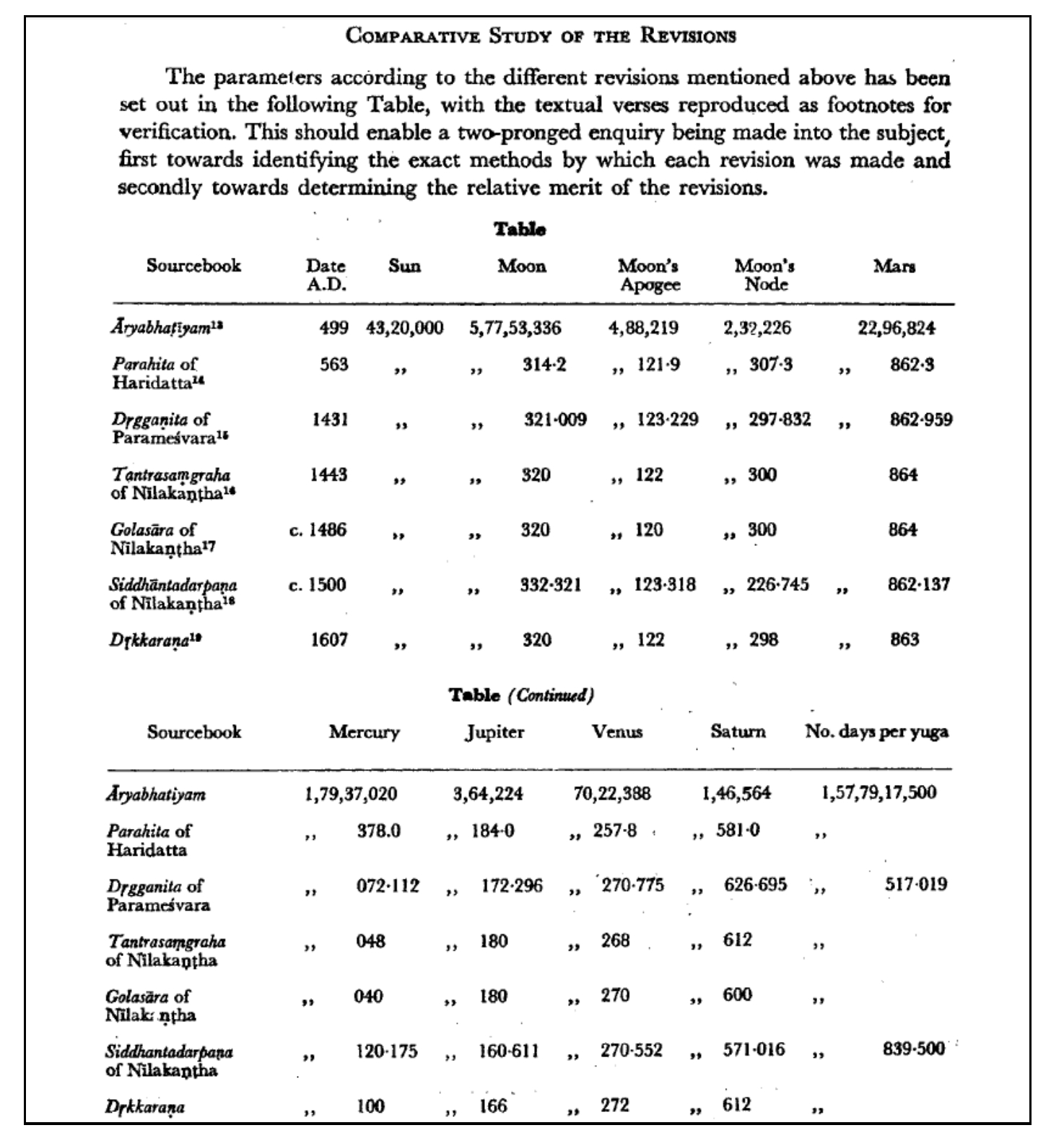}

\chapter{Introduction to the Kalman Filter}
\label{ch2}

\section{Important Features of the Kalman Filter}

\par Due to the seemingly unpretentious fact of splitting the state and measurement equations and switching between the state propagation and its update using the measurements, very interesting outcomes have been shown to be possible. Any amount of deep study and understanding of the state or the measurement equations separately may not be able to comprehend the exciting possibilities and abilities when both are combined together. This is similar to the components of a watch, or the cells in an organism leading respectively to the time keeping ability or life, which do not exist in the individual components. The GPS is another brilliant example of such a synergism. The competence of the Kalman filter is similar to the saying `wholes are more than the sum of their parts' as stated by Minsky \cite{Minsky1988} (1988). It is the above feature that can be called as synergistic, parallel, operator splitting, or a combination of theory and experiment that is the remarkable and profound aspect of the Kalman filter rather than describing it as a sequential least square estimator, or capable of handling time varying states and measurements.

\par As mentioned in the beginning the triplets of change, capture and correct form the Kalman filter. The filter can be viewed or understood from different perspectives. The modeling of the state of a system is subjective (or in other words intuitive) and the system measurements are objective. Generally the knowledge being uncertain (or inaccurately known) and the measurements are inaccurate or corrupted by noise the Kalman filter combines the two to expand the knowledge front. Another way to look at the Kalman filter is that it combines or assimilates the information from two sources namely uncertain system and measurement models in a statistically consistent way. One other way of understanding the Kalman filter is that it matches the model and the measurement and in the process improves both by suppressing the noise in the measurement improves the accuracy of the state and the parameters in it. There could be many ways or criteria of combining the model and the measurements. Each one could give different results but the criterion to accept any result is that the estimates should be meaningful, reasonable, acceptable and useable. Thus one should note that one cannot be at the truth but around the truth. The only way to reach the truth or in other words get to know the absolute source from which the data has come about is to have infinite data together with an algorithm being capable of reaching the truth.

\section{Importance of Proper Data Processing}
\par In this section we discuss the importance of proper processing of the data with a simple example. Consider the estimation of the parameter `a' in the equation
\begin{align}
y = ax
\end{align}
A set of N noisy measurements of x, y are given by
\begin{align*}
x_m = x + w \\
y_m = y + v
\end{align*}
The mean of the random measurement noises are assumed to be zero. In order to estimate `a' one can use different formulae all of which look reasonable such as
\begin{footnotesize}
\begin{align*}
& a   =    \frac{1}{N} \sum (\frac{y_m}{x_m}) &  a = \frac{1}{N}\sum( \frac{y_m}{\sum x_m})\\
& \frac{1}{a} =   \frac{1}{N} \sum (\frac{x_m}{y_m})    & \frac{1}{a}=\frac{1}{N}\sum( \frac{x_m}{\sum y_m})
\end{align*}
\end{footnotesize}
Also the Least square (LS) estimate is $a=\sum(x_my_m)/\sum (x_mx_m)$. Substituting for $x_m$ and $y_m$ into the above and simplifying with further  assumption w $\ll$ x and  v $\ll$ y and expanding in a Taylor's series and truncating one may note that even if N tends to infinity only some of the above provide unbiased estimates assuming that the mean value of `w' and `v' are zero.  Even the LS estimate which balance the basic equation provides unbiased estimate only if the mean value of `w' and `v' equals zero. The above feature shows that any arbitrary way of estimating the parameter in a problem may not in general lead to proper estimates and we need more mathematically sound approaches.

Even in the least squares fit there are many variants depending on whether the departures of the data from the fitted line is measured (i) vertically (the standard exercise in most text books) (ii) horizontally, (iii) bisector of the above two lines, (iv) perpendicularly, and (v) measured both perpendicularly and vertically are possible. However these are dictated by the underlying scientific mechanisms in the various fields of application. A good discussion of the above is available in Isobe et al. \cite{Isobe1990} (1990) and Feigelson and Babu \cite{Feigelson1992} (1992). The above is mentioned to stress the importance of a good physical understanding of the problem which arises from earlier studies or by intuition in newer problems.

In fact such an estimate of obtaining unknown parameters is provided by the Method of Maximum Likelihood Estimation (MMLE) in the statistics pioneered by Gauss (Dover\cite{Dover1963} 1963, reprint of Gauss 1809) and reinvented by Fisher \cite{Fisher1922} (1922) is extensively used in present day estimation theory. If you wish to contribute to estimation theory then read more statistics! Thus one can note that even if an algorithm converges it does not guarantee the result to be correct or even reasonable. This defect can be overcome by considering simulated studies where one knows the true values of the system parameters and the noise statistics. Since the Kalman filter is supposed to provide an estimate together with appropriate uncertainty it is good to generate the same by another procedure that is not filter based at all but does however provide the same. In this report when there is only measurement noise but no process noise we are able to generate the results by using the Newton Raphson minimization technique which helps us to understand the filter operation and its results by comparison with the former. When both the measurement and process noises are present then we are in an unknown territory and have to depend on other ways of having confidence in the filter results. In the present work this has been possible when one introduces the generalized cost function that depends on not only the `innovation' but on the other quantities like the `filtered residue' and the `smoothed residue' as well as those depending on the balance of the governing state equations. One can add another cost function based on the initial state conditions as well.

%
%

\section{Problem Formulation and Extended Kalman Filter Equations}

\par One source of information is the state differential equations and the other source of information that captures the above change are the measurements made on the system. The correction to the state is provided by the measurements based on a proper criterion leading to reduced uncertainty of the state variables. Such a criterion is provided by a probabilistic weighted linear addition of the predicted state and the actual measurement data. The state and measurement variables are related through appropriate functional relationship. Such an update corrective process is repeated at suitable intervals. In fact everything about the state and the measurement equations are to be learnt and estimated from the measurements alone. The above process is described below leads to an optimization problem to be solved which is equivalent to tuning the Kalman filter statistics which are often unknown.

Consider the following nonlinear filtering problem defined for discrete time instants given by k = 1, 2,\ldots N
\begin{align*}
x_k&=f(x_{k-1},\Theta,u_{k-1})+w_k\\
Z_k&=h(x_{k},\Theta)+v_k
\end{align*}
where `x' is the state vector of size $n\times 1$, `u' is the control input and `Z' is the measurement vector of size $m\times 1$. The `f' and `h' are non linear functions of state and measurement equations respectively. The injected process noise, $w_k\sim\mathcal{N}$ ( 0, \textbf{Q}) and the injected measurement noise, $v_k\sim\mathcal{N}$ ( 0, \textbf{R} ) are assumed to be zero mean Additive White Gaussian Noise (AWGN), and are identically and independently distributed (iid). The `$\mathcal{N}$' represents Normal or Gaussian distribution and
\begin{align*}
E\left[w_kw_j^T\right]&=\textbf{Q}\delta(k-j) \text{ \& } E\left[w_k\right]=0   \\
E\left[v_kv_j^T\right]&=\textbf{R}\delta(k-j) \text{ \& } E\left[v_k\right]=0  \\
E\left[w_kv_j^T\right]&=0 \text{ $\forall$ j, k = 1, 2,\ldots N }
\end{align*}
where N is the total number of sampling instants. E$\left[\text{ }\right]$ is the expectation operator, $\delta$ is Kronecker delta function defined as\\
\[ \delta(k-j) = \left\{ \begin{array}{ll}
        0 & \mbox{if $k \neq j$};\\
        1 & \mbox{if $k = j$}.\end{array} \right. \]
The parameter vector `$\Theta$' of size $p \times 1$ is augmented as additional states,
\begin{align*}
\begin{bmatrix}x_k \\ \Theta_k\end{bmatrix}=&
\begin{bmatrix}
f(x_{k-1},\Theta_{k-1},u_{k-1})\\
\Theta_{k-1}
\end{bmatrix}+
\begin{bmatrix}
w_k\\0
\end{bmatrix}
\end{align*}
The non linear filtering problem is now defined as
\begin{align}
\label{e14} X_k&=f(X_{k-1})+w_k\\
\label{e15} Z_k&=h(X_{k})+v_k
\end{align}
where `X' and `w' are respectively the augmented state and process noise vector is of size $(n+p)\times 1$ and thus $w_k\sim \mathcal{N}$ ( 0, $\begin{bmatrix}\textbf{Q} &0\\0&0\end{bmatrix}$ ). The control input `u' and the `hat' symbol for estimates are not shown for brevity. A formal solution to the above problem is the Extended Kalman Filter (EKF) summarised as Brown and Hwang\cite{Brown2012} (2012)
\begin{table}[h]
\begin{center}
\begin{tabular}{  l  l }
\textbf{Initialisation :} & $\mathbf{X_0}=E[X_t]$ \\[10pt]
& $\mathbf{P_0}=E[(X_0-X_t)(X_0-X_t)^T]$ \\[14pt]
\textbf{Predict step :} & $X_{k|k-1}=f(X_{k-1|k-1},u_{k-1})$  \\ [10pt]
& ${P}_{k|k-1}=F_{k-1}{P}_{k-1|k-1}F^T_{k-1}+\textbf{Q}$ \\[14pt]
\textbf{Update step :} & $K_k={P}_{k|k-1}H^T_{k}(H_{k}P_{k|k-1}H^T_{k}+\textbf{R})^{-1}$ \\ [10pt]
& $X_{k|k}=X_{k|k-1}+K_k(Z_{k}-h(X_{k|k-1}))$ \\ [10pt]
& ${P}_{k|k}=(I-K_kH_{k}){P}_{k|k-1}$
\end{tabular}
\end{center}
\end{table}

%
%


where all the symbols have their usual meaning and
\begin{align*}
&\text{True initial state} &:& X_t \\
&\text{Initial state estimate} &:& X_{0|0}=\mathbf{X_0} \\
&\text{Initial state covariance matrix} &:& P_{0|0}= \mathbf{P_0}\\
&\text{State Jacobian matrix}&:&F_{k-1}= \left[\frac{\partial{f}}{\partial{X}}\right]_{X=X_{k-1|k-1}}\\
&\text{Measurement Jacobian matrix}&:&H_{k}= \left[\frac{\partial{h}}{\partial{X}}\right]_{X=X_{k|k-1}}
\end{align*}

The innovation $\left(\mathbf{\nu_k}=Z_k-h(X_{k|k-1})\right)$ following a Gaussian distribution whose probability when maximized leads to the Method of Maximum Likelihood Estimation (MMLE), which is operationally equivalent to minimizing the cost function
\begin{align*}
\mathbf{J}=&\frac{1}{N}\sum \mathbf{\nu_k}(H_kP_{k|k-1}H_k^T+\textbf{R})^{-1}\mathbf{\nu_k}^T\\
=&\mathbf{J}(\mathbf{X_0,P_0,Q,R},\Theta)\\
=&\mathbf{J}(\mathbf{X_0},\Theta,\mathbf{K}(\text{traded for } \mathbf{P_0, Q, R}))
\end{align*}
based on the summation over all the N measurements and thus solving for either $\mathbf{X_0, P_0, Q,}$ \\ \textbf{R}, $\Theta$ or solving for $\mathbf{X_0}$, $\Theta$, K as the case may be. When \textbf{Q} = 0, the MMLE is called as the output error method with the Kalman gain matrix being zero. In the usual Kalman filter implementation generally one does not solve for the statistics $\mathbf{P_0}$, \textbf{Q} and \textbf{R} but tweak manually to obtain acceptable values. The numerical effort of minimizing \textbf{J} cannot be swept under the rug and it has to appear in the estimation of the filter statistics.

The estimation of $\mathbf{X_0}$, $\mathbf{P_0}$, $\Theta$, \textbf{R} and \textbf{Q} in the Kalman filter is known as adaptive filter tuning. The ghost of filter tuning chases every possible formulation or any variant of the Kalman filter be it EKF, Unscented Kalman Filter (UKF), Particle filter (PF), or the Ensemble Kalman Filter (EnKF) or their combinations. The best possible tuning is necessary if one desires to get near optimal solutions. If not properly tuned it is difficult to infer if the performance of the variants of Kalman filter are due to their formulation or filter tuning!

Though Kalman filter has been applied in so many fields of science and technology we are not in a position to estimate the parameters and their uncertainty as denoted by Cramer Rao Bound (CRB).  Even a routine adaptive filtering technique to estimate a constant signal with measurement noise does not seem to exist!

Even if the tunable unknowns are not available or inaccurately known the filter should be able to estimate all of them only from the measurements. Generally the initial estimates are kept not too far from the expected estimates for any algorithm to converge. However for the present approach the initial choice can be over a large range.

There are five steps in the Kalman filter, namely state and covariance propagation with time, Kalman gain calculation and the state and covariance update by incorporating the measurement. In the filter statistics approach all the five steps have to be gone through. However when the constant Kalman gain approach is used only the three steps namely the state propagation, Kalman gain calculation and the state update are necessary.

\par The Kalman filter is not a panacea to obtain better results when compared to simpler techniques of data analysis. The accuracy of the results using Kalman filter depends on its design based on the choice of $\mathbf{X_0}$, $\mathbf{P_0}$, $\Theta$, \textbf{R} and \textbf{Q}. If the above values are  not chosen properly then the filter results can be inferior to that obtained by simpler techniques.

\par There could be other suitable cost functions that one can develop depending on the situation. One cost function called the (Integral of Time multiplied by Absolute Error) ITAE considers the time as a scaling factor. This is meaningful since it is important to ensure a zero error after the filter has converged. This performance index is given by
\begin{align*}
{J_{ITAE}}=\frac{1}{T}\sum a_i|\mathbf{\nu_k}|dt
\end{align*}
where the $a_i$ is suitable weight such as related to the innovation covariance. Another cost function useful to study the effects of inadequate modeling in state estimation problem that is very common in Kalman filter studies  has been proposed and used in rendezvous and docking problem (Philip and Ananthasayanam \cite{Philip2003}  2003),
\begin{align*}
{J_{MODEL}}=\frac{1}{N}\sum (x(i)-x_{ref}(i))^T\mathbf{P}^{-1}(x(i)-x_{ref}(i))	
\end{align*}
with the summation is over all the N time points and the suffix `ref' refers to a desired reference trajectory to be followed and the argument in x (.) denotes the time step or point. The P is the covariance matrix obtained with nominal values for the unknown disturbances. If the variations or a deficiency in the modeling is beyond the statistical fluctuations as denoted by the covariance then the above cost function changes substantially and indicates a degradation of the filter performance. The surprising thing is that when \textbf{R} alone exists in a problem a suitable cost function is considered but when \textbf{Q} also exists most people seen to avoid the cost function. Is this because of the ad hoc approach of choosing the statistics?

\par In the present work we introduce a generalized cost function by an expansion of the usual `innovation' to other quantities generated by the Kalman filter such as the prior and posterior state respectively before and after the measurement is assimilated, the smoothed state after all the measurements are processed. We demonstrate that it is such an approach that decisively indicates the best possible solution in the simulated data analysis and more so in the analysis of real flight data analysis.

\par Some fundamental differences exist between the cost function handled by classical optimization problems and the Kalman filter. The former deals with a cost function that is static, with a fixed model and the number of data, with deterministic unknowns. In contrast the Kalman filter has to grapple with a dynamical cost function due to time varying state and measurement model structure and the parameters therein, with continuously increasing number of measurements and the unknowns are both deterministic quantities and the statistic of probabilistic variables.

\section{Aesthetics, Beauty, and Truth of the Kalman Filter}

The aesthetics of the Kalman filter is to consider only the uncertain estimate and the covariance representing the uncertainty. From a deterministic case to move to probabilistic scenario just only one additional quantity is used for describing the results which is economical.

Also this is in some ways similar to many other problems in science and engineering wherein only the first and second derivatives or moments alone are considered. This is just the reason and fortuitously as well for using velocity and pressure or temperature in equilibrium thermodynamics, which depend on the first and second moments respectively of the distribution function governing the random velocity of the gas molecules. This leads to the consideration of fewer moments or states to describe the dynamics of the gas flow. Imagine what would happen if many higher order moments had become relevant! Even in the equations of motion of classical dynamics only up to linear and angular accelerations, which are second derivatives exist!

Another subtle reasoning can be provided for the above feature. Take a rectangular distribution and with increase in sample size the lower order moments converge faster than the higher order moments. This is because away from the middle the tails control the higher order moments. For a very similar reason, the Boltzmann equation deals only with single particle distribution as against multi particle distribution function.

An analogy can be given from real life. Perhaps it is easy to judge a person early as soft or hard. It would take little more time to find out if how flexible he is in dealing with situations or advice. It would take far more time to understand his way of handling difficult situations in life.

\par The beauty in the Kalman filter is whether it is true or otherwise the many  random variables occurring in a problem are assumed to follow or represented by a multivariate Gaussian distribution, which is completely specified by the mean and covariance. The Gaussian distribution provides an enormous amount of mathematical tractability exactly for linear systems and approximately for nonlinear systems.

\par The truth in the Kalman filter equations is that once it is derived in one  way, it is possible to derive it in a variety of ways with slightly  different assumptions, but leading to similar set of basic equations as  for linear problems. It is interesting to note that the simplest formulation of the Kalman filter is based on minimum amount of a priori knowledge in probability, statistics, and random process providing respectively the Gaussian distribution, linear relationship among the variables, and white noise. The author of each book has his own derivation! If necessary other suitable distributions, nonlinearity, and coloured noise can be introduced later into the filter framework.

\section{Definition and Different ways of Looking at the Kalman Filter}

 We now propose a possible definition of the Kalman filter. The Kalman filter assimilates the measurement information with uncertain system and measurement models based on probabilistic weighted linear addition of the predicted state and the measurement data to adapt both the state and measurement models and their noise statistics in a statistically consistent way. Rao \cite{Rao1987} (1987) had stated that all methods of acquiring knowledge are essentially statistics. Hence the analysis of time dependent process carried out sequentially is contained is the knowledge front by assimilating newer information in a meaningful way is the Kalman filter!

\subsection{Inverse Problem}

\par What is the simplest way to look at the ET? The problem is given the measurements made on the system how can one obtain the structure of both the system and the measurement models? In general due to the redundant amount of data there is no unique solution and this gives rise to a search for best, optimum or even adequate approaches together with subjective assumptions to get a unique solution. Based on previous experience or otherwise one has some feel about a dynamical system. Many systems are deterministic being governed by well known Newton's laws of mechanics or gravity, or laws of electricity, and magnetism and so on. Though their structure is known the parameters in them may not be known with desirable accuracy. If a system is stochastic in nature such as the population, flow in rivers, stock market then suitable stochastic models have to be evolved and the unknown parameters in them estimated.

\par In direct problems given the rules we generate the outcomes and there is nothing more to it. Given a certain scenario today can we say how it had evolved? That is why predicting the future is relatively easy but inferring the past is far more difficult! Given a data how it came to be is far more difficult to answer than answer how it will evolve. You see a certain food stuff, can you say the sequence by which it has been prepared but you can prepare a stuff to reach a certain taste. This is just the reason people patent food items! Again the inverse problem is more difficult than direct problem. If you see a person today can you infer how he had reached the present condition? However knowing him now perhaps you can predict better what he could be later on.

\subsection{Qualitative Modeling}

\par How to qualitatively model the states and the measurements? The system identification according to Zadeh \cite{Zadeh1962} (1962) is the determination on the basis of input, of a system within a specified class of systems, to which the system under test is equivalent. This equivalence implies a loss or error or cost function \textbf{J}. Thus when a qualitatively best, proper or adequate model structure for the state and measurement models are chosen it implies identification. Different model structures can be tested with different criteria and the most acceptable one can be chosen. At times one can model the unknown control input as random process noise. In general for present day activities the algorithms and computing power exist but the physical modeling for example at high angles of attack and sideslip of an aircraft due to their nonlinear, coupled nature with unsteady effects presents a formidable challenge for a flight mechanics specialist to pass on suitable inputs to his control counterpart.

\subsection{Quantitative Estimation}
\label{QE}
\par When to accept the quantitative estimates? The quantitative parameter estimation is defined by Eykhoff\cite{Eykhoff1974} (1974) as the experimental determination of parameters for the above qualitative model. Modeling has to satisfy the conditions of controllability and observability. In simple physical terms these mean that a control has to excite the states and the excited states have to be measured to determine the parameters. For efficient determination of parameters optimal inputs have been proposed. These excite the various modes of the system controlled by the parameters.

\par A simple example of qualitative and quantitative estimation together with consistency checks is provided in  Appendix-\ref{QM}. No matter what you do mathematically in statistics note that there is a manual override and thus it extends to Kalman filter as well! It also shows that some statistical tests due to the level of confidence at which the estimates pass can be deceptive and hence we look for decisive tests and/or criterion in the report.

\subsection{Handling Deterministic State and Measurement Errors}

\par How to handle deterministic state and measurement errors? If there is a random error in the state model or the measurement model it can be handled by including the noise. How to deal with deterministic errors such as an improper structure of the state model or a bias or scale error in the measurement? We can still handle such deterministic errors in the state and measurement equations respectively by adding the process and measurement noise to encompass them.

\subsection{Estimation of Unmodelable Inputs by Noise Modeling}

\par How to handle unmodellable inputs? It is argued subsequently that the effect of unmodellable or unmodelled errors in the state and measurement equations can be offset by adding noise in the model. A more interesting question is can we estimate them as well? The answer is yes and has been done in many cases. The random noise can be modelled as `white' or a more general  Gauss-Markoff process (Gelb \cite{Gelb1974} 1974) of a suitable order thus providing a structure for the random component. The qualitative structure of the random noise modelled with some free parameters whose estimation can provide a quantitative estimate of that noise.

\subsection{Unobservables from Observables}

\par How to estimate the unobservables from the observables? Use intuition or otherwise to connect the unobservables in the state with the measurements. The measurements made on the system are functions of the state and may not always correspond directly to the states and the parameters. The unknown state and parameters may be unobservable and have to be inferred from the measurements. The estimation of the aerodynamic lift, drag, and moment of aerospace vehicles by flight test data analysis is just one example of the above!

\subsection{Expansion of the Scenario}

\par How to get even more knowledge from the measurements? The answer is to improve or in other words put more information into the state model. The expanded state equations contain our understanding that is further improved by the measurement information. As mentioned in the previous section the measurement space can in general be different from the state space. The state equations can be modelled with increasing level of sophistication. In order to appreciate this better consider the radar measurements of an airplane which suffered an accident. The radars provide range, azimuth, and elevation information. If the airplane is treated as a point mass governed by kinematic equations then more accurate estimates of its position, velocity, and perhaps acceleration can be estimated. Next if the state equations are formulated as dynamical equations then the mass, as well as the aerodynamic parameters can be estimated. A further sophisticated model of the airplane's equations of motion helps us to expand the scenario to obtain the forward speed, angles of attack and sideslip, the pitch, yaw, and roll angles, and even the way in which the throttle and the controls have been operated thus facilitating the analysis of the accident and apportion the cause to airframe, pilot, weather or other factors. The expansion of the scenario can be taken as a more sophisticated version of the unobservables from the observables.

\subsection{Deterministic or Probabilistic Approach?}

\par What is the difference between deterministic and probabilistic approach in ET? Gauss as mentioned earlier had been the earliest to utilize the measurements of the position of the asteroid Ceres in the celestial sphere to determine its orbital parameters and predict its position at a later time for sighting it. He used the method of least squares in a deterministic way to handle the more number of redundant measurements. He had provided many clues even for the probabilistic approach. The least squares estimate and the uncertainty are only qualitative. When the characteristics of process and measurement noise denoted respectively by `w' and `v' are specified quantitatively through appropriate probability distributions these get translated after the filter operations through the data into the distribution of the estimates in a quantitative manner. Gauss had understood the deterministic and probabilistic formulations and also that the assumptions in the latter need not always be true.

Before the data is available one can talk of probability. Once the data is available it is deterministic and the entire unknown both deterministic and the statistics of the random variables can all be treated as deterministic unknowns. This is evident by the translation of the problem to minimizing the cost function \textbf{J} which is deterministic. Of course one can use probabilistic approaches to solve a deterministic optimization problem.

\par For theoretical expediency and minimal information the \textbf{Q} and \textbf{R} are generally assumed to be white Gaussian. It is not the validity but the ability of the above that provide acceptable results. There is a charm in doing it that way as between a game of `draught' and `chess' where the latter a few more rules provides panoramic situations. Rules made by humans led them to problems. The basic laws of nature have always been very simple.

\subsection{Handling Numerical Errors by Noise Addition}

\par How to handle numerical errors? In many real time applications propagating the state and covariance equations of the Kalman filter could be highly time consuming. Here one could use less accurate but fast solvers and incur some numerical errors. The less accurate solvers can be interpreted as a modeling deficiency. Once this is accepted then an additional random process noise can offset numerical inaccuracy. However the addition of process noise cannot be a `panacea' for large scale modeling or numerical deficiencies.  Such noise inputs always come at the expense of increased uncertainty of the state or a parameter. One can offset modest amount of numerical errors at the cost of a small increase in the uncertainty of the estimates.

\subsection{Stochastic Corrective Process}

\par Are there other characterizations of the Kalman filter operation? There is a philosophical reason for the sequential approach to be preferable. The evolution of many systems for large times is unpredictable. Typical examples are the population growth, economics, climate, weather, satellite orbit, and space debris. In all the above the dynamics of the system could be changed consciously with time by the society or unmodellable forces or features. Thus additional information at various times regarding the system based on the newer data helps to predict the system better before the next data arrival.

\par As an example a satellite's orbit cannot be predicted accurately for all times to come no matter how accurately one accounts for the model of the earth's gravity field, an atmospheric model, and all other perturbations. Due to the very random nature of unmodellable forces on the satellite its trajectory would depart from the one based on any assumed model after quite sometime. Another classic example is that of weather prediction. No matter how accurate the atmospheric model is one is compelled to take recourse to measurements which is a newer information about the system that has run in parallel in nature without of course solving any governing equations! Thus at various times one is compelled to offset the limitations in the model and input using the measurements. In other words theory and experiment should go hand in hand! Such processes which get corrected at various times have been called as 'Stochastic Corrective Processes' by Narasimha \cite{Narasimha1975} (1975).

\subsection{Data Fusion and Statistical Estimation by Probabilistic Mixing}

\par How are the states and measurements combined? A remarkable feature of present day ET is a variety of information from different sources, rates and accuracies are combined not just algebraically but statistically in an optimal way. This can also be called as data fusion and assimilation (Raol\cite{Raol2010} 2010). Such a fused data is combined with states using suitable weightages to obtain the best possible estimate of $\mathbf{X_0}$, $\mathbf{P_0}$, $\Theta$, \textbf{Q} and {R} all from just the measurement Z only!

\par The Kalman filter was called a mixer by Gelb \cite{Gelb1974} (1974) and the final effect is the filtering of the noise. It is well known that when two independently distributed random variables are added then the variance of their sum is the sum of the individual variances. This means the uncertainty of the new random variable has increased. However if we add them in a weighted manner say half to each one of them then the resulting random variable will have a lower variance than either of the first two. Getting the proper or the so called optimal weight depending on the various individual noise covariances is an important problem in ET and manifests as filter tuning which is a big industry.

\subsection{Optimization in the Time Domain}

\par How to use the formal local filter update equations over the complete data? The earlier sections dealing with propagation from one time to the next and update at a given data point are formal in the sense given an initial $\mathbf{X_0}$, $\mathbf{P_0}$, $\Theta$, \textbf{Q} and {R} the filter can operate on the full data. With improper choice for the above any result can be generated. In order to make sense from the full data, an overall cost function have to be minimized and in the process all the tuning parameters ($\mathbf{X_0}$, $\mathbf{P_0}$, $\Theta$, \textbf{Q} and \textbf{R}) are estimated. This provides a consistent solution valid over the full data. Since the minimization is in general a non linear problem and an iterative processing of the data by the filter equations are needed. The updates for all the above can be carried out at each and every data point or over a window or full data by repeated processing through the data. The optimisation updates based on the sensitivity of the difference between model and measurement if only measurement noise is present is fairly straight forward. The use of Newton Raphson method had been prolific but it turns out if the measurement channels are very few or the data length is very short other techniques have to be utilised. Also probabilistic methods like genetic algorithms have also been successfully applied. After the convergence of \textbf{J} further checks have to be carried out if the initial assumptions regarding the signal and noise are satisfied like if the fit is good, smooth, and the innovations are white and Gaussian. Care has to be taken to see that the above are not statistically deceptive but decisive.

\subsection{Frequency Domain Analysis}

Is a frequency domain approach possible?  Though generally one deals in the time domain by combining or matching of state and measurement it is possible to transform both the above into frequency domain and carry out the analysis. The advantage being in particular for real flight test data, the contamination of the pure airframe excitation, with noise from other sources such as engine, structural vibrations can be looked into in the frequency domain and cut off. A formulation of MMLE was provided in the frequency domain (Klein and Morelli\cite{Klein2006} 2006). The analysis in the frequency domain due to the transformation of the data can lead to changes in the estimates as well as their uncertainties. However it is restricted to linear systems. The earliest filtering technique by Wiener \cite{Wiener1949} (1949) is in the frequency domain but it had many difficulties in practical implementation. The contribution of Kalman in dealing with estimation problems has two aspects. One is the change from a `batch' processing mode to `sequential' mode and the other is to switch from `frequency' domain approach to `time' domain approach. It is felt that based on the sequential least square estimation, such a data processing would have any way come about, but the greater contribution is to move away from frequency to time domain which is more natural and with extensions to general nonlinear systems.

\subsection{Smoothing of the Filter Estimates}

\par What are the best state, measurement and parameter estimates from the filter? During a forward pass of the filter over the complete data only the last point has an estimate based on the complete data. In order to obtain the best possible filter estimates at all data points a backward smoothing can be carried out. This is based on a backward filter pass over the data. Then at every data point by statistically combining the estimates from the forward and the backward filter passes the best estimates can be obtained. After such a smoothing process the state represents the best possible signal content. It is the smoothed states that greatly helped  in the estimation of \textbf{Q} in the present work.

\subsection{Improved States and Measurements}

\par Are the noises in both the states and measurements reduced? The mixing of the complimentary information from the state and measurement helps to obtain better estimates of each as also the parameters in them. After the filter smoothing process, improved state estimate is available and this helps to smooth the measurement. The measurements help to provide improved state variable and their unknown parameters. The Kalman filter can be understood as an advanced model based moving average filter. The usual moving average filters use a simple  polynomial fit over a suitable window size of the data. The Kalman filter using knowledge based state and measurement models is able carry out much better and more accurate signal and noise separation and estimation.

\subsection{Estimation of Process and Measurement Noise }

\par How to estimate \textbf{Q} and \textbf{R}? When the deterministic parts of the state and measurement equations are balanced all over the data what remains is the corresponding noise. However since the filter generates among others predicted, updated, and smoothed state estimates and the corresponding measurements one has to find out which combination of the above helps to provide a suitable `statistic' to obtain an estimate of \textbf{Q} and \textbf{R} if they are also unknown.

\subsection{Correlation of the Innovation Sequence}

\par How to conclude if the estimates are consistent? It is possible to look at the behaviour of the innovation sequence for having tuned the filter statistics $\mathbf{X_0}$, $\mathbf{P_0}$, $\Theta$, \textbf{Q} and \textbf{R} properly. If the filter is optimal then the correlation of the innovation sequence should be `white'. This implies all the information from the data has been extracted and what remains is pure random `white' noise with no information content. This can happen only when the model and the measurement structures are proper, the parameter in them have been obtained after the numerical optimization algorithm has converged properly. Even if any one of the above is not correct then the above sequence would not be `white' within the statistical limits of acceptance. Further an ideal correlation coefficient matrix for the parameter estimates should be an identity matrix. This can happen only with infinite data and with limited data correlations would exist among the estimated parameters and this should be kept in mind in their subsequent use. The `white' noise is the worst data that will fail all modelling, and algorithms in ET and can provide any random result with no meaning. One can refer to Rao\cite{Rao1987} (1987) for some fallacious inferences drawn from random sequences as if they are laws of nature!  If the atmosphere, and earthquakes are truly random then researchers in such fields can stop working!

\subsection{Reversing an `Irreversible' Process}

\par What and how is it finally achieved? Once the signal and noise are mixed they become indistinguishable unless a proper modeling of signal and measurement is made and further numerical effort is put in to separate them. An analogy may be given as mixing of salt and water when the identity of salt is lost. The situation is similar to a signal corrupted by noise. Our understanding is that with evaporation salt alone remains and water vapour goes up. Such a compulsive understanding of the behaviour of salt and water helps to separate them. This is precisely what is demanded in ET, namely model the system and the measurement before processing the data. In the Kalman filter numerical effort is required to propagate the state and the covariance equations (if there is process noise), form a cost function, and optimize it by utilizing various techniques (in general iterative) before the solution is obtained. Thus one can note the mixing of noise with a signal is relatively easy! Thus adding noise is an `irreversible' process and to bring the system back to its original state a lot of effort is required.

\subsection{Some Human Activities from ET Perspective}

How to view some human activities from ET perspective? The advances in science and technology help to improve the cultural style and the living standard of a society provided the wealth is distributed equitably. An inequitable distribution of wealth leads to tensions and turmoil in a society. The cost function for the society need to have proper weightage over food, shelter, health, and leisure for all.

Inventers and innovators in the society must be rewarded.  But it should not be very small or very large imposing burden on the society. In the language of the filter what is the optimum (=reward) Kalman gain? With the availability of new gadgets in sports the rules, regulations, and umpiring will have to change. In cricket these are the stump cameras, and the Hawkeye. The cost function here is not dependent on the gadgets but the immediate excitement, and enjoyment provided to the spectators. The reference to a third umpire in cricket or the challenge by the player in tennis provides a quick and correct decision. The umpiring mistakes are very much reduced.  But if such sport gadgets are not inducted the `status quo' attitude can lead to acrimonious and long drawn disputes. Thus there is a need to sense the change, capture and correct our life.

A fine example of data assimilation is our digestive system extracting the nutrients and rejecting the rest. Accumulating undesirable contents leads to digestive problems or disease. Another fantastic example of assimilation is the evolution of life where the important experiences of the earlier generations are assimilated and compactly coded in the DNA. The spiritualistic state corresponds to system and the materialistic world corresponds to observations. Buddha leading a princely life observed the sufferings due to decay, disease, and death. He assimilated such observations and changed his way of living discarding wine, women, and wealth.

For estimating or controlling a variable it is not always necessary or feasible to handle the same variable. The removal of violence and terrorism could perhaps be achieved more effectively by following a sattvic way of life. The physical, personal, psychological, and professional suffering of people can be overcome not necessarily by direct actions but by remote measures subtly connected to the issues. Such perspectives of some activities in ET (Ananthasayanam and Bharadwaj \cite{MRA2011} 2011) are provided in Table - \ref{tbcape1}.

\section{Tuning of the Kalman Filter Statistics}

\par How does one achieve all the above stated objectives in designing the Kalman filter for any application? It is not known to many that the enthusiasm which followed soon after the Kalman filter was introduced was damped since the noise statistics had to be provided to design the filter. Obviously the effort to be put in minimizing \textbf{J} cannot be escaped!  In relation to its importance to obtain as close to an optimal solution as is possible the corresponding effort does not appear to have been put in the field. Generally one manually tweaks the statistics to reach acceptable results instead of tuning properly to get even better results. Thus the filter tuning has not matured to a level for routine use. Gauss had a relatively ideal situation with a good system model and only the measurements had noise. Kalman when he proposed the filter required the statistics of the process and the measurement noise to be known and dealt with only state estimation. The adaptive approach or filter tuning tries to obtain the filter statistics $\mathbf{P_0}$, \textbf{Q} and \textbf{R} by using the filter operating over the measurement data.

\par The filter tuning varies from ad hoc, through heuristic to rigorous methods. Generally the tuning is manual but adaptive processes are needed to obtain better results. The ad hoc quick fix solutions are such as limiting P from going to zero, or add \textbf{Q} to increase P before calculating the gain, multiply P by a factor to limit K all have obviously limitations in handling involved problems or scenarios. In the fading memory filter the estimates based on the current set is averaged with the previous estimate, with a weighting parameter be it for \textbf{R} or \textbf{Q}. All the above are arbitrary and can lead to inaccurate results. The third rigorous approach could be time consuming to the extent of solving the whole problem. Exact solutions are very hard, approximate choice can lead to inappropriate results but the middle path of heuristic approaches are quite appealing and is the present work. We derive in this report a fairly general adaptive approach to tune the Kalman filter statistics for any system. The present recursive filter recipe helps to achieve statistical equilibrium for all the unknowns together with their internal consistency checks.

\section{Important Qualitative Features of the Filter Statistics}

\par Should the $\mathbf{P_0}$ = \textbf{Q} = 0 then the filter will not learn anything from the measurements all of which will be ignored. The \textbf{R} is fairly objective and can be determined from the measured data. The $\mathbf{P_0}$ is tricky and generally the off diagonal elements are set to zero and the diagonal elements are set to large values but however their relative values are crucial for an optimum filter operation. The $\mathbf{P_0}$ controls the handover from the initial transient to \textbf{Q} for steady state filter behaviour and has to be chosen carefully. The $\mathbf{P_0}$ is important like for a child with deficient or overdose of nutrition when young will have some shortfall even with appropriate nutritious food he/she may have in later life. The \textbf{Q} is notoriously difficult but it helps to inject uncertainty into the state equations to assist the filter to learn from the measurements and it also controls the steady state filter response. Even when there is no state noise but only the measurement noise in the data, starting with some initial estimate $\mathbf{X_0}$ and $\Theta$ which is somewhat far away from the true values in order to assist the filter to learn from the measurements a non zero \textbf{Q} has to be injected into the state equations since the effect of $\mathbf{P_0}$ will fade away quickly. A large value of \textbf{Q} will lead to a short transient with large steady state uncertainty of the estimates and vice versa for small \textbf{Q}. In the present work since both $\mathbf{X_0}$ and $\mathbf{P_0}$ are simultaneously tuned for data without process noise the requirement for injecting \textbf{Q} does not arise.

\par Thus though \textbf{Q} is considered notorious it is the life line of the Kalman filter doing good work all the time. Grossly misunderstood just like good people thought to be bad for a long time! Some classic examples for such systems are the GPS receiver clocks, satellite, trajectory of aircraft, missiles and re-entry vehicles. These are handled by using the kinematic relations between the position, velocity and acceleration all assumed to be driven by white Gaussian noise \textbf{Q} of suitable magnitude to enable the filter to track these systems. Generally the filter parameters are tuned off line using simulated data and subsequently used for on line and real time applications perhaps with efficient numerical procedures.

\par It is only by the \textbf{Q} that an analyst can handle the unmodelled or unmodellable errors  and even account for numerical errors! The \textbf{Q} is also helpful to track systems whose dynamical equations are not known. The process noise can estimate, account or offset for some deficiency, inaccuracy, or error in the following namely the initial conditions, system and measurement model equations, control or external input, measurement noise statistics, the numerical state and covariance propagation or update operations.
\par Similar to the different derivation of the filter equations there could be different methods of implementing the filter for any practical situation. One could work with time varying full, diagonal, or constant matrices \textbf{Q} and R, or work with constant Kalman gain matrix, or with important Kalman gain matrix elements. The constant \textbf{Q} and R matrices approach is generally preferable and the constant gain  approach is more easily implementable though not necessarily optimal.

\section{Constant Gain Approach}

\par In many problems after the initial transients the Kalman gain matrix tends to a non-zero constant value due to finite amount of data or modeling errors. It is useful to trade the filter statistics to the constant Kalman gains to minimize the above \textbf{J} by using an optimization algorithm assuming the innovation covariance to be a constant. The advantage of using constant gains is that the covariance equations need not be propagated and updated thus enormously saving computational time. The gains can be chosen using a suitable cost function based on the normalized innovation over a period of time just as for the filter statistics.

\par For the constant gain approach any suitable optimization algorithm can be utilized to minimize the cost function. It has been found that the simple genetic algorithm are quite effective in handling a variety of aerospace problems such as the docking and rendezvous (Philip and Ananthasayanam \cite{Philip2003} 2003), the evolution of the space debris (Ananthasayanam et al.\cite{MRA2006} 2006), re-entry of a space debris object (Anilkumar et al. \cite{Anilkumar2007} 2007), integrating the MEMS-INS and GPS (Helen Basil et al. \cite{Helen2004} 2004), and in estimating the total electron content in the atmosphere (Anandraj et al. \cite{Anandraj2005} 2005). Based on the above successful applications we suggest the Kalman gain approach in huge data assimilation problems.

\section{Kalman Filter and some of its Variants}

\par The simplest formulation of a Kalman filter is when the state and measurement equations are linear. For linear systems during the evolution and the update the normal distribution is maintained, but with changed mean and covariance.  For nonlinear systems even if the initial distribution is normal, it gets distorted after propagation and so a suitable local approximation or quasi linearization has to be made.

\par In the Extended Kalman filter (EKF) the nonlinear systems and/or measurements equations (Smith et al. \cite{Smith1962} 1962) are approximated by first order Taylor series expansion about the previous updated state. The probability density is approximated by a Gaussian, which may distort the true structure and at times could lead to the  divergence between the filter prediction and the measurements.

\par In the Unscented Kalman filter or UKF (Julier et al. \cite{Julier1997} 1997) approach instead of linearizing the functions, a set of chosen points are propagated through the nonlinear transformation. These points are so chosen such that the mean, covariance, and possibly also higher order moments, match with the propagated distribution.

\par The Interactive Multiple Model (IMM) was introduced to handle rapid system dynamics (Blom et al.\cite{Henk1988} 1988), since the gain cannot change as rapidly as the system dynamics. Here additional model state equations describing velocity, acceleration, jerk and so on are introduced. Here each filter corresponding to the dynamic model computes the states in parallel using the given measurements and the responses from individual filters are combined based on their mode probabilities to form the response from the IMM.  This interaction requires the specification of the transition probability matrix $p_{ij}$  that the model `i' is at the current observation time given that model `j' was at  the previous observation time as well as a sojourn time `$\tau_i$' to be specified by the analyst to calculate suitable weights for each filter's output.

\par The particle filtering (Gordon \cite{Gordon1993} 1993) is a Monte Carlo technique for state estimation that can handle nonlinear models together with non Gaussian noise. Here the state probability density is approximated by using point particles having positive weights. Based on the initial distribution the weights are chosen and then the particles are propagated following the system dynamics together with the state noise. Then using the measurement their weights are adjusted and normalized among all the particles. The particles that can track the measurements gain weight and the ones far away lose their weights. However, after a while all, but one weight, will become zero leading to degeneracy. A resampling scheme is introduced to solve the degeneracy problem that discard the particles with small weights and focus on the particles with more significant weights.

\par For large size systems, such as those occurring in geophysical studies  maintaining the covariance matrix computationally being difficult, in the ensemble KF (EnKF) for large problems (Evensen \cite{Evensen2003} 2003), the estimate and the covariance matrix are replaced by the sample covariance from a large number of ensemble members similar to a particle in the particle filter. Each member of the ensemble is propagated including the process noise and later updated using a so-called virtual observation.

One may note the evolution of the variants of the Kalman filter possesses some similarities as it progressed to handle simple, complex, to massive problems as in many other fields such as fluid dynamics or structural mechanics. In these cases commencing from simple geometries one obtains closed form analytical solutions (as in the linear KF wherein the gains can be pre computed to process the data as and when they arrive) followed by numerical calculations in involved cases (as in EKF). When the geometry is complex and the boundary conditions are involved it becomes necessary to discretise and form cells over appropriate space and time (as in PF) to obtain the solution. Further when massively complex geometries and boundary conditions occur other innovative approaches (as in EnKF) have been developed. An extensive bibliography of the non linear estimation is provided by Georgios \cite{Georgios2001} (2001) and an excellent review of non linear filters is given by Fred Daum \cite{Daum2005} (2005).

\begin{landscape}
\begin{table}[h]
\caption{Perspectives of Some Activities in Terms of Estimation Theory}
\label{tbcape1}
\begin{center}
\begin{tabular}{|c| c| c| c| c|c|c| }

\hline

No	& Description  & \makecell{Estimation \\Theory} & Astronomy &	\makecell{Society and\\ its evolution}	& \makecell{Intellectual\\ property rights} & 	\makecell{Cricket or\\ Tennis}

\\ \hline
1	& Change	& System model	& \makecell{Predict the \\ position of \\ celestial objects}	&
\makecell{Living standards\\ and culture} &	Inventions	& \makecell{Existing Rules \\and umpiring}

\\ \hline
2 &	Capture	& \makecell{Measurement model}	& \makecell{Observations with\\ Unavoidable errors}	& \makecell{Scientific and \\technological\\ advances}  &	\makecell{Impact on \\society}	& \makecell{Camera and \\Hawkeye}

\\ \hline
3	& Correct	& \makecell{The above estimates\\ are combined}	& \makecell{Combine the above\\ to get improved\\ position of objects} 	& \makecell{Changing standard \\ and culture} &	\makecell{Awards, patents,\\ and copyrights}	& Third umpire

\\ \hline
4 &	\makecell{Nature of cost\\ or its criteria}	& \makecell{Highly subjective\\ but cast subjectively\\ reasonably} 	& \makecell{Mean square error\\ between model\\ and measurements}	& \makecell{Many but\\ should be\\ reasonable}& \makecell{Proper recognition,\\ rewards and\\ returns}	& \makecell{Refer to the\\ third umpire or\\ player challenge}

\\ \hline
5	& Cost function	& \makecell{Reduce the\\ difference between\\ model and \\measurements} 	& \makecell{Minimize the \\above error}	& \makecell{Improved standard\\ of living and\\ happiness}	& \makecell{Encourage inventions\\ without burdening \\society}	& \makecell{More number \\ of correct\\ decisions}

\\ \hline
6	& \makecell{Effect of proper\\ Correction}	& \makecell{Better than\\ either the model\\ and Measurement}	& \makecell{Improves estimated \\position of \\celestial objects}	& \makecell{Improved and \\Equitable living\\ for all} &	\makecell{Encourages \\further inventions}	& \makecell{Increased excitement\\ and enjoyment}

\\ \hline
7	& \makecell{Effect of improper\\ Correction}	& \makecell{Can make it worse\\ than model and \\the Measurement}	& \makecell{Celestial objects \\can be lost \\for later tracking} 	& \makecell{Unequal change\\ can lead to\\ tension and turmoil}	& \makecell{Exploitative\\ economics leads to\\ disparity in society}	& \makecell{Incorrect decisions\\ spoils excitement\\ and enjoyment}

\\ \hline

\end{tabular}
\end{center}
\end{table}
\end{landscape}

\chapter{Kalman Filter Tuning : Earlier and the Present Approach}
\label{ch3}

\section{Introduction to Kalman filter Tuning }

\par The estimation of $\mathbf{X_0}$, $\mathbf{P_0}$, $\Theta$, \textbf{R} and \textbf{Q} in the Kalman filter is known as the problem of adaptive filter tuning. The ghost of filter tuning chases every variant of the Kalman filter which can at best be minimized but not completely ignored if one desires to get near optimal solutions. Further it becomes difficult for one to infer if the performance of the variants of Kalman filter are due to their  formulation or filter tuning! It should be remarked that in the best spirit of the estimation theory in particular the recursive Kalman filter approach even if $\mathbf{X_0}$, $\mathbf{P_0}$, $\Theta$, \textbf{R} and \textbf{Q} namely the initial states, their covariances, parameters in the state and measurement equations, the measurement and state process noise covariances are not available or inaccurately known the filter should still have the ability to estimate all the above from the `observables' that are measured and commencing not too far from the proper estimates for the algorithm to converge. One would like to have the initial choice of all the unknowns should not be very critical. The filter should be self consistent in estimating all the unknowns. Generally one tune the filter statistics off line using simulated data and later use it to process real data on line or even in real time. The Table-\ref{tbcape2} discusses the triplets (Ananthasayanam and Bharadwaj \cite{MRA2011} 2011) occurring in the Kalman filter.

\par The present work shows that there is a compulsive proper choice of $\mathbf{X_0}$ and $\mathbf{P_0}$ in order to derive the best possible state and parameter estimates and their uncertainties using the filter. For both the states and parameters the choice of $\mathbf{X_0}$ appears to be relatively weak for the filter estimates but the choice of $\mathbf{P_0}$ is very important in particular for the parameter uncertainty represented by CRB. Generally even with not quite a good tuning of the filter statistics the estimates could be near the optimum but lead to large variation in the uncertainty represented by the CRB. However it may be cautioned that for data with only measurement noise if a Newton Raphson (NR) technique is used to minimize the cost function the effect of the initial unknown state can affect the estimates but can be handled if this is also treated as an unknown. In the case of the filter by the process of smoothing the unknown initial state can be also estimated.

\par As we will see later the elements of $\mathbf{P_0}$ have to be chosen not just low or high but in such a way they lead towards optimum filter estimates and the CRB. The choice of suitable $\mathbf{X_0}$, and $\mathbf{P_0}$ reminds one of the bitter struggles in statistics in specifying the `prior' for the Bayesian approach which had been studied enormously.

\section{Some Simple Choices for Initial $\mathbf{X_0}$ for States and Parameters}

\par There are many `ad hoc' methods to make the filter operate satisfactorily. Since some of the states are generally measured either the first or the average of the first few measurements can be taken as the initial value $\mathbf{X_0}$ for the state. The parameters are used as augmented states in the EKF route. Since either some computational or experimental results are available, these can be set as initial $\mathbf{X_0}$ for the parameter values. At times if possible a least square parameter estimates based on balancing the governing differential equations using the appropriate measurements can also be used as start up values.

\section{Importance of Initial $\mathbf{P_0}$ for States and Parameters}

\par One might wonder as to where is the necessity for tuning $\mathbf{P_0}$ lies, when in principle it can be arbitrary. The answer to this lies in the following. For example if $\mathbf{P_0}$ is set equal to zero (for a very confident choice for the initial estimates) then the filter ignores and learns nothing from the measurements. If $\mathbf{P_0}$ is extremely large (a pessimistic choice for the initial values), then the filter believes the measurements much more and provides very little weightage or ignores the state model values leading to large fluctuations in the state and parameter estimates along with large final uncertainty. Thus there appears to be a proper choice for $\mathbf{P_0}$ which is neither zero nor infinity to provide proper estimates and uncertainty.

\par This $\mathbf{P_0}$ is one of the important tuning parameters as stressed by very few like Maybeck \cite{Maybeck1979} (1979), Candy \cite{Candy1986} (1986), and Gemson \cite{Gemson1991} (1991) but most people seem to treat the above casually. Usually one assumes a guess $\mathbf{P_0}$ which tends to become very low after some data points and in order to make the filter learn from the subsequent measurements introduce an additional \textbf{Q} by trial and error into the state equations. This finally leads to some estimates and uncertainties. The former usually may be close but the latter generally away from the correct values. The peculiar situation is one has introduced \textbf{Q} even when there is no model structure uncertainty. Only to learn from the measurements the \textbf{Q} was introduced. We looked for a recursive procedure that provides a proper $\mathbf{P_0}$ without any \textbf{Q} in such cases and it turns out to be so as will see subsequently.
\par The important point is that the initial state and parameter $\mathbf{P_0}$ can affect the final covariance (P$_{N|N})$ from the filter operation. This can become crucial in certain state estimation problems such as impact point estimation and its uncertainty for target tracking. In such cases the results for the uncertainty from improper or inaccurate tuning can be highly deceptive. Even in parameter estimation problems the estimates and the uncertainties that are attached can be important in the design of control systems.

\section{Choice of Initial \textbf{R}}
\par Usually a good initial estimate for \textbf{R} can be obtained from the calibration of the measuring instrument and generally it is assumed to be constant over the data length. In the NR procedure of minimizing the cost function based on data without process noise even \textbf{R} can be treated as an unknown and estimated along with other unknown quantities.  One usually assumes an initial value for \textbf{R} in the filter which could be close to zero and later adaptively update and estimate it.

\section{Choice of Initial \textbf{Q}}
\par In principle the \textbf{Q} should reflect the uncertainty in the assumed state model or any `unmodellable' feature of the state or even unknown random state input. The \textbf{Q} along with the initial $\mathbf{P_0}$ plays a very important role in the filter operation without divergence. The value of \textbf{Q} should be small enough to retain the learning potential from the measurement but not large to increase the uncertainty so that the filter estimates become useless.  One usually assumes an initial value for \textbf{Q} in the filter which could be close to zero and later adaptively update and estimate it.

\section{Simultaneous Estimation of \textbf{R} and \textbf{Q}}

\par When the data contains the effect of both the measurement and process noise it becomes far more difficult for analysis. In general minimization of the cost function by treating \textbf{R} and \textbf{Q} as unknowns during which the simultaneous update and estimation of \textbf{R} and \textbf{Q} appears to be difficult as reported by many researchers in the field. For a given \textbf{R} changing the sequence of measurement noise makes the dynamics noisy means more blurred. In contrast the varying sequence of process noise alters differently every time the dynamical behaviour of the system making it stochastic meaning drift randomly. The interesting point is the filter by tracking the drifted dynamical behaviour even with large \textbf{Q} estimates the parameters controlling the original dynamics of the system without the effect of \textbf{R} and \textbf{Q}. Since \textbf{R} and \textbf{Q} occur respectively in the measurement and state equations their effects are negatively correlated. The \textbf{Q} represents the average rate of change of the state, tracking ability by indirectly controlling the rate at which the old data are forgotten, and affects the estimation accuracy (Bohilin 1976). The effect of \textbf{R} is opposite to that of \textbf{Q}. Thus during simultaneous recursive estimation if the statistics for estimating them are not properly chosen then \textbf{R} is over estimated and \textbf{Q} is under estimated and vice versa. This is just the reason due to which Gemson \cite{Gemson1991} (1991) in his approach had to update \textbf{R} and \textbf{Q} alternately.

\subsection{Tuning Filter Statistics with only \textbf{R}}

\par In the spirit of recursive filtering operations we have sought to tune all the filter statistics by repeated iterative processing of the data and nowhere used any optimization procedures. When only measurement noise is present in the data it is possible to get both the estimates and the CRB by minimizing the cost function \textbf{J} formed by the difference of the estimated model parameters and the actual measurement. The NR procedure carries out the above task very efficiently for simple to involved systems (Ananthasayanam et al. \cite{MRA2001} 2001). It is the results from the above NR procedure that served as an anchor for tuning the filter parameters to get the closest possible estimates and the CRB from the present recursive filter passes through the data. It appears to be not feasible to exactly reproduce the NR results from the filter for each and every data but sufficiently close such that the over an ensemble of data there is very little difference.

\subsection{Tuning Filter Statistics with both \textbf{R} and \textbf{Q}}
\par However when process noise is also present in the data we have to look for consistency based on a comparison between the injected and estimated values of \textbf{R} and \textbf{Q}. Further the additional cost functions proposed in our work based on balancing the state and measurement equations helped to obtain confidence in the results. When the filter operates through the data it generates prior, post, and smoothed state estimates and their covariances which help to generate candidate `statistic' to estimate both \textbf{R} and \textbf{Q}. From these one can also generate additional cost functions to see how best the state equations are balanced. Then all the above state estimates and covariances can be transformed into measurement space to be compared once again with the actual measurements and its covariance to generate candidate `statistic' to estimate \textbf{R}. These help to form more cost functions to see how best the measurement equations are balanced. It is necessary to choose from among the above `statistics' the proper combination of `statistics' for simultaneously estimating \textbf{R} and \textbf{Q}. Any improper combination does not lead to proper filter operation and even if it does leads to inappropriate results as is the case in the earlier approaches to solve the filter tuning problem.

\par The subsequent sections after a review of earlier approaches propose the present method. The Chapter \ref{ch4} demonstrates the results using the simulated data followed by real data in Chapter \ref{ch5} based on a comparison of the present results with earlier approaches.
\section{Earlier Approaches for Tuning the Kalman Filter Statistics}
\subsection{Mathematical Approaches and Desirable Features}
Most mathematical treatments implicitly assume unlimited data. However we have to deal with finite data and wish to get the solution within a few filter passes. Further most such treatments do not provide sensitivity among the input and output variables in a problem which could lead to insensitive results and conclusions. The Appendix-\ref{QM} using the OEM approach for a simple curve fitting provides an example for such inferences.

\par In any field of science or technology one needs at times stability and at other times sensitivity. For example in the Newton Raphson optimization technique there is stability in the procedure commencing from different reasonable initial values converging to the same result for a given measurement data. Should the measurement data change then the converged parameters also change(!) thus showing sensitivity. We have established such a `stability' and `sensitivity' for the present reference recursive recipe.

\subsection{Review of Earlier Adaptive Kalman Filtering Approaches}
One brute force method for tuning filter statistics is to carry out an optimization exercise to solve for the statistics based on minimizing a suitable cost function over thousands of combinations of the unknowns. These would be too unwieldy requiring such a massive numerical exercise to be carried out for each and every new problem scenario. We should thus look for elegant approaches which help tune the filter with reasonable numerical effort. In particular in EKF if the unknown noise covariances are incorrectly specified biased estimates can arise (Ljung \cite{Ljung1979} 1979, Ljungquist and Balchen \cite{Ljung1994} 1994). Even when the system parameters are known, if an inaccurate description of the noise statistics are used the filter may give poor estimates, or even diverge.

\par There are broadly four approaches for adaptive filtering namely Bayesian, Maximum Likelihood, Covariance Matching and Correlation Techniques (Mehra \cite{Mehra1970} 1970).

\subsubsection{Bayesian Method}
\par Every update in a Kalman filter is obviously a Bayesian update. The various approaches have been called differently based on what and how they handle such as Bayesian, ML based on the innovation, filter statistics, and the autocorrelation of the innovation to adaptively estimate or account for the unknown covariances. The work of Alspach \cite{Alspach1974} (1974) deals with a bank of autonomous Kalman filter run with a range of Kalman gains. Each one stores a running sum of the square of the residuals. Subsequently it is possible to obtain the estimates of the unknowns based on a weighted sum over the grid points of the gain. Hilborn and Lainiotis \cite{Hilborn1969} (1969) show that a Bayesian optimal adaptive estimation system converges to the average performance of an (unrealizable) optimal system operating with knowledge of the parameters which are unknown to a Bayesian adaptive/learning system. Thus an adaptive system learns the true parameters based on assimilating the information from the data.

\subsubsection{Maximum Likelihood (ML) and Expectation Maximization (EM)}

\par The ML and EM methods need the specification of the probability distribution (usually a Gaussian) followed by the difference between the measurement and the model prediction called innovation or prediction error. The maximum likelihood methods (Kashyap \cite{Kashyap1970} 1970, Bohlin \cite{Bohlin1976} 1976) maximize the likelihood function based on the innovations containing the unknown covariances \textbf{Q} and \textbf{R}. Usually time consuming gradient based numerical optimisation or other schemes that reprocess the data several times are required to estimate the unknown covariances.

\par In order to avoid nonlinear optimization in the ML approach Shumway and Stoffer \cite{Shumway2000} (2000) proposed an iterative method using the expectation maximization (EM) technique. The EM consists of expectation and maximization steps. First the states are estimated using an initial guess of the unknown parameters based on a Kalman smoother. The unknown parameters are next estimated by ML method. This process of estimating the states using the Kalman smoother and optimizing the parameters using is repeated until parameter convergence.

\par Bavdekar et al. \cite{Bavdekar2011} (2011) used both a direct ML method and an EM method for nonlinear systems, based on the extended Kalman filter. Their first approach directly optimizes the likelihood function of the innovation sequence (a product of the likelihoods for each innovation since the innovations are white for an optimal estimator) generated by the EKF using a standard constrained nonlinear programming strategy such as sequential quadratic programming. The EM method maximizes the conditional density function of the complete data set of the states and measurements. When the parameters are known the extended EM algorithm provides the estimates for the noise covariances based on algebraic relations without the need for any derivative calculation.

\par Recently Zagrobelny and Rawlings \cite{Megan2014} (2014) similarly proposed a ML method to identify \textbf{Q} and \textbf{R} assuming the parameters are known. By writing the outputs in terms of the process and measurement noises, they form a normal distribution for the sequence of measurements. The variance of this distribution is a function of the unknown noise covariances, and the likelihood is optimized with respect to these covariances. Simulations are used to compare the ML method to several existing methods like the ALS method, an alternate ML method based on the innovations, and an EM method.

\subsubsection{Covariance Matching}

\par During the filter pass over the data a number of random variables such as pre, post, and smoothed states arise and these transformed to the measurements along with actual measurements many more statistics are available. Combining these one can form innovation, residue, smoothed residues and so on. As mentioned earlier among the five filter equations two refer to the sample and the other three refer to the ensemble characteristics. If the filter is performing well then the sample statistics formed by the above should be internally consistent with their ensemble properties also provided by the filter. These can be matched by running many Monte Carlo simulations. However taking sample statistics at various time instants during a filter pass one can form equations connecting the many random variables, their combinations and their covariances to estimate the unknown covariances. The different approaches of Myers and Tapley \cite{MT1976} (1976), Gemson \cite{Gemson1991} (1991), Mohamed and Schwarz \cite{MS1999} (1999), Bavdekar et al. \cite{Bavdekar2011} (2011), and the present approach are examples for the covariance matching. Different statistics are chosen either over a window or the full data length, after each pass for estimating \textbf{R} and \textbf{Q}. These estimates could be used in subsequent passes as is necessary. Myers and Tapley \cite{MT1976} (1976) approach using the innovation (available before update) can at times make the \textbf{R} lose its positive definiteness. This is because \textbf{R} depends on the difference of two matrices which could at times become negative. In order to overcome such an eventuality they proposed a remedy of simply turning the estimate as positive! Jazwinski \cite{Jazwinski1970} (1970) and  Mohamed and Schwarz \cite{MS1999} (1999) suggested a more stable statistic based on the residue (available after update) for estimating \textbf{R} which is the sum of two matrices. The \textbf{R} estimates are generally more accurate than \textbf{Q}. Simultaneous estimation of \textbf{R} and \textbf{Q }is not easy perhaps as they negatively affect the filter performance. These covariance matching techniques are preferred due to their simplicity and speed despite being suboptimal. Gemson \cite{Gemson1991} (1991) and Gemson and Ananthasayanam \cite{Gemson1998} (1998) showed the importance of $\mathbf{P_0}$ for parameter estimation. Similarly many statistics are available and are utilized for estimating \textbf{Q}.

\subsubsection{The Correlation Technique}

\par The innovation theorem of Kailath \cite{Kailath1970} (1970) states that the innovation sequence is zero mean white Gaussian. Kailath further stated that if any gain other than the optimal is used then estimates will be suboptimal and the innovation mean will be non zero and the autocorrelation will not follow the Kronecker delta function. Also if the covariance of the innovations are not as expected they are indicative of the choice of any or all of the system matrices as well as the covariances are incorrect.

\par The correlation technique pioneered by Mehra \cite{Mehra1970,Mehra1972} (1970, 1972) and Carew and Belanger \cite{Carew1973} (1973) and Belanger \cite{Belanger1974} (1974) is the earliest and highly cited method for determining the unknown covariances. Mehra's approach is based on the properties of the innovations process that must be white and Gaussian. Starting from an assumed value for the unknown \textbf{R} and \textbf{Q} an initial estimate for the steady state Kalman gain is obtained. This sequence is checked to see whether the particular Kalman gain implemented generates a statistically acceptable white noise sequence. However the Kalman gain can take correct value even when \textbf{R} and \textbf{Q} are far away from their true values. This is because different combinations of \textbf{R} and \textbf{Q} can lead to the same gain.

\par Later Neethling and Young \cite{Neethling1974} (1974) noted large covariances from Mehra's approach due to some other deficiencies. The approach of Oussalah and De Schutter \cite{Oussalah2000} (2001) improves Mehra's \cite{Mehra1970} (1970) and Carrew's and Bellanger's (1973) approaches by incorporating information about the quality of the innovation estimates leading to a weighted least squares methodology instead of the earlier least squares methodology. The weights are determined using a Bhattacharyya distance criterion between the ideal probability and the distribution referring to the current first and second order statistics of the autocorrelation functions. The latter helps to generate a convergent sequence to the steady state filter, which after some manipulations allows one to determine the values of the noise statistics \textbf{Q} and \textbf{R}.

\par
\begin{figure}[h]
\includegraphics[width=5in,height=4in]{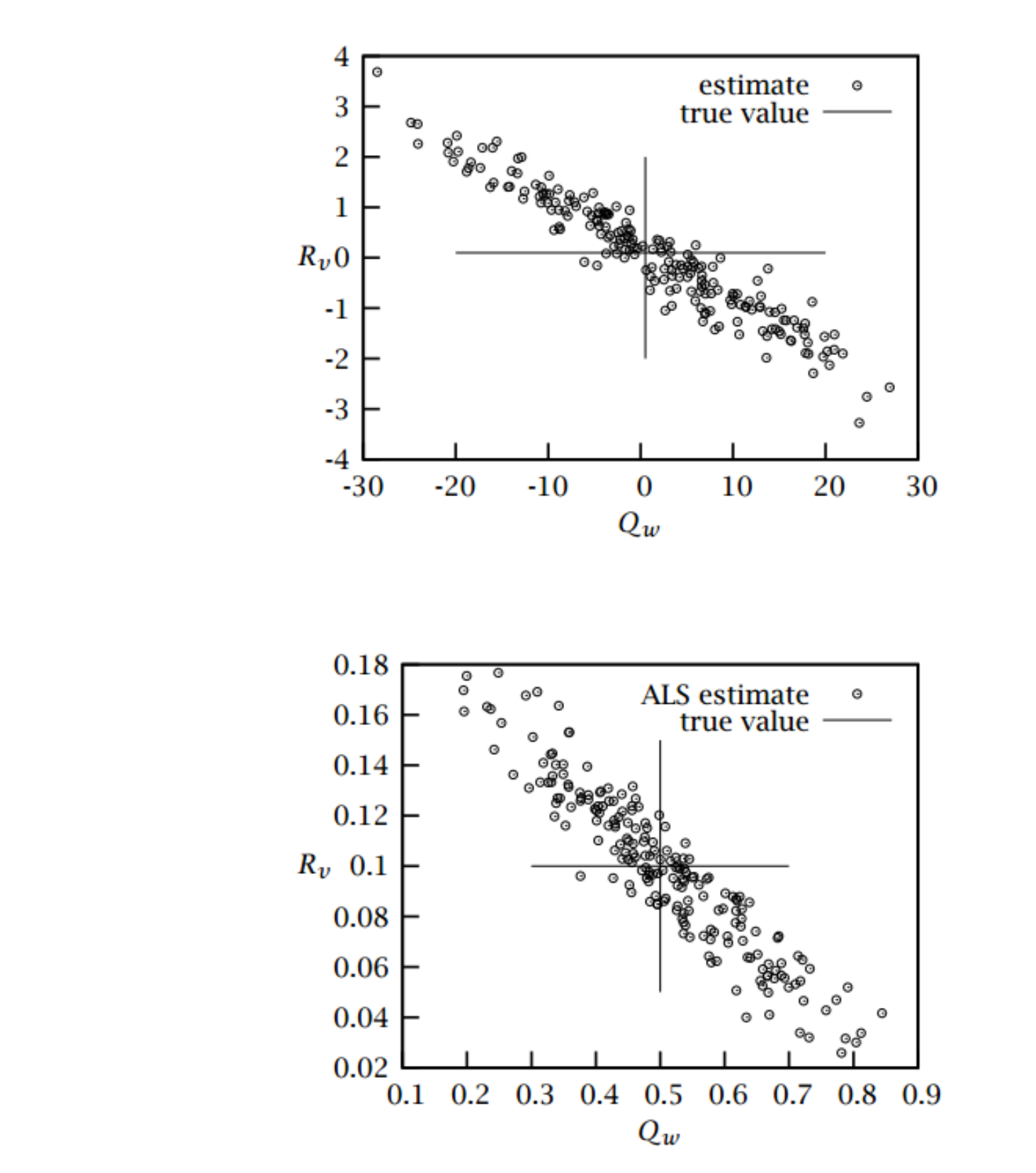}
\caption{Estimates of $Q_w$ and $R_v$ (\textbf{Q} and \textbf{R} in our notation) using }
\caption*{Mehra's method (top) and ALS method (bottom)}
\caption*{Credit: Odelson et al. \cite{Odelson2006} (2006)}
\label{od}
\end{figure}

 The latter work of Odelson et al. \cite{Odelson2006} (2006) showed that based on some counter examples the mathematical conditions regarding the system and measurement matrices are not sufficient and not necessary in Mehra's \cite{Mehra1970} work (1970). Also they showed that the variance estimates of Mehra are larger. They proposed a method called constrained Autocovariance Least Squares (ALS) method corrected the above and obtained a much smaller and better estimates and none negative as shown in Fig. \ref{od} for a typical case.

\subsubsection{Some Approaches in an Airplane Flight Test Data Analysis}

\par In an airplane flight test data analysis many approaches have been suggested to determine the parameters and the noise covariances. For the sake of brevity we will only briefly mention them as natural formulation of Schultz \cite{Schultz1976} (1976), innovation, a combined formulation which takes the advantage of natural and innovation called MMLE3 of Maine and Iliff (1980), the combined formulation of Ishimoto \cite{Ishimoto1997} (1997) to solve the practical problems in natural and innovation formulations. A more detailed discussion on aircraft flight test data analysis can be found in Klein and Morelli \cite{Klein2006} (2006) and Jategaonkar \cite{Jategaonkar2006} (2006).

\par One may ask the question as to why there are so many formulations for solving an optimization problem. The  reason is the unknowns do not occur in a simple way in the cost function, and there are many transformed variables with which one tries to solve for the basic unknowns, the size and the required compatibility conditions among the transformed variables lead to the many difficulties not found in the classical optimization problems. An analogy may be given as when one tries simple difference schemes to solve a problem in Computational Fluid Dynamics one could invariably find himself in distress. A similar problem generally arises in using standard optimization  techniques in tuning the Kalman filter for the purpose of system identification and estimation.

\subsubsection{Some Other Approaches}

\par Valappil and Georgakis \cite{VG2000} (2000) used the available information about the model parametric uncertainties and translate this information to the process noise covariance \textbf{Q}. They propose two methods called linearized and Monte Carlo approaches. They also assume that \textbf{R} is readily available from the measuring instrument characteristics. The Monte Carlo simulations are run with parameter values sampled from the assumed normal distribution, with means and covariances. Then the difference between the nominal and the random dynamical state trajectory over many simulations is taken to provide the process noise at any time instant to be used in the filter equations.

\par Some attempts have been made like Powell \cite{Powell2002} (2002) using the simplex method, Oshman and Shaviv  \cite{Oshman2000} (2000) using the genetic algorithm, and controlled random search (Anilkumar \cite{Anilkumar2000} 2000). However when the dimension, nonlinearlty, and the range of search space become large these could become computationally prohibitive and could lead to local minima.

\par Manika Saha et al. \cite{Manika2014} (2014) felt that $\mathbf{X_0}$ and $\mathbf{P_0}$ are not critical since they affect only the initial transient filter performance and $\mathbf{P_0}$ has to be decided by designers since \textbf{P} changes as the filter operation proceeds reaching a steady state if only the system dynamics does not substantially change and a suitable choice of \textbf{R} can be easily made. Thus for a chosen suitable values for $\mathbf{X_0}$, $\mathbf{P_0}$, and \textbf{R} they identified the critical \textbf{Q} by forming two metrics based on the innovation covariance. These vary from zero to the number of measurements and vice versa as \textbf{Q} changes from zero to infinity they proposed that \textbf{Q} can be chosen around the intersection point of these two cost functions.

\par Lau and Lin \cite{Lau2011} (2011) also discuss the limitations of simulated annealing and particle optimization techniques for filter tuning.  One can summarize that deterministic or probabilistic optimization approaches do not appear to be efficient for solving the filter tuning problem. Hence we tried to see if a recursive filtering approach would work and fortuitously it had worked and will be demonstrated subsequently.

\subsection{Earlier versus the Present Approach for Filter Tuning}

\par Much of the earlier research work have in fact and rightly so concentrated their effort in using simulated data to tune the filter off line for obtaining the statistics to be used later for on line and real time applications. While this is commendable one important feature has been forgotten in such approaches. Many variations in the methods used could lead to convergence and one could feel happy about the same. The convergence of any technique even in simulation studies is no guarantee for a proper solution to the problem. Even the simple case of a linear fit to a set of data many variants tend to different results. It is like a necessary condition but not a sufficient one. It is just the reason that we still lean on simulation studies but wherein exact solution is available to the analyst. Hence presently the filter methods have been applied firstly to some very simple cases such as a constant signal, a ramp function, spring-mass-damper system, longitudinal, and lateral motion of an airplane for which when only the measurement noise exists (\textbf{R} $>$ 0) and the process noise does not exist (\textbf{Q} = 0). For such a situation exact reference results are derivable by using the Newton Raphson technique and these can be compared with filter generated results. Subsequently when the process noise in included in the system under study one is in an uncharted territory and for filter tuning it is necessary to look for the consistency based on a comparison between the injected measurement and process noise values. Further many other cost functions based on balancing the states and measurement equations that are introduced help to move from deceptive to decisive solutions.

\par There are five basic filter operations namely the state and covariance propagation, Kalman gain evaluation, and the state and covariance updates. The state propagation and update refers to sample values and the covariance propagation, its update, and the Kalman gain refer to the population characteristics. The estimation of \textbf{R} and \textbf{Q} is possible provided one utilizes appropriate choice of the estimation `statistics' based on many quantities that arise in the filter operation like the pre and post filter states as well as the ones derivable from the measurements such as the innovation, residue, smoothed and their covariances. The behaviour of the ensemble characteristics of the samples should be consistent with population characteristics.

\par The above helped to guide us in the choice of appropriate initial filter statistics namely $\mathbf{X_0, P_0,}\Theta$, \textbf{Q} and \textbf{R}. In particular after the first filter pass (using some initial statistics) through the data which is a must it has guided the way the filter statistics have to be set from the second iteration onwards. It helps to answer the questions like what should be the $\mathbf{P_0}$ for the states and the parameters be derived from the value at the end of the pass to be used in the next iteration, should they be full, diagonal, or even zero, whether the \textbf{Q} has to be injected into the state and/or the augmented parameter states. Since the reference NR estimates of the parameters, the Cramer Rao bound (CRB), and the measurement noise are available it has been possible to settle such above questions and in particular the CRB played a crucial role to guide the choices for the above. Generally the Kalman filter provides fairly acceptable estimates for the parameters but unless the tuning is good it is very difficult to match the CRBs from the EKF with the NR values.

\par There are many quantities that occur at various times during the filter pass through the data. The present work has shown that a simultaneous update and convergence of \textbf{R} and \textbf{Q} towards proper values are possible provided one utilizes appropriate choice of the estimation `statistics' among the many quantities that arise in the filter operation like the pre and post filter states as well as the ones derivable from the measurements such as the innovation, filtered residue, smoothed residue and their covariances.

\section{Present Approach for Tuning Filter Statistics}

\par Fundamentally the Estimation Theory (ET) is an optimization problem. Hence a suitable cost function \textbf{J} has to be chosen. In general there are many estimation algorithms which have desirable characteristics (Beck and Arnold \cite{Beck1977} 1977, Goodwin and Payne \cite{Goodwin1977} 1977, Sorenson \cite{Sorenson1980} 1980) that can be utilised to solve an estimation problem. Essentially there are two elements in ET (i) Defining a cost function, (ii) Adopting a suitable algorithm to minimize the cost function. The Likelihood cost (L) for normally distributed error (e) (Bohn \cite{Bohn2000} 2000) is given by

\begin{align}
L(\Theta)=&\frac{1}{N}\sum_{k=1}^N \frac{1}{2}(e_k)^TA_k^{-1}(e_k)+log(det(A_k))
\end{align}
where A is the error covariance matrix and det(A) represent determinant of matrix A. It may be noted that the unknown parameters ($\Theta$) occurs implicitly and not explicitly in the cost function L. Since the filter provides many quantities it is possible to have many more terms in the cost function which is a function of errors in the initial state and parameter estimates, process noise driving the system and the measurement noise introduced by the measurement system, each weighted appropriately through suitable weighting covariance matrices. Thus the new cost function \textbf{J} in a weighted least square sense accounts for (i) A priori knowledge about the initial estimates, (ii) Balancing the measurement equations. (iii) Balancing the system equations. One can call this as generalized MLE (\textbf{J}),
\begin{align}
\mathbf{J = J0+ J1 + J2 + J3 +J4 + J5 + J6 + J7 + J8}
\end{align}
whose terms are defined as
\begin{align*}
\mathbf{J0}=&\frac{1}{2}(\mathbf{X_0}-X_t)^T\mathbf{P_0}^{-1}(\mathbf{X_0}-X_t)\\
\mathbf{J1}=&\frac{1}{N}\sum_{k=1}^N (Z_k-h(X_{k|k-1}))^TS1_k^{-1}(Z_k-h(X_{k|k-1}))\\
\mathbf{J2}=&\frac{1}{N}\sum_{k=1}^N (Z_k-h(X_{k|k}))^TS2_k^{-1}(Z_k-h(X_{k|k}))\\
\mathbf{J3}=&\frac{1}{N}\sum_{k=1}^N (Z_k-h(X_{k|N}))^TS3_k^{-1}(Z_k-h(X_{k|N})) \\
\mathbf{J4}=&\frac{1}{N}\sum_{k=1}^N (Z_k-h(Xd_{k|N}))^T(Z_k-h(Xd_{k|N}))\\
\mathbf{J5}=&\frac{1}{N}\sum_{k=1}^N (Z_k-h(X_{k|k-1}))^TS1_k^{-1}(Z_k-h(X_{k|k-1}))+log(det(S1_k))\\
\mathbf{J6}=&\frac{1}{N}\sum_{k=1}^N   w1_{k|N}^TW1_k^{-1}  w1_{k|N}\\
\mathbf{J7}=&\frac{1}{N}\sum_{k=1}^N   w2_{k|N}^TW2_k^{-1}  w2_{k|N}\\
\mathbf{J8}=&\frac{1}{N}\sum_{k=1}^N   w3_{k|k}^TW3_k^{-1}  w3_{k|k}
\end{align*}
The `S' and `W' are functions of the second order moments given by
\begin{align*}
S1_k=&H_{k}P_{k|k-1}H_{k}^T+\textbf{R}\\
S2_k=&-H_{k|k}P_{k|k}H_{k|k}^T+\textbf{R}\\
S3_k=&-H_{k|N}P_{k|k-1}H_{k|N}^T+\textbf{R}\\
W1_k=&-P_{k|N}-F_{k-1|N}P_{k-1|k-1}F_{k-1|N}^T+P_{k,k-1|N}F_{k-1|N}^T+P_{k,k-1|N}^TF_{k-1|N}+\textbf{Q}\\
W2_k=&-P_{k|N}-Fd_{k-1|N}P_{k-1|k-1}Fd_{k-1|N}^T+P_{k,k-1|N}Fd_{k-1|N}^T+P_{k,k-1|N}^TFd_{k-1|N}+\textbf{Q}\\
W3_k=&P_{k|k-1}-P_{k|k}
\end{align*}
If the initial states are known then J0 is not necessary but if they are unknown, their estimate and covariance can be obtained respectively by the smoothed estimates $\mathbf{X_{0|N}}$ and  $\mathbf{P_{0|N}}$.  The cost \textbf{J1, J2} and \textbf{J3} are expected to tend towards the number of measurements ($m$). The cost \textbf{J6, J7} and \textbf{J8} defined for states with process noise are expected to tend towards the number of states (n). The cost \textbf{J4} is expected to tend towards the trace of \textbf{R} for \textbf{Q} = 0 case and \textbf{J5} is the negative log likelihood function.

\subsection{Choice of $\mathbf{X_0}$}
\label{RTS}
\par Commencing from an assumed reasonable initial choice for $\mathbf{X_0}$, $\mathbf{P_0}$, $\Theta$, \textbf{R} and \textbf{Q} the first filter pass through the data is made. Then a backward smoothing procedure is carried out. The Rauch Tung and Striebel (RTS \cite{RTS1965} 1965) and Modified Bryson Frasier (MBF \cite{Bierman1977} 1977) smoothing gave almost the same numerical estimates, the RTS was used as a standard smoothing procedure throughout the present work. The smoothing leads to the best possible state and parameter estimates as well their covariances. It may be noted after smoothing the state estimates and their covariances change but not that of the parameters. We next describe how the above choices are updated for further filter passes through the data to eventually converge which denotes statistical equilibrium. If the exact value of $\mathbf{X_0}$ if not given it can be obtained by the smoothed estimates. In such probabilistic approaches any number of unknowns are only determined in a probabilistic sense and thus contain uncertainty due to the percolation of all the noise effects over all the unknowns.  The RTS smoothing equations for discrete time instants k = N-1, N-2,\ldots 0 are given by
\begin{align*}
K_{k|N}=&{P}_{k|k}F_{k}{P}_{k+1|k}^{-1}\\
X_{k|N}=&{X}_{k|k}+K_{k|N}({X}_{k+1|N}-{X}_{k+1|k})\\
P_{k|N}=&{P}_{k|k}+K_{k|N}({P}_{k+1|N}-{P}_{k+1|k})K_{k|N}^T
\end{align*}
where $K_{k|N},{X}_{k|N}$ and $P_{k|N}$ is the smoothed gain, smoothed state estimate and smoothed state covariance matrix respectively. All the other variables have their usual meaning which are obtained from the EKF. After an estimate of the initial state (${X}_{0|N}$) if one uses the smoothed initial covariance ($P_{0|N}$) for tuning the filter then the parameter CRBs are not properly estimated but has to be scaled up  as seen in the simulated case studies in Chapter \ref{ch4}.

\subsection{Choice of $\mathbf{P_0}$}
\label{choiceP0}
\par The choice of $\mathbf{P_0}$ for the next filter pass is very tricky. If one were to take the smoothed initial state covariance ($P_{0|N}$) and use it as the $\mathbf{P_0}$ for the next pass then the final covariance keep on decreasing with further filter passes and eventually tend towards zero. We know that such a fact is unrealistic. In order to remedy the above behaviour the final covariance at the end of the pass was scaled up by the number of data points (N) and used at the beginning of the next pass. The only heuristic reasoning that can be provided from statistics is based on the fact that the mean from a sample has an uncertainty P that keeps decreasing with sample size as P/N where P is the population variance. Since in the filter steps the estimates and its update refer to the sample and the other covariance propagation, its update, and the calculation of the Kalman gain refer to the ensemble characteristics before every filter pass we carry out the above scale up method to obtain the $\mathbf{P_0}$ for the next filter pass. The importance of scaling up is discussed in Appendix-\ref{X0P0}. We also propagated backwards the final covariance using the estimated parameters and it turned out that there was not much of a difference with the $\mathbf{P_0}$ that was used in the forward pass. Of course the above process does not end and we have to further modify the above $\mathbf{P_0}$ to obtain the best possible CRB after some passes. The scaled up $\mathbf{P_0}$ is a full matrix. Many changes such as using only the diagonal elements and many more variations were tried out. Finally the reference $\mathbf{P_0}$ to obtain the proper CRB for the parameter estimates turns out to have all the elements are zero except the diagonal elements corresponding to the parameters. If all the elements of the parameter covariances were included and the state covariances set to zero, it did not make much of a difference in the final results. The scaled up $\mathbf{P_0}$ (Shyam \cite{Shyam2014} 2014) is given by
\begin{align}
\label{P2} \mathbf{P_0}=N\times P_{N|N}
\end{align}
With the above changes, the surprising feature is to set the initial covariance as zero even for the states which are measured. The usual recommendation (Mehra \cite{Mehra1970} 1970) in such cases is to set $\mathbf{P_0}$ = \textbf{R}. Even by using the IIM (Inverse of Information Matrix) approach of Gemson \cite{Gemson1991} (1991) obtained the same estimate as \textbf{R} for $\mathbf{P_0}$. The IIM is given by
\begin{align}
\label{P1}  \mathbf{P_0}=\left[\frac{1}{N}\sum_{k=1}^{N} F_{k-1}^TH_k^TR^{-1}H_kF_{k-1} \right]^{-1}
\end{align}

\subsection{Estimation of \textbf{R} and \textbf{Q} using EM Method}
\par The expression for \textbf{R} and \textbf{Q} using the Expectation Maximisation (EM) method extended to a non-linear system by Bavdekar et al. \cite{Bavdekar2011} (2011) is given by,

\begin{align}
\label{max1} \textbf{R}&= \frac{1}{N}\sum_{k=1}^N E\left[v_kv_k^T|Z_N\right]\\
\label{max2} \textbf{Q}&= \frac{1}{N}\sum_{k=1}^N E\left[w_kw_k^T|Z_N\right]
\end{align}

\subsubsection{Estimation of \textbf{R}}
\par Consider the measurement noise, $v_k=Z_k-h(X_k)$ which can be approximated using first order Taylor series expansion around the smoothed estimate, $X_{k|N}$ given by
\begin{align*}
&v_k \approx Z_k-h(X_{k|N})-H_{k|N} \tilde X_{k|N}
\end{align*}
where $H_{k|N}=\frac{\partial{h}}{\partial{X}}_{|X=X_{k|N}}$ and $\tilde X_{k|N}=X_k-X_{k|N}$
\begin{align}
\nonumber &v_kv_k^T=Z_kZ_k^T-Z_kh^T(X_{k|N})-Z_k\tilde X_{k|N}^T H_{k|N}^T-h(X_{k|N})Z_k^T+h(X_{k|N})h^T(X_{k|N})\\
\nonumber &+h(X_{k|N})\tilde X_{k|N}^T H_{k|N}^T-H_{k|N}\tilde X_{k|N}Z_k^T+H_{k|N}\tilde X_{k|N}h^T(X_{k|N})+H_{k|N}\tilde X_{k|N}\tilde X_{k|N}^T H_{k|N}^T
\end{align}
We know that
\begin{align*}
&E[\tilde X_{k|N}]=E[X_k-X_{k|N}]=0\\
&E[\tilde X_{k|N} \tilde X_{k|N}^T]=E[(X_k-X_{k|N})(X_k-X_{k|N})^T]=P_{k|N}
\end{align*}
Thus the conditional expectation is given by
\begin{align*}
E\left[v_kv_k^T|Z_N\right]=Z_kZ_k^T-Z_kh^T(X_{k|N})-h(X_{k|N})Z_k^T+h(X_{k|N})h^T(X_{k|N})+H_{k|N}P_{k|N} H_{k|N}^T
\end{align*}
Rearranging the above terms and using Eq-\ref{max1}, we get
\begin{align}
\label{R1}   \textbf{R}=\frac{1}{N}\sum_{k=1}^{N}\left\lbrace(Z_k-h(X_{k|N}))(Z_k-h(X_{k|N}))^T+H_{k|N}P_{k|N} H_{k|N}^T\right\rbrace
\end{align}

\subsubsection{Estimation of \textbf{Q}}
\par Consider the process noise, $w_k=X_k-f(X_{k-1})$ which can be approximated using first order Taylor series expansion around the smoothed estimate, $X_{k-1|N}$ given by
\begin{align*}
w_k \approx X_k-f(X_{k-1|N})-F_{k-1|N} \tilde X_{k-1|N}
\end{align*}
where $F_{k-1|N}=\frac{\partial{f}}{\partial{X}}_{|X=X_{k-1|N}}$ and $\tilde X_{k-1|N}=X_{k-1}-X_{k-1|N}$.
\begin{align}
\nonumber w_kw_k^T&=X_kX_k^T-X_kf^T(X_{k-1|N})-f(X_{k-1|N})X_k^T+f(X_{k-1|N})f^T(X_{k-1|N})\\
\nonumber &-X_k\tilde X_{k-1|N}^T F_{k-1|N}^T+f(X_{k-1|N})\tilde X_{k-1|N}^TF_{k-1|N}^T-F_{k-1|N}\tilde X_{k-1|N}X_k^T\\
\label{e11} &+F_{k-1|N}\tilde X_{k-1|N}f(X_{k-1|N})^T+F_{k-1|N}\tilde X_{k-1|N}\tilde X_{k-1|N}^T F_{k-1|N}^T
\end{align}
The following results are used in calculating the conditional expectation in Eq-\ref{max2}
\begin{align}
\left.\begin{array}{r@{\;}l}
E[X_kX_k^T|Z_N]&=X_{k|N}X_{k|N}^T+P_{k|N}\\
E[X_kX_{k-1}^T|Z_N]&=X_{k|N}X_{k-1|N}^T+P_{k,k-1|N}
\end{array}\right\}\label{r1}
\end{align}
The lag-one covariance, $P_{k,k-1|N}$ for k = N-1, N-2,\ldots 1 is given by
\begin{align*}
& P_{k,k-1|N}=E[(X_k-X_{k|N})(X_{k-1}-X_{k-1|N})^T]\\
& P_{N,N-1|N}=(I-K_NH_N)F_{N-1}P_{N-1|N-1}\\
& P_{k,k-1|N}=P_{k|k}K_{k-1|N}^T+K_{k|N}(P_{k+1,k|N}-F_{k}P_{k|k})K_{k-1|N}^T
\end{align*}
where $K_{k|N}$ is the smoothed gain at discrete time `k' obtained from the RTS smoothing algorithm-\ref{RTS}. Using Eq-\ref{e11} and Eq-\ref{r1} we get the conditional expectation as
\begin{align*}
&E\left[w_kw_k^T|Z_N\right]=X_{k|N}X_{k|N}^T+P_{k|N}+X_{k|N}f^T(X_{k-1|N})-f(X_{k-1|N})X_{k|N}^T\\
\nonumber &+f(X_{k-1|N})f^T(X_{k-1|N})-P_{k,k-1|N}F_{k-1|N}^T-P_{k,k-1|N}^TF_{k-1|N}+F_{k-1|N}P_{k-1|k-1}F_{k-1|N}^T
\end{align*}
Rearranging the above terms and using Eq-\ref{max2}, we get
\begin{footnotesize}
\begin{align}
 \label{Q1}   \textbf{Q}=\frac{1}{N}\sum_{k=1}^N &\{  w1_{k|N}   w1_{k|N}^T+P_{k|N}+F_{k-1|N}P_{k-1|N}F_{k-1|N}^T-P_{k,k-1|N}F_{k-1|N}^T- P_{k,k-1|N}^TF_{k-1|N}\}
\end{align}
\end{footnotesize}
where $  w1_{k|N}=X_{k|N}-f(X_{k-1|N})$

\subsection{The Present DSDT Method for Estimating \textbf{Q}}
\label{DSDT}
\par We now estimate \textbf{Q} in a different way using the difference between the stochastic and dynamical trajectory (DSDT method) by following the extended EM method. The stochastic trajectory with the process noise can be approximated using the first order Taylor series expansion around a nominal point ($X_n$) as
\begin{align}
\label{e30} X_k&=f(X_{n_{k-1}})+f'(X_{n_{k-1}})(X_{k-1}-X_{n_{k-1}})+w_k
\end{align}
where $f'$ represents partial differentiation operation on f and thus $f'=\frac{\partial f}{\partial X}$. Consider the \\ dynamical trajectory ($Xd$) without the process noise defined as
\begin{align}
\nonumber Xd_{k}&=f(Xd_{k-1})\\
\label{e31} Xd_k&=f(X_{n_{k-1}})+f'(X_{n_{k-1}})(Xd_{k-1}-X_{n_{k-1}})
\end{align}
where and $Xd_0=X_0$. It is assumed that the nominal point ($X_n$) of both the above trajectories are close to the estimated dynamical trajectory ($X_{n_k}\approx Xd_{k|N}$) where $Xd_{k|N}=f(Xd_{k-1|N})$ and $Xd_{0|N}=X_{0|N}$. Subtracting Eq-\ref{e31} from Eq-\ref{e30} we get
\begin{align}
\nonumber X_k-Xd_{k}=&f'(Xd_{k-1|N})(X_{k-1}-Xd_{k-1|N}-Xd_{k-1}+Xd_{k-1|N})+w_k\\
\label{e20} w_k=&X_k-Xd_k-Fd_{k-1|N}(X_{k-1}-Xd_{k-1})
\end{align}
where the dynamical state Jacobian, $Fd_{k-1|N}=\frac{\partial f}{\partial X}_{|X=Xd_{k-1|N}}$.
\begin{align}
\nonumber w_kw_k^T&=X_kX_k^T-X_kXd_k^T-X_kX_{k-1}^TFd_{k-1}^T+X_kXd_{k-1}^TFd_{k-1}^T-Xd_kX_k^T+Xd_kXd_k^T\\
\nonumber &+Xd_kX_{k-1}^TFd_{k-1}^T-Xd_kXd_{k-1}^TFd_{k-1}^T-Fd_{k-1|N}X_{k-1}X_{k}^T+Fd_{k-1|N}X_{k-1}Xd_{k}^T\\
\nonumber &+Fd_{k-1|N}X_{k-1}X_{k-1}^TFd_{k-1}^T-Fd_{k-1|N}X_{k-1}Xd_{k}^TFd_{k-1}^T+Fd_{k-1|N}Xd_{k-1}X_{k}^T\\
\nonumber &-Fd_{k-1|N}Xd_{k-1}Xd_{k}^T-Fd_{k-1|N}Xd_{k-1}X_{k-1}^TF_{k-1|N}^T+Fd_{k-1|N}Xd_{k-1}Xd_{k}^TFd_{k-1}^T
\end{align}
The additional results apart from Eq-\ref{r1} used in calculating the conditional expectation in Eq-\ref{max2} are
\begin{align}
\left.\begin{array}{r@{\;}l}
E[X_kXd_k^T|{Z_N}]&=E[X_k|Z_N]E[Xd_k^T|Z_N]=X_{k|N}Xd_{k|N}\\
E[Xd_kXd_k^T|{Z_N}]&=Xd_{k|N}Xd_{k|N}^T+Pd_{k|N}\\
E[Xd_kXd_{k-1}^T|{Z_N}]&=Xd_{k|N}Xd_{k-1|N}^T+Pd_{k,k-1|N}
\end{array}\right\}\label{r2}
\end{align}
where $Xd_{k|N}=f(Xd_{k-1|N})$ is the predicted dynamical state trajectory without the measurement and process noise using the estimated parameter, $\Theta_{N|N}$. Using Eq-\ref{r1} and Eq-\ref{r2} we get
\begin{align}
\nonumber E\left[w_kw_k^T|Z_N\right]=&  w2_{k|N}  w2_{k|N}^T+P_{k|N}+Fd_{k-1|N}P_{k-1|N}Fd_{k-1|N}-P_{k,k-1|N}Fd_{k-1|N}^T\\
\nonumber &-Fd_{k-1|N}P_{k,k-1|N}^T+Pd_{k|N}+Fd_{k-1|N}Pd_{k-1|N}Fd_{k-1|N}\\
\label{e33}&-Pd_{k,k-1|N}Fd_{k-1|N}^T-Fd_{k-1|N}Pd_{k,k-1|N}^T
\end{align}
where $w2_{k|N}=X_{k|N}-Xd_{k|N}-Fd_{k-1|N}(X_{k-1|N}-Xd_{k-1|N})$. Now consider the following term,
\begin{align}
\nonumber &Xd_k-Xd_{k|N}=f(Xd_{k-1})-f(Xd_{k-1|N})\\
\nonumber &\approx f(Xd_{k-1|N})+Fd_{k-1|N}(Xd_{k-1}-Xd_{k-1|N})-f(Xd_{k-1|N})\\
\label{D1}&\approx Fd_{k-1|N}(Xd_{k-1}-Xd_{k-1|N})
\end{align}
Using Eq-\ref{D1}, we get the covariance of the dynamical trajectory as
\begin{align*}
Pd_{k|N}=&E[(Xd_k-Xd_{k|N})(Xd_k-Xd_{k|N})^T]\\
=&Fd_{k-1|N}Pd_{k-1|N}Fd_{k-1|N}^T
\end{align*}
where $Pd_{0|N}=P_{0|N}$ since $Xd_{0|N}=X_{0|N}$. The lag one covariance of the dynamical trajectory is given by
\begin{align*}
Pd_{k,k-1|N}=&E[(Xd_k-Xd_{k|N})(Xd_{k-1}-Xd_{k-1|N})^T]\\
=&Fd_{k-1|N}Pd_{k-1|N}
\end{align*}
Substituting the value of $Pd_{k,k-1|N}$ and $Pd_{k|N}$ in Eq-\ref{e33} and using Eq-\ref{max2} we get
\begin{footnotesize}
\begin{align}
\label{Q2}   \textbf{Q}=&\frac{1}{N}\sum_{k=1}^N \{  w2_{k|N}  w2_{k|N}^T+P_{k|N}+Fd_{k-1|N}P_{k-1|N}Fd_{k-1|N}^T-P_{k,k-1|N}Fd_{k-1|N}^T-P_{k,k-1|N}^TFd_{k-1|N}\}
\end{align}
\end{footnotesize}
If \textbf{Q} = 0 then X = $Xd$ and assuming that $P_{0|N}\approx$ 0, \textbf{R} can be estimated as
\begin{align}
\label{R3} {\textbf{R}}\approx \frac{1}{N}\sum_{k=1}^{N}{\left\lbrace (Z_k-h(Xd_{k|N}))(Z_k-h(Xd_{k|N}))^T\right\rbrace}
\end{align}

\subsection{Choice of \textbf{R}}
\label{sum1}
The choice of \textbf{R} for the next filter pass can utilize one appropriate among the many that are possible. Bavdekar et al. \cite{Bavdekar2011} (2011) used the smoothed statistic $Z_k-h(X_{k|N})$ for the \textbf{R} estimation using extended EM method given by,
\begin{align*}
\textbf{R}=&\frac{1}{N}\sum_{k=1}^{N}\left\lbrace(Z_k-h(X_{k|N}))(Z_k-h(X_{k|N}))^T+H_{k|N}P_{k|N} H_{k|N}^T\right\rbrace
\end{align*}
We can also use the dynamical trajectory to estimate \textbf{R} exclusively for \textbf{Q}=0 case using the sample covariance of dynamical residue based on Eq-\ref{R3}. The choice of Mohamed and Schwarz (MS) for the \textbf{R} estimation is filtered residue and is given by,
\begin{align}
\label{R4} \textbf{R}&=\frac{1}{N}\sum_{k=1}^{N}{\left\lbrace (Z_k-h(X_{k|k}))(Z_k-h(X_{k|k}))^T+H_{k|k}{P}_{k|k}H_{k|k}^T\right\rbrace}
\end{align}
The choice of Myers and Tapley (MT) for the \textbf{R} estimation is innovation given by,
\begin{align}
\label{R2} {\textbf{R}}&=\frac{1}{N}\sum_{k=1}^{N}{\left\lbrace (Z_k-h(X_{k|k-1}))(Z_k-h(X_{k|k-1}))^T-H_k{P}_{k|k-1}H_k^T\right\rbrace}
\end{align}

Thus the above three equations use respectively smoothed, after update, and before updated states, the measurement and their corresponding covariances. All the measurement noise statistics innovations, filtered residue, smoothed residue and dynamical residue are assumed to be zero mean. We note that the smoothed residue is the best statistic for \textbf{R} estimation for both \textbf{Q} = 0 and \textbf{Q} $>$ 0 case.
\subsection{Choice of \textbf{Q}}
\label{sum2}
The choice of \textbf{Q} for the next filter pass can utilize one appropriate among the many that are possible. Bavdekar et al. \cite{Bavdekar2011}(2011) used the smoothed statistic $X_{k|N}-f(X_{k-1|N})$ for the \textbf{Q} estimation using extended EM method given by,
\begin{align*}
 \textbf{Q}=&\frac{1}{N}\sum_{k=1}^N \left\lbrace  w1_{k|N}   w1_{k|N}^T+P_{k|N}+F_{k-1|N}P_{k-1|N}F_{k-1|N}^T-P_{k,k-1|N}F_{k-1|N}^T-P_{k,k-1|N}^TF_{k-1|N}\right\rbrace
\end{align*}
where $ w1_{k|N}=X_{k|N}-f(X_{k-1|N})$. The present work introduced the DSDT statistic for \textbf{Q} (section-\ref{DSDT}) and is given by
\begin{align*}
\textbf{Q}=&\frac{1}{N}\sum_{k=1}^N \left\lbrace  w2_{k|N}   w2_{k|N}^T+P_{k|N}+Fd_{k-1|N}P_{k-1|N}Fd_{k-1|N}^T-P_{k,k-1|N}Fd_{k-1|N}^T-P_{k,k-1|N}^TFd_{k-1|N}\right\rbrace
\end{align*}
where $w2_{k|N}=X_{k|N}-Xd_{k|N}-Fd_{k-1|N}(X_{k-1|N}-Xd_{k-1|N})$. Mohamed and Schwarz (MS) used innovations and  gain for estimating \textbf{Q} given by,
\begin{align}
\label{Q4} \textbf{Q}&=K_N\left\lbrace \frac{1}{N}\sum_{k=1}^N (Z_k-h(X_{k|k-1}))(Z_k-h(X_{k|k-1}))^T\right\rbrace K_N^T
\end{align}
The choice of Myers and Tapley (MT) for \textbf{Q} is $w3_{k|k}=X_{k|k}-X_{k|k-1}$ and is given by,
\begin{align}
\label{Q3} {\textbf{Q}}&=\frac{1}{N}\sum_{k=1}^{N}{\left\lbrace   w3_{k|k}  w3_{k|k}^T-\left(F_{k-1}{P}_{k|k-1}F_{k-1}^T-P_{k|k}\right)\right\rbrace}
\end{align}
All the process noise statistics, $w1_{k|N},w2_{k|N}$ and $w3_{k|k}$ are assumed to be zero mean. We note that the smoothed statistics $w1_{k|N}$ and $w2_{k|N}$ give identical results and are the best for the \textbf{Q} estimation.

\section{Adaptive Tuning Algorithm and the Reference Recursive Recipe}

The different methods and options for filter tuning forms a part of sensitivity study on different simulated case studies which are discussed below,

\begin{itemize}
\item $\mathbf{P_0}$ can be estimated by Scale up, Inverse of Information Matrix = IIM  or by Smoothing which will have to be scaled up and all of which has to be further modified as in the first option for obtaining proper results.
\item Options for $\mathbf{P_0}$ which can be split as cov ([State-S;Parameter-P]) are, 1 - [0,0;0,\checkmark], 2 -Diagonal matrix, 3 - Full matrix. The checkmark (\checkmark) represent a non-zero value at the indicated position. The `cov (.)' represents covariance matrix.
\item The process noise \textbf{Q} can be estimated by EM, DSDT, MT or MS method.
\item Options for \textbf{Q} = cov ([S;P]) are, 1 - [\checkmark,0;0,0], 2 - Diagonal matrix, 3 - Full matrix.
\item The measurement noise \textbf{R} can be estimated by EM, MT, MS or using Eq - \ref{R3}.
\end{itemize}

The following steps explain the recursive or iterative algorithm for tuning the EKF.

\begin{enumerate}

\item Given the system model and the measurements the first iteration of EKF is carried out with guess values of $\mathbf{X_0}$, $\mathbf{P_0}$, $\Theta$, \textbf{R} and \textbf{Q}.

\item Run the extended RTS smoother using the filtered data to get the smoothed state estimate $X_{k|N}$ and the corresponding smoothed covariance $P_{k|N}$.

\item The $\mathbf{P_0}$ can be estimated by Scale up (Eq - \ref{P2}), IIM (Eq - \ref{P1}) or by smoothing ($P_{0|N}$) which will have to be scaled up and modified using the the first option for obtaining proper results.

\item The measurement and process noise covariance can be estimated by any of the method and options as discussed in section-\ref{sum1} and \ref{sum2}.

\item EKF is run using the estimates of $\mathbf{X_0}$, $\mathbf{P_0}$, $\Theta$, \textbf{Q} and \textbf{R} in the next few iterations until statistical equilibrium is reached.

\item Many simulation runs (say 50) are carried out by repeating the above steps using different injected measurement ($v$) and process noise ($w$) sequences.

\item Different cost functions (\textbf{J1} to \textbf{J8}) are checked for convergence.

\end{enumerate}

For the \textbf{Q} = 0 case the value of \textbf{Q} is set at $10^{-10}$ or lower for all iterations to help the filter that would otherwise generate a pseudo \textbf{Q} and then slowly grind it to zero in hundreds of iterations. For the \textbf{Q} $>$ 0 case if any of the states is known to have \textbf{Q} = 0 then it can be set at $10^{-10}$ or lower. For \textbf{Q} = 0 case one can even estimate \textbf{R} by ignoring the second order terms (assuming P$\rightarrow$ 0). It is of interest to note that for Q $>$ 0 case unless the second order terms of the filter output covariance terms are also included in (i) the estimate for \textbf{R }and \textbf{Q} using the EM option and (ii) the estimate for \textbf{R} using the EM together with \textbf{Q} using the DSDT option the estimates for \textbf{R} and \textbf{Q} do not converge to the proper value. A comparative study is conducted among the different filter tuning methods and the best adaptive recipe is proposed as a reference method as shown below

\begin{table}[h]
\begin{center}
\caption*{\textbf{Reference Recursive Recipe}}{}
\begin{tabular}{| c | c |  }
\hline
\textbf{Q} = 0 &  \textbf{Q} $>$ 0 \\ \hline
\makecell{ $\mathbf{X_0}$ : Given or $X_{0|N}$  \\ $\Theta$ : $\Theta_{N|N}$  \\ $\mathbf{P_0}$ : Scaled up-[0,0;0,\checkmark]\\\textbf{Q} : $10^{-10}$-[\checkmark,0;0,0] \\ \textbf{R} : EM-diag} &
\makecell{$\mathbf{X_0}$ : Given or $X_{0|N}$ \\ $\Theta$ : $\Theta_{N|N}$  \\ $\mathbf{P_0}$ : Scaled up-[0,0;0,\checkmark]\\\textbf{Q} : EM/DSDT-[\checkmark,0;0,0] \\ \textbf{R} : EM-diag} \\ \hline
\end{tabular}
\end{center}
\end{table}

\textbf{Note :} The $\mathbf{X_0}$ in all cases is either given or obtained by the smoother. The smoothed initial estimate which can be split as $X_{0|N}$ =  $(x_{0|N},\Theta_{0|N})^T$ including both state and parameter. Hence the unknown parameters are not shown explicitly as $\Theta$. The estimated parameter is taken as $\Theta_{N|N}$ obtained from $X_{N|N}=[x_{N|N},\Theta_{N|N}]$ with covariance $P_\Theta$ obtained at the end of the final filter pass over the data, $P_{N|N}$ = $[ P_{xx}, P_{x\Theta}; P_{\Theta x}, P_\Theta]$.

\newpage
\begin{table}[h]
\caption{The Triplets Occurring in the Discussion on Kalman Filter}
\label{tbcape2}
\begin{tiny}
\begin{tabular}{c }
\end{tabular}
\end{tiny}
\end{table}

\includegraphics[width=6in,height=8in]{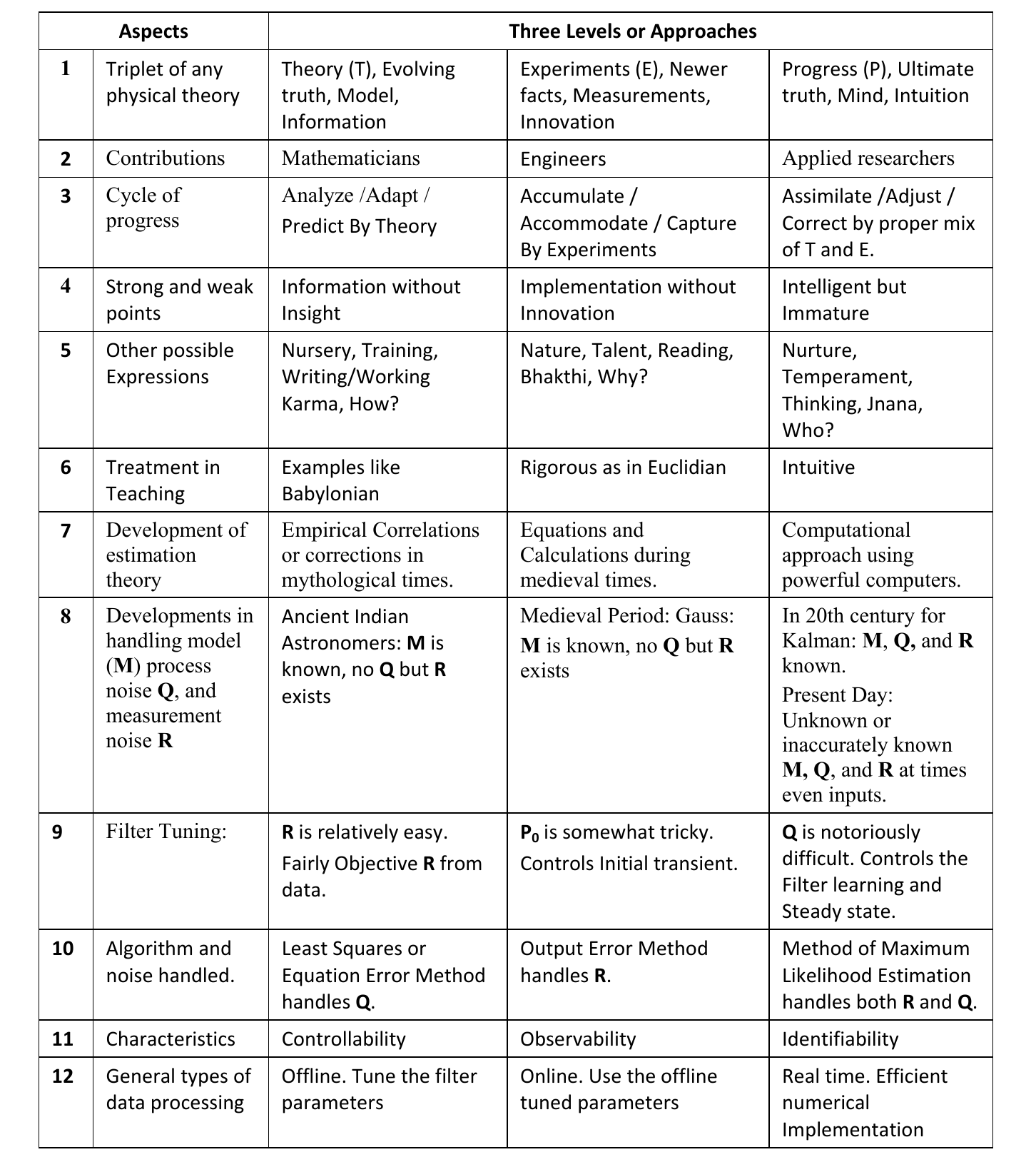}

\afterpage{\null\newpage}
\chapter{Simulated Case Studies}
\label{ch4}
\par In this Chapter we carry out simulation studies commencing from the simplest case having one or two unknown parameters and noise components and move on to more involved case of dynamical systems having around a dozen or more of unknown parameters and noise components. The dynamical systems are (i) a constant signal, (ii) a ramp, (iii) a spring, mass, damper system with a weak non linear spring constant, (iv) the longitudinal motion of an airplane and (v) the lateral directional motion of an airplane. We study all the above systems by including first only the measurement noise \textbf{R} and later include the process noise \textbf{Q} as well. Except in the simplest of cases the governing dynamical equations are integrated using ode23, ode45, or using the appropriately derived transition matrices in the MATLAB\textsuperscript{\textregistered} all of which provided equally accurate state trajectories. Then for studies by including \textbf{R} and \textbf{Q} using the outputs from the default normal random number generator in MATLAB\textsuperscript{\textregistered} the proper noise samples were added to the appropriate state and measurements. The accuracy of this approach in generating the stochastic trajectories with process noise input was also checked against the Van Loan's \cite{VanLoan1978} (1978) procedure. The effect of higher order filters like Second Order Kalman Filter and Iterated Extended Kalman Filter (Jazwinski \cite{Jazwinski1970} 1970) on non linear complex systems (Longitudinal and Lateral dynamics of Aircraft system) are also studied. The detailed numerical inputs and figures used in the simulation of the various case studies are given in Table-\ref{sysdes}.

\par In all simulated system study, the guess value of $\mathbf{P_0}$ chosen is $10^{-1}$ for all states which is assumed to be a diagonal matrix in the first iteration. The guess value of \textbf{Q} chosen is $10^{-1}$ for all states and zero for the augmented parameters. The guess value of \textbf{R} chosen is $2^{-1}$ for all measurement channels. The initial parameters are chosen to be within $\pm 20 \%$ error. A total of N = 100 measurement data are simulated with the time varying from 0 to 10 seconds in very small steps of $\delta$t = 0.1 s. For zero process noise case, the maximum number of iteration is set to 20 over 50 simulations and for non zero process noise case it is set to 100 over 50 simulations for obtaining generally four digit accuracy (though not necessary) in the results as presented in various tables in the report. In the present reference procedure it was noticed that generally even if the initial state covariance, initial process and measurement noise variances are varied over a wide range of powers from -3 to +3 together with the initial parameter values being set to zero one can reach the same estimation results for a given data. Thus there is stability with respect to far off initial conditions but sensitivity in the estimates with different data. Further detailed sensitivity studies were carried out and the results are provided in the various Tables of this Chapter along with brief remarks which are simple and self explanatory. One comment in most tables as unity means near about it. Such studies show that the present reference recursive recipe leads to a non diverging, and consistent filter performance over many simulations and provides better results when compared to earlier approaches.

\section{Simulated Constant Signal System}

\par Consider a constant signal system for discrete time instants k=1,2,\ldots N is given by
\begin{align*}
x_k=\Theta x_{k-1}+w_k
\end{align*}
where $x_0=10$ and $\Theta_{true}=1$ is considered as an unknown parameter. The measurement equation is
\begin{align*}
Z_k=HX_k+v_k
\end{align*}
where the augmented state, $X_k=[x_k,\Theta_k]$, H=[1 0] is the measurement matrix of size $m\times (n+p)$ and $m=n=p=1$. The numerical values of the noise variances are shown in Table-\ref{sysdes}. All the figures are presented for only one run to prevent cluttering.

\subsection{Details of Sensitivity Study}

\par The system under consideration is first solved using the reference recipe. Later other sensitivity studies were carried out using other possible variations for $\mathbf{P_0}$, \textbf{Q} and \textbf{R}. Such studies lead to the conclusion that the reference recipe is just about the best possible and others may not always help the filter operation without divergence and even if they did may not lead to results as good as the reference recipe. The tabulated results are averaged over 50 simulations which includes

\begin{itemize}
\item $\Theta$ ratio is the ratio of EKF estimated parameter ($\Theta_{N|N}$) to that of $\Theta_{true}$.
\item CRB ratio is the ratio of the square root of $CRB$ estimated by the NR-[\ref{NR}] method to that of the square root of parameter covariance ($P_{\Theta}$) estimated by EKF.
\item PCRB ratio is the ratio of the square root of Posterior CRB-[\ref{PCRB}] to that of the square root of EKF estimated state covariance at the last time instant.
\item Consistency check on the estimated parameters is done using the ratio defined as
\begin{align*}
&\text{Consistency ratio = }\frac {\sigma_{  \Theta}}{\text{SIGMA}_{avg}}\\
&\sigma_{  \Theta}=\left[\frac{1}{50}\sum_{s=1}^{50}{(  \Theta^s-\bar \Theta)^2}\right]^{\frac{1}{2}}\\
&\text{SIGMA}_{avg}=\frac{1}{50}\sum_{s=1}^{50}{\sqrt{P^s_{\Theta}}}
\end{align*}
where `s' is the simulation number, $\bar \Theta$ is the sample mean of the estimated parameters over 50 simulations.
\item Spread factor is a measure of percentage spread seen in the estimates using both first and second order moments which is defined as
\begin{equation*}
\text{Spread factor = }\left[\frac{1}{50}\sum_{s=1}^{50}{\sqrt{(\Theta-  \Theta^s)^2 + P_{  \Theta^s}}}\right]\times \frac{100}{|\Theta|}
\end{equation*}

\item \textbf{R} ratio EKF/True is the ratio of the EKF estimated \textbf{R} to that of the true reference value of \textbf{R}.
\item \textbf{R} ratio EKF/NR is the ratio of the EKF estimated \textbf{R} to that of the \textbf{R} estimated by NR method.
\item \textbf{Q} ratio is the ratio of the EKF estimated \textbf{Q} to that of the true \textbf{Q}.
\item The mean ($\mu$) and standard deviation ($\sigma$) of the cost functions (\textbf{J1-J8}) over many simulations.
\end{itemize}

\subsubsection{\textbf{Q} = 0 CASE}
\begin{enumerate}
\item If the \textbf{Q} is allowed to learn it takes thousands of iterations for it to tend to zero. Hence when studying cases with no process noise it is better to set it to a very small value such as $10^{-10}$.
\item If \textbf{R} is estimated using the EM statistic using smoothed data (i) the $\Theta$ and (ii) the \textbf{R} estimate is good. It is the choice of $\mathbf{P_0}$ that controls the behaviour of the CRB.
\item The reference case with scaled up $\mathbf{P_0}$ with zero state covariance gives close CRB results obtainable from the modified Newton Raphson (NR) method (section-\ref{NR}). If a scaled up or diagonal $\mathbf{P_0}$ which have non zero initial state covariance is used then the filter CRB is higher.  For this simple constant signal case, all other possible choices for \textbf{R} such as (i) dynamical residue, (ii) smoothed residue ignoring second order term, (iii) filtered residue ignoring second order term, (iv) innovation ignoring second order term, (v) MT and (vi) MS all give very good filter CRBs.
\item If the IIM is used instead of scaled up $\mathbf{P_0}$ with zero state covariance it gives good CRBs. If IIM with non zero initial state covariance is used then the filter estimated CRB is higher.
\item If the smoothed $\mathbf{P_0}$ is used instead of scaled up $\mathbf{P_0}$ then in all the above three combinations it leads to low filter estimated CRB. This is because the smoothed $\mathbf{P_0}$ is much lower than the scaled up $\mathbf{P_0}$ by a factor of the number of time points (N).
\item During smoothing the parameter covariances do not change but only the state covariance changes. Hence if the smoothed $\mathbf{P_0}$ is scaled up and the state covariance is set to zero then it becomes equivalent to the reference case.
\item The value of the cost functions \textbf{J1, J2} and \textbf{J3} are quite close to unity as is expected, \textbf{J4} is the trace of \textbf{R} and \textbf{J5} is the negative log likelihood cost function.
\end{enumerate}


\subsubsection{\textbf{Q} $>$ 0 CASE}

\begin{figure}[h]
\includegraphics[width=6in,height=4in]{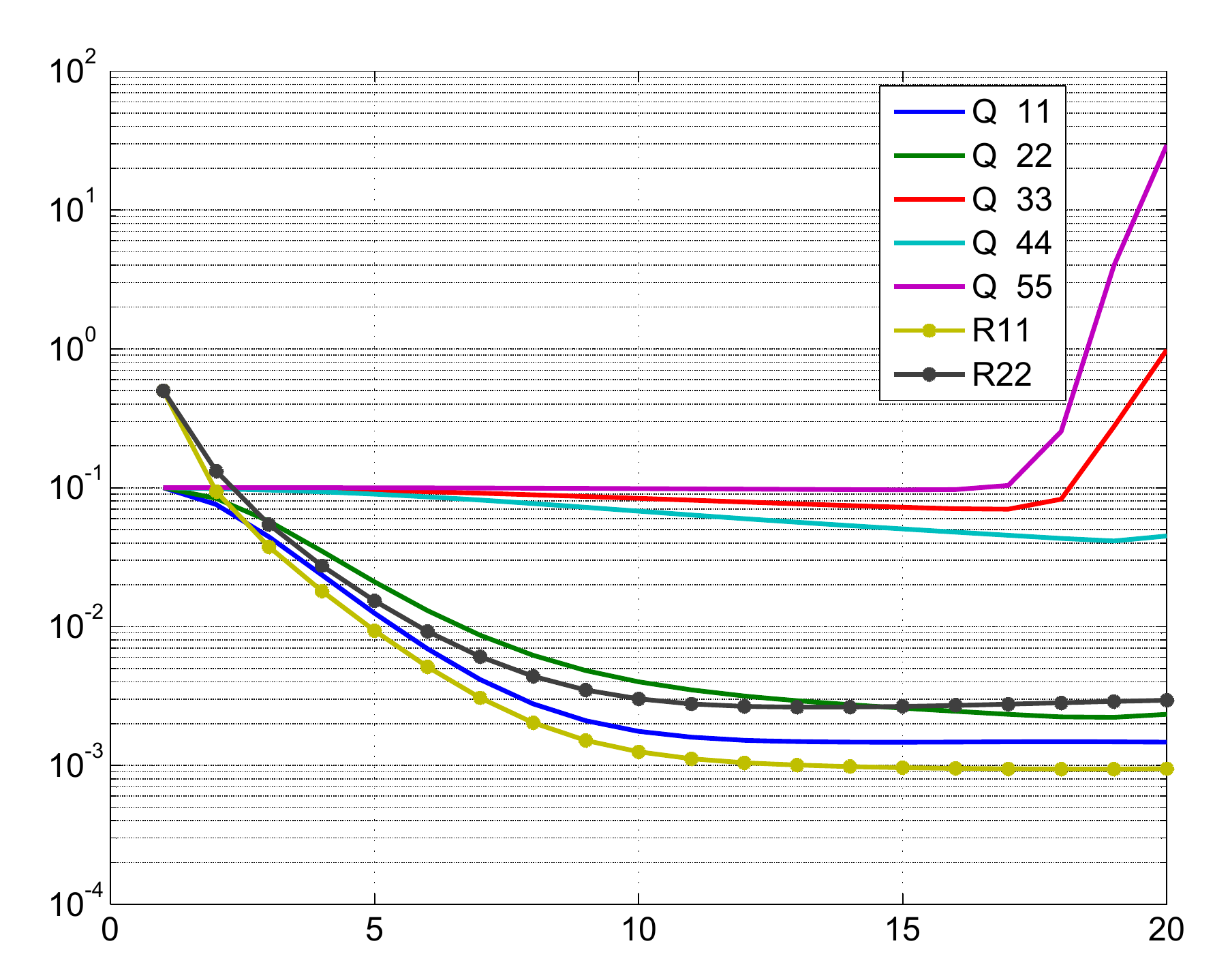}
\caption{Variation of \textbf{Q, R} with iterations using the}
\caption*{procedure suggested by Bavdekar et al. \cite{Bavdekar2011} (2011)}
\label{EM_SMD_Q}
\end{figure}

\begin{figure}[h]
\includegraphics[width=6in,height=4in]{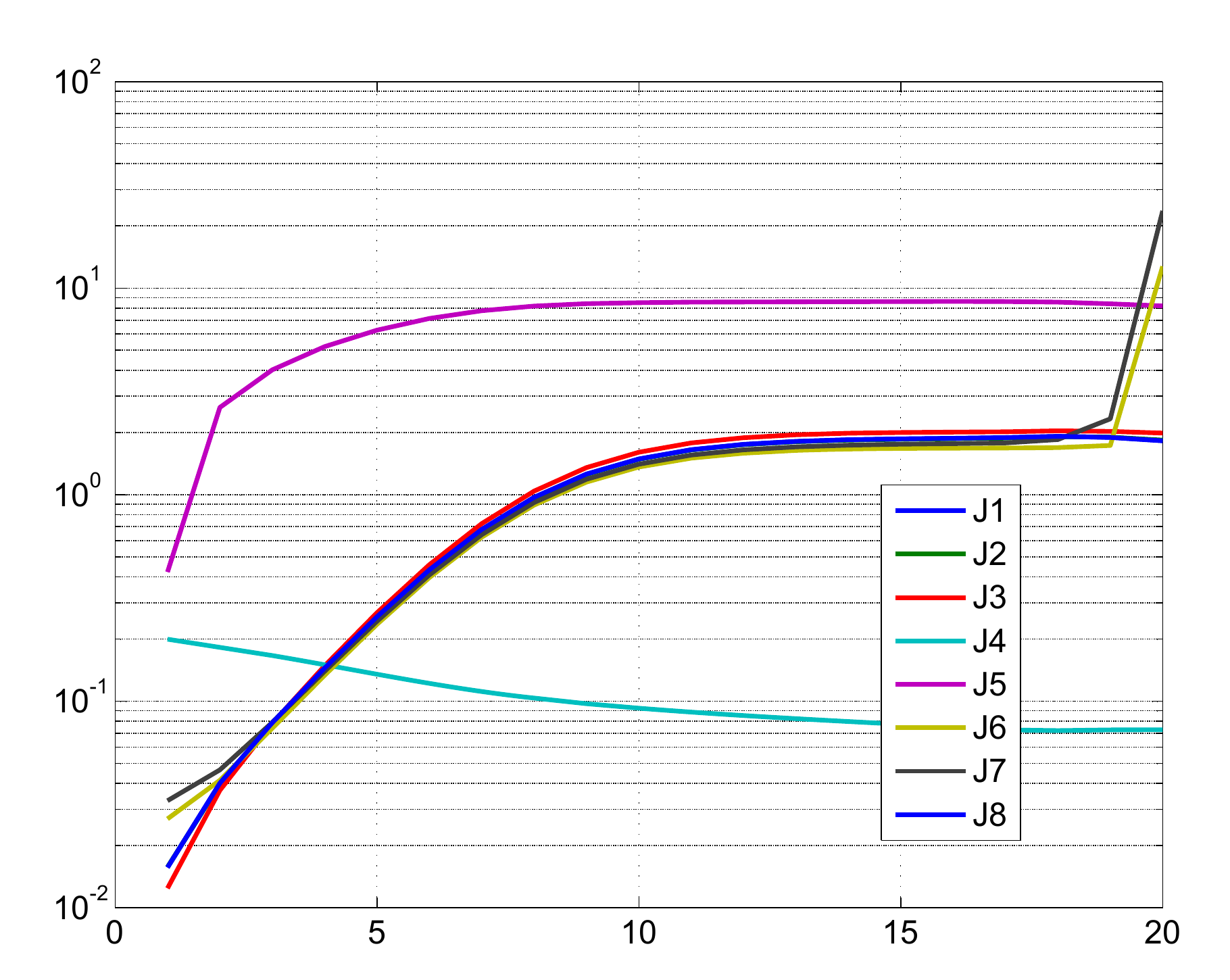}
\caption{Variation of different costs with iterations using the }
\caption*{procedure suggested by Bavdekar et al. \cite{Bavdekar2011} (2011)}
\label{EM_SMD_J}
\end{figure}

\begin{enumerate}
\item Further experiments were carried out for the non zero \textbf{Q} using the reference recursive recipe among many possible studies a few such as (i) \textbf{R} being known, (ii) $\Theta$ known, (ii) \textbf{R} and $\Theta$ and the like as shown in Table-\ref{tbconQ}.
\item The next study is the procedure suggested by Bavdekar et al. \cite{Bavdekar2011} (2011). This had difficuly in providing consistent parameter estimates. The smoothed $\mathbf{P_0}$ used as the parameter covariance tend to decrease with iterations and the consistency ratio becomes greater than one. As the system complexity increases their procedure requires a stopping criterion using the cost function \textbf{J5} which may vary for different simulations. The cause of divergence is the process noise injected into the parameter augmented state equations which increase through the iterations as seen in the Fig. \ref{EM_SMD_Q} and Fig. \ref{EM_SMD_J} referring to the SMD system studied later in the report. It is not desirable to inject process noise into the augmented states corresponding to the parameters in the filter operation. If only diagonal terms are used for $\mathbf{P_0}$, \textbf{Q} and \textbf{R} then no stopping criterion is necessary but again the covariance keep on decreasing through iterations. Next if scaled up $\mathbf{P_0}$ is used and the state covariance is reset to zero then the results come very close to the reference approach. If the DSDT statistics is used for estimating \textbf{Q} then it also provides results very close to the reference case.
\item The next study is by using the MT statistics for \textbf{\textbf{R} and Q}. Simultaneous estimation of \textbf{R, Q} is generally not possible. The diagonal terms can become negative at times which has to be forced to be a positive value. There is no proper convergence even if $\Theta$ is known. If \textbf{R} is assumed to be known then the results using only the later half of the filter outputs are somewhat close to the reference recipe. The procedure of Gemson using the MT statistics for \textbf{Q} and initial \textbf{R} close to the true value led to reasonably acceptable but not as close as the reference results.
\item We next ran the filter using the MS statistics of \textbf{Q} and \textbf{R}. Here there was no systematic filter operation and convergence even if \textbf{R} and/or $\Theta$ is known.  The procedure of Gemson but using the MS statistics for \textbf{Q} and initial \textbf{R} close to the true value led to reasonably acceptable but not as close as the reference results.
\item All the above studies show that even for the case of a constant signal only a proper combination of the choices for $\mathbf{P_0}$, \textbf{Q} and \textbf{R} can lead systematically to a set of good results. The other choices appear to present some difficulties both in the filter operation and providing accurate estimates.
\end{enumerate}

\subsection{Remarks on the Results}

\par
\begin{figure}[h]
\includegraphics[width=6in,height=3.2in]{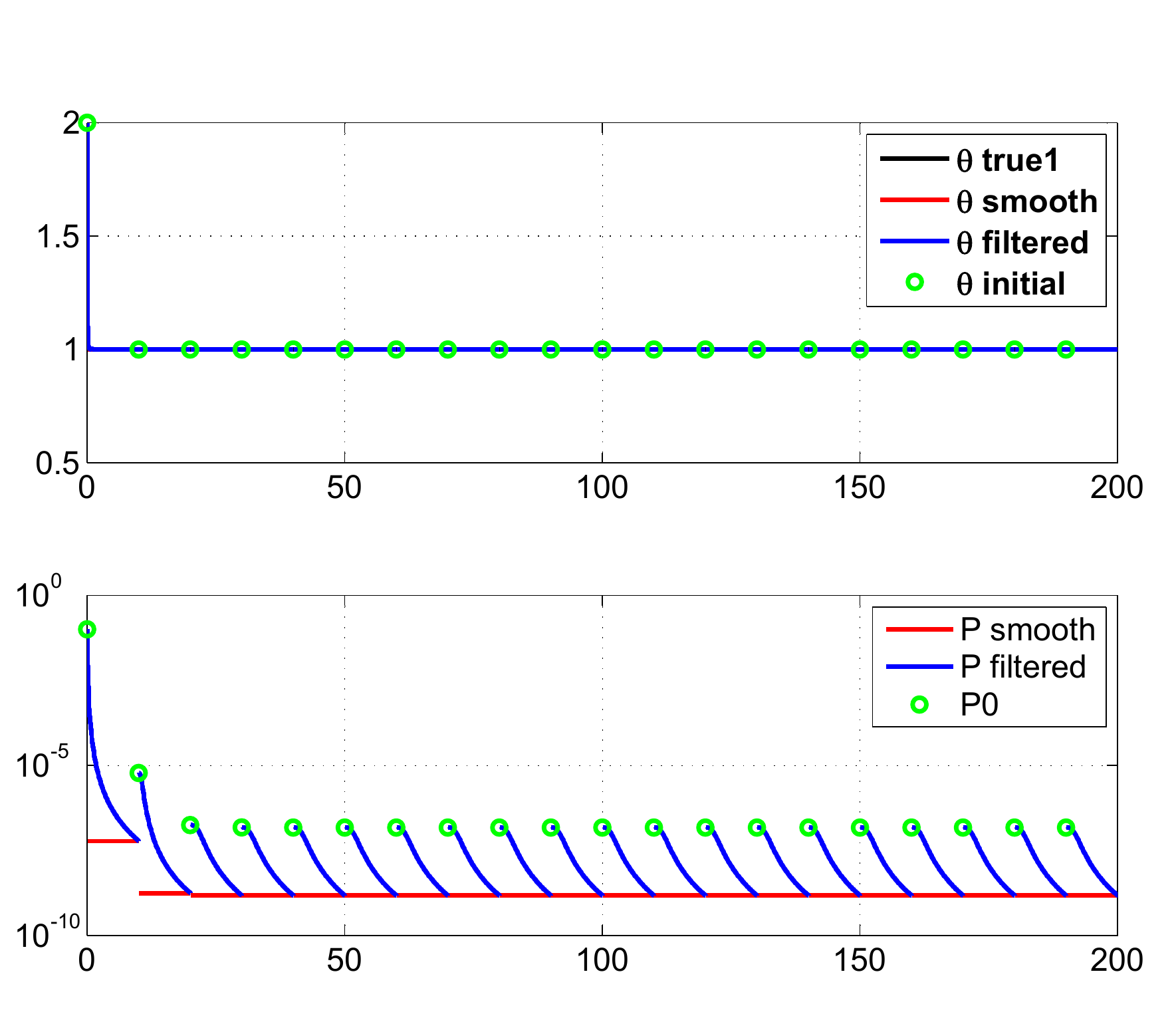}
\caption{The variation of different parameter estimate and their filtered }
\caption*{and smoothed covariances through (with the time cumulatively) the iterations}
\label{con_p1}
\end{figure}

\begin{figure}[h]
\includegraphics[width=6in,height=3.2in]{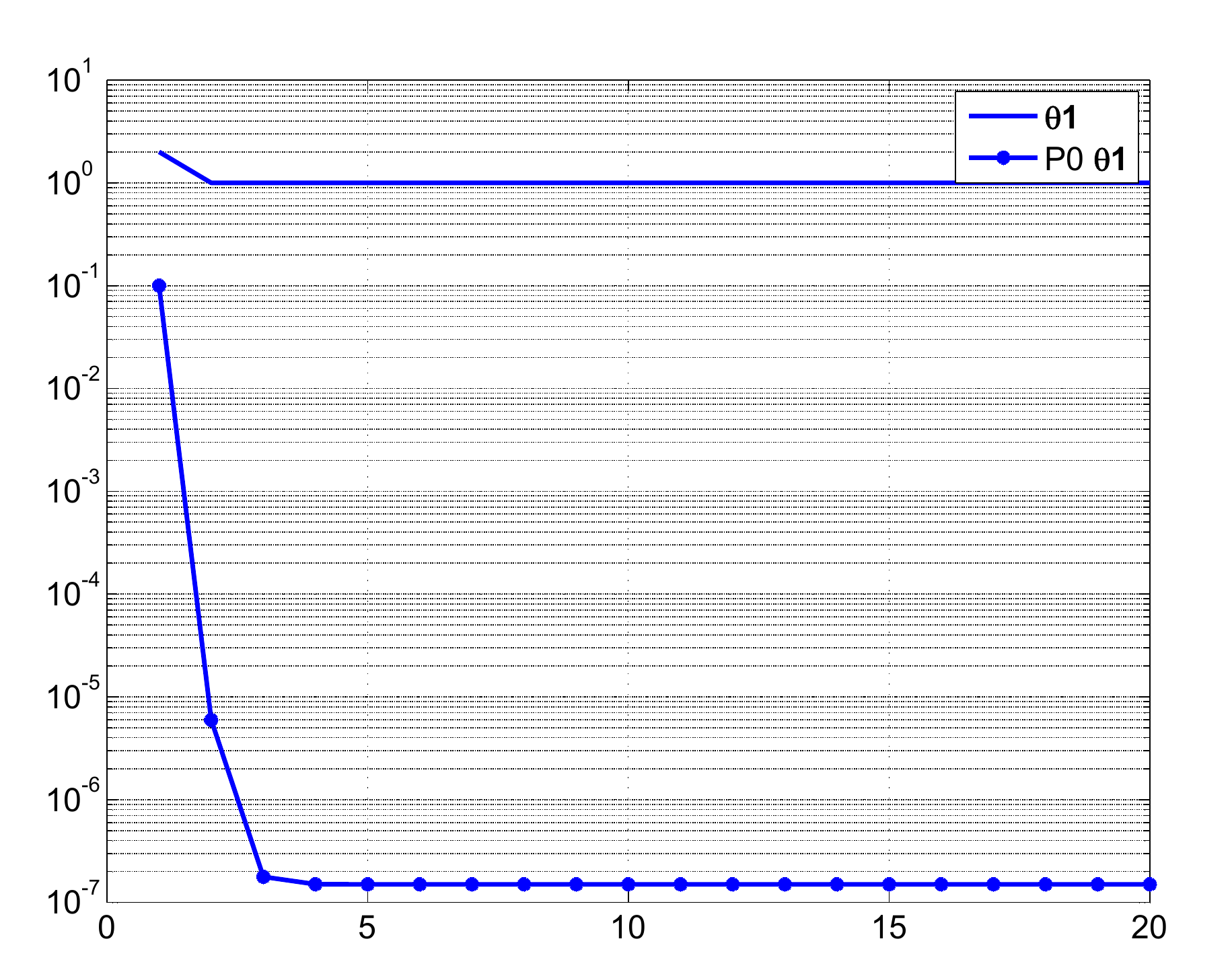}
\caption{Variation of parameter and its initial covariance ($\mathbf{P_0}$) with iterations}
\label{con_P0}
\end{figure}

\begin{figure}[h]
\includegraphics[width=6in,height=4in]{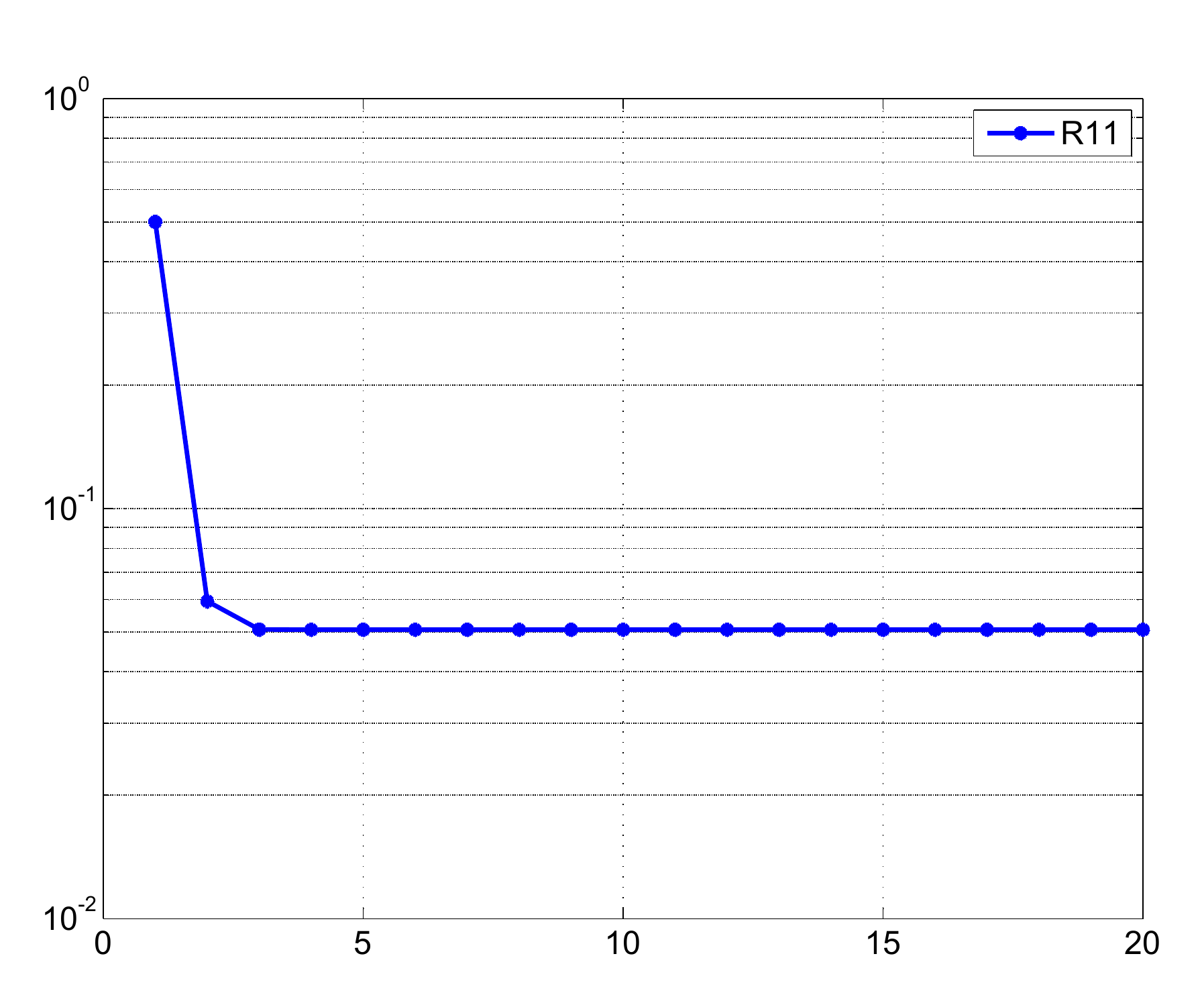}
\caption{Variation of \textbf{R} with iterations}
\label{con_R}
\end{figure}

\begin{figure}[h]
\includegraphics[width=6in,height=4in]{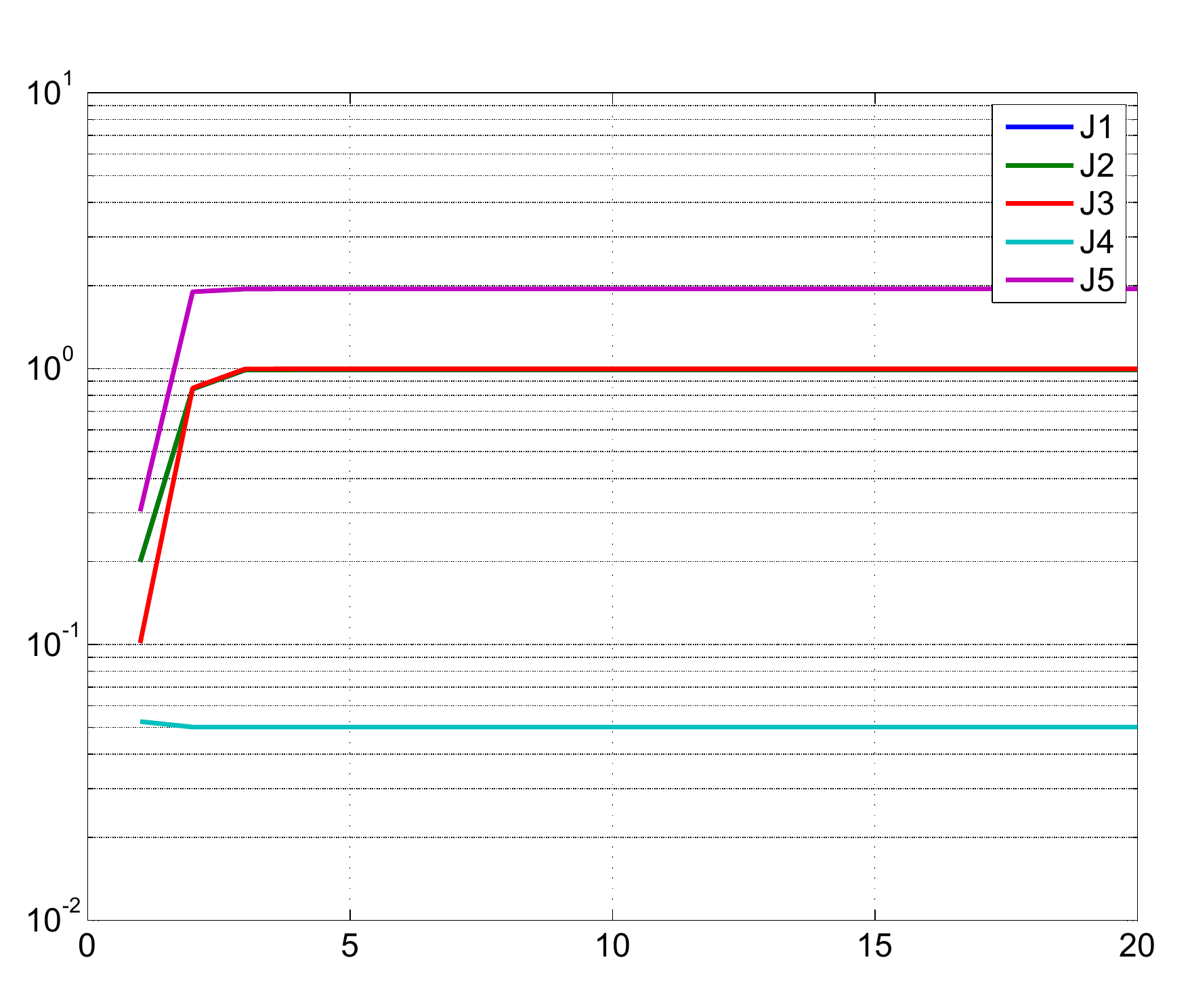}
\caption{Variation of different costs (\textbf{J1-J5}) with iterations}
\label{con_cost}
\end{figure}

\begin{figure}[h]
\includegraphics[width=6in,height=4in]{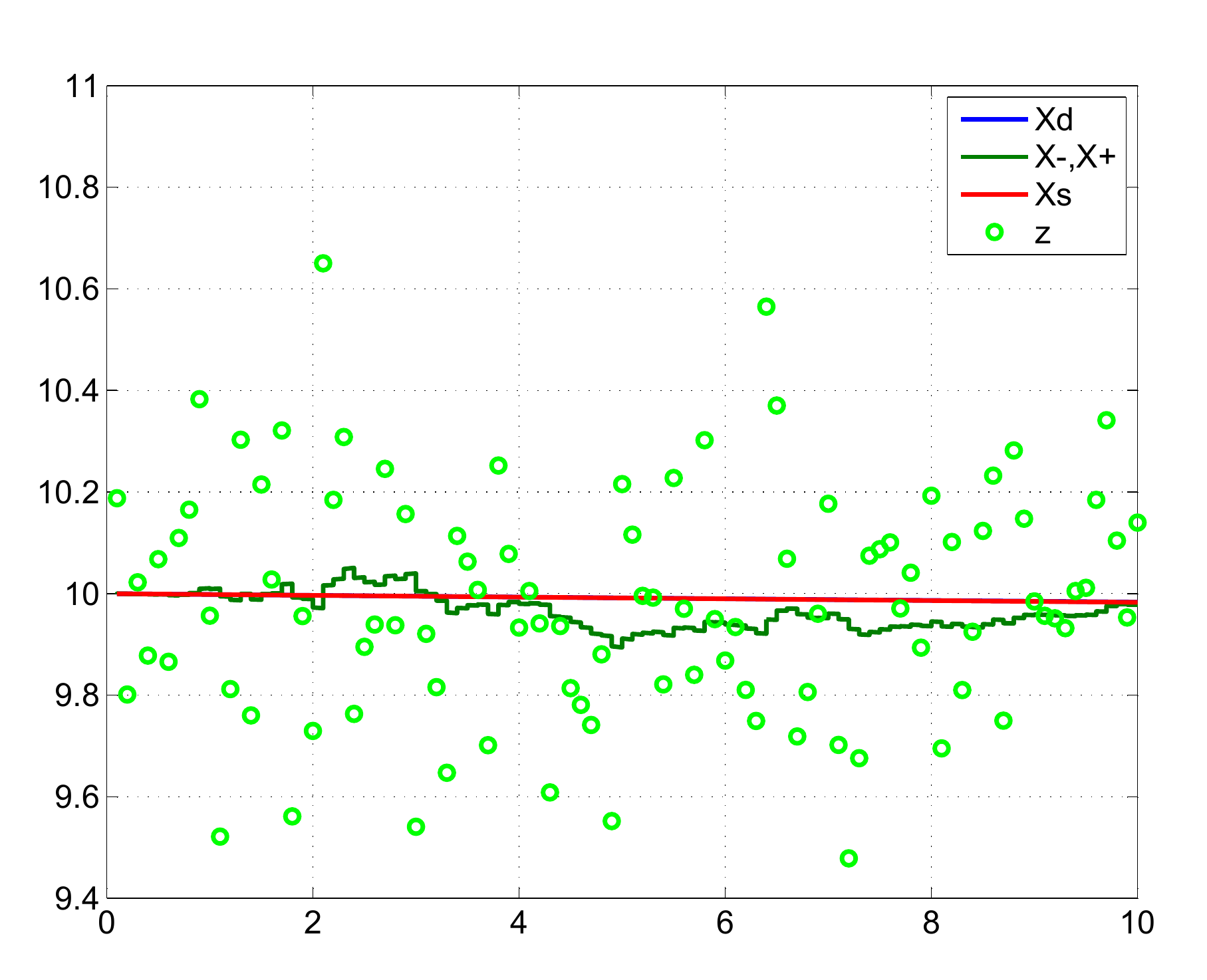}
\caption{Comparison of the predicted dynamics, posterior, smoothed }
\caption*{and the measurement}
\label{con_h1}
\end{figure}

\begin{figure}[h]
\includegraphics[width=6in,height=4in]{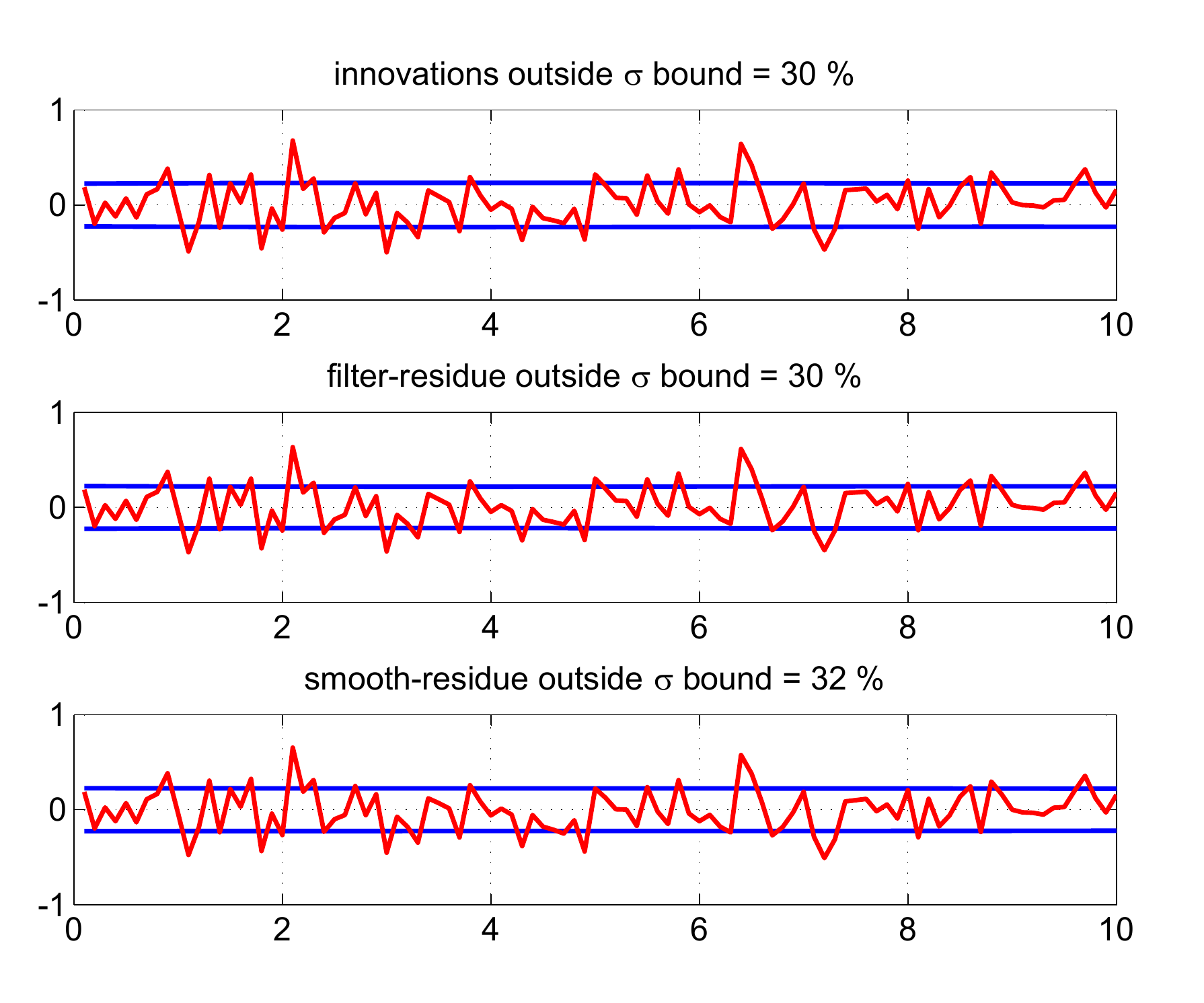}
\caption{The innovations, filtered residue and smoothed residue }
\label{con_innov}
\end{figure}

\begin{figure}[h]
\includegraphics[width=6in,height=4in]{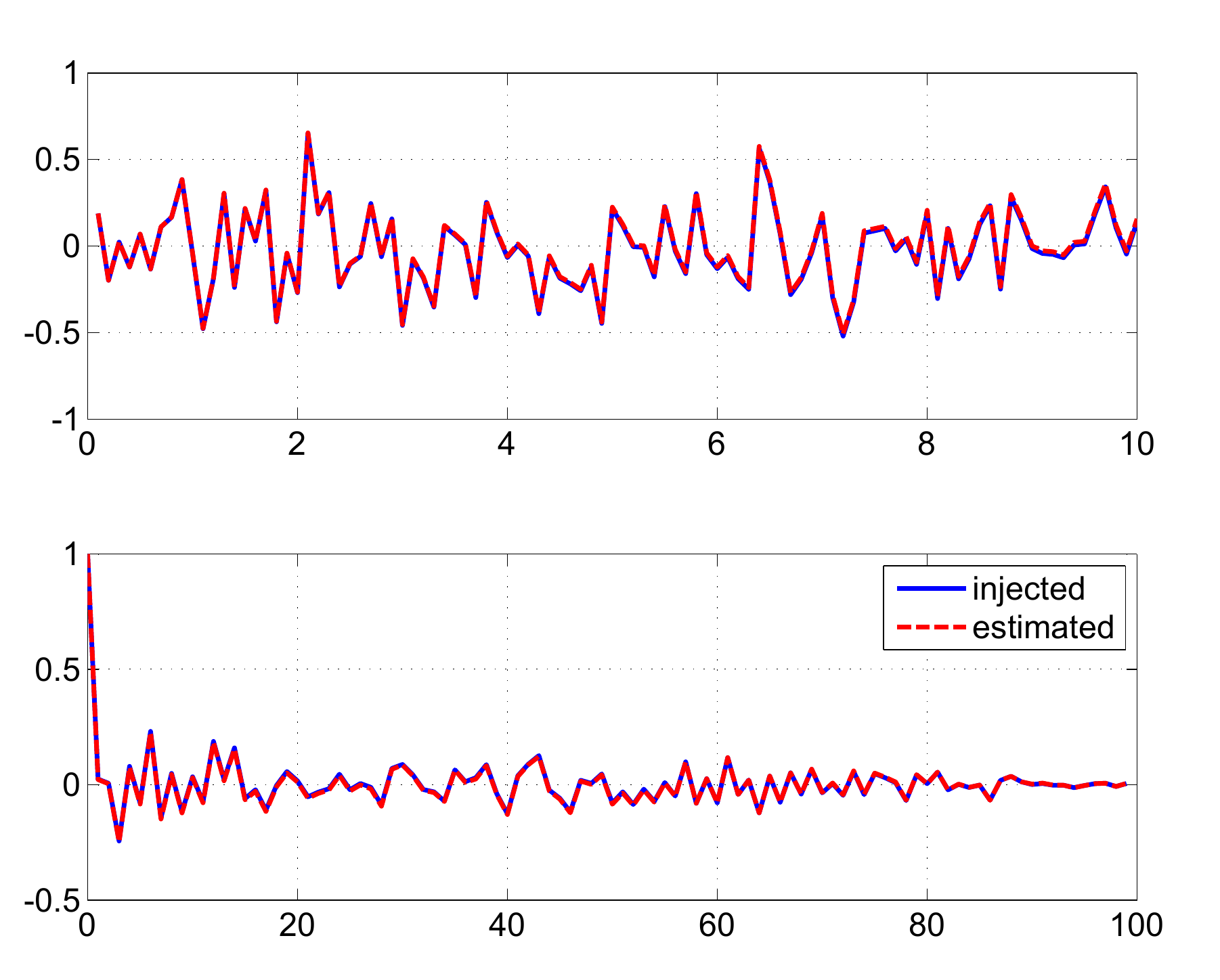}
\caption{Time variation of injected and estimated measurement noise (top) and}
\caption*{their autocorrelation (bottom)}
\label{con_mnoise}
\end{figure}

 We first run the filter assuming \textbf{Q} = 0. It was found that about 20 iterations of the data would suffice. The Fig. \ref{con_p1} shows the various parameter estimates and its corresponding variances through cumulative time instants with iterations. The variation of the estimated initial parameters and their variances through iterations are shown in Fig. \ref{con_P0}. The parameter and the uncertainty reach almost their final estimated values in about 2 and 5 iterations respectively. A similar plot in Fig. \ref{con_R} shows the variation of the estimated measurement noise. The variation of different cost functions (\textbf{J1-J5}) through the iterations is shown in Fig. \ref{con_cost}. The Fig. \ref{con_h1} shows the predicted dynamics, filtered and smoothed estimate at the last iteration. The Fig. \ref{con_innov} show the innovations, filtered residue and smoothed residue together with the square root of their variance ($\pm\sigma$ bound). In the EKF approach most of the quantities are Gaussian or approximated as quasi Gaussian and one would expect all the above quantities are close to being Gaussian and hence around one third of the total sample points to be outside the $\sigma$ bound. The injected and estimated measurement noise distributions during the final iteration shown in Fig. \ref{con_mnoise} indicate that they are close to each other. Their autocorrelations are ideally expected to be close to the Kronecker delta function which provides confidence in the proposed algorithm.

\begin{figure}[h]
\includegraphics[width=6in,height=3.2in]{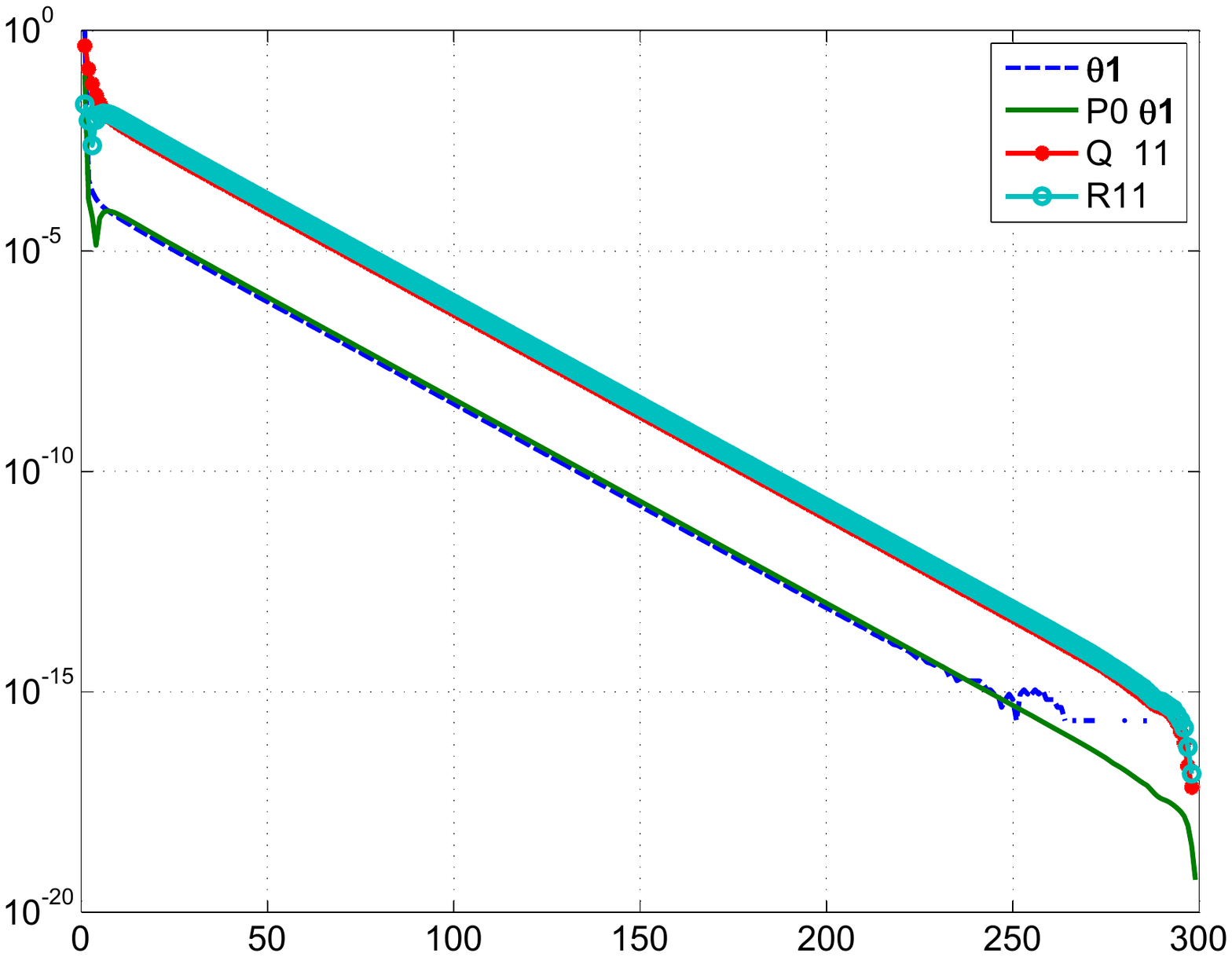}
\caption{The absolute difference between the iterated and final values}
\caption*{with 500 iterations}
\label{con_err}
\end{figure}

\begin{figure}[h]
\includegraphics[width=6in,height=3.2in]{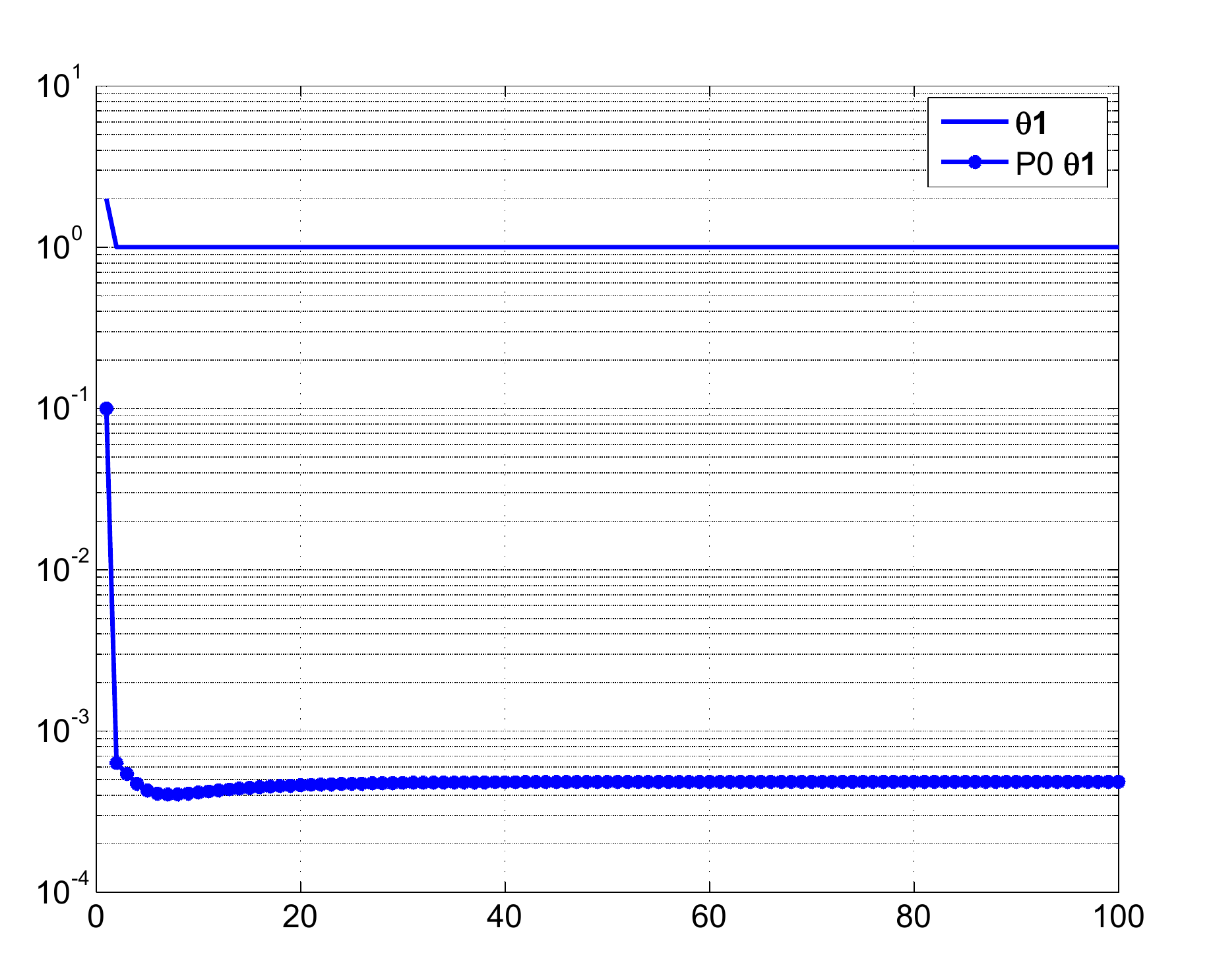}
\caption{Variation of parameter and its initial covariance ($\mathbf{P_0}$) with iterations}
\label{conQ_P0}
\end{figure}

\begin{figure}[h]
\includegraphics[width=6in,height=4in]{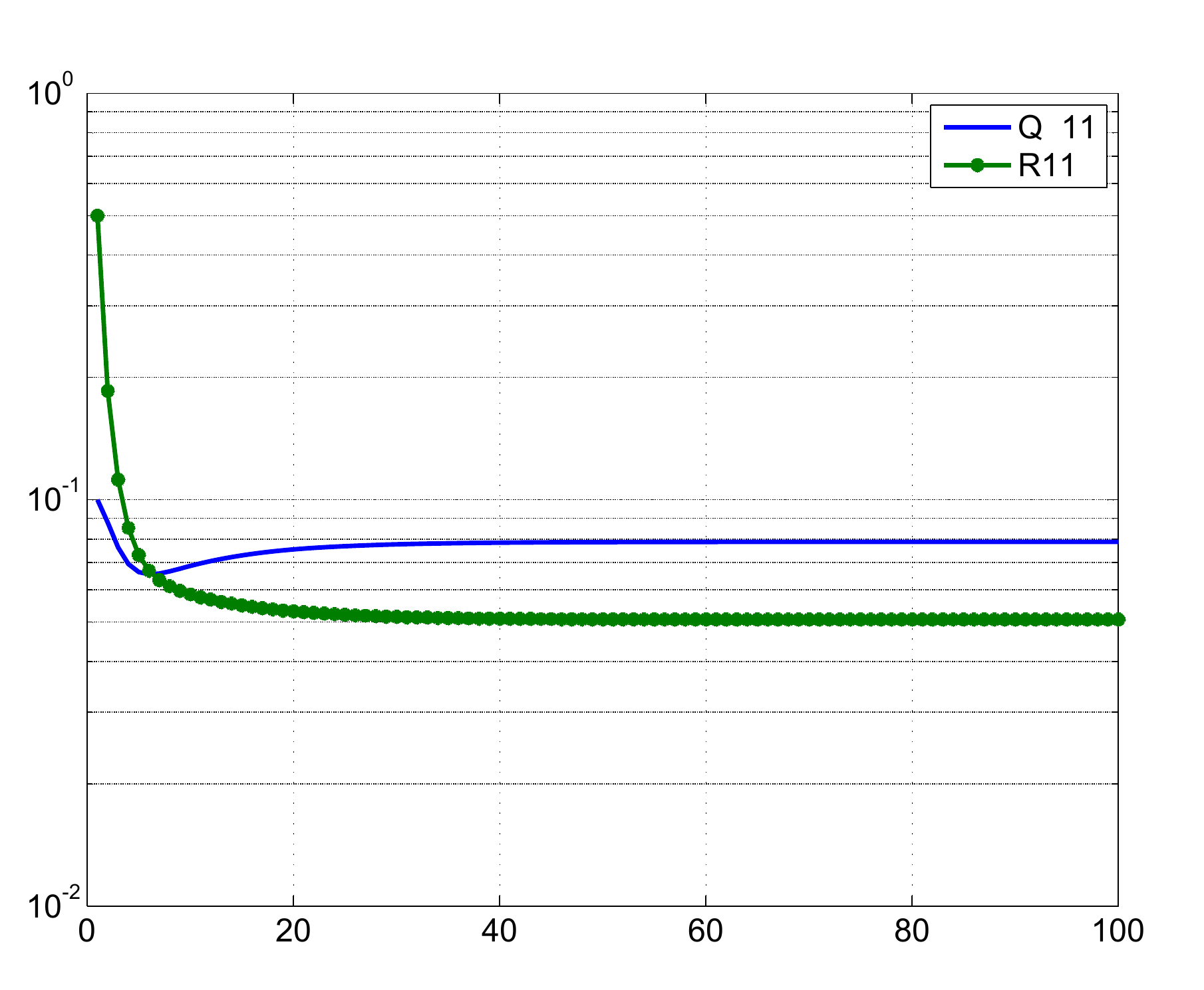}
\caption{Variation of \textbf{Q} and \textbf{R} with iterations}
\label{conQ_R}
\end{figure}

\begin{figure}[h]
\includegraphics[width=6in,height=4in]{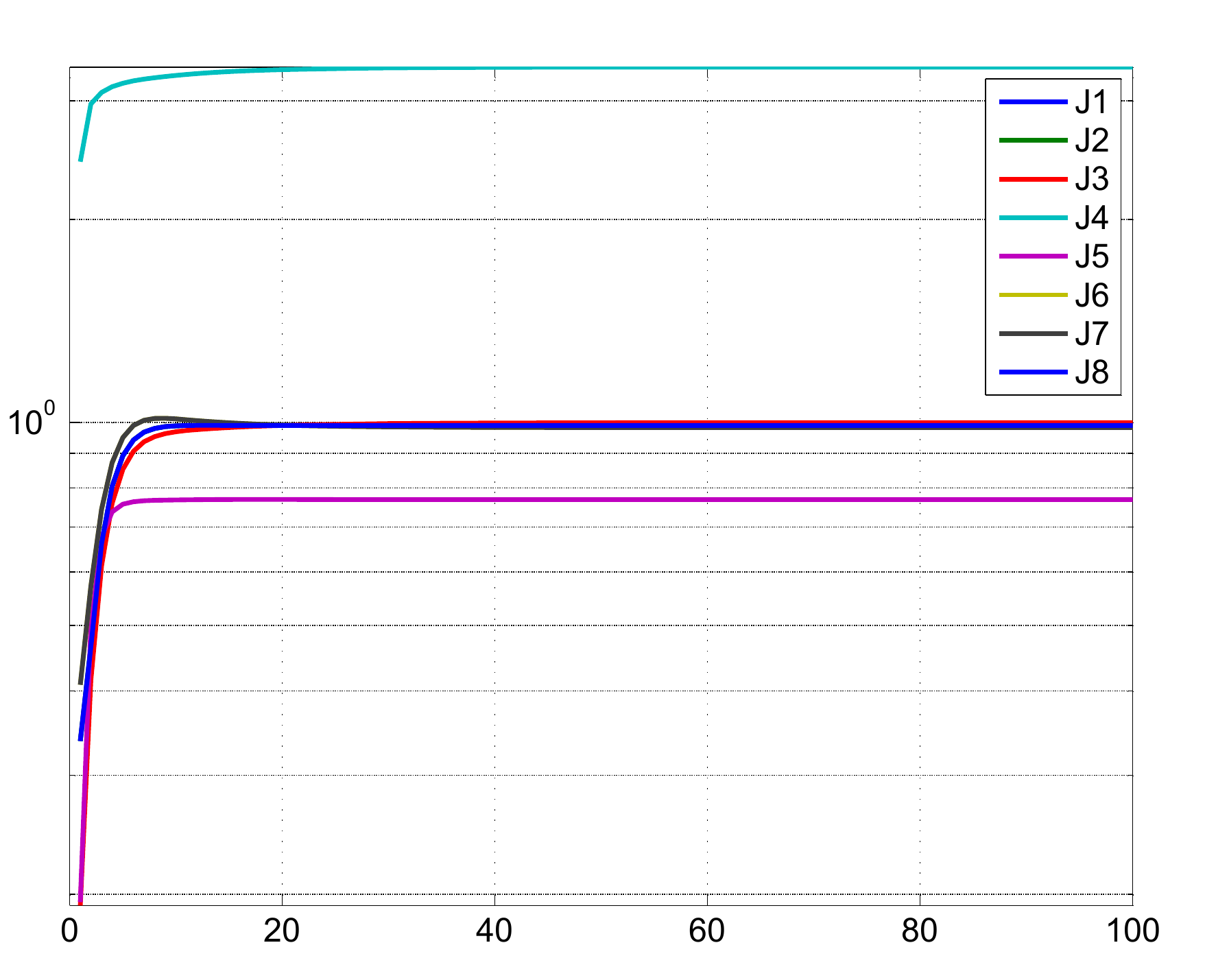}
\caption{Variation of different costs (\textbf{J1-J8}) with iterations}
\label{conQ_J}
\end{figure}

\begin{figure}[h]
\includegraphics[width=6in,height=4in]{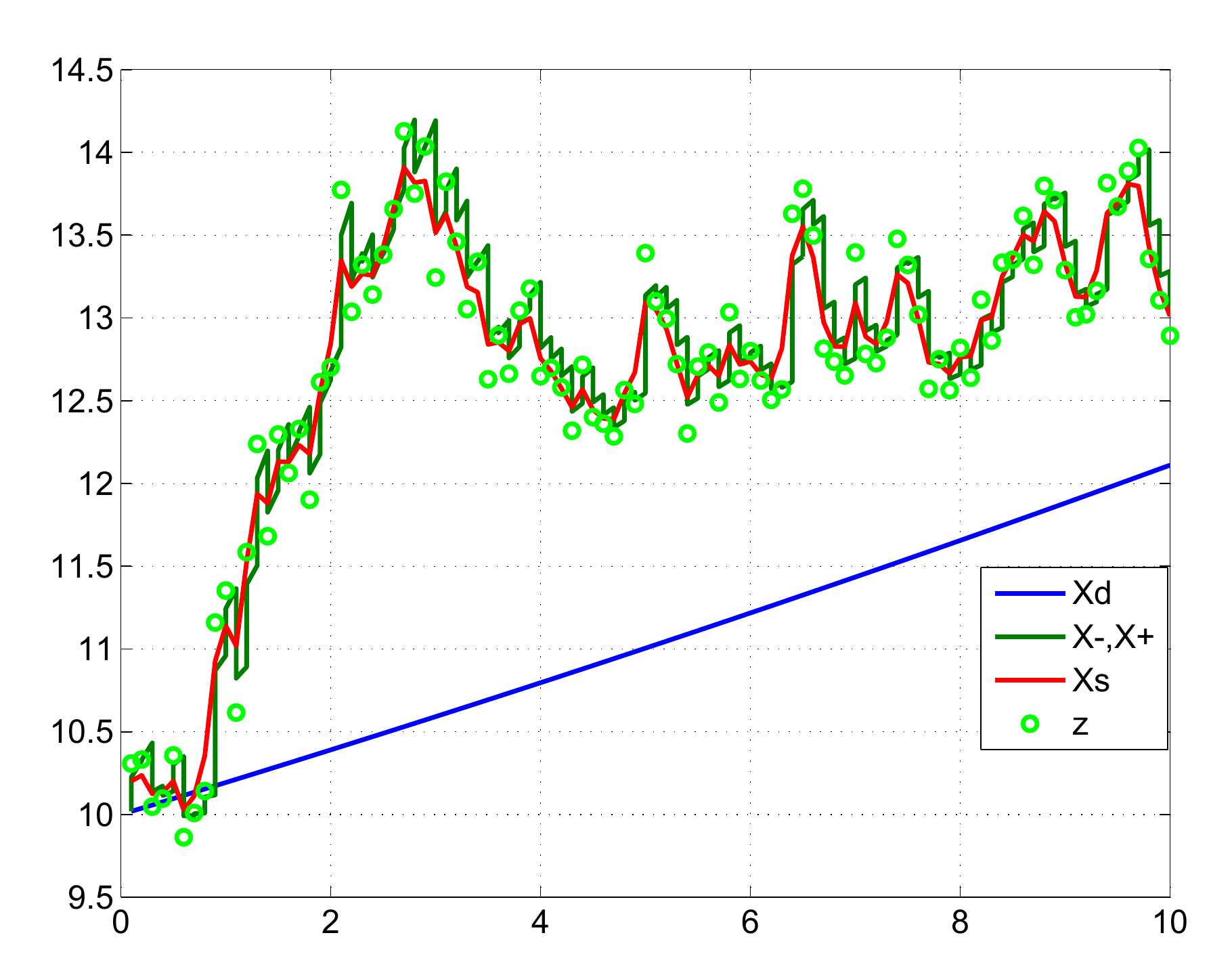}
\caption{Comparison of the predicted dynamics, posterior, smoothed}
\caption*{and the measurement}
\label{conQ_h}
\end{figure}

\begin{figure}[h]
\includegraphics[width=6in,height=4in]{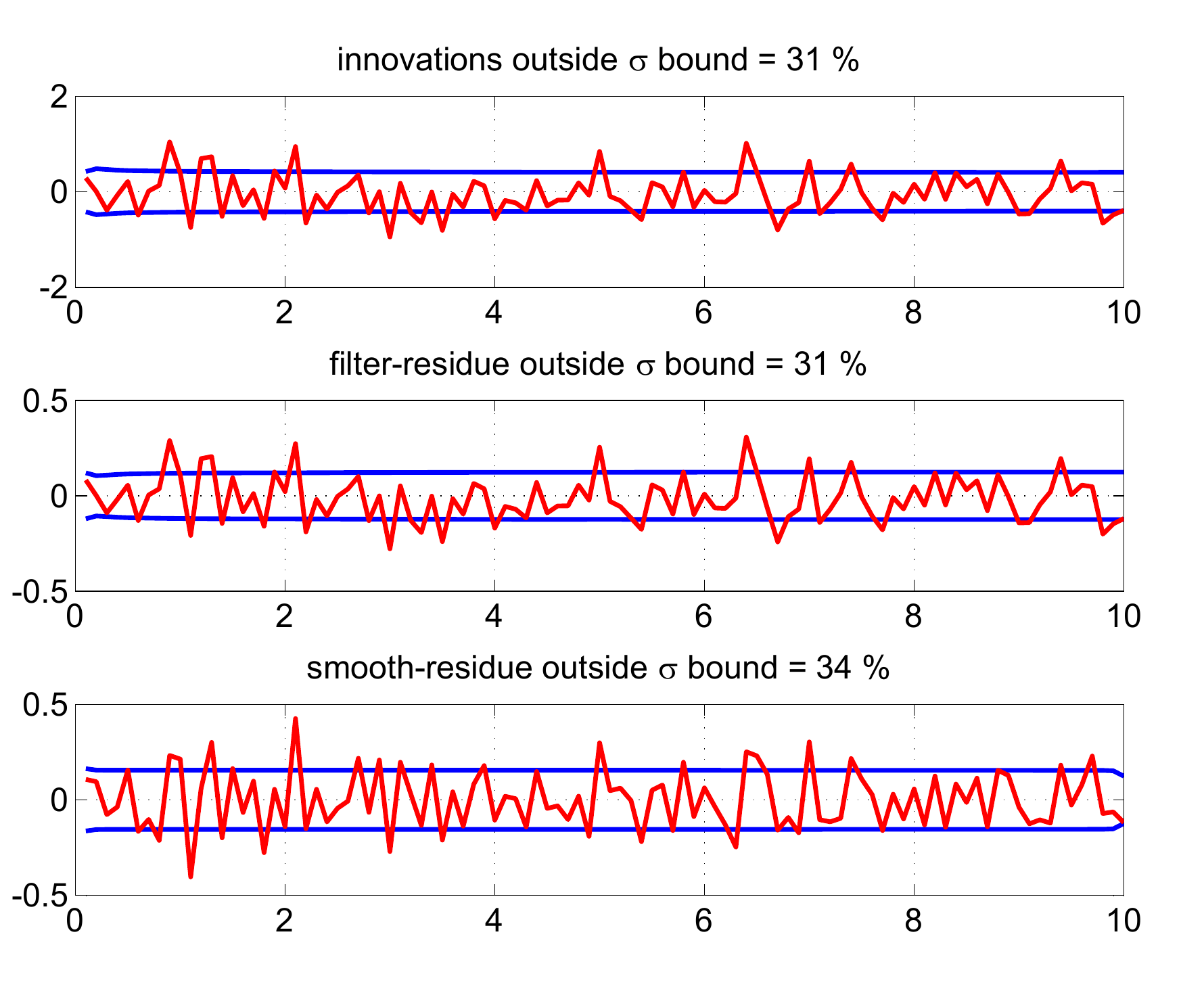}
\caption{The innovations, filtered residue and smoothed residue }
\label{conQ_innov}
\end{figure}

\begin{figure}[h]
\includegraphics[width=6in,height=4in]{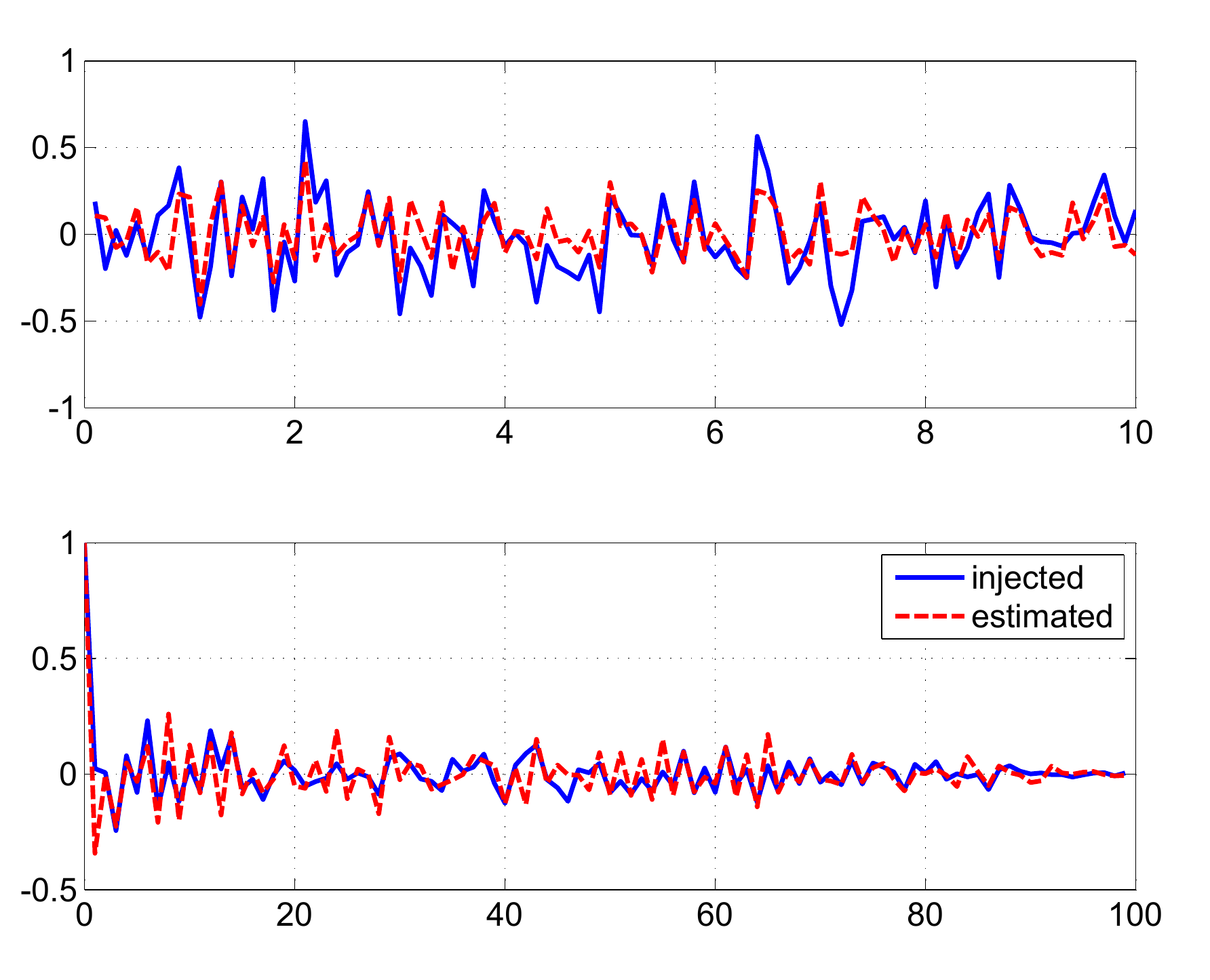}
\caption{Time variation of injected and estimated measurement noise (top) and}
\caption*{their autocorrelation (bottom)}
\label{conQ_mnoise}
\end{figure}

\begin{figure}[h]
\includegraphics[width=6in,height=4in]{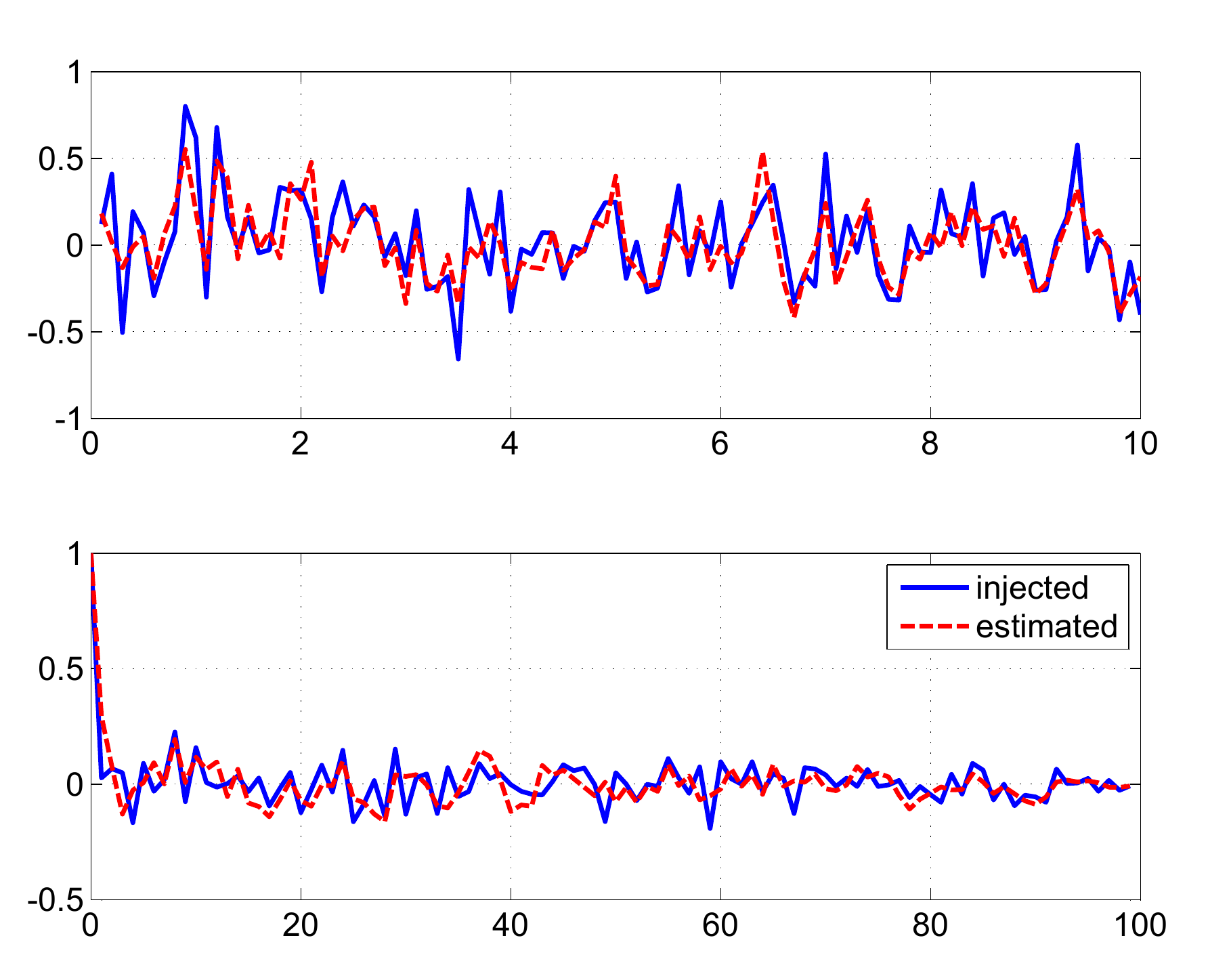}
\caption{Time variation of injected and estimated process noise (top) and}
\caption*{their autocorrelation (bottom)}
\label{conQ_pnoise}
\end{figure}

The next step is to process the data with process noise (\textbf{Q} $>$ 0). The Fig. \ref{con_err} shows the absolute difference between the iterated and final values with iterations which indicates the accuracy level that one needs and it was found that 100 iterations are required. The variation of the estimated initial parameters and their variances through iterations are shown in Fig. \ref{conQ_P0}. The parameter and the uncertainty reach almost their final estimated values in about 5 and 20 iterations respectively. A similar plot in Fig. \ref{conQ_R} shows the variation of the estimated \textbf{R} and \textbf{Q}. The variation of different cost functions through the iterations are shown in Fig. \ref{conQ_J}. The cost functions \textbf{J1-J3} correspond to the number of measurement ($m$=1) and in presence of process noise, \textbf{J6-J8} correspond to the number of states  ($n$=1). The \textbf{J4} in absence of process noise corresponds to the trace of \textbf{R}. The \textbf{J5} is the negative log likelihood function whose absolute value is shown in the plot. There is a mismatch in the predicted dynamics and the measurement as seen in Fig. \ref{conQ_h} indicating the presence of process noise. The subsequent Fig. \ref{conQ_innov} correspond to the earlier Fig. \ref{con_innov} of \textbf{Q} = 0 case. The Fig. \ref{conQ_mnoise} and Fig. \ref{conQ_pnoise} shows respectively the injected and estimated measurement and process noise samples across time during the final iteration.

\begin{landscape}
\begin{table}[h]
\caption{Simulated System Noise inputs and Figures Numbers}{}
\label{sysdes}
\begin{center}
\begin{footnotesize}
\begin{tabular}{| c | c | c | c | c |}
\hline
Simulated system & \makecell{State and process \\ noise variance} & \makecell{Measurement and \\ noise variance} & \makecell{Figures \\ (\textbf{Q} = 0)} & \makecell{Figures \\ (\textbf{Q} $>$ 0) }

\\ \hline

Constant & \makecell{Constant state\\ (0.05)} & \makecell{Constant state \\(0.05)} &

\makecell{Fig. \ref{con_p1},  Fig. \ref{con_P0},\\ Fig. \ref{con_R}, Fig. \ref{con_cost} ,  \\Fig. \ref{con_h1} , Fig. \ref{con_innov} , \\Fig. \ref{con_mnoise} }  &
\makecell{ Fig. \ref{con_err} , Fig. \ref{conQ_P0} , \\Fig. \ref{conQ_R} , Fig. \ref{conQ_J} , \\Fig. \ref{conQ_h} , Fig. \ref{conQ_innov} , \\Fig. \ref{conQ_mnoise} , Fig. \ref{conQ_pnoise}}
 \\ \hline

Ramp &  \makecell{Ramp state\\ (0.05)} & \makecell{Ramp state \\(0.05)} &

\makecell{Fig. \ref{ramp_p1} ,  Fig. \ref{ramp_P0} , \\ Fig. \ref{ramp_R} , Fig. \ref{ramp_cost} , \\ Fig. \ref{ramp_h1} , Fig. \ref{ramp_innov} , \\Fig. \ref{ramp_mnoise} }  &
\makecell{ Fig. \ref{ramp_err} , Fig. \ref{rampQ_P0} , \\Fig. \ref{rampQ_R} , Fig. \ref{rampQ_J} , \\Fig. \ref{rampQ_h} , Fig. \ref{rampQ_innov} , \\Fig. \ref{rampQ_mnoise} , Fig. \ref{rampQ_pnoise} }
\\ \hline

 SMD & \makecell{Displacement (0.001)\\Velocity (0.002)} & \makecell{Displacement (0.001)\\Velocity (0.004)} &

\makecell{Fig. \ref{smd_p1}-\ref{smd_p3} ,  Fig. \ref{smd_P0} , \\
 Fig. \ref{smd_R} , Fig. \ref{smd_cost} ,\\
 Fig. \ref{smd_h1}-\ref{smd_h2} , Fig. \ref{smd_innov1}-\ref{smd_innov2} ,\\
 Fig. \ref{smd_mnoise1}-\ref{smd_mnoise2}
}
&
\makecell{ Fig. \ref{smd_err} , Fig. \ref{smdQ_P0} , \\Fig. \ref{smdQ_R} , Fig. \ref{smdQ_J} ,\\Fig. \ref{smdQ_h1}--\ref{smdQ_h2} , Fig. \ref{smdQ_innov1}-\ref{smdQ_innov2} , \\Fig. \ref{smdQ_mnoise1}-\ref{smdQ_mnoise2} , Fig. \ref{smdQ_pnoise1}-\ref{smdQ_pnoise2} }
 \\
\hline

\end{tabular}
\end{footnotesize}
\end{center}
\end{table}
\end{landscape}

\begin{landscape}
\begin{table}[h]
\subsection{Constant System Tables (\textbf{Q} = 0) }
\vspace{14pt}
\caption{Sensitivity Study : (\textbf{Q} = 0) : Constant system.\\ No. of iterations=20, No. of simulations=50.}{}
\label{tbcon}
\begin{center}
\begin{footnotesize}
\begin{tabular}{| c | c | c | c | c | c | c | c | c | c | c | c |}
\hline
Study &
$\Theta$ ratio &
\makecell{CRB ratio\\NR/EKF} &
\makecell{Consistency\\ ratio-EKF}&
\makecell{Consistency\\ ratio-NR}&
\makecell{Spread\\factor\\EKF} &
\makecell{Spread\\factor\\NR} &
\makecell{\textbf{R} ratio\\EKF/True} & \makecell{\textbf{R} ratio\\EKF/NR} &
\makecell{$\mu$\ of \\\textbf{J1-J5}} &
\makecell{$\sigma$ of \\\textbf{J1-J5}} &
Remarks
\\ \hline

\multicolumn{12}{|c|}{Reference adaptive EKF used for non zero process noise case} \\ \hline
\makecell{$\mathbf{P_0}$ : Scaled up-[0,0;0,\checkmark]\\\textbf{Q} : EM-[\checkmark,0;0,0] \\ \textbf{R} : EM-diag}
&\multicolumn{11}{c|}{Extremely slow convergence of \textbf{Q} taking more than 100 iterations as seen in Fig. \ref{EM_slow}.} \\ \hline

\multicolumn{12}{|c|}{Proposed reference adaptive EKF for zero process noise case, \textbf{Q} is diagonal matrix fixed to $10^{-10}$ for all states} \\ \specialrule{.2em}{.0em}{.0em}

\multicolumn{1}{!{\vrule width 2pt}c|}{
\makecell{$\mathbf{P_0}$ : Scaled up-[0,0;0,\checkmark]\\\textbf{Q} : $10^{-10}$-[\checkmark,0;0,0] \\ \textbf{R} : EM-diag} } &
 1.0000 &   1.0042  &
 0.9613 &   0.9577  &
 0.0051 &   0.0051  &
 0.9761 &   1.0013  &
\makecell{ 0.9901  \\  0.9901  \\  0.9998  \\  0.0483  \\ -1.9933 }&
\makecell{0.0000  \\  0.0000  \\  0.0012 \\   0.0065 \\   0.1399 } &
\multicolumn{1}{|c!{\vrule width 2pt}}{
\makecell{$\sigma_{  \Theta}$=\\3.6257e-05\\SIGMA$_{avg}$=\\3.7718e-05}} \\ \specialrule{.2em}{.0em}{.0em}


\multicolumn{12}{|c|}{Using different method and options for $\mathbf{P_0}$ } \\
\end{tabular}
\end{footnotesize}
\end{center}
\end{table}
\end{landscape}

\addtocounter{table}{-1}
\begin{landscape}
\begin{table}[h]
\caption{Sensitivity Study : (\textbf{Q} = 0) : Constant system.\\ No. of iterations=20, No. of simulations=50.}{}
\begin{center}
\begin{footnotesize}
\begin{tabular}{|c| c| c| c| c| c| c|c|c|c|c|c|}
\hline
Study &
$\Theta$ ratio &
\makecell{CRB ratio\\NR/EKF} &
\makecell{Consistency\\ ratio-EKF}&
\makecell{Consistency\\ ratio-NR}&
\makecell{Spread\\factor\\EKF} &
\makecell{Spread\\factor\\NR} &
\makecell{\textbf{R} ratio\\EKF/True} & \makecell{\textbf{R} ratio\\EKF/NR} &
\makecell{$\mu$\ of \\\textbf{J1-J5}} &
\makecell{$\sigma$ of \\\textbf{J1-J5}} &
Remarks
\\ \hline


\makecell{$\mathbf{P_0}$ : IIM-[0,0;0,\checkmark]\\\textbf{Q} : $10^{-10}$-[\checkmark,0;0,0] \\ \textbf{R} : EM-diag}  &
 1.0000 &   1.0003  &
  1.0186 &   0.9577  &
  0.0052 &   0.0051  &
 0.9770 &   1.0022  &
\makecell{  0.9900 \\   0.9900  \\  0.9998 \\   0.0485  \\ -1.8054  }&
\makecell{   0.0000   \\ 0.0000 \\   0.0012  \\  0.0066 \\   0.1378 } &
Unity CRB ratio\\ \hline


\makecell{$\mathbf{P_0}$ : IIM-diag\\\textbf{Q} : $10^{-10}$-[\checkmark,0;0,0] \\ \textbf{R} : EM-diag}  &
 1.0000 &   0.4939  &
 0.7709 &   0.9577  &
 0.0094 &   0.0051  &
 0.9811 &   1.0064  &
\makecell{0.9800  \\  0.9800 \\   0.9996  \\  0.0484  \\ -1.6845}&
\makecell{ 0.0000  \\  0.0000 \\   0.0011  \\  0.0067  \\  0.1383 }&
\makecell{Higher filter CRB\\ than the NR}\\ \hline


\makecell{$\mathbf{P_0}$ : IIM-full\\\textbf{Q} : $10^{-10}$-[\checkmark,0;0,0] \\ \textbf{R} : EM-diag}  &
 1.0000 &   0.4986  &
 0.7793 &   0.9577  &
 0.0093 &   0.0051  &
 0.9963 &   1.0220  &
\makecell{ 0.9709  \\  0.9709  \\  0.9892  \\  0.0487  \\ -1.8111}&
\makecell{ 0.0688  \\  0.0688  \\  0.0710  \\  0.0070  \\  0.1417  }&
\makecell{Higher filter CRB\\ than that of the NR}\\
\end{tabular}
\end{footnotesize}
\end{center}
\end{table}
\end{landscape}

\addtocounter{table}{-1}
\begin{landscape}
\begin{table}[h]
\caption{Sensitivity Study : (\textbf{Q} = 0) : Constant system.\\ No. of iterations=20, No. of simulations=50.}{}
\begin{center}
\begin{footnotesize}
\begin{tabular}{|c| c| c| c| c| c| c|c|c|c|c|c|}
\hline
Study &
$\Theta$ ratio &
\makecell{CRB ratio\\NR/EKF} &
\makecell{Consistency\\ ratio-EKF}&
\makecell{Consistency\\ ratio-NR}&
\makecell{Spread\\factor\\EKF} &
\makecell{Spread\\factor\\NR} &
\makecell{\textbf{R} ratio\\EKF/True} & \makecell{\textbf{R} ratio\\EKF/NR} &
\makecell{$\mu$\ of \\\textbf{J1-J5}} &
\makecell{$\sigma$ of \\\textbf{J1-J5}} &
Remarks
\\ \hline

\multicolumn{12}{|c|}{Using different methods for estimating \textbf{R}} \\ \hline

\makecell{$\mathbf{P_0}$ : Scaled up-[0,0;0,\checkmark]\\\textbf{Q} : $10^{-10}$-[\checkmark,0;0,0] \\ \textbf{R} : diag \textbf{R}*}  &
 1.0000 &   1.0048  &
  0.9619 &   0.9577  &
 0.0051 &   0.0051  &
 0.9749 &   1.0000  &
\makecell{  0.9914  \\  0.9914  \\  1.0012 \\   0.0483 \\  -1.9933 }&
\makecell{ 0.0025  \\  0.0025 \\   0.0028  \\  0.0065  \\  0.1399 }&
\makecell{*Using dynamical residue\\ as per Eq-\ref{R3}}\\ \hline


\makecell{$\mathbf{P_0}$ : Scaled up-[0,0;0,\checkmark]\\\textbf{Q} : $10^{-10}$-[\checkmark,0;0,0] \\\textbf{R} : diag \textbf{R}*}  &
 1.0000 &    1.0092  &
 0.9660 &   0.9577  &
 0.0051 &   0.0051  &
 0.9664 &   0.9913  &
\makecell{1.0000  \\  1.0000 \\   1.0098  \\  0.0483 \\  -1.9932}&
\makecell{ 0.0000 \\   0.0000 \\   0.0013  \\  0.0065  \\  0.1399   }&
\makecell{*Using smoothed residue\\ignoring $2^{nd}$ order\\terms in Eq-\ref{R1}}\\
\end{tabular}
\end{footnotesize}
\end{center}
\end{table}
\end{landscape}


\begin{landscape}
\begin{table}[h]
\subsection{Constant System Tables (\textbf{Q} $>$ 0) }
\vspace{14pt}
\caption{Sensitivity Study : (\textbf{Q} $>$ 0) : Constant system.\\ No. of iterations=100, No. of simulations=50.}{}
\label{tbconQ}
\begin{center}
\begin{footnotesize}
\begin{tabular}{|c| c| c| c| c| c| c|c|c|c|c|}
\hline
Study &
$\Theta$ ratio &
\makecell{PCRB\\ ratio} &
\makecell{Consistency\\ ratio-EKF}&
\makecell{Spread\\factor\\EKF} &
\makecell{\textbf{R} ratio\\EKF/True} &
\makecell{\textbf{Q} ratio\\EKF/True}  &
\makecell{$\mu$\ of \\\textbf{J1-J8}} &
\makecell{$\sigma$ of \\\textbf{J1-J8}} &
Remarks
\\ \hline

\multicolumn{10}{|c|}{Proposed reference adaptive EKF for non zero process noise} \\ \specialrule{.2em}{.0em}{.0em}

\multicolumn{1}{!{\vrule width 2pt}c|}{
\makecell{$\mathbf{P_0}$ : Scaled up-[0,0;0,\checkmark]\\\textbf{Q} : EM-[\checkmark,0;0,0] \\\textbf{R} : EM-diag}}  &
 1.0000 &  1.0275 &
 1.0270 &  0.2906 &
 0.9348 &  1.0300 &
\makecell{ 0.9898 \\   0.9898\\    1.0003\\    0.9880 \\   -1.0448 \\   0.9778 \\   0.9778 \\ 0.9898} &
\makecell{  0.0018  \\  0.0018 \\   0.0019 \\   0.8767  \\  0.1262  \\  0.0089  \\  0.0089 \\ 0.0018} &
\multicolumn{1}{|c!{\vrule width 2pt}}{
\makecell{$\sigma_{\Theta}$= 0.00222\\SIGMA$_{avg}$= 0.00216}} \\ \specialrule{.2em}{.0em}{.0em}

\multicolumn{1}{!{\vrule width 2pt}c|}{
\makecell{$\mathbf{P_0}$ : Scaled up-[0,0;0,\checkmark]\\\textbf{Q} : EM-[\checkmark,0;0,0] \\\textbf{R} : Known}}  &
 1.0000 &   0.9936 &
  1.0371 &  0.2907 &
 NA &   0.9986 &
\makecell{  0.9846 \\   0.9846 \\   0.9873 \\   0.9791  \\ -1.0368 \\   0.9783 \\   0.9783\\ 0.9846} &
\makecell{  0.0586  \\  0.0586 \\   0.1059   \\ 0.8448 \\   0.1247  \\  0.0057   \\ 0.0057\\ 0.0586} &
\multicolumn{1}{|c!{\vrule width 2pt}}{
\makecell{\textbf{R} is known}} \\ \specialrule{.2em}{.0em}{.0em}

\end{tabular}
\end{footnotesize}
\end{center}
\end{table}
\end{landscape}

\addtocounter{table}{-1}
\begin{landscape}
\begin{table}[h]
\caption{Sensitivity Study : (\textbf{Q} $>$ 0) : Constant system.\\ No. of iterations=100, No. of simulations=50.}{}
\begin{center}
\begin{footnotesize}
\begin{tabular}{|c| c| c| c| c| c| c|c|c|c|c|}
\hline
Study &
$\Theta$ ratio &
\makecell{PCRB\\ ratio} &
\makecell{Consistency\\ ratio-EKF}&
\makecell{Spread\\factor\\EKF} &
\makecell{\textbf{R} ratio\\EKF/True} &
\makecell{\textbf{Q} ratio\\EKF/True}  &
\makecell{$\mu$\ of \\\textbf{J1-J8}} &
\makecell{$\sigma$ of \\\textbf{J1-J8}} &
Remarks
\\ \specialrule{.2em}{.0em}{.0em}

\multicolumn{1}{!{\vrule width 2pt}c|}{
\makecell{$\mathbf{P_0}$ : Zero\\\textbf{Q} : EM-diag \\\textbf{R} : EM-diag}}  &
\makecell{NA} &
   1.0309 &
NA & NA &
0.9355 &  1.0280  &
\makecell{ 1.0000 \\   1.0000 \\   1.0006 \\   2.5377  \\ -1.0804  \\  1.0001 \\   1.0001 \\1.0000} &
\makecell{ 0.0001  \\  0.0001  \\  0.0051  \\  3.5893   \\  0.1198  \\  0.0067  \\  0.0067 \\ 0.0001} &
\multicolumn{1}{|c!{\vrule width 2pt}}{
\makecell{The parameters are \\assumed to be known}} \\ \specialrule{.2em}{.0em}{.0em}

\multicolumn{1}{!{\vrule width 2pt}c|}{
\makecell{$\mathbf{P_0}$ : Zero\\\textbf{Q} : EM-diag \\\textbf{R} : Known}}  &
\makecell{NA} &
 0.9963 &
NA & NA &
 NA &  0.9989 &
\makecell{ 0.9945 \\   0.9945  \\  0.9869  \\  2.5377 \\  -1.0718  \\  1.0004  \\  1.0004 \\0.9945} &
\makecell{ 0.0618  \\  0.0618 \\   0.1119 \\   3.5893  \\  0.1180  \\  0.0051  \\  0.0051 \\ 0.0618} &
\multicolumn{1}{|c!{\vrule width 2pt}}{
\makecell{The parameters and \textbf{R} are \\assumed to be known}} \\ \specialrule{.2em}{.0em}{.0em}

\end{tabular}
\end{footnotesize}
\end{center}
\end{table}
\end{landscape}

\addtocounter{table}{-1}
\begin{landscape}
\begin{table}[h]
\caption{Sensitivity Study : (\textbf{Q} $>$ 0) : Constant system.\\ No. of iterations=100, No. of simulations=50.}{}
\begin{center}
\begin{footnotesize}
\begin{tabular}{|c| c| c| c| c| c| c|c|c|c|c|}
\hline
Study &
$\Theta$ ratio &
\makecell{PCRB\\ ratio} &
\makecell{Consistency\\ ratio-EKF}&
\makecell{Spread\\factor\\EKF} &
\makecell{\textbf{R} ratio\\EKF/True} &
\makecell{\textbf{Q} ratio\\EKF/True}  &
\makecell{$\mu$\ of \\\textbf{J1-J8}} &
\makecell{$\sigma$ of \\\textbf{J1-J8}} &
Remarks
\\ \hline

\multicolumn{10}{|c|}{Extended EM algorithm (Bavdekar et al. \cite{Bavdekar2011}(2011))} \\ \hline
\makecell{$\mathbf{P_0}$ : Smoothed-full\\\textbf{Q} : EM-full \\\textbf{R} : EM-full}  & \multicolumn{9}{c|}{\makecell{There is no consistency in the parameter estimates since the smoothed $\mathbf{P_0}$ simply decreases through iterations.}}  \\ \hline

\multicolumn{10}{|c|}{Modification-1} \\ \hline

\makecell{$\mathbf{P_0}$ : Smoothed-diag\\\textbf{Q} : EM-diag \\\textbf{R} : EM-diag}   &
 1.0000 &  1.0313 &
 10.8083 &  0.1680 &
 0.9745 &   0.9400 &
\makecell{0.9993 \\   0.9993  \\  1.0004  \\  1.0079 \\  -1.1020  \\  1.0041 \\   1.0041 \\ 0.9993} &
\makecell{ 0.0017  \\  0.0017  \\  0.0017 \\   0.8743 \\   0.1278  \\  0.0030 \\   0.0030 \\ 0.0017} &
\makecell{No stoping condition. \\The low spread factor is deceptive\\since there is no consistency  in the\\ parameter estimates with its covariance}\\ \hline

\multicolumn{10}{|c|}{Modification-2} \\ \hline

\makecell{$\mathbf{P_0}$ : Scaled up-[0,0;0,\checkmark]\\\textbf{Q} : EM-diag \\\textbf{R} : EM-diag}   &
 1.0000 &  1.0275 &
  1.0270 &  0.2906 &
0.9348 &  1.0300 &
\makecell{0.9898 \\   0.9898 \\   1.0003  \\  0.9880 \\  -1.0448  \\  0.9778 \\   0.9778 \\ 0.9898} &
\makecell{0.0018 \\   0.0018  \\  0.0019  \\  0.8767  \\  0.1262  \\  0.0089  \\  0.0089 \\ 0.0018} &
\makecell{The slight differences from \\the reference case is due to the small \textbf{Q} \\that is generated for the parameters}\\ \hline

\multicolumn{10}{|c|}{Modification-3} \\
\end{tabular}
\end{footnotesize}
\end{center}
\end{table}
\end{landscape}

\addtocounter{table}{-1}
\begin{landscape}
\begin{table}[H]
\caption{Sensitivity Study : (\textbf{Q} $>$ 0) : Constant system.\\ No. of iterations=100, No. of simulations=50.}{}
\begin{center}
\begin{footnotesize}
\begin{tabular}{|c| c| c| c| c| c| c|c|c|c|}
\hline
Study &
$\Theta$ ratio &
\makecell{PCRB\\ ratio} &
\makecell{Consistency\\ ratio-EKF}&
\makecell{Spread\\factor\\EKF} &
\makecell{\textbf{R} ratio\\EKF/True} &
\makecell{\textbf{Q} ratio\\EKF/True}  &
\makecell{$\mu$\ of \\\textbf{J1-J8}} &
\makecell{$\sigma$ of \\\textbf{J1-J8}} &
Remarks
\\ \hline

\multicolumn{10}{|c|}{MS method as suggested by Mohamed and Schwarz} \\ \hline
\makecell{$\mathbf{P_0}$ : Scaled up-[0,0;0,\checkmark]\\\textbf{Q} : MS-diag \\\textbf{R} : MS-diag}  & \multicolumn{9}{c|}{\makecell{\textbf{R} can converge to a very large value and the cost functions diverge as seen in Fig. \ref{MS_con_Q},\ref{MS_con_J}. \\ There is no convergence in cost even if $\Theta$ is known.}}\\ \hline

\multicolumn{10}{|c|}{MS method with \textbf{Q} being estimated using only last N/2 samples} \\ \hline
\makecell{$\mathbf{P_0}$ : zero\\\textbf{Q} : MS-diag \\\textbf{R} : known} &
 NA &  1.0100 &
 NA &  NA &
 NA &  0.9927 &
\makecell{1.0354  \\  1.0354  \\  1.0071 \\   2.5377 \\  -1.0504  \\  1.0983  \\  1.0983 \\ 1.0354} &
\makecell{ 0.1629  \\  0.1629   \\ 0.1304  \\  3.5893  \\  0.1170 \\   0.2726 \\   0.2726 \\ 0.1629} &
\makecell{\textbf{R}, $\Theta$ is known}\\ \hline

\multicolumn{10}{|c|}{MS method assuming \textbf{R} is known and \textbf{Q} is estimated using only last N/2 samples} \\ \hline
\makecell{$\mathbf{P_0}$ : Scaled up-[0,0;0,\checkmark]\\\textbf{Q} : MS-[\checkmark,0;0,0] \\\textbf{R} : Known} &
 1.0000 &  1.0173 &
  0.9643 &   0.3096 &
 0.8703 &  1.1872 &
\makecell{0.9837  \\  0.9837  \\  1.0067  \\  0.9703 \\  -0.9896 \\   0.9709\\    0.9709 \\ 0.9837} &
\makecell{ 0.1474  \\  0.1474 \\   0.1313  \\  0.8348  \\  0.1242  \\  0.1866   \\ 0.1866 \\ 0.1474} &
\makecell{Stable operation when \\\textbf{R} is known}\\ \hline

\multicolumn{10}{|c|}{ Gemson's approach using MS method} \\
\end{tabular}
\end{footnotesize}
\end{center}
\end{table}
\end{landscape}

\addtocounter{table}{-1}
\begin{landscape}
\begin{table}[h]
\caption{Sensitivity Study : (\textbf{Q} $>$ 0) : Constant system.\\ No. of iterations=100, No. of simulations=50.}{}
\begin{center}
\begin{footnotesize}
\begin{tabular}{|c| c| c| c| c| c| c|c|c|c|}
\hline
Study &
$\Theta$ ratio &
\makecell{PCRB\\ ratio} &
\makecell{Consistency\\ ratio-EKF}&
\makecell{Spread\\factor\\EKF} &
\makecell{\textbf{R} ratio\\EKF/True} &
\makecell{\textbf{Q} ratio\\EKF/True}  &
\makecell{$\mu$\ of \\\textbf{J1-J8}} &
\makecell{$\sigma$ of \\\textbf{J1-J8}} &
Remarks
\\ \hline

\multicolumn{10}{|c|}{Covariance matching as suggested by Myers and Tapley (MT)} \\ \hline
\makecell{$\mathbf{P_0}$ : Scaled up-[0,0;0,\checkmark]\\\textbf{Q} : MT-diag \\\textbf{R} : MT-diag}  & \multicolumn{9}{c|}{\makecell{Positive definiteness of \textbf{Q,R} is not gauranteed and the cost functions can diverge as seen in Fig. \ref{MT_con_Q},-\ref{MT_con_J}.\\ There is no proper convergence even if $\Theta$ is known.}} \\ \hline

\multicolumn{10}{|c|}{MT method with \textbf{Q} being estimated using only last N/2 samples} \\ \hline
\makecell{$\mathbf{P_0}$ : zero\\\textbf{Q} : MT-diag \\\textbf{R} : Known} &
 NA &  1.0087 &
 NA &  NA &
 NA &  0.9933 &
\makecell{1.0310 \\   1.0310  \\  1.0068   \\ 2.5377 \\  -1.0527 \\   1.0832  \\  1.0832 \\ 1.0310} &
\makecell{ 0.1634 \\   0.1634   \\ 0.1383 \\   3.5893 \\   0.1182  \\  0.2476  \\  0.2476\\ 0.1634} &
\makecell{\textbf{R}, $\Theta$ is known}\\ \hline

\multicolumn{10}{|c|}{MT method assuming \textbf{R} is known and \textbf{Q} is estimated using only last N/2 samples} \\ \hline
\makecell{$\mathbf{P_0}$ : Scaled up-[0,0;0,\checkmark]\\\textbf{Q} : MT-[\checkmark,0;0,0] \\\textbf{R} : Known}  &
  1.0000 &  1.0053 &
 1.0583 &  0.2888 &
NA &  0.9884 &
\makecell{ 1.0184 \\   1.0184  \\  1.0066  \\  0.9770  \\ -1.0205   \\ 1.0500 \\   1.0500 \\ 1.0184} &
\makecell{ 0.1518 \\   0.1518 \\   0.1300 \\   0.8339  \\  0.1258 \\   0.2179  \\  0.2179 \\0.1518} &
\makecell{Stable Operation}\\ \hline

\multicolumn{10}{|c|}{ Gemson's approach using MT method} \\
\hline

\end{tabular}
\end{footnotesize}
\end{center}
\end{table}
\end{landscape}


\clearpage
\subsection{Constant System Figures (\textbf{Q} = 0) }

\begin{figure}[h]
\includegraphics[width=6in,height=4in]{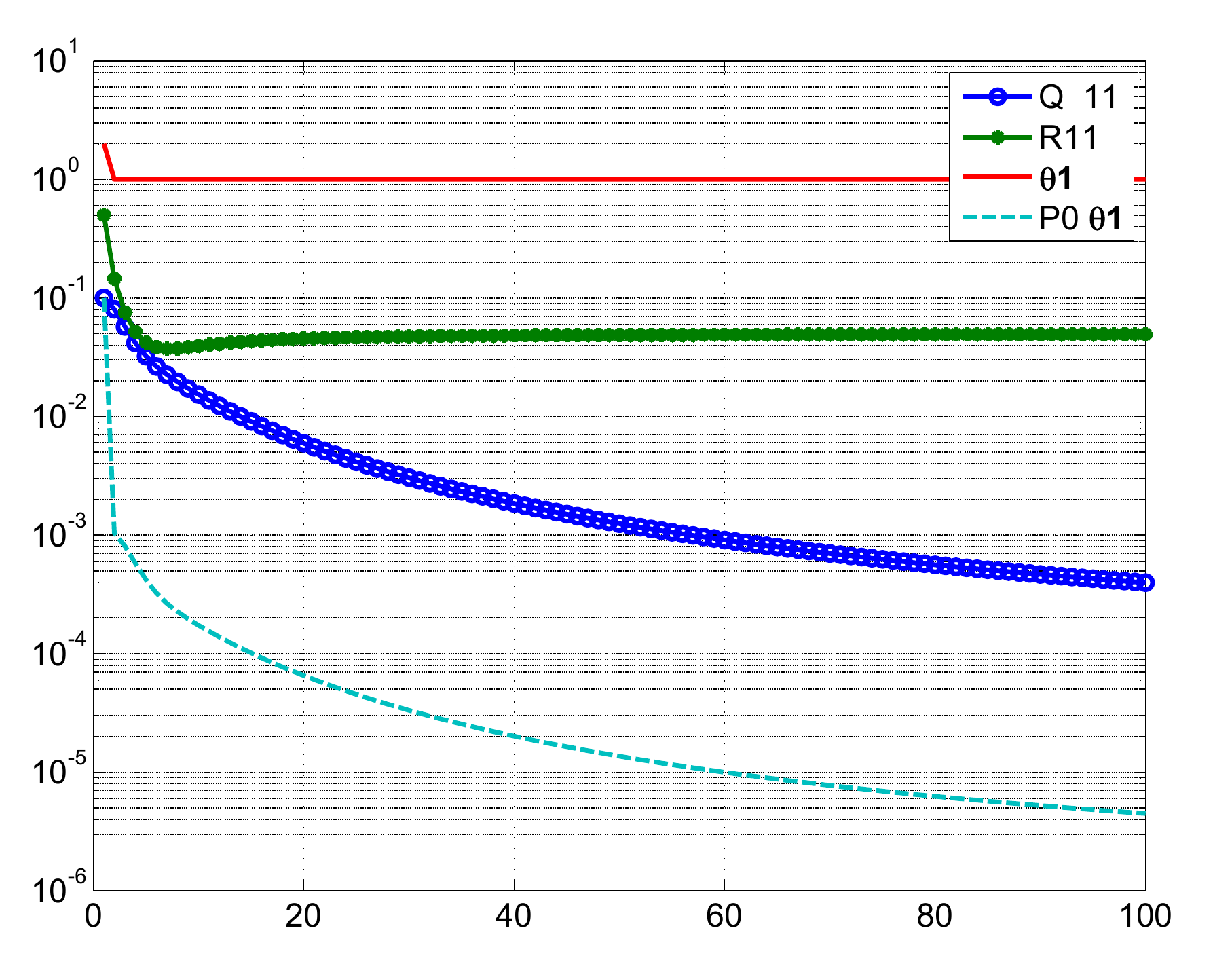}
\caption{Variation of different estimates with iterations using}
\caption*{Reference EKF (\textbf{Q} $>$ 0)}
\label{EM_slow}
\end{figure}

\clearpage
\subsection{Constant System Figures (\textbf{Q} $>$ 0) }

\begin{figure}[h]
\includegraphics[width=6in,height=4in]{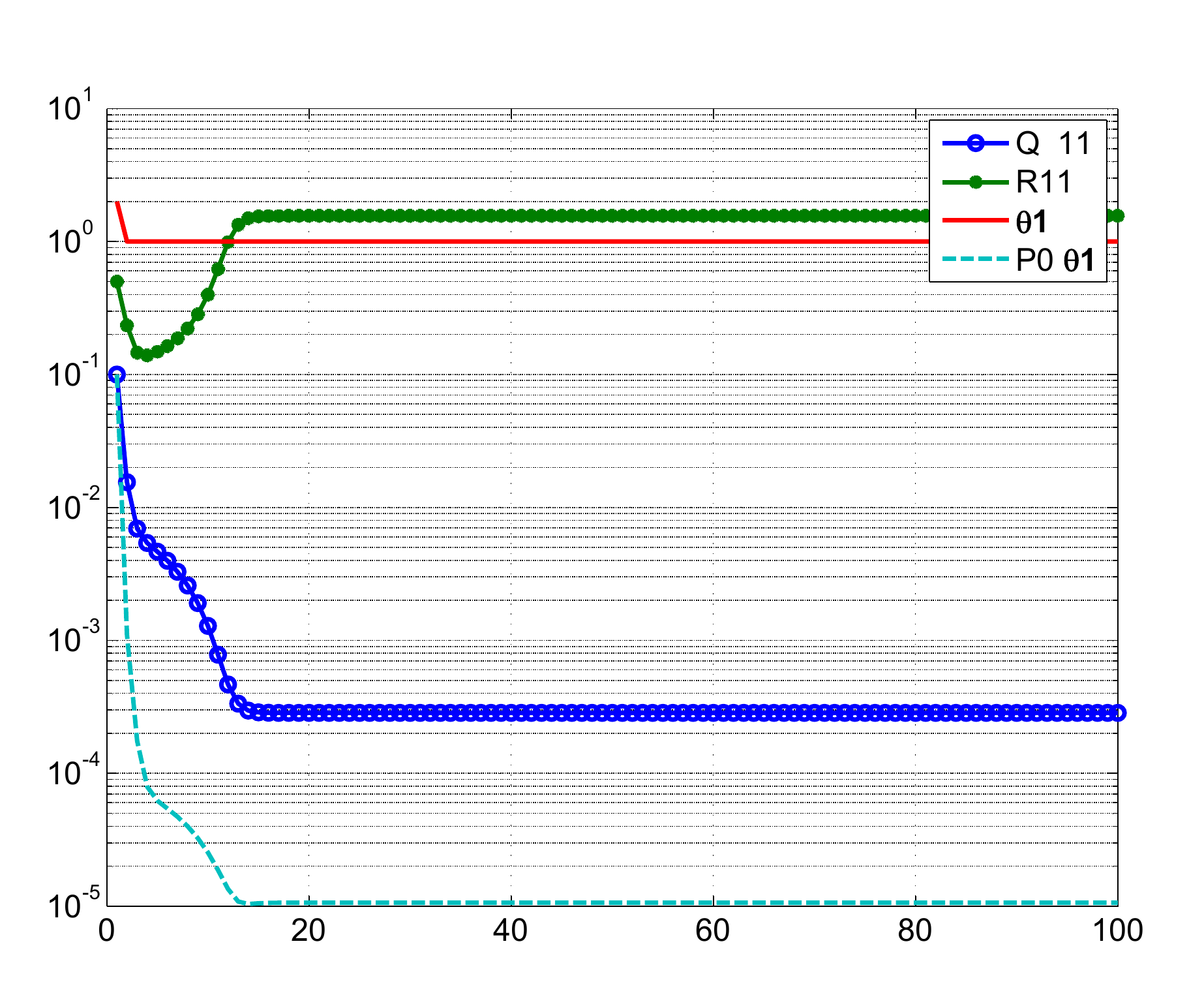}
\caption{Variation of different estimates with iterations using MS method}
\label{MS_con_Q}
\end{figure}

\begin{figure}[h]
\includegraphics[width=6in,height=4in]{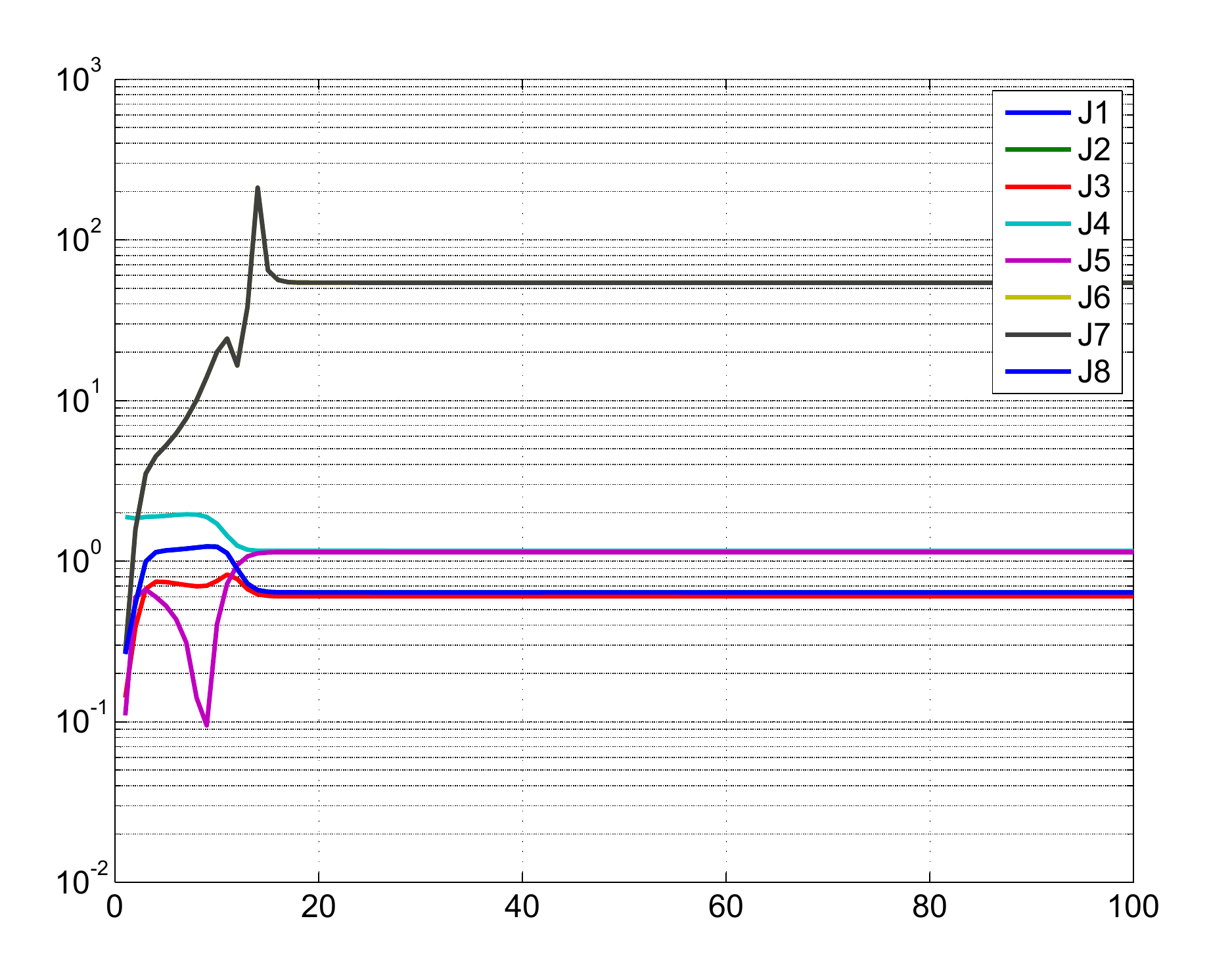}
\caption{Variation of different costs with iterations using MS method}
\label{MS_con_J}
\end{figure}

\begin{figure}[h]
\includegraphics[width=6in,height=4in]{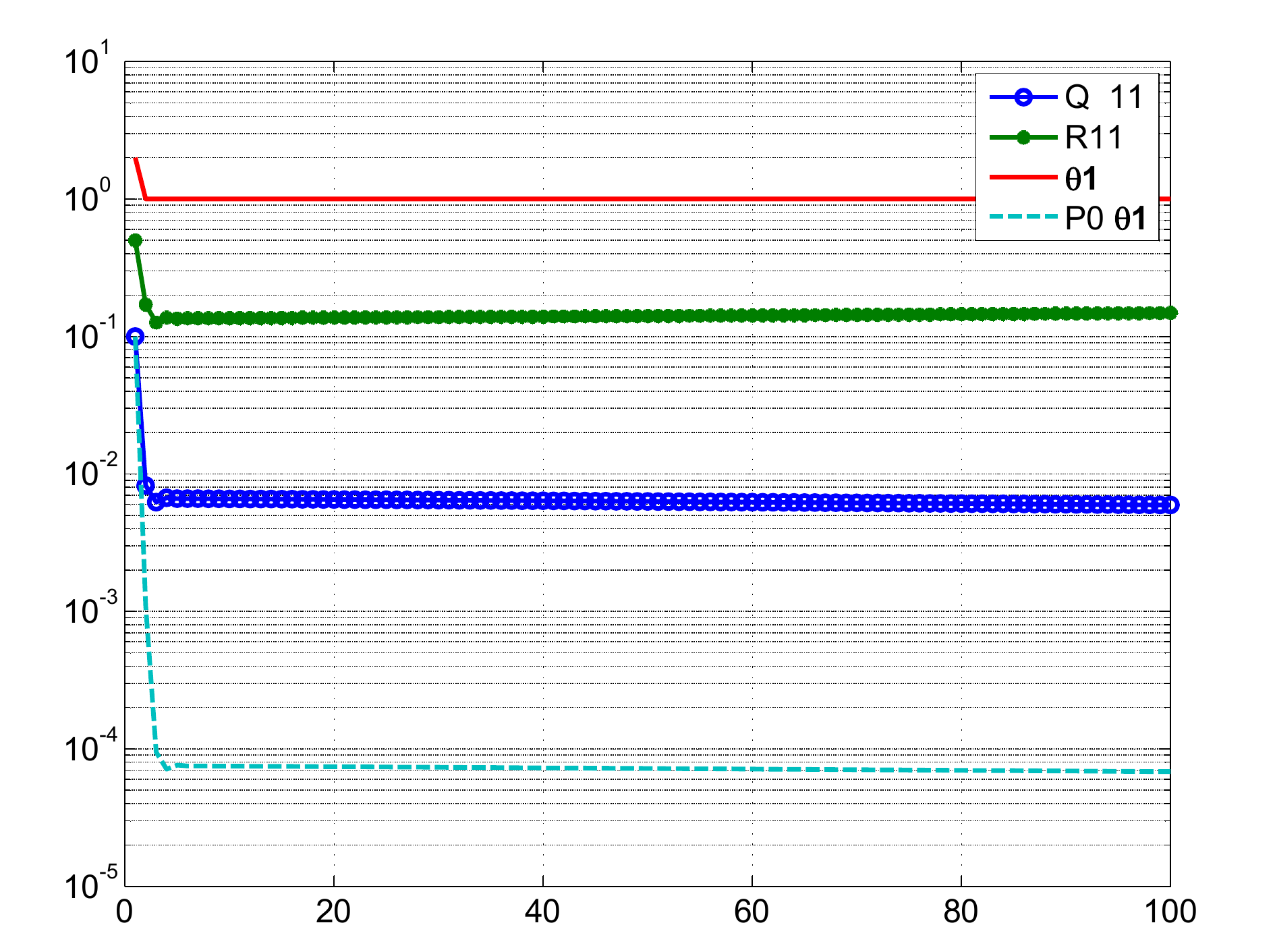}
\caption{Variation of different estimates with iterations using MT method}
\label{MT_con_Q}
\end{figure}

\begin{figure}[h]
\includegraphics[width=6in,height=4in]{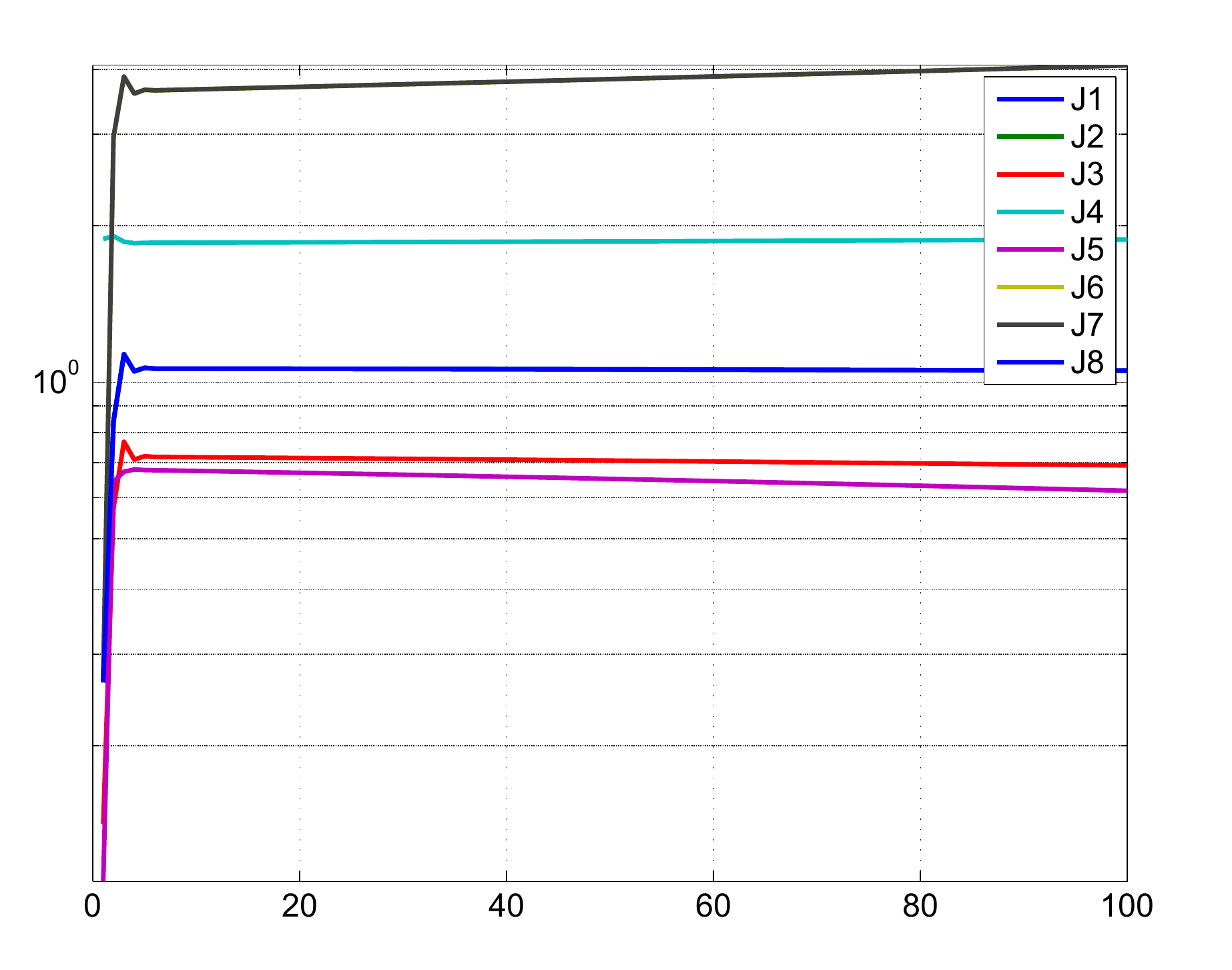}
\caption{Variation of different costs with iterations using MT method}
\label{MT_con_J}
\end{figure}

\clearpage
\section{Simulated Ramp System}

\par Consider a straight line (ramp) defined by, x=at+b where the unknowns are parameter `a' which is the slope and parameter `b' which is the intercept. The discretized state and measurement equations of a ramp signal system is given by
\begin{align*}
x_k=x_{k-1}+\Theta \delta t+w_k
\end{align*}
where $b=x_0=10$,$\delta t$=0.1sec is the sampling time period and $a=\Theta_{true}=2$ is considered as an unknown parameter. The measurement equation is given by
\begin{align*}
Z_k=HX_k+v_k
\end{align*}
where the augmented state, $X_k=[x_k,\Theta_k]$, H=[1 0] is the measurement matrix of size $m\times (n+p)$ where $m=n=p=1$. The numerical values of the noise variances are shown in Table-\ref{sysdes}. All the figures are presented for only one simulation run to prevent cluttering.

\subsection{Remarks on the Results}

\par
\begin{figure}[h]
\includegraphics[width=6in,height=3.2in]{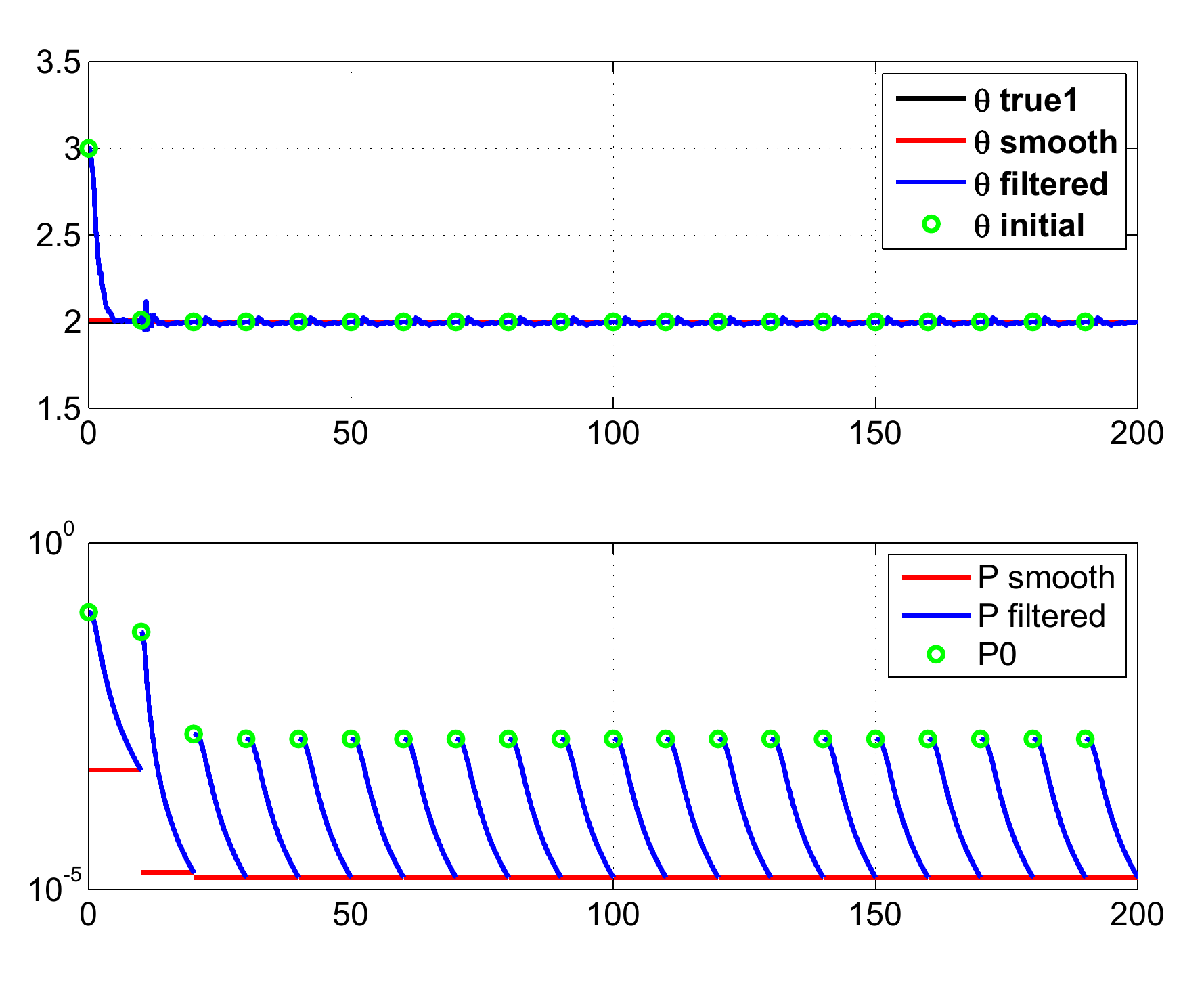}
\caption{The variation of different parameter estimate and their filtered and }
\caption*{smoothed covariances through (with the time cumulatively) the iterations}
\label{ramp_p1}
\end{figure}

\begin{figure}[h]
\includegraphics[width=6in,height=3.2in]{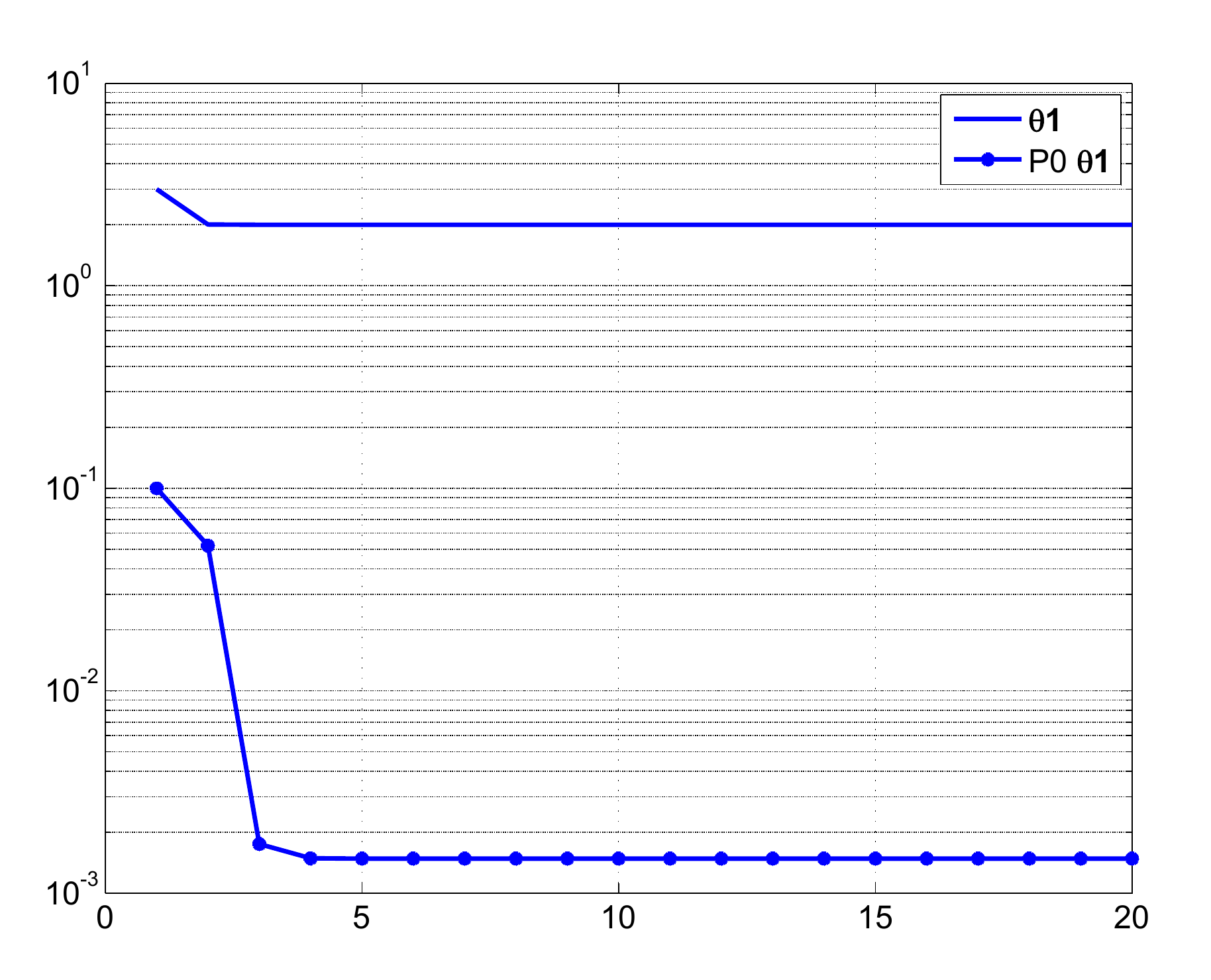}
\caption{Variation of parameter and its initial covariance ($\mathbf{P_0}$) with iterations}
\label{ramp_P0}
\end{figure}

\begin{figure}[h]
\includegraphics[width=6in,height=4in]{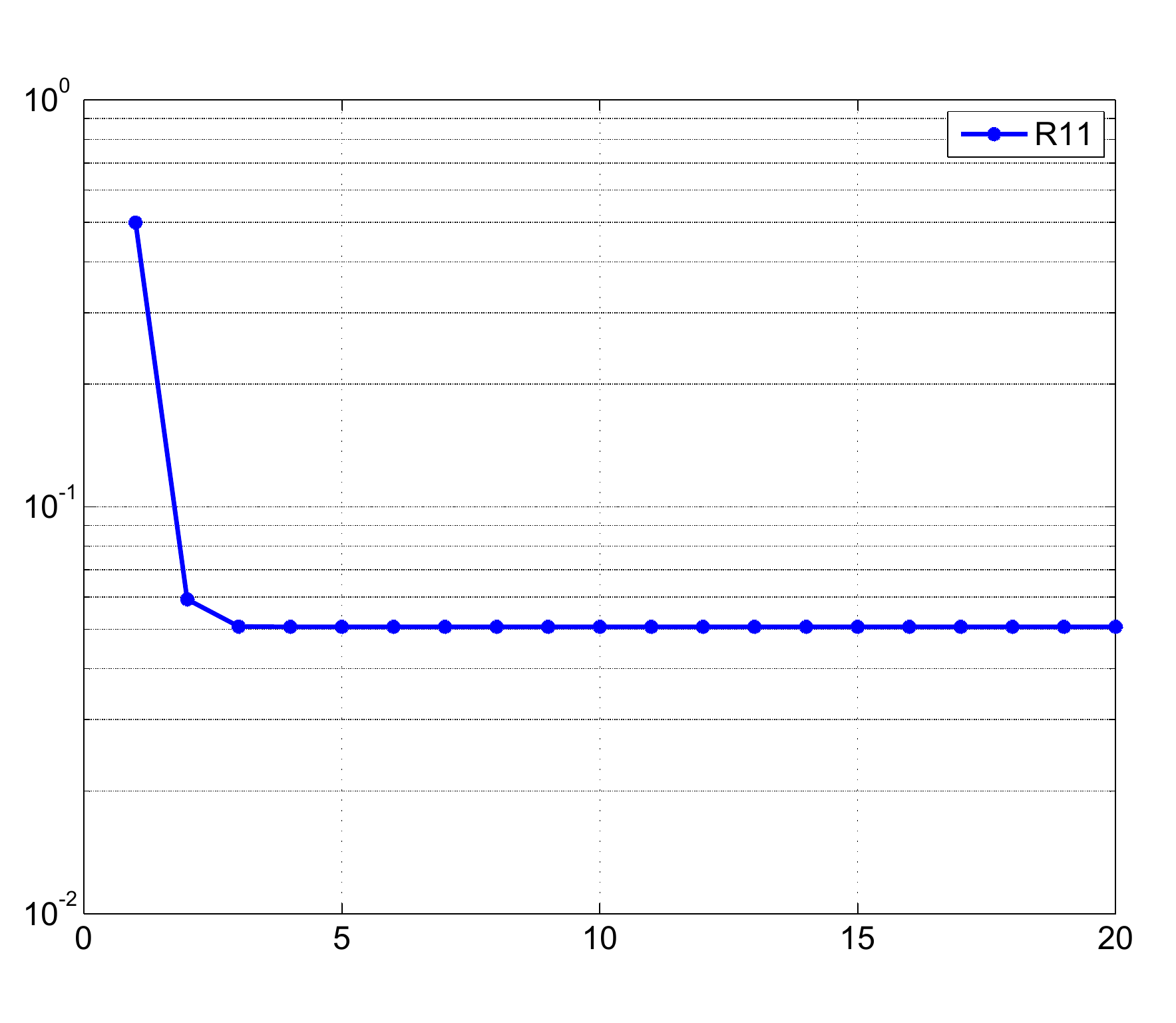}
\caption{Variation of \textbf{R} with iterations}
\label{ramp_R}
\end{figure}

\begin{figure}[h]
\includegraphics[width=6in,height=4in]{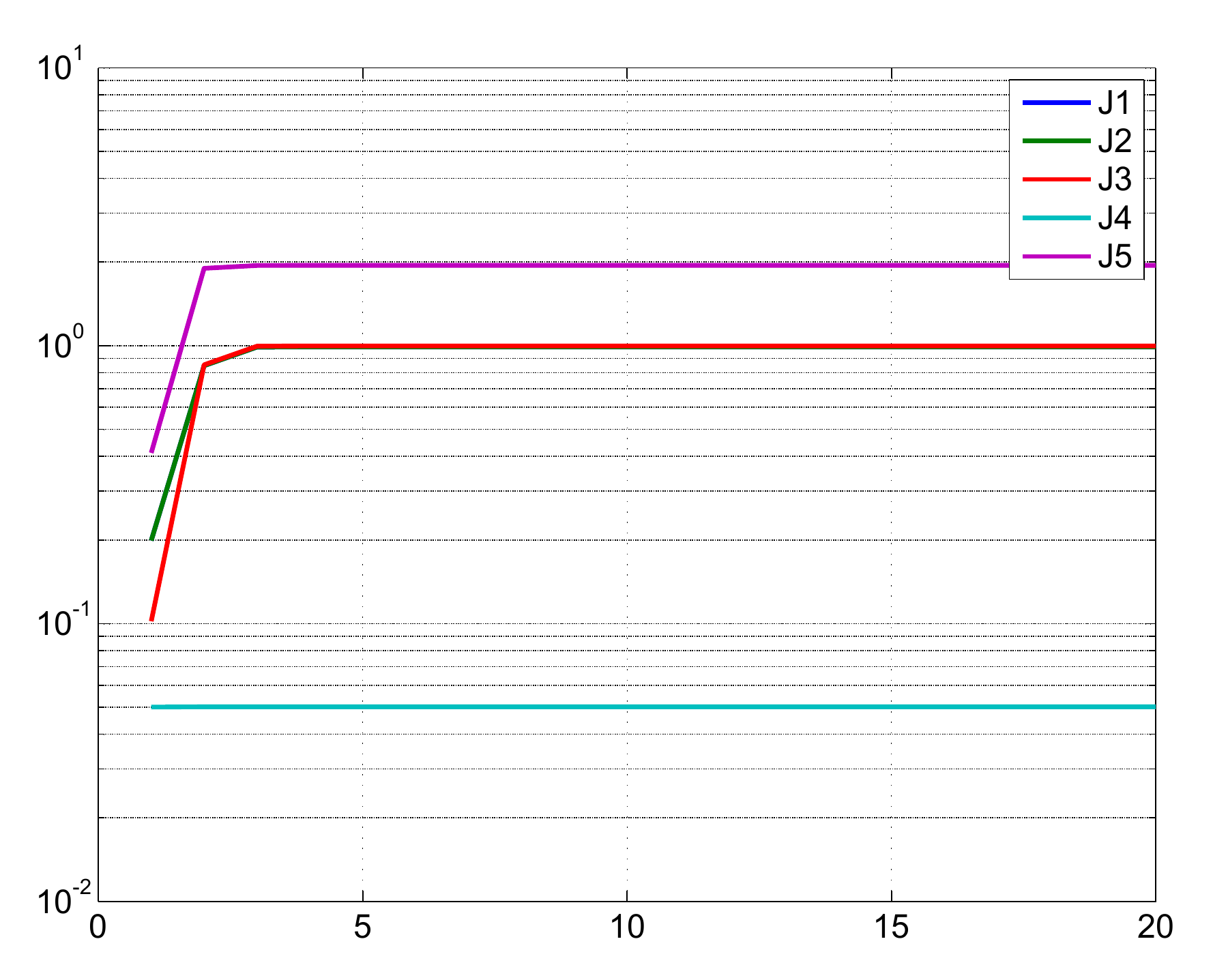}
\caption{Variation of different costs (\textbf{J1-J5}) with iterations}
\label{ramp_cost}
\end{figure}

\begin{figure}[h]
\includegraphics[width=6in,height=4in]{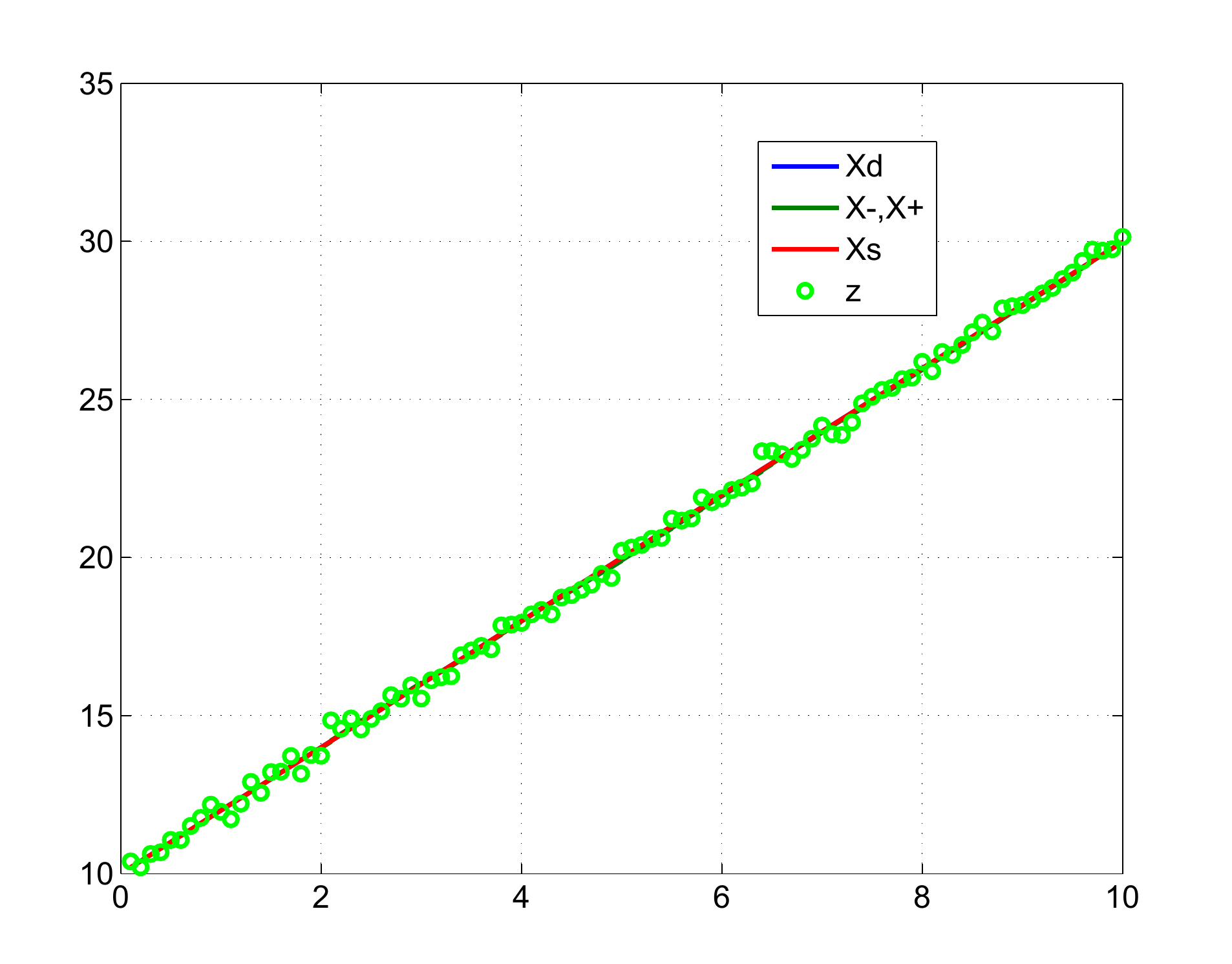}
\caption{Comparison of the predicted dynamics, posterior, smoothed}
\caption*{and the measurement}
\label{ramp_h1}
\end{figure}

\begin{figure}[h]
\includegraphics[width=6in,height=4in]{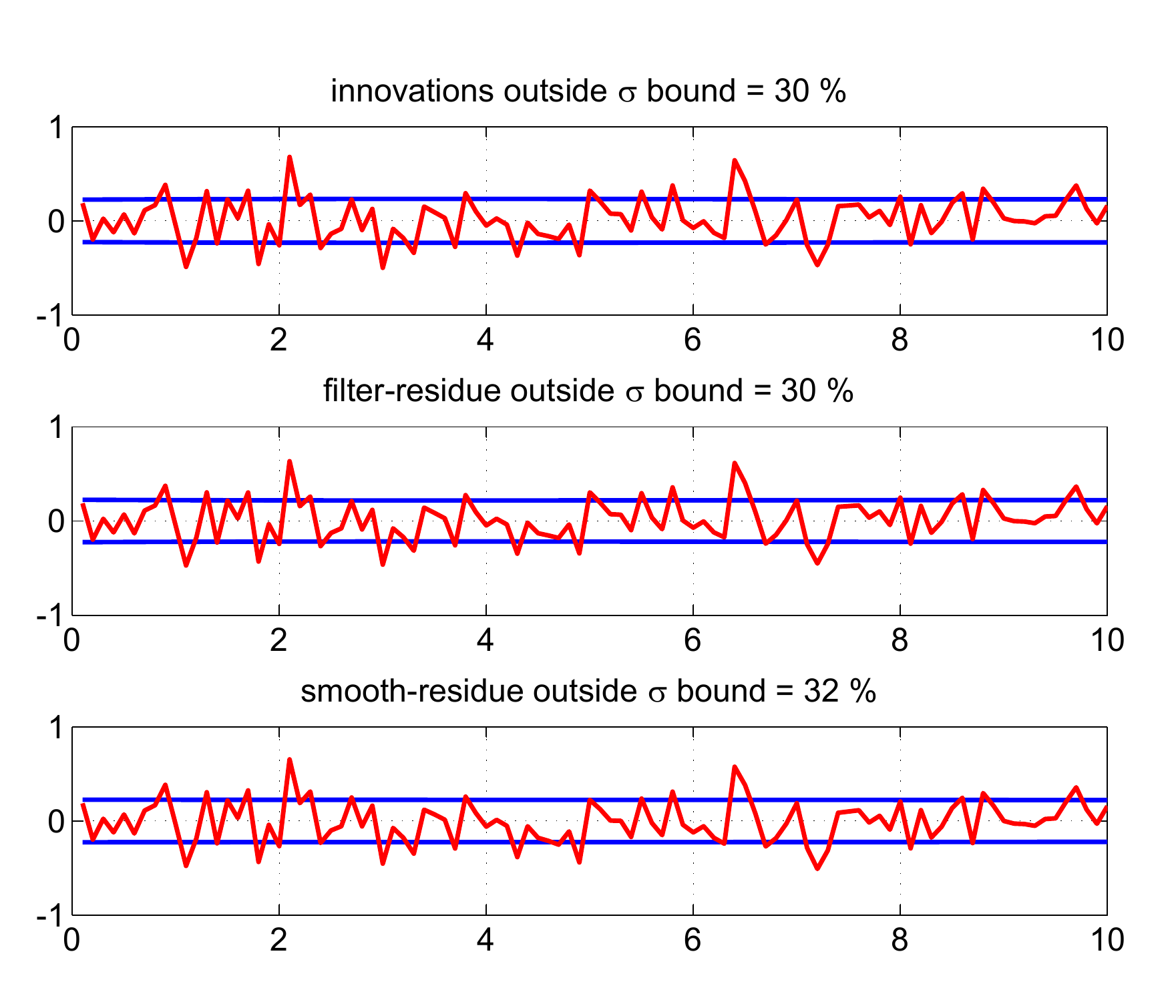}
\caption{The innovations, filtered residue and smoothed residue }
\label{ramp_innov}
\end{figure}

\begin{figure}[h]
\includegraphics[width=6in,height=4in]{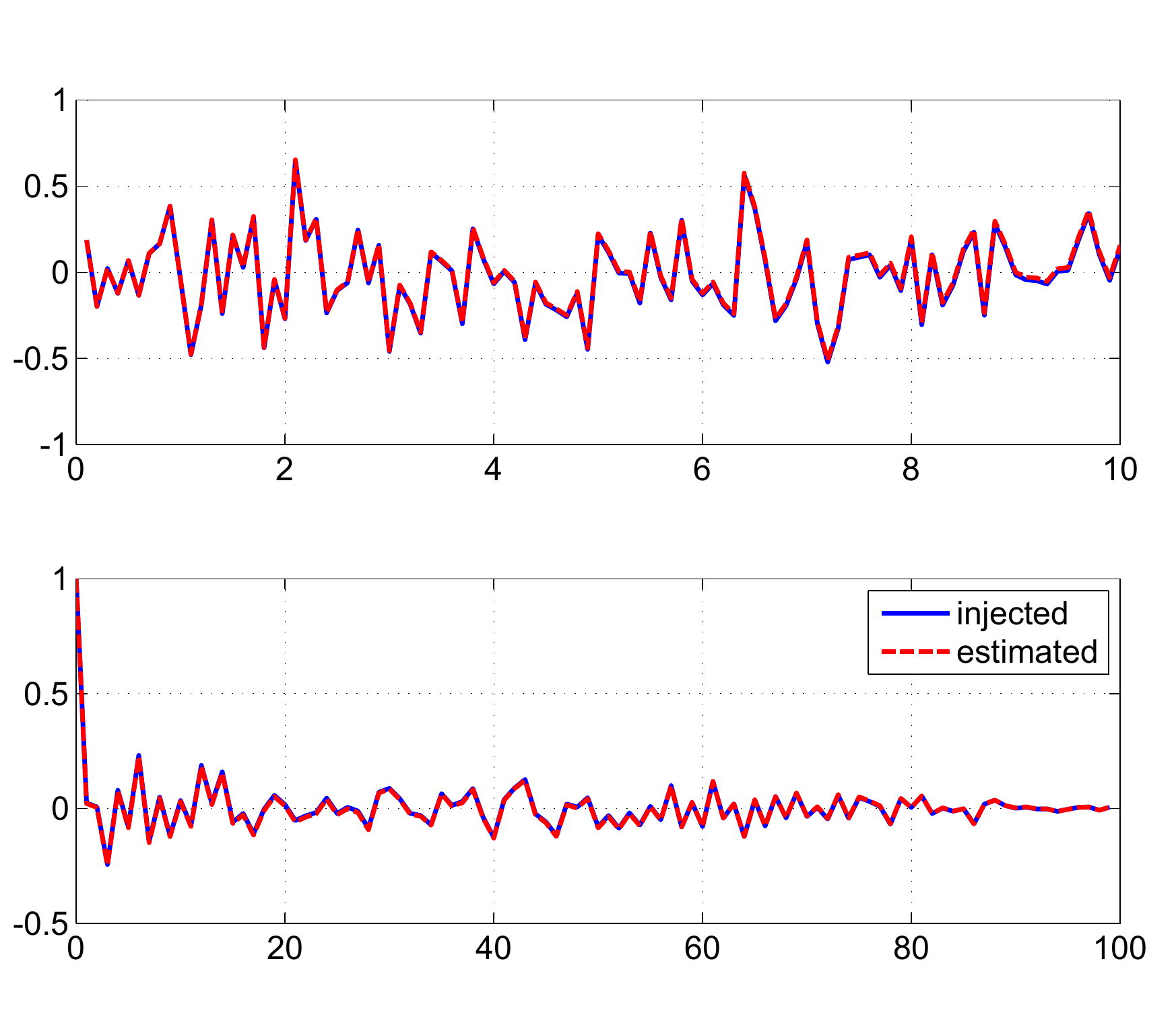}
\caption{Time variation of injected and estimated measurement noise (top) and}
\caption*{their autocorrelation (bottom)}
\label{ramp_mnoise}
\end{figure}

 We first run the filter assuming \textbf{Q} = 0. It was found that about 20 iterations of the data would suffice. The Fig. \ref{ramp_p1}-\ref{ramp_mnoise} refer to the \textbf{Q} = 0 case. The Fig. \ref{ramp_p1} shows the various parameter estimates and its corresponding variances through cumulative time instants with iterations. The variation of the estimated initial parameters and their variances through iterations are shown in Fig. \ref{ramp_P0}. The parameter and the uncertainty reach almost their final estimated values in about 2 and 5 iterations respectively. A similar plot in Fig. \ref{ramp_R} shows the variation of the estimated measurement noise. The variation of different cost functions (\textbf{J1-J5}) through the iterations is shown in Fig. \ref{ramp_cost}. The Fig. \ref{ramp_h1} shows the predicted dynamics, filtered and smoothed estimate at the last iteration. The Fig. \ref{ramp_innov} show the innovations, filtered residue and smoothed residue together with the square root of their variance ($\pm\sigma$ bound). In the EKF approach most of the quantities are Gaussian or approximated as quasi Gaussian and one would expect all the above quantities are close to being Gaussian and hence around one third of the total sample points to be outside the $\sigma$ bound. The injected and estimated measurement noise distributions during the final iteration shown in Fig. \ref{ramp_mnoise} indicate that they are close to each other. Further even their autocorrelations are ideally expected to be close to the Kronecker delta function which provides confidence in the proposed filter algorithm.

\par
\begin{figure}[h]
\includegraphics[width=6in,height=3.2in]{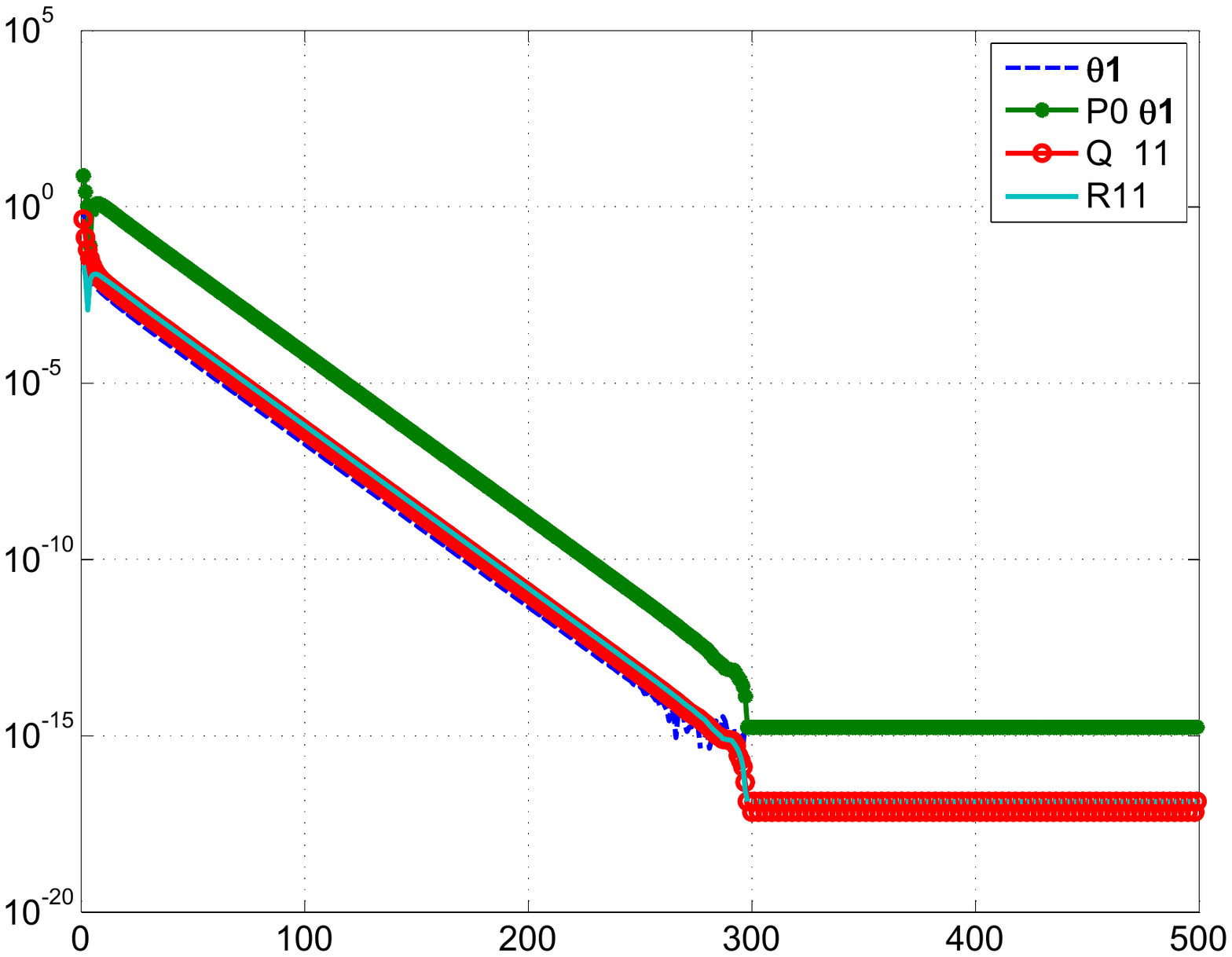}
\caption{The absolute difference between the iterated and final values}
\caption*{with 500 iterations}
\label{ramp_err}
\end{figure}

\begin{figure}[h]
\includegraphics[width=6in,height=3.2in]{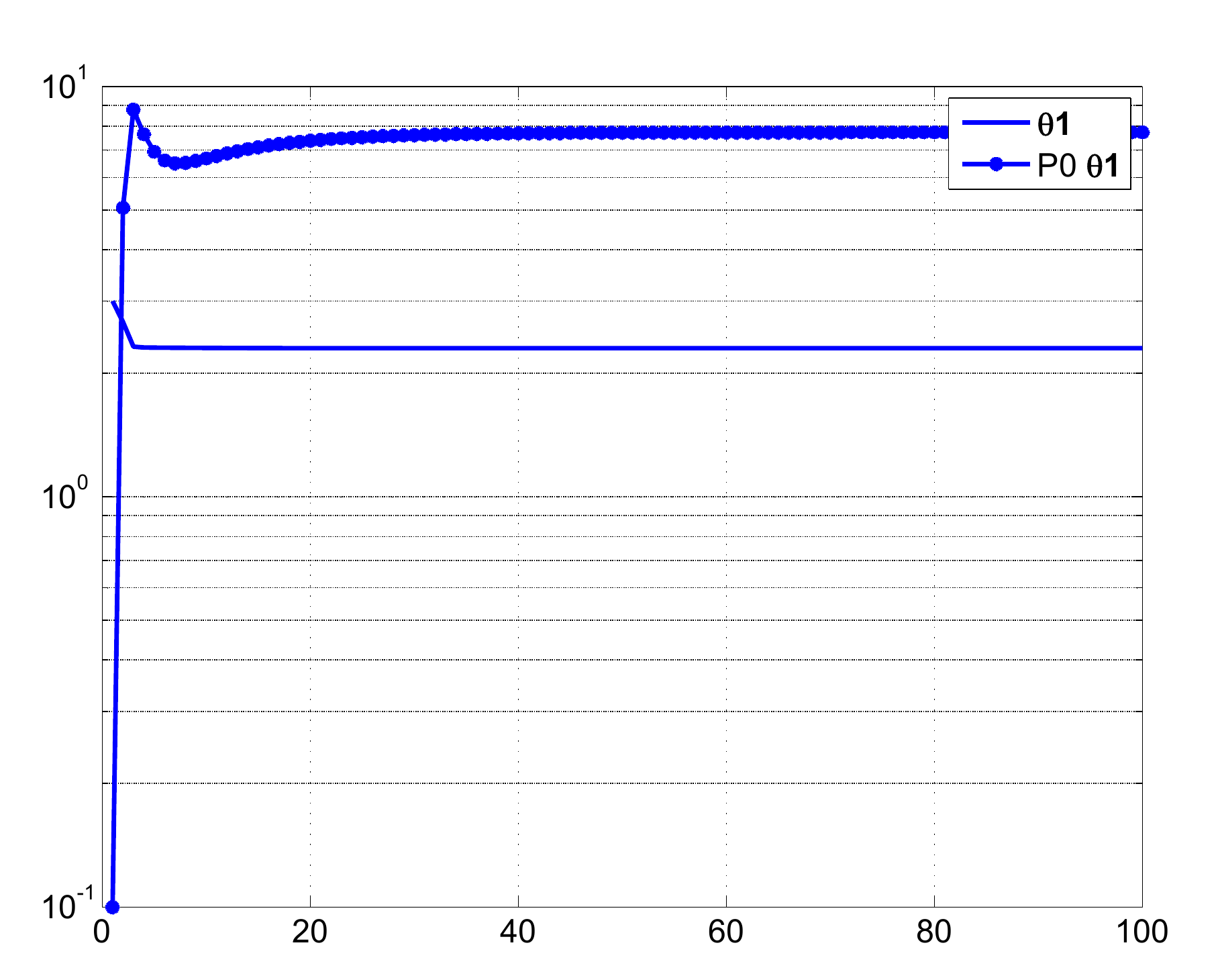}
\caption{Variation of parameter and its initial covariance ($\mathbf{P_0}$) with iterations}
\label{rampQ_P0}
\end{figure}

\begin{figure}[h]
\includegraphics[width=6in,height=4in]{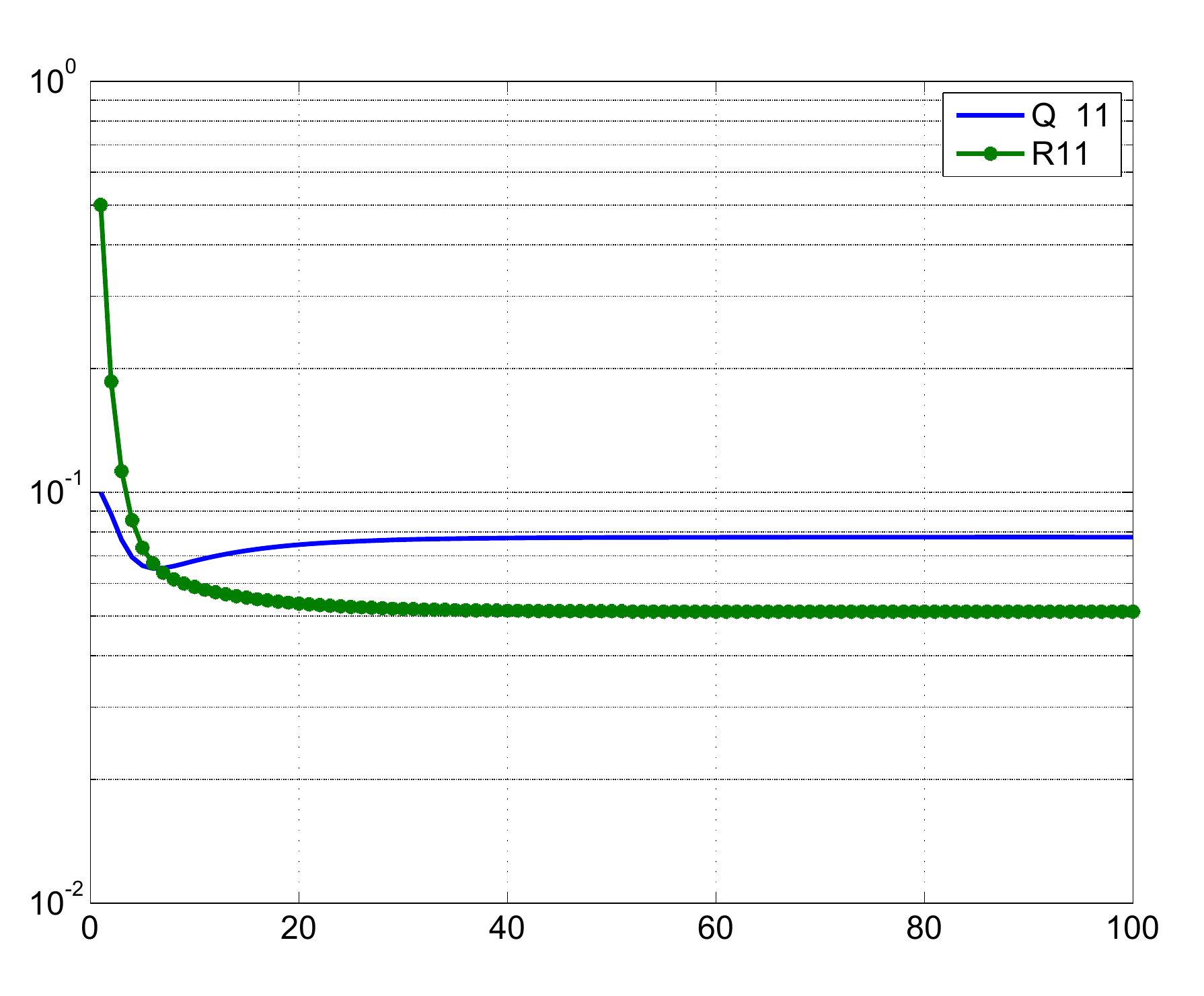}
\caption{Variation of \textbf{Q} and \textbf{R} with iterations}
\label{rampQ_R}
\end{figure}

\begin{figure}[h]
\includegraphics[width=6in,height=4in]{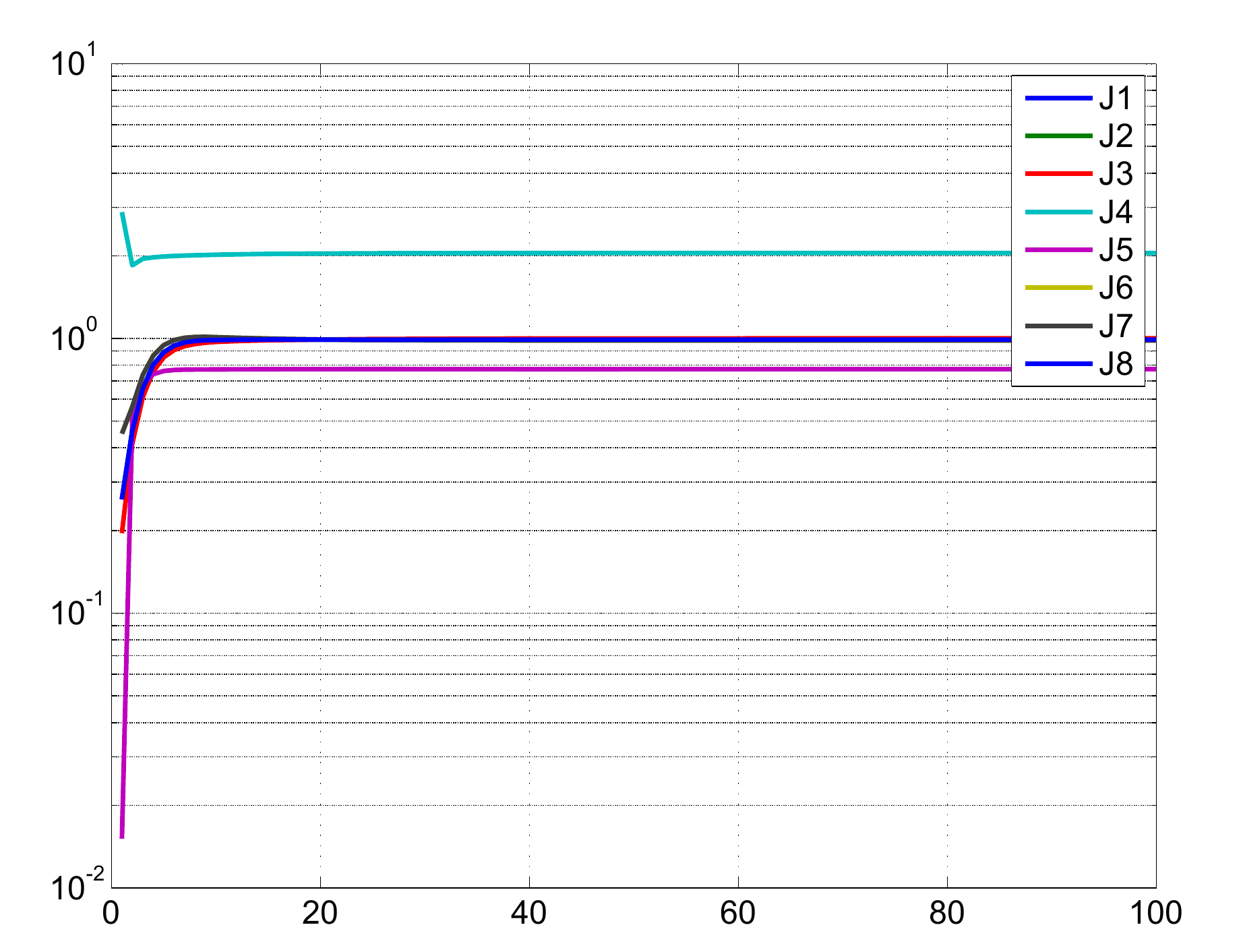}
\caption{Variation of different costs (\textbf{J1-J8}) with iterations}
\label{rampQ_J}
\end{figure}

\begin{figure}[h]
\includegraphics[width=6in,height=4in]{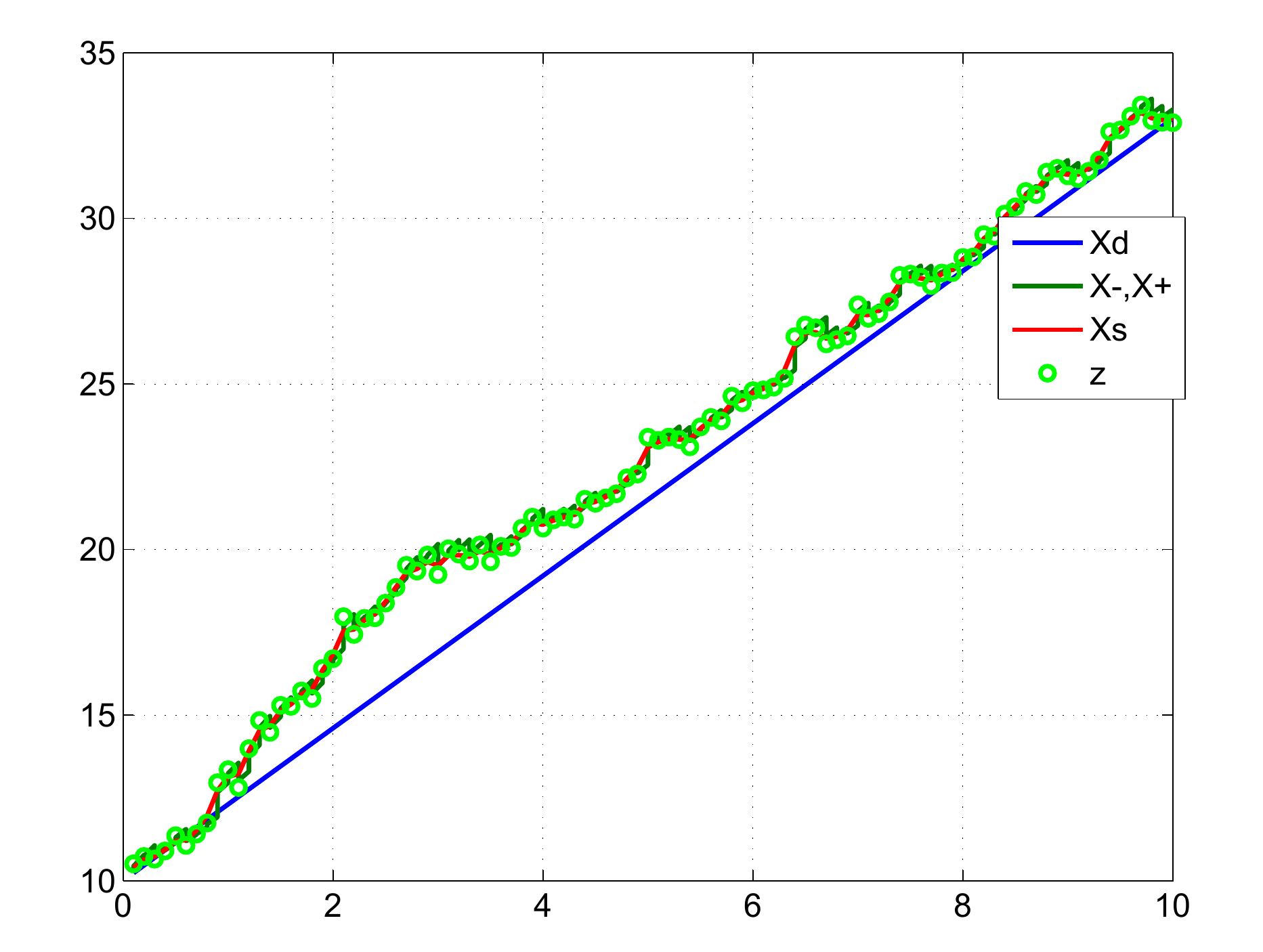}
\caption{Comparison of the predicted dynamics, posterior, smoothed}
\caption*{and the measurement}
\label{rampQ_h}
\end{figure}

\begin{figure}[h]
\includegraphics[width=6in,height=4in]{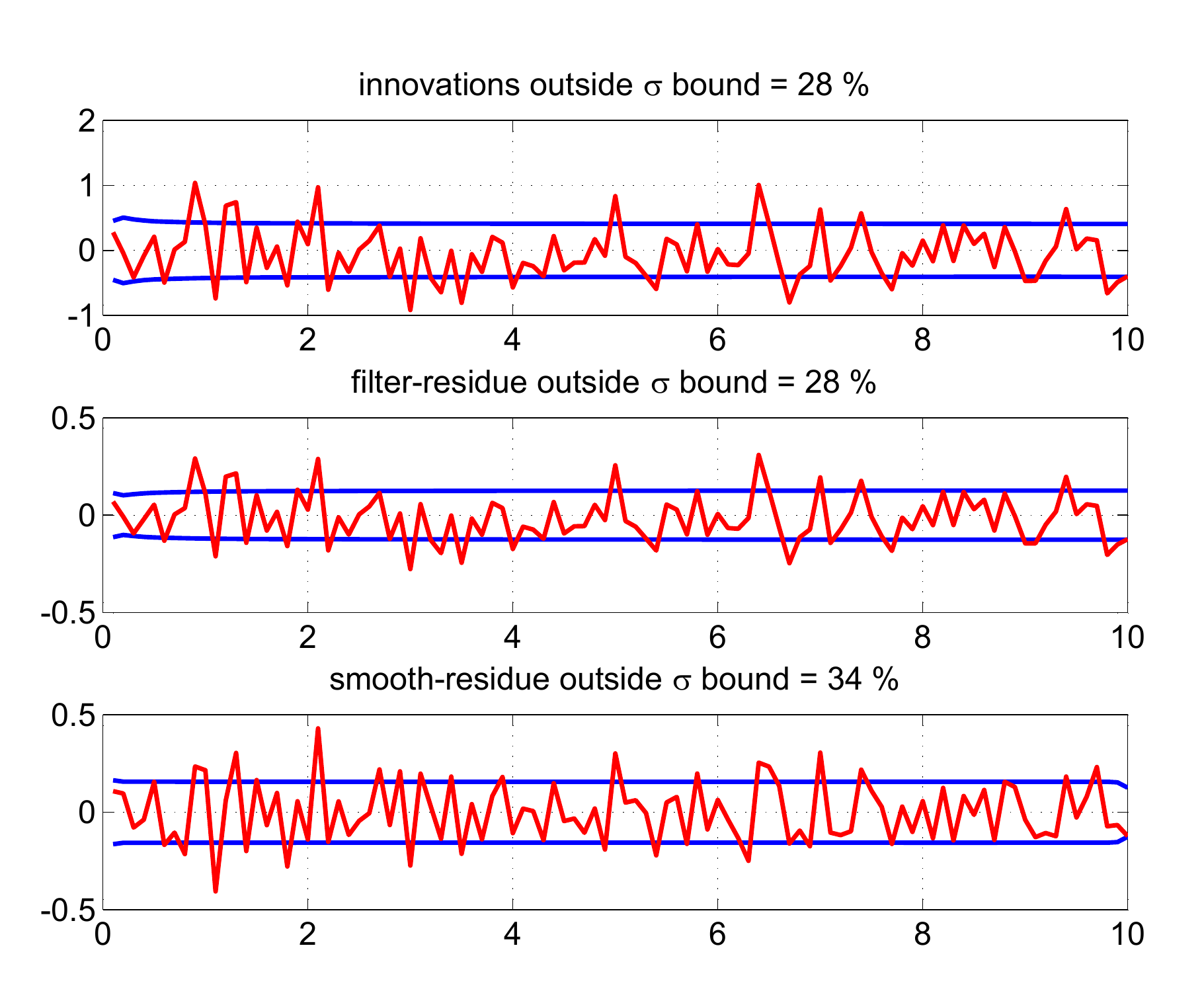}
\caption{The innovations, filtered residue and smoothed residue }
\label{rampQ_innov}
\end{figure}

\begin{figure}[h]
\includegraphics[width=6in,height=4in]{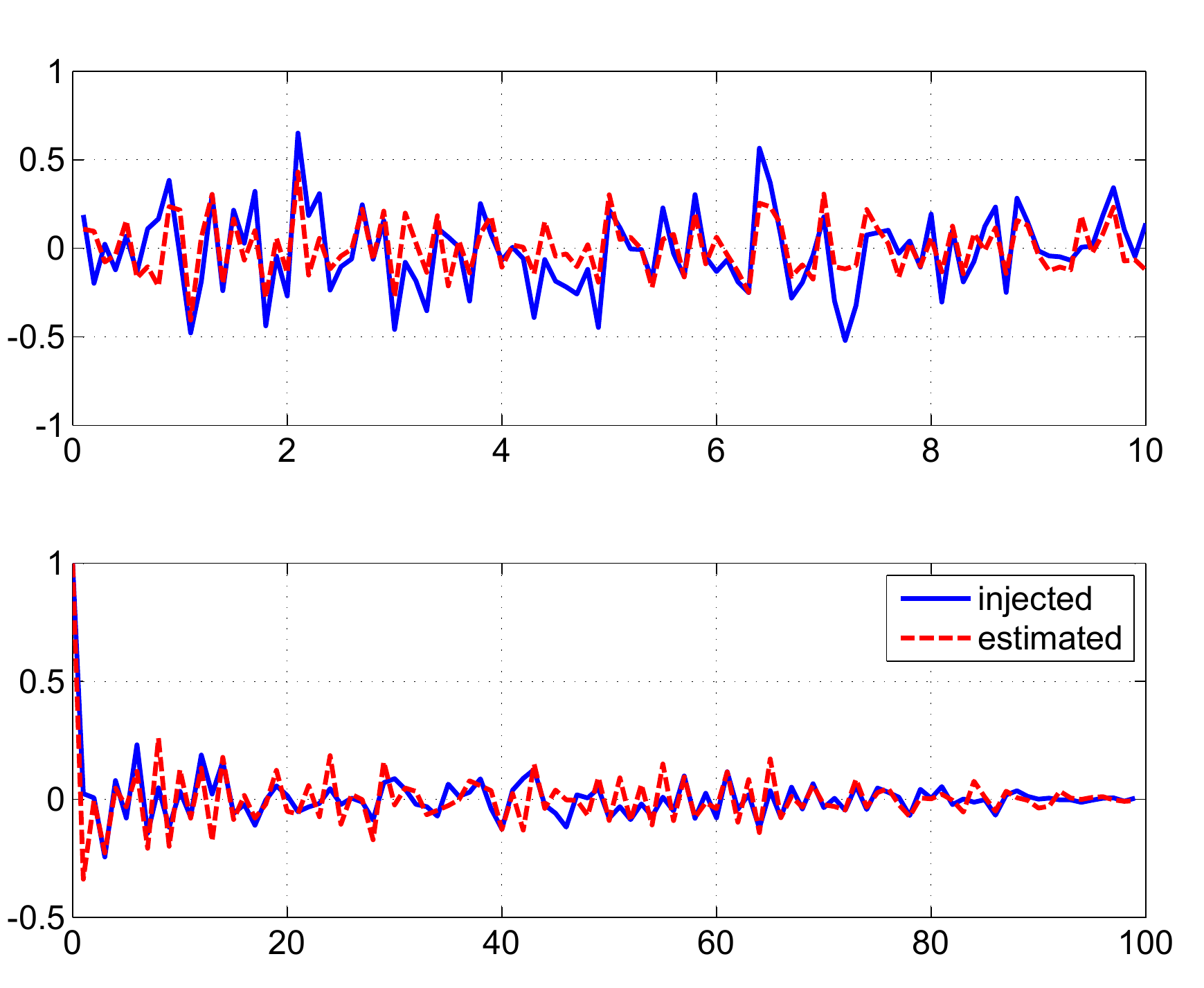}
\caption{Time variation of injected and estimated measurement noise (top) and}
\caption*{their autocorrelation (bottom)}
\label{rampQ_mnoise}
\end{figure}

\begin{figure}[h]
\includegraphics[width=6in,height=4in]{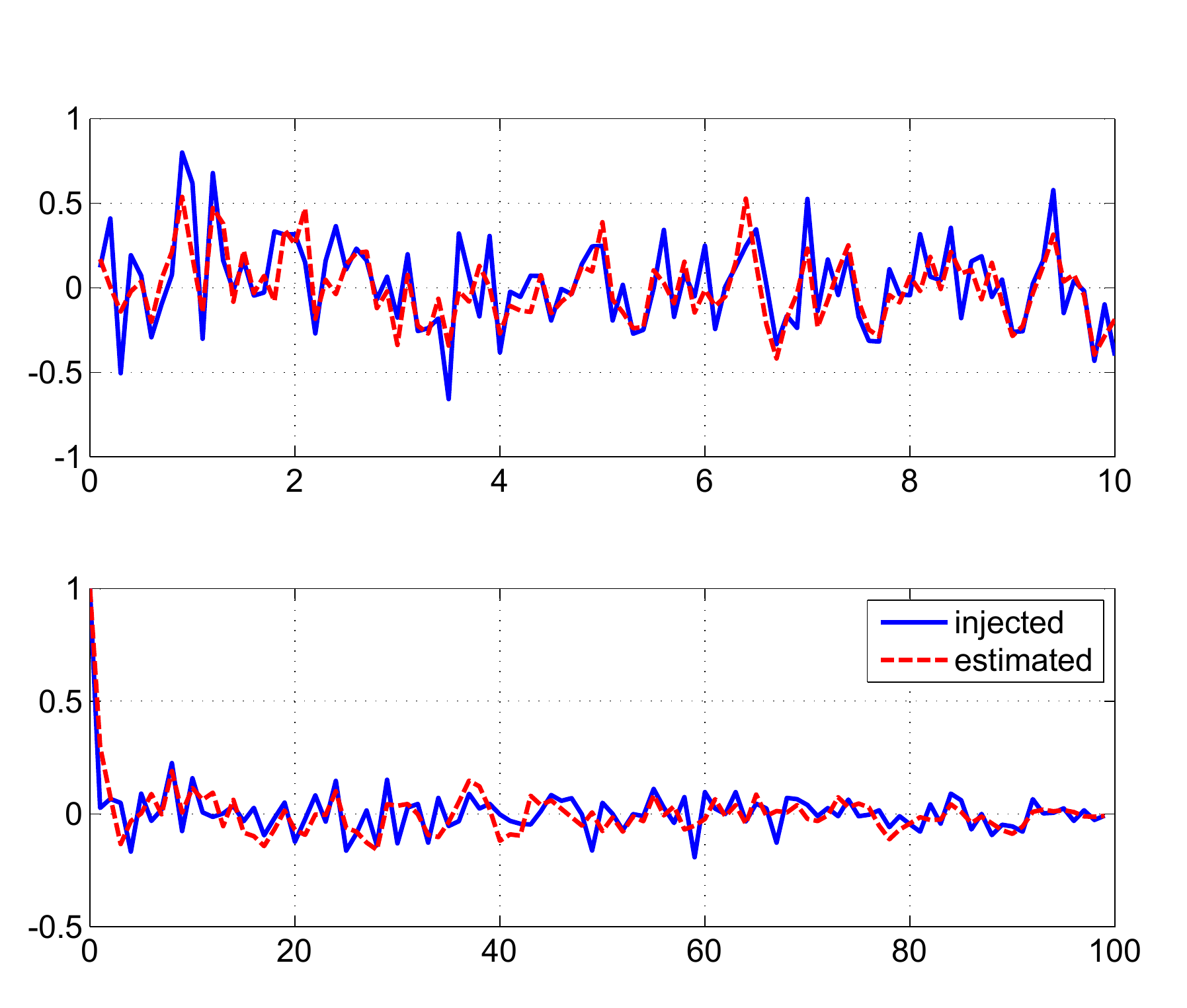}
\caption{Time variation of injected and estimated process noise (top) and}
\caption*{their autocorrelation (bottom)}
\label{rampQ_pnoise}
\end{figure}

 The next step is to process the data with process noise (\textbf{Q} $>$ 0). The Fig. \ref{ramp_err}-\ref{rampQ_pnoise} refer to the \textbf{Q} $>$ 0 case. The Fig. \ref{ramp_err} shows the absolute difference between the iterated and final values with iterations which indicates the accuracy level that one needs and it was found that 100 iterations are required. The variation of the estimated initial parameters and their variances through iterations are shown in Fig. \ref{rampQ_P0}. The parameter and the uncertainty reach almost their final estimated values in about 5 and 20 iterations respectively. A similar plot in Fig. \ref{rampQ_R} shows the variation of the estimated measurement and process noise. The variation of different cost functions (\textbf{J1-J8}) through the iterations are shown in Fig. \ref{rampQ_J}. The cost functions \textbf{J1-J3} correspond to the number of measurement ($m$=1) and in presence of process noise, \textbf{J6-J8} correspond to the number of states ($n$=1). The \textbf{J4} in absence of process noise corresponds to the trace of the measurement noise \textbf{R}. The \textbf{J5} is the negative log likelihood cost whose absolute value is shown in the plot. There is a mismatch in the predicted dynamics and the measurement as seen in Fig. \ref{rampQ_h} indicating the presence of process noise. The subsequent Fig. \ref{rampQ_innov} correspond to the earlier Fig. \ref{ramp_innov} of \textbf{Q} = 0 case. The Fig. \ref{rampQ_mnoise} and Fig. \ref{rampQ_pnoise} shows respectively the injected and estimated measurement and process noise samples across time during the final iteration.

\begin{landscape}
\begin{table}[h]
\subsection{Ramp System Tables (\textbf{Q} = 0) }
\vspace{14pt}
\caption{Sensitivity Study : (\textbf{Q} = 0) : Ramp system.\\ No. of iterations=20, No. of simulations=50.}{}
\label{tbramp}
\begin{center}
\begin{footnotesize}
\begin{tabular}{| c | c | c | c | c | c | c | c | c | c | c | c |}
\hline
Study &
$\Theta$ ratio &
\makecell{CRB ratio\\NR/EKF} &
\makecell{Consistency\\ ratio-EKF}&
\makecell{Consistency\\ ratio-NR}&
\makecell{Spread\\factor\\EKF} &
\makecell{Spread\\factor\\NR} &
\makecell{\textbf{R} ratio\\EKF/True} & \makecell{\textbf{R} ratio\\EKF/NR} &
\makecell{$\mu$\ of \\\textbf{J1-J5}} &
\makecell{$\sigma$ of \\\textbf{J1-J5}} &
Remarks
\\ \hline

\multicolumn{12}{|c|}{Reference adaptive EKF used for non zero process noise case} \\ \hline
\makecell{$\mathbf{P_0}$ : Scaled up-[0,0;0,\checkmark]\\\textbf{Q} : EM-[\checkmark,0;0,0] \\ \textbf{R} : EM-diag}
&\multicolumn{11}{c|}{Extremely slow convergence of \textbf{Q} taking more than 100 iterations.} \\ \hline

\multicolumn{12}{|c|}{Proposed reference adaptive EKF for zero process noise case, \textbf{Q} is diagonal matrix fixed to $10^-10$ for all states} \\ \specialrule{.2em}{.0em}{.0em}

\multicolumn{1}{!{\vrule width 2pt}c|}{
\makecell{$\mathbf{P_0}$ : Scaled up-[0,0;0,\checkmark]\\\textbf{Q} : $10^{-10}$-[\checkmark,0;0,0] \\ \textbf{R} : EM-diag} } &
 0.9996 &   1.0044 &
0.9623 &   0.9581 &
   0.2538 &      0.2544 &
  0.9761 &   1.0013  &
\makecell{ 0.9901 \\   0.9901  \\  0.9998   \\ 0.0483 \\  -1.9933 }&
\makecell{ 0.0000  \\  0.0000  \\  0.0012  \\  0.0065   \\ 0.1399   } &
\multicolumn{1}{|c!{\vrule width 2pt}}{
\makecell{$\sigma_{  \Theta}$= 0.0036\\SIGMA$_{avg}$= 0.0038}} \\ \specialrule{.2em}{.0em}{.0em}


\multicolumn{12}{|c|}{Using different method and options for $\mathbf{P_0}$ } \\
\hline

\end{tabular}
\end{footnotesize}
\end{center}
\end{table}
\end{landscape}

\addtocounter{table}{-1}
\begin{landscape}
\begin{table}[h]
\caption{Sensitivity Study : (\textbf{Q} = 0) : Ramp system.\\ No. of iterations=20, No. of simulations=50.}{}
\begin{center}
\begin{footnotesize}
\begin{tabular}{|c| c| c| c| c| c| c|c|c|c|c|c|}
\hline
Study &
$\Theta$ ratio &
\makecell{CRB ratio\\NR/EKF} &
\makecell{Consistency\\ ratio-EKF}&
\makecell{Consistency\\ ratio-NR}&
\makecell{Spread\\factor\\EKF} &
\makecell{Spread\\factor\\NR} &
\makecell{\textbf{R} ratio\\EKF/True} & \makecell{\textbf{R} ratio\\EKF/NR} &
\makecell{$\mu$\ of \\\textbf{J1-J5}} &
\makecell{$\sigma$ of \\\textbf{J1-J5}} &
Remarks
\\ \hline

\makecell{$\mathbf{P_0}$ : IIM-diag*\\\textbf{Q} : $10^{-10}$-[\checkmark,0;0,0] \\ \textbf{R} : EM-diag}  &
 0.9996 &     0.5024  &
 0.7839 &     0.9581  &
 0.4656 &     0.2544  &
 0.9803 &     1.0056  &
\makecell{ 0.9804  \\  0.9804  \\  0.9996   \\ 0.0481 \\  -1.9010 }&
\makecell{ 0.0000  \\  0.0000 \\   0.0011 \\   0.0065  \\  0.1404} &
\makecell{IIM taken only for dynamic \\states and unity $\mathbf{P_0}$ \\ for parameters}\\ \hline


\makecell{$\mathbf{P_0}$ : Smoothed-[0,0;0,\checkmark]\\\textbf{Q} : $10^{-10}$-[\checkmark,0;0,0] \\ \textbf{R} : EM-diag}  &
 0.9996 &      4.3583  &
 4.1745 &     0.9581  &
 0.1599 &     0.2544  &
 0.9670 &     0.9919  &
\makecell{0.9994 \\   0.9994  \\  0.9999  \\  0.0483  \\ -2.0388 }&
\makecell{ 0.0000 \\   0.0000  \\  0.0001  \\  0.0065  \\  0.1399 }&
\makecell{Lower filter CRB\\ than that of the NR}\\ \hline


\makecell{$\mathbf{P_0}$ : Smoothed-diag\\\textbf{Q} : $10^{-10}$-[\checkmark,0;0,0] \\ \textbf{R} : EM-diag}  &
0.9997 &     3.9208  &
6.0621 &     0.9581  &
0.2353 &     0.2544  &
0.9624 &      0.9872  &
\makecell{ 0.9986  \\  0.9986  \\  0.9999  \\  0.0481 \\  -2.0438}&
\makecell{0.0000 \\   0.0000  \\  0.0001  \\  0.0065  \\  0.1418  }&
\makecell{Lower filter CRB\\ than that of the NR}\\
\end{tabular}
\end{footnotesize}
\end{center}
\end{table}
\end{landscape}

\addtocounter{table}{-1}
\begin{landscape}
\begin{table}[h]
\caption{Sensitivity Study : (\textbf{Q} = 0) : Ramp system.\\ No. of iterations=20, No. of simulations=50.}{}
\begin{center}
\begin{footnotesize}
\begin{tabular}{|c| c| c| c| c| c| c|c|c|c|c|c|}
\hline
Study &
$\Theta$ ratio &
\makecell{CRB ratio\\NR/EKF} &
\makecell{Consistency\\ ratio-EKF}&
\makecell{Consistency\\ ratio-NR}&
\makecell{Spread\\factor\\EKF} &
\makecell{Spread\\factor\\NR} &
\makecell{\textbf{R} ratio\\EKF/True} & \makecell{\textbf{R} ratio\\EKF/NR} &
\makecell{$\mu$\ of \\\textbf{J1-J5}} &
\makecell{$\sigma$ of \\\textbf{J1-J5}} &
Remarks
\\ \hline

\multicolumn{12}{|c|}{Using different methods for estimating \textbf{R}} \\ \hline

\makecell{$\mathbf{P_0}$ : Scaled up-[0,0;0,\checkmark]\\\textbf{Q} : $10^{-10}$-[\checkmark,0;0,0] \\ \textbf{R} : diag \textbf{R}*}  &
 0.9996 &     1.0050  &
 0.9629 &     0.9581  &
  0.2537 &     0.2544  &
 0.9749 &      1.0000  &
\makecell{ 0.9914  \\  0.9914  \\  1.0012  \\  0.0483 \\  -1.9933 }&
\makecell{ 0.0026  \\  0.0026  \\  0.0029  \\  0.0065 \\   0.1399 }&
\makecell{*Using dynamical residue\\ as per Eq-\ref{R3}}\\ \hline


\makecell{$\mathbf{P_0}$ : Scaled up-[0,0;0,\checkmark]\\\textbf{Q} : $10^{-10}$-[\checkmark,0;0,0] \\\textbf{R} : diag \textbf{R}*}  &
 0.9996 &     1.0093  &
 0.9670 &     0.9581  &
 0.2530 &     0.2544  &
 0.9664 &     0.9913  &
\makecell{ 1.0000  \\  1.0000 \\   1.0098  \\  0.0483  \\ -1.9932}&
\makecell{   0.0000   \\ 0.0000  \\  0.0013   \\ 0.0065 \\   0.1399    }&
\makecell{*Using smoothed residue\\ignoring $2^{nd}$ order\\terms in Eq-\ref{R1}}\\
\end{tabular}
\end{footnotesize}
\end{center}
\end{table}
\end{landscape}

\begin{landscape}
\subsection{Ramp System Tables (\textbf{Q} $>$ 0) }
\vspace{14pt}
\begin{table}[h]
\caption{Sensitivity Study : \textbf{Q}$>0$ : Ramp system.\\ No. of iterations=100, No. of simulations=50.}{}
\label{rampQ}
\begin{center}
\begin{footnotesize}
\begin{tabular}{|c| c| c| c| c| c| c|c|c|c|c|}
\hline
Study &
$\Theta$ ratio &
\makecell{PCRB\\ ratio} &
\makecell{Consistency\\ ratio-EKF}&
\makecell{Spread\\factor\\EKF} &
\makecell{\textbf{R} ratio\\EKF/True} &
\makecell{\textbf{Q} ratio\\EKF/True}  &
\makecell{$\mu$\ of \\\textbf{J1-J8}} &
\makecell{$\sigma$ of \\\textbf{J1-J8}} &
Remarks
\\ \hline

\multicolumn{10}{|c|}{Proposed reference adaptive EKF for non zero process noise} \\ \specialrule{.2em}{.0em}{.0em}

\multicolumn{1}{!{\vrule width 2pt}c|}{
\makecell{$\mathbf{P_0}$ : Scaled up-[0,0;0,\checkmark]\\\textbf{Q} : EM-[\checkmark,0;0,0] \\\textbf{R} : EM-diag}}  &
 1.0220 &  1.0275 &
 0.9981 &  14.9505 &
 0.9362 &  1.0279 &
\makecell{0.9898  \\  0.9898 \\   1.0003  \\  0.8370  \\ -1.0451 \\   0.9778  \\  0.9778 \\ 0.9898} &
\makecell{  0.0018 \\   0.0018  \\  0.0018 \\   0.5799 \\   0.1263  \\  0.0090  \\  0.0090 \\ 0.0018} &
\multicolumn{1}{|c!{\vrule width 2pt}}{
\makecell{$\sigma_{  \Theta}$= 0.2219\\SIGMA$_{avg}$=  0.2224}} \\ \specialrule{.2em}{.0em}{.0em}

\multicolumn{1}{!{\vrule width 2pt}c|}{
\makecell{$\mathbf{P_0}$ : Scaled up-[0,0;0,\checkmark]\\\textbf{Q} : EM-[\checkmark,0;0,0] \\\textbf{R} : EM-diag}}  &
 1.0221 &  0.9938 &
 1.0070 &  14.9282 &
 NA &  0.9979 &
\makecell{0.9848 \\   0.9848 \\   0.9876 \\   0.8365 \\  -1.0369  \\  0.9783 \\   0.9783\\ 0.9848} &
\makecell{0.0590  \\  0.0590 \\   0.1065  \\  0.5726 \\   0.1245 \\   0.0056  \\  0.0056\\ 0.0590} &
\multicolumn{1}{|c!{\vrule width 2pt}}{
\makecell{\textbf{R} is known}} \\ \specialrule{.2em}{.0em}{.0em}

\end{tabular}
\end{footnotesize}
\end{center}
\end{table}
\end{landscape}

\addtocounter{table}{-1}
\begin{landscape}
\begin{table}[h]
\caption{Sensitivity Study : \textbf{Q}$>0$ : Ramp system.\\ No. of iterations=100, No. of simulations=50.}{}
\begin{center}
\begin{footnotesize}
\begin{tabular}{|c| c| c| c| c| c| c|c|c|c|c|}
\hline
Study &
$\Theta$ ratio &
\makecell{PCRB\\ ratio} &
\makecell{Consistency\\ ratio-EKF}&
\makecell{Spread\\factor\\EKF} &
\makecell{\textbf{R} ratio\\EKF/True} &
\makecell{\textbf{Q} ratio\\EKF/True}  &
\makecell{$\mu$\ of \\\textbf{J1-J8}} &
\makecell{$\sigma$ of \\\textbf{J1-J8}} &
Remarks
 \\ \specialrule{.2em}{.0em}{.0em}

\multicolumn{1}{!{\vrule width 2pt}c|}{
\makecell{$\mathbf{P_0}$ : Zero\\\textbf{Q} : EM-diag \\\textbf{R} : EM-diag}}  &
\makecell{NA} &
 1.0309 &
NA & NA &
 0.9355 &   1.0280 &
\makecell{ 1.0000  \\  1.0000 \\   1.0006  \\  2.5377 \\  -1.0804 \\   1.0001 \\   1.0001 \\ 1.0000} &
\makecell{ 0.0001\\    0.0001\\    0.0051   \\ 3.5893  \\  0.1198  \\  0.0067  \\  0.0067 \\ 0.0001} &
\multicolumn{1}{|c!{\vrule width 2pt}}{
\makecell{The parameters are \\assumed to be known}} \\ \specialrule{.2em}{.0em}{.0em}

\multicolumn{1}{!{\vrule width 2pt}c|}{
\makecell{$\mathbf{P_0}$ : Zero\\\textbf{Q} : EM-diag \\\textbf{R} : Known}}  &
\makecell{NA} &  0.9963 &
NA & NA &
0.9670 &   0.9989 &
\makecell{0.9945  \\  0.9945  \\  0.9869  \\  2.5377 \\  -1.0718  \\  1.0004  \\  1.0004 \\ 0.9945} &
\makecell{ 0.0618 \\   0.0618  \\  0.1119  \\  3.5893  \\  0.1180  \\  0.0051  \\  0.0051 \\ 0.0618} &
\multicolumn{1}{|c!{\vrule width 2pt}}{
\makecell{The parameters  and \textbf{R} are \\assumed to be known}} \\ \specialrule{.2em}{.0em}{.0em}

\end{tabular}
\end{footnotesize}
\end{center}
\end{table}
\end{landscape}

\addtocounter{table}{-1}
\begin{landscape}
\begin{table}[h]
\caption{Sensitivity Study : \textbf{Q} $>$ 0 : Ramp system.\\ No. of iterations=100, No. of simulations=50.}{}
\begin{center}
\begin{footnotesize}
\begin{tabular}{|c| c| c| c| c| c| c|c|c|c|c|}
\hline
Study &
$\Theta$ ratio &
\makecell{PCRB\\ ratio} &
\makecell{Consistency\\ ratio-EKF}&
\makecell{Spread\\factor\\EKF} &
\makecell{\textbf{R} ratio\\EKF/True} &
\makecell{\textbf{Q} ratio\\EKF/True}  &
\makecell{$\mu$\ of \\\textbf{J1-J8}} &
\makecell{$\sigma$ of \\\textbf{J1-J8}} &
Remarks
\\ \hline

\multicolumn{10}{|c|}{Extended EM algorithm (Bavdekar et al. \cite{Bavdekar2011} (2011))} \\ \hline
\makecell{$\mathbf{P_0}$ : Smoothed-full\\\textbf{Q} : EM-full \\\textbf{R} : EM-full}  & \multicolumn{9}{c|}{\makecell{There is no consistency in the parameter estimates since the smoothed $\mathbf{P_0}$ simply decreases through iterations.}}  \\ \hline

\multicolumn{10}{|c|}{Modification-1} \\ \hline

\makecell{$\mathbf{P_0}$ : Smoothed-diag\\\textbf{Q} : EM-diag \\\textbf{R} : EM-diag}   &
1.0273 &  1.0314 &
10.3040 &   8.7111 &
0.9753 &  0.9389&
\makecell{ 0.9993  \\  0.9993  \\  1.0004 \\   0.8505  \\ -1.1022  \\  1.0041  \\  1.0041 \\ 0.9993} &
\makecell{  0.0017  \\  0.0017  \\  0.0017  \\  0.5852 \\   0.1277  \\  0.0030   \\ 0.0030 \\ 0.0017} &
\makecell{No stoping condition.\\ The low spread factor is deceiving\\since there is no consistency in the\\ parameter estimates with its covariance}\\ \hline

\multicolumn{10}{|c|}{Modification-2} \\ \hline

\makecell{$\mathbf{P_0}$ : Scaled up-[0,0;0,\checkmark]\\\textbf{Q} : EM-diag \\\textbf{R} : EM-diag}   &
 1.0220 &   1.0275 &
0.9981 &   14.9505 &
0.9362 &   1.0279 &
\makecell{  0.9898\\    0.9898 \\   1.0003  \\  0.8370  \\ -1.0451 \\   0.9778  \\  0.9778 \\ 0.9898} &
\makecell{0.0018  \\  0.0018 \\   0.0018 \\   0.5799  \\  0.1263  \\  0.0090  \\  0.0090 \\ 0.0018} &
\makecell{The slight differences from \\the reference case is due to the small \textbf{Q} \\that is generated for the parameters}\\ \hline

\multicolumn{10}{|c|}{Modification-3} \\
\end{tabular}
\end{footnotesize}
\end{center}
\end{table}
\end{landscape}

\addtocounter{table}{-1}
\begin{landscape}
\begin{table}[H]
\caption{Sensitivity Study : \textbf{Q} $>$ 0 : Ramp system.\\ No. of iterations=100, No. of simulations=50.}{}
\begin{center}
\begin{footnotesize}
\begin{tabular}{|c| c| c| c| c| c| c|c|c|c|}
\hline
Study &
$\Theta$ ratio &
\makecell{PCRB\\ ratio} &
\makecell{Consistency\\ ratio-EKF}&
\makecell{Spread\\factor\\EKF} &
\makecell{\textbf{R} ratio\\EKF/True} &
\makecell{\textbf{Q} ratio\\EKF/True}  &
\makecell{$\mu$\ of \\\textbf{J1-J8}} &
\makecell{$\sigma$ of \\\textbf{J1-J8}} &
Remarks
\\ \hline

\multicolumn{10}{|c|}{MS method (Mohamed and Schwarz)} \\ \hline
\makecell{$\mathbf{P_0}$ : Scaled up-[0,0;0,\checkmark]\\\textbf{Q} : MS-diag \\\textbf{R} : MS-diag}  & \multicolumn{9}{c|}
{\makecell{\textbf{R} can converge to a very large value and the cost functions diverge as seen in Fig. \ref{MS_ramp_Q},\ref{MS_ramp_J}. \\ There is no convergence in cost even if $\Theta$ is known.}}\\ \hline

\multicolumn{10}{|c|}{MS method with \textbf{Q} being estimated using only last N/2 samples} \\ \hline
\makecell{$\mathbf{P_0}$ : Zero\\\textbf{Q} : MS-diag \\\textbf{R} : Known} &
 NA &  1.0100 &
 NA & NA &
 NA & 0.9927 &
\makecell{ 1.0354  \\  1.0354  \\  1.0071  \\  2.5377 \\  -1.0504  \\  1.0983 \\   1.0983 \\ 1.0354} &
\makecell{0.1629  \\  0.1629 \\   0.1304  \\  3.5893  \\  0.1170   \\ 0.2726   \\ 0.2726 \\ 0.1629} &
\makecell{\textbf{R}, $\Theta$ is known}\\ \hline

\multicolumn{10}{|c|}{MS method assuming \textbf{R} is known and \textbf{Q} is estimated using only last N/2 samples} \\ \hline
\makecell{$\mathbf{P_0}$ : Scaled up-[0,0;0,\checkmark]\\\textbf{Q} : MS-[\checkmark,0;0,0] \\\textbf{R} : Known} &
 1.0221 &  0.9804 &
  0.9599 & 15.4046 &
 NA & 1.0968 &
\makecell{ 0.9447 \\   0.9447  \\  0.9543  \\  0.8383  \\ -1.0321  \\  0.9376  \\  0.9376 \\ 0.9447} &
\makecell{0.0101 \\   0.0101  \\  0.0590  \\  0.5718 \\   0.1237  \\  0.0807 \\   0.0807 \\0.0101} &
\makecell{Stable operation when \\\textbf{R} is known}\\ \hline

\multicolumn{10}{|c|}{ Gemson's approach using MS method} \\
\end{tabular}
\end{footnotesize}
\end{center}
\end{table}
\end{landscape}

\addtocounter{table}{-1}
\begin{landscape}
\begin{table}[h]
\caption{Sensitivity Study : \textbf{Q} $>$ 0 : Ramp system.\\ No. of iterations=100, No. of simulations=50.}{}
\begin{center}
\begin{footnotesize}
\begin{tabular}{|c| c| c| c| c| c| c|c|c|c|}
\hline
Study &
$\Theta$ ratio &
\makecell{PCRB\\ ratio} &
\makecell{Consistency\\ ratio-EKF}&
\makecell{Spread\\factor\\EKF} &
\makecell{\textbf{R} ratio\\EKF/True} &
\makecell{\textbf{Q} ratio\\EKF/True}  &
\makecell{$\mu$\ of \\\textbf{J1-J8}} &
\makecell{$\sigma$ of \\\textbf{J1-J8}} &
Remarks
\\ \hline

\multicolumn{10}{|c|}{MT method (Myers and Tapley)} \\ \hline
\makecell{$\mathbf{P_0}$ : Scaled up-[0,0;0,\checkmark]\\\textbf{Q} : MT-diag \\\textbf{R} : MT-diag}  & \multicolumn{9}{c|}{\makecell{Positive definiteness of \textbf{Q, R} is not gauranteed and the cost functions can diverge as seen in Fig. \ref{MT_ramp_Q},-\ref{MT_ramp_J}.\\ There is no proper convergence even if $\Theta$ is known.}} \\ \hline

\multicolumn{10}{|c|}{MT method with \textbf{Q} being estimated using only last N/2 samples} \\ \hline
\makecell{$\mathbf{P_0}$ : Zero\\\textbf{Q} : MT-diag \\\textbf{R} : Known} &
 NA &   1.0087 &
 NA & NA &
 NA & 0.9933 &
\makecell{ 1.0310  \\  1.0310  \\  1.0068  \\  2.5377  \\ -1.0527   \\ 1.0832  \\  1.0832 \\1.0310} &
\makecell{0.1634  \\  0.1634  \\  0.1383 \\   3.5893  \\  0.1182  \\  0.2476  \\  0.2476 \\ 0.1634} &
\makecell{\textbf{R}, $\Theta$ is known}\\ \hline

\multicolumn{10}{|c|}{MT method assuming \textbf{R} is known and \textbf{Q} is estimated using only last N/2 samples} \\ \hline
\makecell{$\mathbf{P_0}$ : Scaled up-[0,0;0,\checkmark]\\\textbf{Q} : MT-[\checkmark,0;0,0] \\\textbf{R} : Known}  &
1.0222 & 1.0049 &
1.0261 &  14.8499 &
0.9362 &  0.9900 &
\makecell{ 1.0171 \\   1.0171  \\  1.0058  \\  0.8358 \\  -1.0209  \\  1.0478 \\   1.0478 \\ 1.0171} &
\makecell{ 0.1507   \\ 0.1507  \\  0.1296 \\   0.5691  \\  0.1255  \\  0.2155  \\  0.2155 \\ 0.1507} &
\makecell{Stable Operation}\\ \hline

\multicolumn{10}{|c|}{ Gemson's approach using MT method} \\
\end{tabular}
\end{footnotesize}
\end{center}
\end{table}
\end{landscape}

\clearpage
\subsection{Ramp System Figures (\textbf{Q} = 0) }

\vspace{17pt}

\clearpage
\subsection{Ramp System Figures (\textbf{Q} $>$ 0) }

\begin{figure}[h]
\includegraphics[width=6in,height=4in]{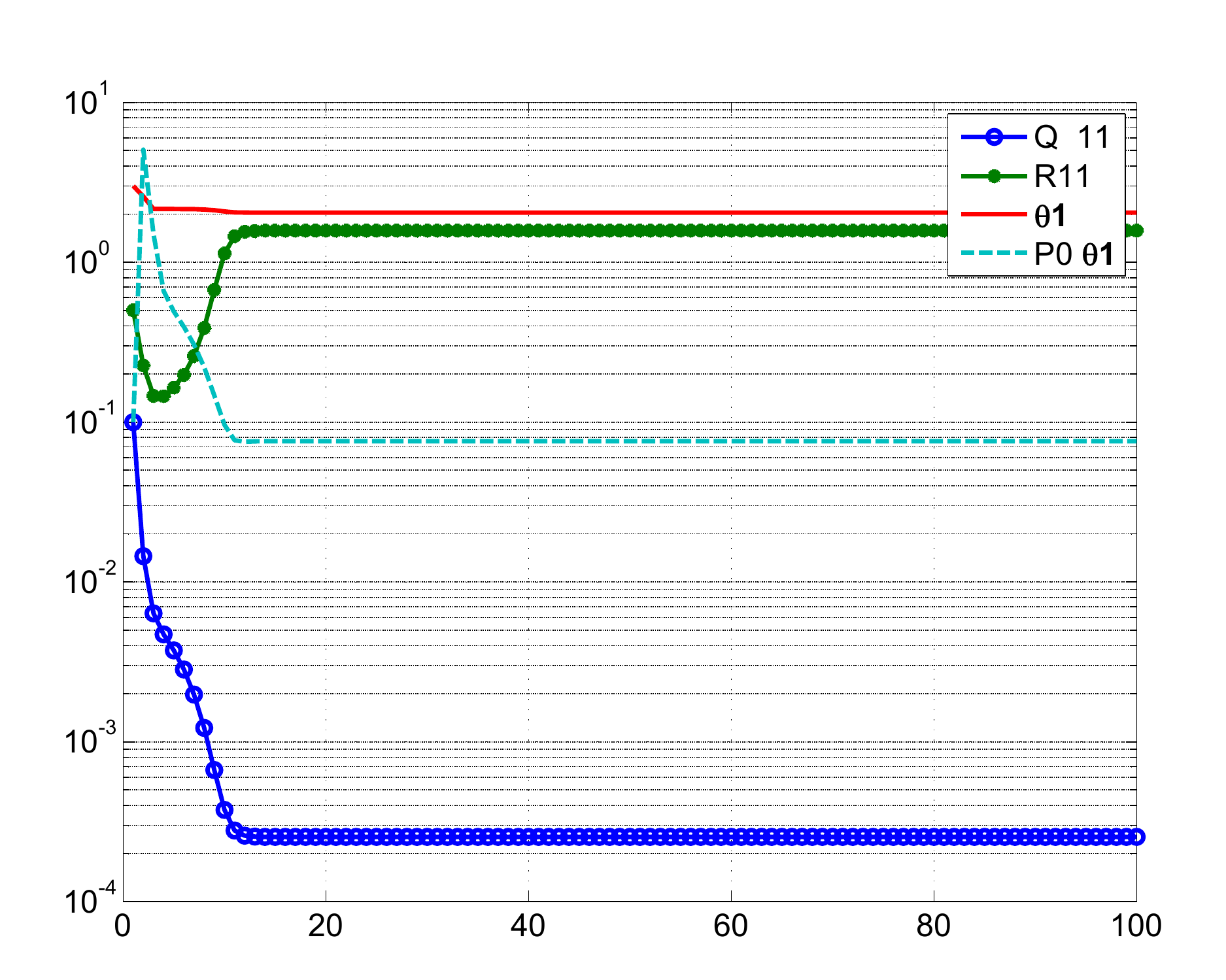}
\caption{Variation of different estimates with iterations using MS method}
\label{MS_ramp_Q}
\end{figure}

\begin{figure}[h]
\includegraphics[width=6in,height=4in]{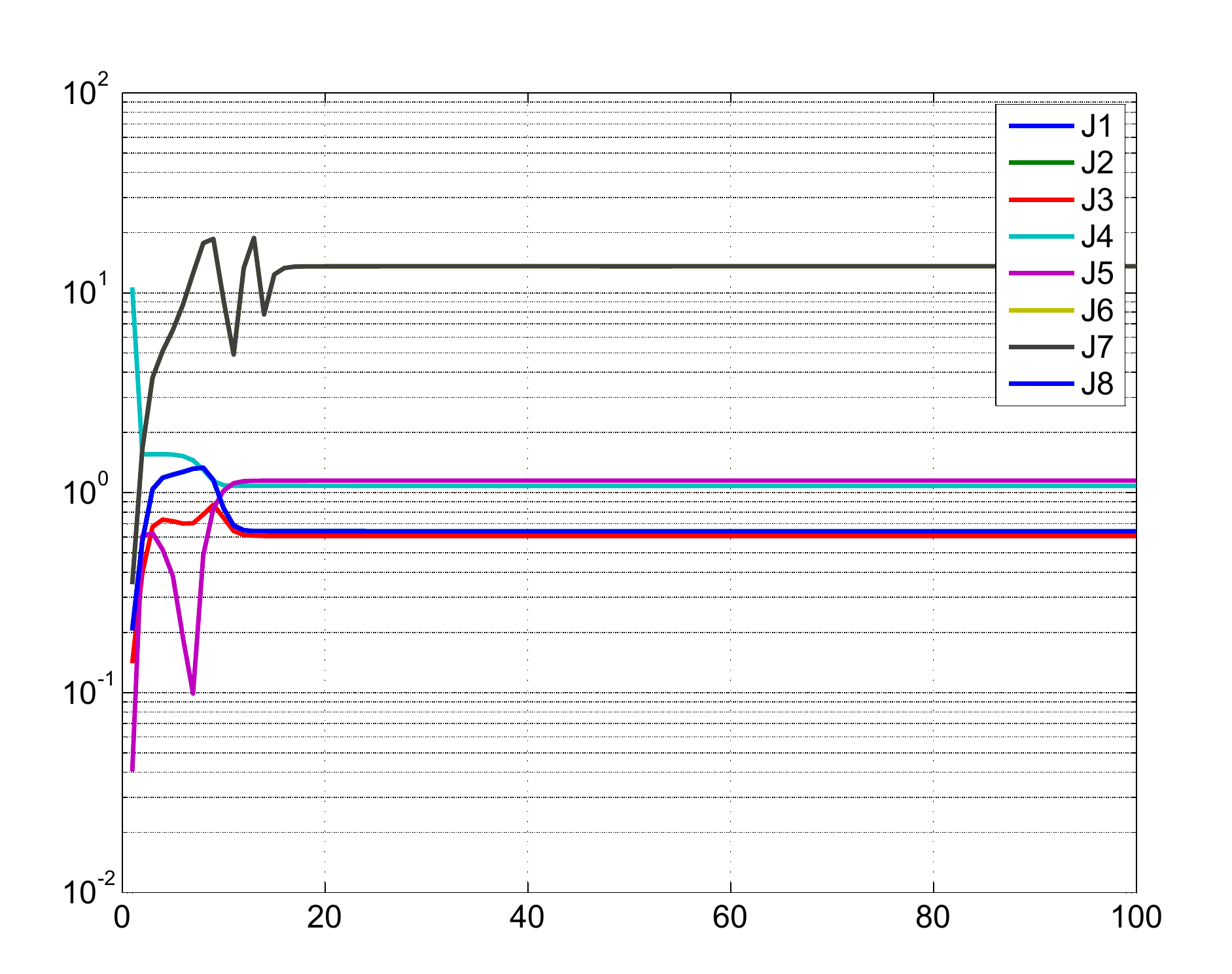}
\caption{Variation of different costs with iterations using MS method}
\label{MS_ramp_J}
\end{figure}

\begin{figure}[h]
\includegraphics[width=6in,height=4in]{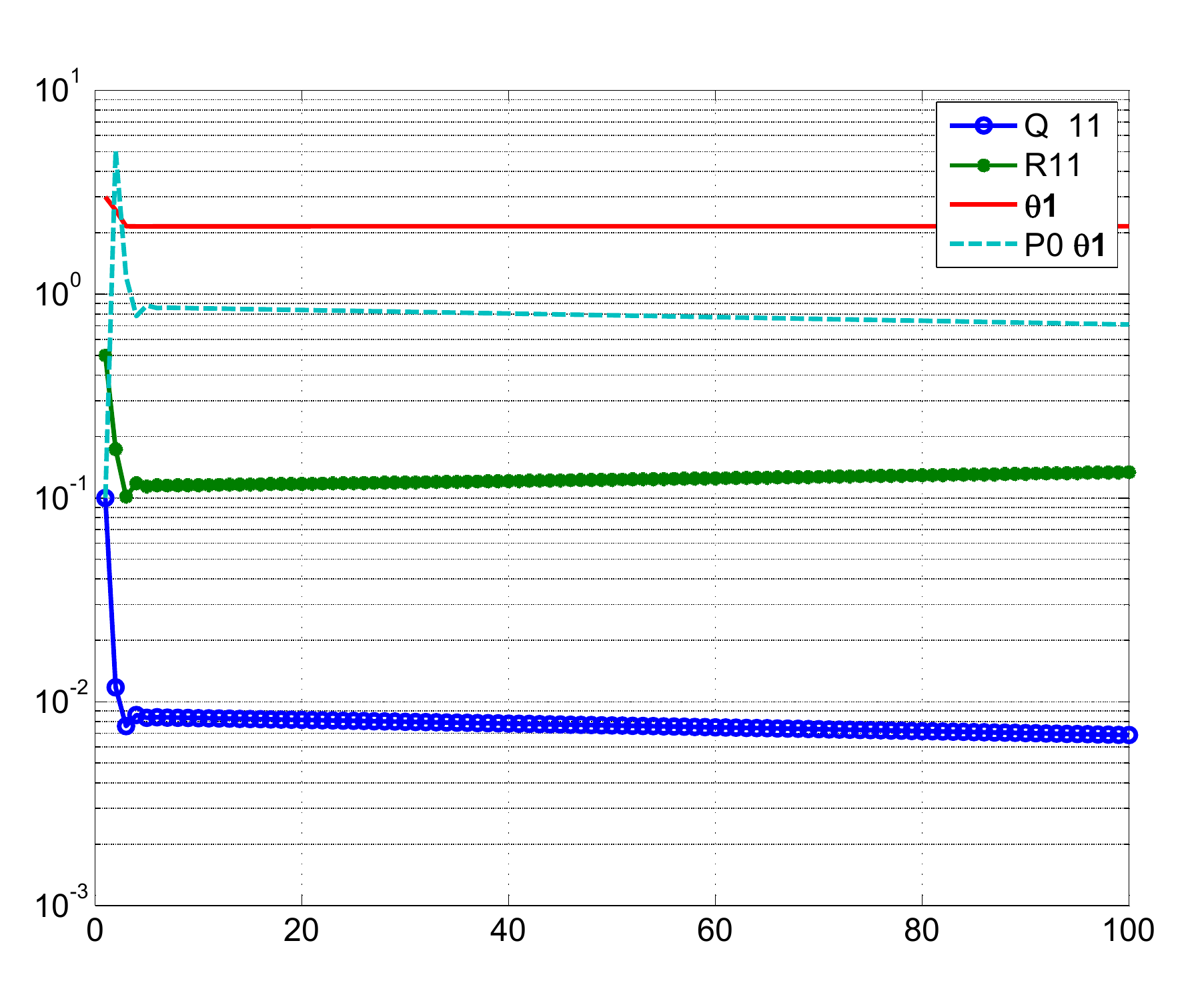}
\caption{Variation of different estimates with iterations using MT method}
\label{MT_ramp_Q}
\end{figure}

\begin{figure}[h]
\includegraphics[width=6in,height=4in]{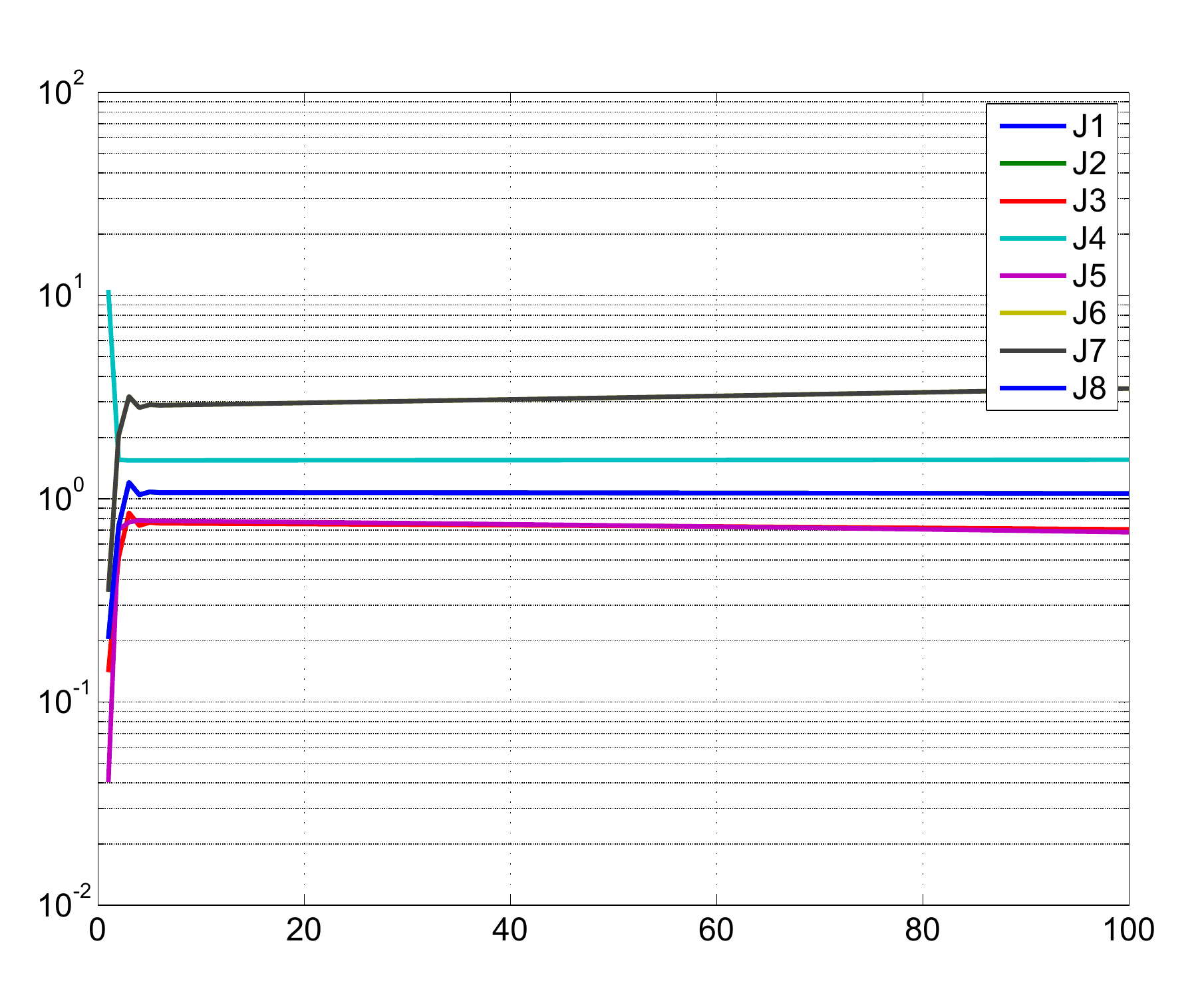}
\caption{Variation of different costs with iterations using MT method}
\label{MT_ramp_J}
\end{figure}


\clearpage
\section{Simulated Spring, Mass, and Damper System}

\par A spring mass damper (SMD) system with weak non-linearity in the continuous time (t) state space form is given by
\begin{align*}
\dot{x}_1(t)&=x_2(t)\\
\dot{x}_2(t)&=-\Theta_1x_1(t)-\Theta_2x_2(t)-\Theta_3x_1^3(t)
\end{align*}
where $x_1$, $x_2$ is the displacement and velocity state with initial condition 1,0 respectively. $\dot{x}$ represents differentiation with respect to time (t). The unknown parameter vector is $\Theta=[\Theta_1,\Theta_2,\Theta_3]^T$ with the true values being $\Theta_{true}=(4,0.4,0.6)^T$. The $\Theta_3$ is a weak parameter since its value do not affect the system dynamics much. The complete state vector, X=$[x_1,x_2,\Theta_1,\Theta_2,\Theta_3]^T$ of size $(n+p)\times 1$. The measurement equation is given by
\begin{align*}
Z_k=HX_k+v_k
\end{align*}
where H=$\begin{bmatrix}
1 & 0 & 0 & 0 & 0 \\ 1 & 0 & 0 & 0 & 0
\end{bmatrix}$ is the measurement matrix of size $m\times (n+p)$ where $m=n=2$ and $p=3$. The numerical values of the noise variances are shown in Table-\ref{sysdes}. All the figures are presented for only one simulation run to prevent cluttering.

\subsection{Remarks on the Results}

\par
\begin{figure}[h]
\includegraphics[width=6in,height=3.0in]{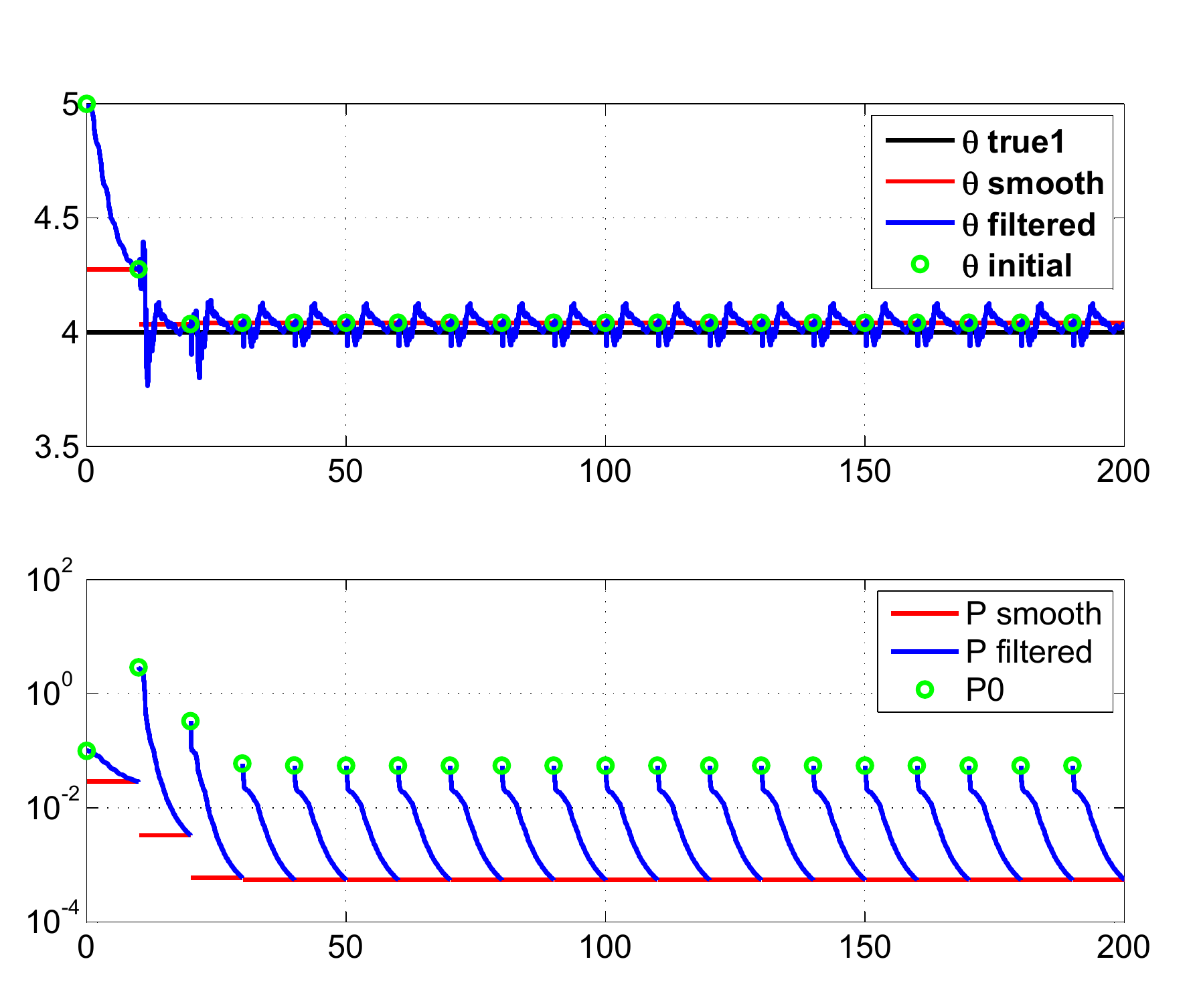}
\caption{The variation of parameter estimate 1 and their filtered and }
\caption*{smoothed covariances through (with the time cumulatively) the iterations}
\label{smd_p1}
\end{figure}

\begin{figure}[h]
\includegraphics[width=6in,height=3.0in]{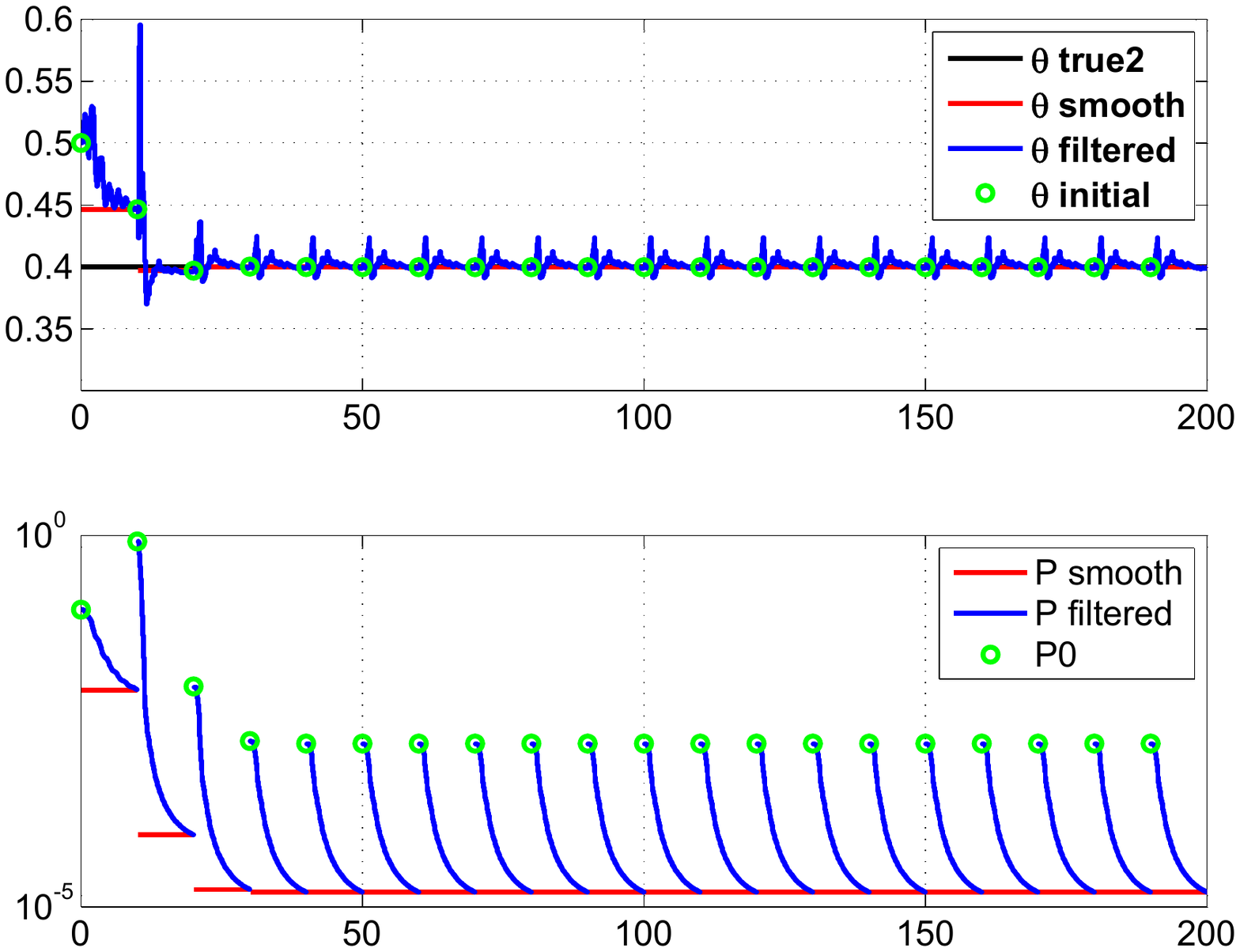}
\caption{The variation of parameter estimate 2 and their filtered and }
\caption*{smoothed covariances through (with the time cumulatively) the iterations}
\label{smd_p2}
\end{figure}

\begin{figure}[h]
\includegraphics[width=6in,height=4in]{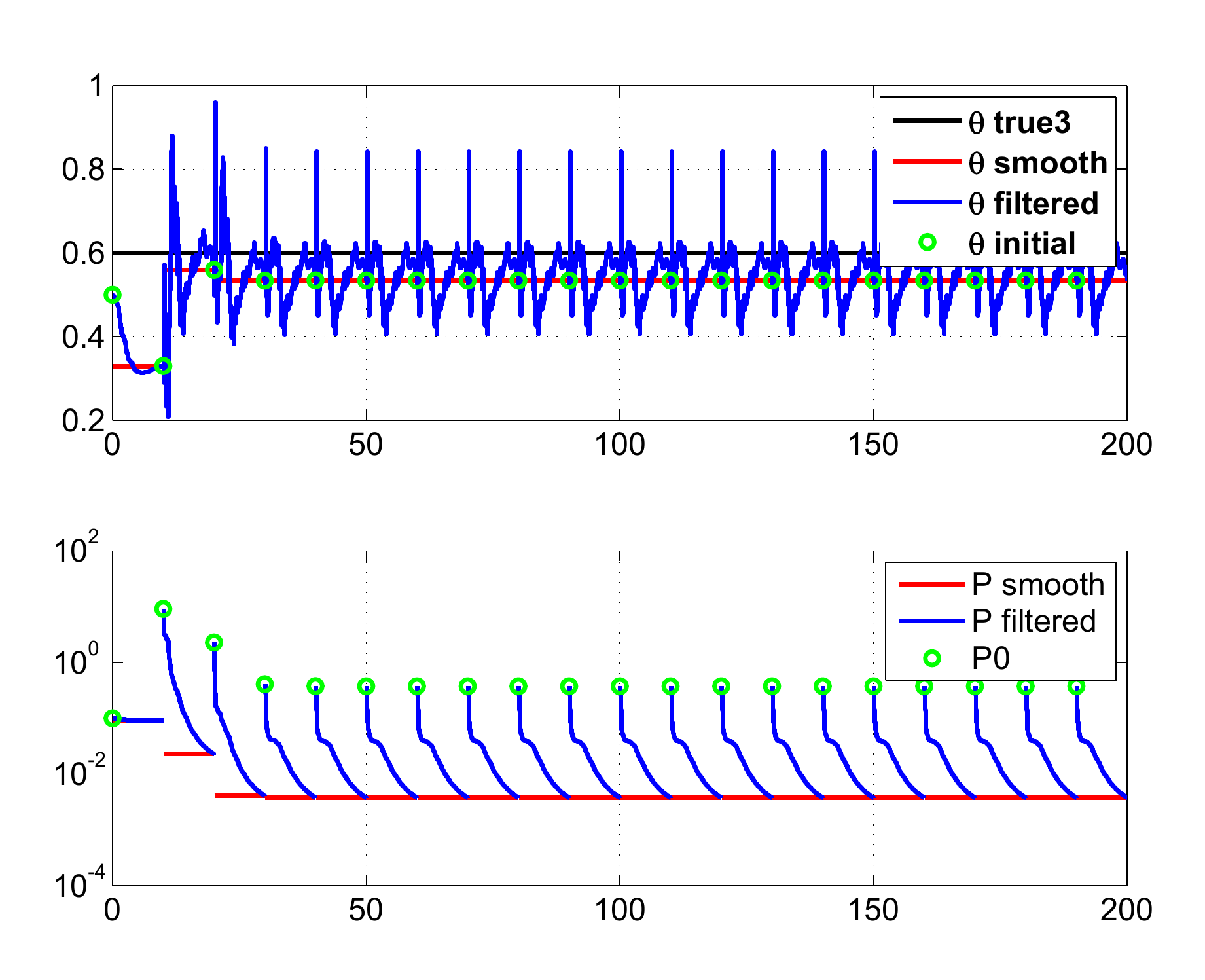}
\caption{The variation of parameter estimate 3 and their filtered and }
\caption*{smoothed covariances through (with the time cumulatively) the iterations}
\label{smd_p3}
\end{figure}

\begin{figure}[h]
\includegraphics[width=6in,height=4in]{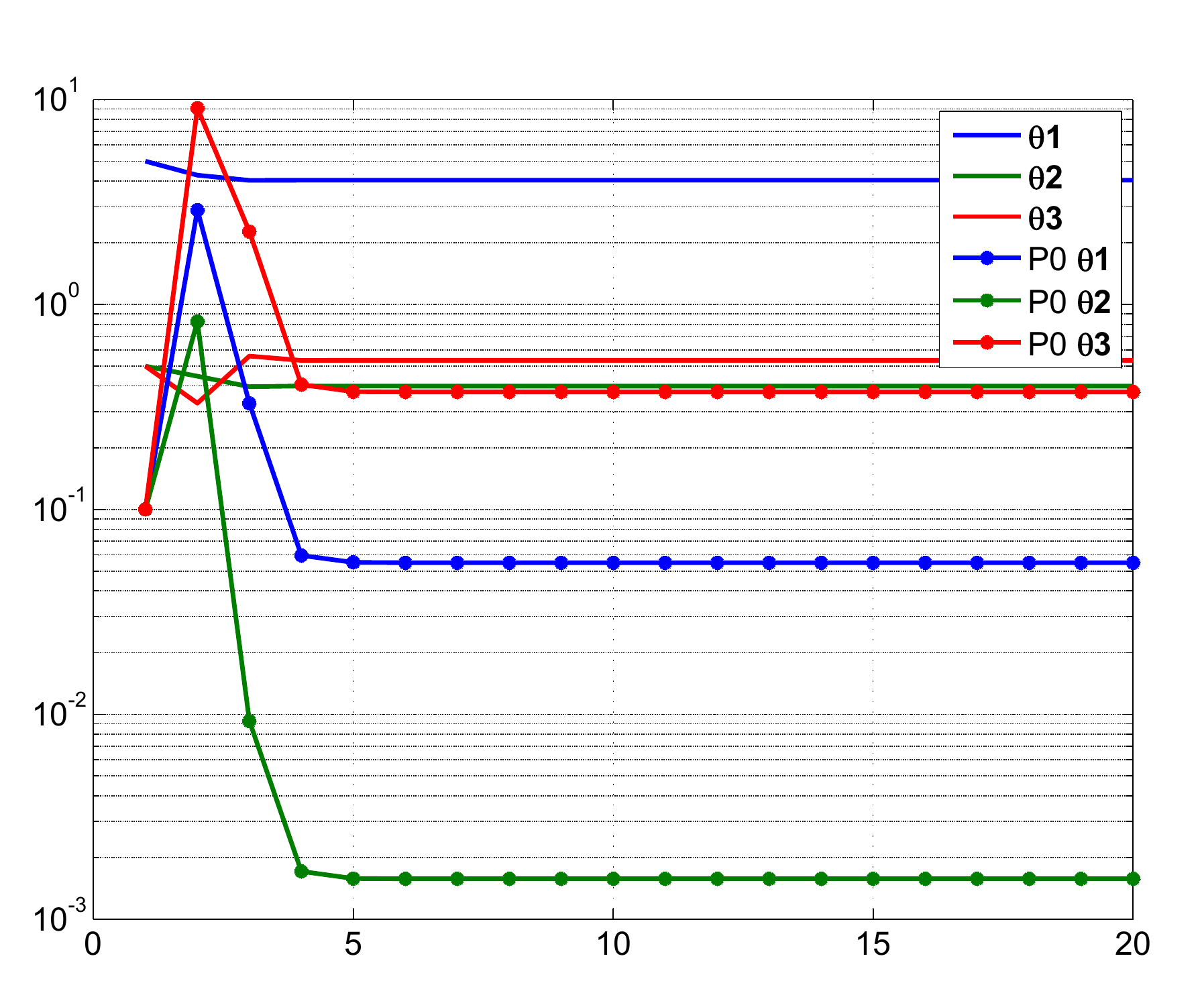}
\caption{Variation of parameter and its initial covariance ($\mathbf{P_0}$) with iterations}
\label{smd_P0}
\end{figure}

\begin{figure}[h]
\includegraphics[width=6in,height=4in]{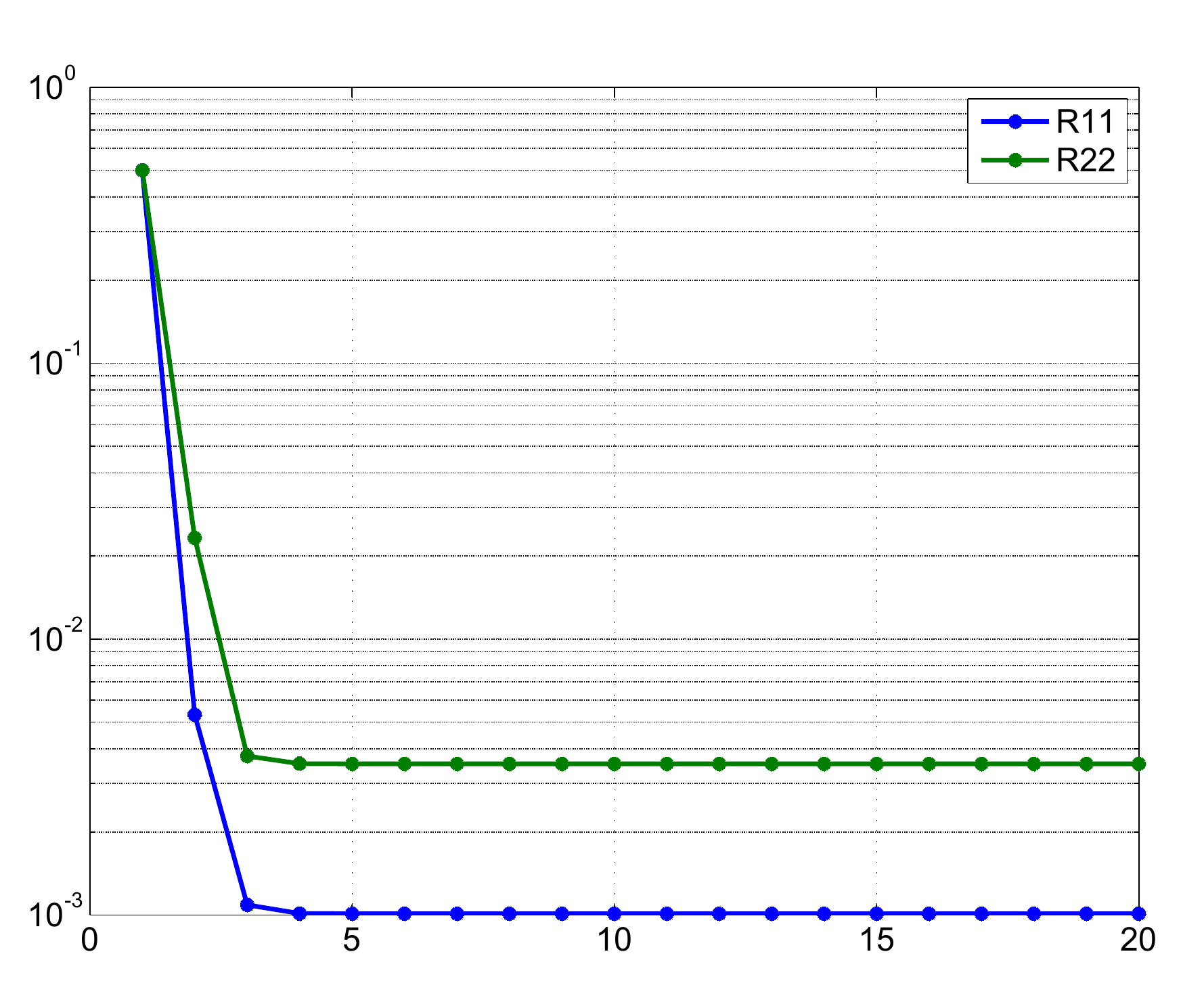}
\caption{Variation of \textbf{R} with iterations}
\label{smd_R}
\end{figure}

\begin{figure}[h]
\includegraphics[width=6in,height=4in]{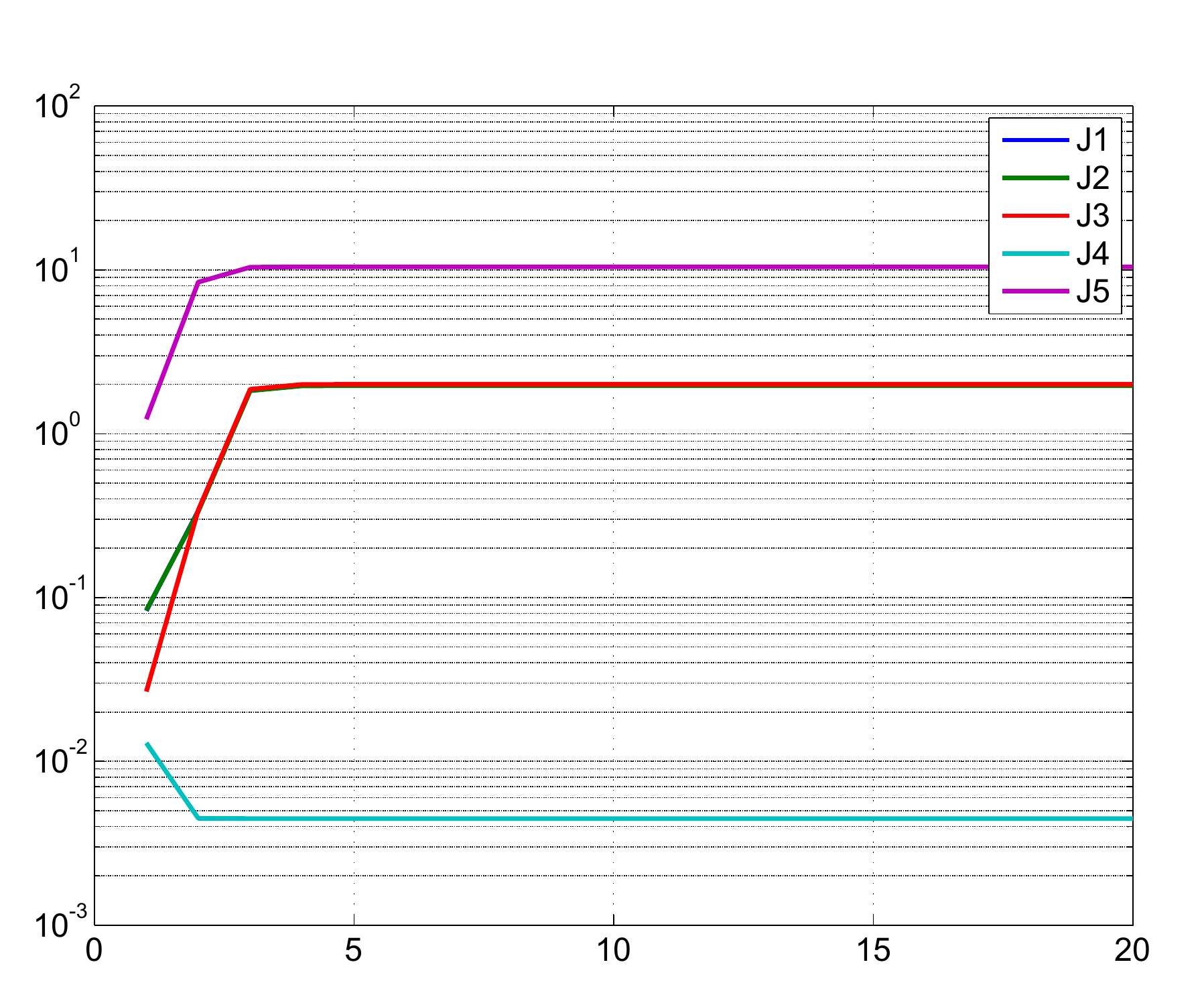}
\caption{Variation of different costs (\textbf{J1-J5}) with iterations}
\label{smd_cost}
\end{figure}

\begin{figure}[h]
\includegraphics[width=6in,height=4in]{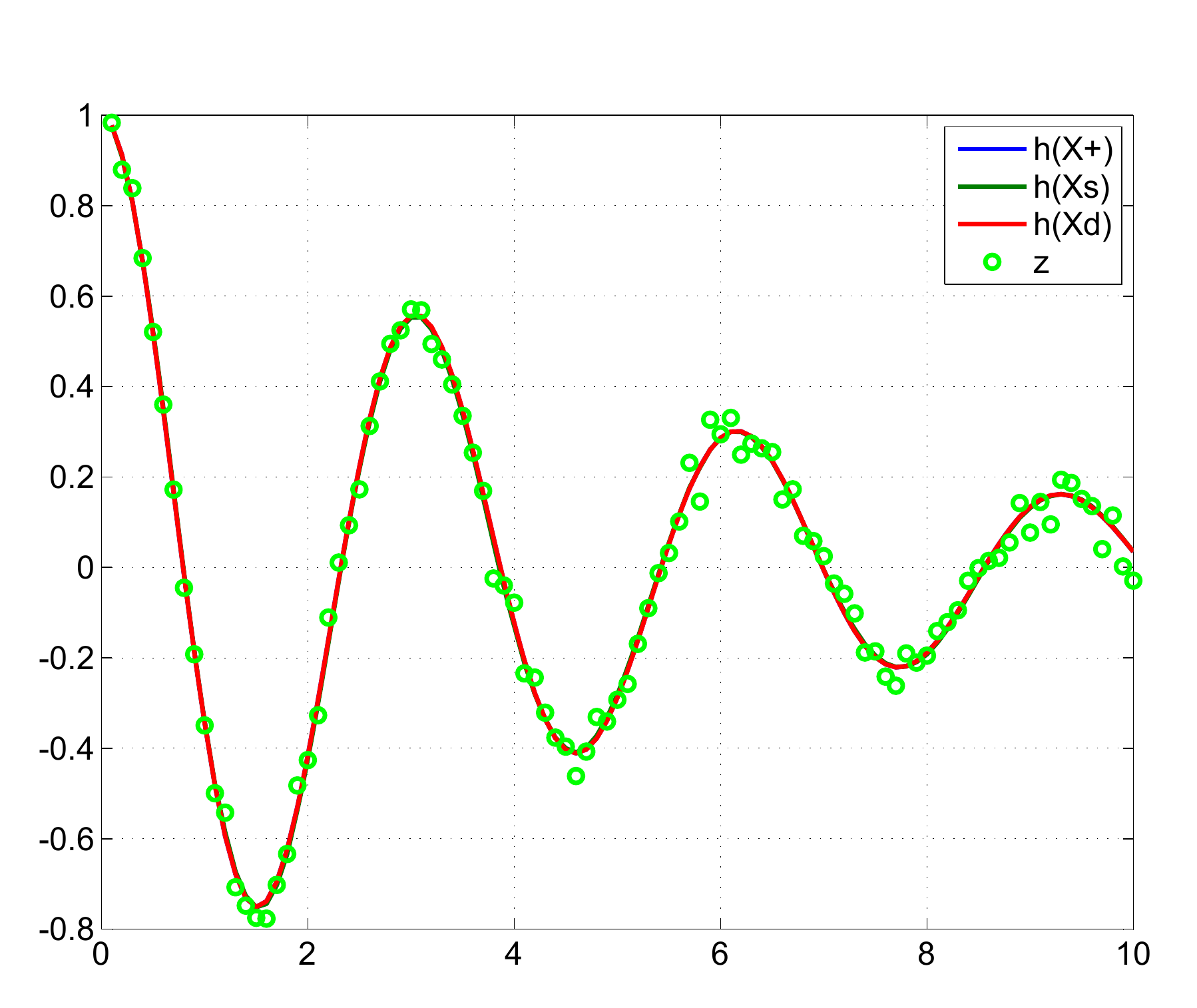}
\caption{Comparison of the predicted dynamics, posterior, smoothed}
\caption*{and the measurement 1 (displacement) }
\label{smd_h1}
\end{figure}

\begin{figure}[h]
\includegraphics[width=6in,height=4in]{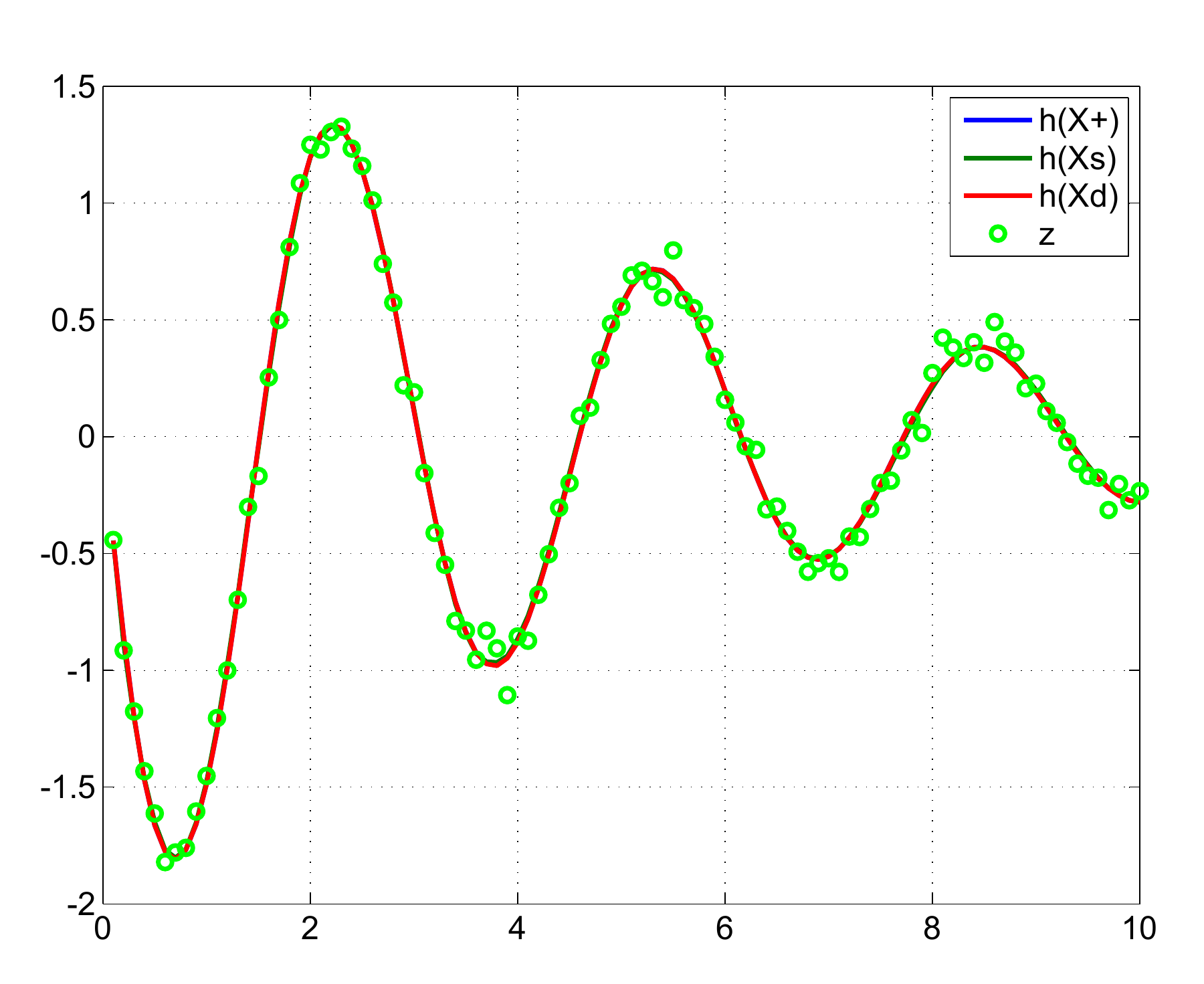}
\caption{Comparison of the predicted dynamics, posterior, smoothed}
\caption*{and the measurement 2 (velocity) }
\label{smd_h2}
\end{figure}

\begin{figure}[h]
\includegraphics[width=6in,height=4in]{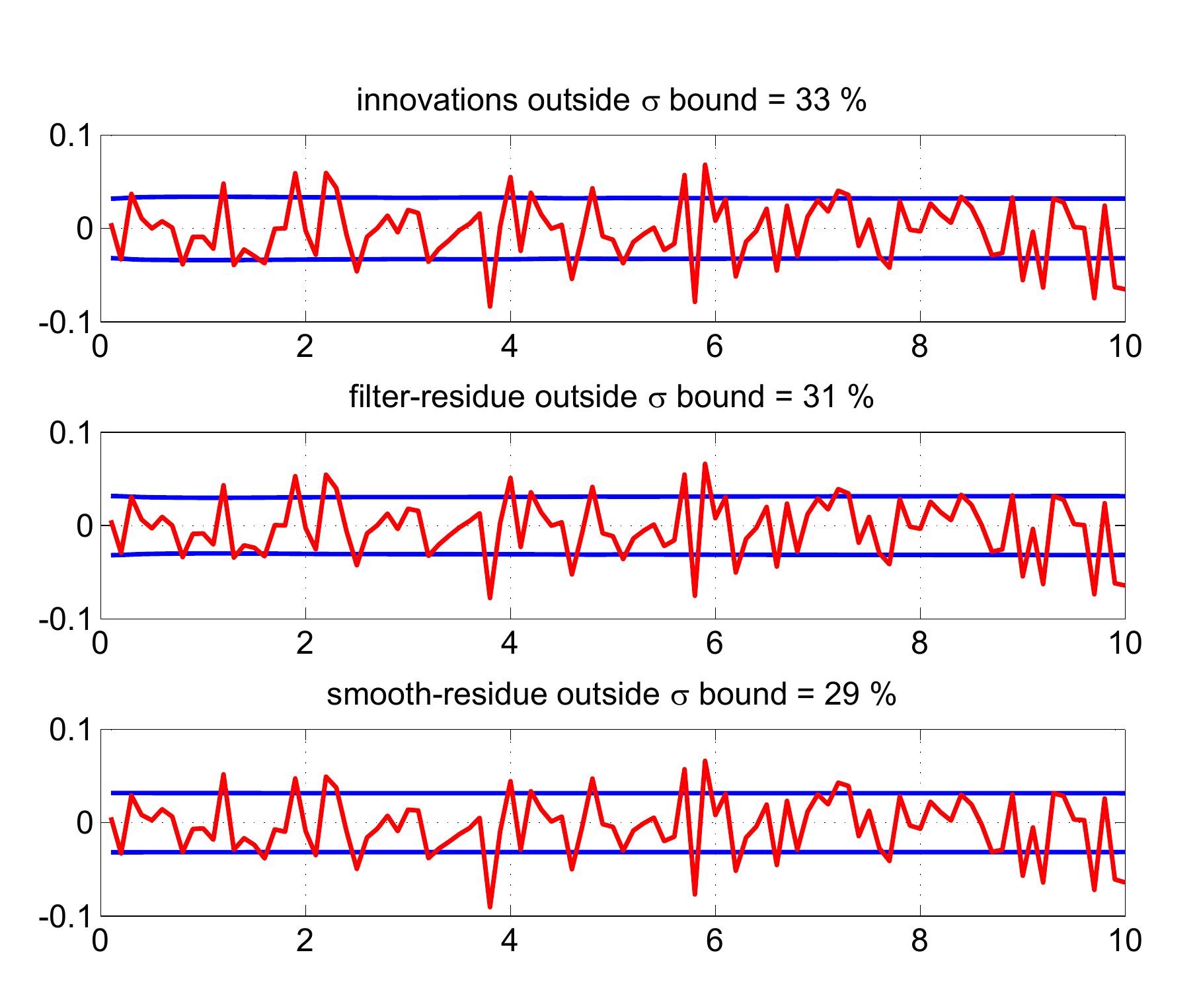}
\caption{The innovations, filtered residue and smoothed residue}
\caption*{corresponding to measurement 1 (displacement) }
\label{smd_innov1}
\end{figure}

\begin{figure}[h]
\includegraphics[width=6in,height=4in]{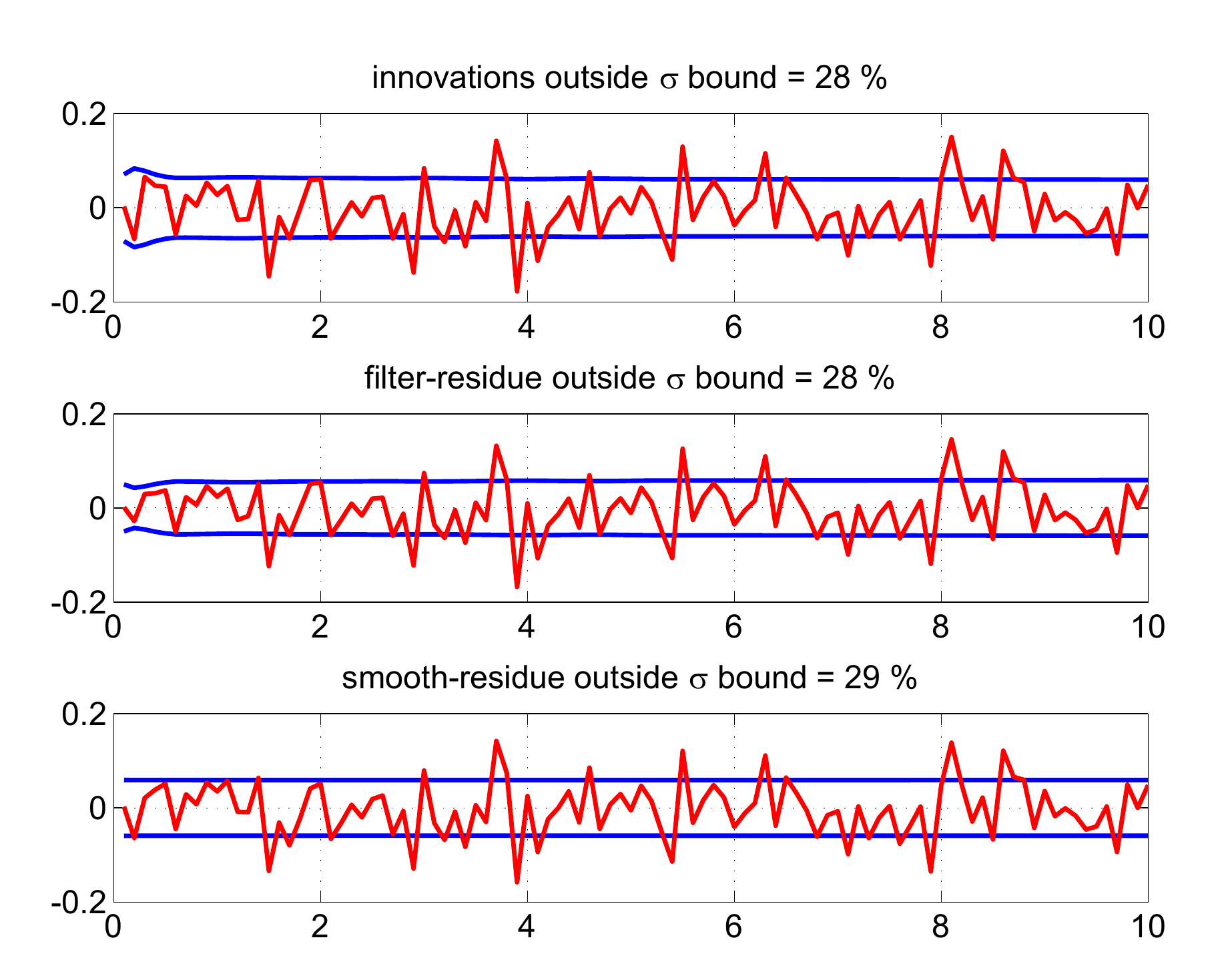}
\caption{The innovations, filtered residue and smoothed residue}
\caption*{corresponding to measurement 2 (velocity)}
\label{smd_innov2}
\end{figure}

\begin{figure}[h]
\includegraphics[width=6in,height=4in]{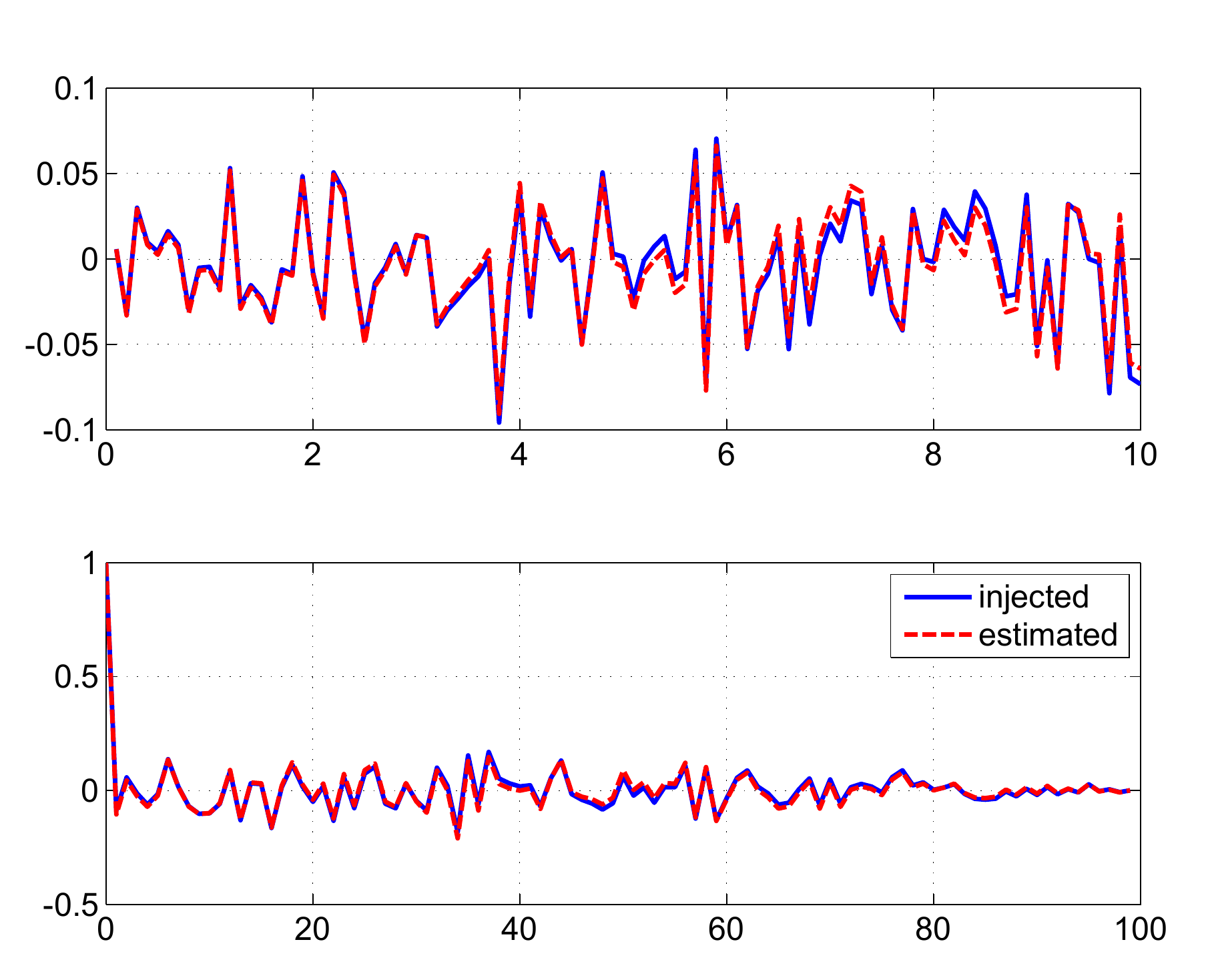}
\caption{Time variation of injected and estimated measurement noise (top) and}
\caption*{their autocorrelation (bottom) for measurement 1 (displacement)}
\label{smd_mnoise1}
\end{figure}

\begin{figure}[h]
\includegraphics[width=6in,height=4in]{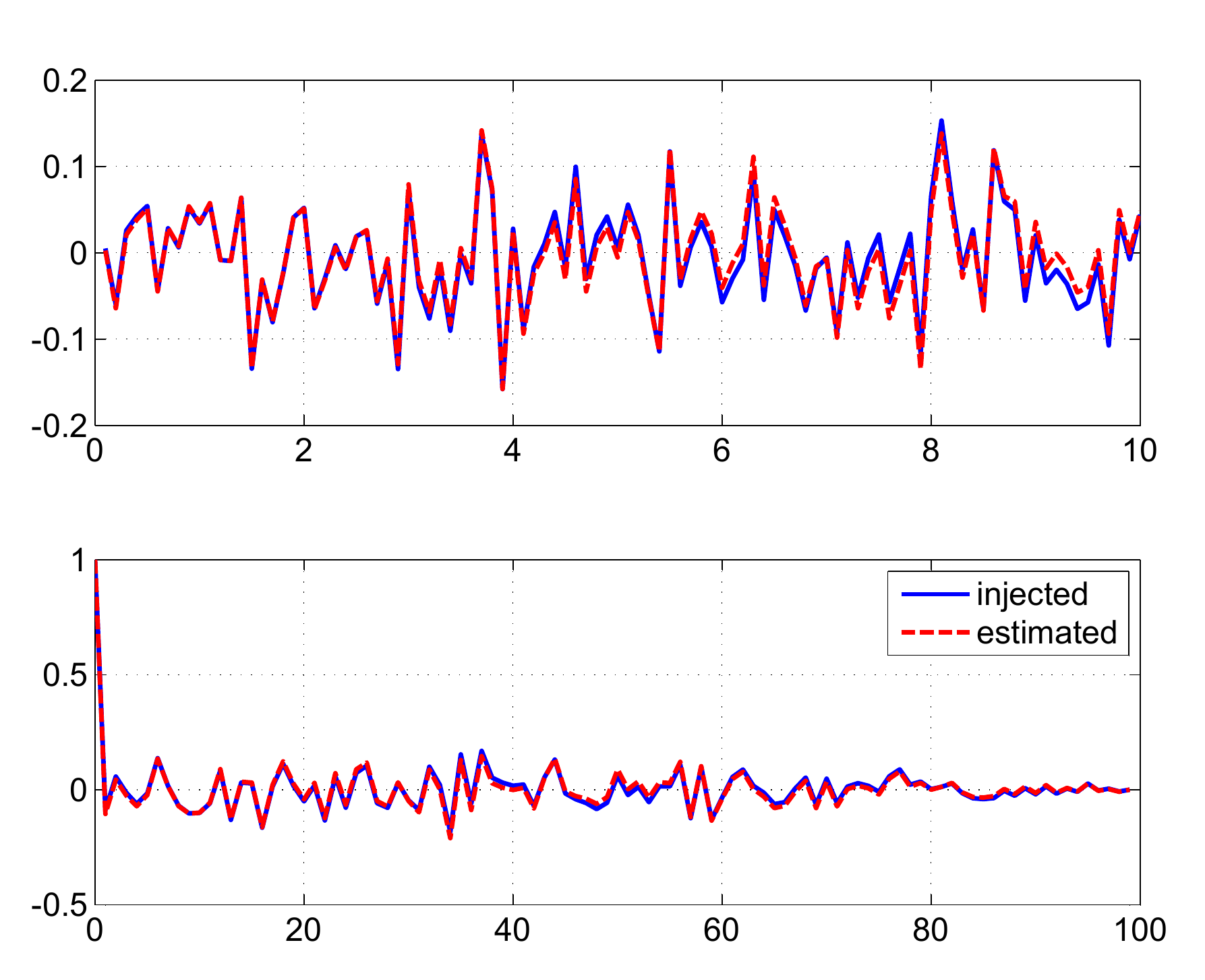}
\caption{Time variation of injected and estimated measurement noise (top) and}
\caption*{their autocorrelation (bottom) for measurement 2 (velocity)}
\label{smd_mnoise2}
\end{figure}

 We first run the filter assuming \textbf{Q} = 0. It was found that about 20 iterations of the data would suffice. The Fig. \ref{smd_p1}-\ref{smd_mnoise2} refer to the \textbf{Q} = 0 case. The Fig. \ref{smd_p1}-\ref{smd_p3} shows the various parameter estimates and its corresponding variances through cumulative time instants with iterations. The variation of the estimated initial parameters and their variances through iterations are shown in Fig. \ref{smd_P0}. The parameter and the uncertainty reach almost their final estimated values in about 2 and 5 iterations respectively. A similar plot in Fig. \ref{smd_R} shows the variation of the estimated measurement noise. The variation of different cost functions (\textbf{J1-J5}) through the iterations is shown in Fig. \ref{smd_cost}. The Fig. \ref{smd_h1}-\ref{smd_h2} shows the predicted dynamics, filtered and smoothed estimate at the last iteration. The Fig. \ref{smd_innov1}-\ref{smd_innov2} show the innovations, filtered residue and smoothed residue together with the square root of their variance ($\pm\sigma$ bound). In the EKF approach most of the quantities are Gaussian or approximated as quasi Gaussian and one would expect all the above quantities are close to being Gaussian and hence around one third of the total sample points to be outside the $\sigma$ bound. The injected and estimated measurement noise distributions during the final iteration shown in Fig. \ref{smd_mnoise1}-\ref{smd_mnoise2} indicate that they are close to each other. Further even their autocorrelations are ideally expected to be close to the Kronecker delta function which provides confidence in the proposed filter algorithm.\par

\par
\begin{figure}[h]
\includegraphics[width=6in,height=3.2in]{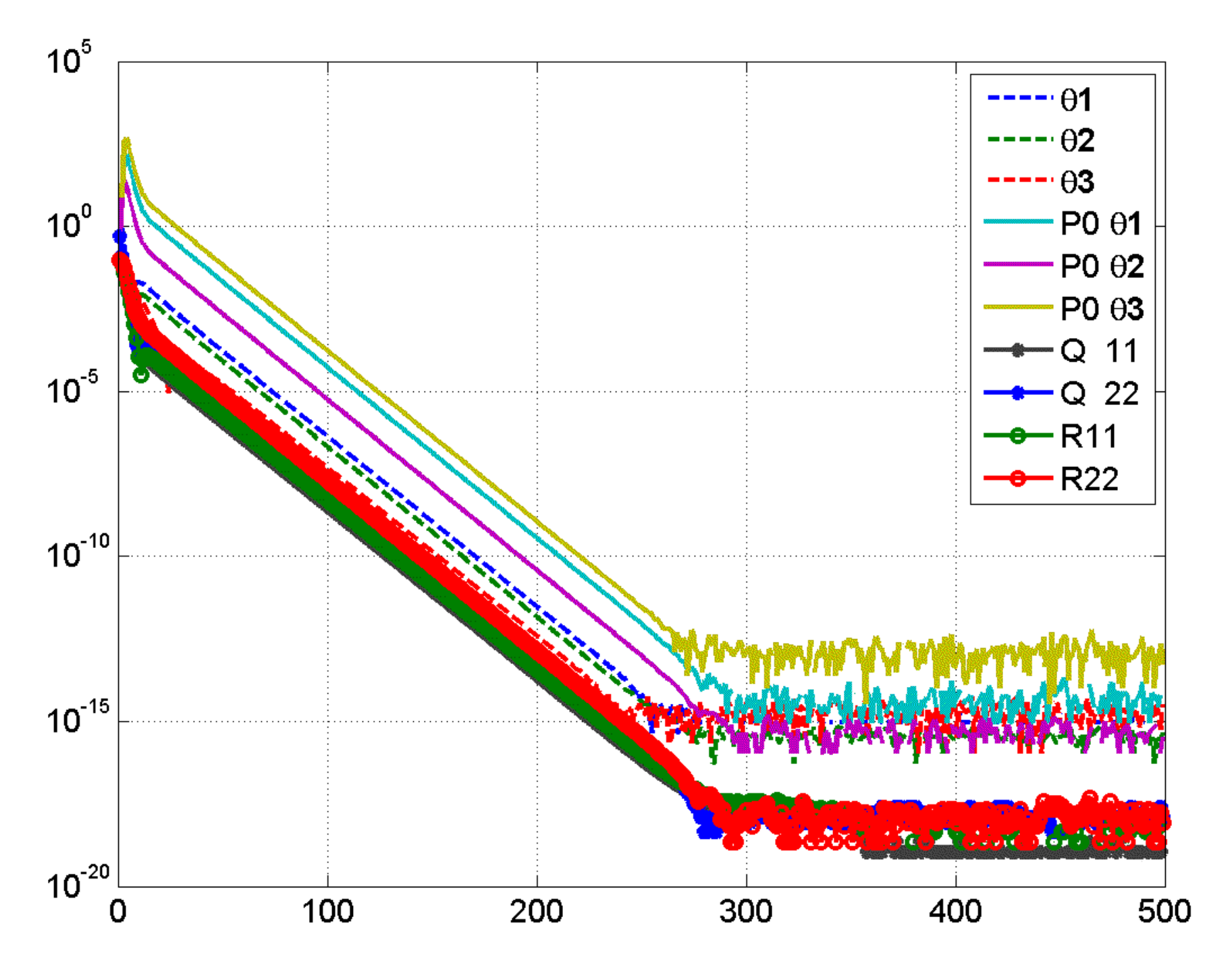}
\caption{The absolute difference between the iterated and final values}
\caption*{with 500 iterations}
\label{smd_err}
\end{figure}

\begin{figure}[h]
\includegraphics[width=6in,height=3.2in]{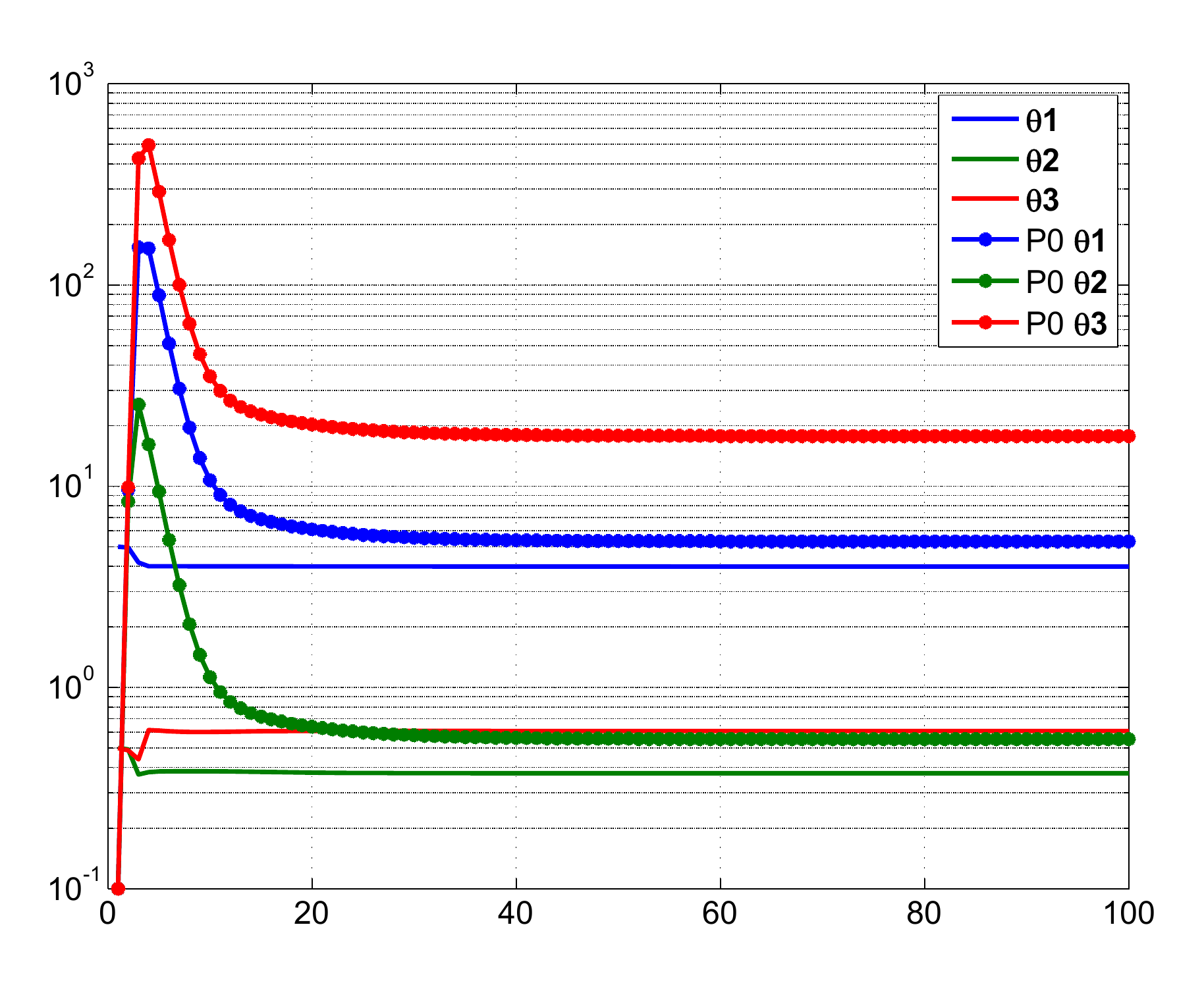}
\caption{Variation of parameter and its initial covariance ($\mathbf{P_0}$) with iterations}
\label{smdQ_P0}
\end{figure}

\begin{figure}[h]
\includegraphics[width=6in,height=4in]{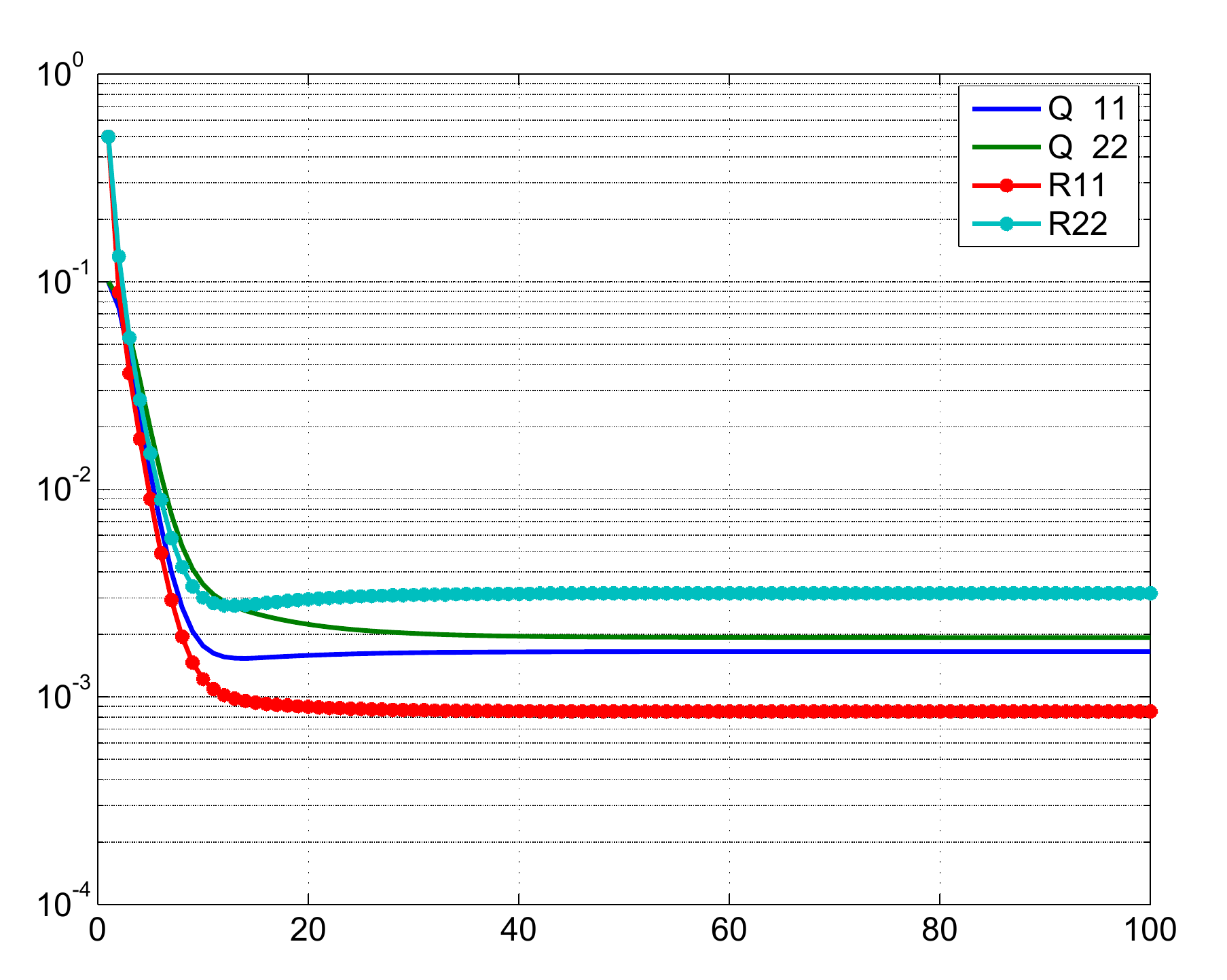}
\caption{Variation of\textbf{Q} and \textbf{R} with iterations}
\label{smdQ_R}
\end{figure}

\begin{figure}[h]
\includegraphics[width=6in,height=3.4in]{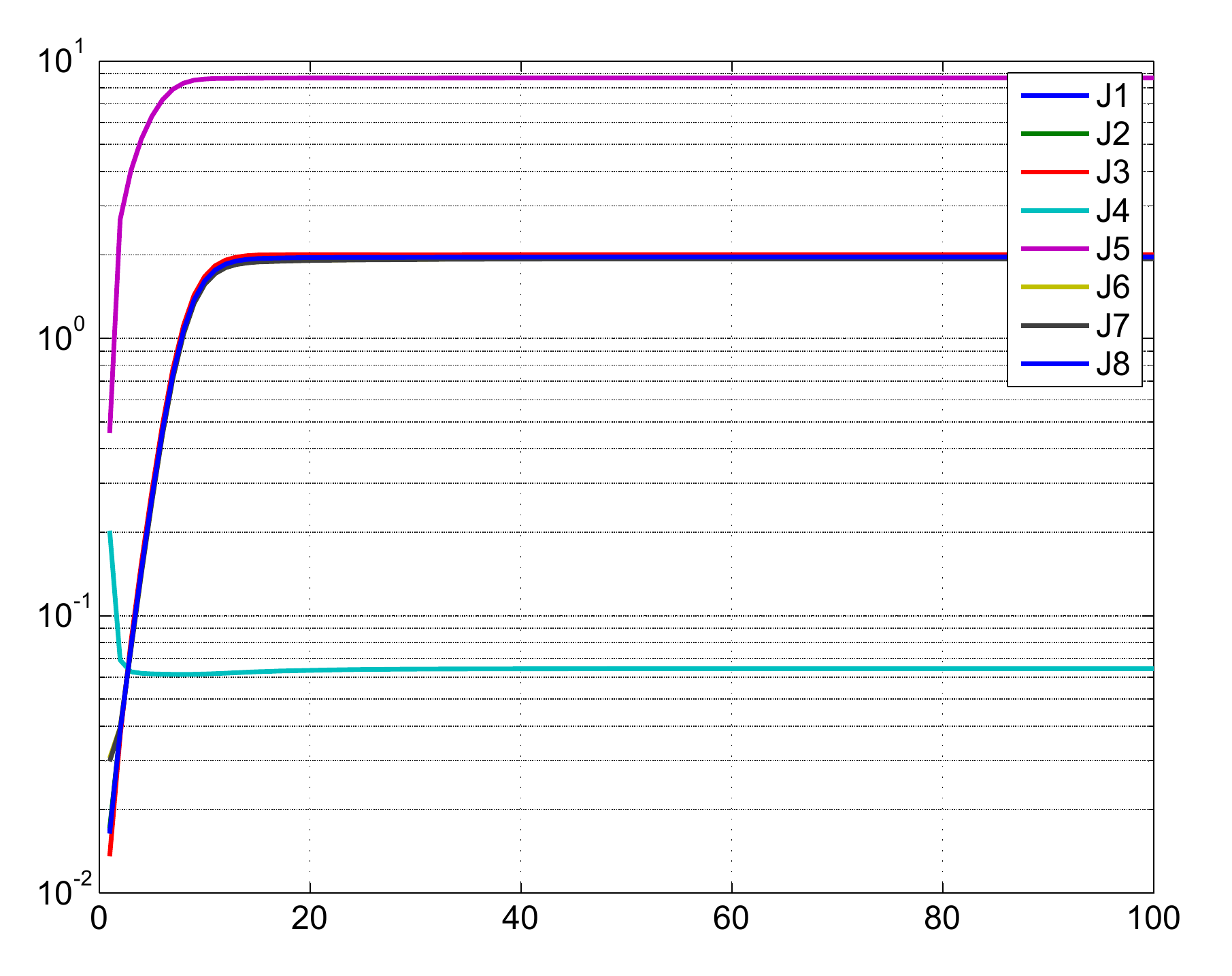}
\caption{Variation of different costs (\textbf{J1-J8}) with iterations}
\label{smdQ_J}
\end{figure}

\begin{figure}[h]
\includegraphics[width=6in,height=4in]{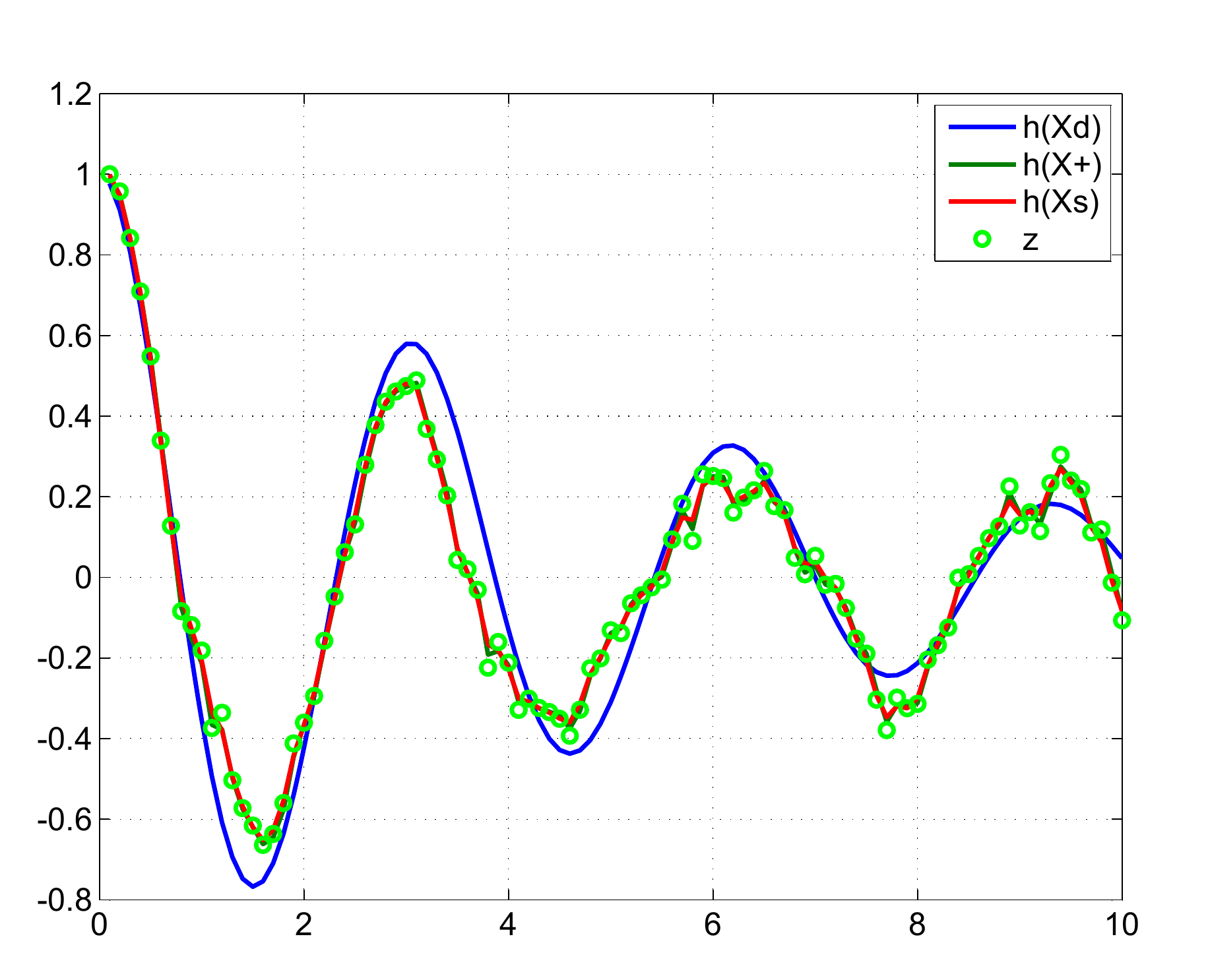}
\caption{Comparison of the predicted dynamics, posterior, smoothed}
\caption*{and the measurement 1 (displacement) }
\label{smdQ_h1}
\end{figure}

\begin{figure}[h]
\includegraphics[width=6in,height=4in]{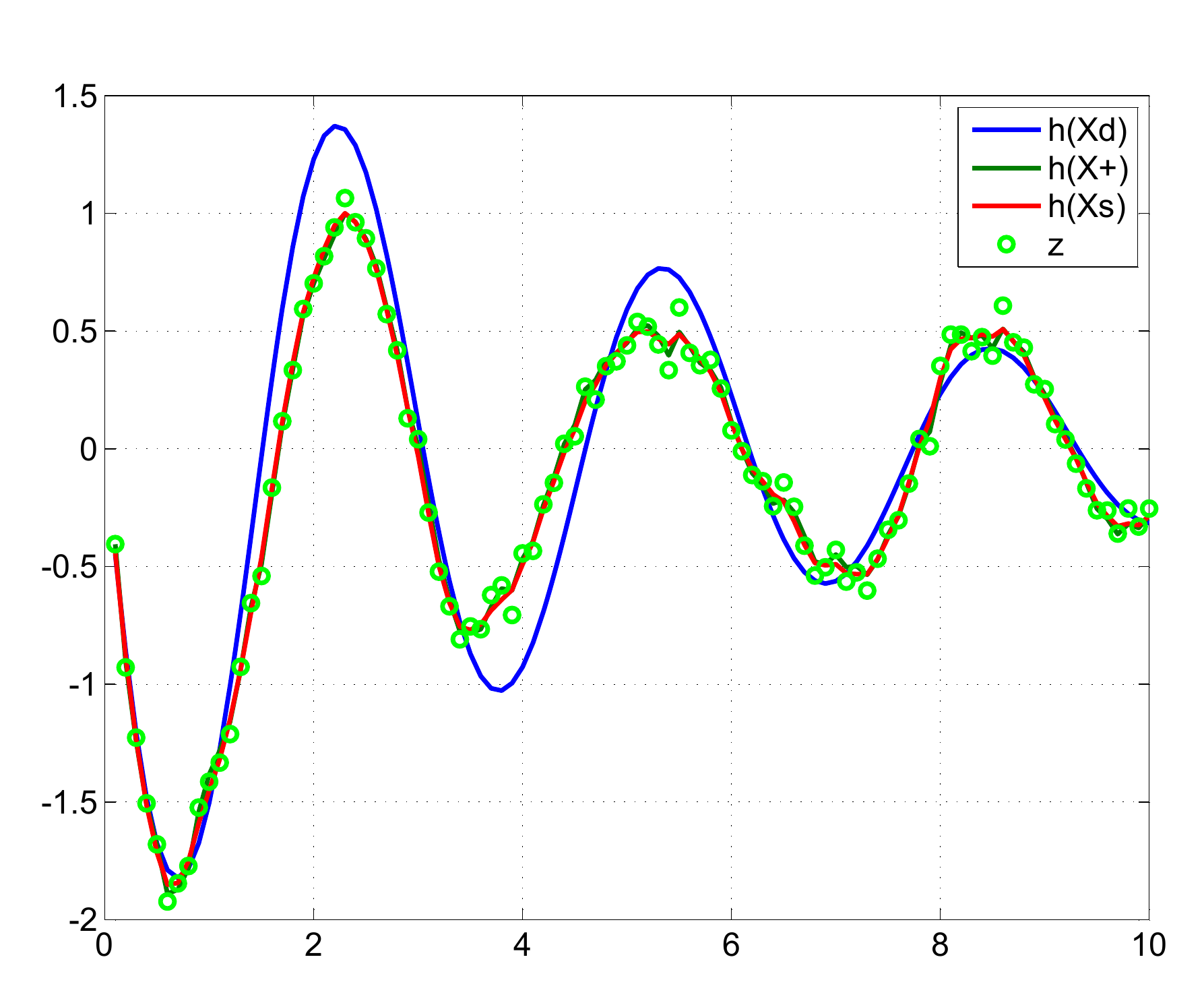}
\caption{Comparison of the predicted dynamics, posterior, smoothed}
\caption*{and the measurement 2 (velocity) }
\label{smdQ_h2}
\end{figure}

\begin{figure}[h]
\includegraphics[width=6in,height=4in]{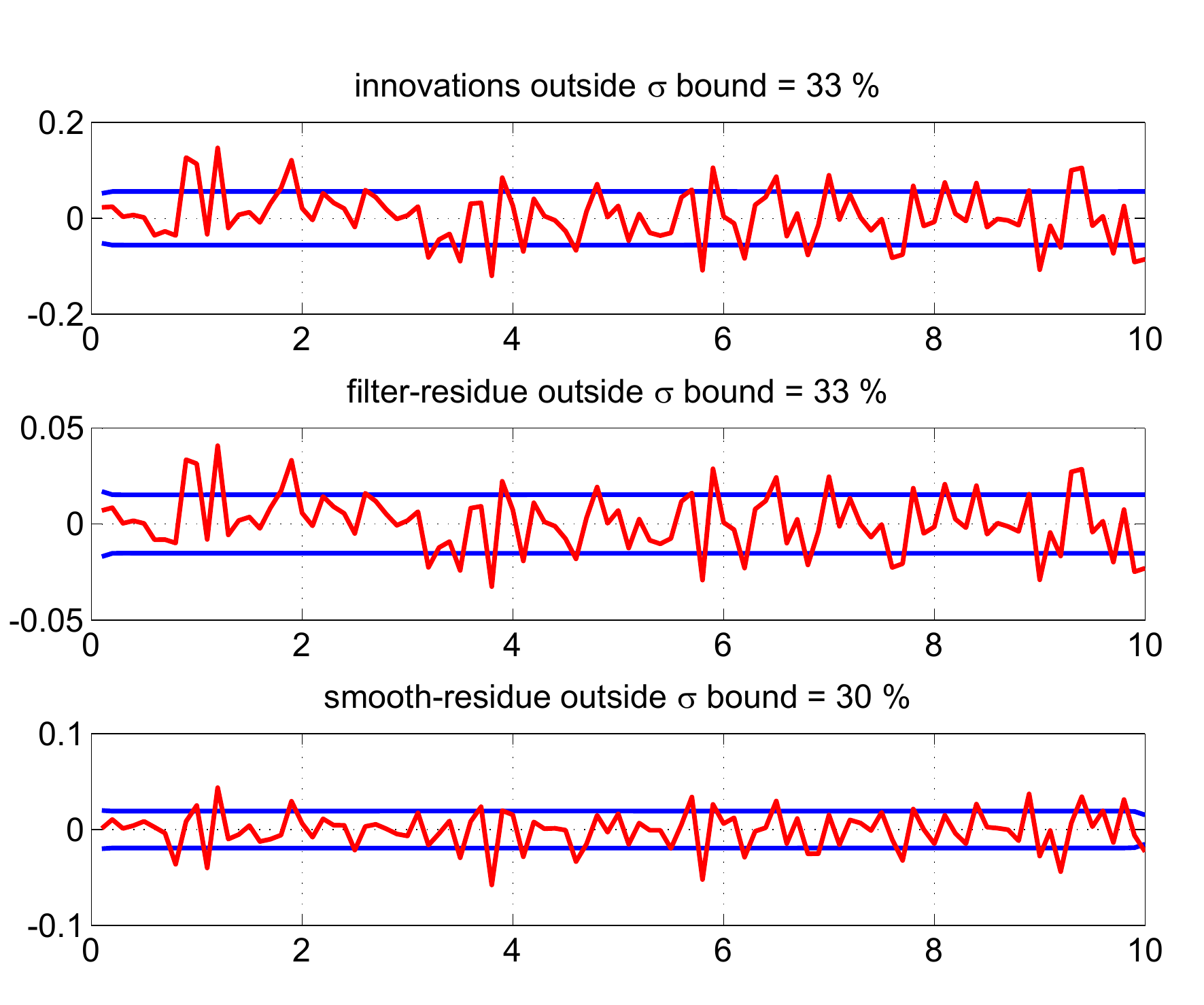}
\caption{The innovations, filtered residue and smoothed residue}
\caption*{corresponding to measurement 1 (displacement) }
\label{smdQ_innov1}
\end{figure}

\begin{figure}[h]
\includegraphics[width=6in,height=4in]{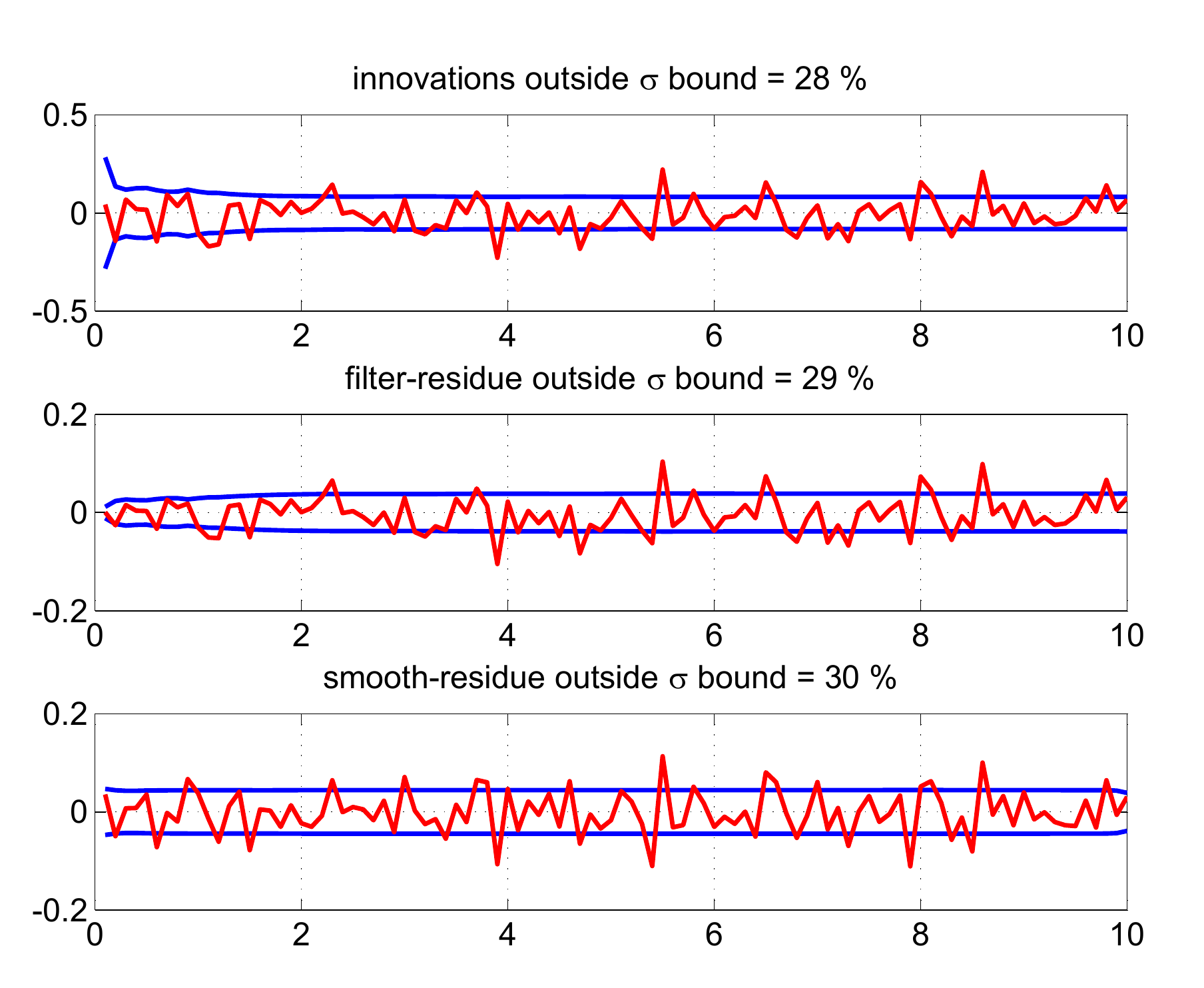}
\caption{The innovations, filtered residue and smoothed residue}
\caption*{corresponding to measurement 2 (velocity)}
\label{smdQ_innov2}
\end{figure}

\begin{figure}[h]
\includegraphics[width=6in,height=4in]{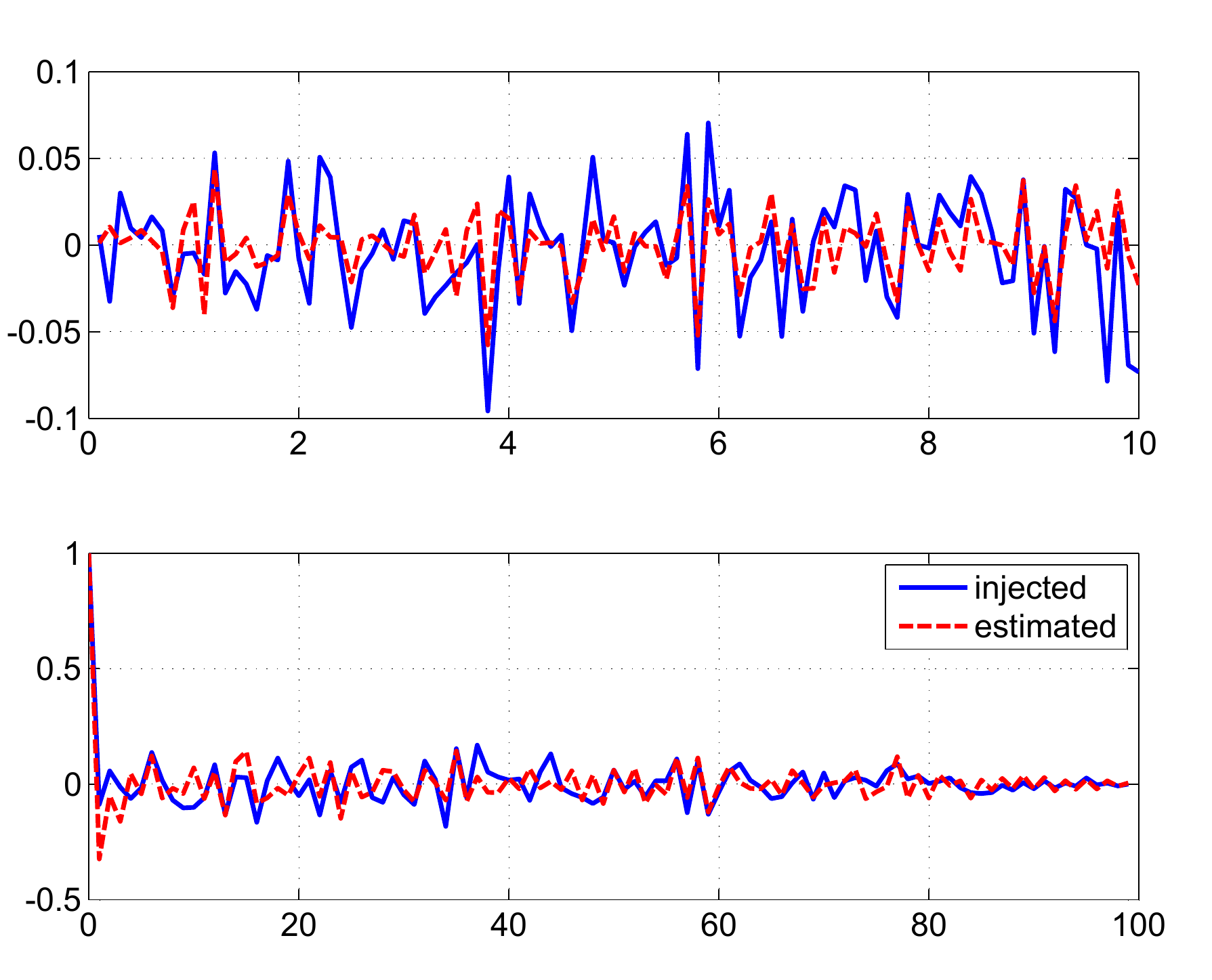}
\caption{Time variation of injected and estimated measurement noise (top) and}
\caption*{their autocorrelation (bottom) for measurement 1 (displacement)}
\label{smdQ_mnoise1}
\end{figure}

\begin{figure}[h]
\includegraphics[width=6in,height=4in]{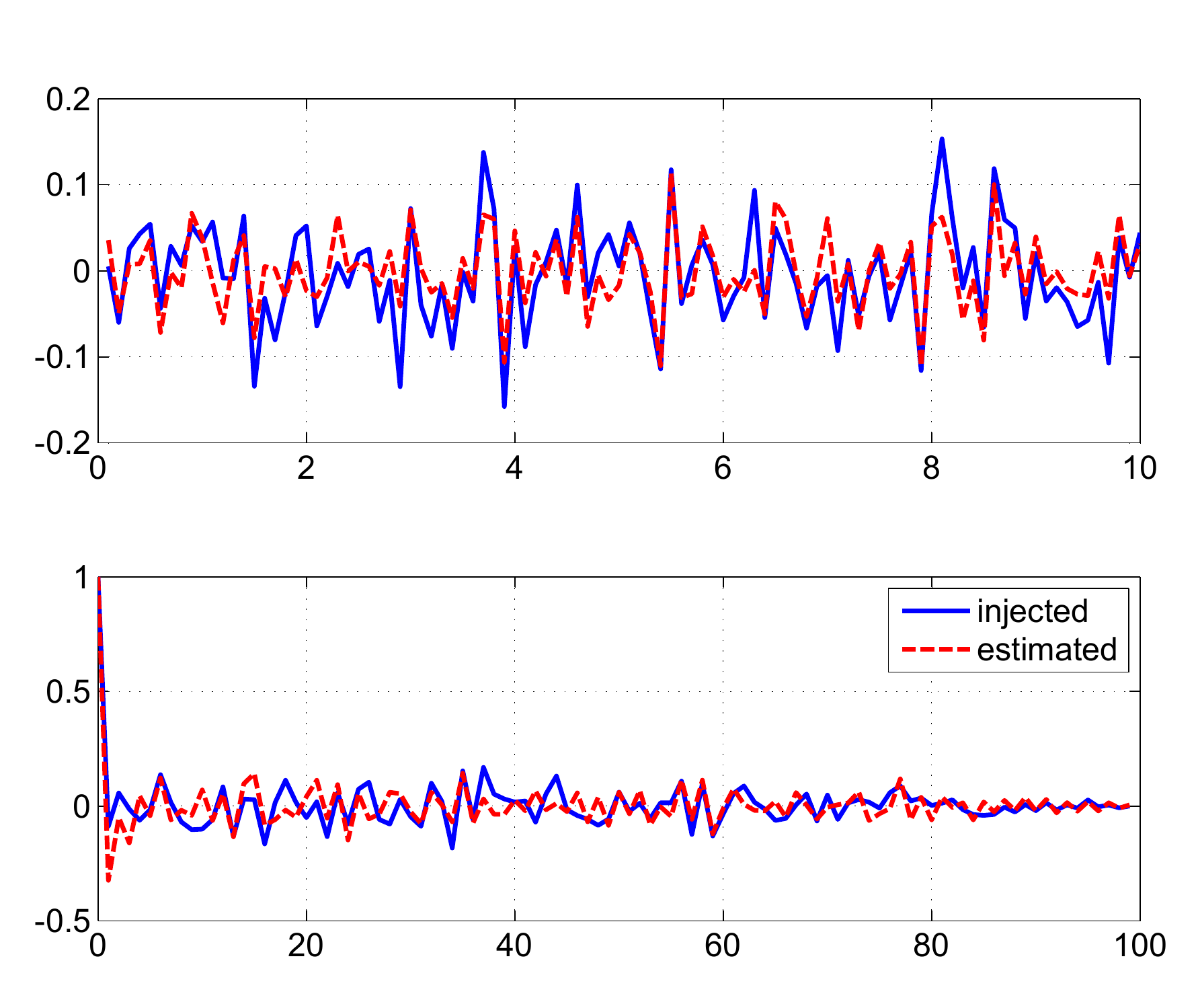}
\caption{Time variation of injected and estimated measurement noise (top) and}
\caption*{their autocorrelation (bottom) for measurement 2 (velocity)}
\label{smdQ_mnoise2}
\end{figure}

\begin{figure}[h]
\includegraphics[width=6in,height=4in]{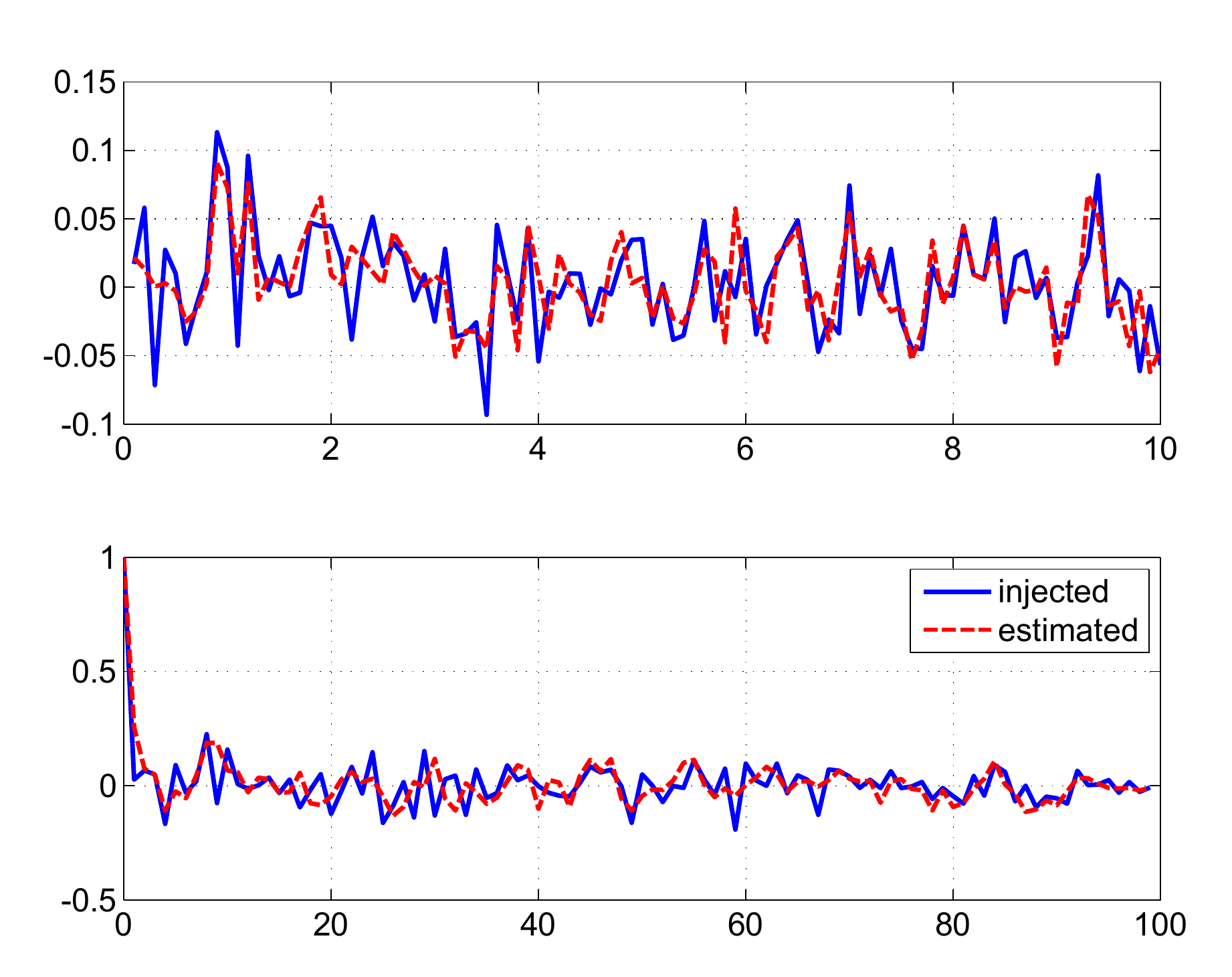}
\caption{Time variation of injected and estimated process noise (top) and}
\caption*{their autocorrelation (bottom) for state 1 (displacement)}
\label{smdQ_pnoise1}
\end{figure}

\begin{figure}[h]
\includegraphics[width=6in,height=4in]{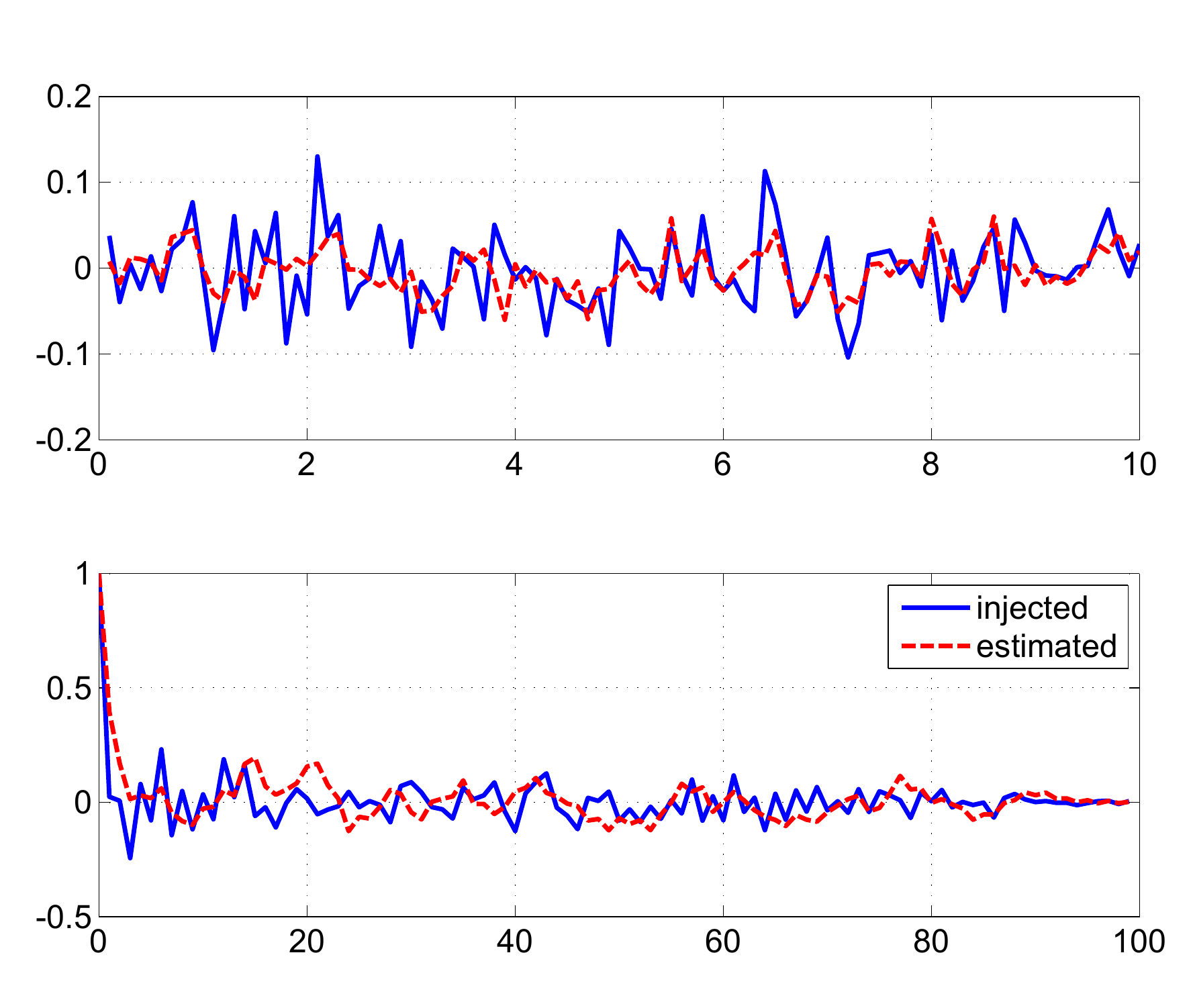}
\caption{Time variation of injected and estimated process noise (top) and}
\caption*{their autocorrelation (bottom) for state 2 (velocity)}
\label{smdQ_pnoise2}
\end{figure}

 The next step is to process the data with process noise (\textbf{Q} $>$ 0). The Fig. \ref{smd_err}-\ref{smdQ_pnoise2} refer to the \textbf{Q} $>$ 0 case. The Fig. \ref{smd_err} shows the absolute difference between the iterated and final values with iterations which indicates the accuracy level that one needs and it was found that 100 iterations are required. The variation of the estimated initial parameters and their variances through iterations are shown in Fig. \ref{smdQ_P0}. The parameter and the uncertainty reach almost their final estimated values in about 5 and 20 iterations respectively. A similar plot in Fig. \ref{smdQ_R} shows the variation of the estimated measurement and process noise. The variation of different cost functions through the iterations are shown in Fig. \ref{smdQ_J}. The cost functions \textbf{J1-J3} correspond to the number of measurement ($m$=2) and in presence of process noise, \textbf{J6-J8} correspond to the number of states ($n$=2). The \textbf{J4} in absence of process noise corresponds to the trace of the measurement noise \textbf{R}. The \textbf{J5} is the negative log likelihood cost whose absolute value is shown in the plot. There is a mismatch in the predicted dynamics and the measurement as seen in Fig. \ref{smdQ_h1}-\ref{smdQ_h2} indicating the presence of process noise. The subsequent Fig. \ref{smdQ_innov1}-\ref{smdQ_innov2} correspond to the earlier Fig. \ref{smd_innov1}-\ref{smd_innov2} of \textbf{Q} = 0 case. The Fig.  \ref{smdQ_mnoise1}-\ref{smdQ_mnoise2} and Fig. \ref{smdQ_pnoise1}-\ref{smdQ_pnoise2} shows respectively the injected and estimated measurement and process noise samples across time during the final iteration.


\begin{landscape}
\begin{table}[h]
\subsection{SMD System Tables (\textbf{Q} = 0) }
\vspace{14pt}
\caption{Sensitivity Study : (\textbf{Q} = 0) : SMD system.\\ No. of iterations=20, No. of simulations=50.}{}
\label{tbsmd}
\begin{center}
\begin{footnotesize}
\begin{tabular}{| c | c | c | c | c | c | c | c | c | c | c | c |}
\hline
Study &
$\Theta$ ratio &
\makecell{CRB ratio\\NR/EKF} &
\makecell{Consistency\\ ratio-EKF}&
\makecell{Consistency\\ ratio-NR}&
\makecell{Spread\\factor\\EKF} &
\makecell{Spread\\factor\\NR} &
\makecell{\textbf{R} ratio\\EKF/True} & \makecell{\textbf{R} ratio\\EKF/NR} &
\makecell{$\mu$\ of \\\textbf{J1-J5}} &
\makecell{$\sigma$ of \\\textbf{J1-J5}} &
Remarks
\\ \hline

\multicolumn{12}{|c|}{Reference adaptive EKF used for non zero process noise case} \\ \hline
\makecell{$\mathbf{P_0}$ : Scaled up-[0,0;0,\checkmark]\\\textbf{Q} : EM-[\checkmark,0;0,0] \\ \textbf{R} : EM-diag}
&\multicolumn{11}{c|}{Extremely slow convergence of \textbf{Q} taking more than 100 iterations.} \\ \hline

\multicolumn{12}{|c|}{Proposed reference adaptive EKF for zero process noise case, \textbf{Q} is diagonal matrix fixed to $10^{-10}$ for all states} \\ \specialrule{.2em}{.0em}{.0em}

\multicolumn{1}{!{\vrule width 2pt}c|}{
\makecell{$\mathbf{P_0}$ : Scaled up-[0,0;0,\checkmark]\\\textbf{Q} : $10^{-10}$-[\checkmark,0;0,0] \\ \textbf{R} : EM-diag \\ RTS-Smoothing} } &
\makecell{ 1.0007  \\  1.0000  \\  0.9869} & \makecell{ 0.9952   \\ 1.0171  \\  1.0173} &
\makecell{ 1.0140  \\  1.3764 \\   1.0919} & \makecell{  1.0533 \\   1.3138 \\   1.0817} &
\makecell{0.8128  \\  1.5488 \\  14.5838} & \makecell{0.8126  \\  1.5079 \\  14.5843} &
\makecell{ 0.9807  \\  0.9743} & \makecell{ 1.0139 \\  1.0128 } &
\makecell{1.9704\\    1.9702  \\  1.9999 \\   0.0048 \\ -10.3911}&
\makecell{ 0.0512  \\  0.0512  \\  0.0013  \\  0.0008  \\  0.2243  } &
\multicolumn{1}{|c!{\vrule width 2pt}}{
\makecell{ $\sigma_{  \Theta}$= \\  0.0240, 0.0055, 0.0670 \\
$SIGMA_{avg}$= \\  0.0236, 0.0040, 0.0614
}}
\\
\hline

\end{tabular}
\end{footnotesize}
\end{center}
\end{table}
\end{landscape}

\addtocounter{table}{-1}
\begin{landscape}
\begin{table}[h]
\caption{Sensitivity Study : (\textbf{Q} = 0) : SMD system.\\ No. of iterations=20, No. of simulations=50.}{}
\begin{center}
\begin{footnotesize}
\begin{tabular}{|c| c| c| c| c| c| c|c|c|c|c|c|}
\hline
Study &
$\Theta$ ratio &
\makecell{CRB ratio\\NR/EKF} &
\makecell{Consistency\\ ratio-EKF}&
\makecell{Consistency\\ ratio-NR}&
\makecell{Spread\\factor\\EKF} &
\makecell{Spread\\factor\\NR} &
\makecell{\textbf{R} ratio\\EKF/True} & \makecell{\textbf{R} ratio\\EKF/NR} &
\makecell{$\mu$\ of \\\textbf{J1-J5}} &
\makecell{$\sigma$ of \\\textbf{J1-J5}} &
Remarks
\\ \hline

\makecell{$\mathbf{P_0}$ : IIM-[0,0;0,\checkmark]\\\textbf{Q} : $10^{-10}$-[\checkmark,0;0,0] \\ \textbf{R} : EM-diag}  &
\makecell{ 1.0014  \\  0.9990  \\  0.9667} & \makecell{  0.9934  \\  1.0058  \\  1.0212} &
\makecell{ 1.1296  \\  2.2338   \\ 1.3654} & \makecell{1.0533  \\  1.3138 \\   1.0817} &
\makecell{0.8997  \\  2.3360 \\  17.3808} & \makecell{0.8126  \\  1.5079 \\  14.5843}&
\makecell{ 0.9894 \\   0.9816} & \makecell{ 1.0228 \\   1.0204 } &
\makecell{  1.9690 \\   1.9710   \\ 1.9998 \\   0.0050  \\-10.2078 }&
\makecell{0.0776  \\  0.0776 \\   0.0015 \\   0.0009 \\   0.2349 }&
Unity CRB ratio\\ \hline

\makecell{$\mathbf{P_0}$ : IIM-diag\\\textbf{Q} : $10^{-10}$-[\checkmark,0;0,0] \\ \textbf{R} : EM-diag}  &
\makecell{ 0.9995  \\  1.0023 \\   1.0375} & \makecell{0.8007 \\   0.7215  \\  0.5713} &
\makecell{ 1.0476  \\  1.5883  \\  1.1881} & \makecell{1.0533  \\  1.3138 \\   1.0817} &
\makecell{  1.0205 \\   2.4209  \\ 26.8883} & \makecell{0.8126  \\  1.5079 \\  14.5843} &
\makecell{0.9828   \\ 0.9718} & \makecell{  1.0160 \\   1.0102 } &
\makecell{1.9530  \\  1.9542  \\  1.9998 \\   0.0049 \\ -10.1918}&
\makecell{ 0.0655  \\  0.0655  \\  0.0049  \\  0.0009   \\ 0.2403 }&
\makecell{Higher filter CRB\\ than that of the NR}\\
\hline

\end{tabular}
\end{footnotesize}
\end{center}
\end{table}
\end{landscape}

\addtocounter{table}{-1}
\begin{landscape}
\begin{table}[h]
\caption{Sensitivity Study : (\textbf{Q} = 0) : SMD system.\\ No. of iterations=20, No. of simulations=50.}{}
\begin{center}
\begin{footnotesize}
\begin{tabular}{|c| c| c| c| c| c| c|c|c|c|c|c|}
\hline
Study &
$\Theta$ ratio &
\makecell{CRB ratio\\NR/EKF} &
\makecell{Consistency\\ ratio-EKF}&
\makecell{Consistency\\ ratio-NR}&
\makecell{Spread\\factor\\EKF} &
\makecell{Spread\\factor\\NR} &
\makecell{\textbf{R} ratio\\EKF/True} & \makecell{\textbf{R} ratio\\EKF/NR} &
\makecell{$\mu$\ of \\\textbf{J1-J5}} &
\makecell{$\sigma$ of \\\textbf{J1-J5}} &
Remarks
\\ \hline

\multicolumn{12}{|c|}{Using different methods for estimating \textbf{R}} \\ \hline

\makecell{$\mathbf{P_0}$ : Scaled up-[0,0;0,\checkmark]\\\textbf{Q} : $10^{-10}$-[\checkmark,0;0,0] \\ \textbf{R} : diag \textbf{R}*}  &
\makecell{1.0007  \\  1.0000  \\  0.9868} & \makecell{ 1.0015 \\   1.0234  \\  1.0237} &
\makecell{ 1.0191 \\   1.3877 \\   1.0984} & \makecell{1.0533  \\  1.3138 \\   1.0817} &
\makecell{0.8094  \\  1.5462 \\  14.5332} & \makecell{0.8126  \\  1.5079 \\  14.5843} &
\makecell{0.9699  \\  0.9615} &  \makecell{1.0027 \\   0.9994} &
\makecell{1.9950  \\  1.9948 \\   2.0248  \\  0.0048 \\ -10.3911 }&
\makecell{0.0589 \\   0.0589 \\   0.0271 \\   0.0008 \\   0.2244 }&
\makecell{*Using dynamical residue\\ as per Eq-\ref{R3}}\\ \hline

\makecell{$\mathbf{P_0}$ : Scaled up-[0,0;0,\checkmark]\\\textbf{Q} : $10^{-10}$-[\checkmark,0;0,0] \\\textbf{R} : diag \textbf{R}*}  &
\makecell{1.0007  \\  1.0000  \\  0.9869} & \makecell{ 1.0028  \\  1.0248  \\  1.0250} &
\makecell{ 1.0218   \\ 1.3873 \\   1.1004}  & \makecell{1.0533  \\  1.3138 \\   1.0817} &
\makecell{ 0.8094  \\  1.5436 \\  14.5273} & \makecell{0.8126  \\  1.5079 \\  14.5843}&
\makecell{ 0.9665  \\  0.9598} & \makecell{ 0.9992 \\   0.9977} &
\makecell{ 2.0001  \\  1.9999  \\  2.0300  \\  0.0048\\  -10.3913}&
\makecell{  0.0522  \\  0.0522  \\  0.0037 \\   0.0008  \\  0.2243  }&
\makecell{*Using smoothed residue\\ignoring $2^{nd}$ order\\terms in Eq-\ref{R1}}\\
\hline

\end{tabular}
\end{footnotesize}
\end{center}
\end{table}
\end{landscape}


\begin{landscape}
\begin{table}[h]
\subsection{SMD System Tables (\textbf{Q} $>$ 0) }
\vspace{14pt}
\caption{Sensitivity Study : (\textbf{Q} $>$ 0) : SMD system.\\ No. of iterations=100, No. of simulations=50.}{}
\label{tbsmdQ}
\begin{center}
\begin{footnotesize}
\begin{tabular}{|c| c| c| c| c| c| c|c|c|c|c|}
\hline
Study &
$\Theta$ ratio &
\makecell{PCRB\\ ratio} &
\makecell{Consistency\\ ratio-EKF}&
\makecell{Spread\\factor\\EKF} &
\makecell{\textbf{R} ratio\\EKF/True} &
\makecell{\textbf{Q} ratio\\EKF/True}  &
\makecell{$\mu$\ of \\\textbf{J1-J8}} &
\makecell{$\sigma$ of \\\textbf{J1-J8}} &
Remarks
\\ \hline

\multicolumn{10}{|c|}{Proposed reference adaptive EKF for non zero process noise} \\ \specialrule{.2em}{.0em}{.0em}

\multicolumn{1}{!{\vrule width 2pt}c|}{
\makecell{$\mathbf{P_0}$ : Scaled up-[0,0;0,\checkmark]\\\textbf{Q} : EM-[\checkmark,0;0,0] \\\textbf{R} : EM-diag\\RTS-Smoothing}}  &
\makecell{ 0.9933 \\   1.0201 \\   1.0764} &
\makecell{ 1.0179  \\  1.0286} &
\makecell{1.0327  \\  1.0772  \\  0.9128} & \makecell{7.9434 \\  23.5655 \\  98.4334} &
\makecell{  0.9450  \\  0.9135} &
\makecell{ 1.0648 \\   1.0922} & \makecell{  1.9650\\1.9700\\1.9982\\ 0.0709\\-8.7886\\1.9439\\1.9533\\1.9585} & \makecell{ 0.0224 \\   0.0217\\    0.0091 \\  0.0396 \\  0.2281 \\ 0.0422 \\   0.0618 \\   0.0373} &
\multicolumn{1}{|c!{\vrule width 2pt}}{
\makecell{$\sigma_{  \Theta}$= \\ 0.2418, 0.0720, 0.4089 \\
$SIGMA_{avg}$= \\ 0.2342, 0.0668, 0.4480}} \\ \specialrule{.2em}{.0em}{.0em}

\multicolumn{1}{!{\vrule width 2pt}c|}{
\makecell{$\mathbf{P_0}$ : Scaled up-[0,0;0,\checkmark]\\\textbf{Q} : EM-[\checkmark,0;0,0] \\\textbf{R} : EM-diag\\MBF-Smoothing}}  &
\makecell{  0.9933  \\  1.0203 \\   1.0764} &
\makecell{ 1.0178  \\  1.0309} &
\makecell{   1.0314  \\  1.0762  \\  0.9113} & \makecell{ 7.9494 \\  23.5909 \\  98.4953} &
\makecell{ 0.9445 \\   0.9116} & \makecell{  1.0648  \\  1.0965} &
\makecell{1.9656 \\   1.9707  \\  1.9994  \\  0.0709 \\  -8.7886   \\ 1.9441  \\  1.9535 \\ 1.9592} &
\makecell{ 0.0221   \\ 0.0219  \\  0.0029  \\  0.0396  \\  0.2281  \\  0.0429 \\   0.0623 \\0.0366} &
\multicolumn{1}{|c!{\vrule width 2pt}}{
\makecell{$\sigma_{  \Theta}$= \\  0.2418, 0.0720,  0.4087 \\
$SIGMA_{avg}$= \\ 0.2345, 0.0669,  0.4485}} \\
\end{tabular}
\end{footnotesize}
\end{center}
\end{table}
\end{landscape}

\addtocounter{table}{-1}
\begin{landscape}
\begin{table}[h]
\caption{Sensitivity Study : (\textbf{Q} $>$ 0) : SMD system.\\ No. of iterations=100, No. of simulations=50.}{}
\begin{center}
\begin{footnotesize}
\begin{tabular}{|c| c| c| c| c| c| c|c|c|c|}
\hline
Study &
$\Theta$ ratio &
\makecell{PCRB\\ ratio} &
\makecell{Consistency\\ ratio-EKF}&
\makecell{Spread\\factor\\EKF} &
\makecell{\textbf{R} ratio\\EKF/True} &
\makecell{\textbf{Q} ratio\\EKF/True}  &
\makecell{$\mu$\ of \\\textbf{J1-J8}} &
\makecell{$\sigma$ of \\\textbf{J1-J8}} &
Remarks
\\ \specialrule{.2em}{.0em}{.0em}

\multicolumn{1}{!{\vrule width 2pt}c|}{
\makecell{$\mathbf{P_0}$ : Scaled up-[0,0;0,\checkmark]\\\textbf{Q} : EM-[\checkmark,0;0,0] \\\textbf{R} : Known}}  &
\makecell{ 0.9942  \\  1.0143  \\  1.0705} &
\makecell{ 0.9914 \\   0.9926} &
\makecell{  1.0052  \\  1.0596 \\   0.9034} &
\makecell{7.8903 \\  23.2708 \\  97.1068} &
\makecell{NA} &
\makecell{ 1.0343 \\   1.0112} &
\makecell{  1.9489 \\   1.9516   \\ 1.9670  \\  0.0691 \\  -8.7675  \\  1.9335 \\   1.9395\\ 1.9456} &
\makecell{   0.1121  \\  0.1136 \\   0.1862  \\  0.0357  \\  0.2236  \\  0.0350  \\  0.0451\\ 0.1118} &
\multicolumn{1}{|c!{\vrule width 2pt}}{
\makecell{\textbf{R} is known}} \\ \specialrule{.2em}{.0em}{.0em}

\multicolumn{1}{!{\vrule width 2pt}c|}{
\makecell{$\mathbf{P_0}$ : Zero\\\textbf{Q} : EM-diag \\\textbf{R} : EM-diag}}  &
\makecell{NA} &
\makecell{1.0136 \\   1.0214} &
\makecell{NA} & \makecell{NA} &
\makecell{ 0.9484  \\  0.9248} & \makecell{1.0582 \\   1.0601} &
\makecell{ 1.9997 \\   2.0011  \\  1.9981 \\   0.0819 \\  -8.8857 \\   1.9978 \\    2.0067 \\ 1.9982} &
\makecell{ 0.0216  \\  0.0216  \\  0.0083  \\  0.0508 \\   0.2244  \\  0.0158  \\  0.0403 \\0.0246} &
\multicolumn{1}{|c!{\vrule width 2pt}}{
\makecell{The parameters are \\assumed to be known}} \\
\end{tabular}
\end{footnotesize}
\end{center}
\end{table}
\end{landscape}

\addtocounter{table}{-1}
\begin{landscape}
\begin{table}[h]
\caption{Sensitivity Study : (\textbf{Q} $>$ 0) : SMD system.\\ No. of iterations=100, No. of simulations=50.}{}
\begin{center}
\begin{footnotesize}
\begin{tabular}{|c| c| c| c| c| c| c|c|c|c|}
\hline
Study &
$\Theta$ ratio &
\makecell{PCRB\\ ratio} &
\makecell{Consistency\\ ratio-EKF}&
\makecell{Spread\\factor\\EKF} &
\makecell{\textbf{R} ratio\\EKF/True} &
\makecell{\textbf{Q} ratio\\EKF/True}  &
\makecell{$\mu$\ of \\\textbf{J1-J8}} &
\makecell{$\sigma$ of \\\textbf{J1-J8}} &
Remarks
\\ \hline

\multicolumn{10}{|c|}{Extended EM algorithm (Bavdekar et al. \cite{Bavdekar2011} (2011)} \\ \hline
\makecell{$\mathbf{P_0}$ : Smoothed-full\\\textbf{Q} : EM-full \\\textbf{R} : EM-full}  & \multicolumn{9}{c|}{\makecell{Cost functions diverges after few iterations.\\There is a need for a precise stopping condition without which unity ratios cannot be achieved.}}  \\ \hline

\multicolumn{10}{|c|}{Modification-1} \\ \hline

\makecell{$\mathbf{P_0}$ : Smoothed-diag\\\textbf{Q} : EM-diag \\\textbf{R} : EM-diag}   &
\makecell{  1.0210  \\  1.0018 \\   0.6806} & \makecell{ 1.0164  \\  1.0473} &
\makecell{ 13.3560 \\  10.0828 \\  13.6464} & \makecell{ 3.9686 \\  13.6349  \\ 51.6473} &
\makecell{ 0.9502 \\   0.9703} & \makecell{1.0286 \\   0.8676} &
\makecell{ 1.9988\\ 1.9996\\ 1.9998\\0.0656\\-8.9374 \\ 2.0031\\ 2.0027\\1.9976} & \makecell{0.0241\\0.0241\\0.0076  \\  0.0325 \\   0.2290  \\  0.0147   \\ 0.0239\\ 0.0282 }& \makecell{No stoping condition.\\ The low spread factor is deceiving\\since there is no consistency in the\\ parameter estimates with its covariance.}\\ \hline

\multicolumn{10}{|c|}{Modification-2} \\
\hline

\end{tabular}
\end{footnotesize}
\end{center}
\end{table}
\end{landscape}

\addtocounter{table}{-1}
\begin{landscape}
\begin{table}[h]
\caption{Sensitivity Study : (\textbf{Q} $>$ 0) : SMD system.\\ No. of iterations=100, No. of simulations=50.}{}
\begin{center}
\begin{footnotesize}
\begin{tabular}{|c| c| c| c| c| c| c|c|c|c|}
\hline
Study &
$\Theta$ ratio &
\makecell{PCRB\\ ratio} &
\makecell{Consistency\\ ratio-EKF}&
\makecell{Spread\\factor\\EKF} &
\makecell{\textbf{R} ratio\\EKF/True} &
\makecell{\textbf{Q} ratio\\EKF/True}  &
\makecell{$\mu$\ of \\\textbf{J1-J8}} &
\makecell{$\sigma$ of \\\textbf{J1-J8}} &
Remarks
\\ \hline

\multicolumn{10}{|c|}{MS method (Mohamed and Schwarz)} \\ \hline
\makecell{$\mathbf{P_0}$ : Scaled up-[0,0;0,\checkmark]\\\textbf{Q} : MS-diag \\\textbf{R} : MS-diag}  & \multicolumn{9}{c|}
{\makecell{\textbf{R} can converge to a very large value and the cost functions diverge as seen in Fig. \ref{MS_SMD_Q},\ref{MS_SMD_J}.\\ There is no convergence in cost even if $\Theta$ is known.}}\\ \hline

\multicolumn{10}{|c|}{MS method with \textbf{Q} being estimated using last N/2 samples} \\ \hline
\makecell{$\mathbf{P_0}$ : Zero\\\textbf{Q} : MS-diag \\\textbf{R} : MS-diag}  &
NA & \makecell{ 1.0084 \\   1.0437} &
NA & NA & NA &
\makecell{0.9510  \\  0.8739} &
\makecell{2.1281  \\  2.1309  \\  2.0729  \\  0.0819  \\ -8.8300  \\  2.3141  \\  2.3369 \\ 2.1230} &
\makecell{ 0.3269  \\  0.3290  \\  0.2541 \\   0.0508  \\  0.2363  \\  0.6473  \\  0.6726 \\ 0.3213} &
\makecell{\textbf{R}, $\Theta$ is known} \\ \hline

\multicolumn{10}{|c|}{MS method assuming \textbf{R} is known and \textbf{Q} estimated using last N/2 samples} \\ \hline
\makecell{$\mathbf{P_0}$ : Scaled up-[0,0;0,\checkmark]\\\textbf{Q} : MS-[\checkmark,0;0,0] \\\textbf{R} : Known} &
\makecell{0.9926 \\   1.0258 \\   1.0883} & \makecell{0.9989  \\  0.9756} &
\makecell{ 1.0005  \\  1.0762  \\  0.8684} & \makecell{ 8.1555 \\  24.2527 \\  99.2637} &
NA & \makecell{1.0085 \\   1.1562} &
\makecell{ 1.9341 \\   1.9419 \\   1.9466 \\   0.0698 \\  -8.7546 \\   1.9178 \\   1.9220\\1.9260} &
\makecell{ 0.0842  \\  0.0902  \\  0.1191  \\  0.0381  \\  0.2185 \\   0.2016 \\   0.2007\\0.0836} &
\makecell{Stable operation when \\\textbf{R} is known}\\
\hline

\end{tabular}
\end{footnotesize}
\end{center}
\end{table}
\end{landscape}

\addtocounter{table}{-1}
\begin{landscape}
\begin{table}[H]
\caption{Sensitivity Study : (\textbf{Q} $>$ 0) : SMD system.\\ No. of iterations=100, No. of simulations=50.}{}
\begin{center}
\begin{footnotesize}
\begin{tabular}{|c| c| c| c| c| c| c|c|c|c|}
\hline
Study &
$\Theta$ ratio &
\makecell{PCRB\\ ratio} &
\makecell{Consistency\\ ratio-EKF}&
\makecell{Spread\\factor\\EKF} &
\makecell{\textbf{R} ratio\\EKF/True} &
\makecell{\textbf{Q} ratio\\EKF/True}  &
\makecell{$\mu$\ of \\\textbf{J1-J8}} &
\makecell{$\sigma$ of \\\textbf{J1-J8}} &
Remarks
\\ \hline

\multicolumn{10}{|c|}{Covariance matching method  suggested by Myers and Tapley (MT)} \\ \hline
\makecell{$\mathbf{P_0}$ : Scaled up-[0,0;0,\checkmark]\\\textbf{Q} : MT-diag \\\textbf{R} : MT-diag}  & \multicolumn{9}{c|}{\makecell{Positive definiteness of \textbf{Q, R} is not gauranteed and the cost functions can diverge as seen in Fig. \ref{MT_SMD_Q},-\ref{MT_SMD_J}.\\ There is no proper convergence even if $\Theta$ is known.}} \\ \hline

\multicolumn{10}{|c|}{MT method with \textbf{Q} being estimated using only last N/2 samples   } \\ \hline
\makecell{$\mathbf{P_0}$ : Zero\\\textbf{Q} : MS-diag \\\textbf{R} : MS-diag}  &
NA & \makecell{1.0058  \\  1.0145} &
NA & NA & NA &
\makecell{ 0.9716 \\   1.0205} &
\makecell{2.0620 \\   2.0643 \\   2.0266  \\  0.0819  \\ -8.8372  \\  2.2149  \\  2.7665 \\ 2.0569} &
\makecell{0.3138  \\  0.3153 \\   0.2537  \\  0.0508  \\  0.2300 \\   0.6658\\    4.2955\\ 0.3092} &
\makecell{\textbf{R}, $\Theta$ is known} \\ \hline

\multicolumn{10}{|c|}{MT method assuming \textbf{R} is known and \textbf{Q} is estimated using only last N/2 samples} \\ \hline
\makecell{$\mathbf{P_0}$ : Scaled up-[0,0;0,\checkmark]\\\textbf{Q} : MT-[\checkmark,0;0,0] \\\textbf{R} : Known}  &
\makecell{0.9915  \\  1.0148  \\  1.1158}  & \makecell{ 1.0091 \\   1.0223} &
\makecell{  1.1623  \\  1.1677 \\   1.1151} & \makecell{ 7.9841 \\  23.0947 \\ 100.5373} &
NA & \makecell{0.9940 \\   1.0208} &
\makecell{2.0247 \\   2.0656 \\   2.0439 \\   0.0698 \\  -8.7304 \\   2.0215  \\  2.0197 \\ 1.9923} &
\makecell{ 0.2809  \\  0.2941 \\   0.2493  \\  0.0382 \\   0.2290  \\  0.4419  \\  0.4271\\0.2666} &
\makecell{Stable operation}\\
\hline

\end{tabular}
\end{footnotesize}
\end{center}
\end{table}
\end{landscape}

\clearpage
\subsection{SMD System Figures (\textbf{Q} = 0) }

\vspace{14pt}

\clearpage
\subsection{SMD System Figures (\textbf{Q} $>$ 0) }

\begin{figure}[h]
\includegraphics[width=6in,height=4in]{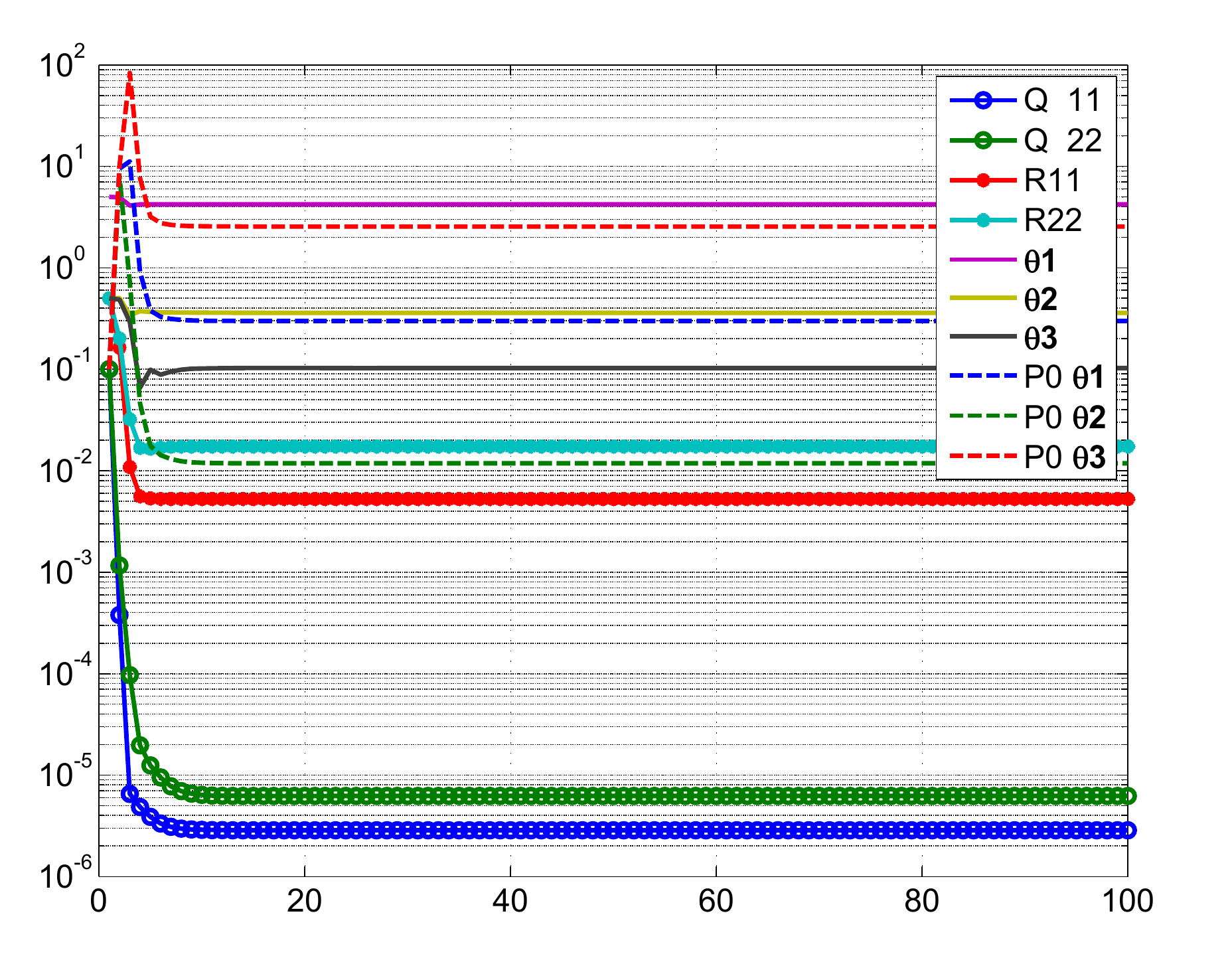}
\caption{Variation of different estimates with iterations using MS method}
\label{MS_SMD_Q}
\end{figure}

\begin{figure}[h]
\includegraphics[width=6in,height=4in]{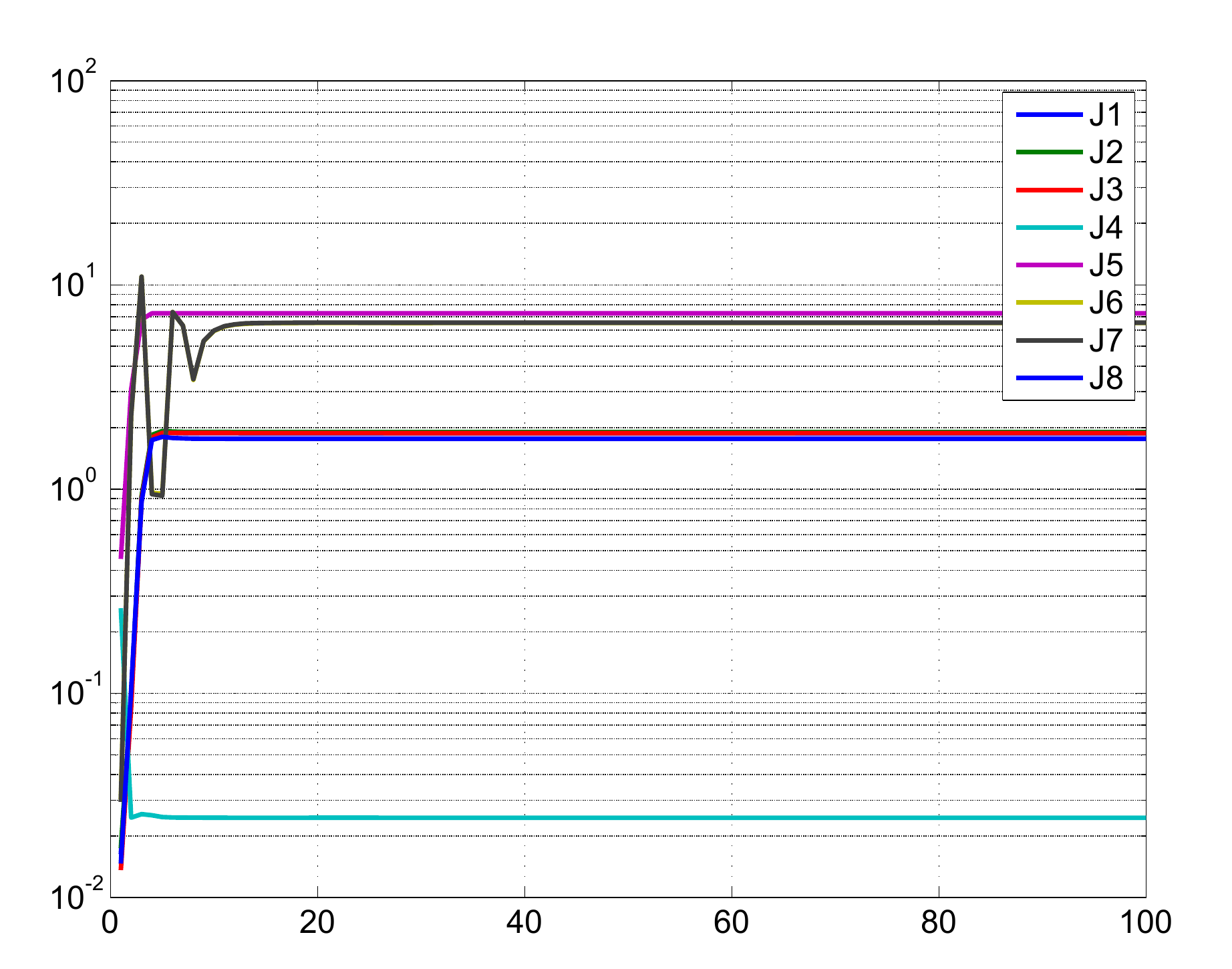}
\caption{Variation of different costs with iterations using MS method}
\label{MS_SMD_J}
\end{figure}

\begin{figure}[h]
\includegraphics[width=6in,height=4in]{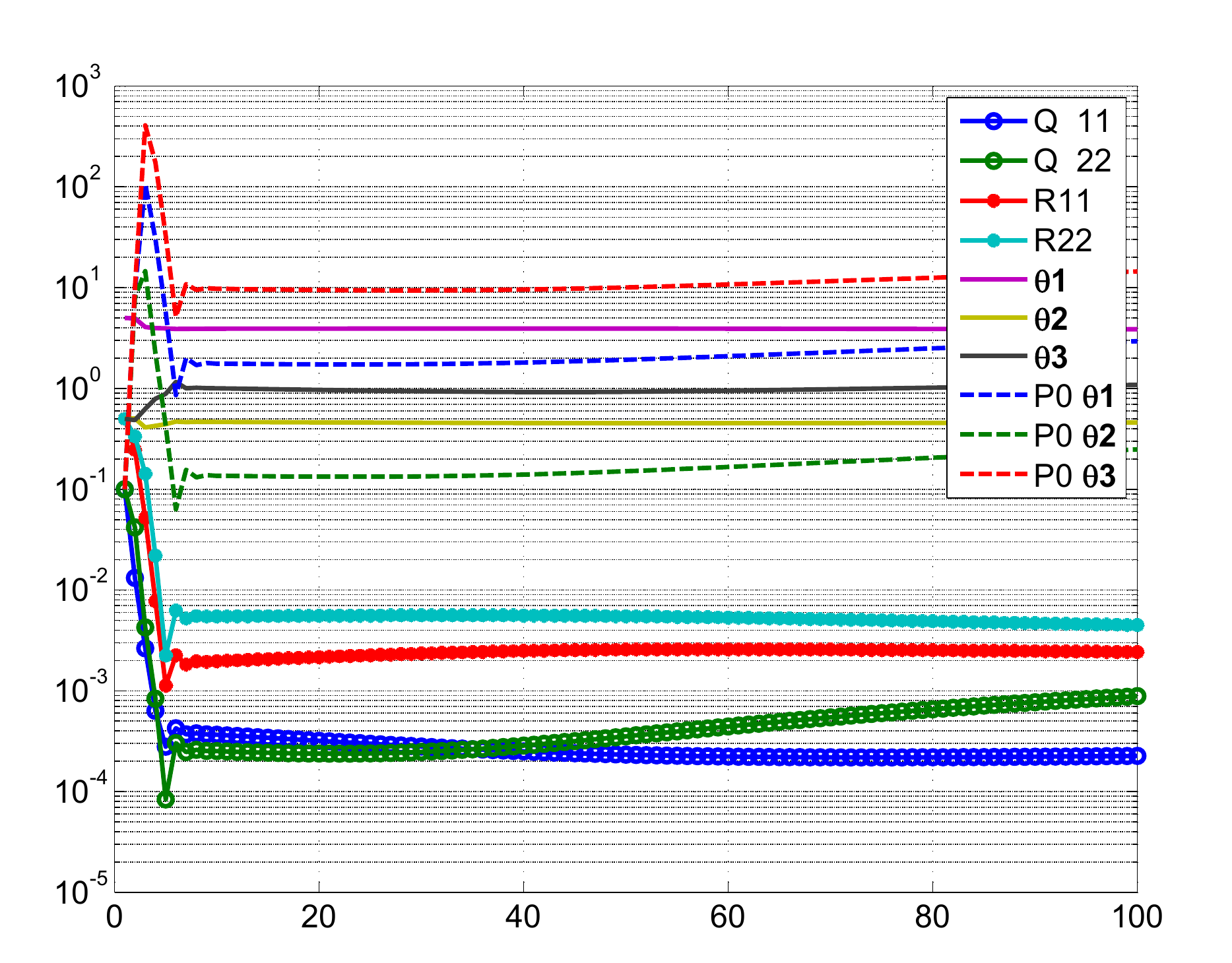}
\caption{Variation of different estimates with iterations using MT method}
\label{MT_SMD_Q}
\end{figure}

\begin{figure}[h]
\includegraphics[width=6in,height=4in]{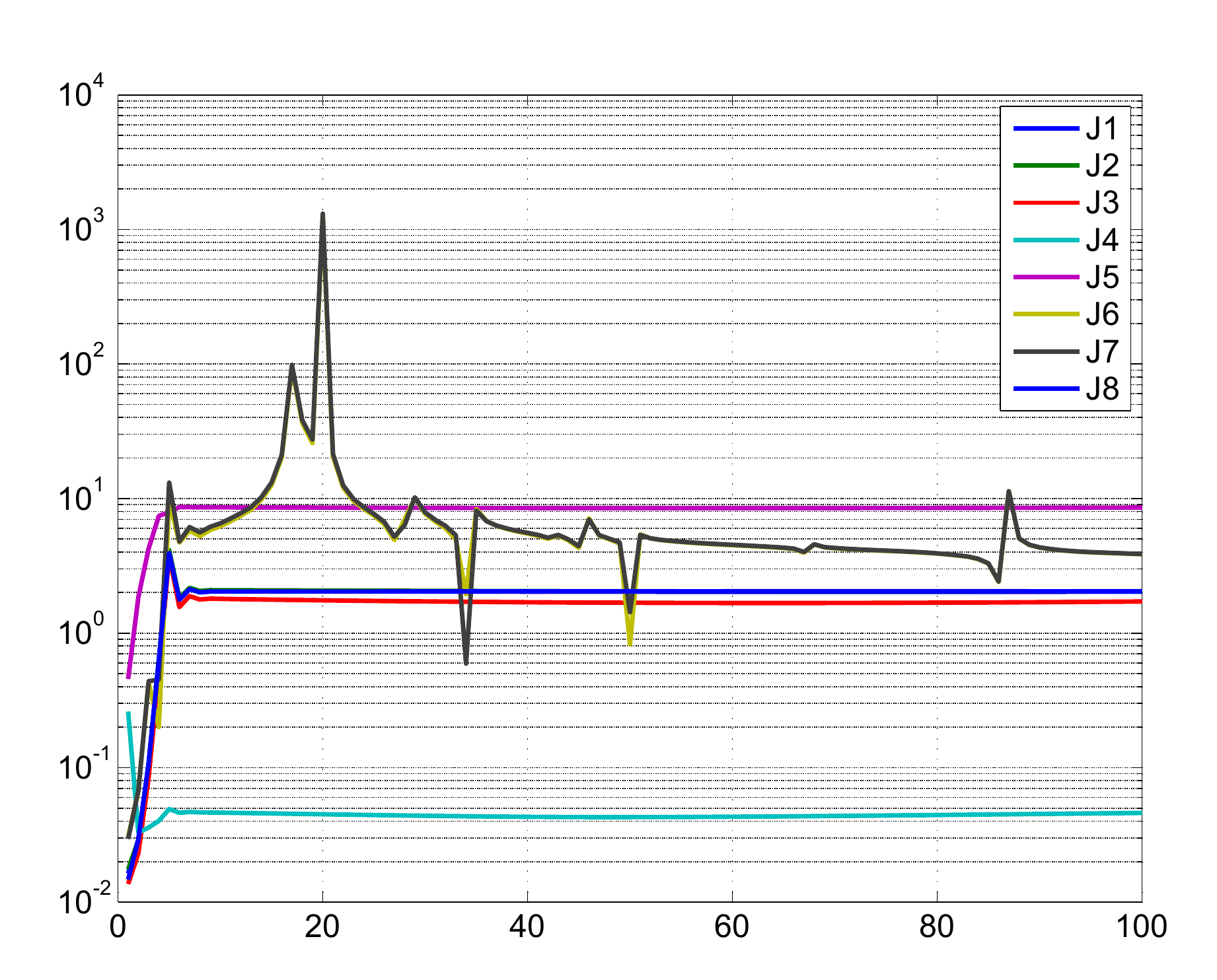}
\caption{Variation of different costs with iterations using MT method}
\label{MT_SMD_J}
\end{figure}

\clearpage
\section{Simulated Longitudinal Motion of Aircraft}

\par
\begin{figure}[h]
\includegraphics[width=6in,height=3in]{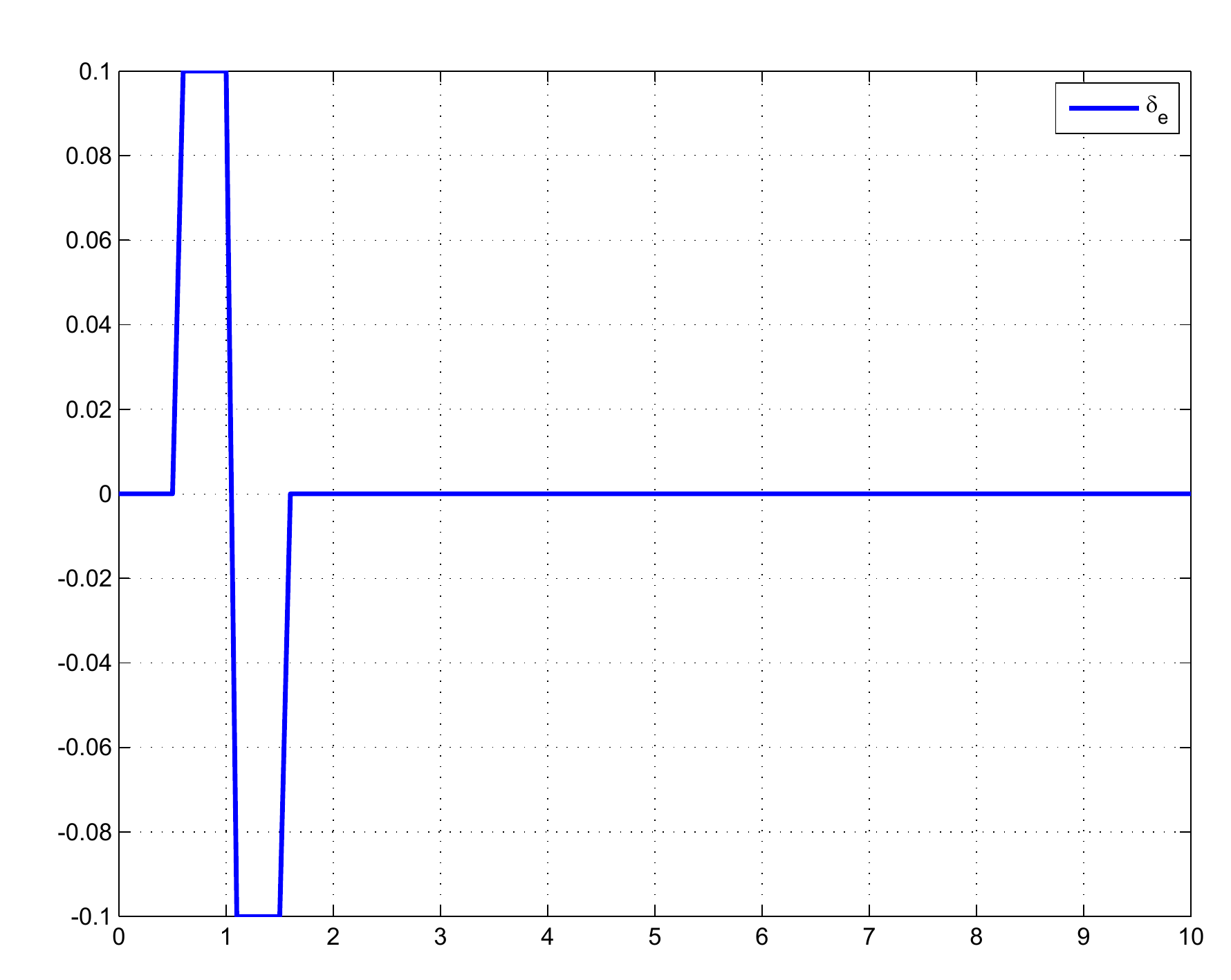}
\caption{Control input of simulated Longitudinal aircraft system}
\label{input_lon}
\end{figure}

 Consider the Longitudinal motion of an aircraft system excited by control input as shown in Fig. \ref{input_lon}. The state equations ($n=4$) for the angle of attack ($\alpha$), pitch rate (q), pitch angle ($\theta$) and downward velocity (V) are respectively given by
\begin{align*}
\dot{\alpha}=&Z_\alpha\alpha+q-0.0021\theta+Z_{\delta_e}\delta_e\\
\dot{q}=&m_\alpha\alpha+m_qq+m_{\delta_e}\delta_e\\
\dot{\theta}=&q\\
\dot{V}=&15.67\alpha-32.16\theta+8.354\delta_e
\end{align*}
All the states have zero initial conditions. The measured quantities are indicated with subscript `m' which includes the angle of attack ($\alpha$), pitch rate (q), pitch angle ($\theta$), downward velocity (V) and normal acceleration ($a_z$). The measurement equations ($m$=5) are given by
\begin{align*}
\alpha_m=&\alpha+v_\alpha \\
q_m=&q+v_q \\
\theta_m=&\theta+v_\theta \\
V_m=&V+v_V \\
a_{z_m}=&U_0\times(Z_\alpha\alpha+Z_{\delta_e}\delta_e)+v_{a_z}
\end{align*}

The unknown parameter ($p=5$) set is $\Theta=(Z_\alpha,m_\alpha,m_q,z_{\delta_e},m_{\delta_e})^T$ with the true values being $\Theta_{true}=(-0.42,-3.7943,-0.3632,-0.006489,-6.2807)^T$ and $U_0$ = 100. The time indices are not shown for brevity in the state and measurement equations. The numerical values of the noise variances are shown in Table-\ref{sysdes}. All the figures are presented for only one simulation run to prevent cluttering.

\subsection{Remarks on the Results}

\par
\begin{figure}[h]
\includegraphics[width=6in,height=3.0in]{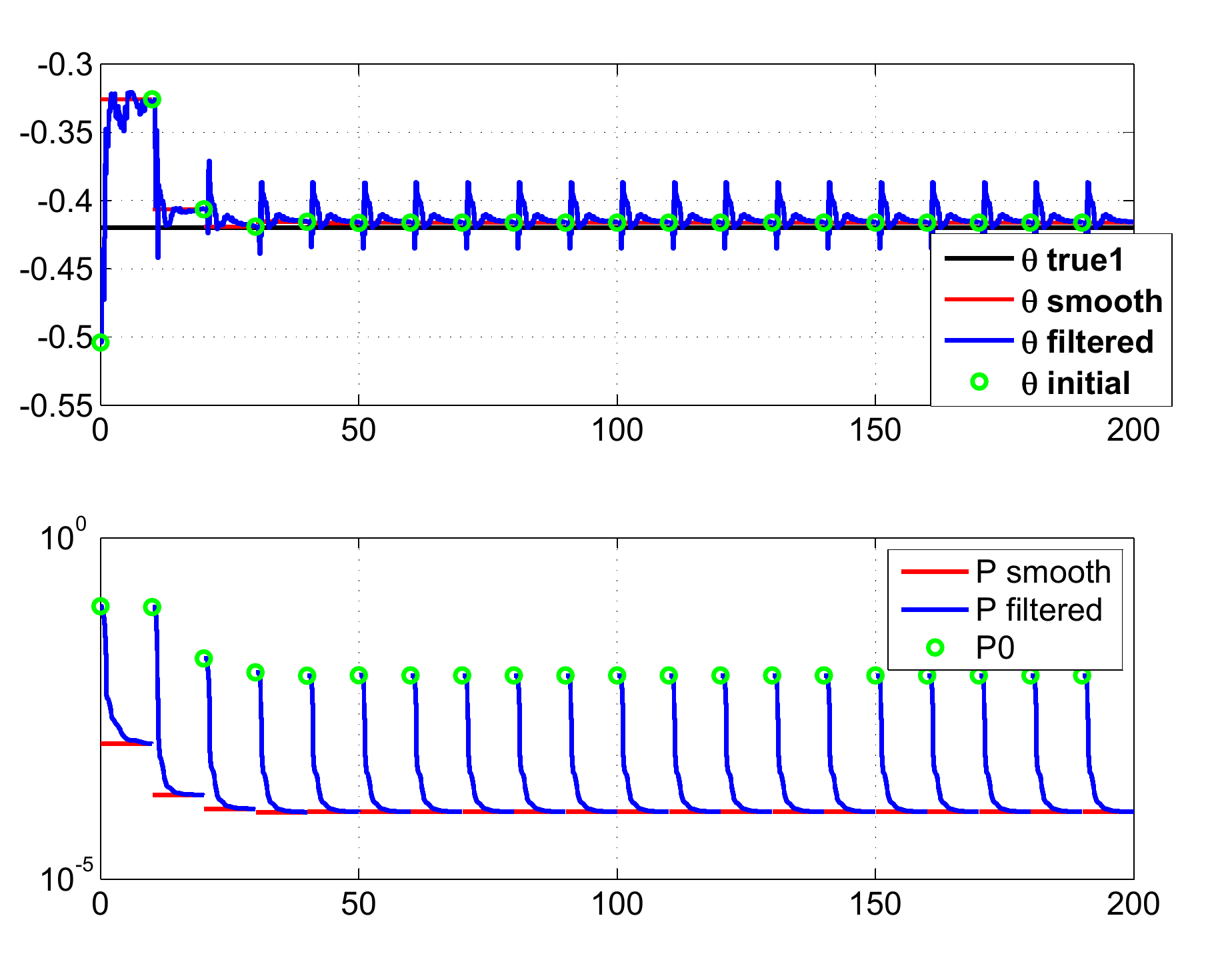}
\caption{The variation of parameter estimate 1 and their filtered and}
\caption*{smoothed covariances through (with the time cumulatively) the iterations}
\label{lon_p1}
\end{figure}

\begin{figure}[h]
\includegraphics[width=6in,height=4in]{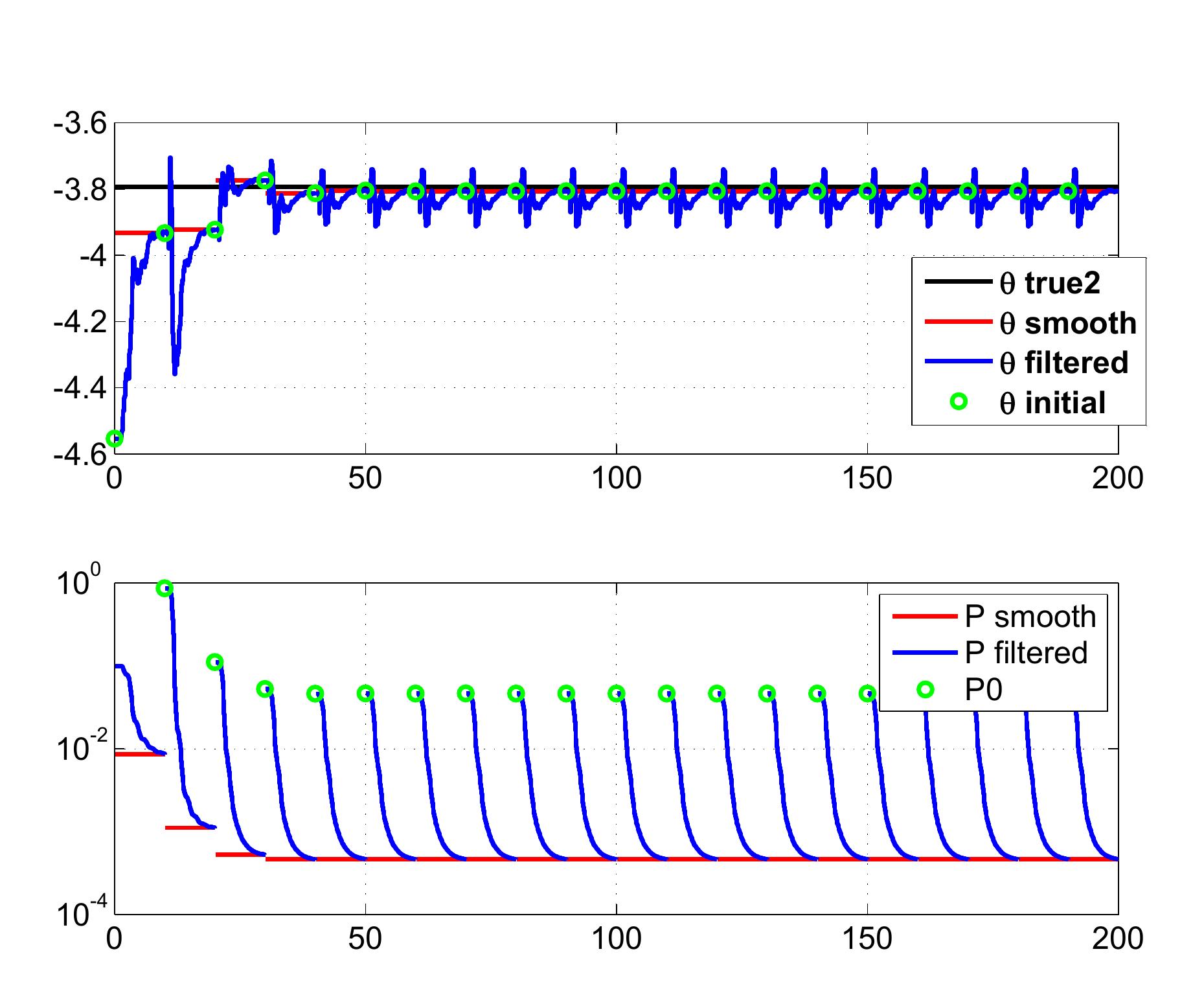}
\caption{The variation of parameter estimate 2 and their filtered and}
\caption*{smoothed covariances through (with the time cumulatively) the iterations}
\label{lon_p2}
\end{figure}

\begin{figure}[h]
\includegraphics[width=6in,height=4in]{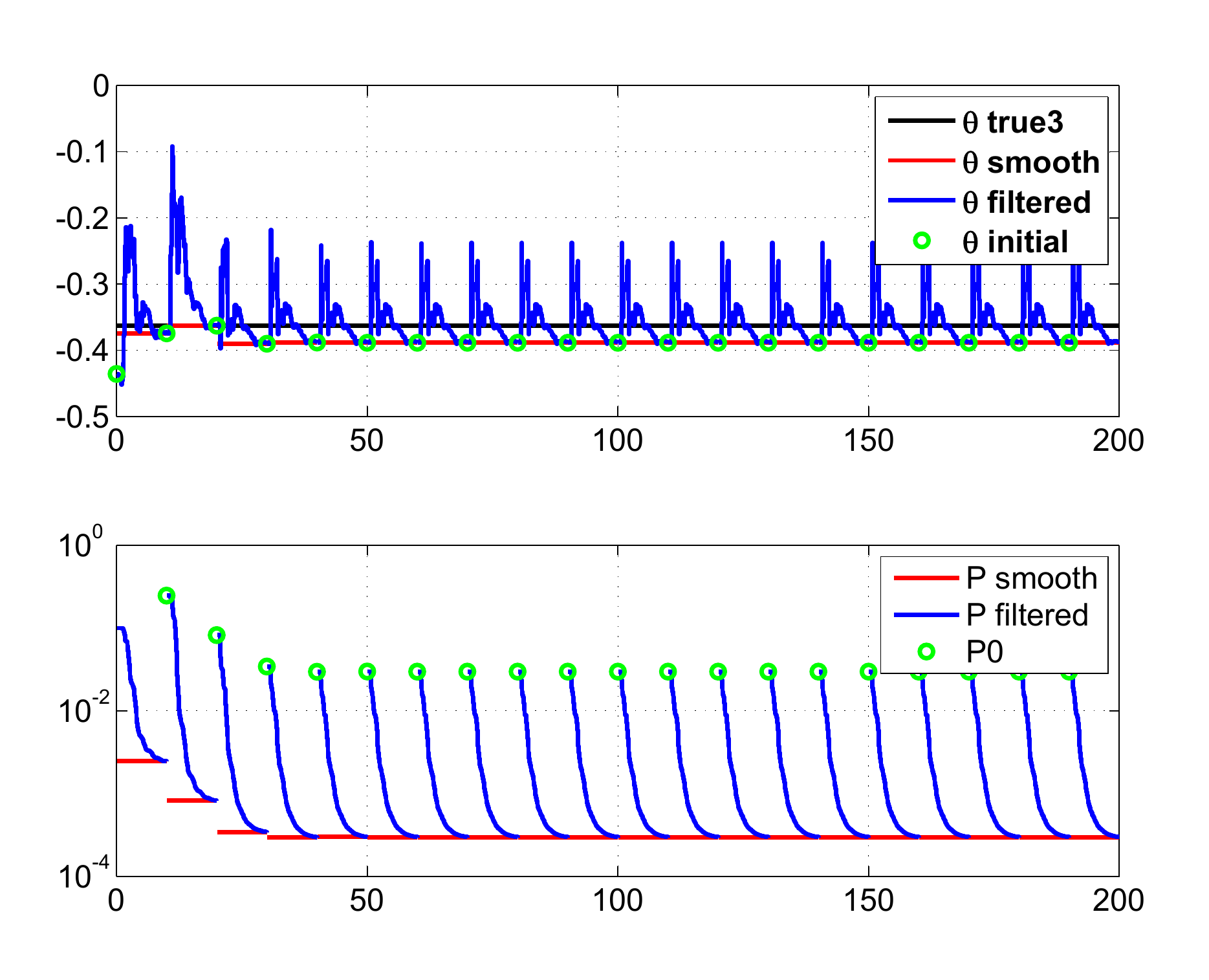}
\caption{The variation of parameter estimate 3 and their filtered and}
\caption*{smoothed covariances through (with the time cumulatively) the iterations}
\label{lon_p3}
\end{figure}

\begin{figure}[h]
\includegraphics[width=6in,height=4in]{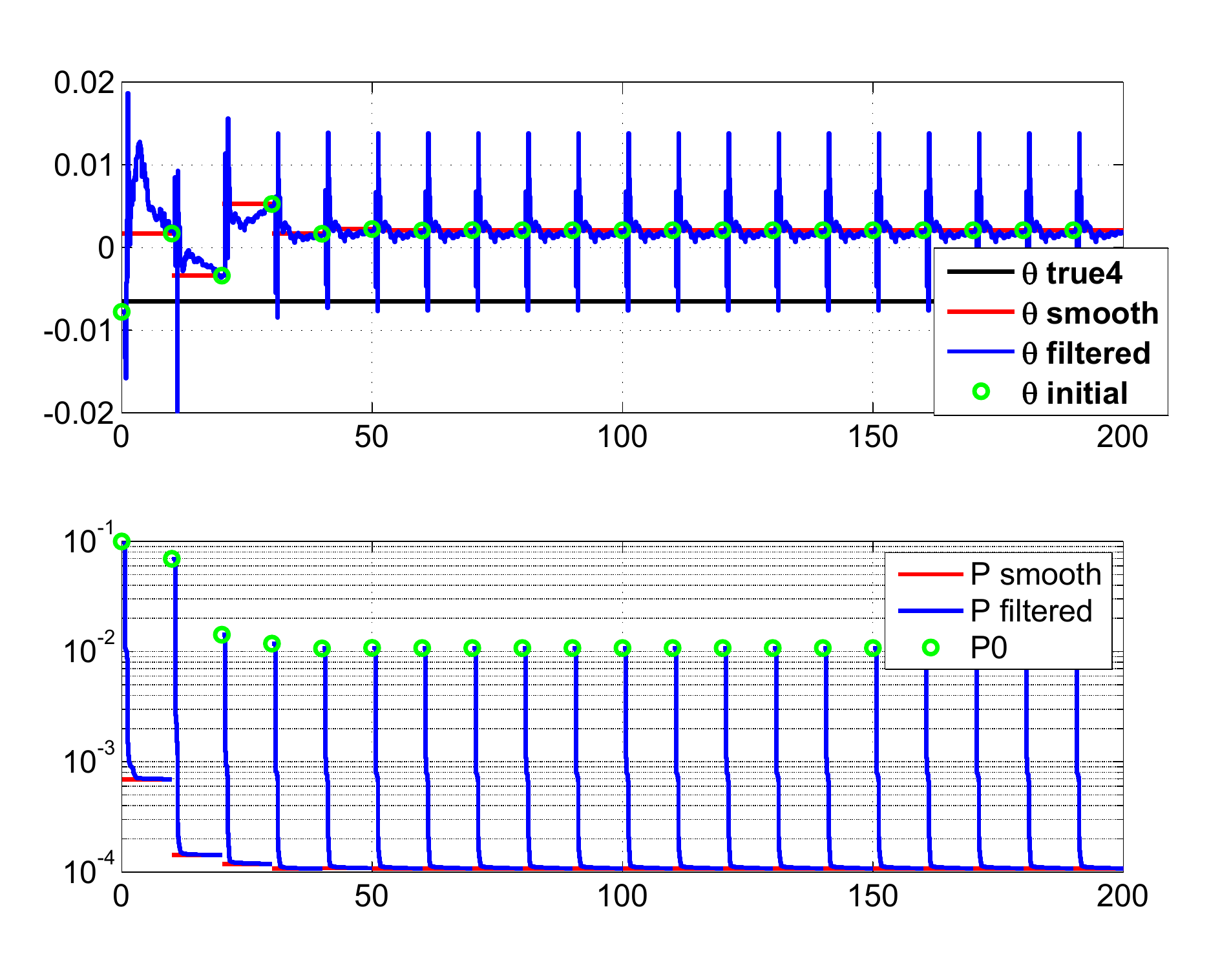}
\caption{The variation of parameter estimate 4 and their filtered and}
\caption*{smoothed covariances through (with the time cumulatively) the iterations}
\label{lon_p4}
\end{figure}

\begin{figure}[h]
\includegraphics[width=6in,height=4in]{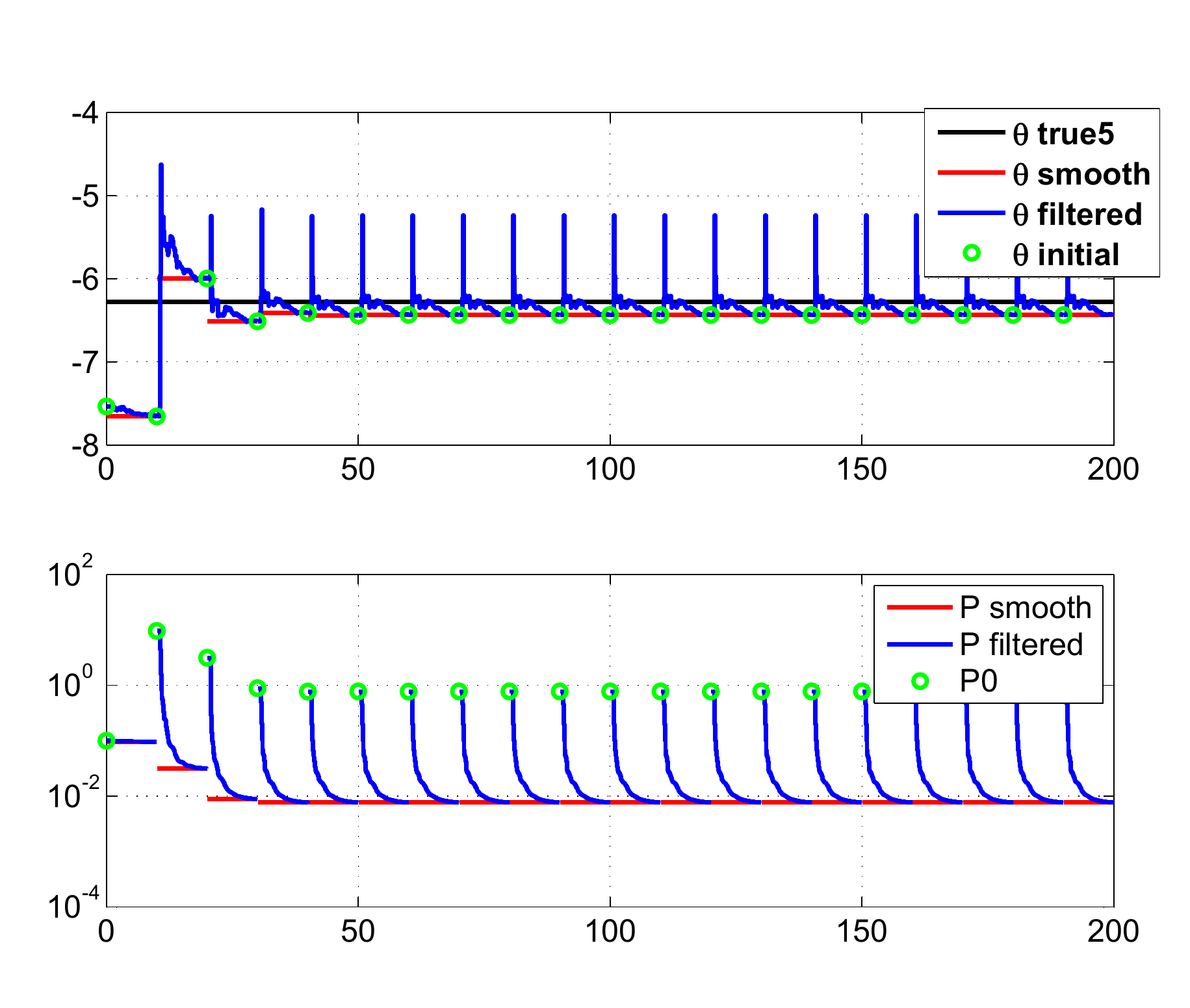}
\caption{The variation of parameter estimate 5 and their filtered and}
\caption*{smoothed covariances through (with the time cumulatively) the iterations}
\label{lon_p5}
\end{figure}

\begin{figure}[h]
\includegraphics[width=6in,height=4in]{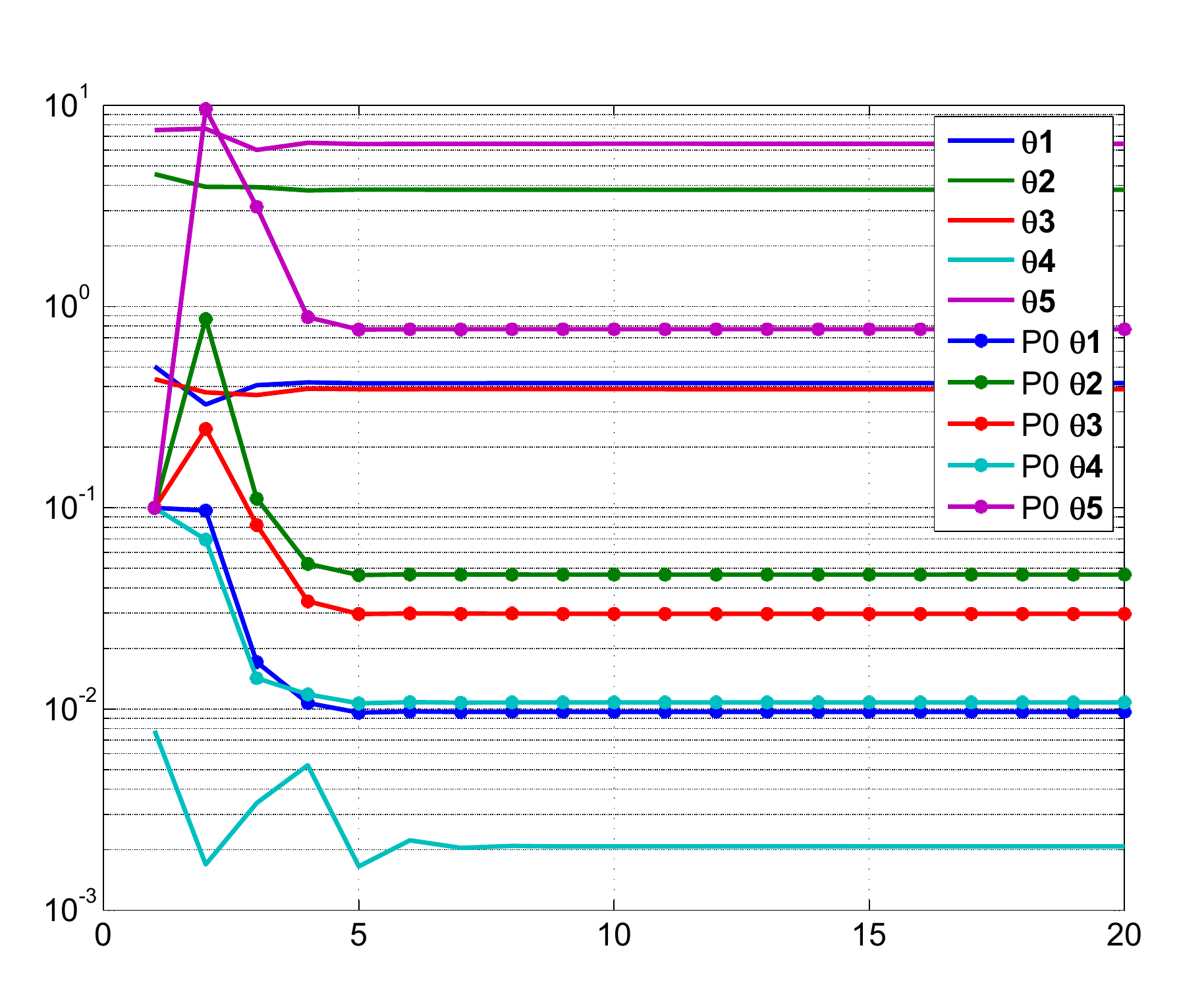}
\caption{Variation of parameter and its initial covariance ($\mathbf{P_0}$) with iterations}
\label{lon_P0}
\end{figure}

\begin{figure}[h]
\includegraphics[width=6in,height=4in]{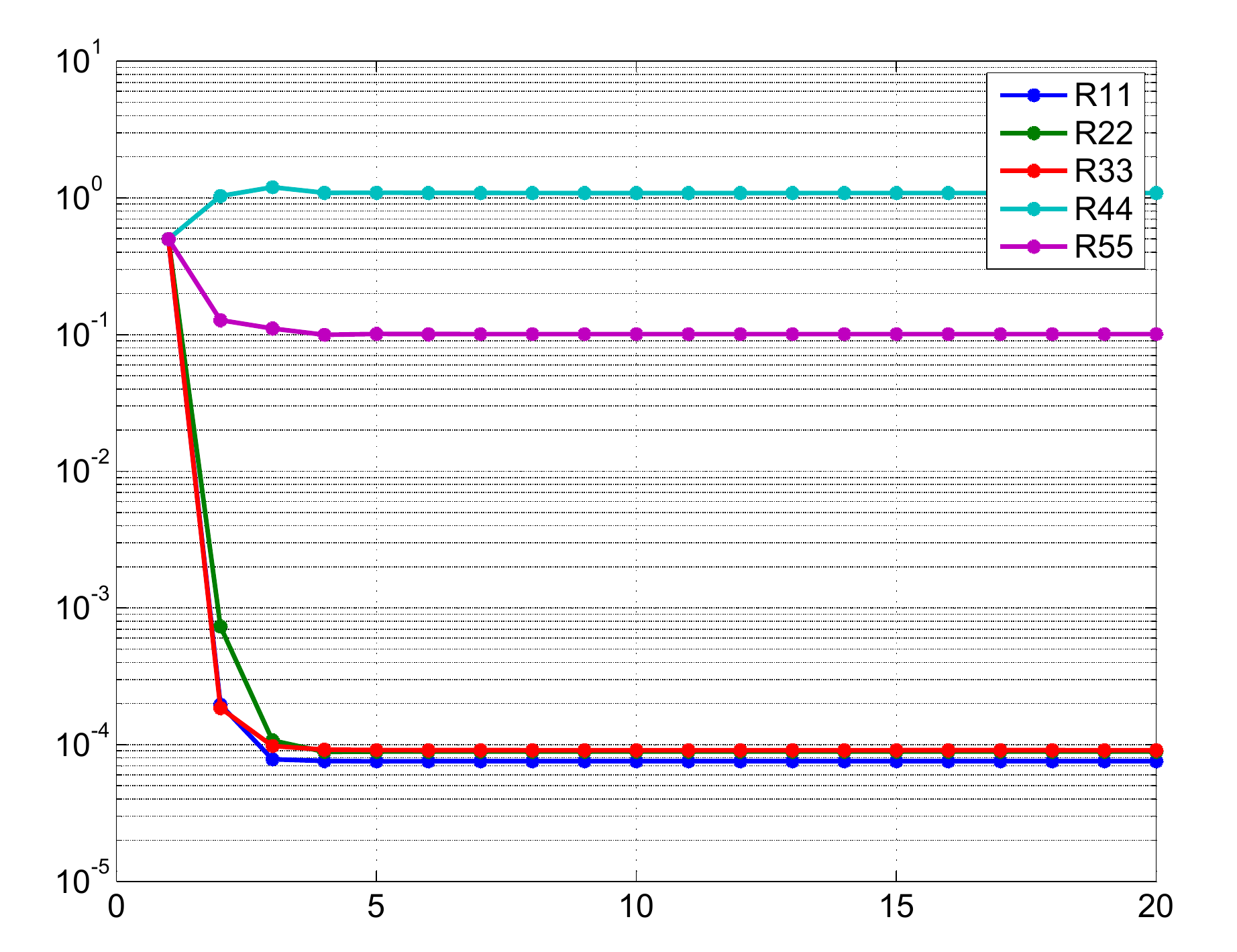}
\caption{Variation of \textbf{R} with iterations}
\label{lon_R}
\end{figure}

\begin{figure}[h]
\includegraphics[width=6in,height=4in]{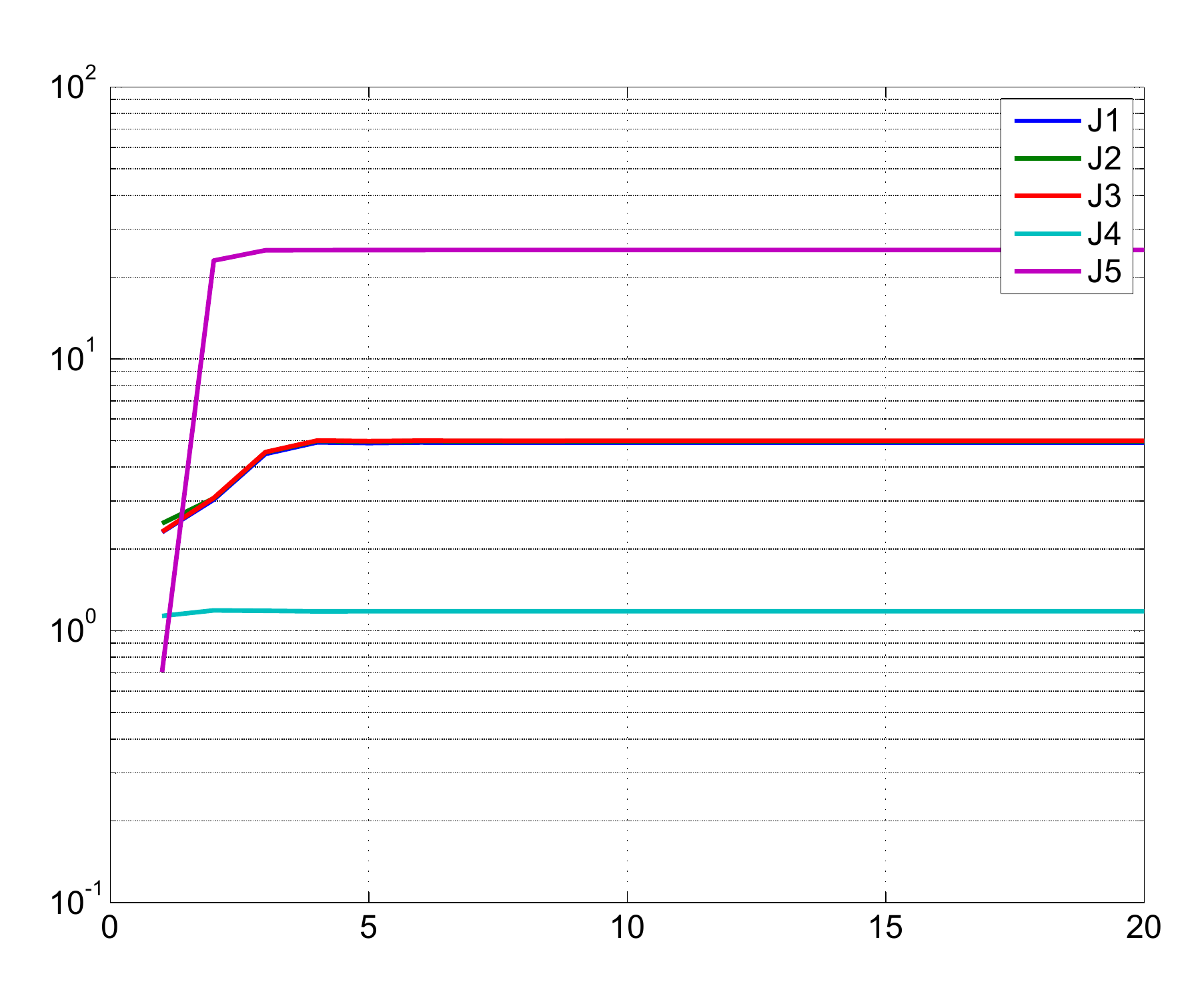}
\caption{Variation of different costs (\textbf{J1-J5}) with iterations}
\label{lon_cost}
\end{figure}

\begin{figure}[h]
\includegraphics[width=6in,height=4in]{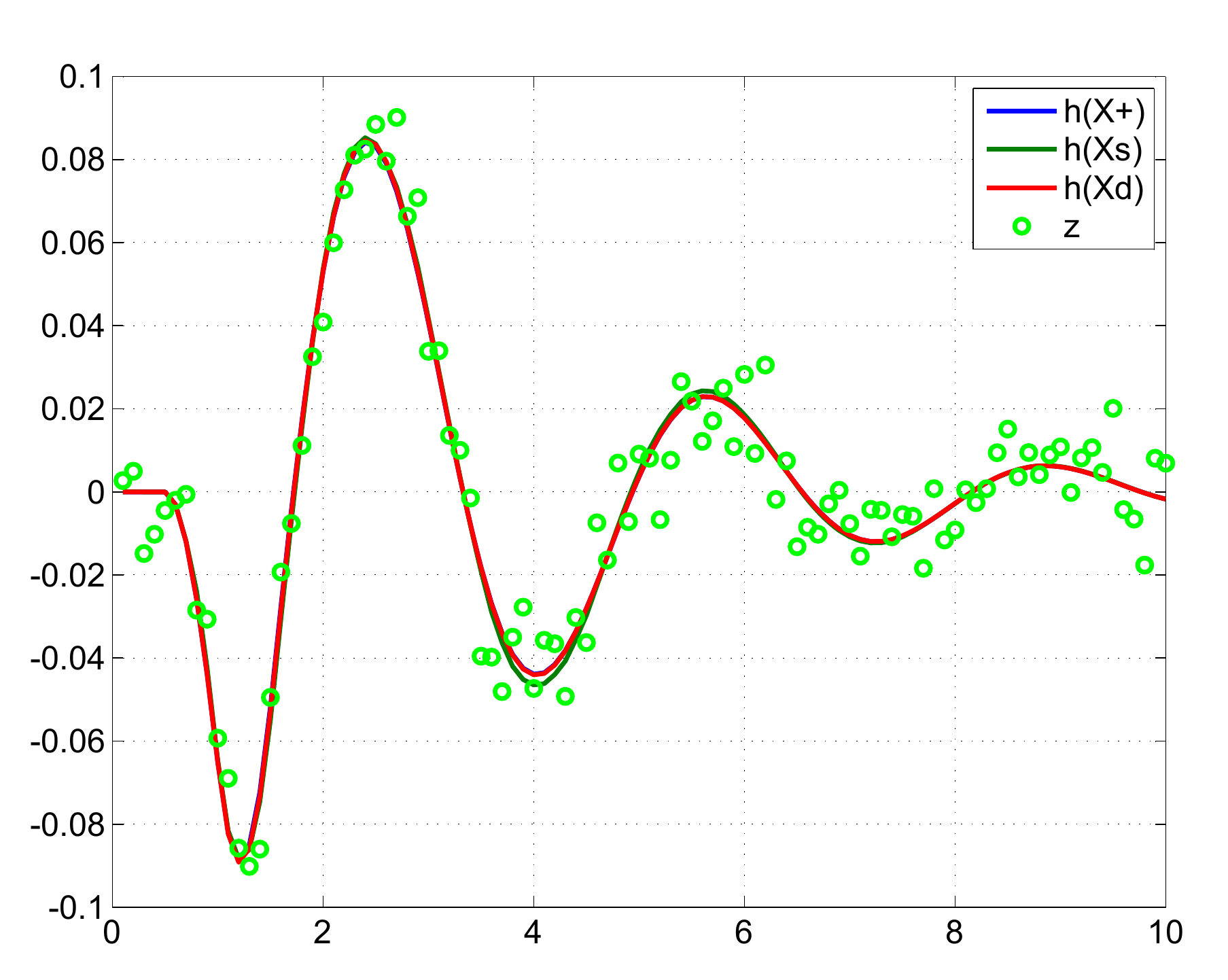}
\caption{Comparison of the predicted dynamics, posterior, smoothed}
\caption*{and the measurement 1 }
\label{lon_h1}
\end{figure}

\begin{figure}[h]
\includegraphics[width=6in,height=4in]{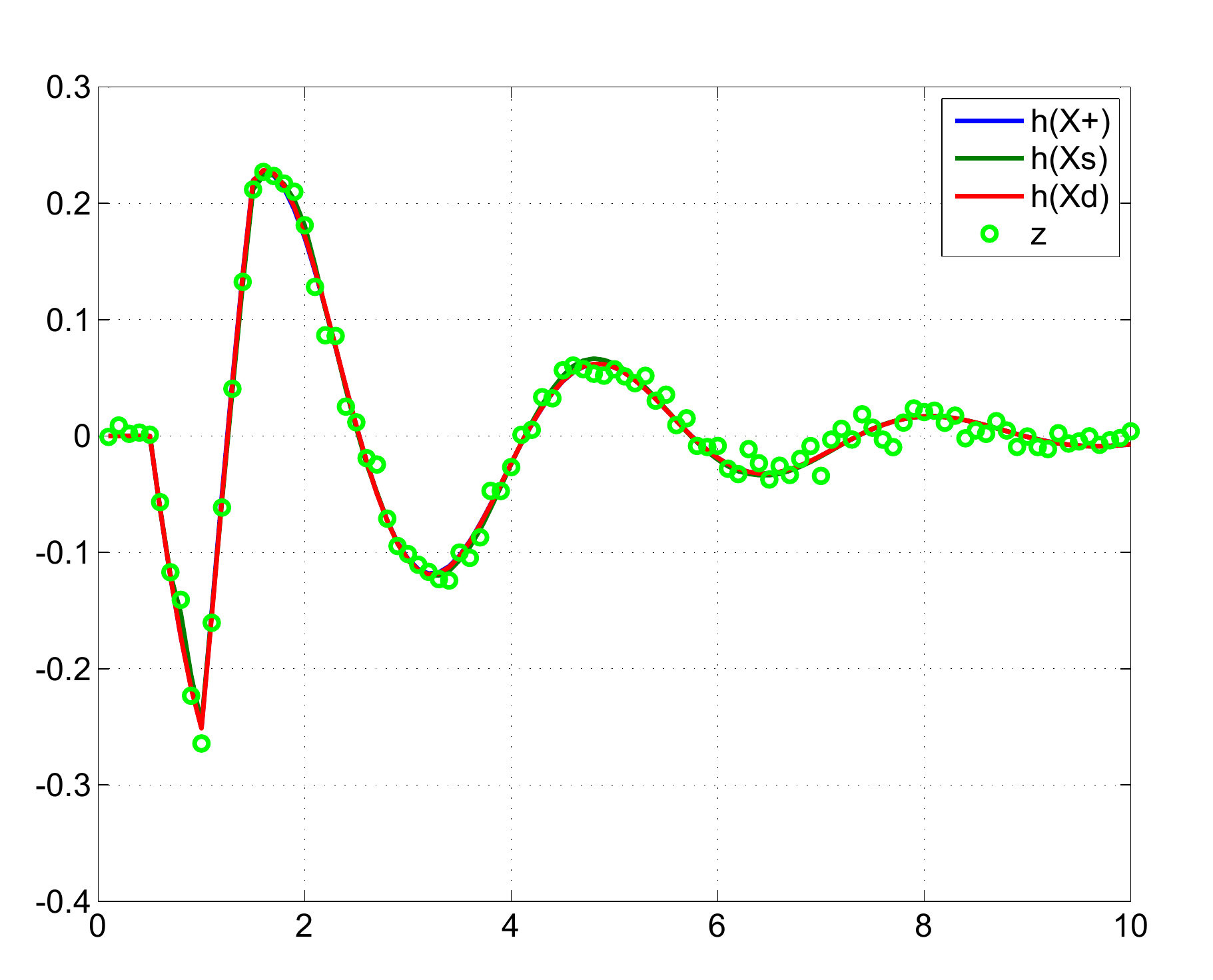}
\caption{Comparison of the predicted dynamics, posterior, smoothed}
\caption*{and the measurement 2}
\label{lon_h2}
\end{figure}

\begin{figure}[h]
\includegraphics[width=6in,height=4in]{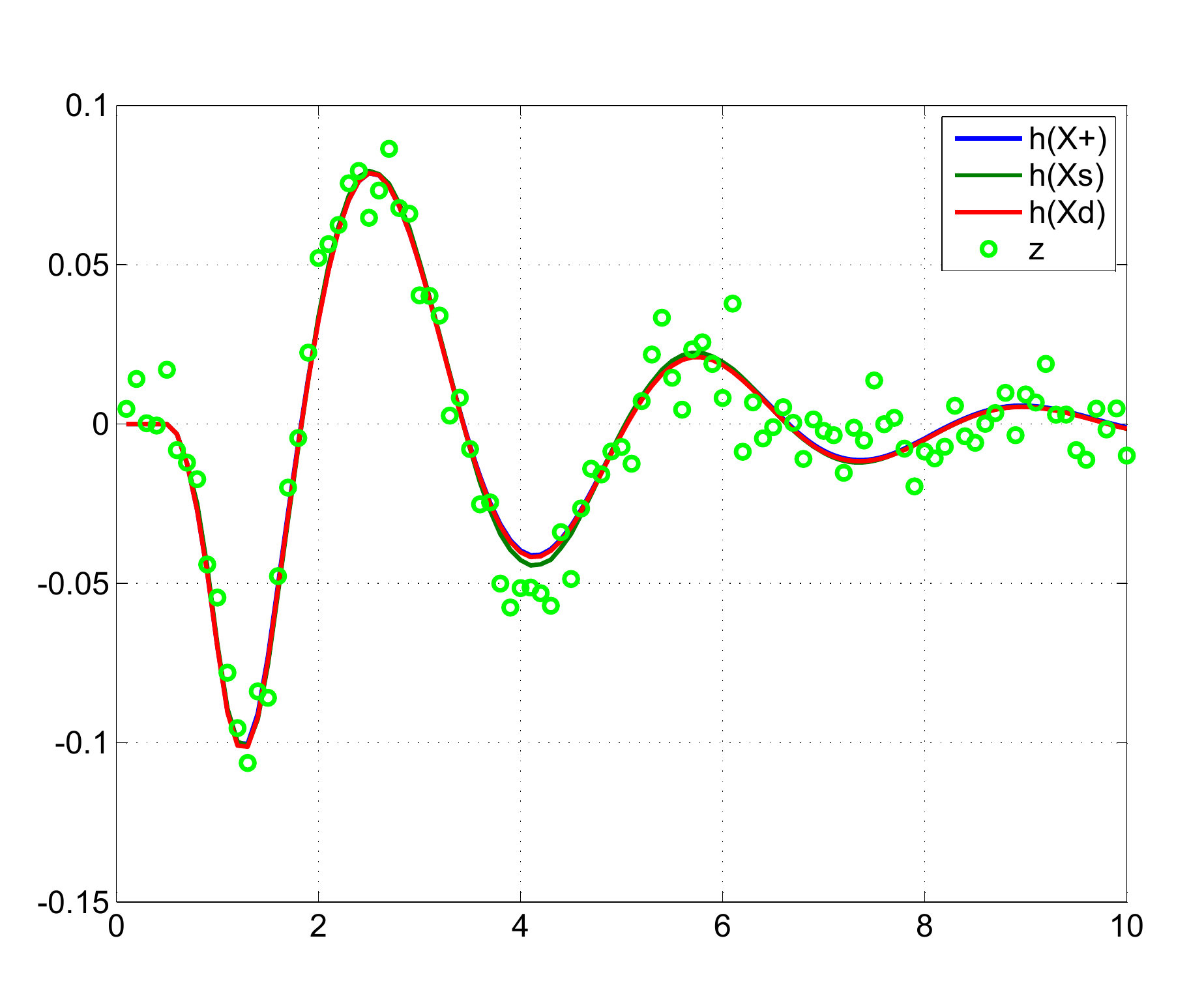}
\caption{Comparison of the predicted dynamics, posterior, smoothed}
\caption*{and the measurement 3 }
\label{lon_h3}
\end{figure}

\begin{figure}[h]
\includegraphics[width=6in,height=4in]{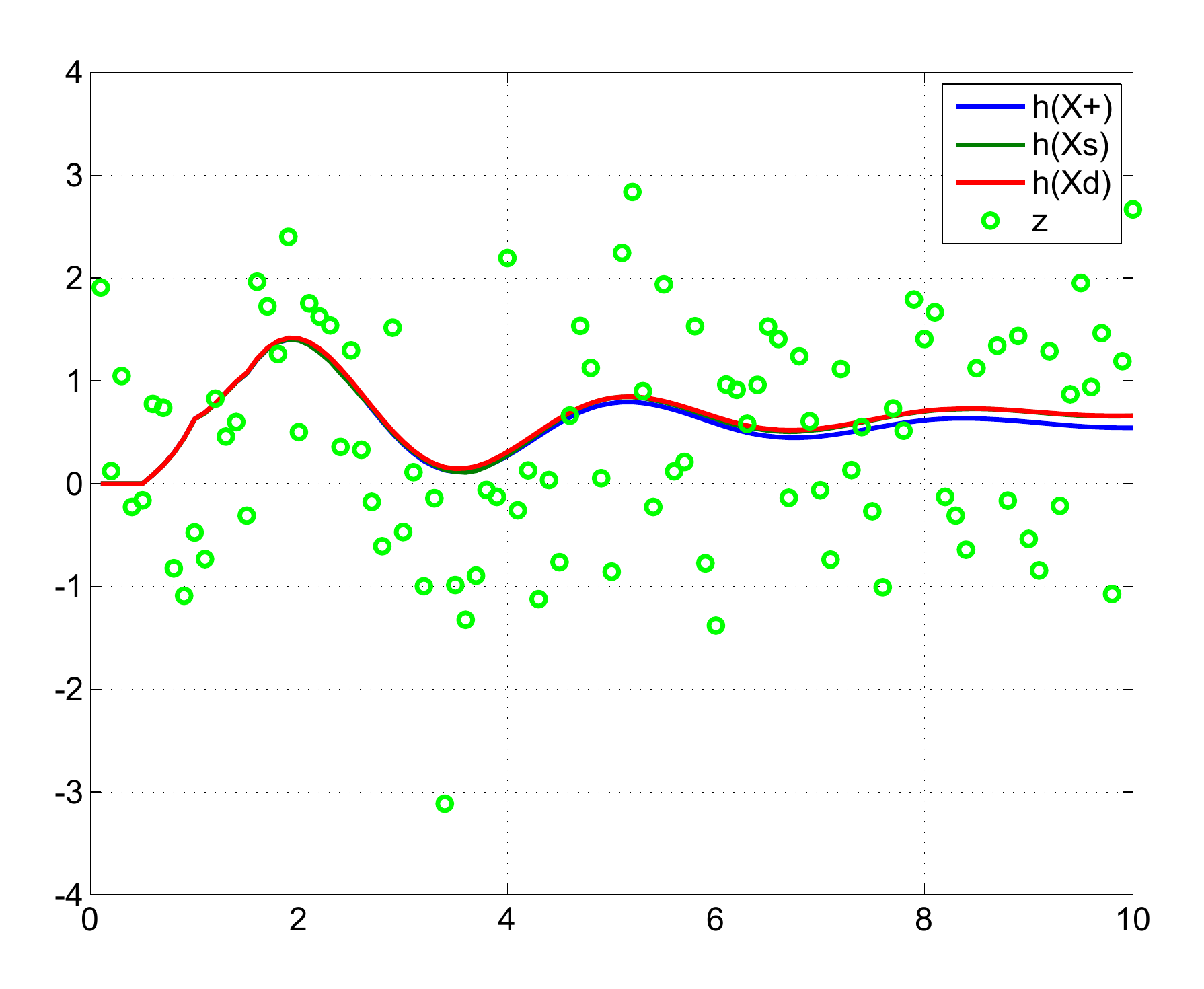}
\caption{Comparison of the predicted dynamics, posterior, smoothed}
\caption*{and the measurement 4}
\label{lon_h4}
\end{figure}

\begin{figure}[h]
\includegraphics[width=6in,height=4in]{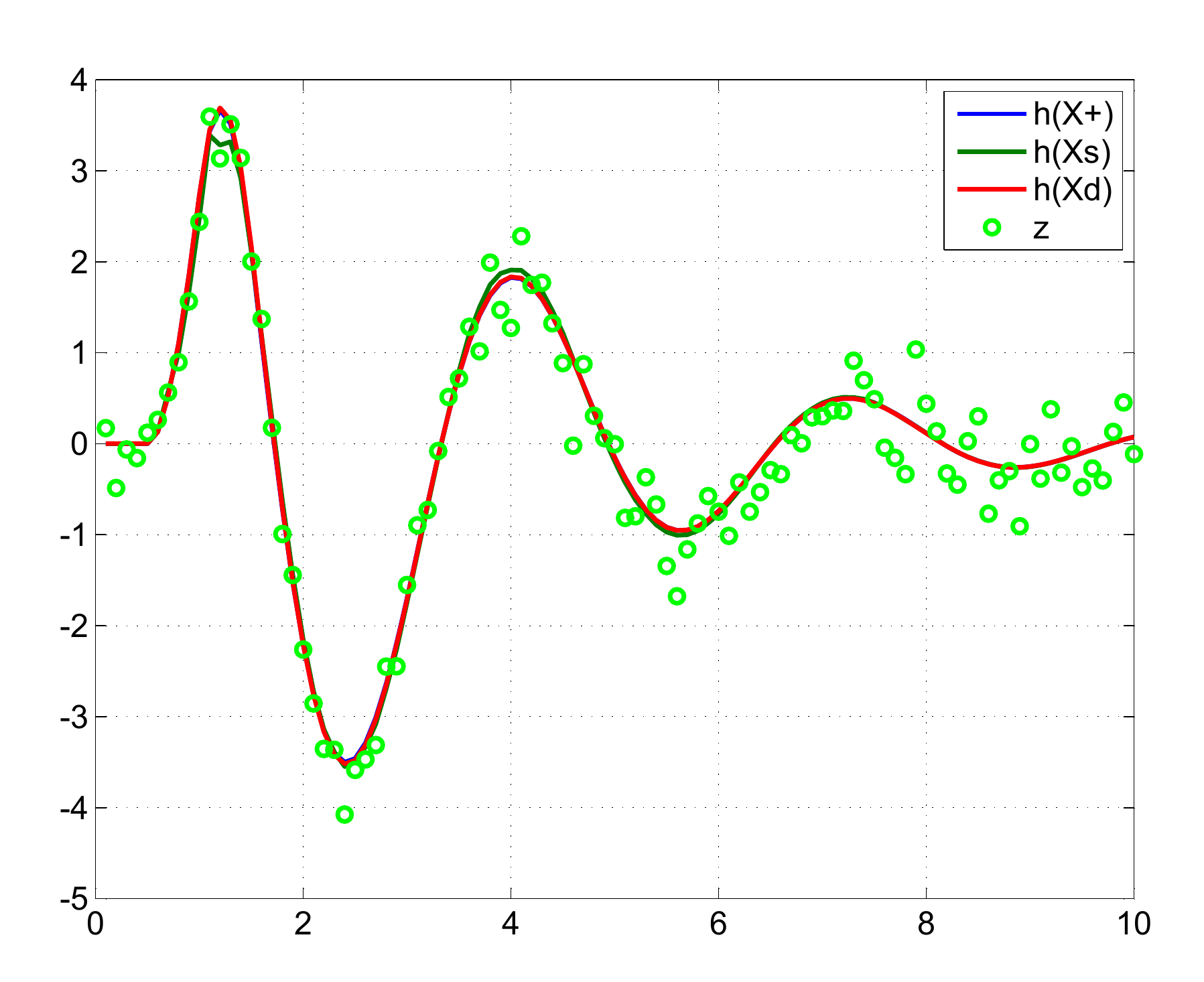}
\caption{Comparison of the predicted dynamics, posterior, smoothed}
\caption*{and the measurement 5}
\label{lon_h5}
\end{figure}

\begin{figure}[h]
\includegraphics[width=6in,height=4in]{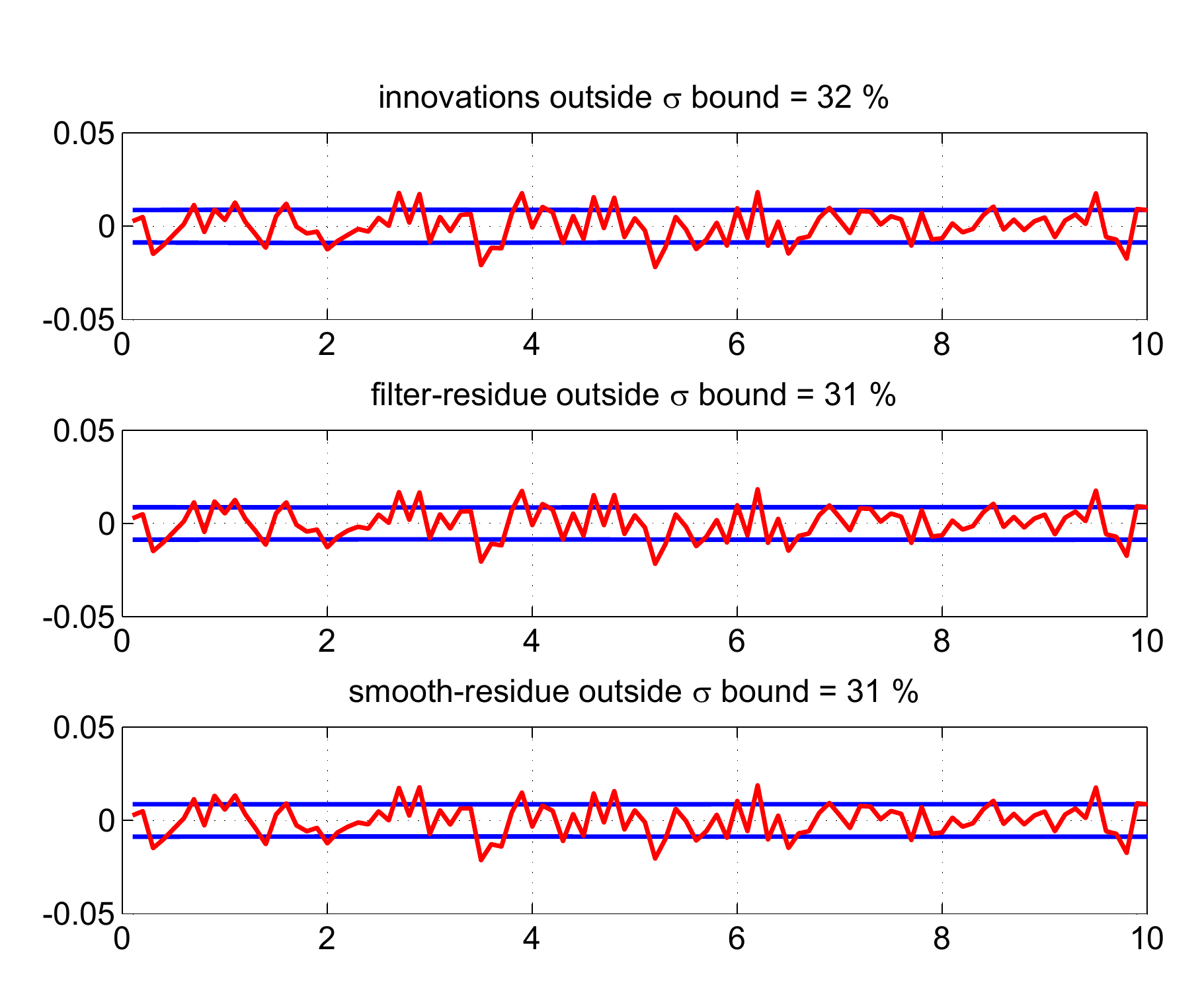}
\caption{The innovations, filtered residue and smoothed residue}
\caption*{corresponding to measurement 1 }
\label{lon_innov1}
\end{figure}

\begin{figure}[h]
\includegraphics[width=6in,height=4in]{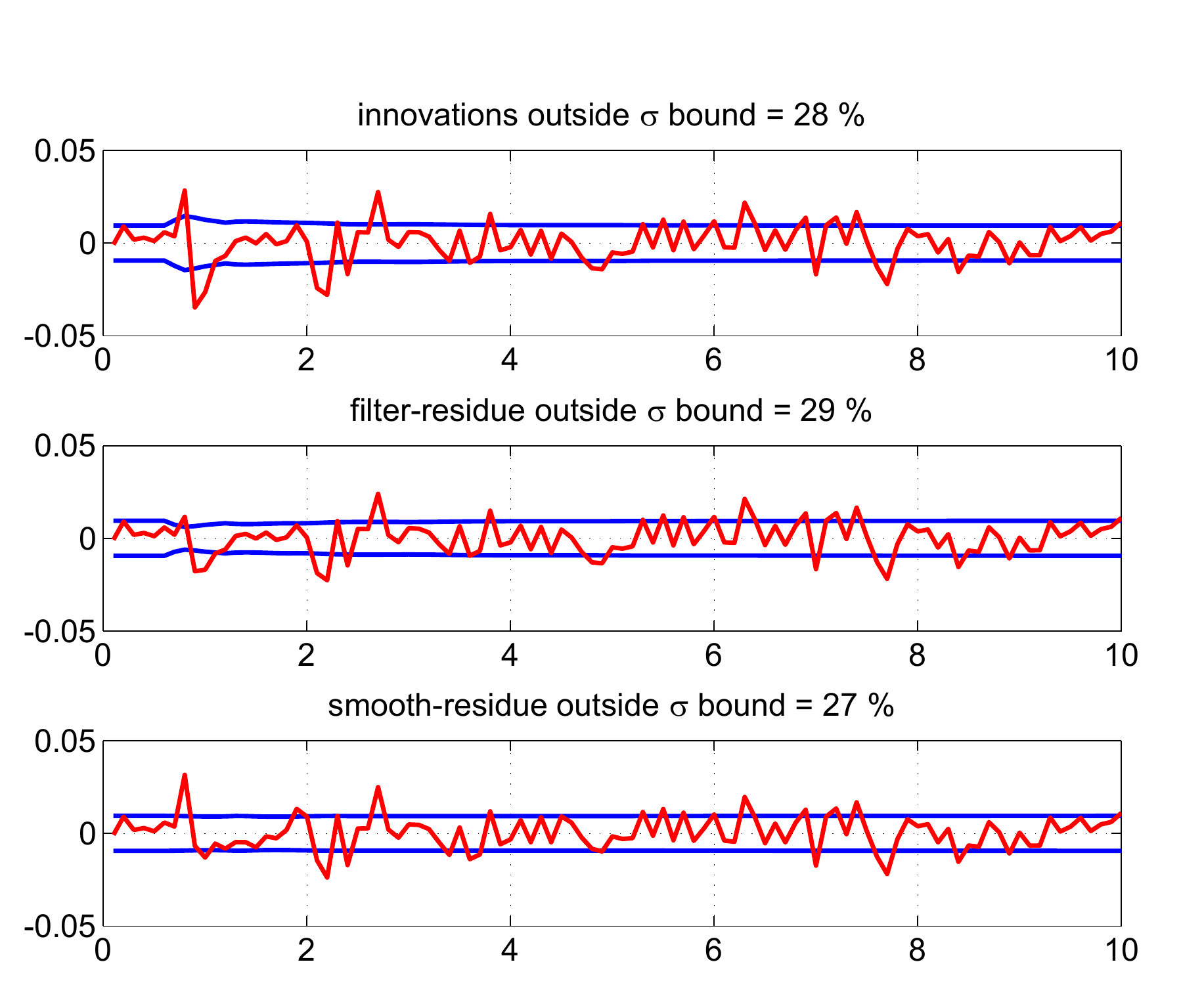}
\caption{The innovations, filtered residue and smoothed residue}
\caption*{corresponding to measurement 2}
\label{lon_innov2}
\end{figure}

\begin{figure}[h]
\includegraphics[width=6in,height=4in]{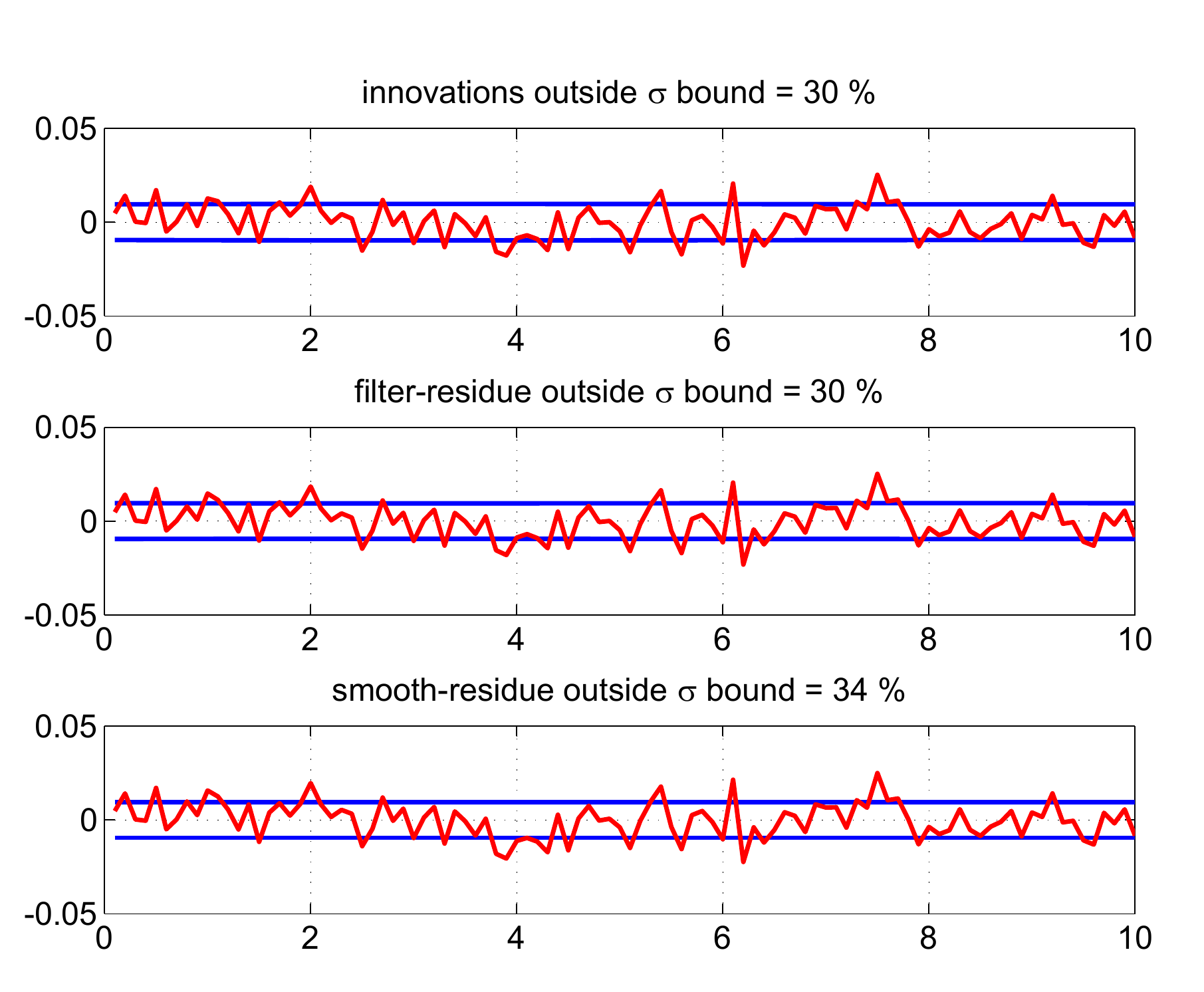}
\caption{The innovations, filtered residue and smoothed residue}
\caption*{corresponding to measurement 3 }
\label{lon_innov3}
\end{figure}

\begin{figure}[h]
\includegraphics[width=6in,height=4in]{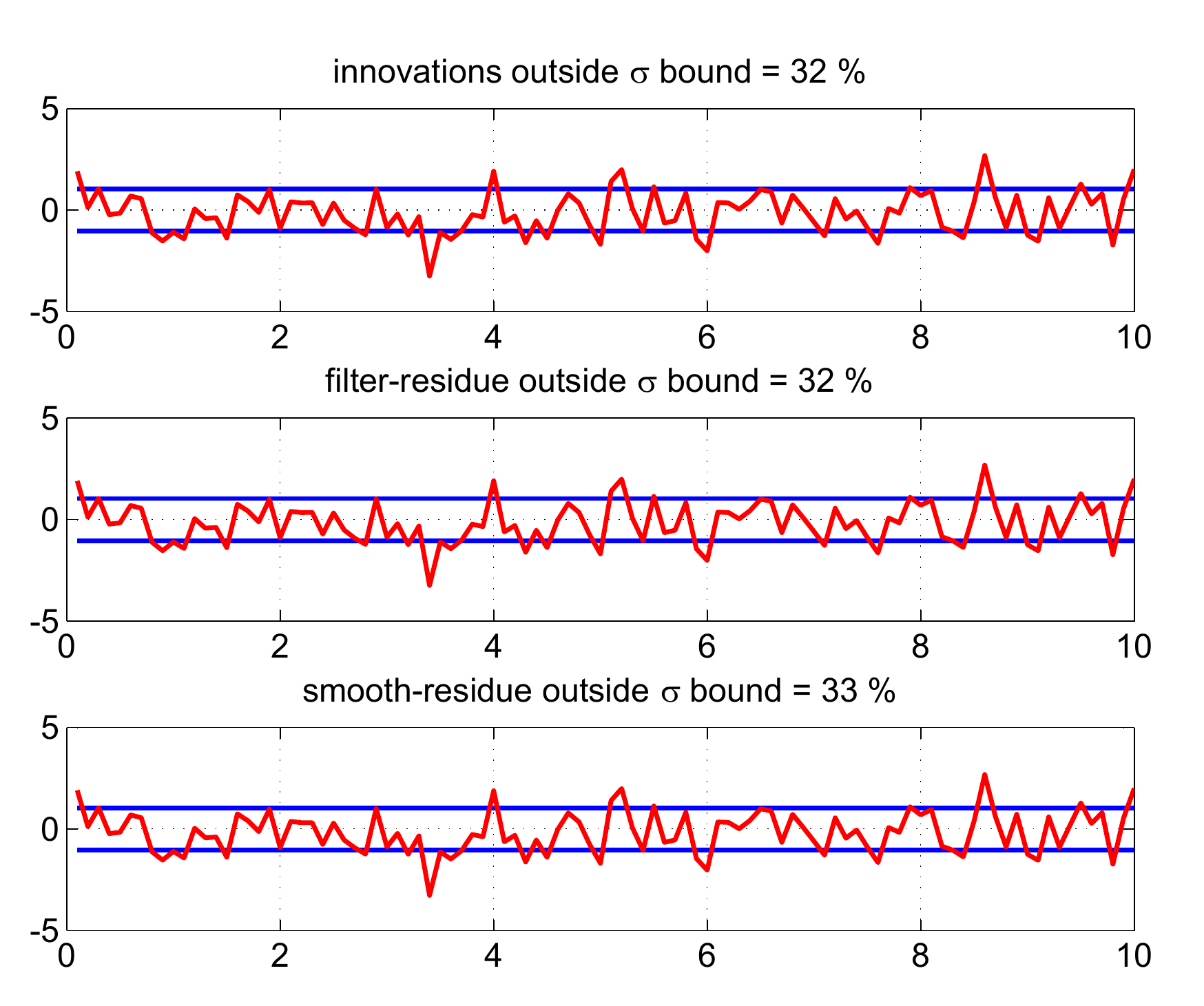}
\caption{The innovations, filtered residue and smoothed residue}
\caption*{corresponding to measurement 4}
\label{lon_innov4}
\end{figure}

\begin{figure}[h]
\includegraphics[width=6in,height=4in]{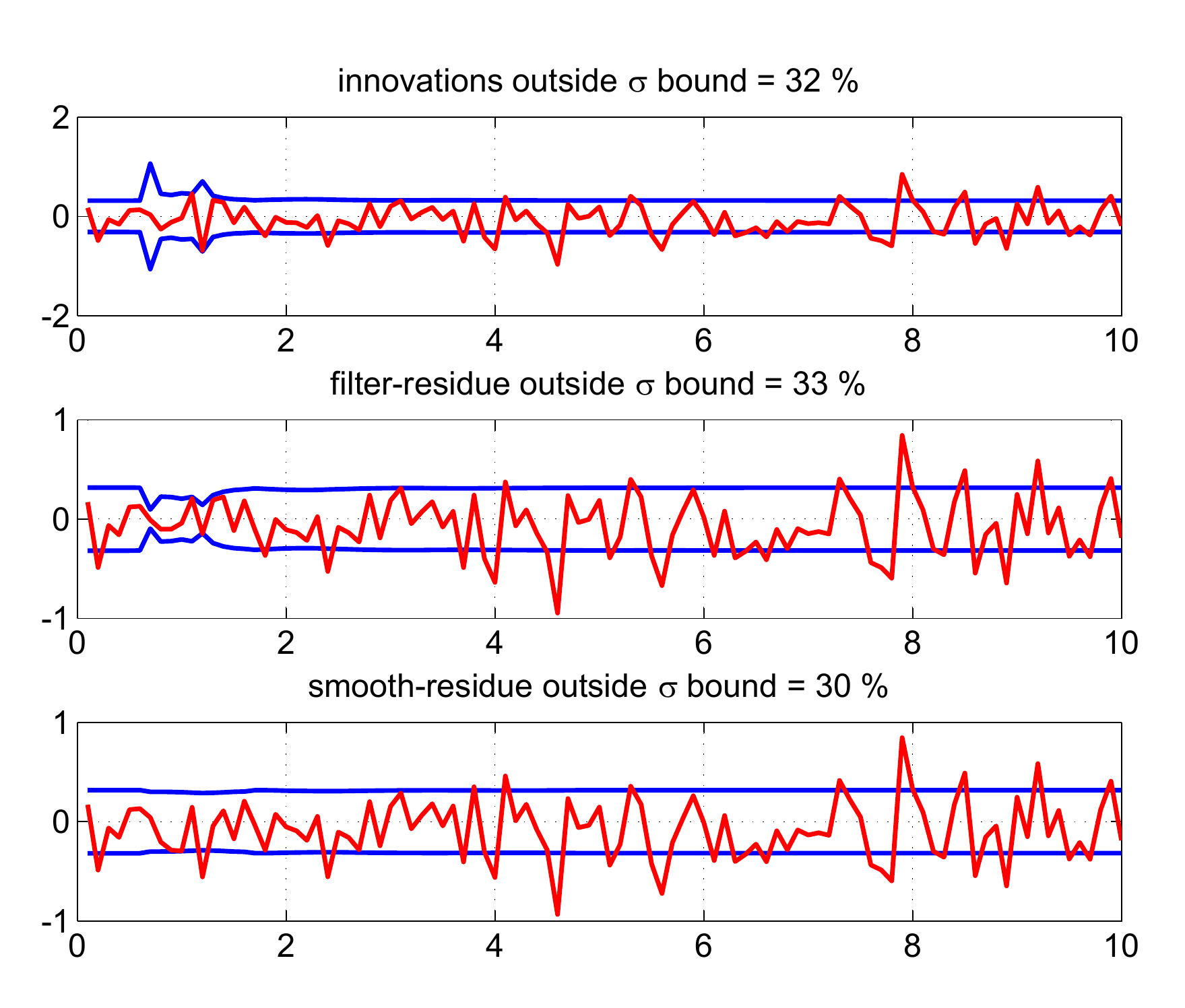}
\caption{The innovations, filtered residue and smoothed residue}
\caption*{corresponding to measurement 5}
\label{lon_innov5}
\end{figure}

\begin{figure}[h]
\includegraphics[width=6in,height=4in]{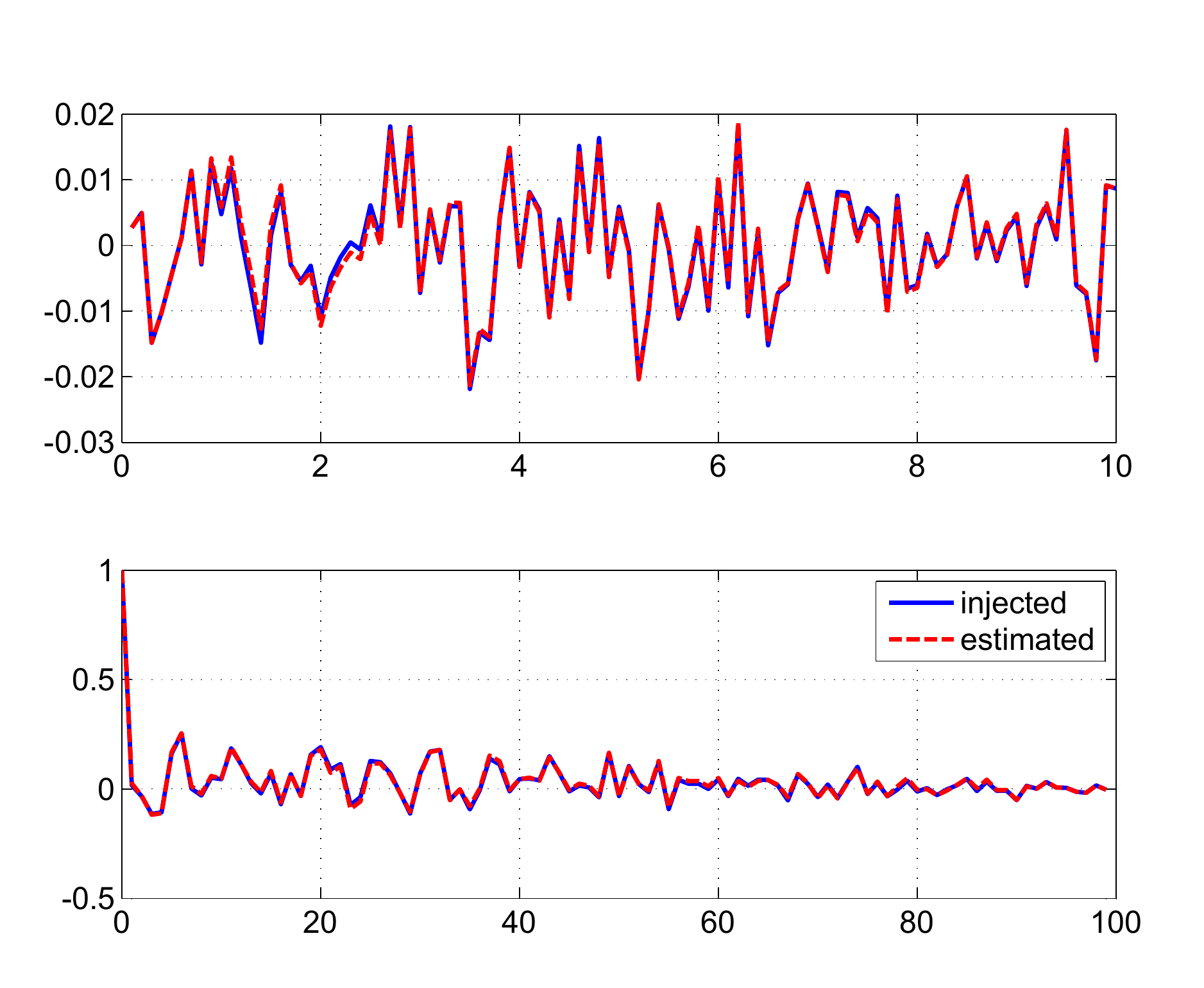}
\caption{Time variation of injected and estimated measurement noise (top) and}
\caption*{their autocorrelation (bottom) for measurement 1}
\label{lon_mnoise1}
\end{figure}

\begin{figure}[h]
\includegraphics[width=6in,height=4in]{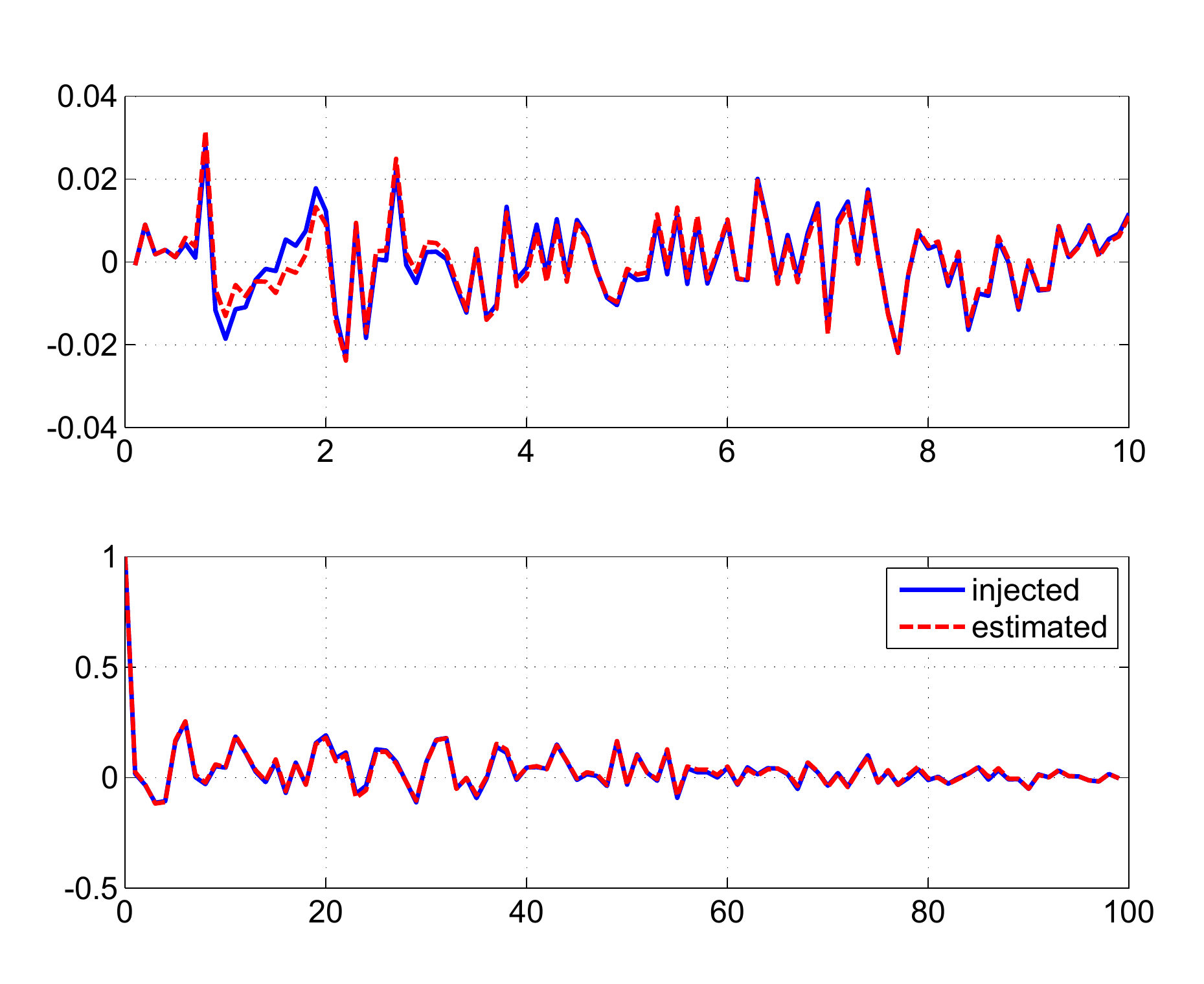}
\caption{Time variation of injected and estimated measurement noise (top) and}
\caption*{their autocorrelation (bottom) for measurement 2}
\label{lon_mnoise2}
\end{figure}

\begin{figure}[h]
\includegraphics[width=6in,height=4in]{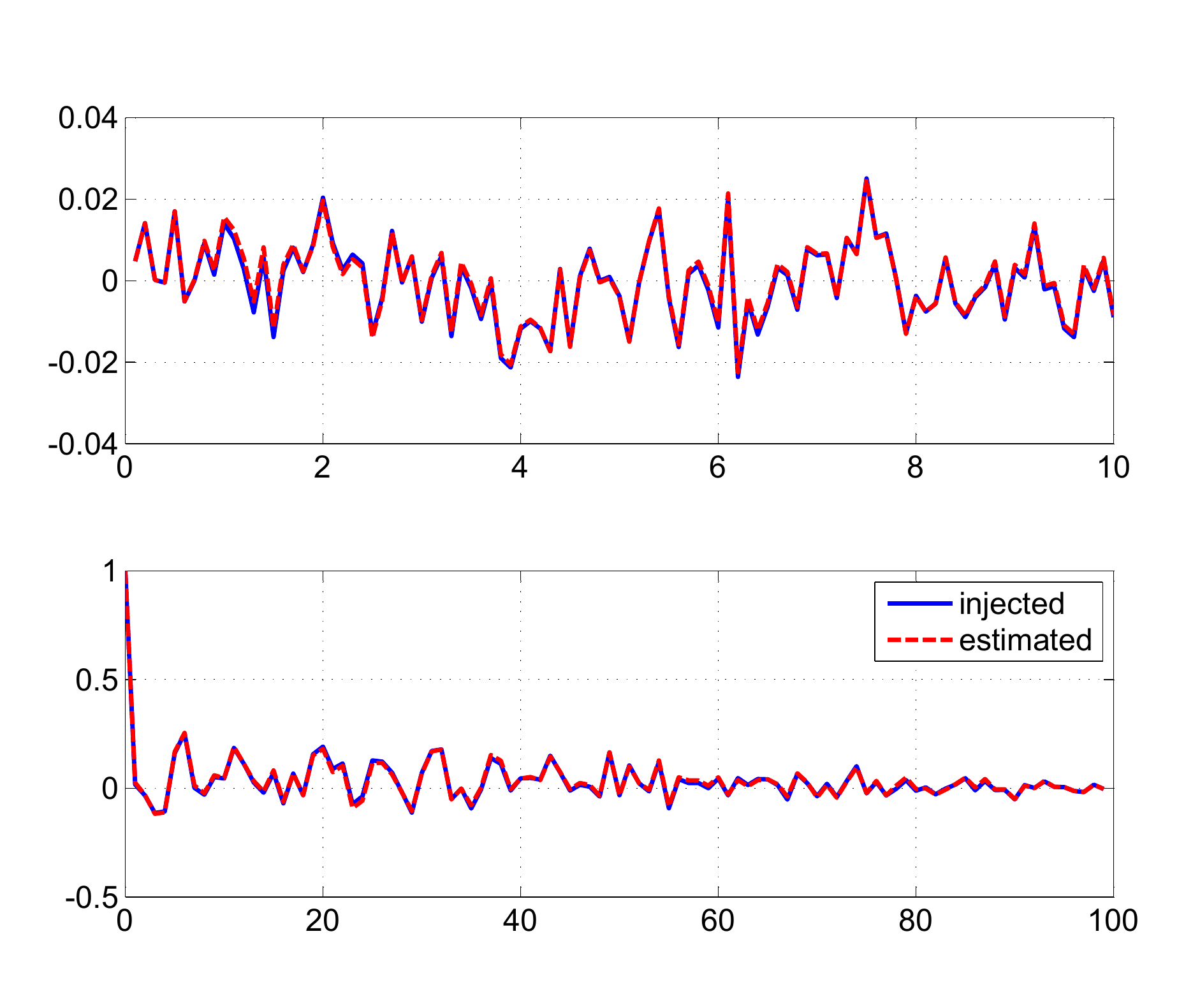}
\caption{Time variation of injected and estimated measurement noise (top) and}
\caption*{their autocorrelation (bottom) for measurement 3}
\label{lon_mnoise3}
\end{figure}

\begin{figure}[h]
\includegraphics[width=6in,height=4in]{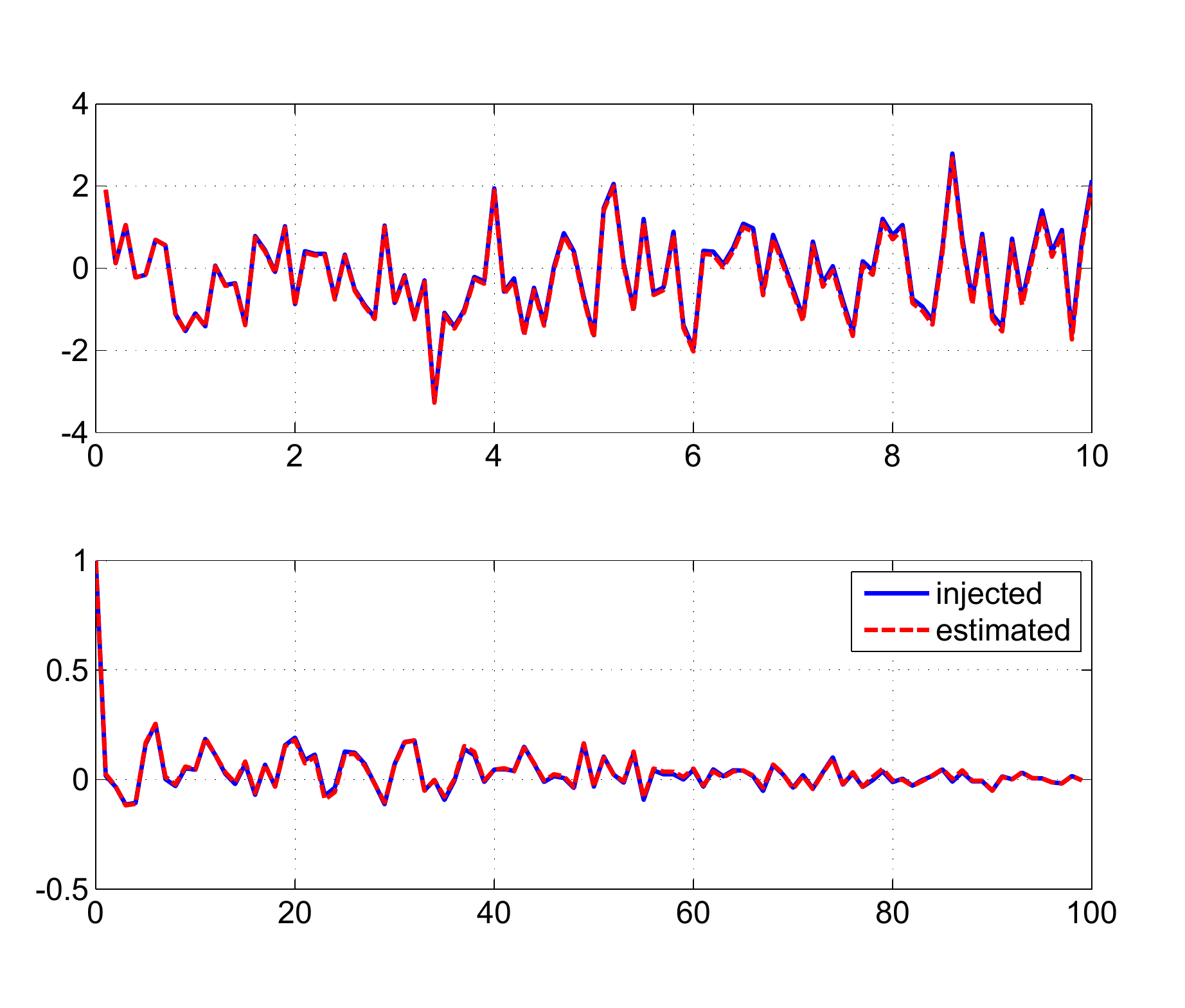}
\caption{Time variation of injected and estimated measurement noise (top) and}
\caption*{their autocorrelation (bottom) for measurement 4}
\label{lon_mnoise4}
\end{figure}

\begin{figure}[h]
\includegraphics[width=6in,height=4in]{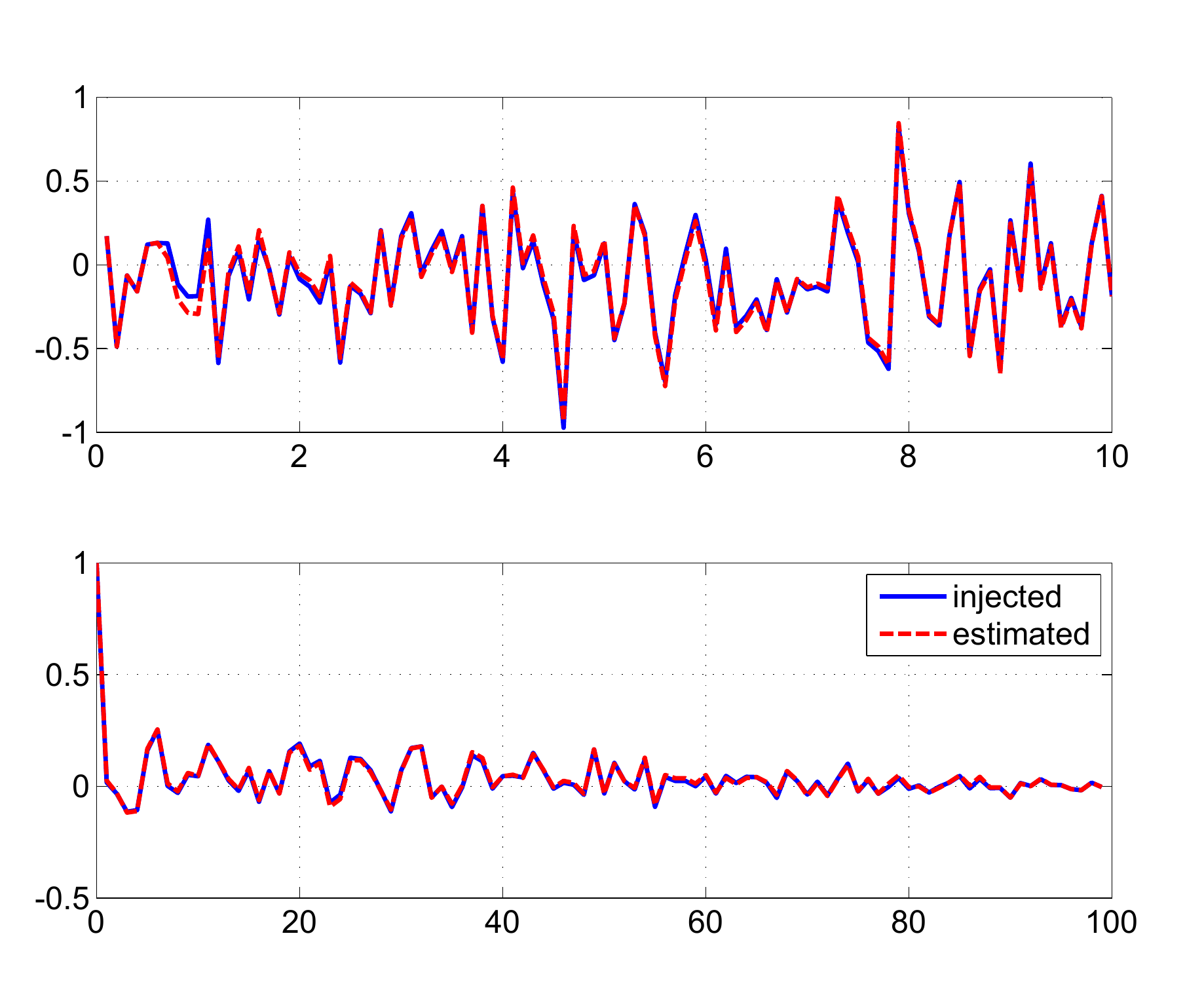}
\caption{Time variation of injected and estimated measurement noise (top) and}
\caption*{their autocorrelation (bottom) for measurement 5}
\label{lon_mnoise5}
\end{figure}

 We first run the filter assuming \textbf{Q} = 0. It was found that about 20 iterations of the data would suffice. The Fig. \ref{lon_p1}-\ref{lon_mnoise5} refer to the \textbf{Q} = 0 case. The Fig. \ref{lon_p1}-\ref{lon_p5} shows the various parameter estimates and its corresponding variances through cumulative time instants with iterations. The variation of the estimated initial parameters and their variances through iterations are shown in Fig. \ref{lon_P0}. The parameter and the uncertainty reach almost their final estimated values in about 2 and 5 iterations respectively. A similar plot in Fig. \ref{lon_R} shows the variation of the estimated measurement noise. The variation of different cost functions (\textbf{J1-J5}) through the iterations is shown in Fig. \ref{lon_cost}. The Fig. \ref{lon_h1}-\ref{lon_h5} shows the predicted dynamics, filtered and smoothed estimate at the last iteration. The Fig. \ref{lon_innov1}-\ref{lon_innov5} show the innovations, filtered residue and smoothed residue together with the square root of their variance ($\pm\sigma$ bound). In the EKF approach most of the quantities are Gaussian or approximated as quasi Gaussian and one would expect all the above quantities are close to being Gaussian and hence around one third of the total sample points to be outside the $\sigma$ bound. The injected and estimated measurement noise distributions during the final iteration shown in Fig. \ref{lon_mnoise1}-\ref{lon_mnoise5} indicate that they are close to each other. Further even their autocorrelations are ideally expected to be close to the Kronecker delta function which provides confidence in the proposed filter algorithm.\par

\begin{figure}[h]
\includegraphics[width=6in,height=3.2in]{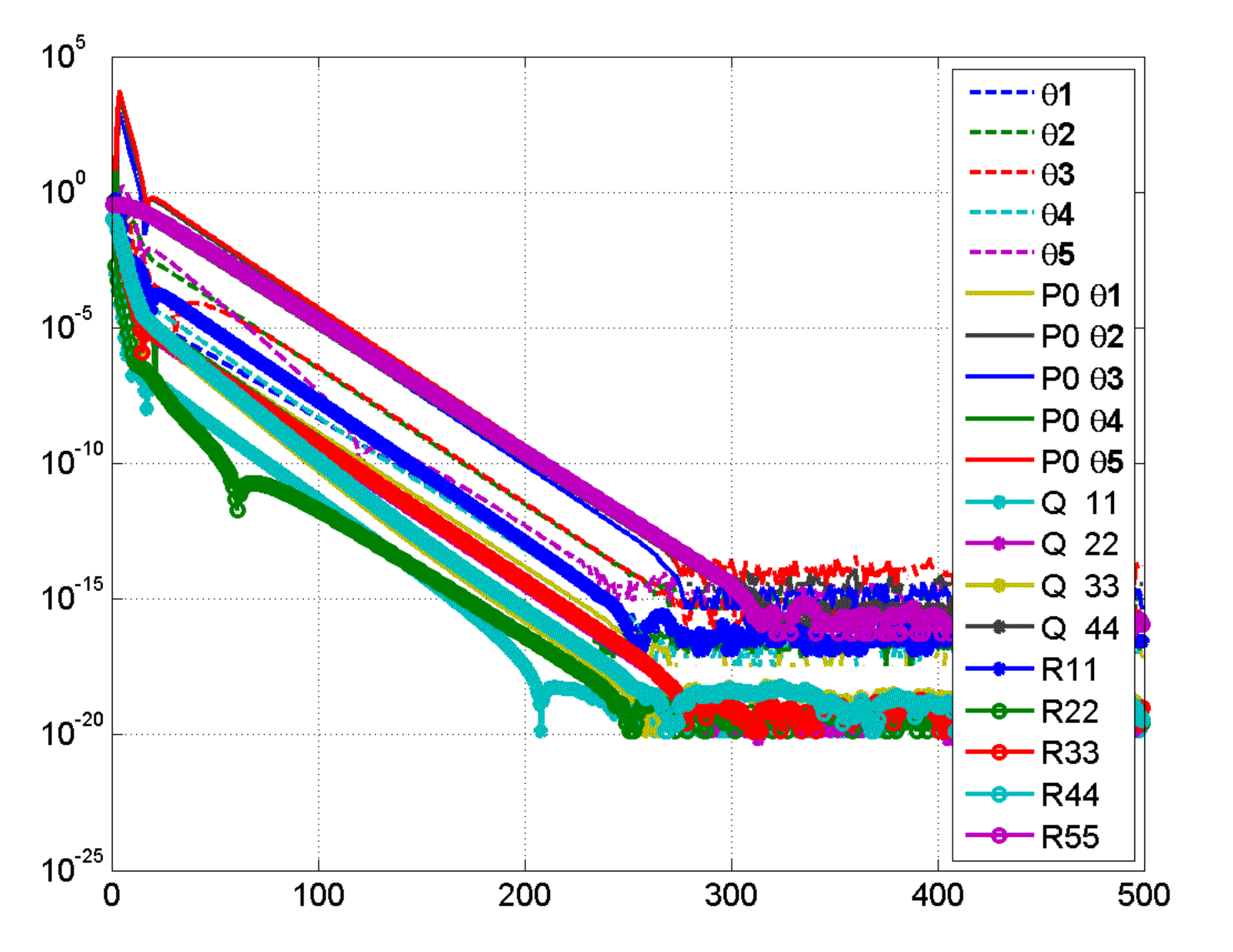}
\caption{The absolute difference between the iterated and final values}
\caption*{with 500 iterations}
\label{lon_err}
\end{figure}

\begin{figure}[h]
\includegraphics[width=6in,height=3.2in]{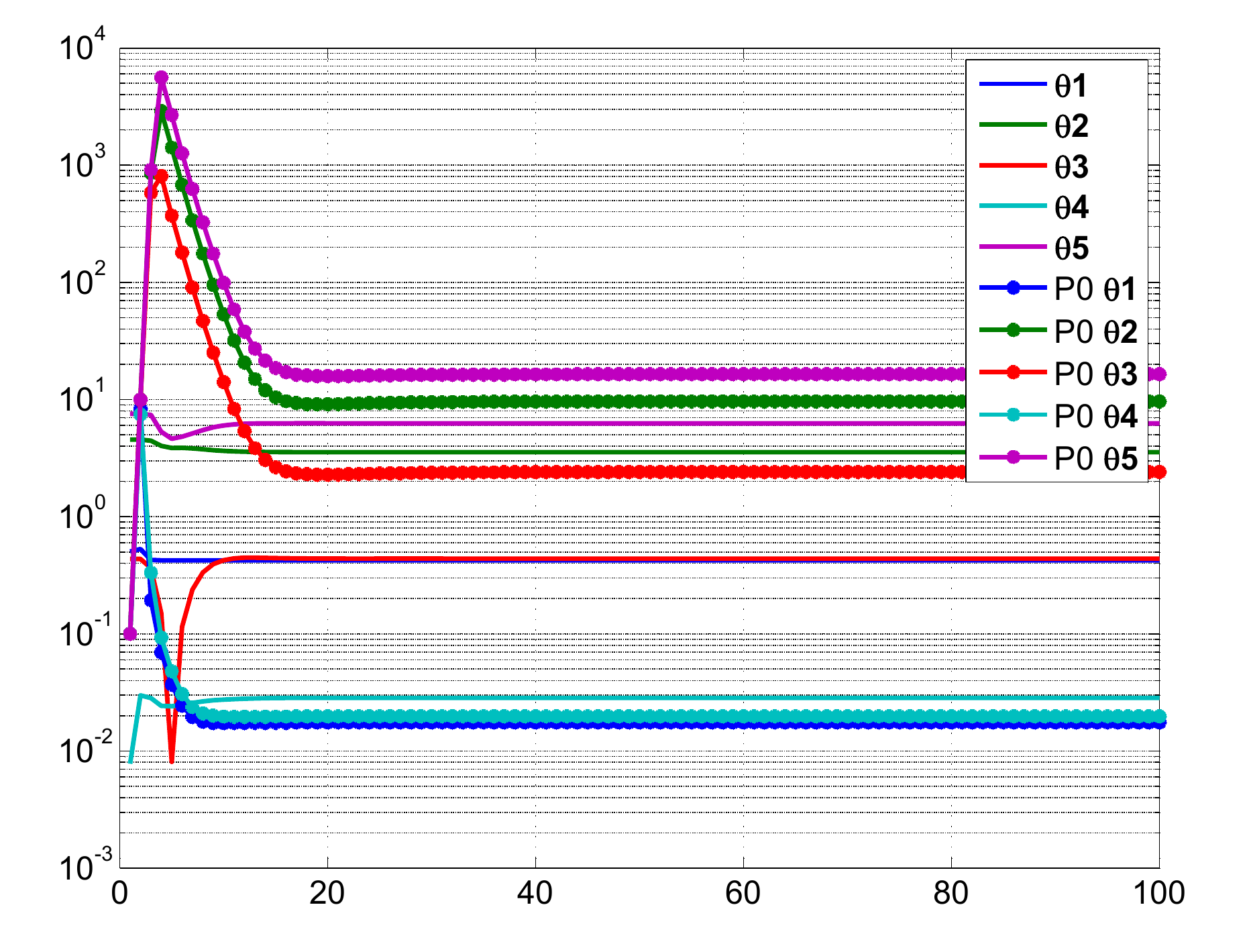}
\caption{Variation of parameter and its initial covariance (P0) with iterations}
\label{lonQ_P0}
\end{figure}

\begin{figure}[h]
\includegraphics[width=6in,height=4in]{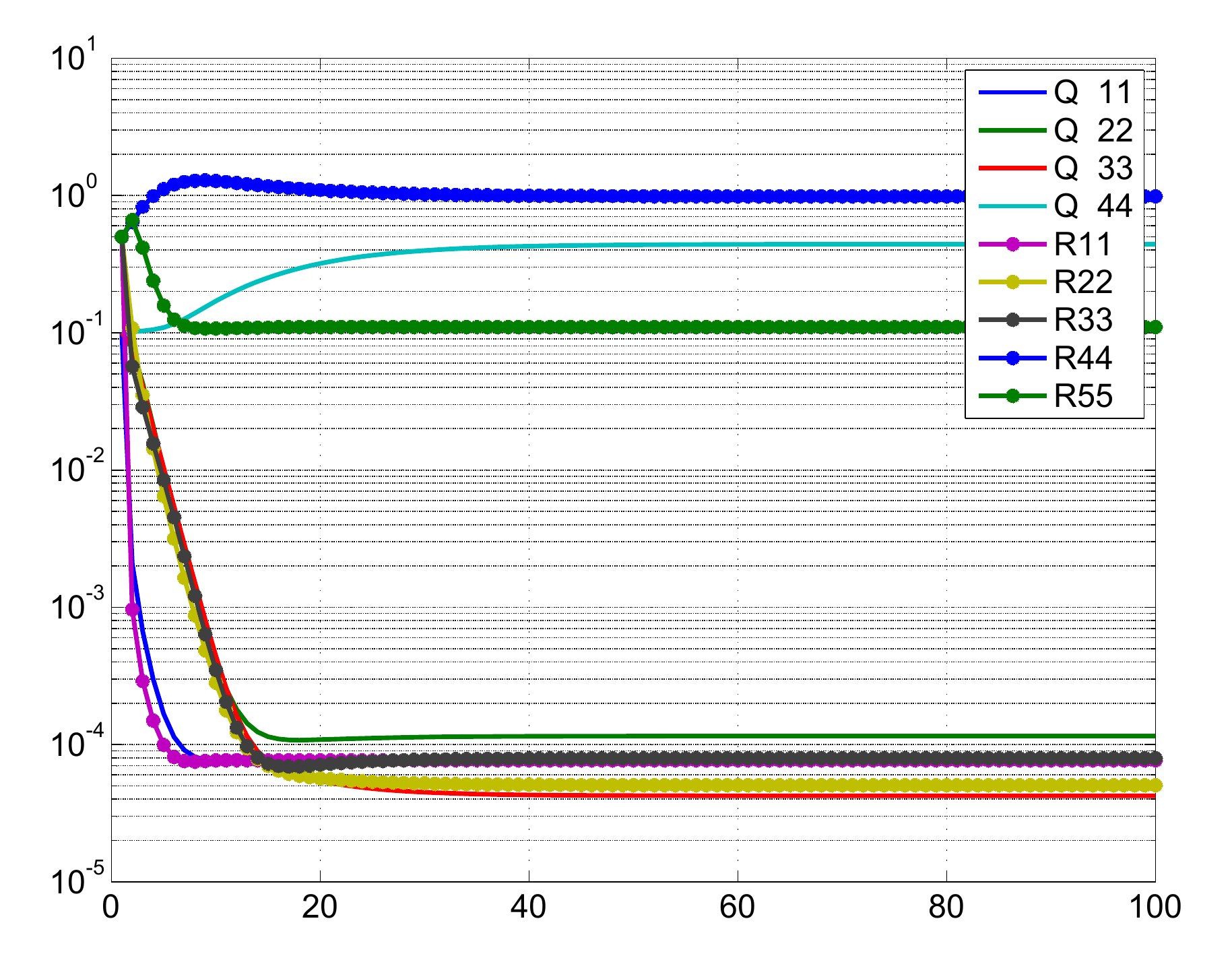}
\caption{Variation of \textbf{Q} and \textbf{R} with iterations}
\label{lonQ_R}
\end{figure}

\begin{figure}[h]
\includegraphics[width=6in,height=3.4in]{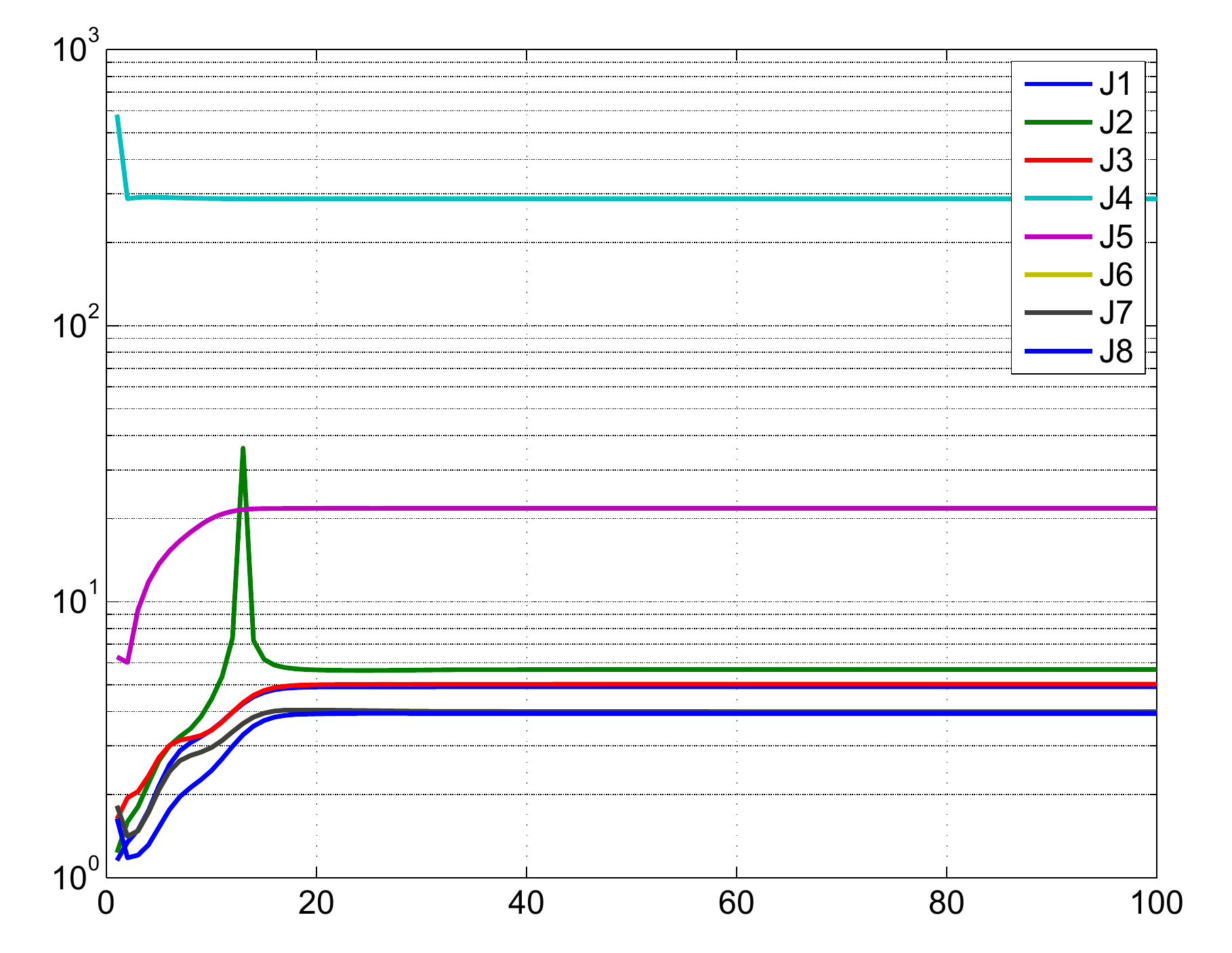}
\caption{Variation of different costs (\textbf{J1-J8}) with iterations}
\label{lonQ_J}
\end{figure}

\begin{figure}[h]
\includegraphics[width=6in,height=4in]{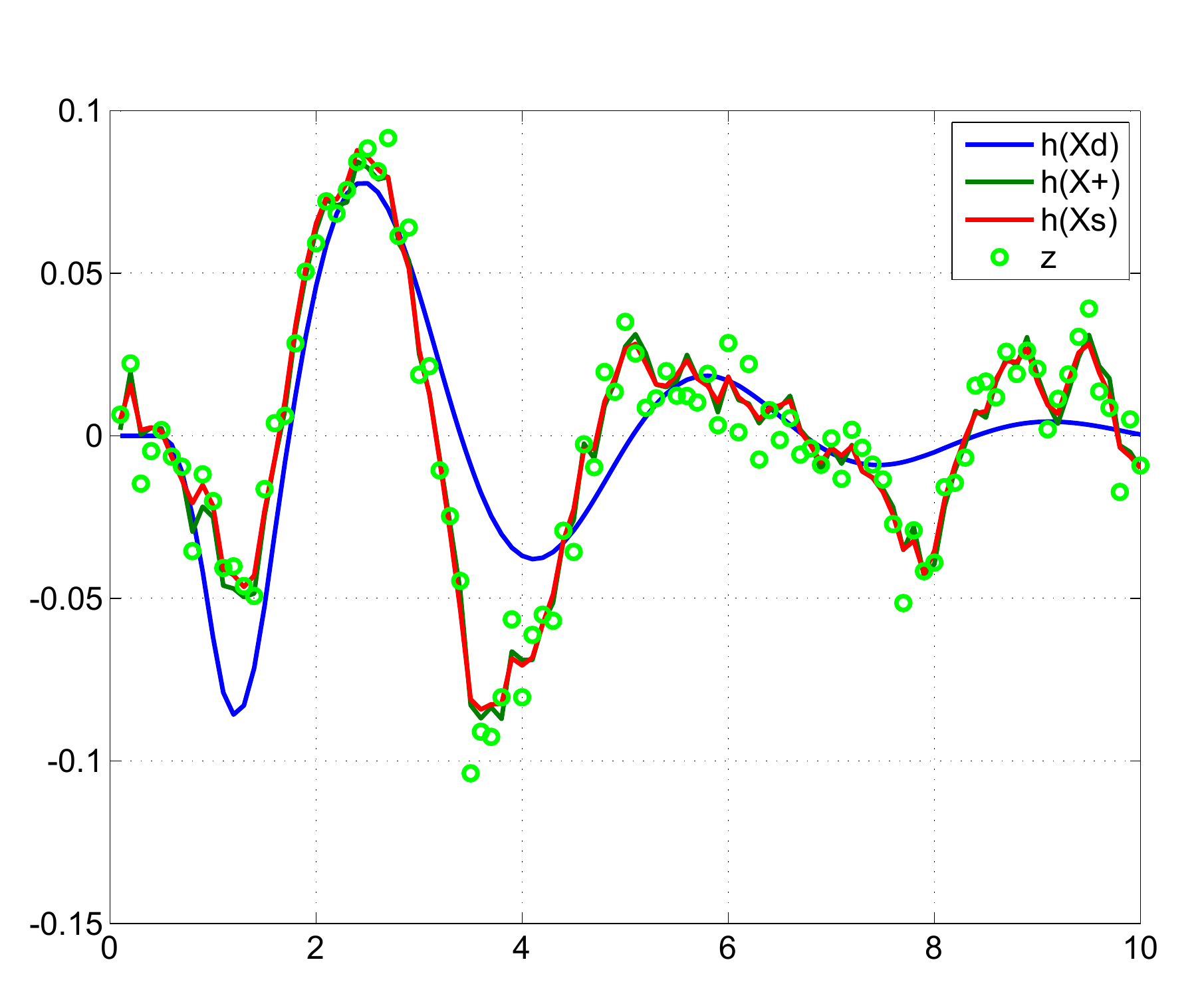}
\caption{Comparison of the predicted dynamics, posterior, smoothed}
\caption*{and the measurement 1 }
\label{lonQ_h1}
\end{figure}

\begin{figure}[h]
\includegraphics[width=6in,height=4in]{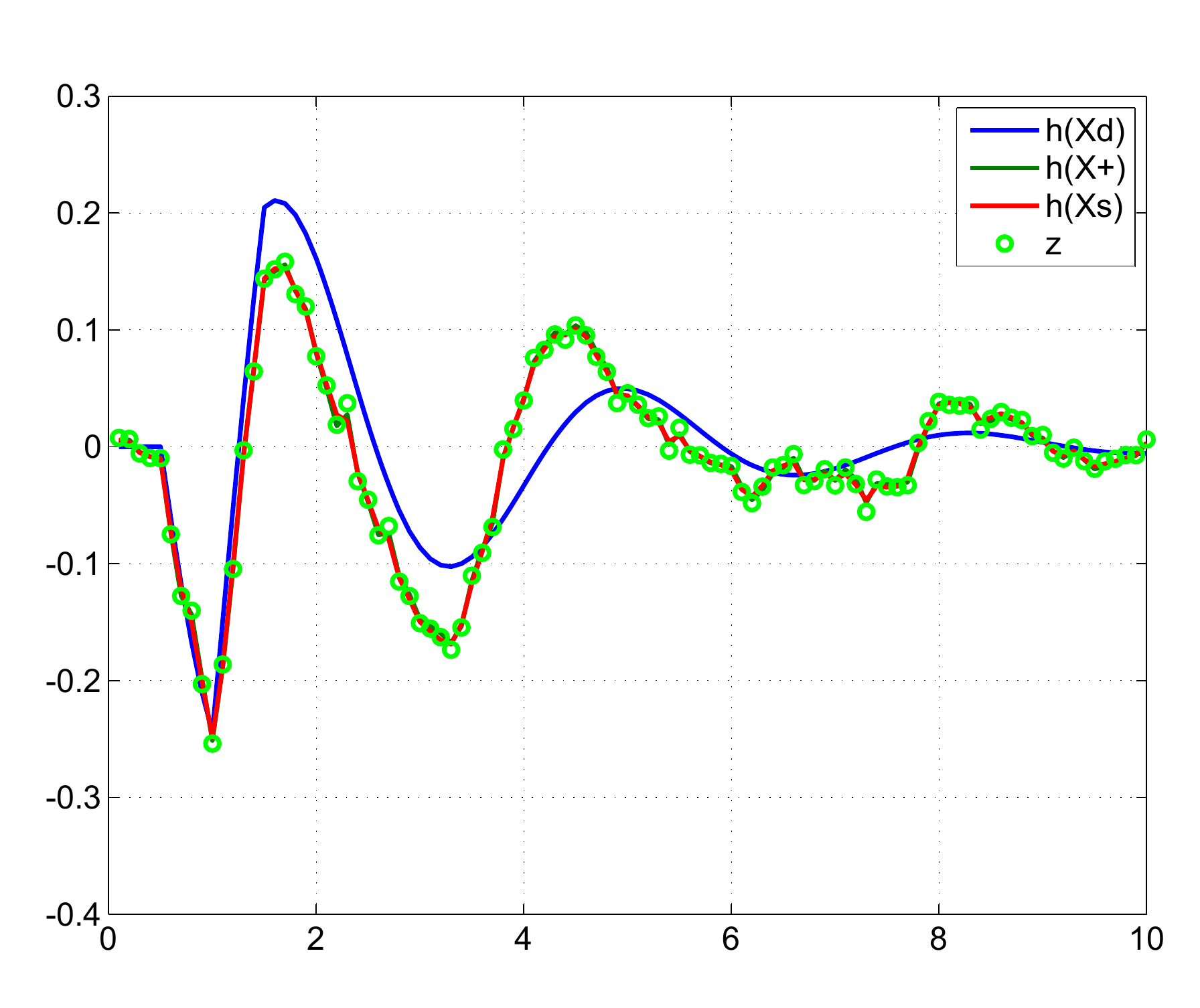}
\caption{Comparison of the predicted dynamics, posterior, smoothed}
\caption*{and the measurement 2}
\label{lonQ_h2}
\end{figure}

\begin{figure}[h]
\includegraphics[width=6in,height=4in]{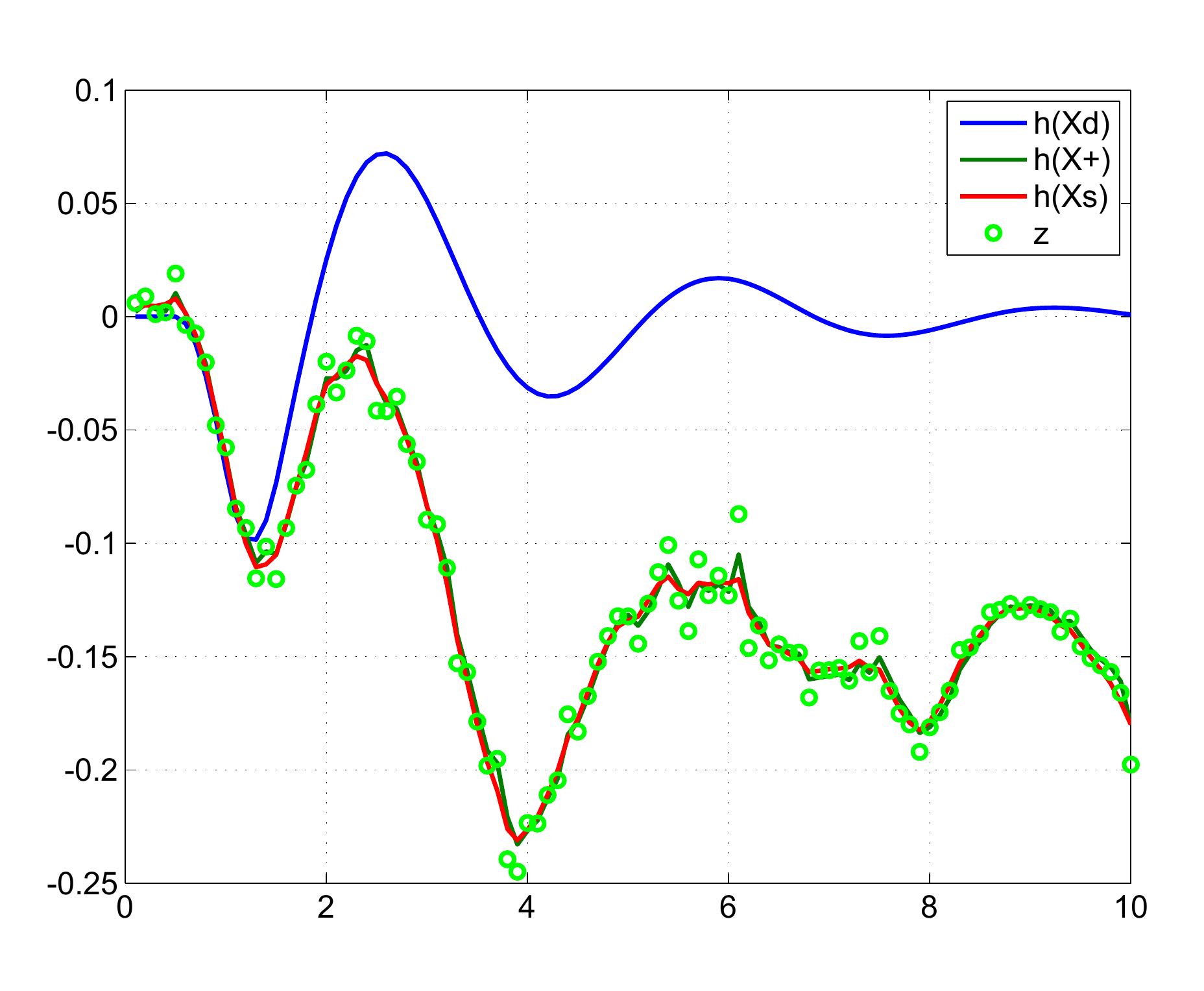}
\caption{Comparison of the predicted dynamics, posterior, smoothed}
\caption*{and the measurement 3 }
\label{lonQ_h3}
\end{figure}

\begin{figure}[h]
\includegraphics[width=6in,height=4in]{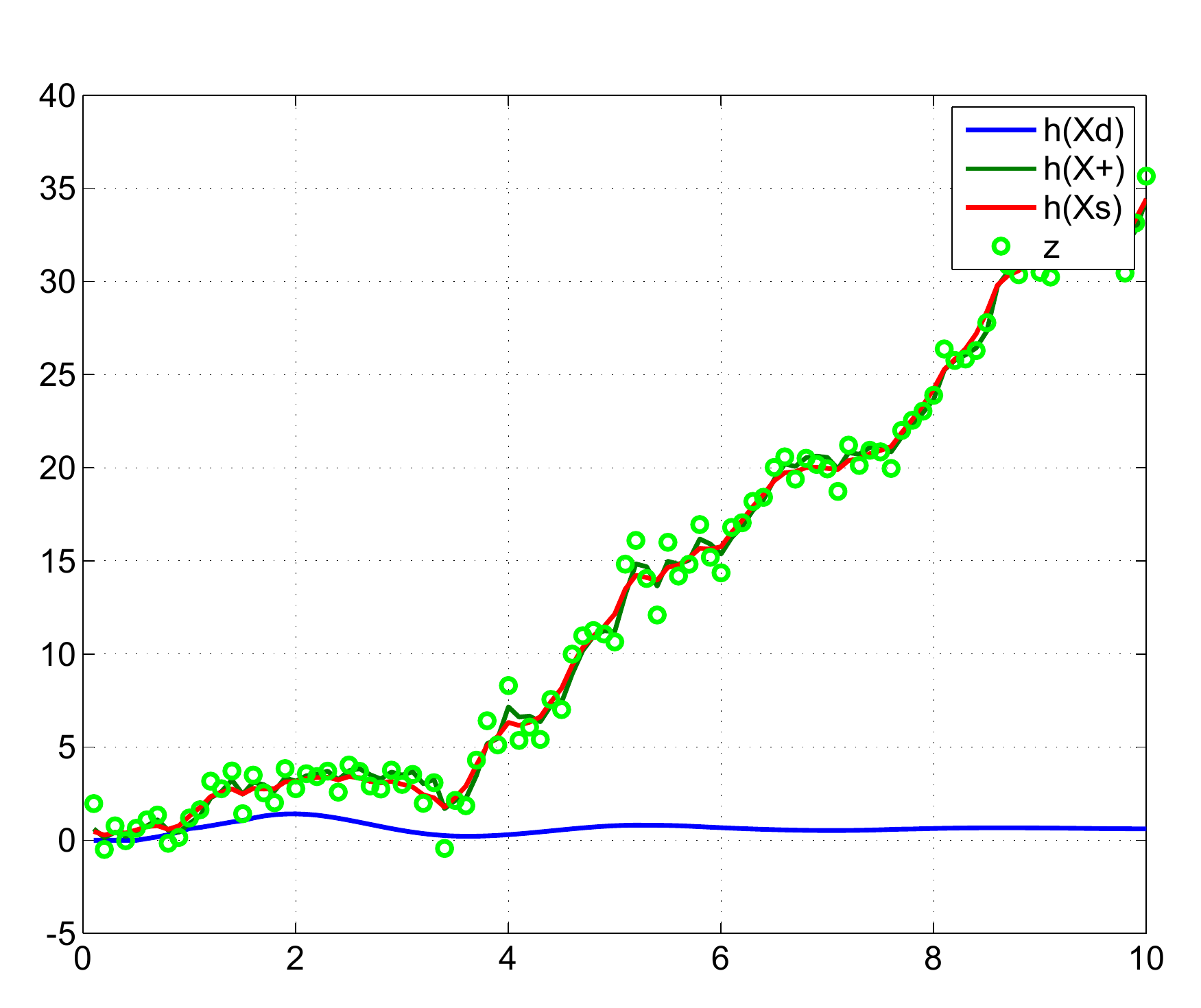}
\caption{Comparison of the predicted dynamics, posterior, smoothed}
\caption*{and the measurement 4}
\label{lonQ_h4}
\end{figure}

\begin{figure}[h]
\includegraphics[width=6in,height=4in]{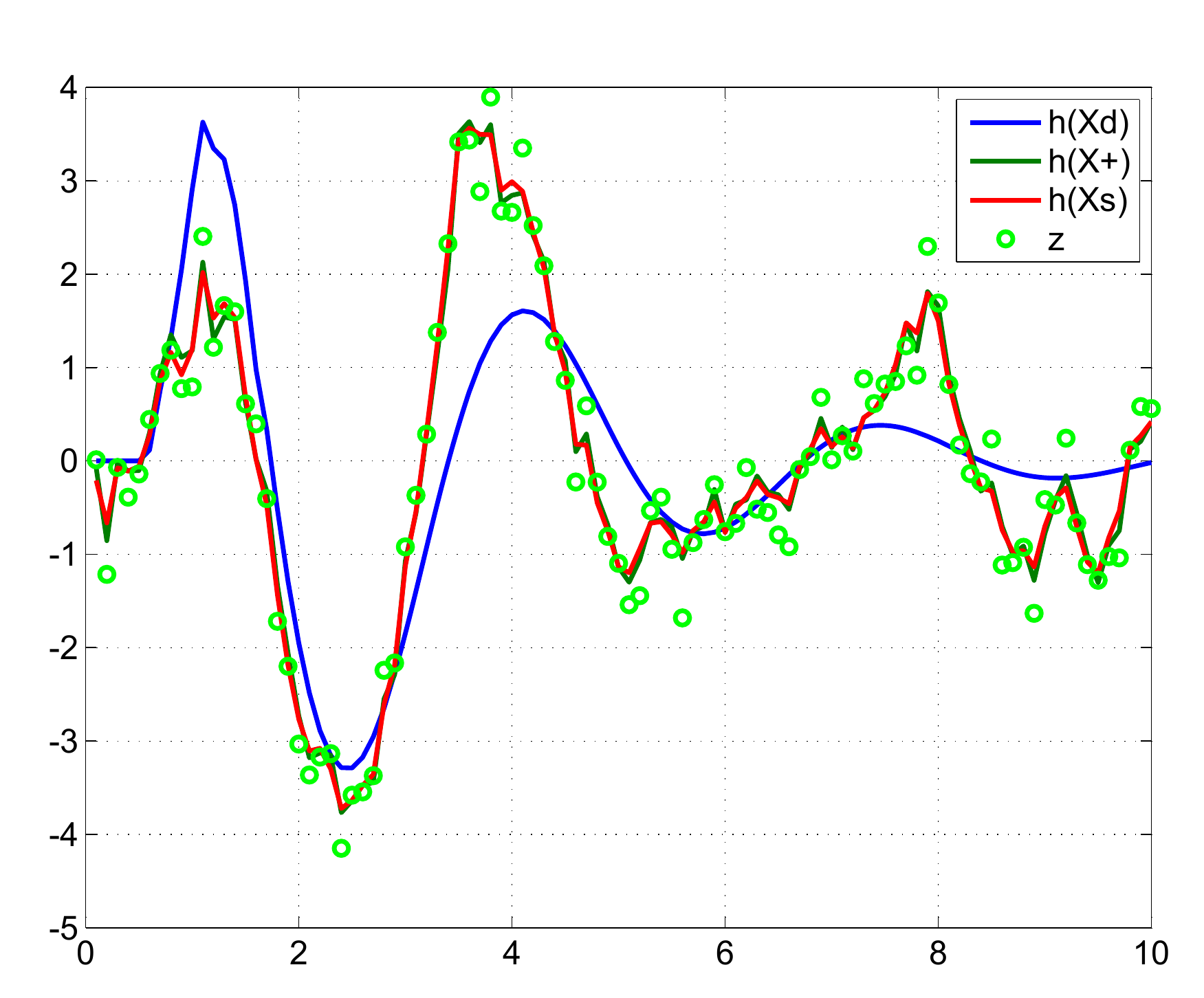}
\caption{Comparison of the predicted dynamics, posterior, smoothed}
\caption*{and the measurement 5}
\label{lonQ_h5}
\end{figure}

\begin{figure}[h]
\includegraphics[width=6in,height=4in]{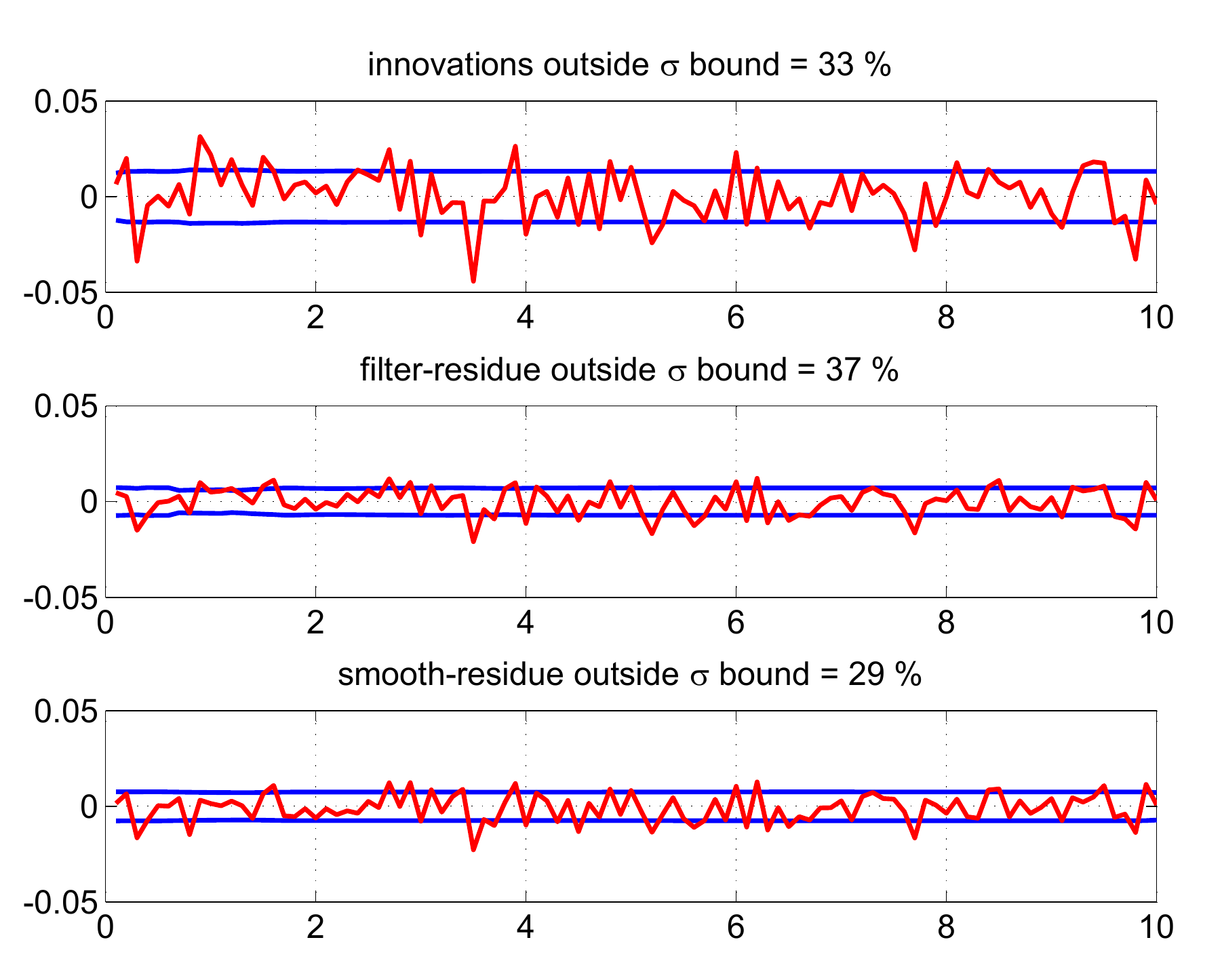}
\caption{The innovations, filtered residue and smoothed residue}
\caption*{corresponding to measurement 1 }
\label{lonQ_innov1}
\end{figure}

\begin{figure}[h]
\includegraphics[width=6in,height=4in]{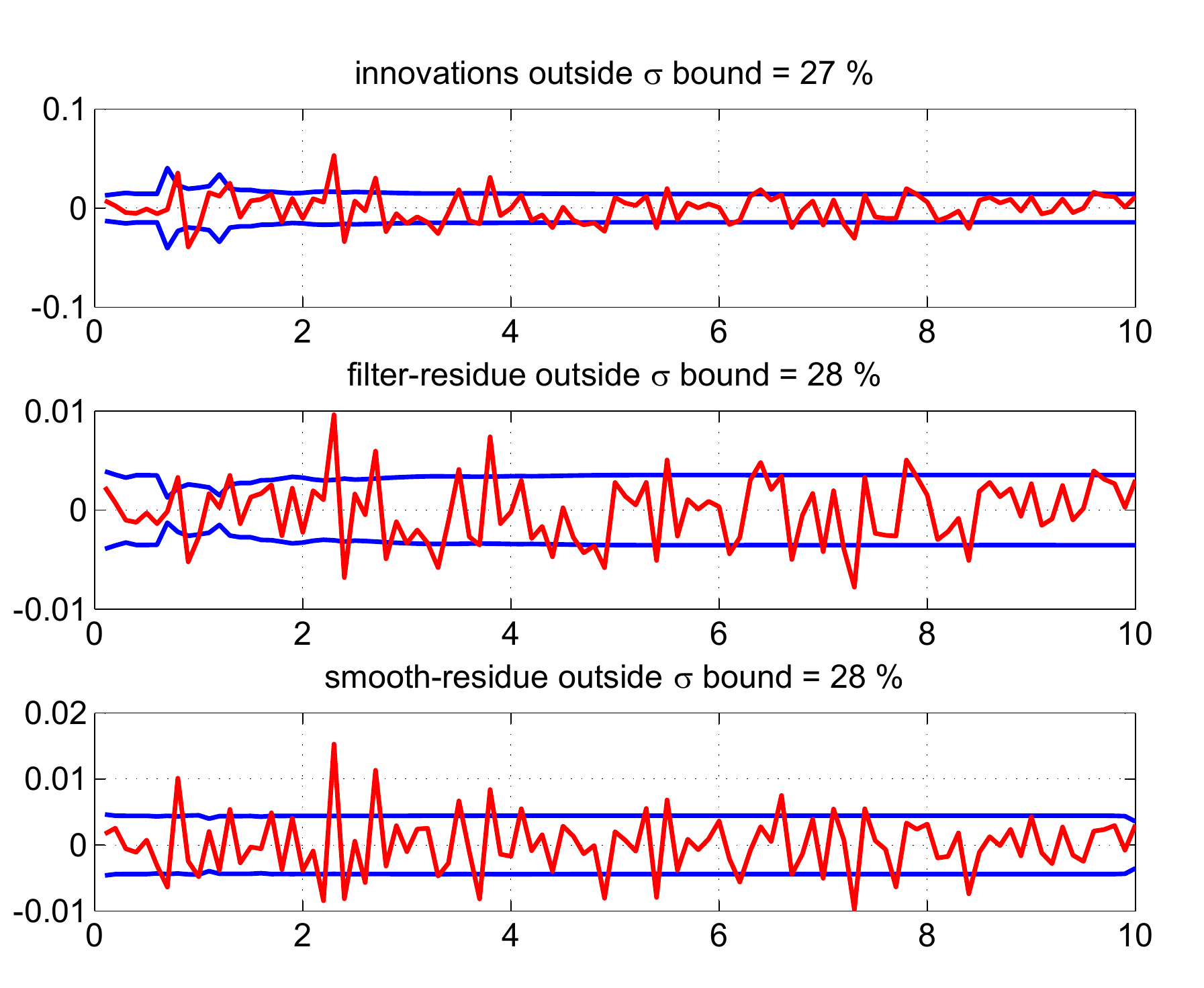}
\caption{The innovations, filtered residue and smoothed residue}
\caption*{corresponding to measurement 2}
\label{lonQ_innov2}
\end{figure}

\begin{figure}[h]
\includegraphics[width=6in,height=4in]{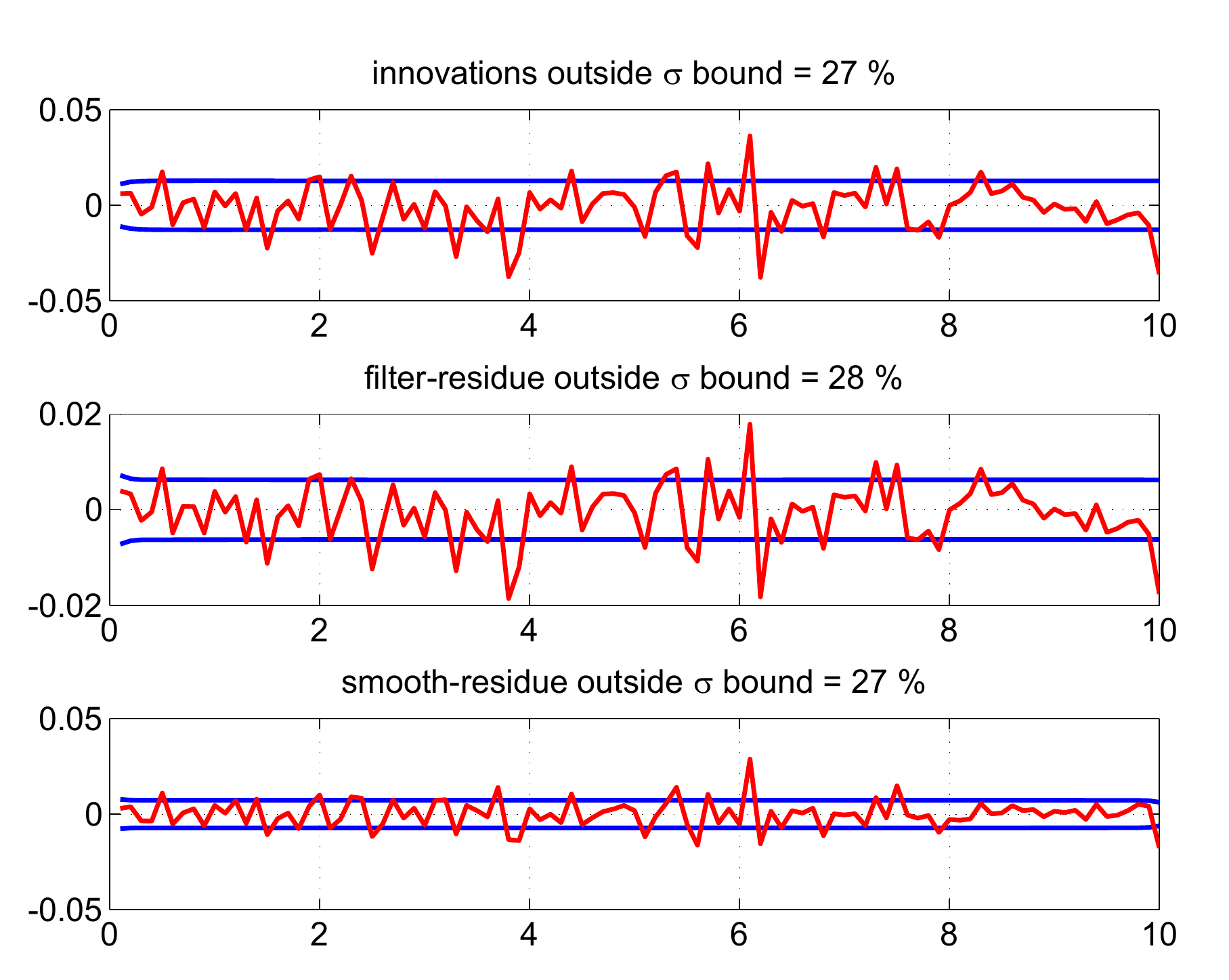}
\caption{The innovations, filtered residue and smoothed residue }
\caption*{corresponding to measurement 3 }
\label{lonQ_innov3}
\end{figure}

\begin{figure}[h]
\includegraphics[width=6in,height=4in]{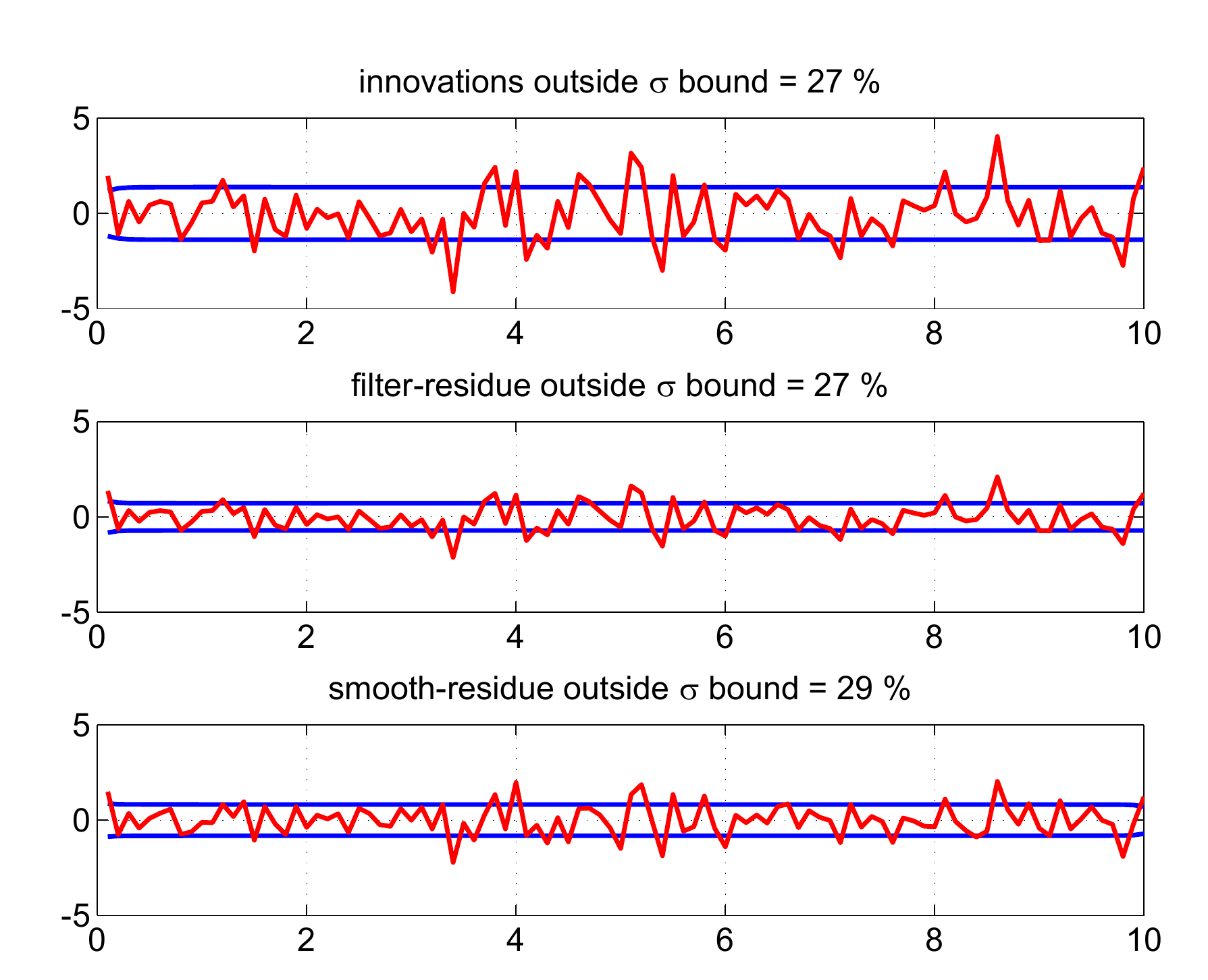}
\caption{The innovations, filtered residue and smoothed residue}
\caption*{corresponding to measurement 4}
\label{lonQ_innov4}
\end{figure}

\begin{figure}[h]
\includegraphics[width=6in,height=4in]{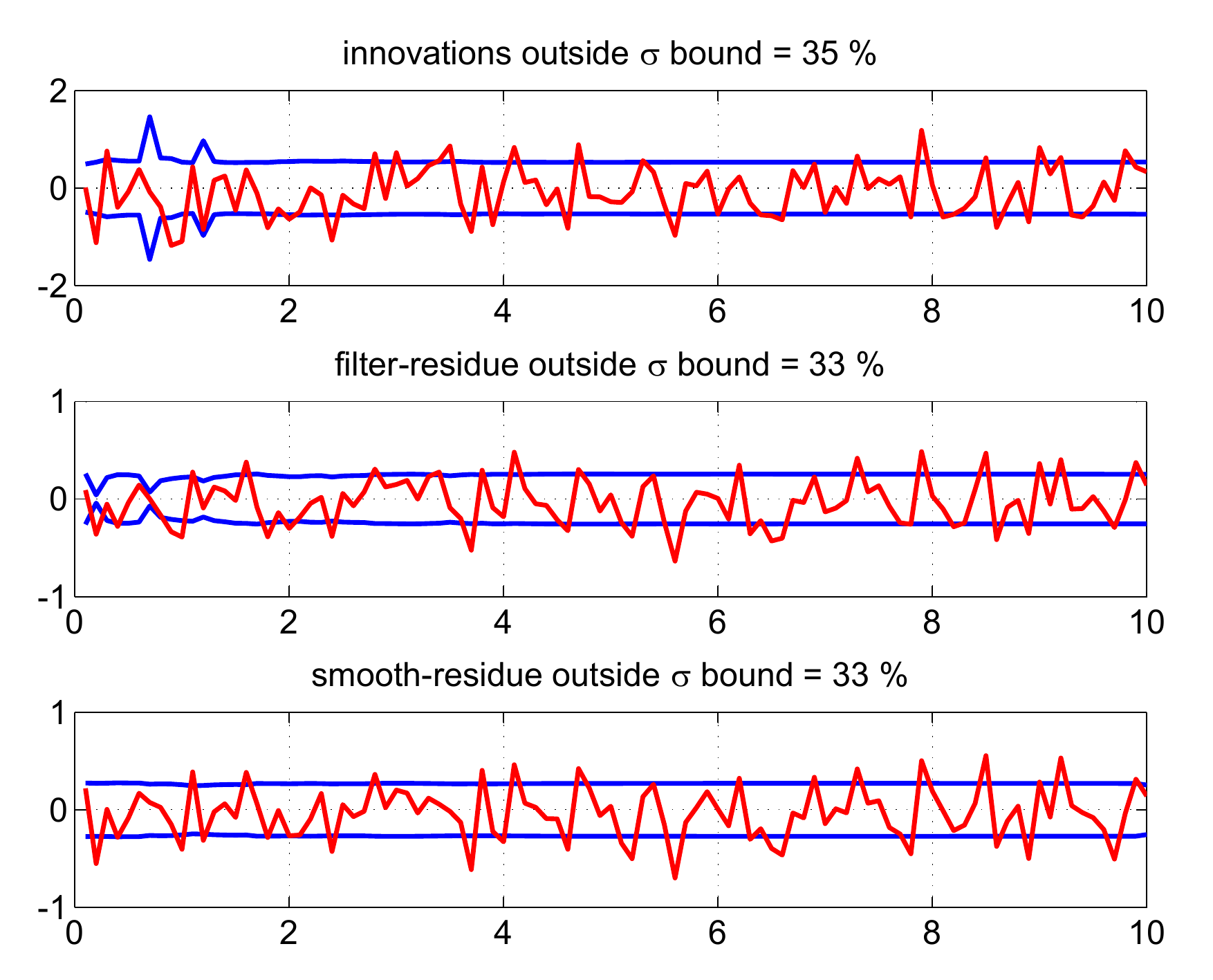}
\caption{The innovations, filtered residue and smoothed residue}
\caption*{corresponding to measurement 5}
\label{lonQ_innov5}
\end{figure}

\begin{figure}[h]
\includegraphics[width=6in,height=4in]{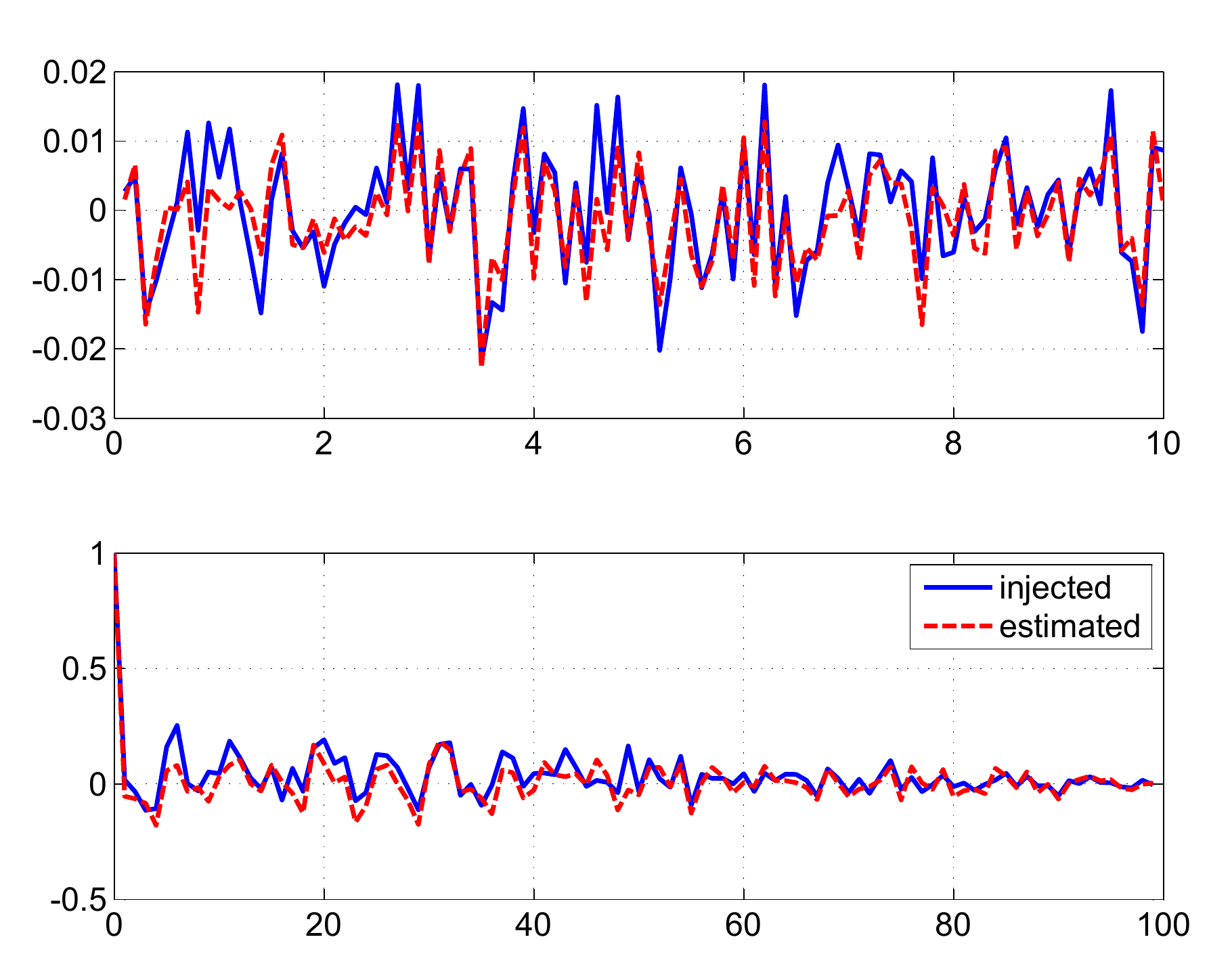}
\caption{Time variation of injected and estimated measurement noise (top) and}
\caption*{their autocorrelation (bottom) for measurement 1}
\label{lonQ_mnoise1}
\end{figure}

\begin{figure}[h]
\includegraphics[width=6in,height=4in]{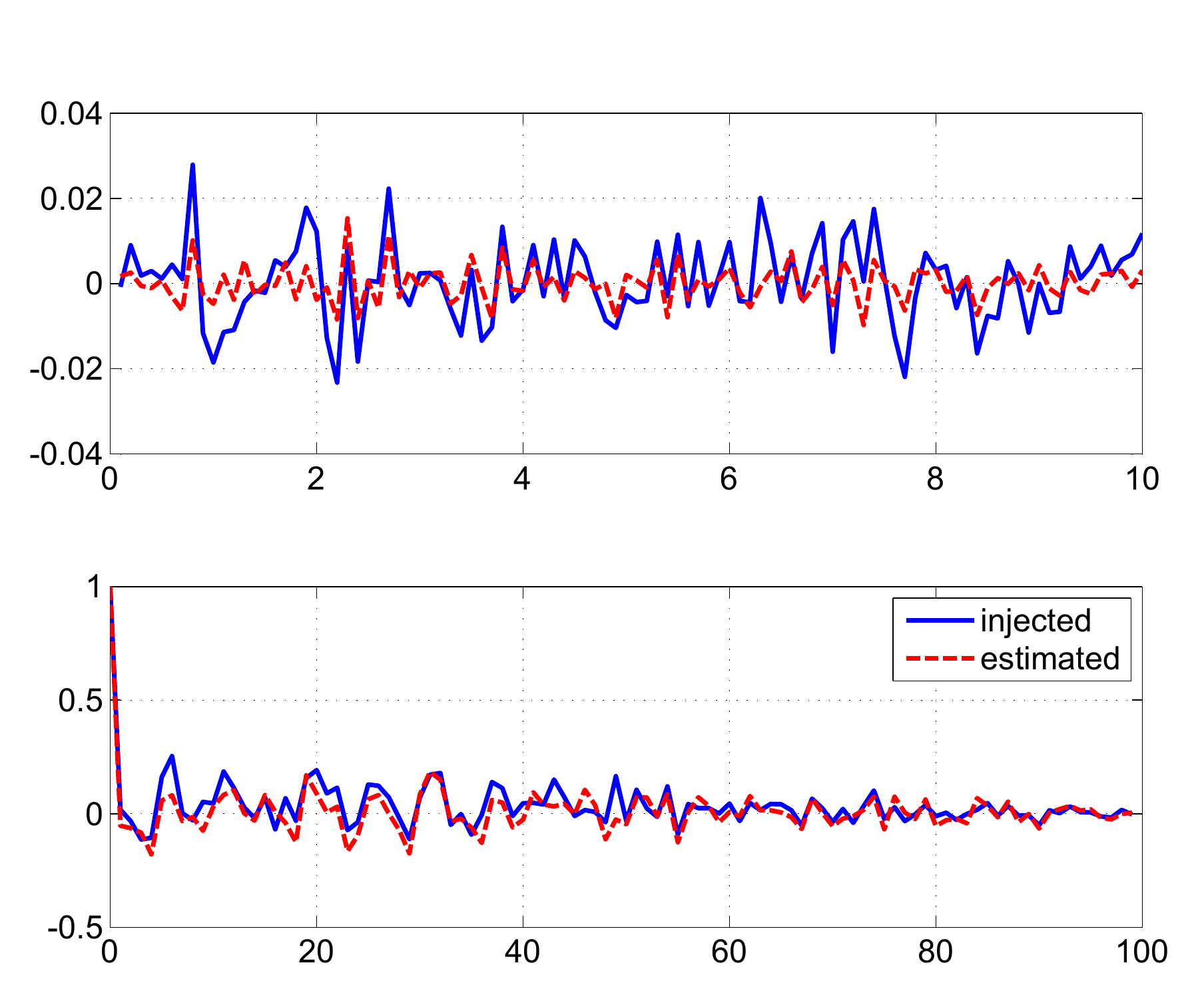}
\caption{Time variation of injected and estimated measurement noise (top) and}
\caption*{their autocorrelation (bottom) for measurement 2}
\label{lonQ_mnoise2}
\end{figure}

\begin{figure}[h]
\includegraphics[width=6in,height=4in]{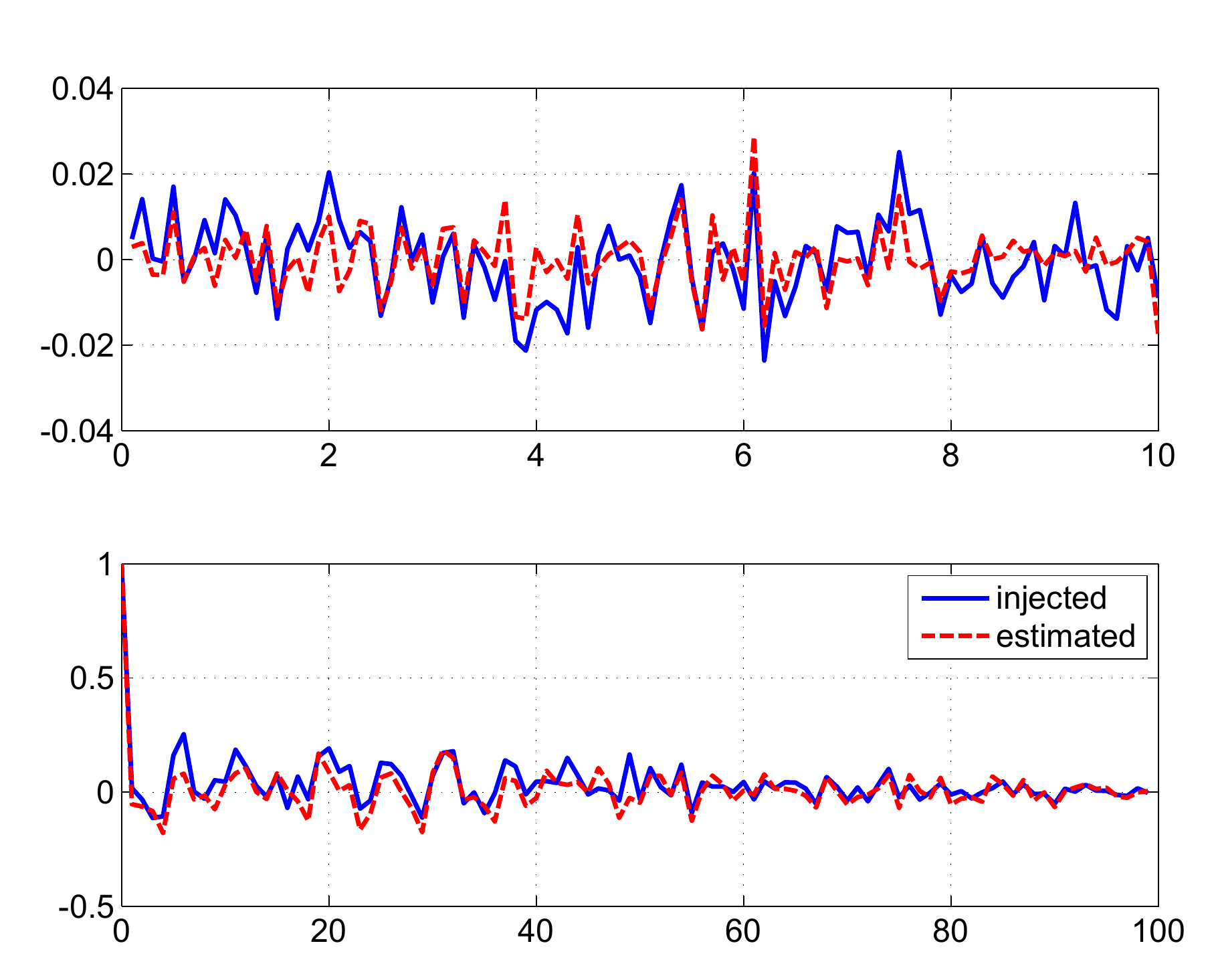}
\caption{Time variation of injected and estimated measurement noise (top) and}
\caption*{their autocorrelation (bottom) for measurement 3}
\label{lonQ_mnoise3}
\end{figure}

\begin{figure}[h]
\includegraphics[width=6in,height=4in]{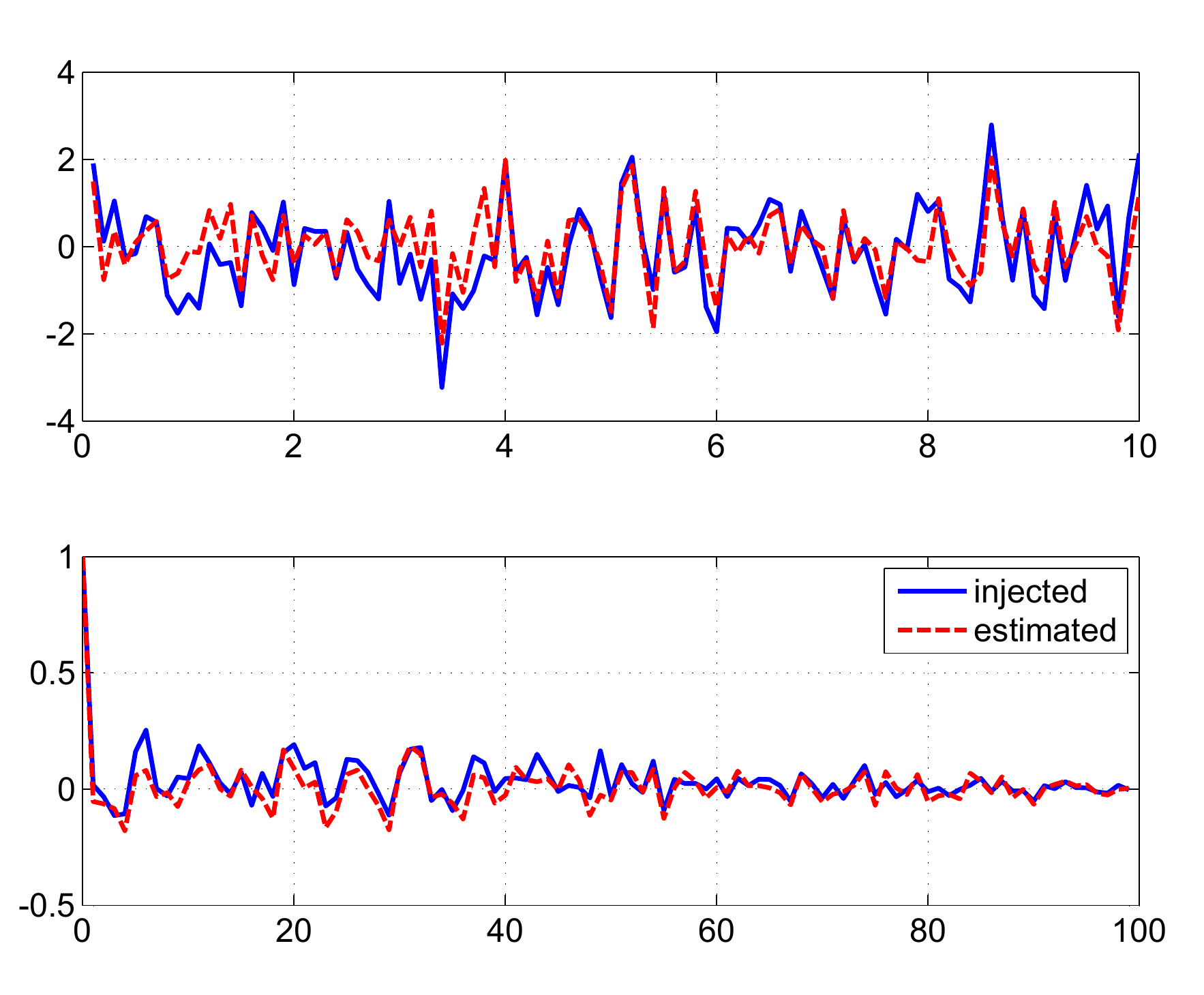}
\caption{Time variation of injected and estimated measurement noise (top) and}
\caption*{their autocorrelation (bottom) for measurement 4}
\label{lonQ_mnoise4}
\end{figure}

\begin{figure}[h]
\includegraphics[width=6in,height=4in]{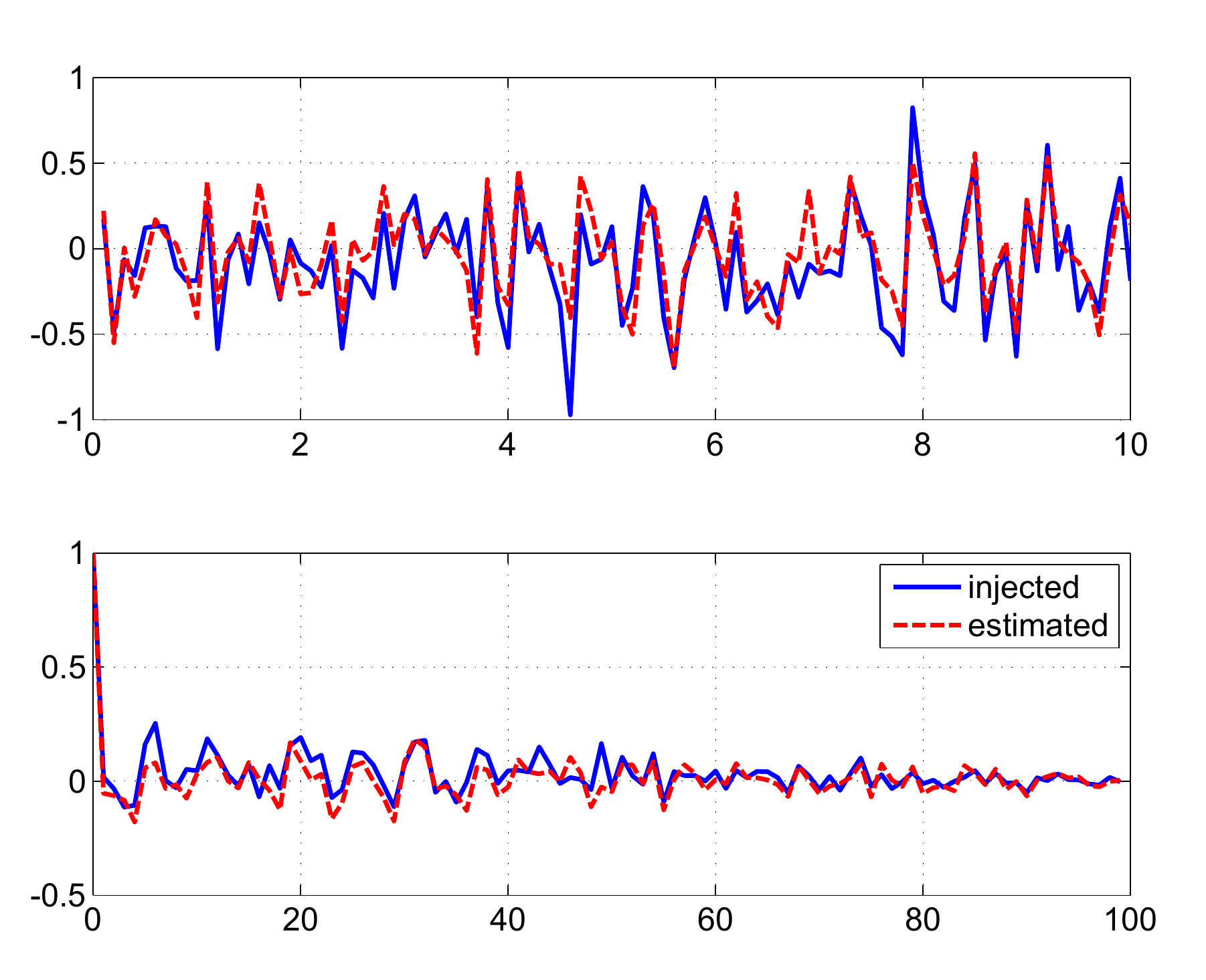}
\caption{Time variation of injected and estimated measurement noise (top) and}
\caption*{their autocorrelation (bottom) for measurement 5}
\label{lonQ_mnoise5}
\end{figure}

\begin{figure}[h]
\includegraphics[width=6in,height=4in]{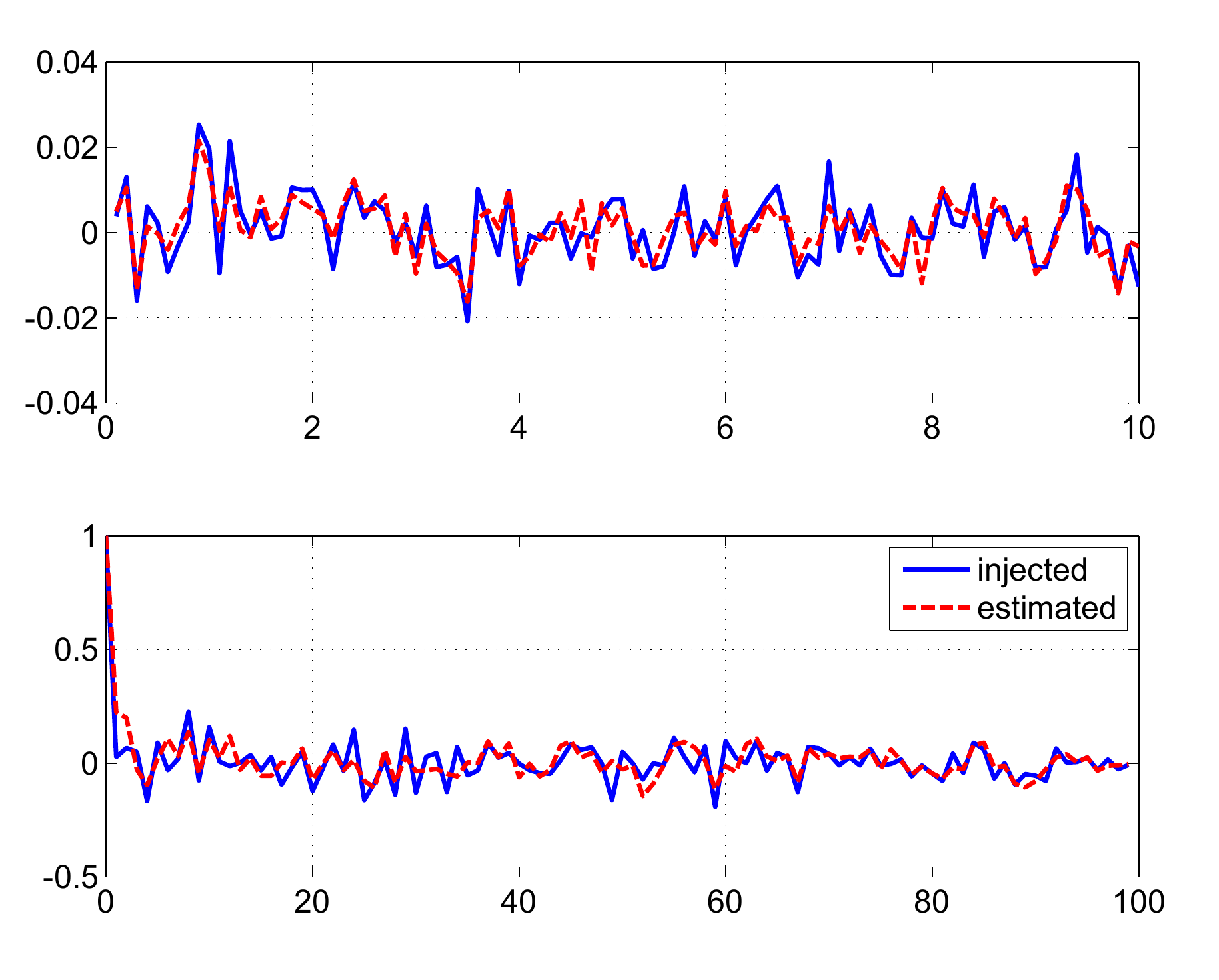}
\caption{Time variation of injected and estimated process noise (top) and}
\caption*{their autocorrelation (bottom) for state 1}
\label{lonQ_pnoise1}
\end{figure}

\begin{figure}[h]
\includegraphics[width=6in,height=4in]{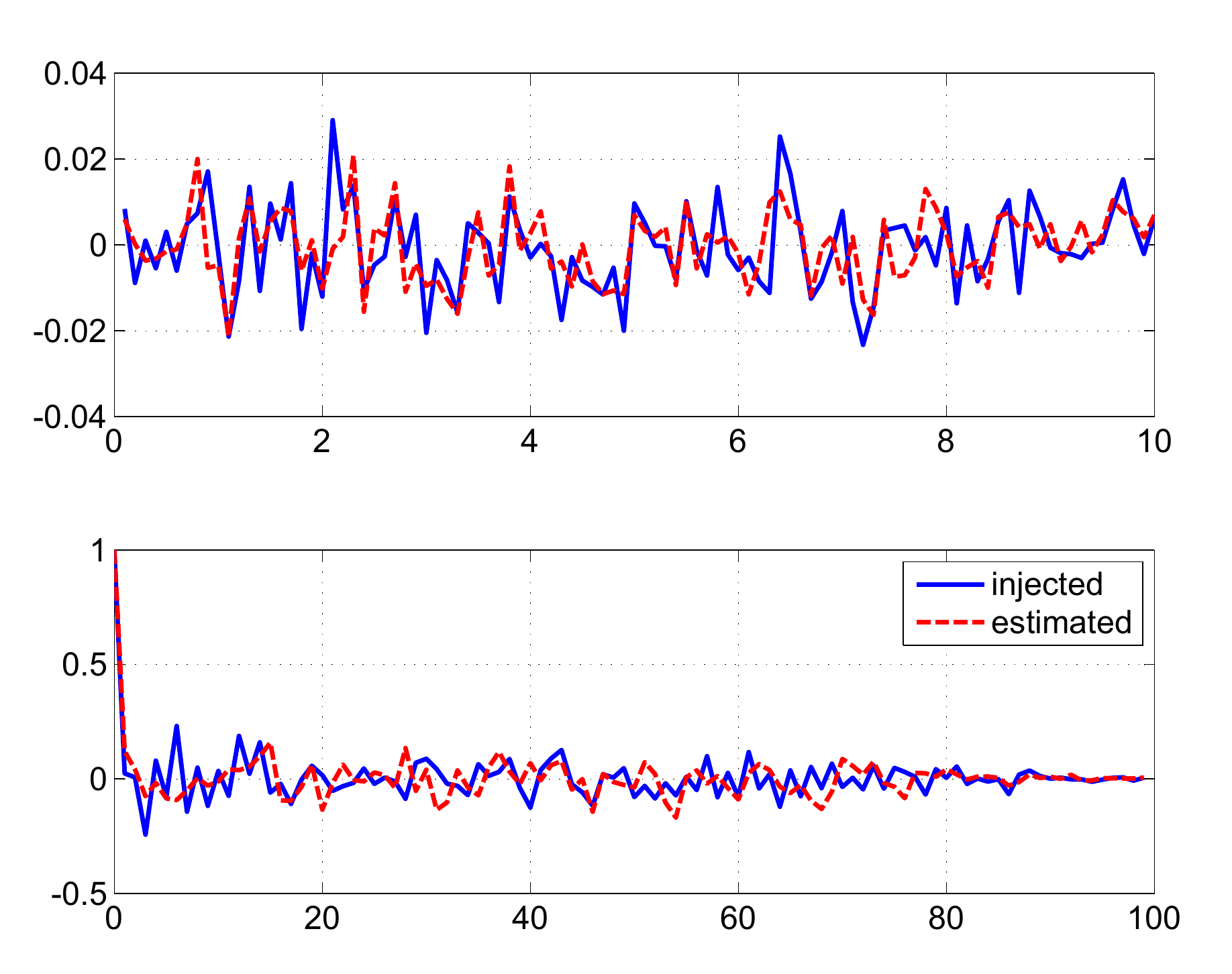}
\caption{Time variation of injected and estimated process noise (top) and}
\caption*{their autocorrelation (bottom) for state 2}
\label{lonQ_pnoise2}
\end{figure}

\begin{figure}[h]
\includegraphics[width=6in,height=4in]{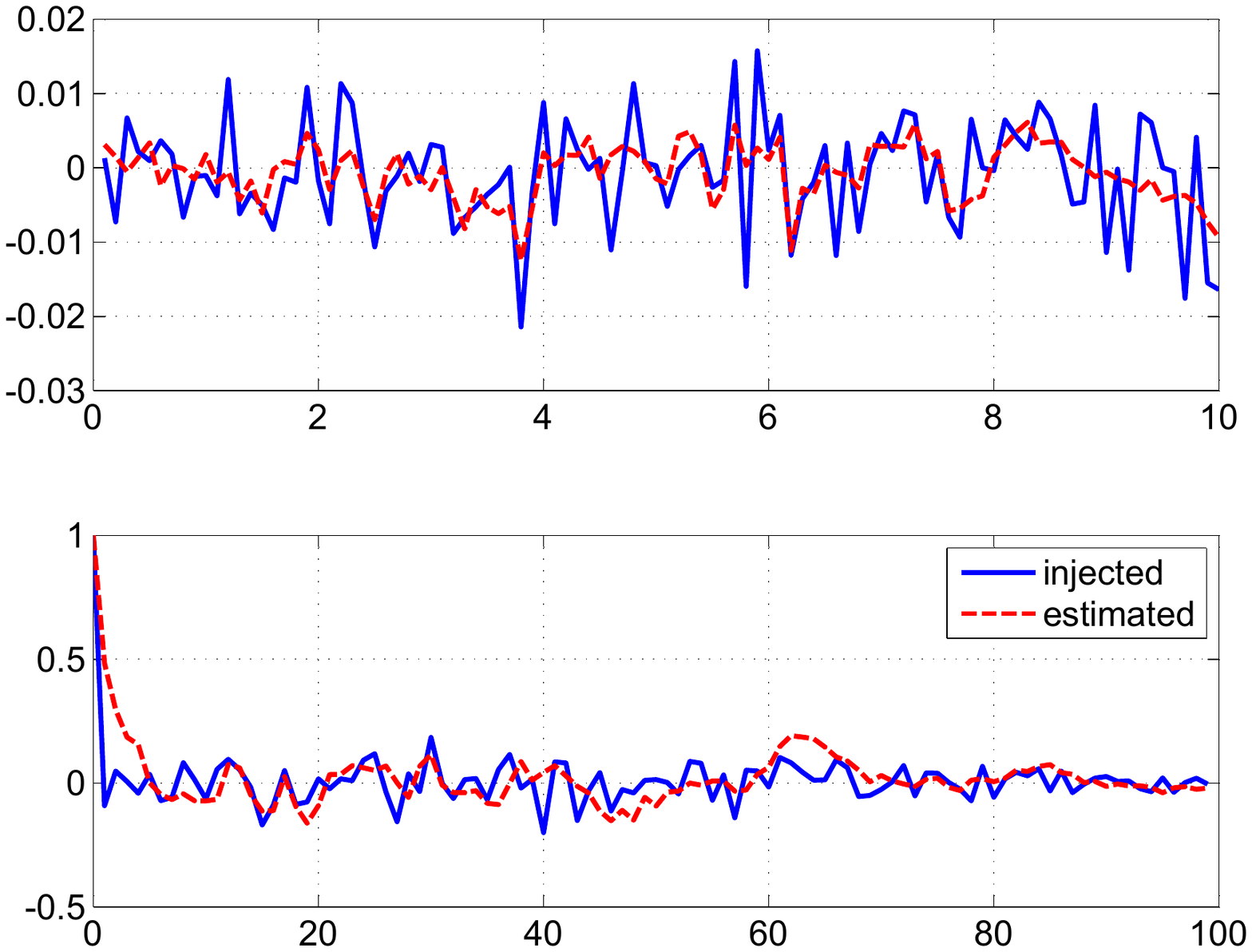}
\caption{Time variation of injected and estimated process noise (top) and}
\caption*{their autocorrelation (bottom) for state 3}
\label{lonQ_pnoise3}
\end{figure}

\begin{figure}[h]
\includegraphics[width=6in,height=4in]{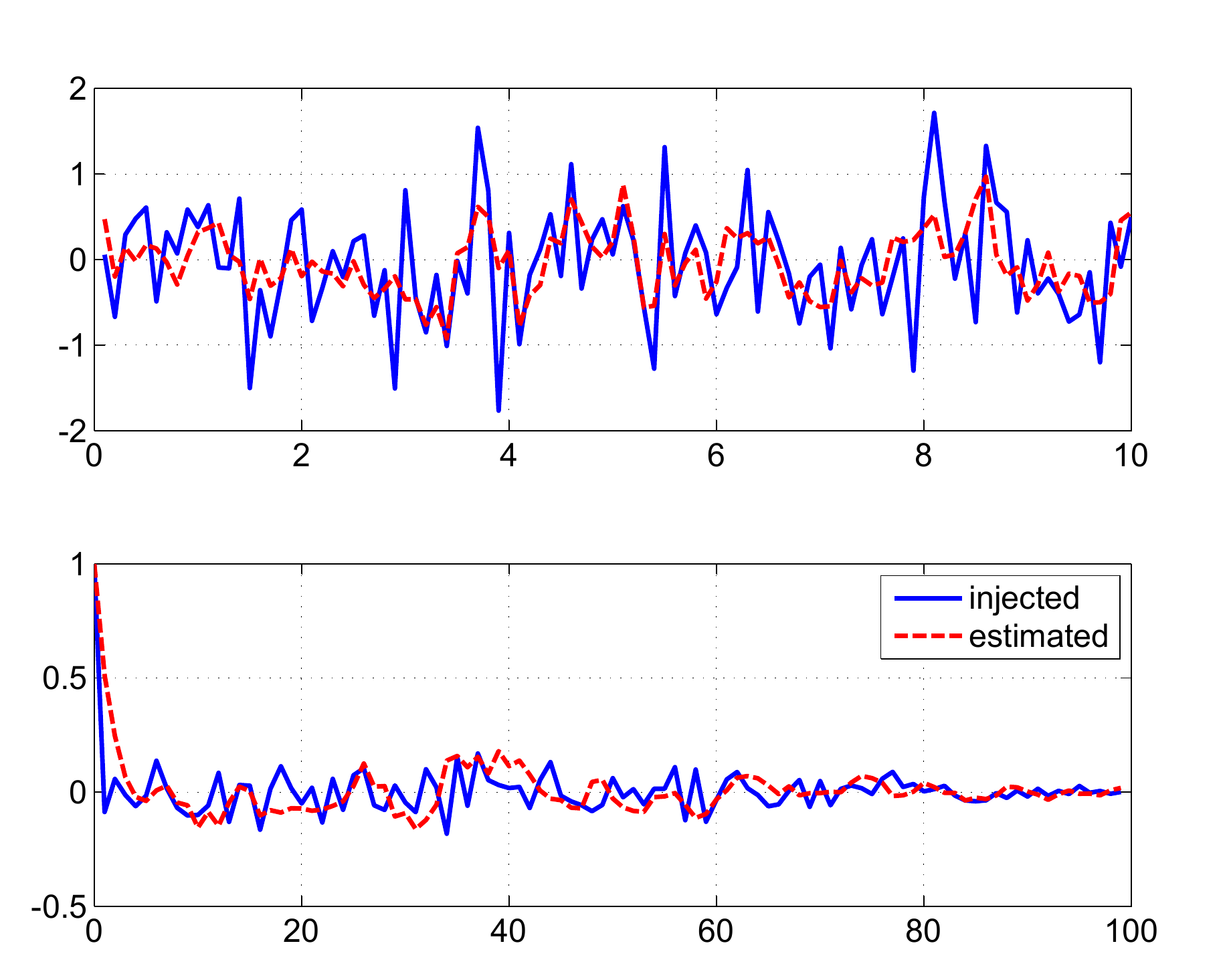}
\caption{Time variation of injected and estimated process noise (top) and}
\caption*{their autocorrelation (bottom) for state 4}
\label{lonQ_pnoise4}
\end{figure}

The next step is to process the data with process noise (\textbf{Q} $>$ 0). The Fig. \ref{lon_err}-\ref{lonQ_pnoise2} refer to the \textbf{Q} $>$ 0 case. The Fig. \ref{lon_err} shows the absolute difference between the iterated and final values with iterations which indicates the accuracy level that one needs and it was found that 100 iterations are required. The variation of the estimated initial parameters and their variances through iterations are shown in Fig. \ref{lonQ_P0}. The parameter and the uncertainty reach almost their final estimated values in about 5 and 20 iterations respectively. A similar plot in Fig. \ref{lonQ_R} shows the variation of the estimated measurement and process noise. The variation of different cost functions through the iterations are shown in Fig. \ref{lonQ_J}. The cost functions \textbf{J1-J3} correspond to the number of measurement ($m$=5) and in presence of process noise, \textbf{J6-J8} correspond to the number of states ($n$=4). The \textbf{J4} in absence of process noise corresponds to the trace of the measurement noise \textbf{R}. The \textbf{J5} is the negative log likelihood cost whose absolute value is shown in the plot. There is a mismatch in the predicted dynamics and the measurement as seen in Fig. \ref{lonQ_h1}-\ref{lonQ_h5} indicating the presence of process noise. The subsequent Fig. \ref{lonQ_innov1}-\ref{lonQ_innov5} correspond to the earlier Fig. \ref{lon_innov1}-\ref{lon_innov5} of \textbf{Q} = 0 case. The Fig.  \ref{lonQ_mnoise1}-\ref{lonQ_mnoise5} and Fig. \ref{lonQ_pnoise1}-\ref{lonQ_pnoise4} shows respectively the injected and estimated measurement and process noise samples across time during the final iteration.


\begin{landscape}
\begin{table}[h]
\subsection{Longitudinal Motion of Aircraft System Tables (\textbf{Q} = 0) }
\vspace{14pt}
\caption{Sensitivity Study : (\textbf{Q} = 0) : Longitudinal Aircraft system.\\ No. of iterations=20, No. of simulations=50.}{}
\label{tblon}
\begin{center}
\begin{footnotesize}
\begin{tabular}{| c | c | c | c | c | c | c | c | c | c |}
\hline
Study &
$\Theta$ ratio &
\makecell{CRB ratio\\NR/EKF} &
\makecell{Consistency\\ ratio-EKF}&
\makecell{Spread\\factor\\EKF} &
\makecell{\textbf{R} ratio\\EKF/True} & \makecell{\textbf{R} ratio\\EKF/NR} &
\makecell{$\mu$\ of \\\textbf{J1-J5}} &
\makecell{$\sigma$ of \\\textbf{J1-J5}} &
Remarks
\\ \hline

\multicolumn{10}{|c|}{Reference adaptive EKF used for non zero process noise case} \\ \hline
\makecell{$\mathbf{P_0}$ : Scaled up-[0,0;0,\checkmark]\\\textbf{Q} : EM-[\checkmark,0;0,0] \\ \textbf{R} : EM-diag}
&\multicolumn{9}{c|}{Extremely slow convergence of \textbf{Q} taking more than 100 iterations.} \\ \hline

\multicolumn{10}{|c|}{\makecell{Proposed reference adaptive EKF for zero process noise case, \textbf{Q} is diagonal matrix fixed to $10^{-10}$ for all states \\
Consistency ratio NR=   1.1320,    1.0367,    0.9417,    1.1147,    0.9322,
\\
Spread Factor NR= 3.5427,    0.8768,    6.5086,  233.9268,    1.9185}} \\
\hline

\end{tabular}
\end{footnotesize}
\end{center}
\end{table}
\end{landscape}

\begin{landscape}
\begin{table}[h]
\addtocounter{table}{-1}
\caption{Sensitivity Study : (\textbf{Q} = 0) : Longitudinal Aircraft system.\\ No. of iterations=20, No. of simulations=50.}{}
\begin{center}
\begin{footnotesize}
\begin{tabular}{| c | c | c | c | c | c | c | c | c | c |}
\hline
Study &
$\Theta$ ratio &
\makecell{CRB ratio\\NR/EKF} &
\makecell{Consistency\\ ratio-EKF}&
\makecell{Spread\\factor\\EKF} &
\makecell{\textbf{R} ratio\\EKF/True} & \makecell{\textbf{R} ratio\\EKF/NR} &
\makecell{$\mu$\ of \\\textbf{J1-J5}} &
\makecell{$\sigma$ of \\\textbf{J1-J5}} &
Remarks
\\ \hline

\makecell{$\mathbf{P_0}$ : IIM-[0,0;0,\checkmark]\\\textbf{Q} : $10^{-10}$-[\checkmark,0;0,0] \\ \textbf{R} : EM-diag}  &
\makecell{  1.0043  \\  1.0005  \\  0.9865 \\   0.9496  \\  1.0000} &
\makecell{ 0.9974  \\  1.0207 \\   0.9433 \\   1.0112 \\   0.9233} &
\makecell{ 1.2091  \\  2.6391  \\  1.0989  \\  1.1716 \\   1.0806} &
\makecell{3.6684  \\  1.5370  \\  7.3022\\  236.3277  \\  2.1750} &
\makecell{ 1.0132  \\  1.0569  \\  0.9890  \\  1.0271 \\   0.9942} &
\makecell{1.0082  \\  1.0518  \\  1.0210  \\  1.0507  \\  1.0418} &
\makecell{  4.9498  \\  4.9532  \\  5.0001  \\  1.0732 \\ -24.6034 } &
\makecell{ 0.1284  \\  0.1245 \\   0.0095 \\   0.1525  \\  0.3909 } &
\makecell{Unity CRB ratio} \\ \hline

\makecell{$\mathbf{P_0}$ : IIM-diag\\\textbf{Q} : $10^{-10}$-[\checkmark,0;0,0] \\ \textbf{R} : EM-diag}  &
\makecell{  1.0060  \\  0.9969  \\  0.9929  \\  0.7579  \\  1.0091} &
\makecell{ 0.9856  \\  0.9120 \\   0.9345 \\   0.9957  \\  0.8027} &
\makecell{  1.1812  \\  2.2915  \\  1.0050 \\   1.1052 \\   1.1132} &
\makecell{3.6970  \\  1.5035  \\  7.0722\\  233.7360  \\  2.6459} &
\makecell{ 1.0227 \\   1.0539 \\   0.9800  \\  0.9781  \\  1.0041} &
\makecell{ 1.0176  \\  1.0489  \\  1.0116 \\   1.0007  \\  1.0522} &
\makecell{  4.9116  \\  4.9132  \\  5.0017  \\  1.1739 \\ -24.5304 } &
\makecell{  0.1555  \\  0.1497  \\  0.0135 \\   0.2161 \\   0.3722  } &
\makecell{Unity CRB ratio} \\
\hline

\end{tabular}
\end{footnotesize}
\end{center}
\end{table}
\end{landscape}

\begin{landscape}
\begin{table}[h]
\addtocounter{table}{-1}
\caption{Sensitivity Study : (\textbf{Q} = 0) : Longitudinal Aircraft system.\\ No. of iterations=20, No. of simulations=50.}{}
\begin{center}
\begin{footnotesize}
\begin{tabular}{| c | c | c | c | c | c | c | c | c | c |}
\hline
Study &
$\Theta$ ratio &
\makecell{CRB ratio\\NR/EKF} &
\makecell{Consistency\\ ratio-EKF}&
\makecell{Spread\\factor\\EKF} &
\makecell{\textbf{R} ratio\\EKF/True} & \makecell{\textbf{R} ratio\\EKF/NR} &
\makecell{$\mu$\ of \\\textbf{J1-J5}} &
\makecell{$\sigma$ of \\\textbf{J1-J5}} &
Remarks
\\ \hline

\makecell{$\mathbf{P_0}$ : Smoothed-[0,0;0,\checkmark]\\\textbf{Q} : $10^{-10}$-[\checkmark,0;0,0] \\ \textbf{R} : EM-diag}  &
\makecell{  1.0023  \\  1.0011  \\  0.9943 \\   0.8240 \\   0.9980} &
\makecell{ 4.4108 \\   4.4306 \\   4.1146 \\   4.4680 \\   4.0746} &
\makecell{5.0072  \\  4.6519  \\  4.1359  \\  4.9952 \\   4.0162} &
\makecell{  2.4008 \\   0.5516  \\  4.4445 \\ 160.2168  \\  1.2668} &
\makecell{ 1.0032   \\ 1.0074  \\  0.9684  \\  0.9767  \\  0.9528} &
\makecell{  0.9982  \\  1.0026 \\   0.9997  \\  0.9991 \\   0.9984} &
\makecell{  4.9971 \\   4.9971 \\   4.9997  \\  1.0722\\  -25.0816 } &
\makecell{ 0.0014  \\  0.0014  \\  0.0005 \\   0.1532  \\  0.3498 } &
\makecell{Lower filter CRB\\ than that of the NR}\\ \hline

\makecell{$\mathbf{P_0}$ : Smoothed-diag\\\textbf{Q} : $10^{-10}$-[\checkmark,0;0,0] \\ \textbf{R} : EM-diag}  &
\makecell{ 0.9962 \\   0.9973 \\   1.0225  \\  0.6821 \\   1.0149} &
\makecell{4.9542   \\ 4.6358  \\  6.4682 \\   4.5531 \\   5.8173} &
\makecell{  5.5242 \\   5.3890  \\  6.3287  \\  5.1484  \\  6.3994} &
\makecell{ 2.3346  \\  0.6331 \\   4.4581\\  163.9454  \\  1.8093} &
\makecell{0.9992 \\   1.0085 \\   0.9650  \\  0.9655 \\   0.9492} &
\makecell{0.9942  \\  1.0037  \\  0.9962 \\   0.9878 \\   0.9946} &
\makecell{  4.9935  \\  4.9935  \\  4.9993 \\   1.0585 \\ -25.1038} &
\makecell{   0.0025 \\   0.0025 \\   0.0008  \\  0.1515  \\  0.3557   } &
\makecell{Lower filter CRB\\ than that of the NR}\\
\hline

\end{tabular}
\end{footnotesize}
\end{center}
\end{table}
\end{landscape}

\begin{landscape}
\begin{table}[h]
\addtocounter{table}{-1}
\caption{Sensitivity Study : (\textbf{Q} = 0) : Longitudinal Aircraft system.\\ No. of iterations=20, No. of simulations=50.}{}
\begin{center}
\begin{footnotesize}
%
\end{footnotesize}
\end{center}
\end{table}
\end{landscape}

\begin{landscape}
\begin{table}[h]
\subsection{Longitudinal Motion of Aircraft System Tables (\textbf{Q} $>$ 0) }
\vspace{14pt}
\caption{Sensitivity Study : (\textbf{Q} $>$ 0) : Longitudinal Aircraft system.\\ No. of iterations=100, No. of simulations=50.}{}
\label{tblonQ}
\begin{center}
\begin{footnotesize}
%
\end{footnotesize}
\end{center}
\end{table}
\end{landscape}

\begin{landscape}
\begin{table}[h]
\addtocounter{table}{-1}
\caption{Sensitivity Study : (\textbf{Q} $>$ 0) : Longitudinal Aircraft system.\\ No. of iterations=100, No. of simulations=50.}{}
\begin{center}
\begin{footnotesize}
%
\end{footnotesize}
\end{center}
\end{table}
\end{landscape}

\begin{landscape}
\begin{table}[H]
\addtocounter{table}{-1}
\caption{Sensitivity Study : \textbf{Q}$>0$ : Longitudinal Aircraft system.\\ No. of iterations=100, No. of simulations=50.}{}
\begin{center}
\begin{footnotesize}
%
\end{footnotesize}
\end{center}
\end{table}
\end{landscape}

\begin{landscape}
\begin{table}[H]
\addtocounter{table}{-1}
\caption{Sensitivity Study : \textbf{Q} $>$ 0 : Longitudinal Aircraft system.\\ No. of iterations=100, No. of simulations=50.}{}
\begin{center}
\begin{footnotesize}
%
\end{footnotesize}
\end{center}
\end{table}
\end{landscape}


\clearpage
\subsection{Longitudinal Motion of Aircraft System Figures (\textbf{Q} = 0) }

\clearpage
\subsection{Longitudinal Motion of Aircraft System Figures (\textbf{Q} $>$ 0) }

\vspace{14pt}

\clearpage
\section{Simulated Lateral Motion of Aircraft System}

\par
\begin{figure}[h]
\includegraphics[width=6in,height=3in]{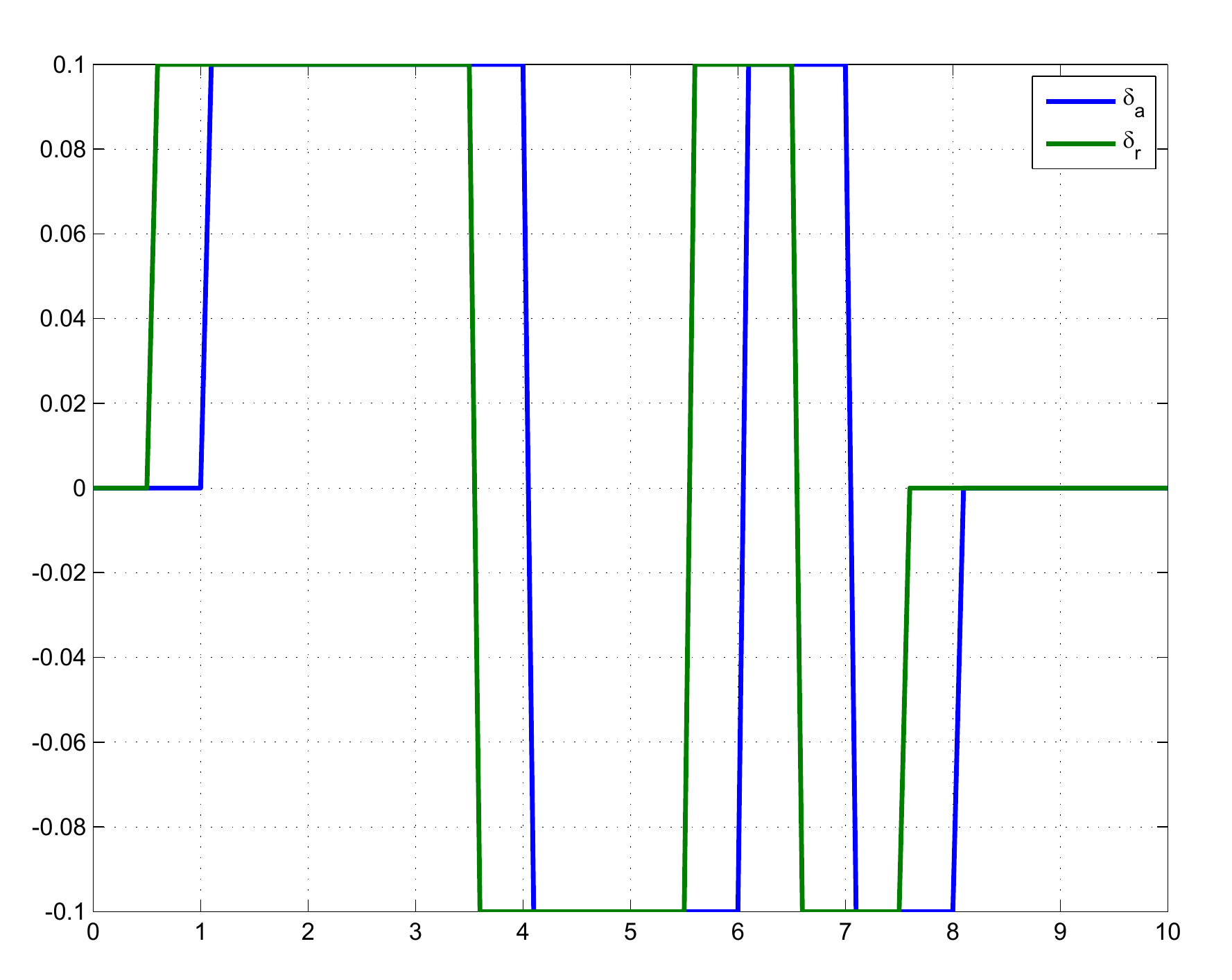}
\caption{Control input of simulated Lateral aircraft system}
\label{input_lat}
\end{figure}

 Consider the Lateral motion of an aircraft system excited by control input as shown in Fig. \ref{input_lat}. The state equations ($n=4$) for the angle of sideslip ($\beta$), roll rate (p), roll angle ($\phi$) and yaw rate (r) are respectively given by
\begin{align*}
\dot{\beta}&=Y_{\beta}\beta+Y_pp+Y_\phi\phi-r+Y_{\delta_a}\delta_a +Y_{\delta_r}\delta_r\\
\dot{p}&=L_{\beta}\beta+L_pp+L_rr+L_{\delta_a}\delta_a +L_{\delta_r}\delta_r\\
\dot\phi&=p\\
\dot{r}&=N_{\beta}\beta+N_pp+N_rr+N_{\delta_a}\delta_a +N_{\delta_r}\delta_r
\end{align*}
All the states have zero initial conditions. The measured quantites are indicated with subscript `m' which includes the angle of sideslip ($\beta$), roll rate (p), roll angle ($\phi$), yaw rate (r), lateral acceleration ($a_y$), roll acceleration ($\dot p$) and yaw acceleration ($\dot r$). The measurement equations ($m=7$) are given by
\begin{align*}
{\beta_m}&=\beta+v_\beta\\
{p_m}&=p+v_p\\
{\phi_m}&=\phi+v_\phi\\
{r_m}&=r+v_r\\
{a_{y_m}}&=U_0\times(Y_{\beta}\beta+Y_{\delta_a}\delta_a +Y_{\delta_r}\delta_r)+v_{a_y} \\
{\dot{p}_m}&=L_{\beta}\beta+L_pp+L_rr+L_{\delta_a}\delta_a +L_{\delta_r}\delta_r+v_p\\
{\dot{r}_m}&=N_{\beta}\beta+N_pp+N_rr+N_{\delta_a}\delta_a +N_{\delta_r}\delta_r+v_r
\end{align*}

The unknown parameter set($p$=15) is $\Theta=(Y_\beta,Y_p,Y_r,L_\beta,L_p,L_r,N_\beta,N_p,N_r,Y_{\delta_a},Y_{\delta_r},L_{\delta_a},L_{\delta_r},\\N_{\delta_a},N_{\delta_r})^T$ with the true values being $(-0.18,-0.00278,0.14,-0.097,-5.82,1.782,0.0084,\\-0.665,-0.712,-0.00447,0.02657,16.434,0.434,-0.428,-2.824)^T$ and $U_0$ = 100. The numerical values of the noise variances are shown in Table-\ref{sysdes}. All the figures are presented for only one simulation run to prevent cluttering.

\subsection{Remarks on the Results}

\par
\begin{figure}[h]
\includegraphics[width=6in,height=3.0in]{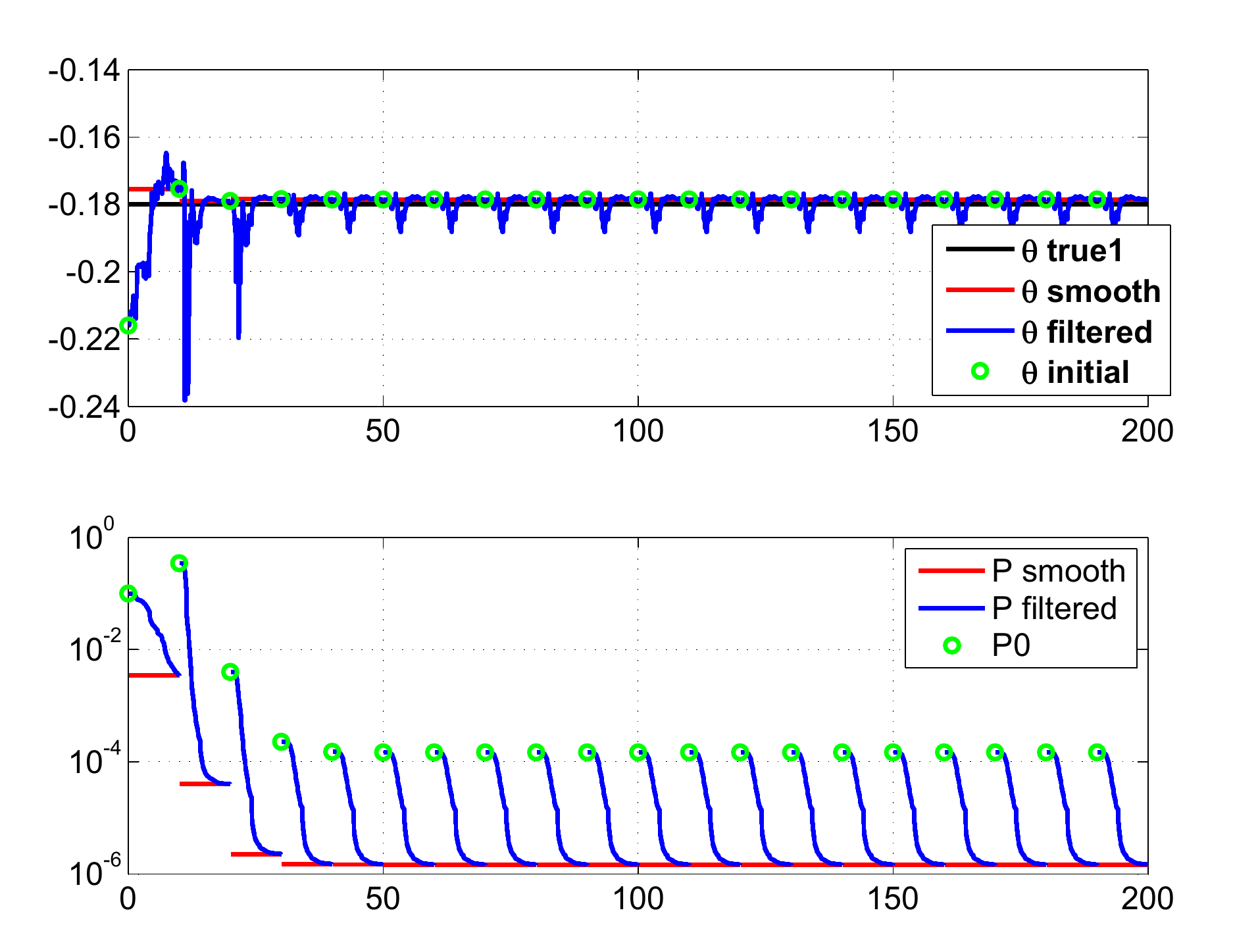}
\caption{The variation of parameter estimate 1 and their filtered and}
\caption*{smoothed covariances through (with the time cumulatively) the iterations}
\label{lat_p1}
\end{figure}

\begin{figure}[h]
\includegraphics[width=6in,height=4in]{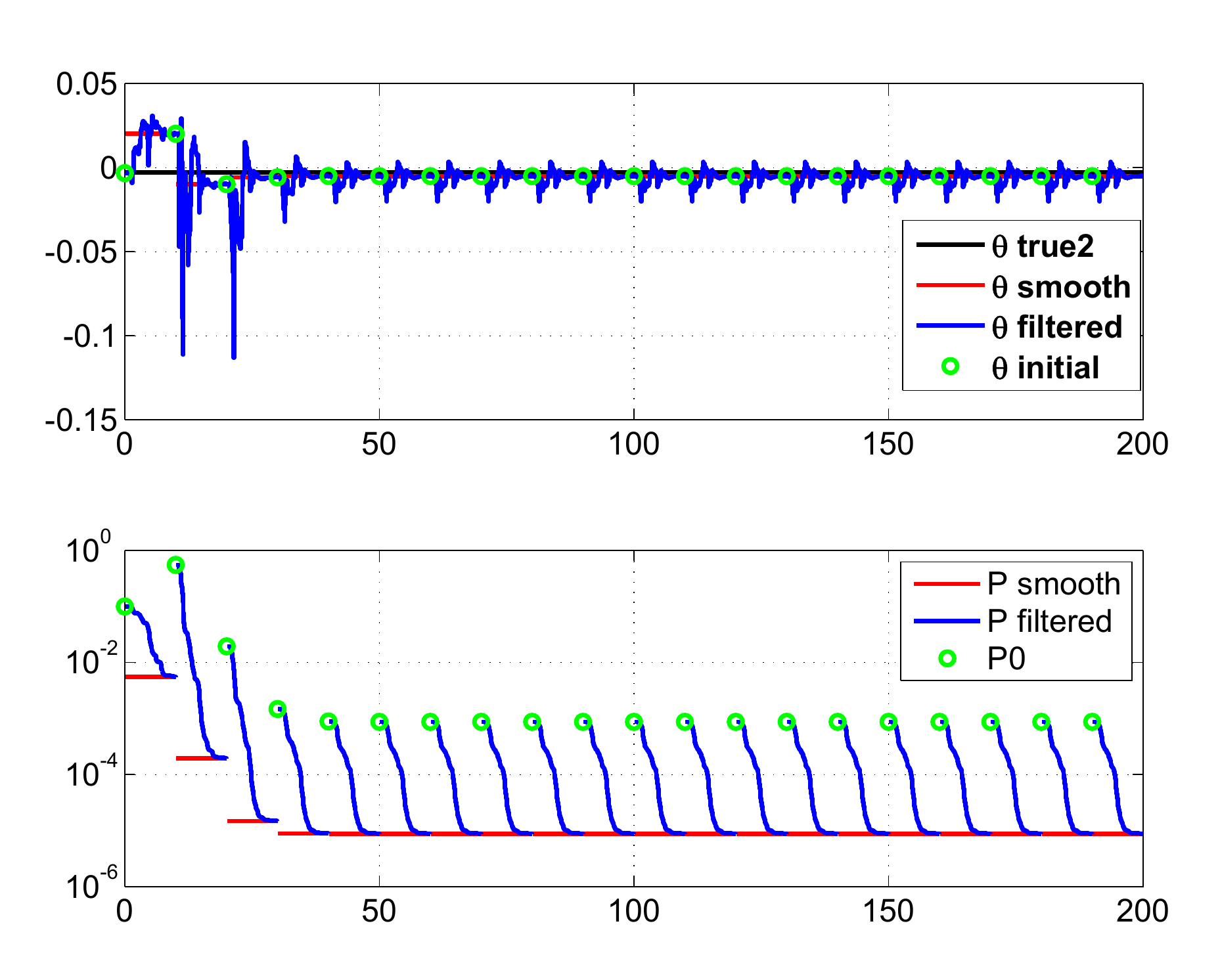}
\caption{The variation of parameter estimate 2 and their filtered and}
\caption*{smoothed covariances through (with the time cumulatively) the iterations}
\label{lat_p2}
\end{figure}

\begin{figure}[h]
\includegraphics[width=6in,height=4in]{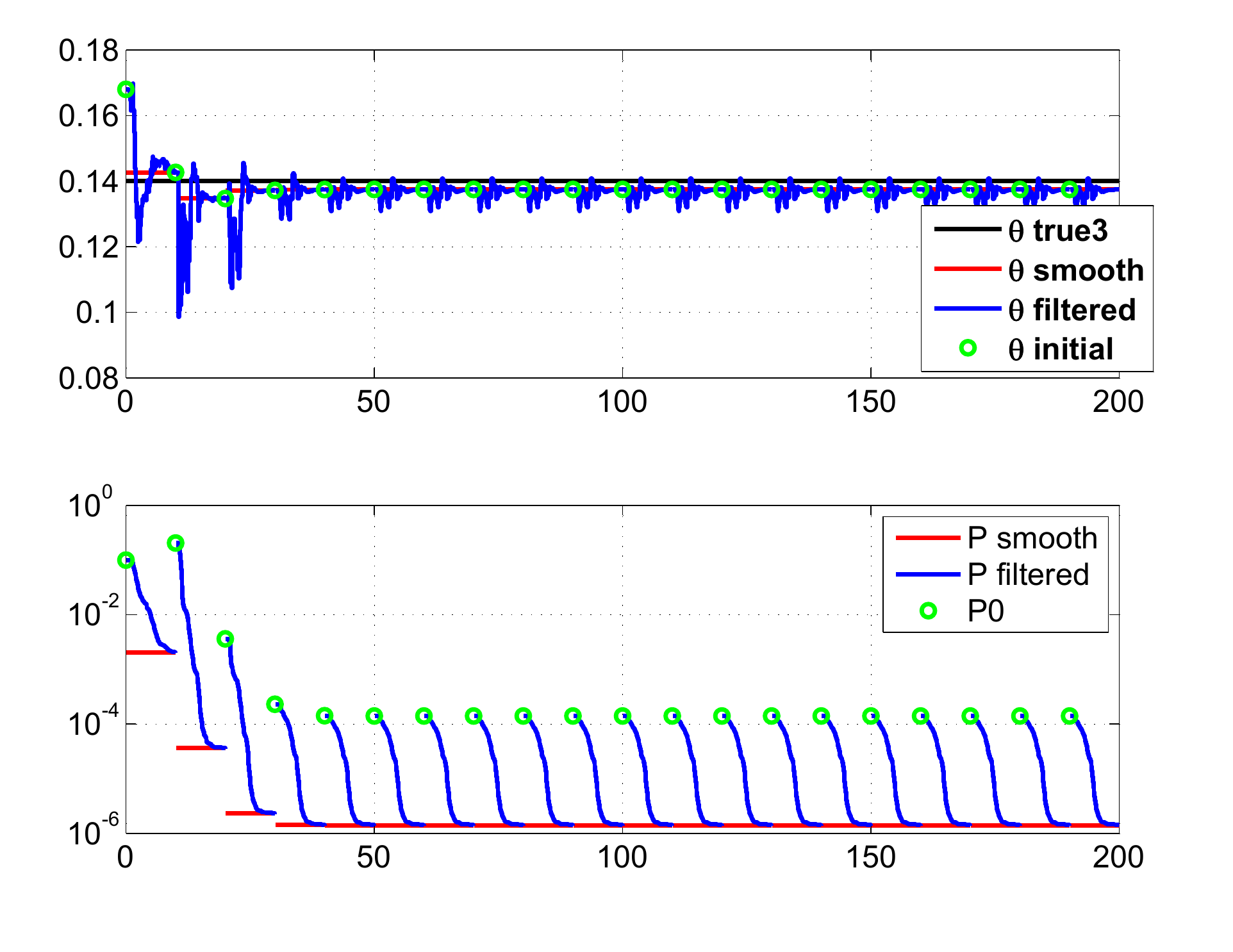}
\caption{The variation of parameter estimate 3 and their filtered and}
\caption*{smoothed covariances through (with the time cumulatively) the iterations}
\label{lat_p3}
\end{figure}

\begin{figure}[h]
\includegraphics[width=6in,height=4in]{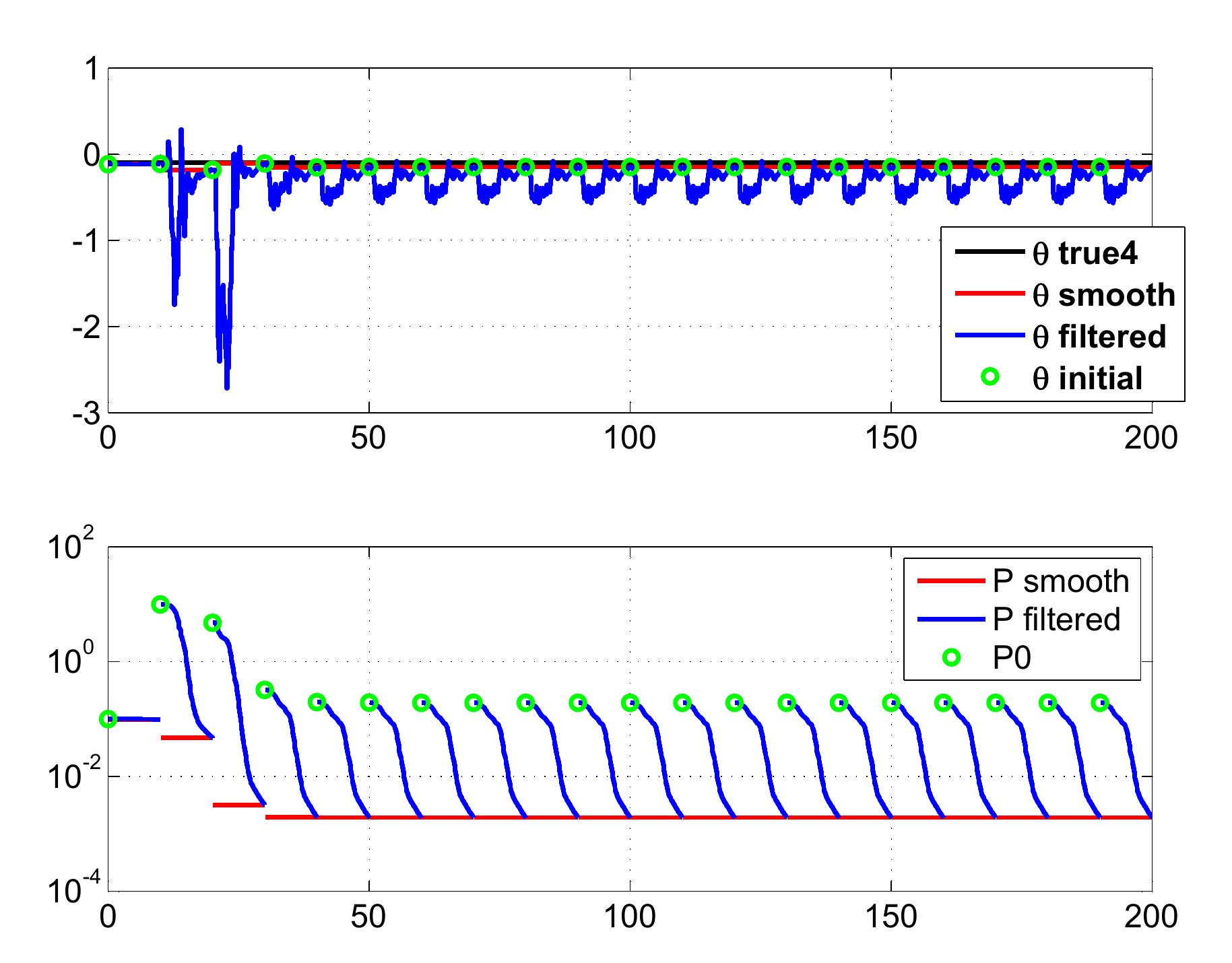}
\caption{The variation of parameter estimate 4 and their filtered and}
\caption*{smoothed covariances through (with the time cumulatively) the iterations}
\label{lat_p4}
\end{figure}

\begin{figure}[h]
\includegraphics[width=6in,height=4in]{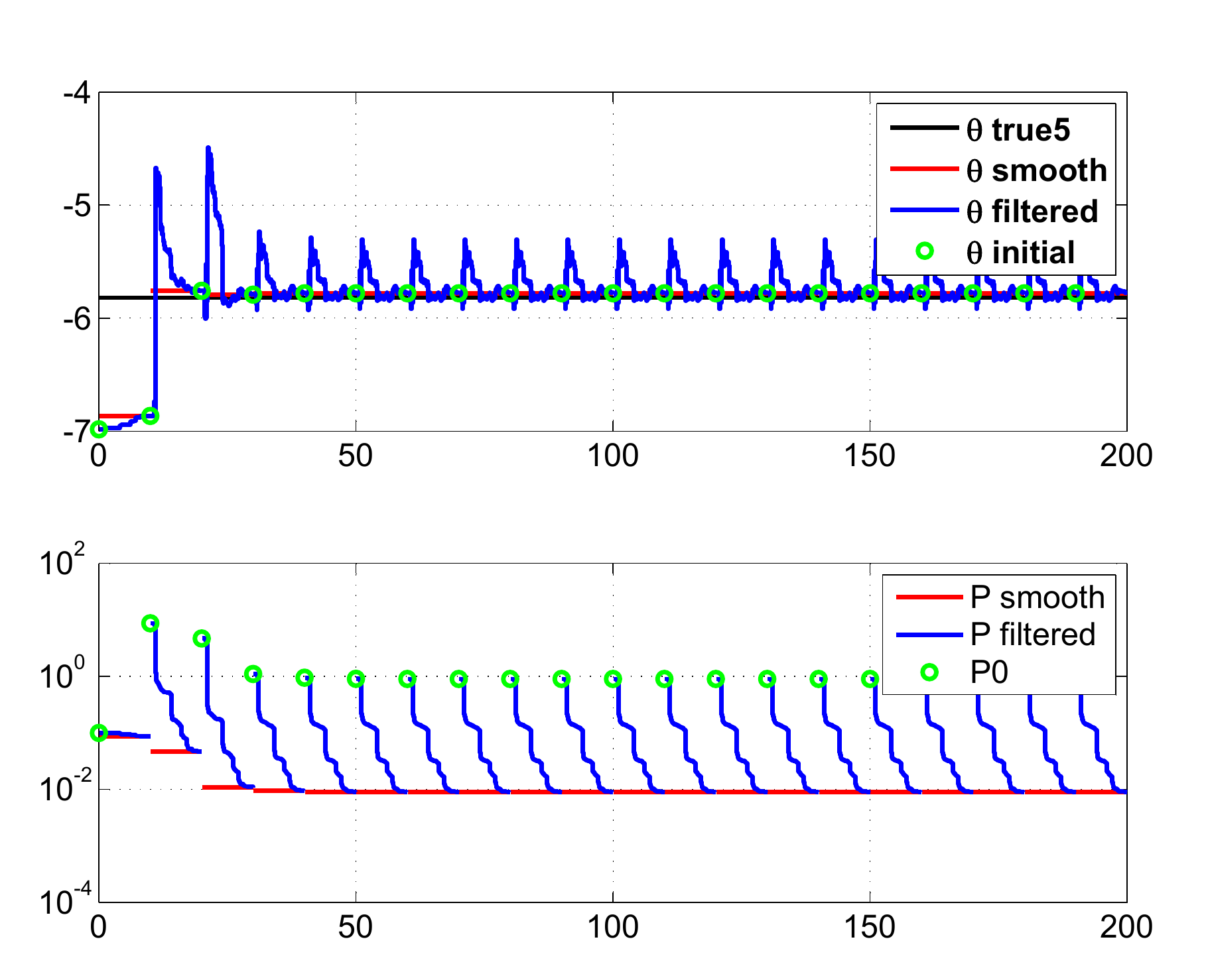}
\caption{The variation of parameter estimate 5 and their filtered and}
\caption*{smoothed covariances through (with the time cumulatively) the iterations}
\label{lat_p5}
\end{figure}

\begin{figure}[h]
\includegraphics[width=6in,height=4in]{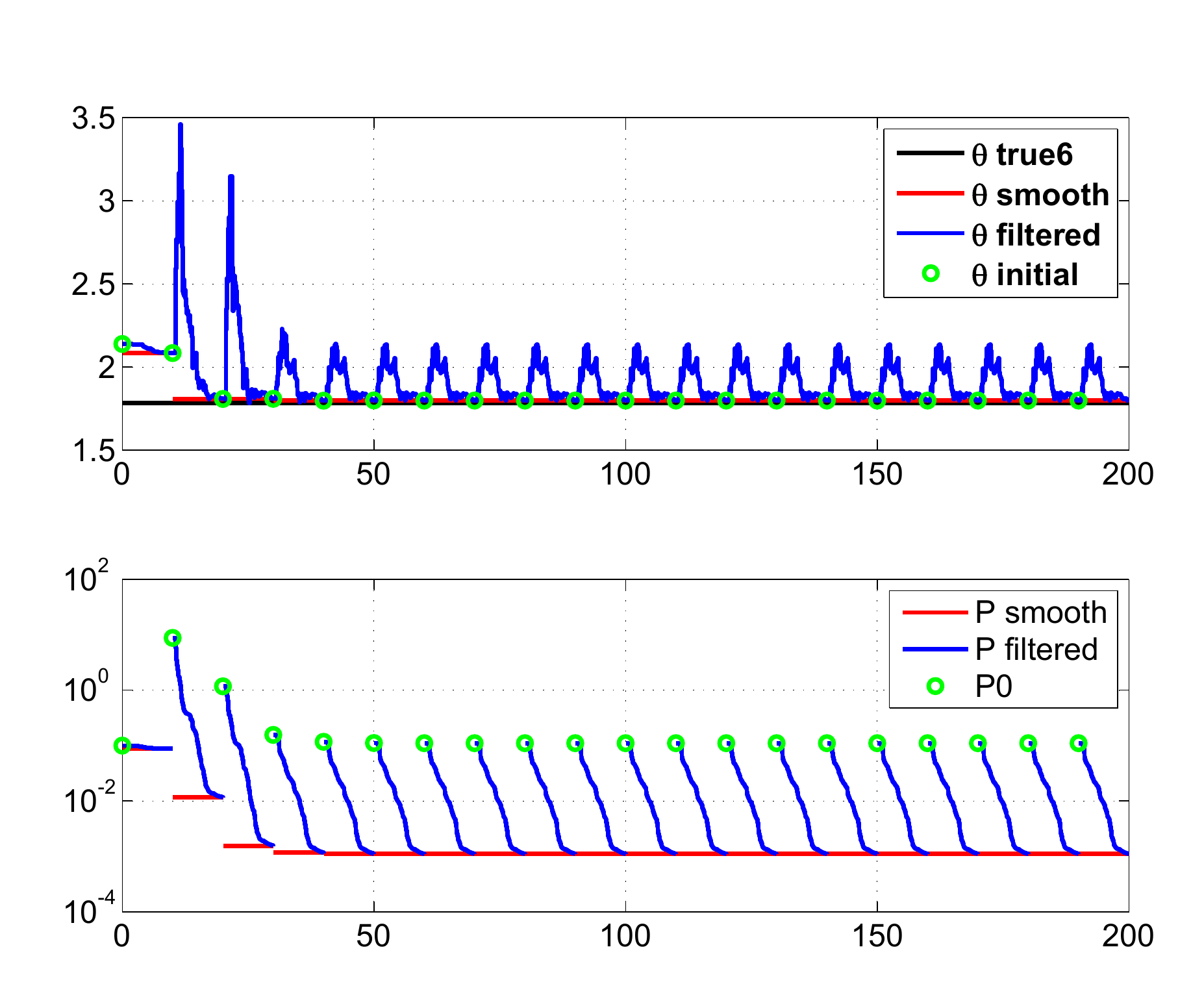}
\caption{The variation of parameter estimate 6 and their filtered and}
\caption*{smoothed covariances through (with the time cumulatively) the iterations}
\label{lat_p6}
\end{figure}

\begin{figure}[h]
\includegraphics[width=6in,height=4in]{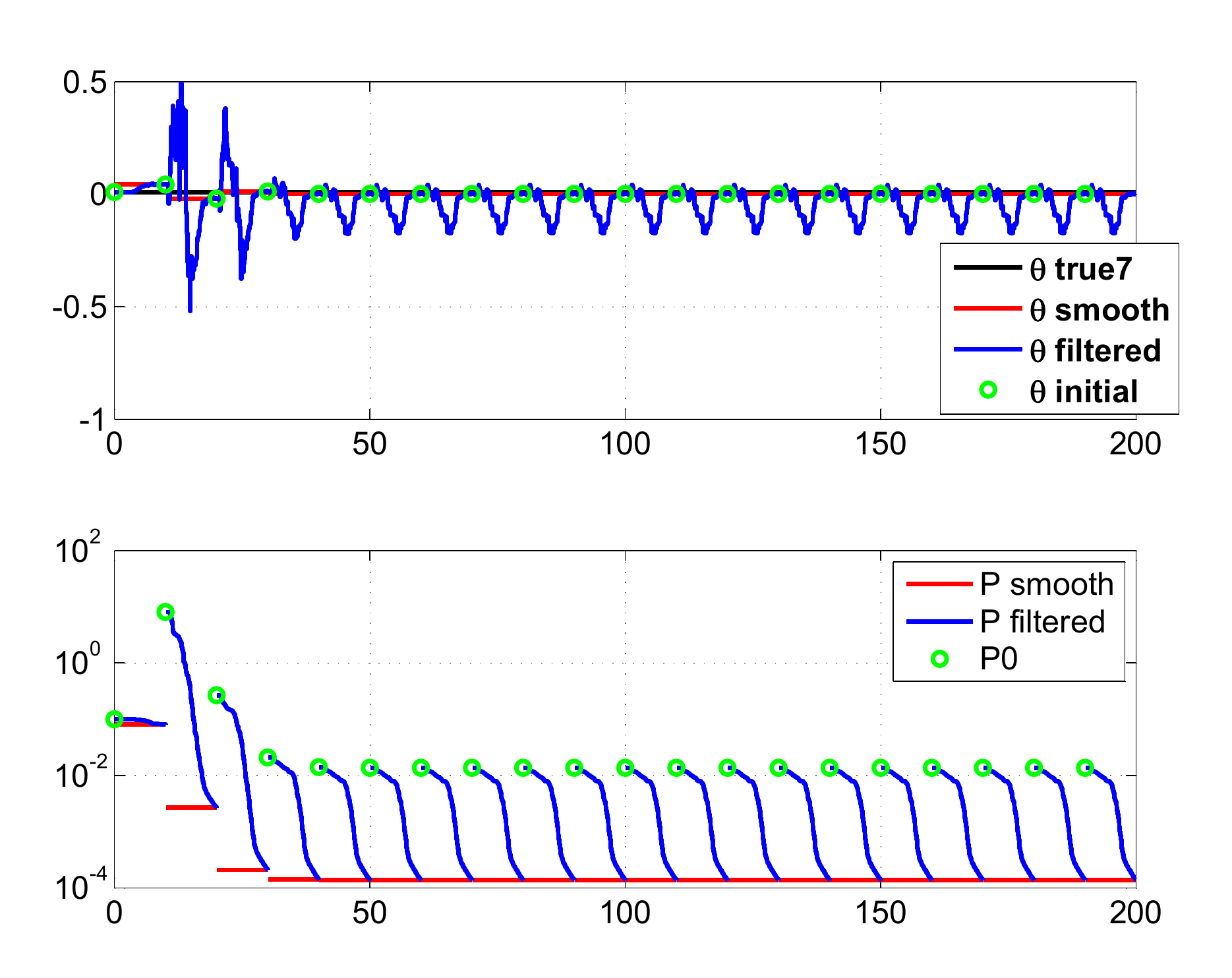}
\caption{The variation of parameter estimate 7 and their filtered and}
\caption*{smoothed covariances through (with the time cumulatively) the iterations}
\label{lat_p7}
\end{figure}

\begin{figure}[h]
\includegraphics[width=6in,height=4in]{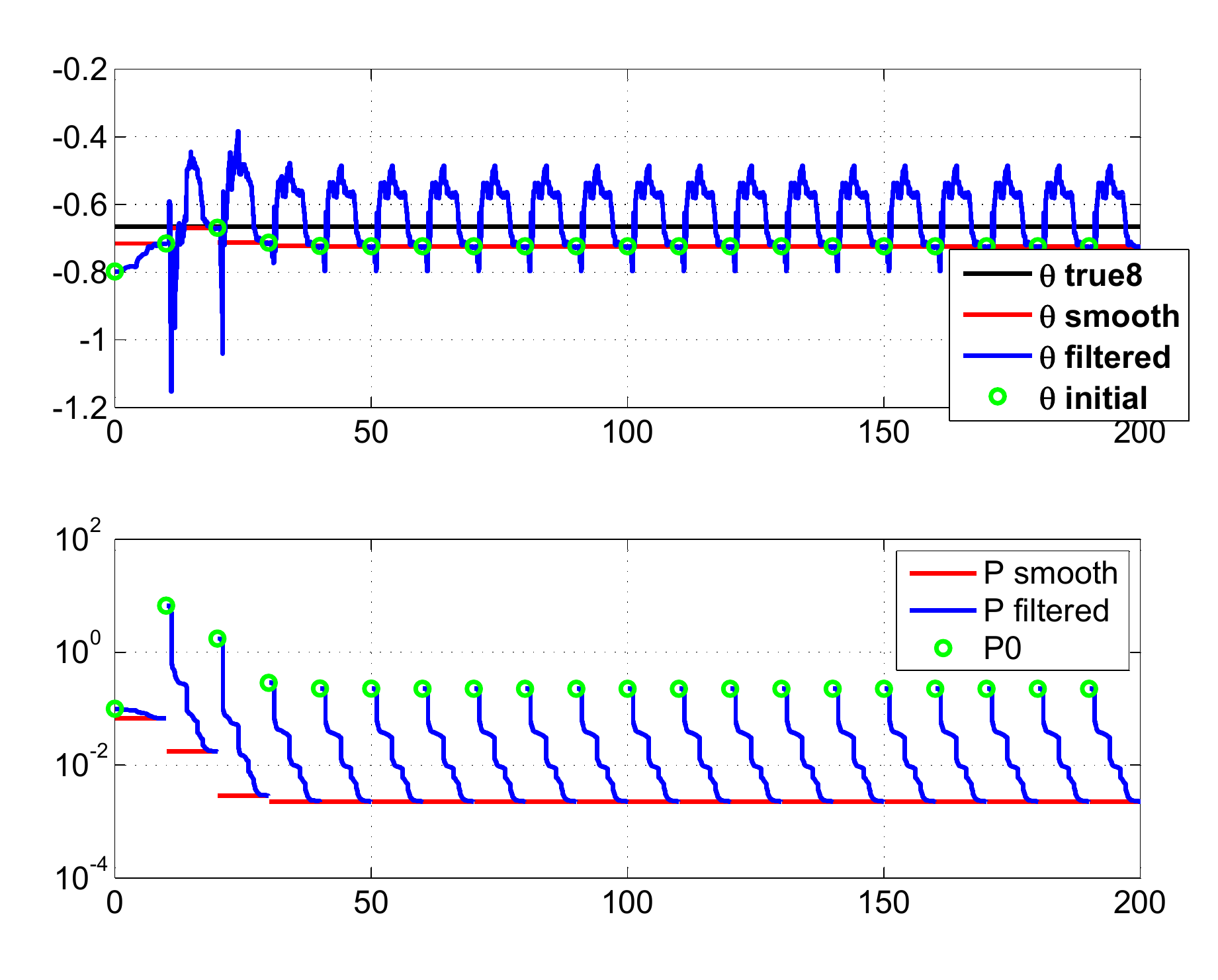}
\caption{The variation of parameter estimate 8 and their filtered and}
\caption*{smoothed covariances through (with the time cumulatively) the iterations}
\label{lat_p8}
\end{figure}

\begin{figure}[h]
\includegraphics[width=6in,height=4in]{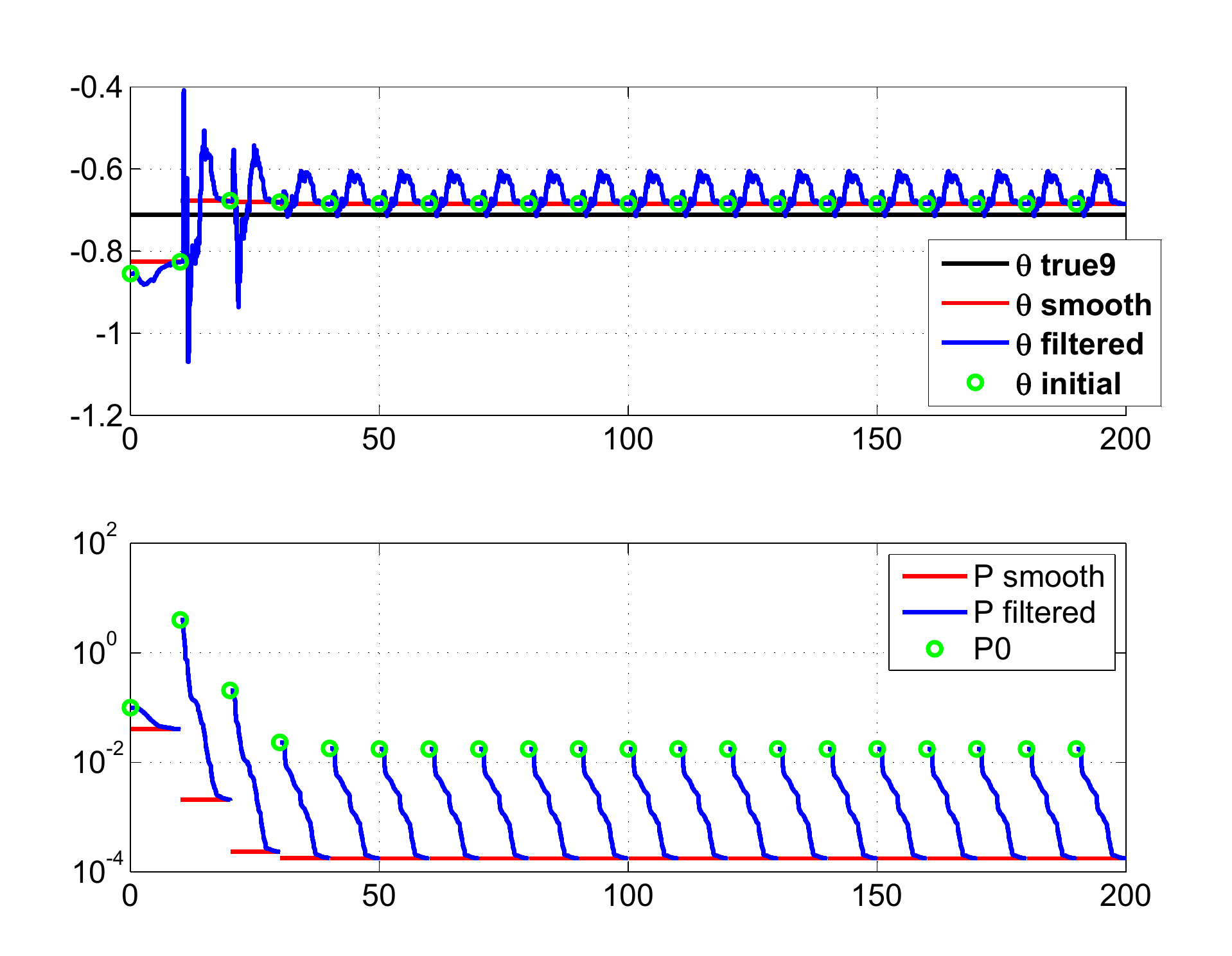}
\caption{The variation of parameter estimate 9 and their filtered and}
\caption*{smoothed covariances through (with the time cumulatively) the iterations}
\label{lat_p9}
\end{figure}

\begin{figure}[h]
\includegraphics[width=6in,height=4in]{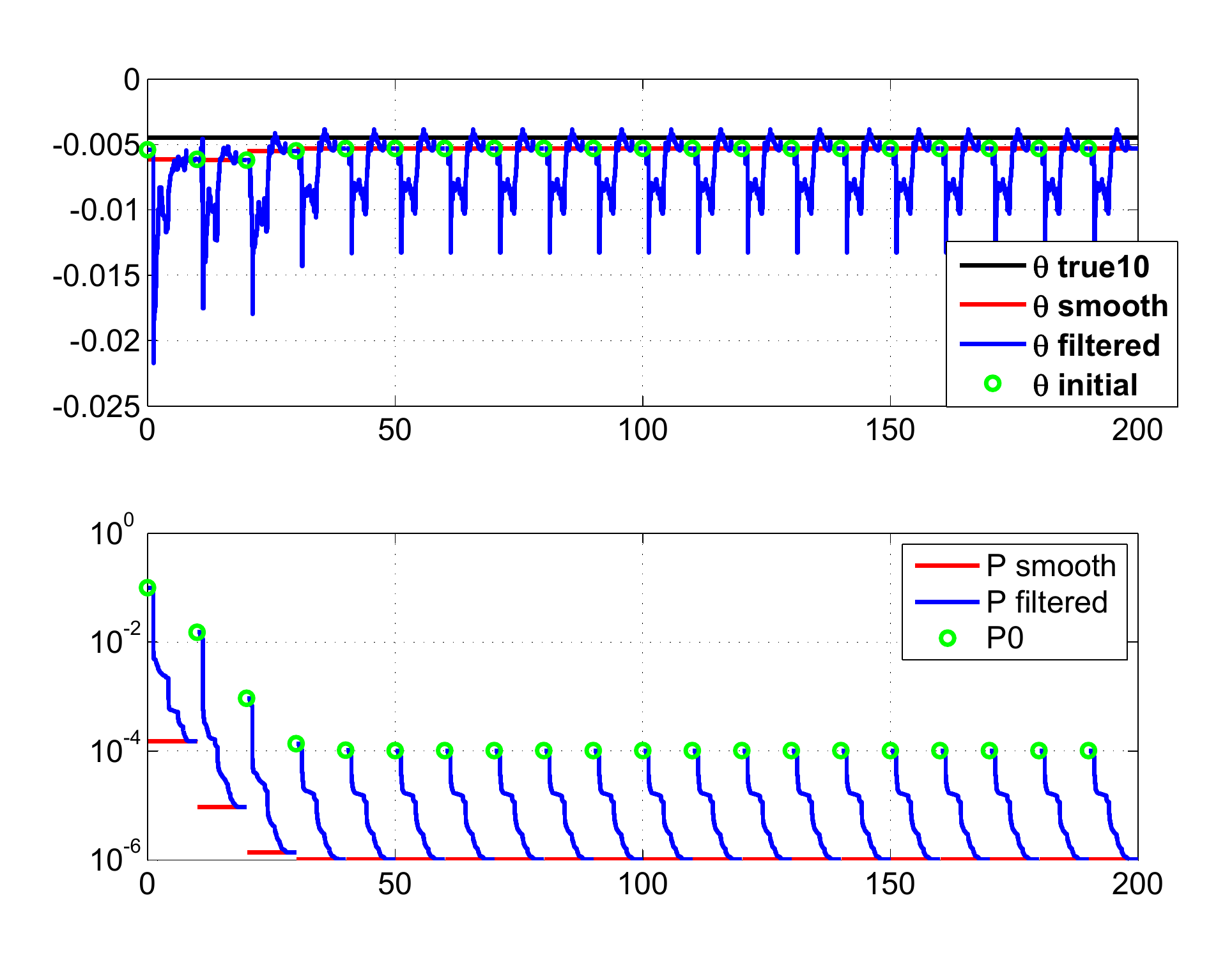}
\caption{The variation of parameter estimate 10 and their filtered and}
\caption*{smoothed covariances through (with the time cumulatively) the iterations}
\label{lat_p10}
\end{figure}

\begin{figure}[h]
\includegraphics[width=6in,height=4in]{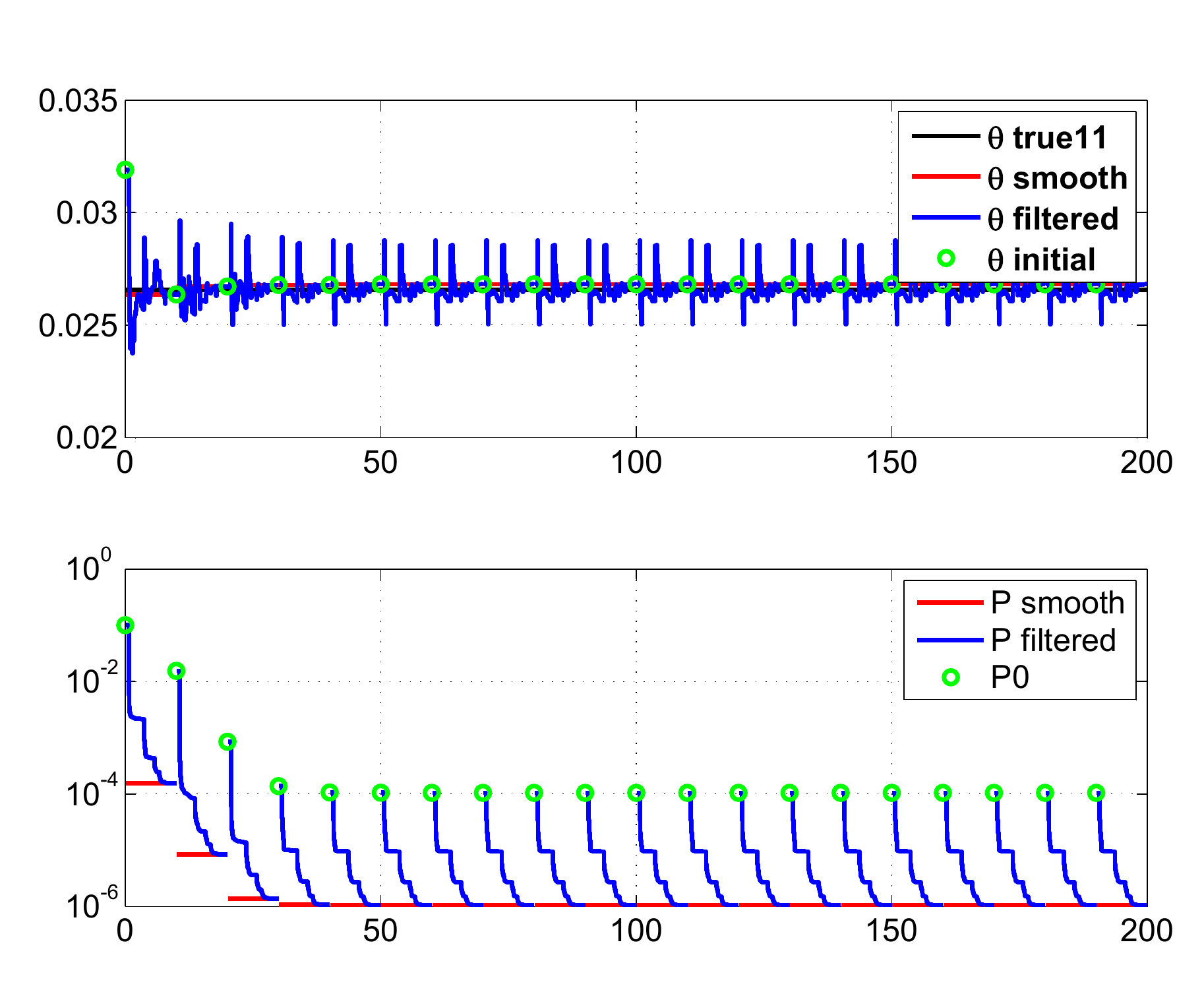}
\caption{The variation of parameter estimate 11 and their filtered and}
\caption*{smoothed covariances through (with the time cumulatively) the iterations}
\label{lat_p11}
\end{figure}

\begin{figure}[h]
\includegraphics[width=6in,height=4in]{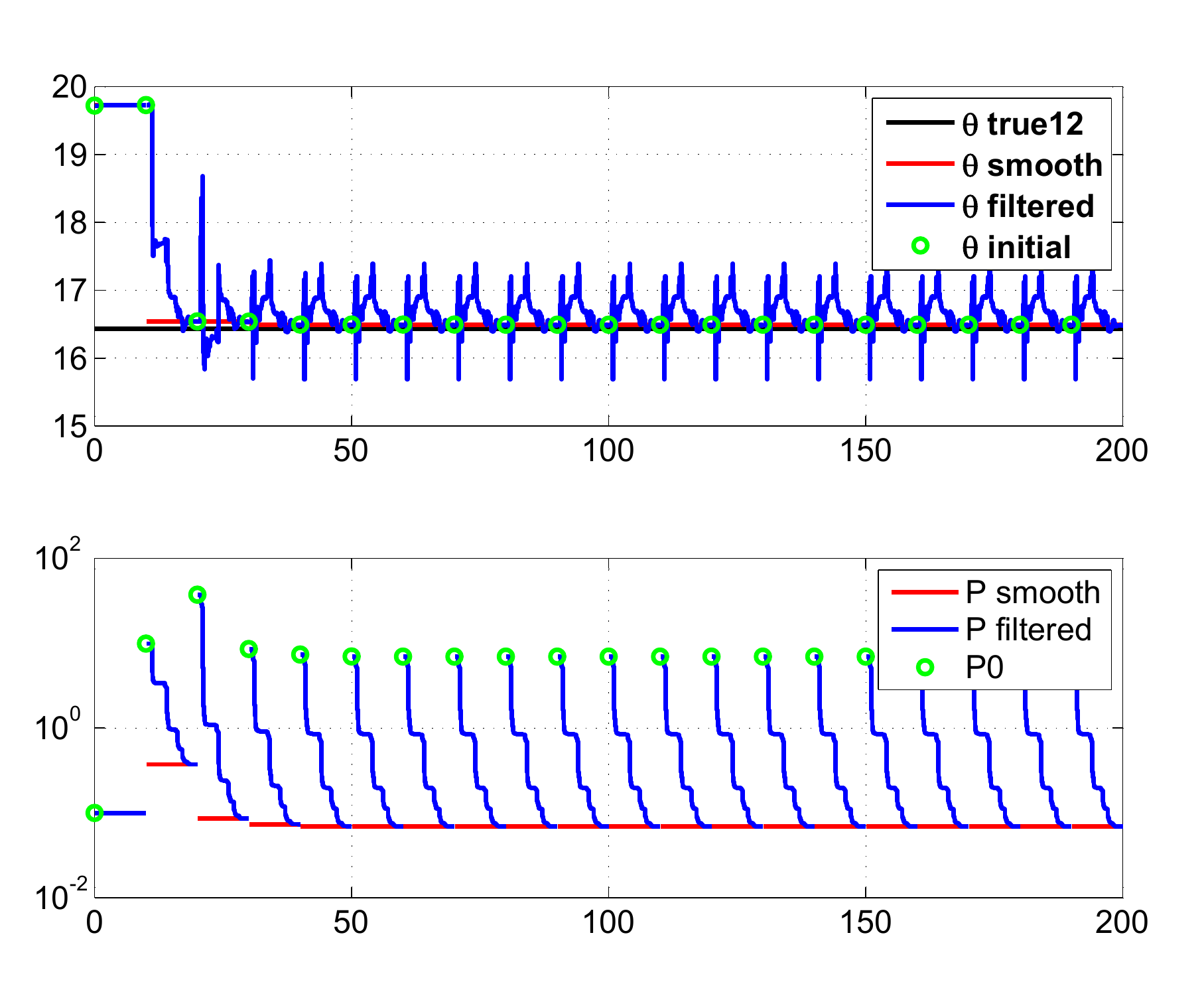}
\caption{The variation of parameter estimate 12 and their filtered and}
\caption*{smoothed covariances through (with the time cumulatively) the iterations}
\label{lat_p12}
\end{figure}

\begin{figure}[h]
\includegraphics[width=6in,height=4in]{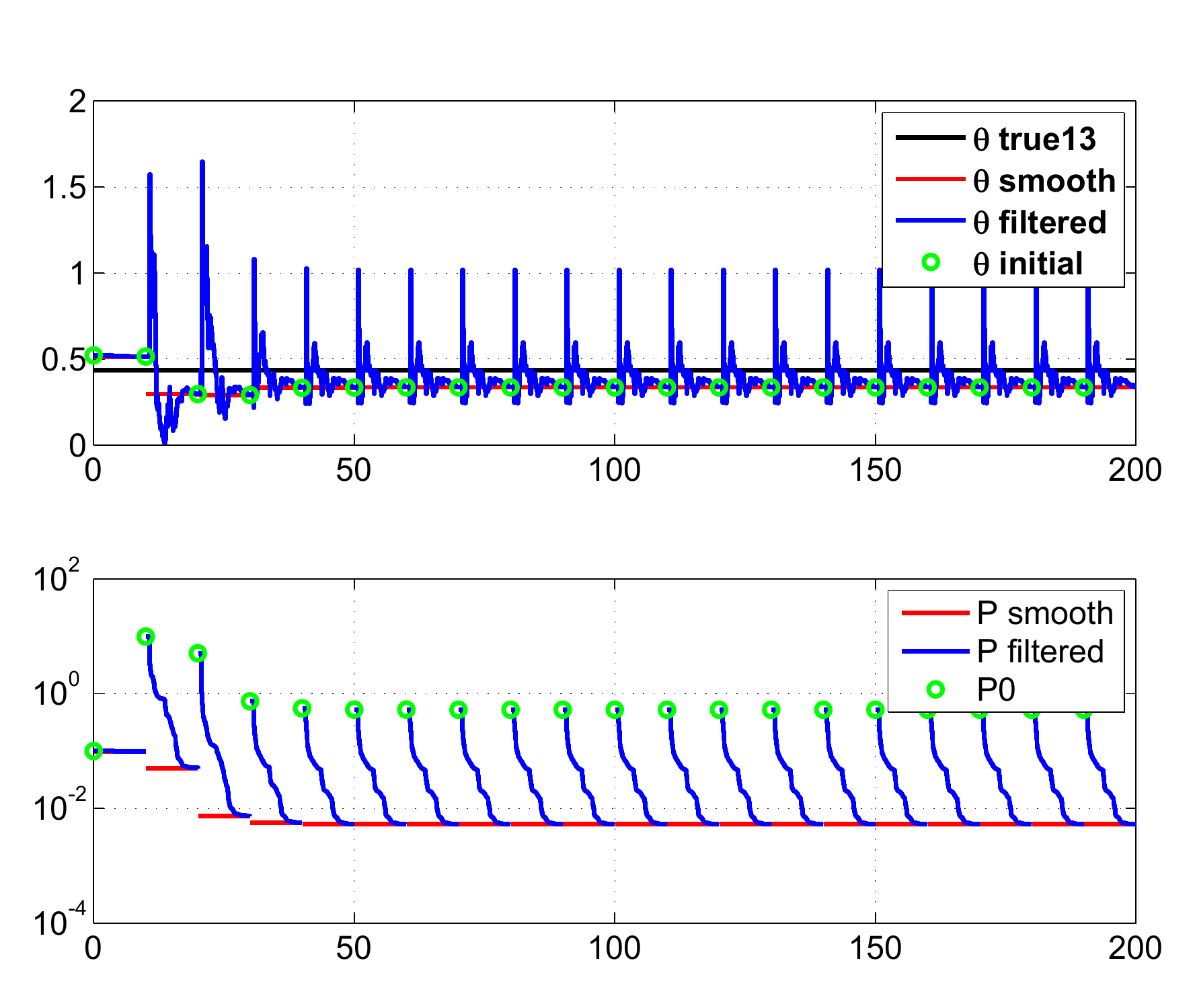}
\caption{The variation of parameter estimate 13 and their filtered and}
\caption*{smoothed covariances through (with the time cumulatively) the iterations}
\label{lat_p13}
\end{figure}

\begin{figure}[h]
\includegraphics[width=6in,height=4in]{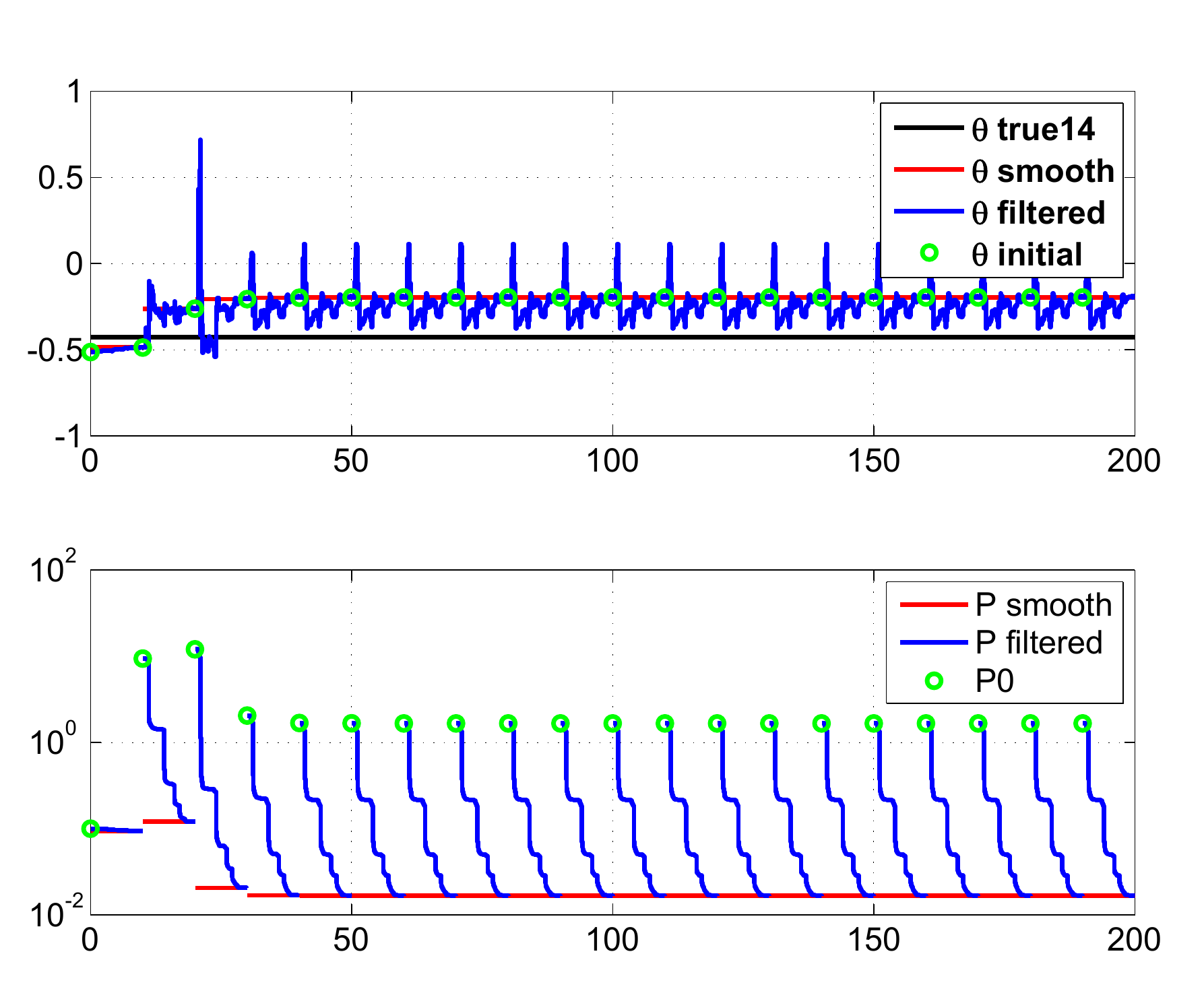}
\caption{The variation of parameter estimate 14 and their filtered and}
\caption*{smoothed covariances through (with the time cumulatively) the iterations}
\label{lat_p14}
\end{figure}

\begin{figure}[h]
\includegraphics[width=6in,height=4in]{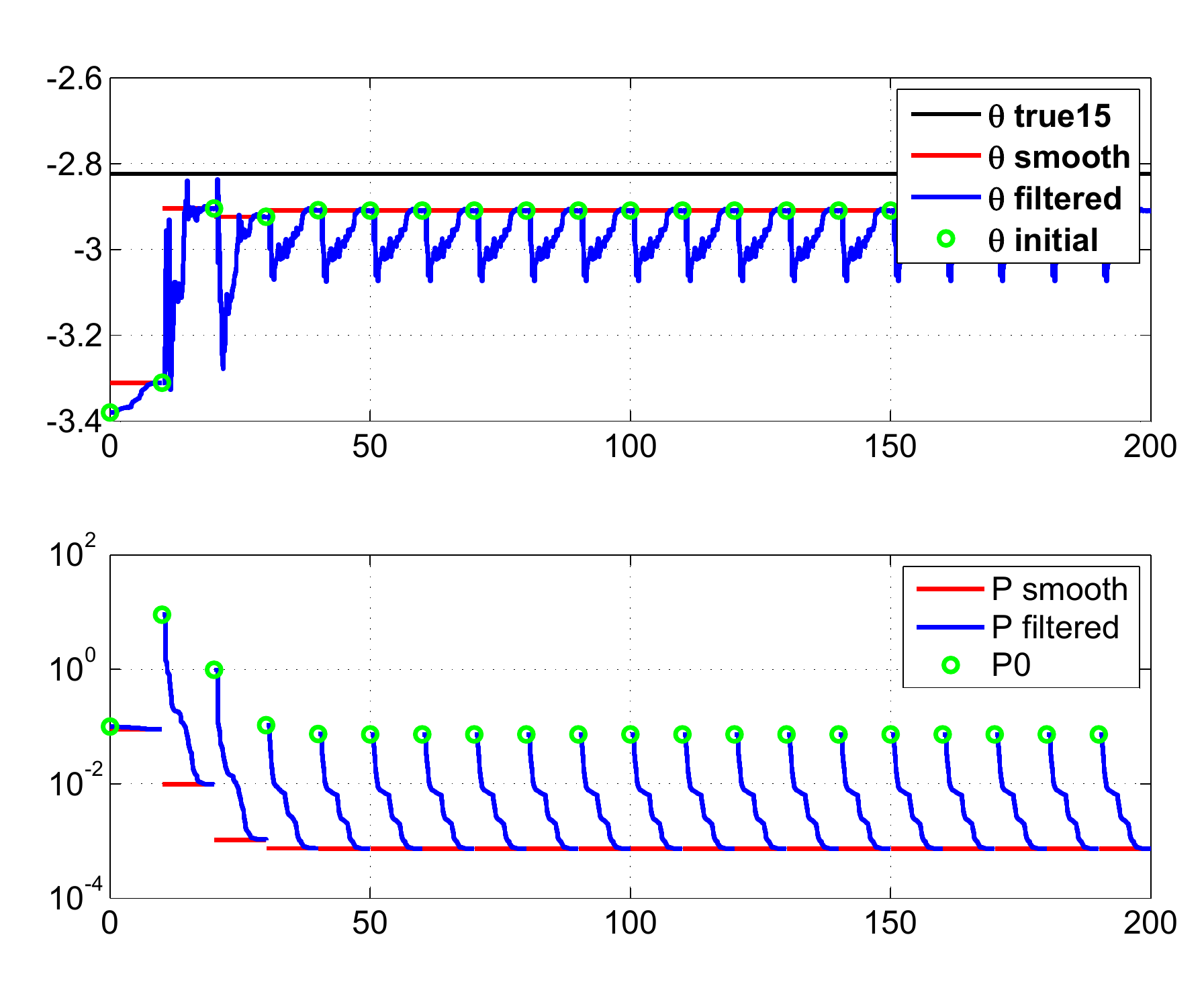}
\caption{The variation of parameter estimate 15 and their filtered and}
\caption*{smoothed covariances through (with the time cumulatively) the iterations}
\label{lat_p15}
\end{figure}

\begin{figure}[h]
\includegraphics[width=6in,height=4in]{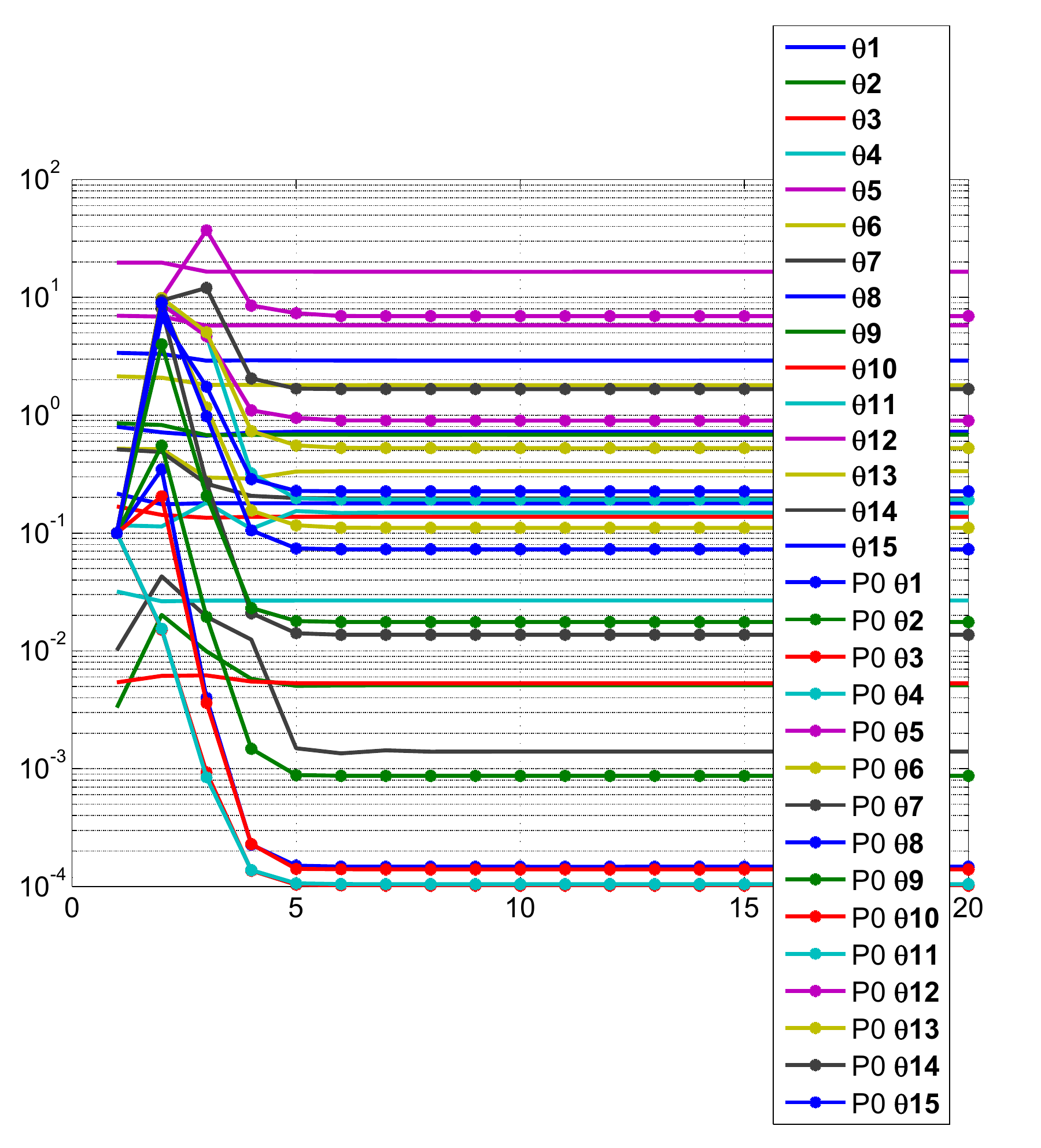}
\caption{Variation of parameter and its initial covariance ($\mathbf{P_0}$) with iterations}
\label{lat_P0}
\end{figure}

\begin{figure}[h]
\includegraphics[width=6in,height=4in]{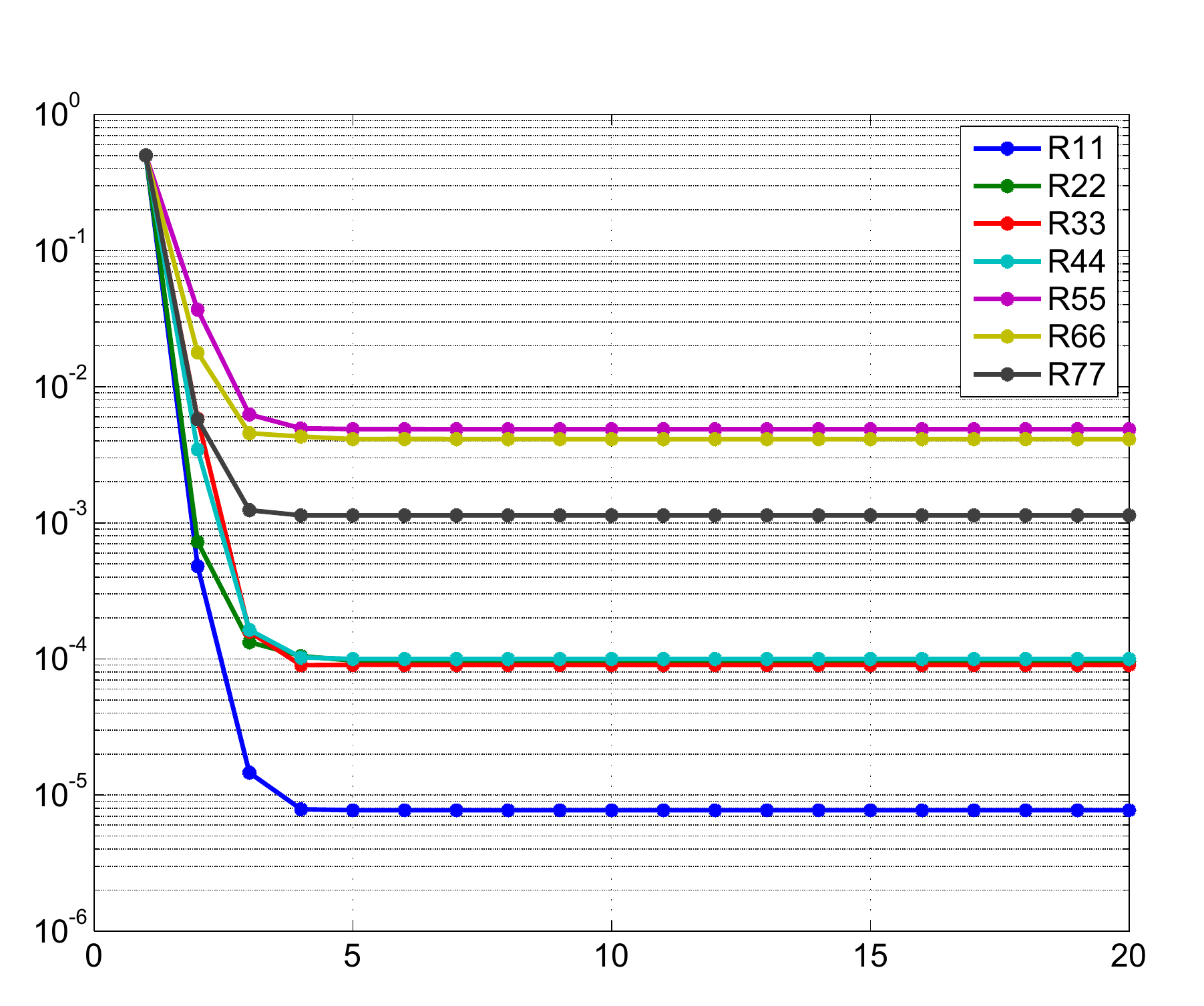}
\caption{Variation of \textbf{R} with iterations}
\label{lat_R}
\end{figure}

\begin{figure}[h]
\includegraphics[width=6in,height=4in]{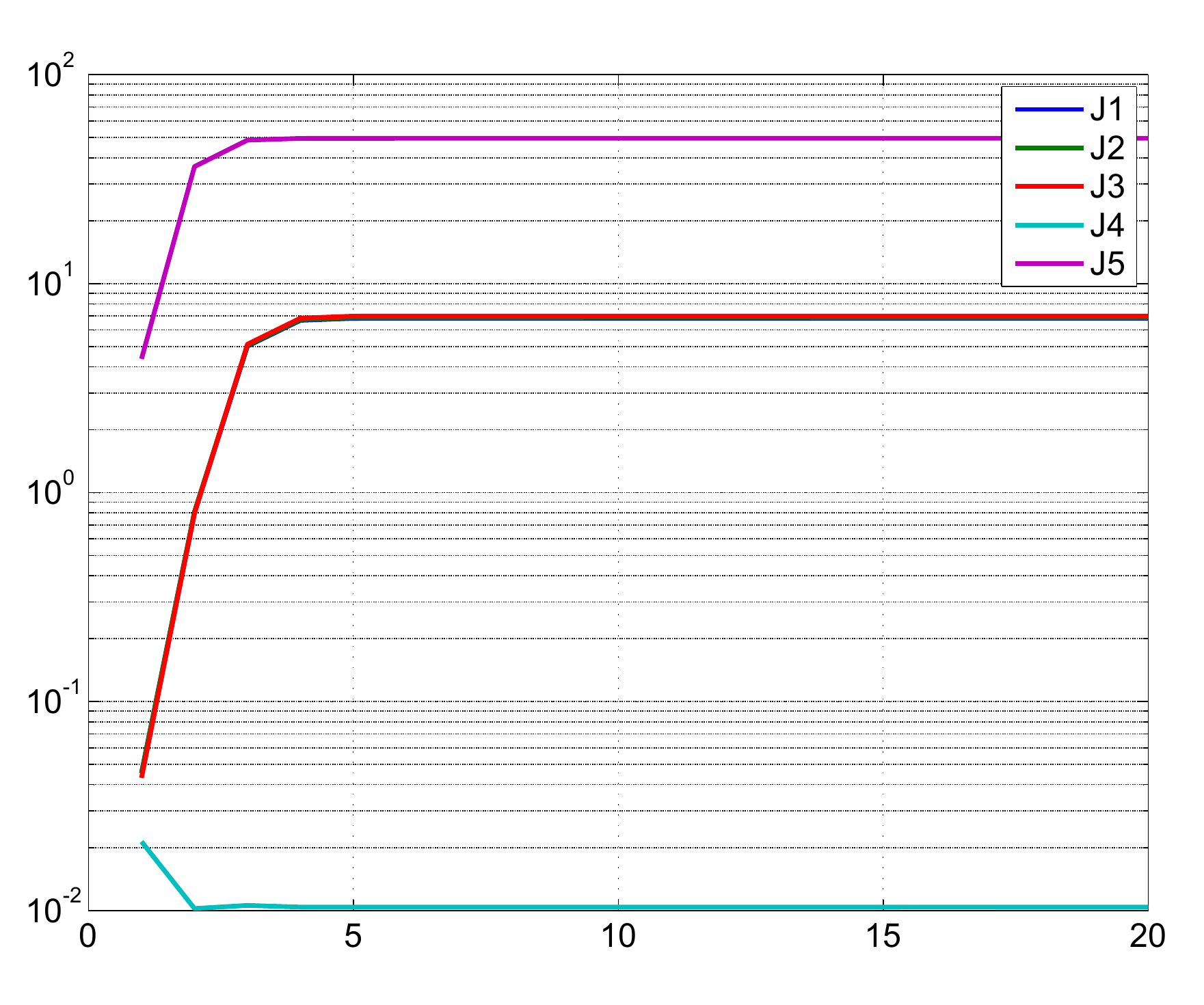}
\caption{Variation of different costs (\textbf{J1-J5}) with iterations}
\label{lat_cost}
\end{figure}

\begin{figure}[h]
\includegraphics[width=6in,height=4in]{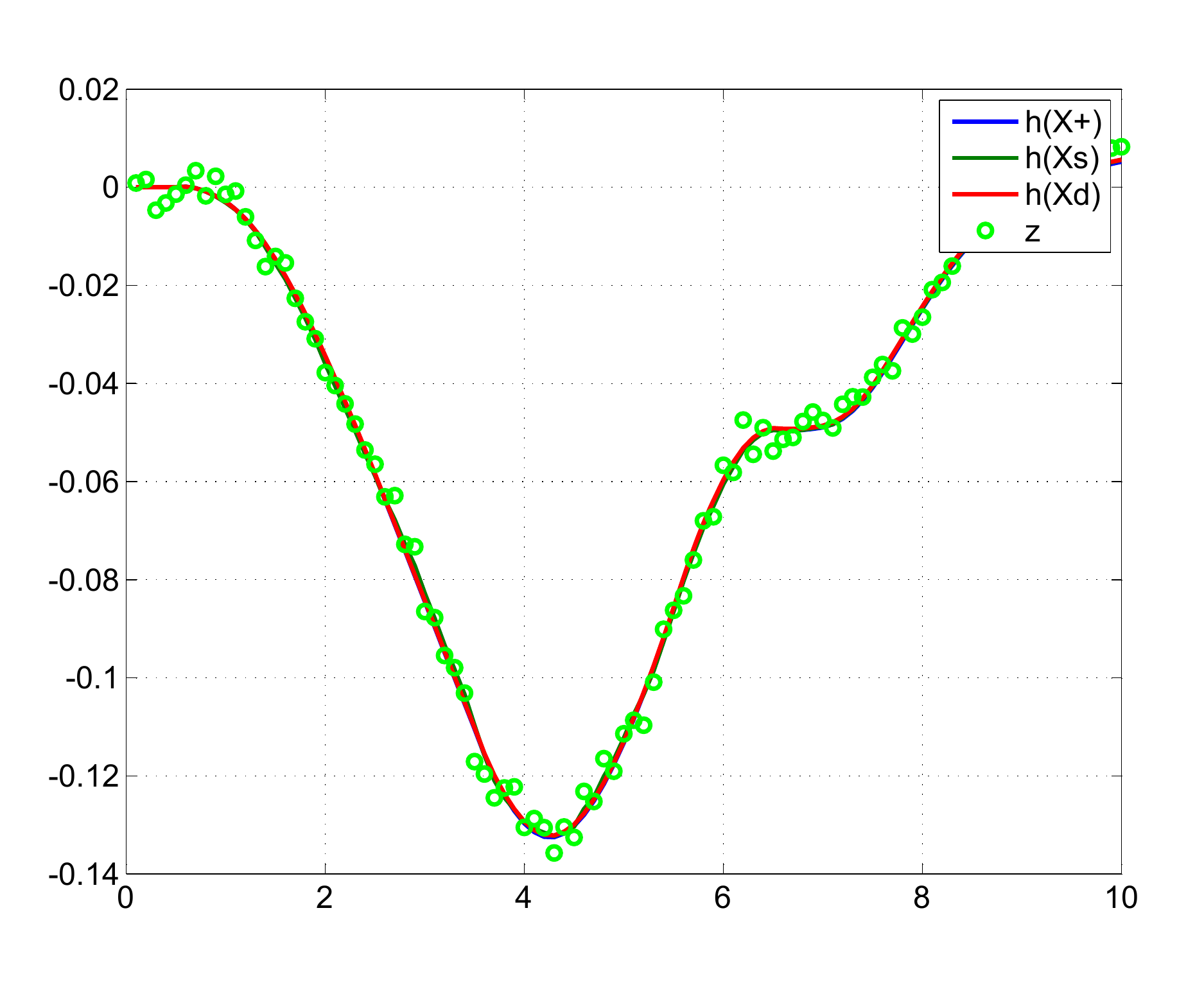}
\caption{Comparison of the predicted dynamics, posterior, smoothed}
\caption*{and the measurement 1 }
\label{lat_h1}
\end{figure}

\begin{figure}[h]
\includegraphics[width=6in,height=4in]{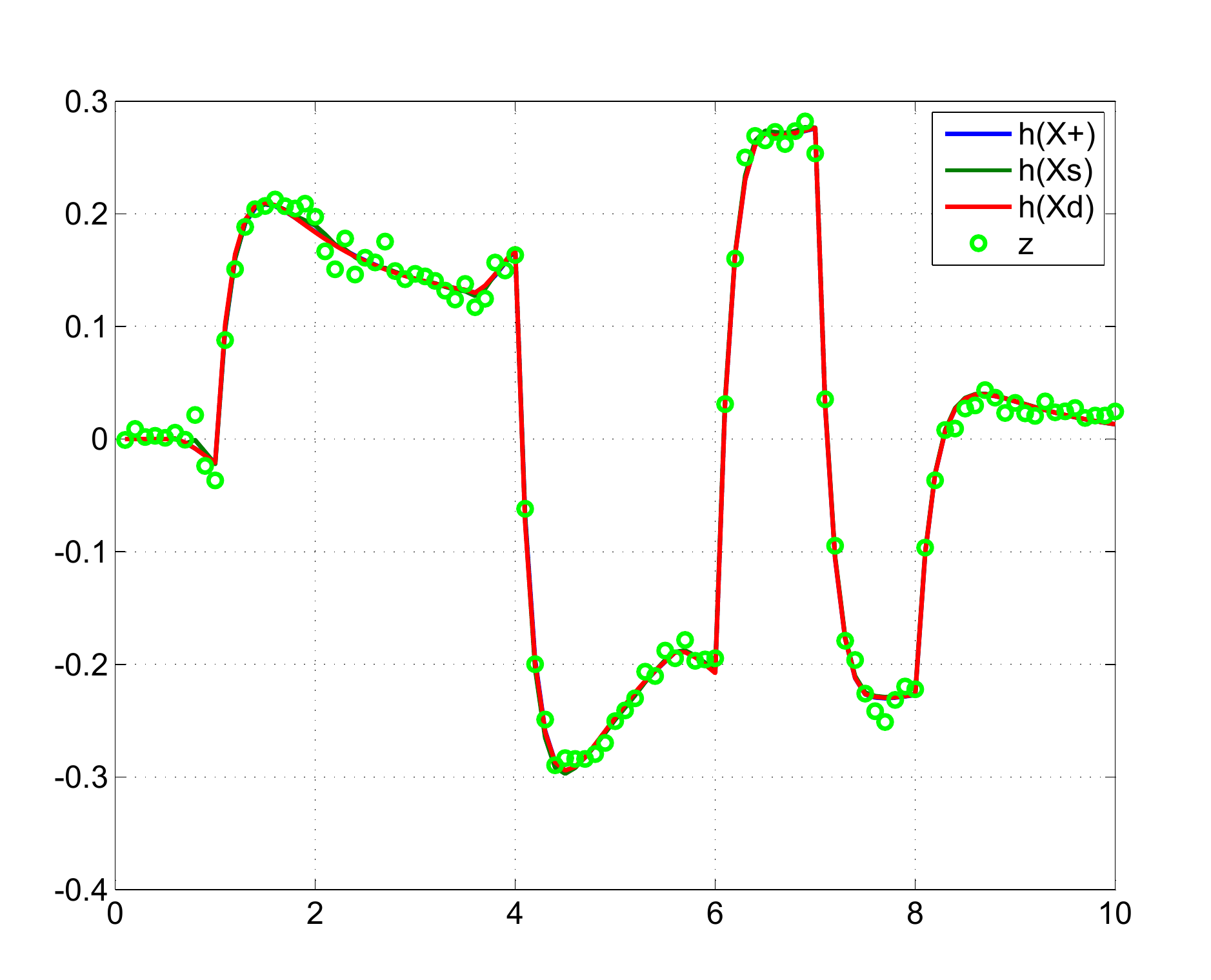}
\caption{Comparison of the predicted dynamics, posterior, smoothed}
\caption*{and the measurement 2}
\label{lat_h2}
\end{figure}

\begin{figure}[h]
\includegraphics[width=6in,height=4in]{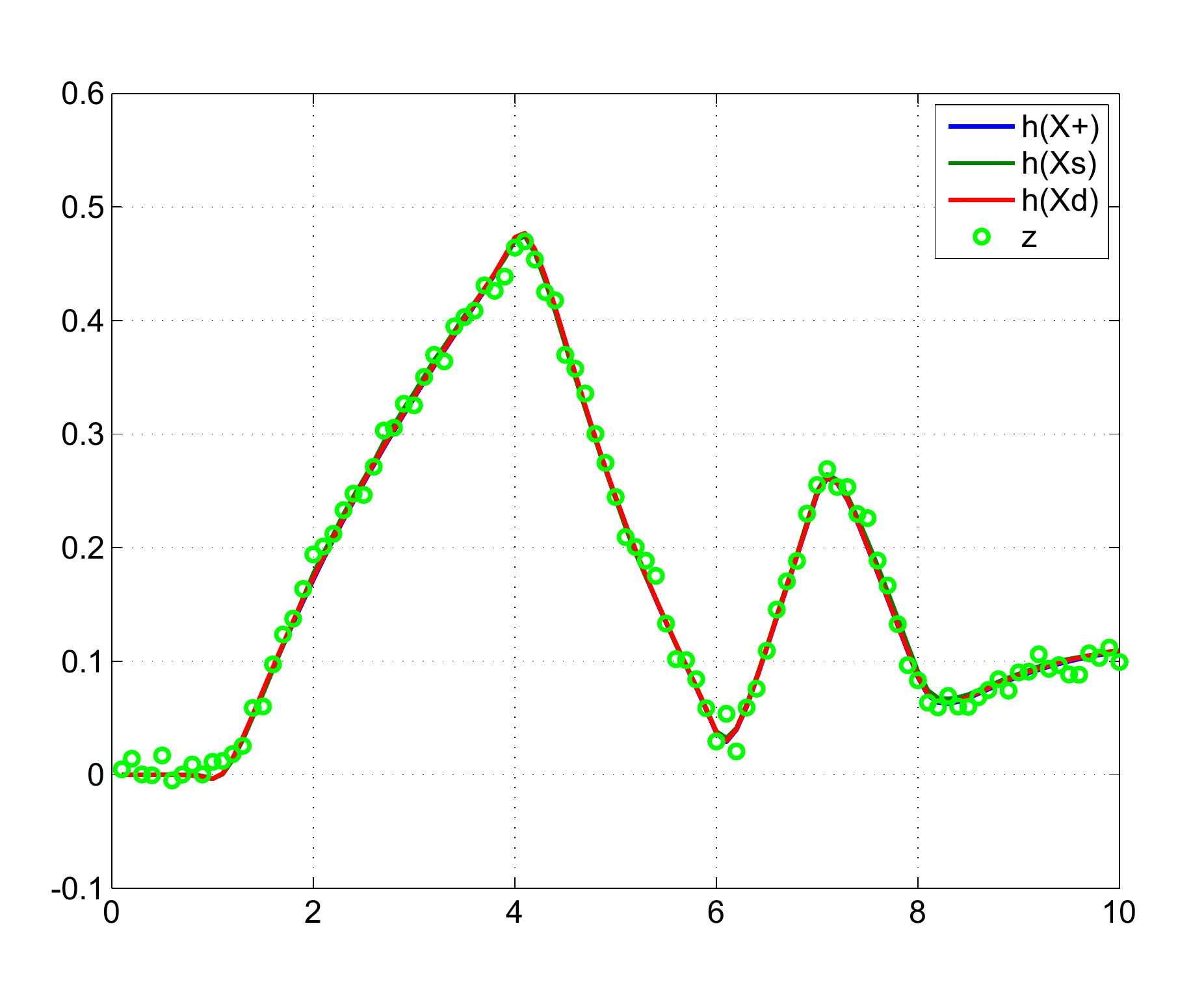}
\caption{Comparison of the predicted dynamics, posterior, smoothed}
\caption*{and the measurement 3 }
\label{lat_h3}
\end{figure}

\begin{figure}[h]
\includegraphics[width=6in,height=4in]{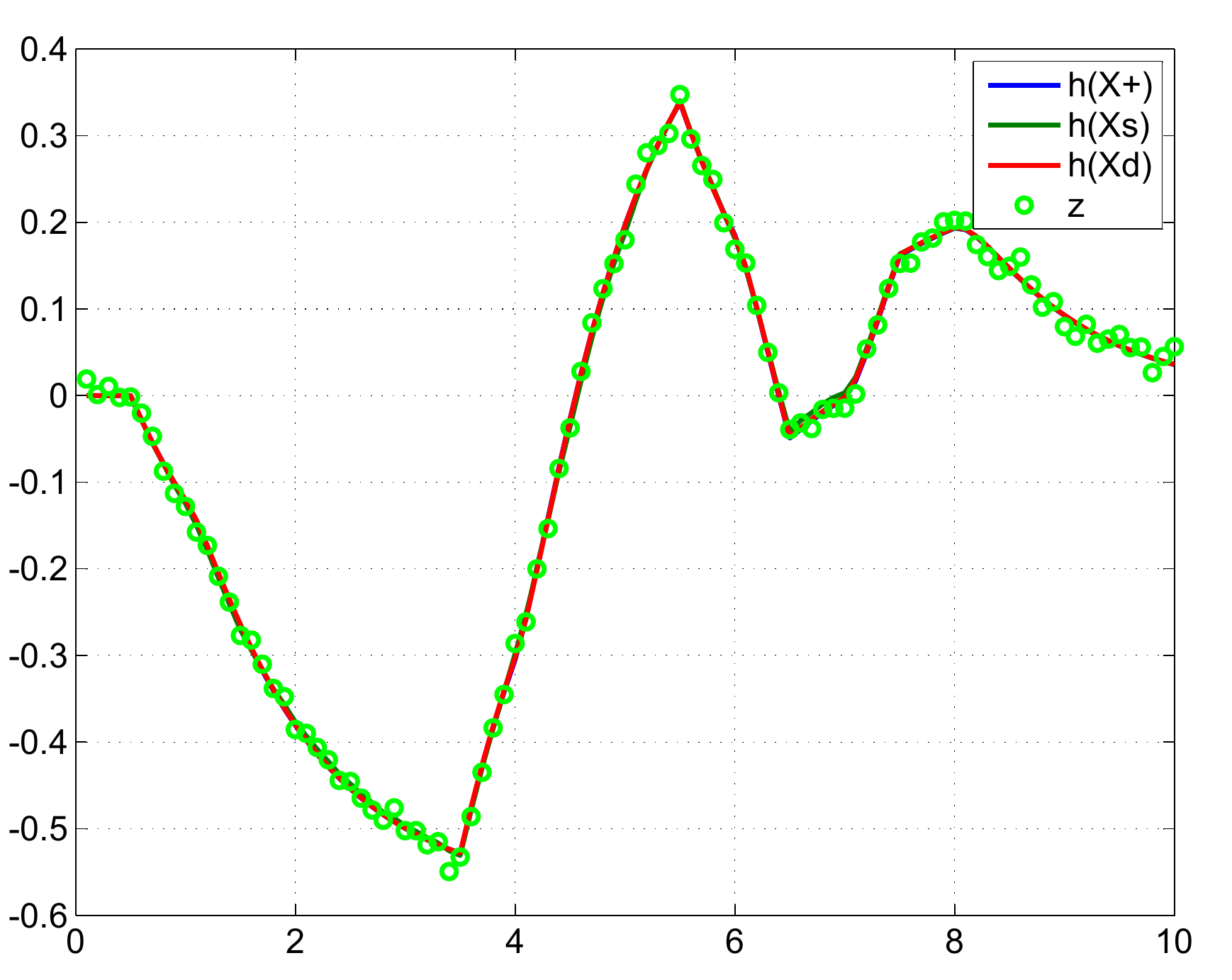}
\caption{Comparison of the predicted dynamics, posterior, smoothed}
\caption*{and the measurement 4}
\label{lat_h4}
\end{figure}

\begin{figure}[h]
\includegraphics[width=6in,height=4in]{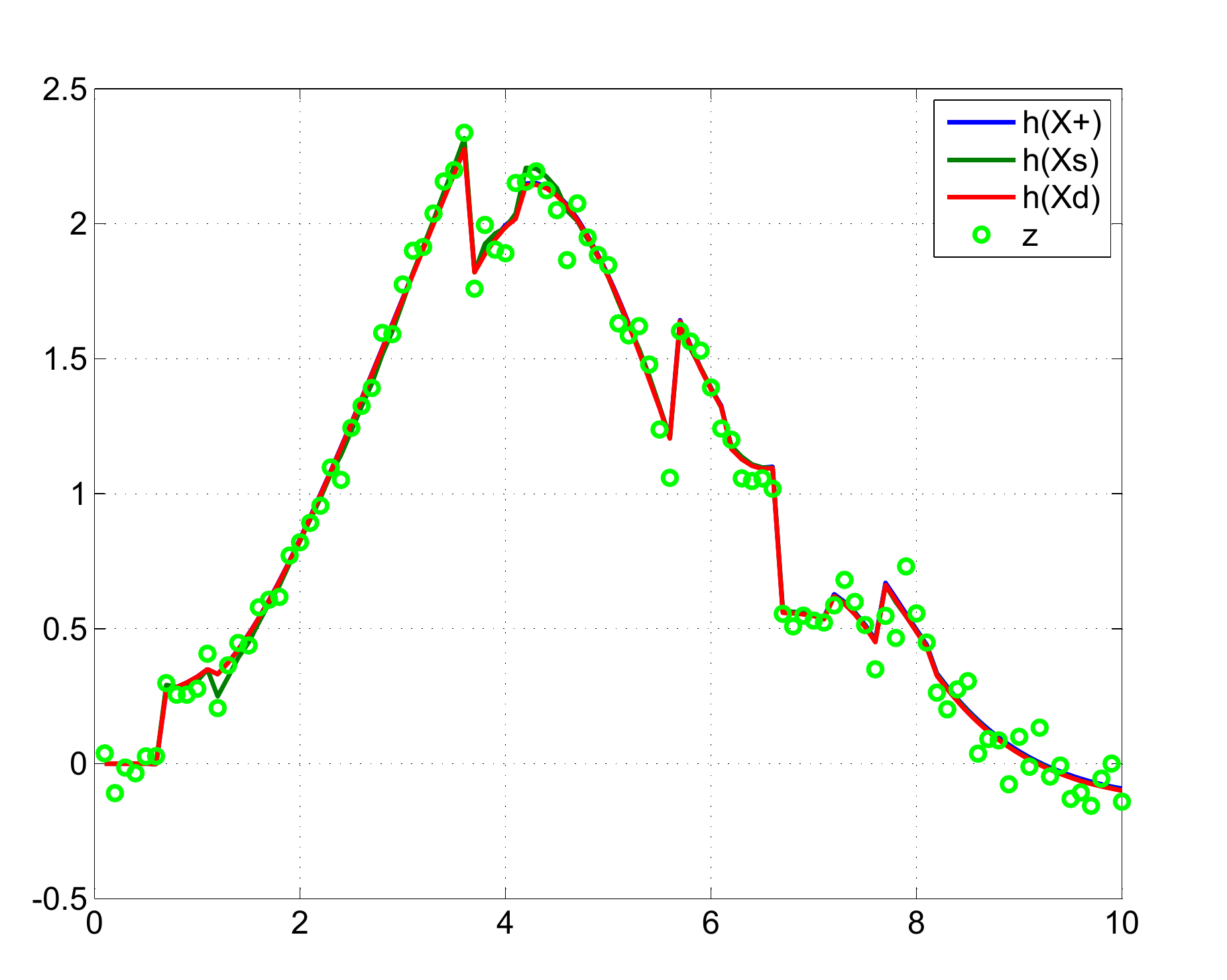}
\caption{Comparison of the predicted dynamics, posterior, smoothed}
\caption*{and the measurement 5}
\label{lat_h5}
\end{figure}

\begin{figure}[h]
\includegraphics[width=6in,height=4in]{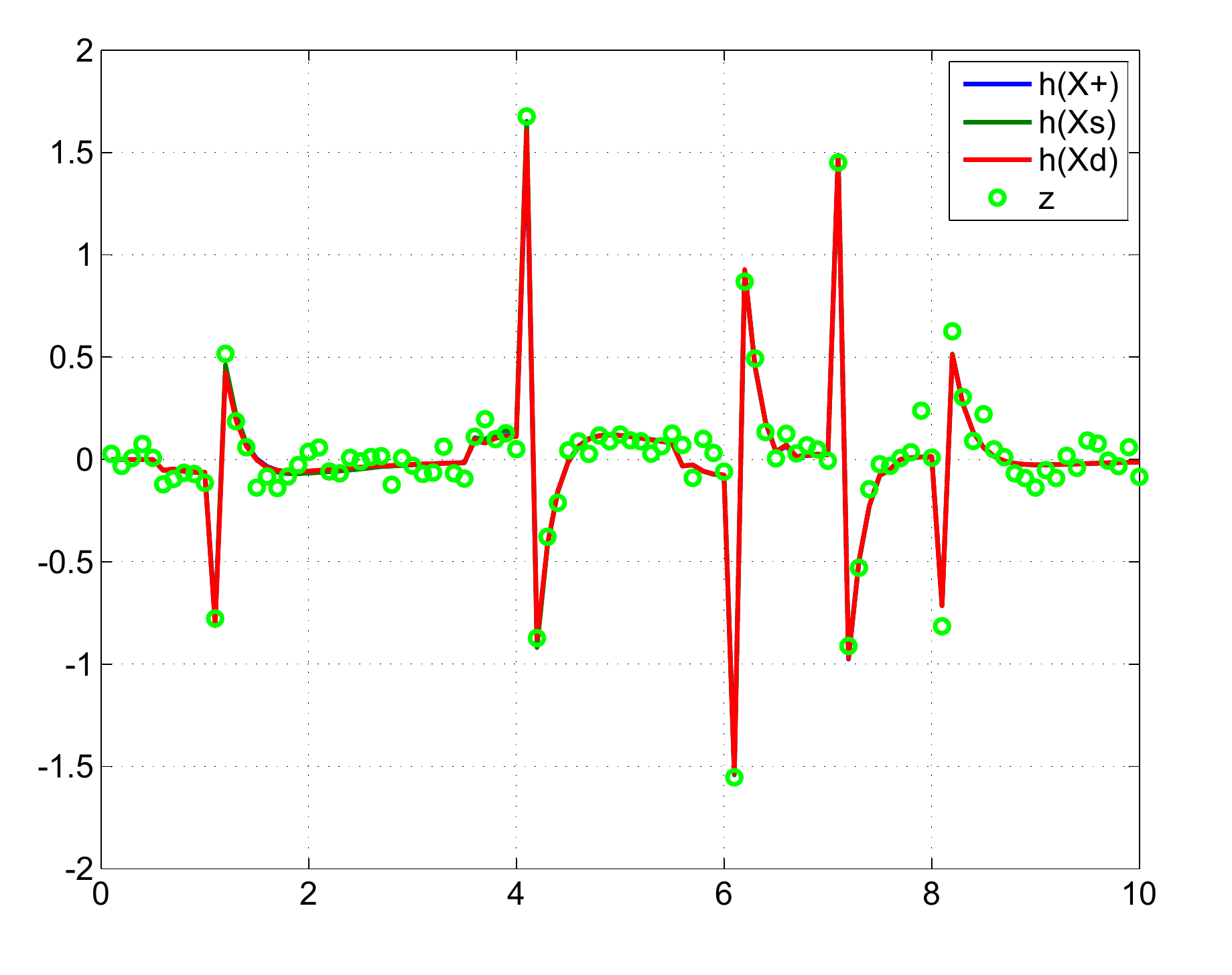}
\caption{Comparison of the predicted dynamics, posterior, smoothed}
\caption*{and the measurement 6}
\label{lat_h6}
\end{figure}

\begin{figure}[h]
\includegraphics[width=6in,height=4in]{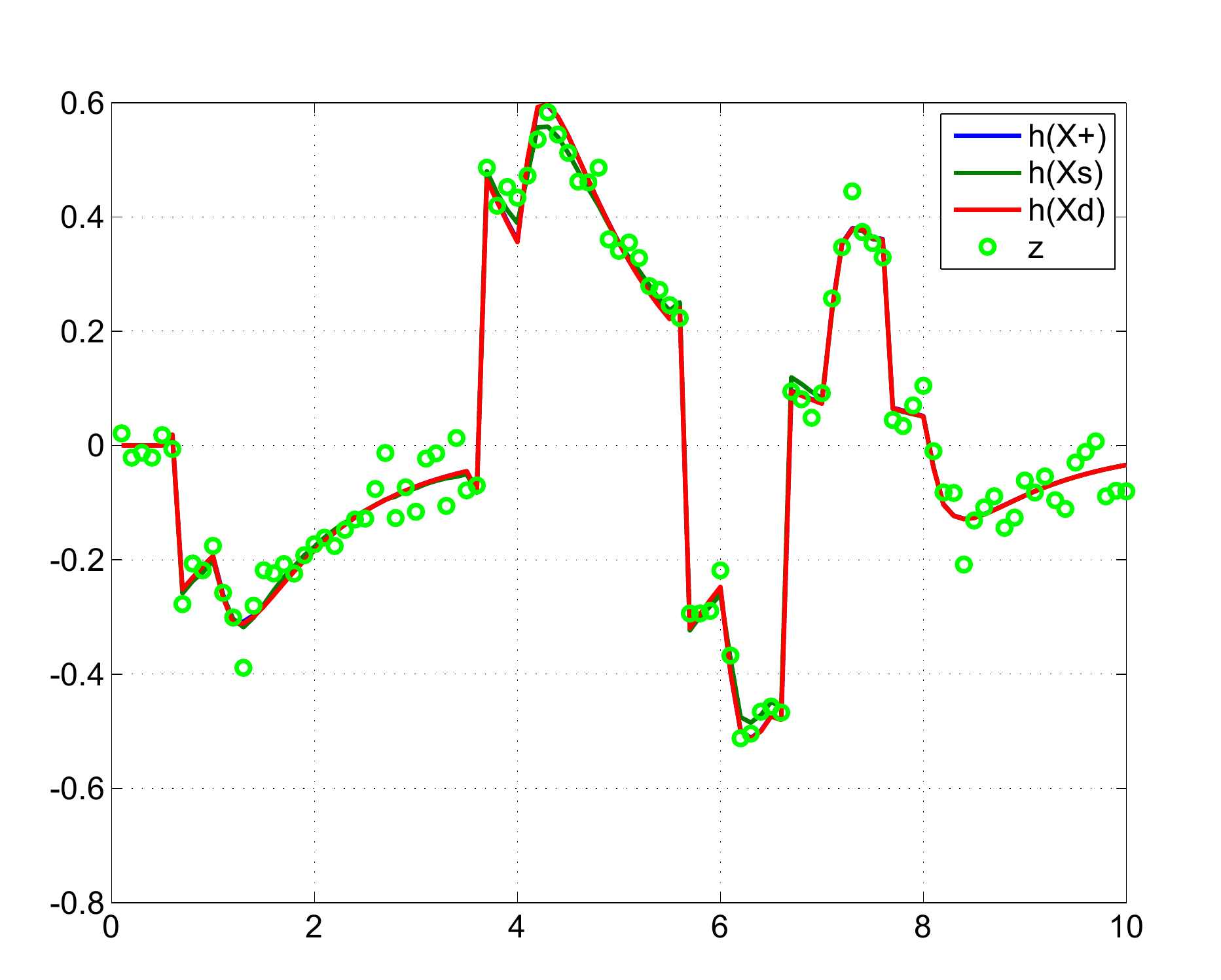}
\caption{Comparison of the predicted dynamics, posterior, smoothed}
\caption*{and the measurement 7}
\label{lat_h7}
\end{figure}

\begin{figure}[h]
\includegraphics[width=6in,height=4in]{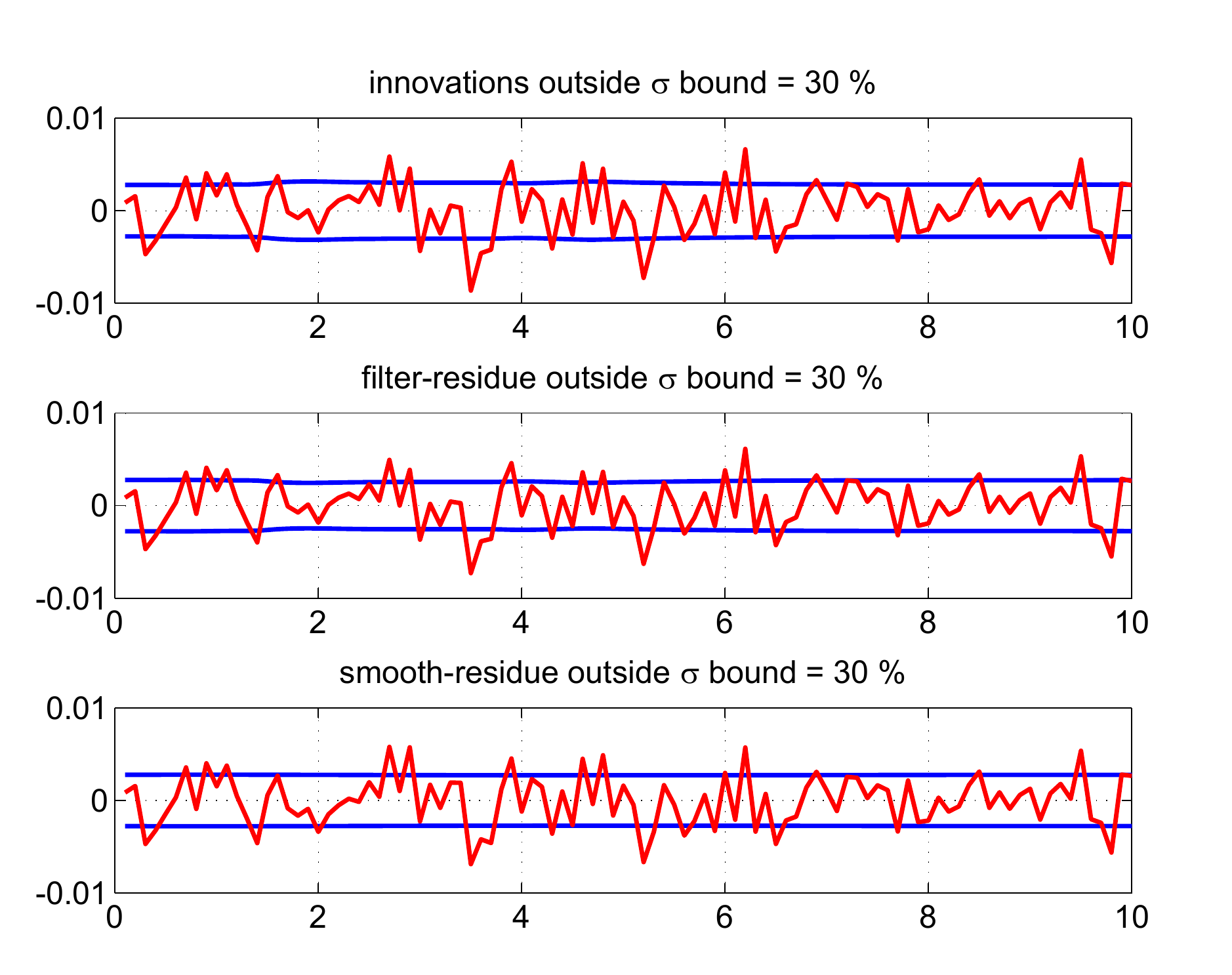}
\caption{The innovations, filtered residue and smoothed residue}
\caption*{corresponding to measurement 1 }
\label{lat_innov1}
\end{figure}

\begin{figure}[h]
\includegraphics[width=6in,height=4in]{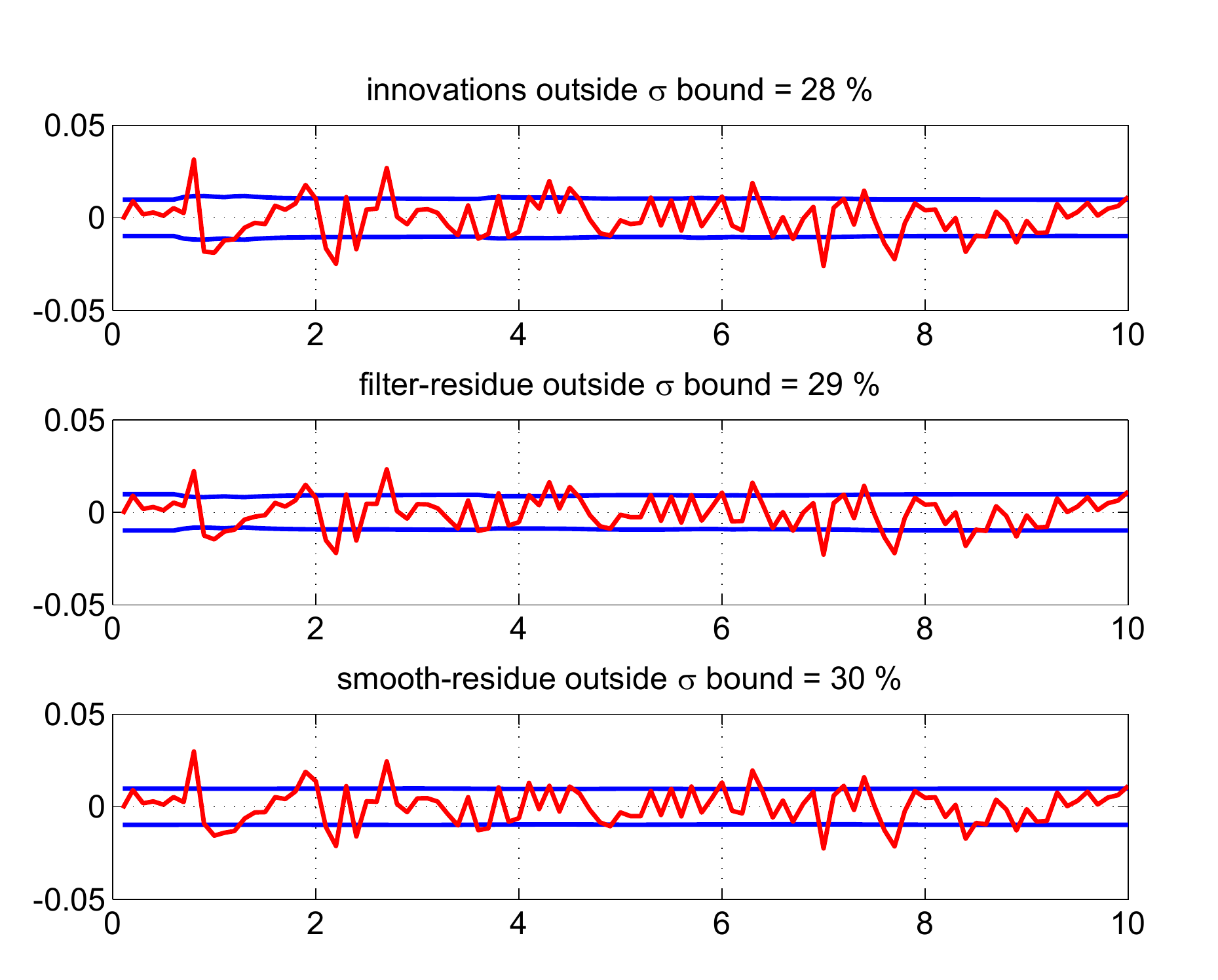}
\caption{The innovations, filtered residue and smoothed residue}
\caption*{corresponding to measurement 2}
\label{lat_innov2}
\end{figure}

\begin{figure}[h]
\includegraphics[width=6in,height=4in]{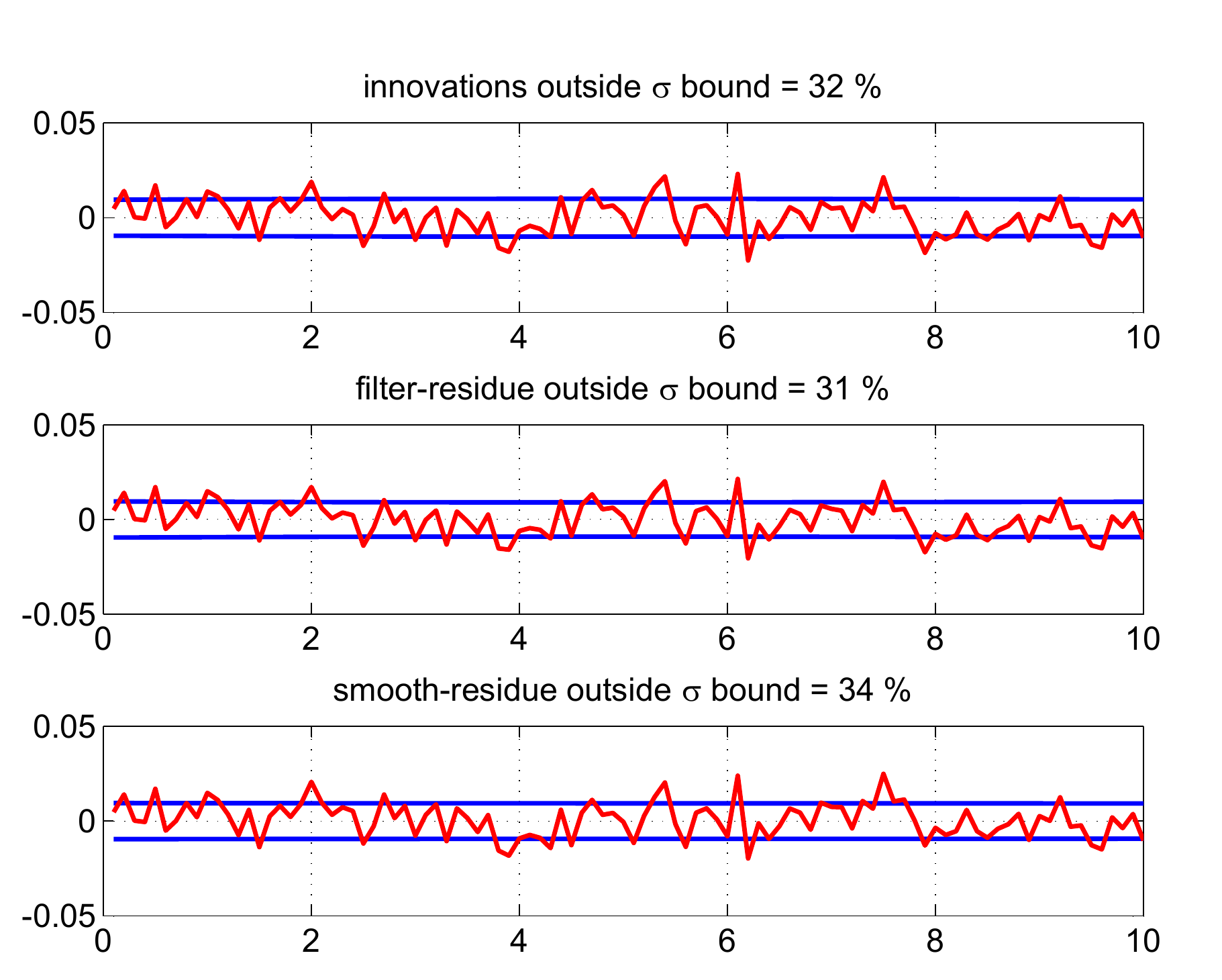}
\caption{The innovations, filtered residue and smoothed residue}
\caption*{corresponding to measurement 3 }
\label{lat_innov3}
\end{figure}

\begin{figure}[h]
\includegraphics[width=6in,height=4in]{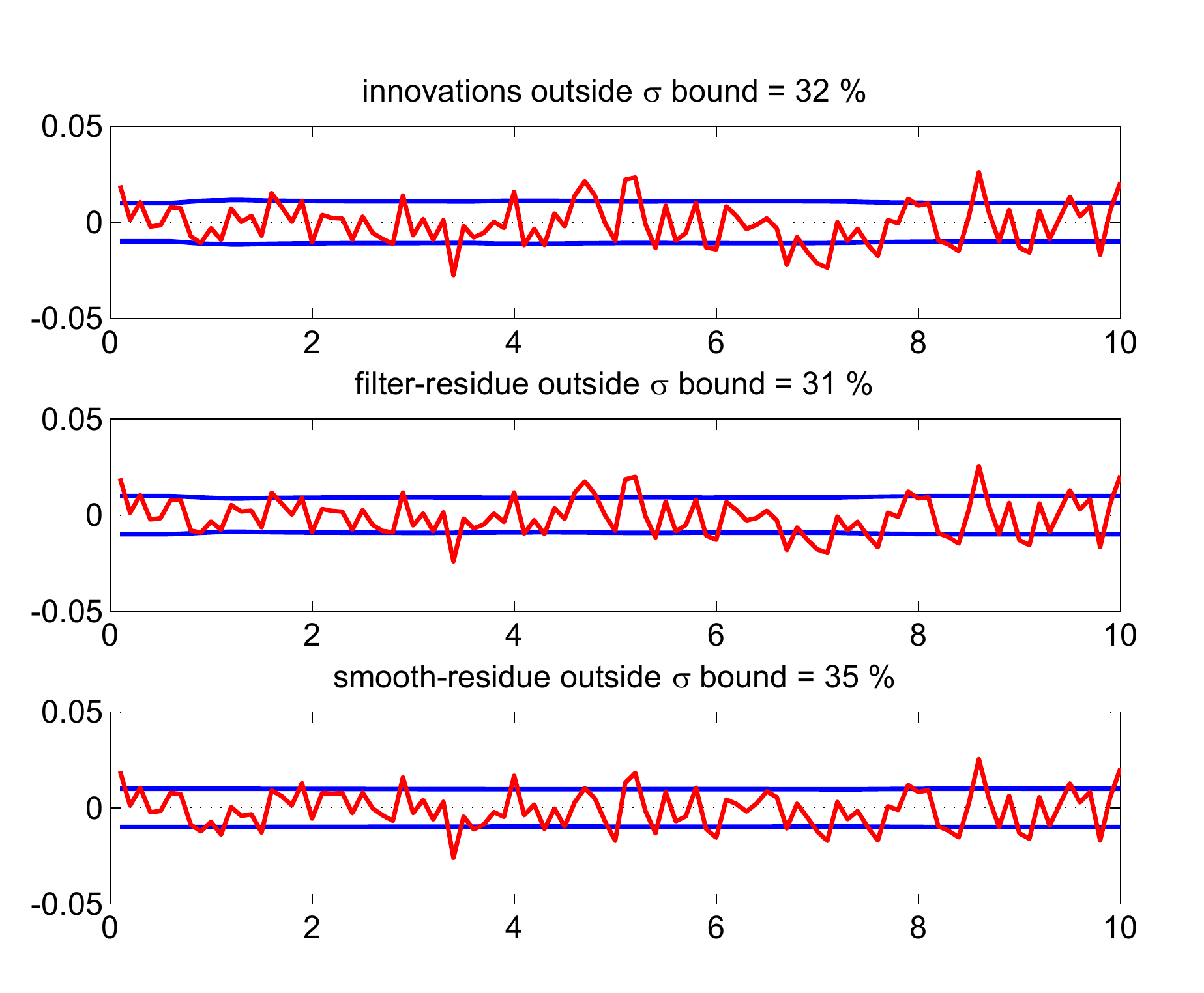}
\caption{The innovations, filtered residue and smoothed residue}
\caption*{corresponding to measurement 4}
\label{lat_innov4}
\end{figure}

\begin{figure}[h]
\includegraphics[width=6in,height=4in]{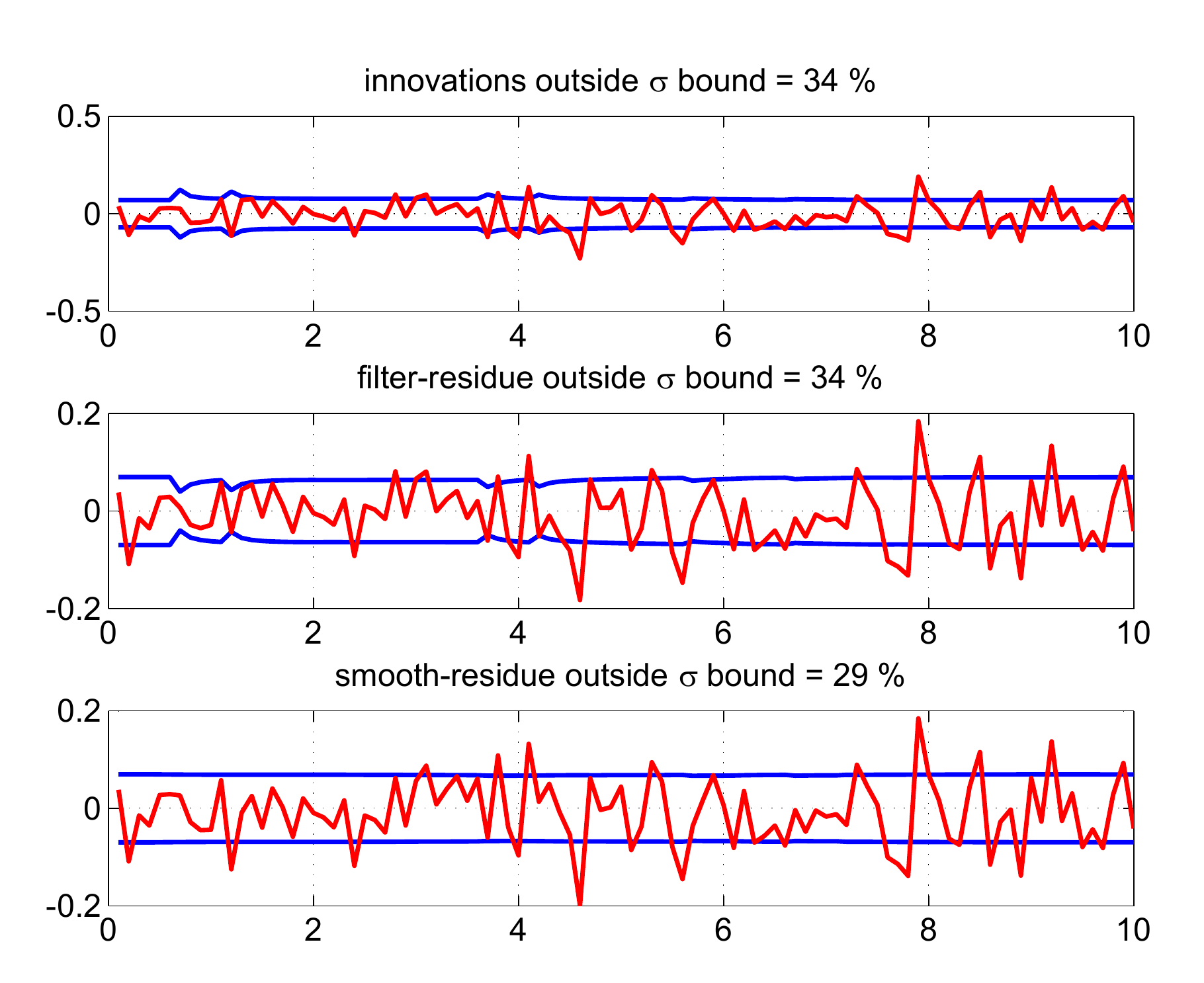}
\caption{The innovations, filtered residue and smoothed residue}
\caption*{corresponding to measurement 5}
\label{lat_innov5}
\end{figure}

\begin{figure}[h]
\includegraphics[width=6in,height=4in]{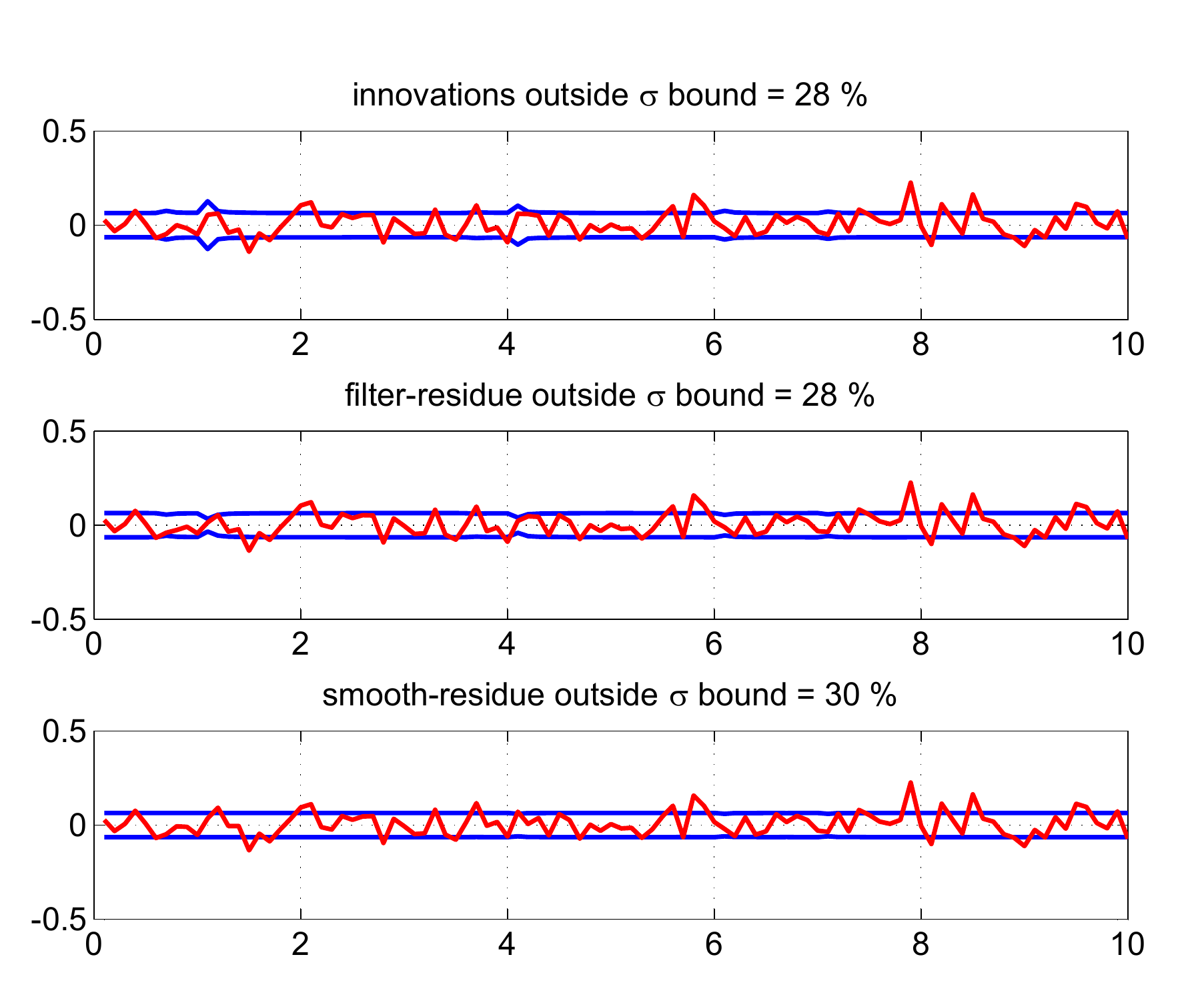}
\caption{The innovations, filtered residue and smoothed residue}
\caption*{corresponding to measurement 6}
\label{lat_innov6}
\end{figure}

\begin{figure}[h]
\includegraphics[width=6in,height=4in]{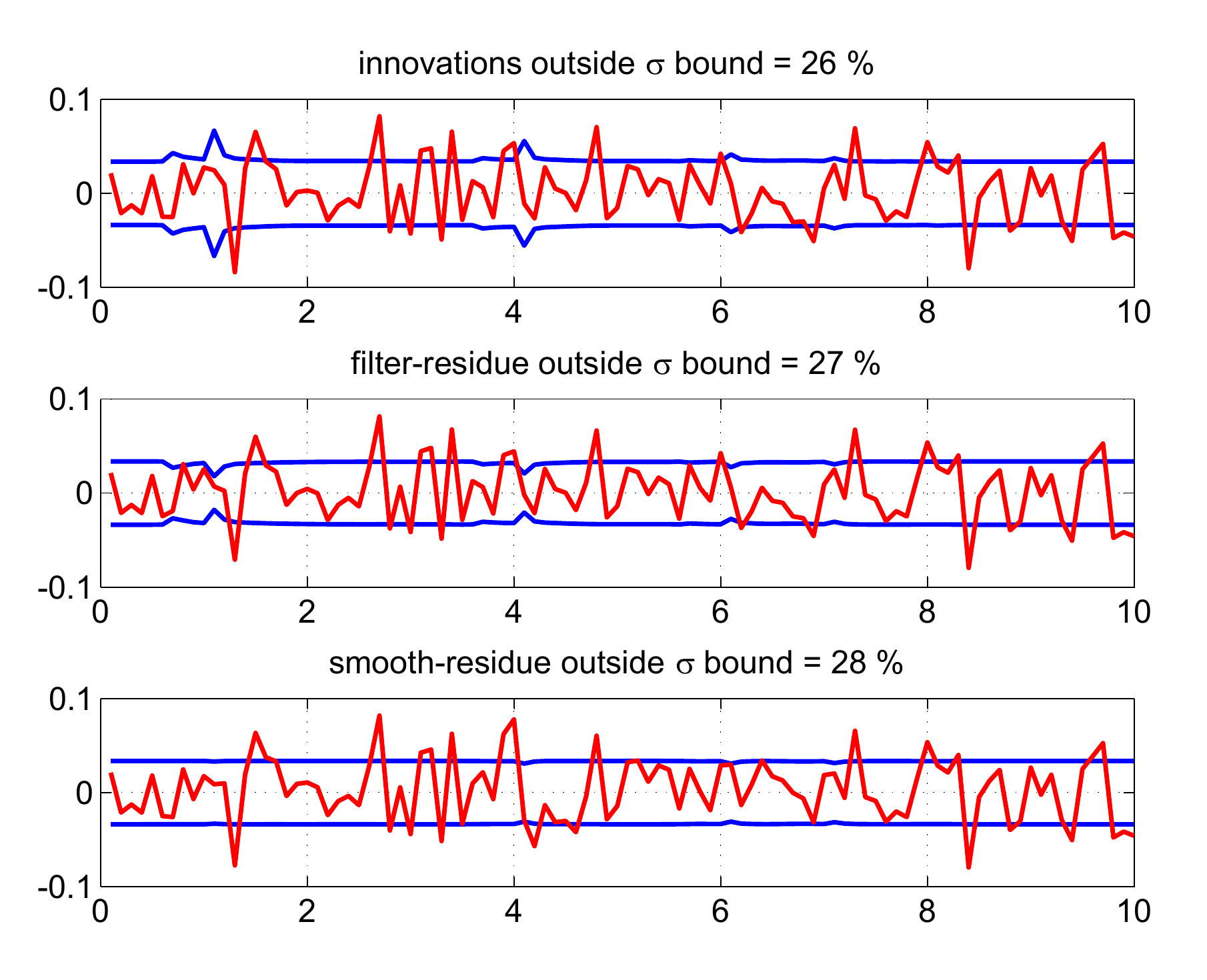}
\caption{The innovations, filtered residue and smoothed residue}
\caption*{corresponding to measurement 7}
\label{lat_innov7}
\end{figure}

\begin{figure}[h]
\includegraphics[width=6in,height=4in]{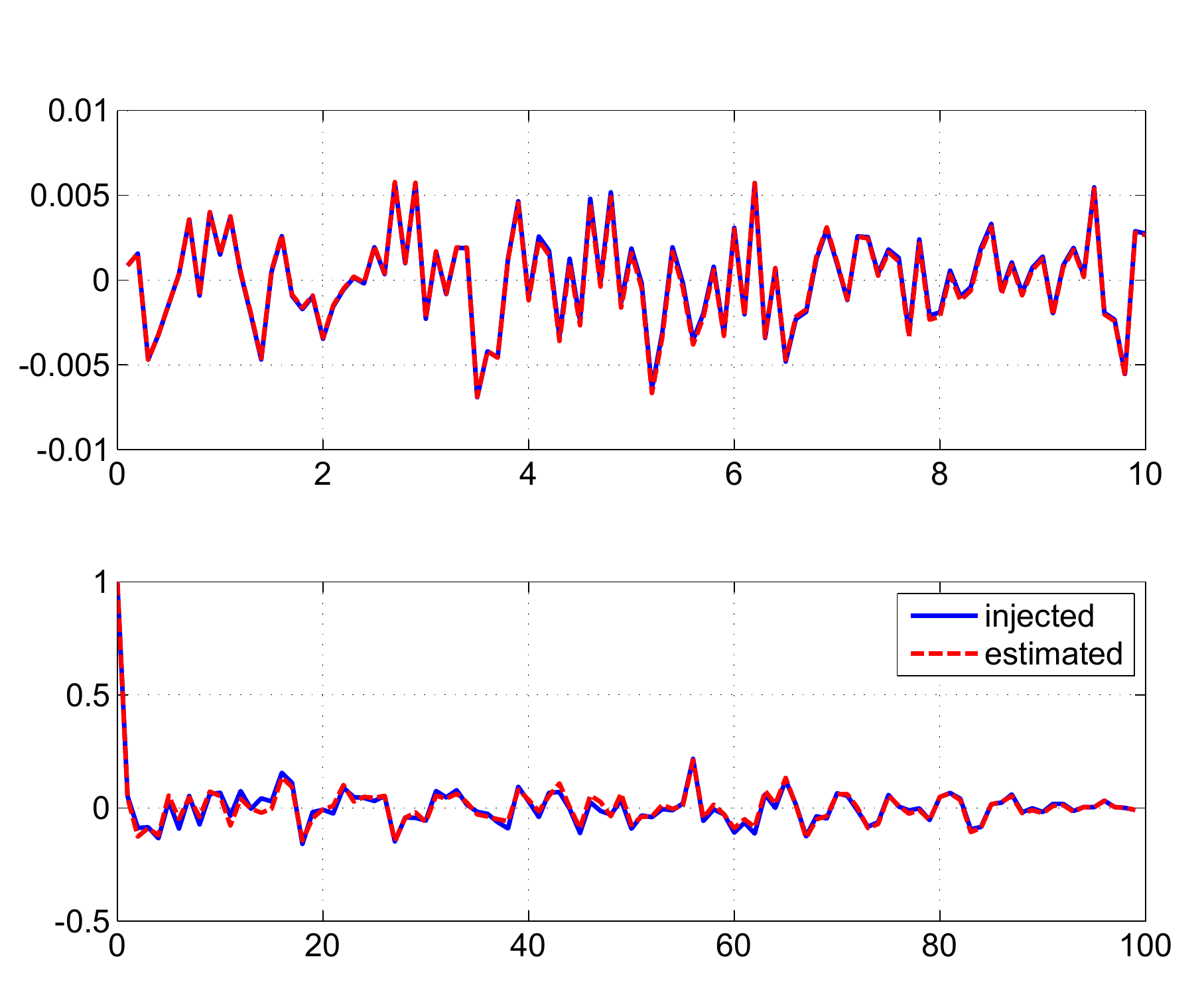}
\caption{Time variation of injected and estimated measurement noise (top) and}
\caption*{their autocorrelation (bottom) for measurement 1}
\label{lat_mnoise1}
\end{figure}

\begin{figure}[h]
\includegraphics[width=6in,height=4in]{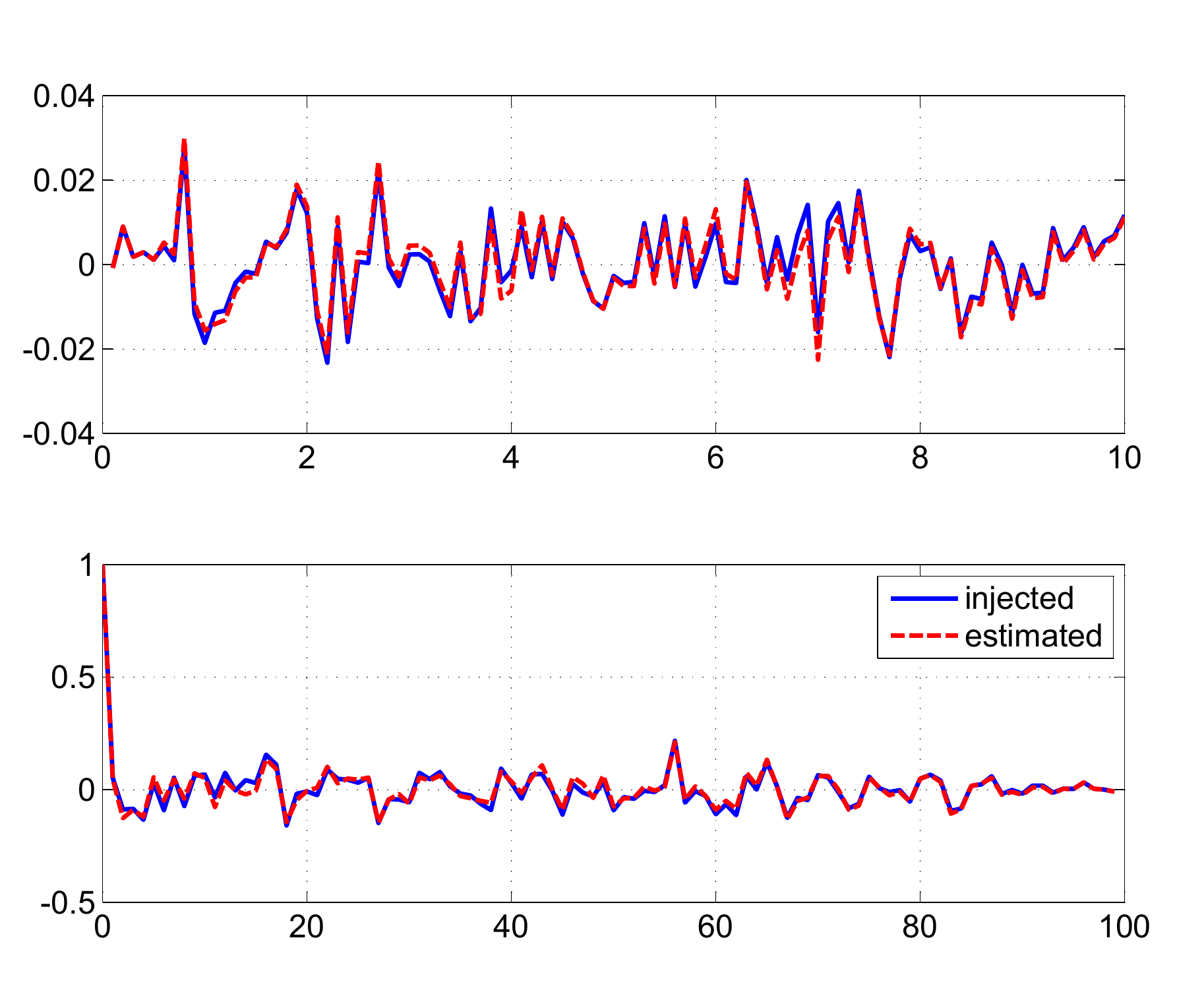}
\caption{Time variation of injected and estimated measurement noise (top) and}
\caption*{their autocorrelation (bottom) for measurement 2}
\label{lat_mnoise2}
\end{figure}

\begin{figure}[h]
\includegraphics[width=6in,height=4in]{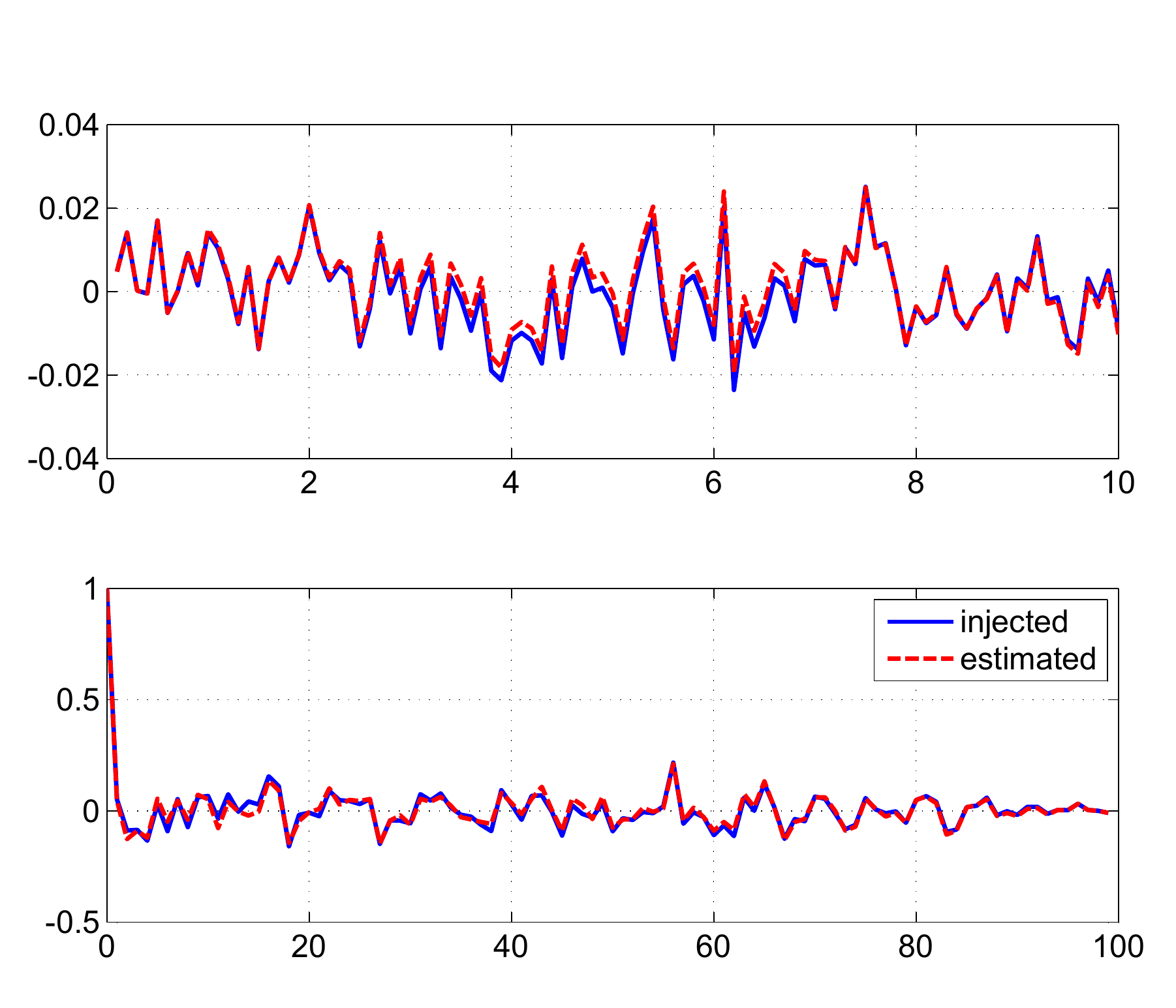}
\caption{Time variation of injected and estimated measurement noise (top) and}
\caption*{their autocorrelation (bottom) for measurement 3}
\label{lat_mnoise3}
\end{figure}

\begin{figure}[h]
\includegraphics[width=6in,height=3in]{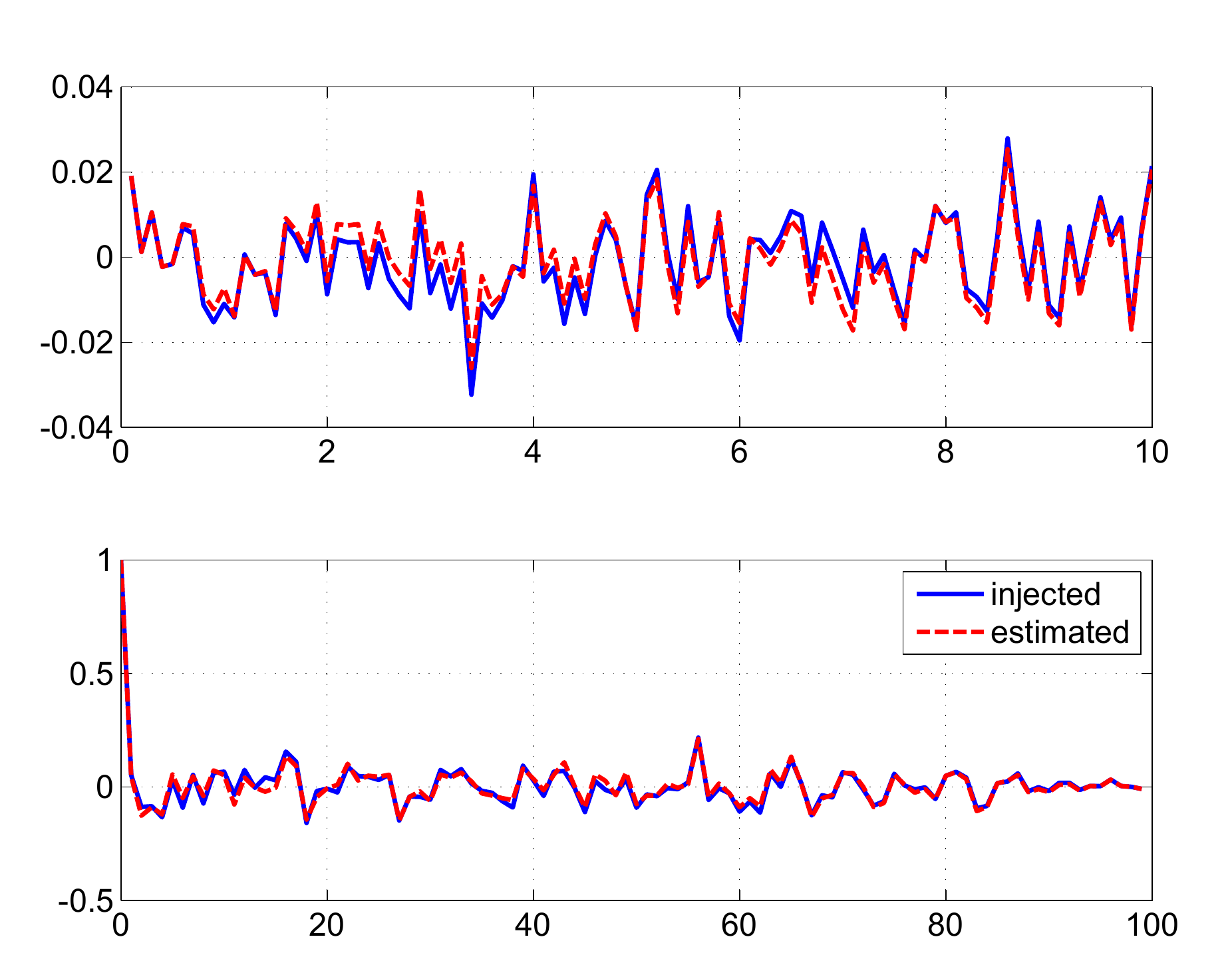}
\caption{Time variation of injected and estimated measurement noise (top) and}
\caption*{their autocorrelation (bottom) for measurement 4}
\label{lat_mnoise4}
\end{figure}

\begin{figure}[h]
\includegraphics[width=6in,height=3in]{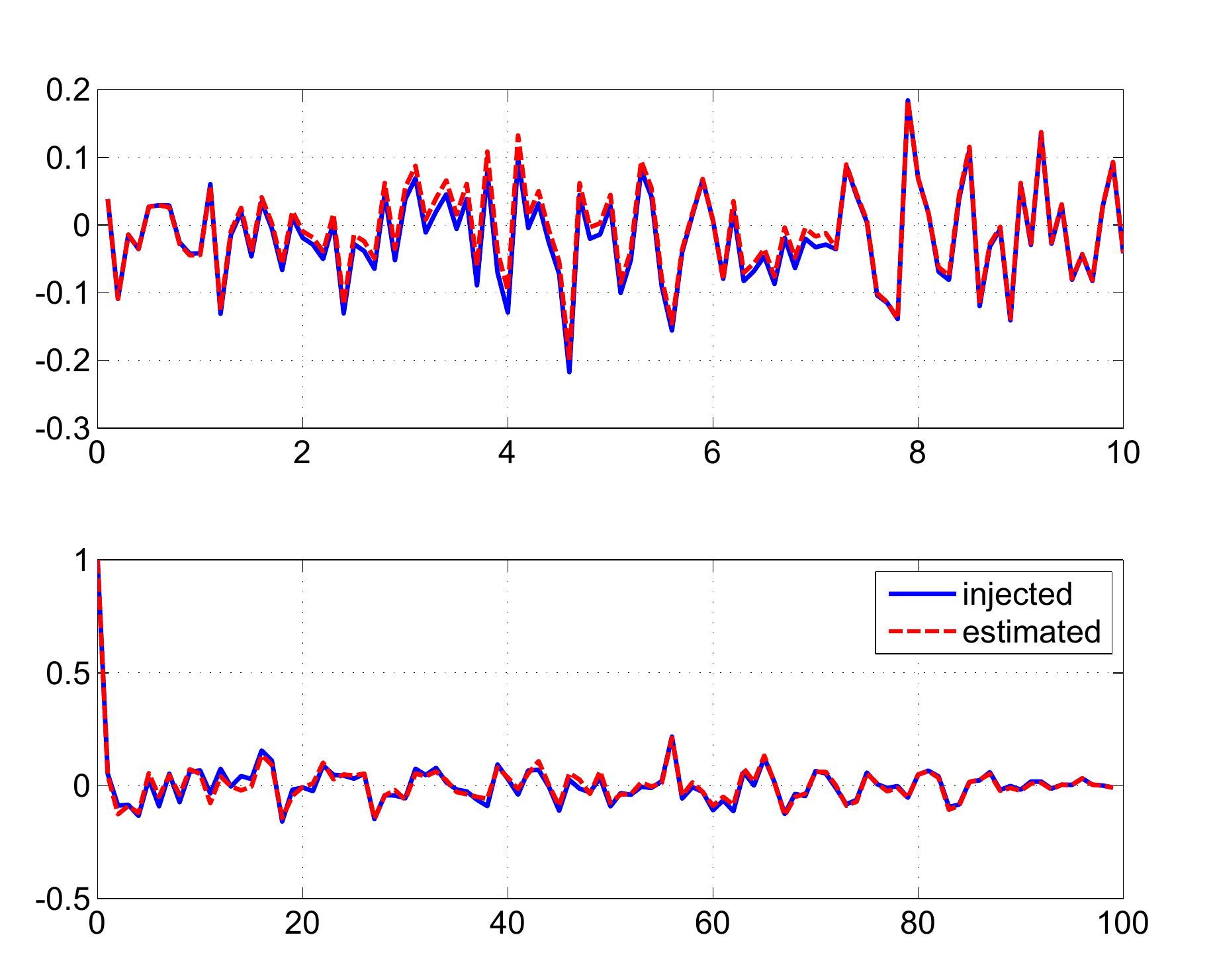}
\caption{Time variation of injected and estimated measurement noise (top) and}
\caption*{their autocorrelation (bottom) for measurement 5}
\label{lat_mnoise5}
\end{figure}

\begin{figure}[h]
\includegraphics[width=6in,height=4in]{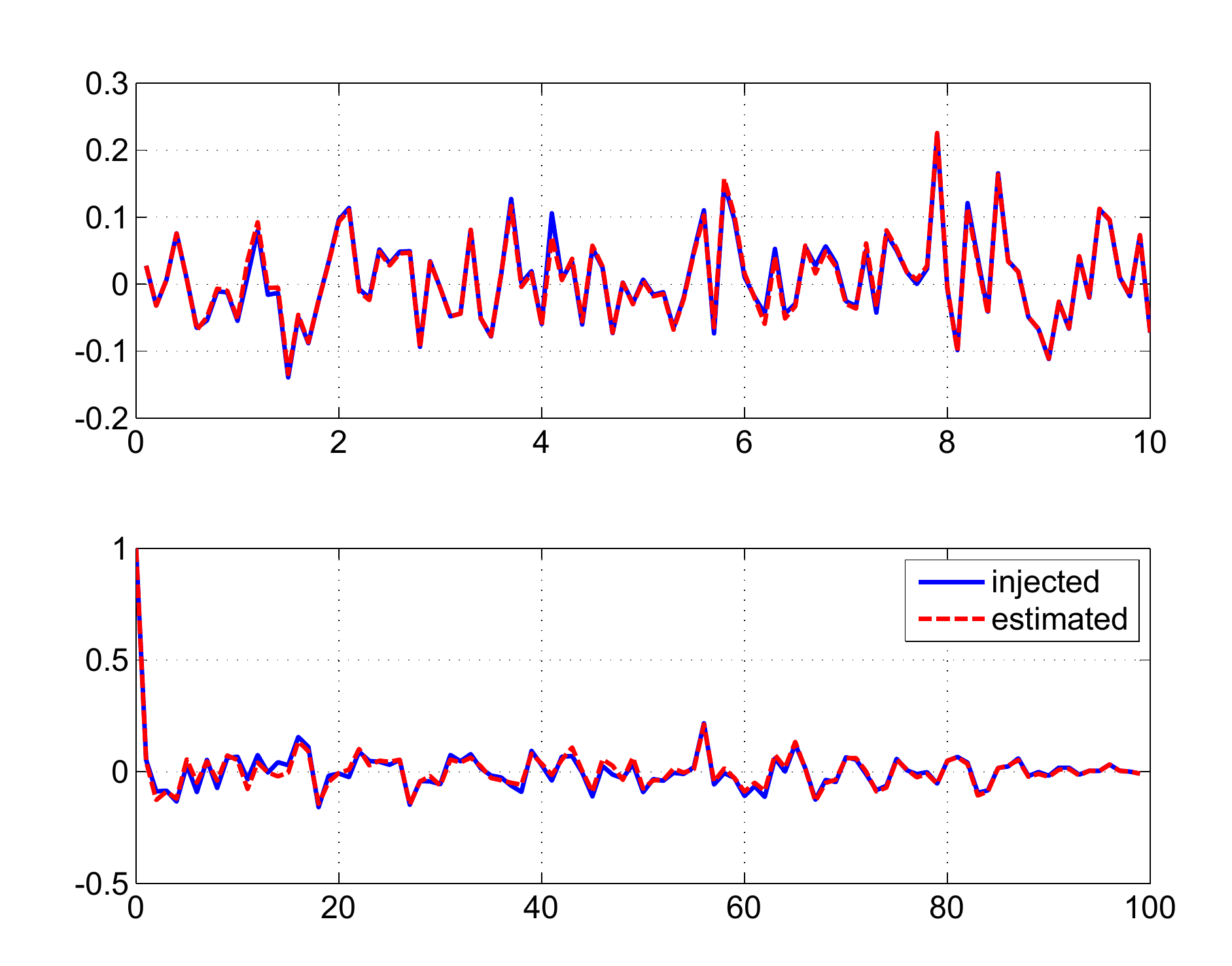}
\caption{Time variation of injected and estimated measurement noise (top) and}
\caption*{their autocorrelation (bottom) for measurement 6}
\label{lat_mnoise6}
\end{figure}

\begin{figure}[h]
\includegraphics[width=6in,height=4in]{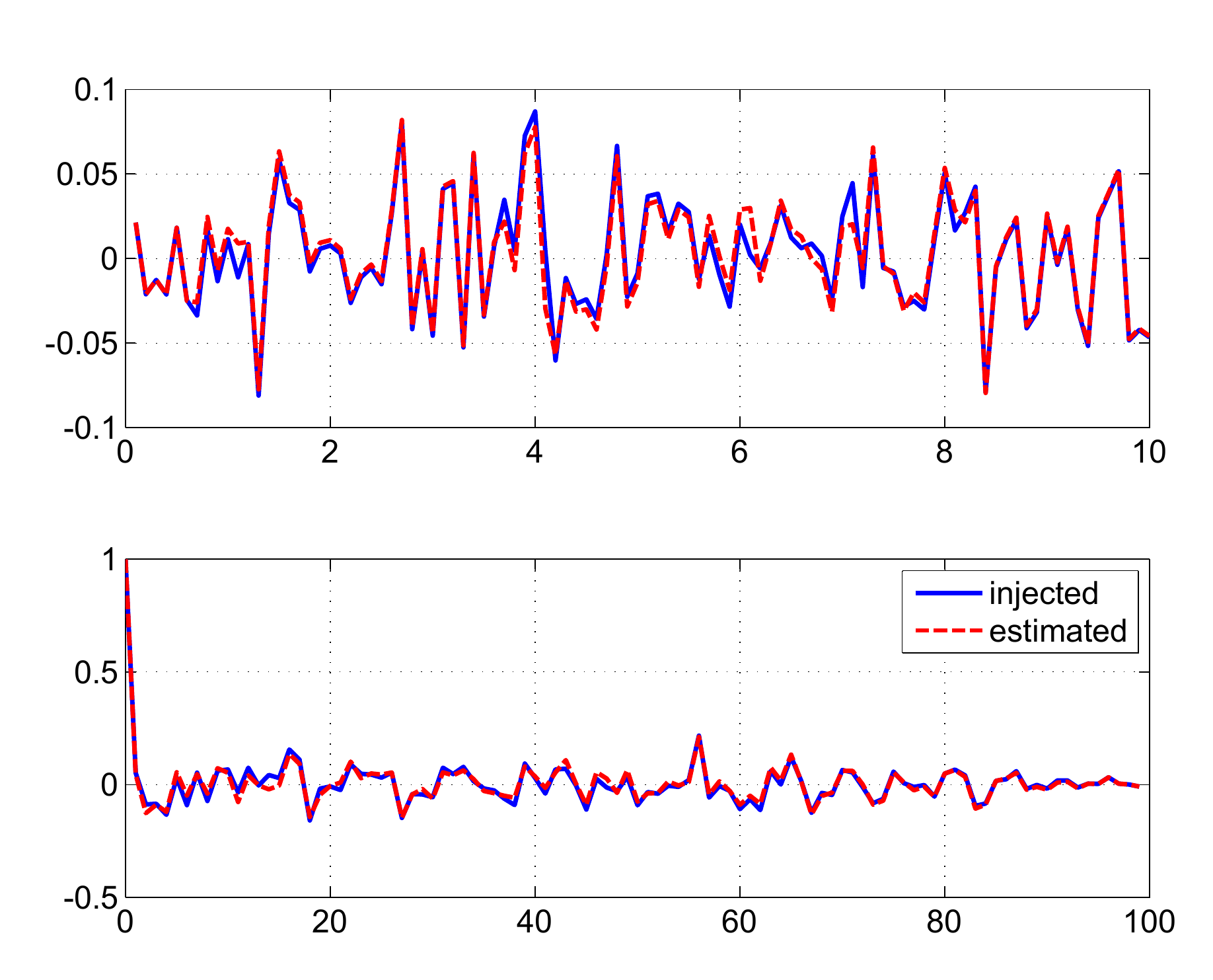}
\caption{Time variation of injected and estimated measurement noise (top) and}
\caption*{their autocorrelation (bottom) for measurement 7}
\label{lat_mnoise7}
\end{figure}

 We first run the filter assuming \textbf{Q} = 0. It was found that about 20 iterations of the data would suffice. The Fig. \ref{lat_p1}-\ref{lat_p15} shows the various parameter estimates and its corresponding variances through cumulative time instants with iterations. The variation of the estimated initial parameters and their variances through iterations are shown in Fig. \ref{lat_P0}. The parameter and the uncertainty reach almost their final estimated values in about 2 and 5 iterations respectively. Similar plots in Fig. \ref{lat_R} and Fig. \ref{lat_cost} shows the variation of the estimated \textbf{R} and cost functions (\textbf{J1-J5}) through the iterations. The Fig. \ref{lat_h1}-\ref{lat_h7} shows the predicted dynamics, filtered and smoothed estimate at the last iteration. The Fig. \ref{lat_innov1}-\ref{lat_innov7} show the innovations, filtered residue and smoothed residue together with the square root of their variance ($\pm\sigma$ bound). In the EKF approach most of the quantities are Gaussian or approximated as quasi Gaussian and one would expect all the above quantities are close to being Gaussian and hence around one third of the total sample points to be outside the $\sigma$ bound. The injected and estimated measurement noise distributions during the final iteration shown in Fig. \ref{lat_mnoise1}-\ref{lat_mnoise7} indicate that they are close to each other. Further their autocorrelations are ideally expected to be close to the Kronecker delta function which provides confidence in the proposed  algorithm.

\begin{figure}[h]
\includegraphics[width=6in,height=3.2in]{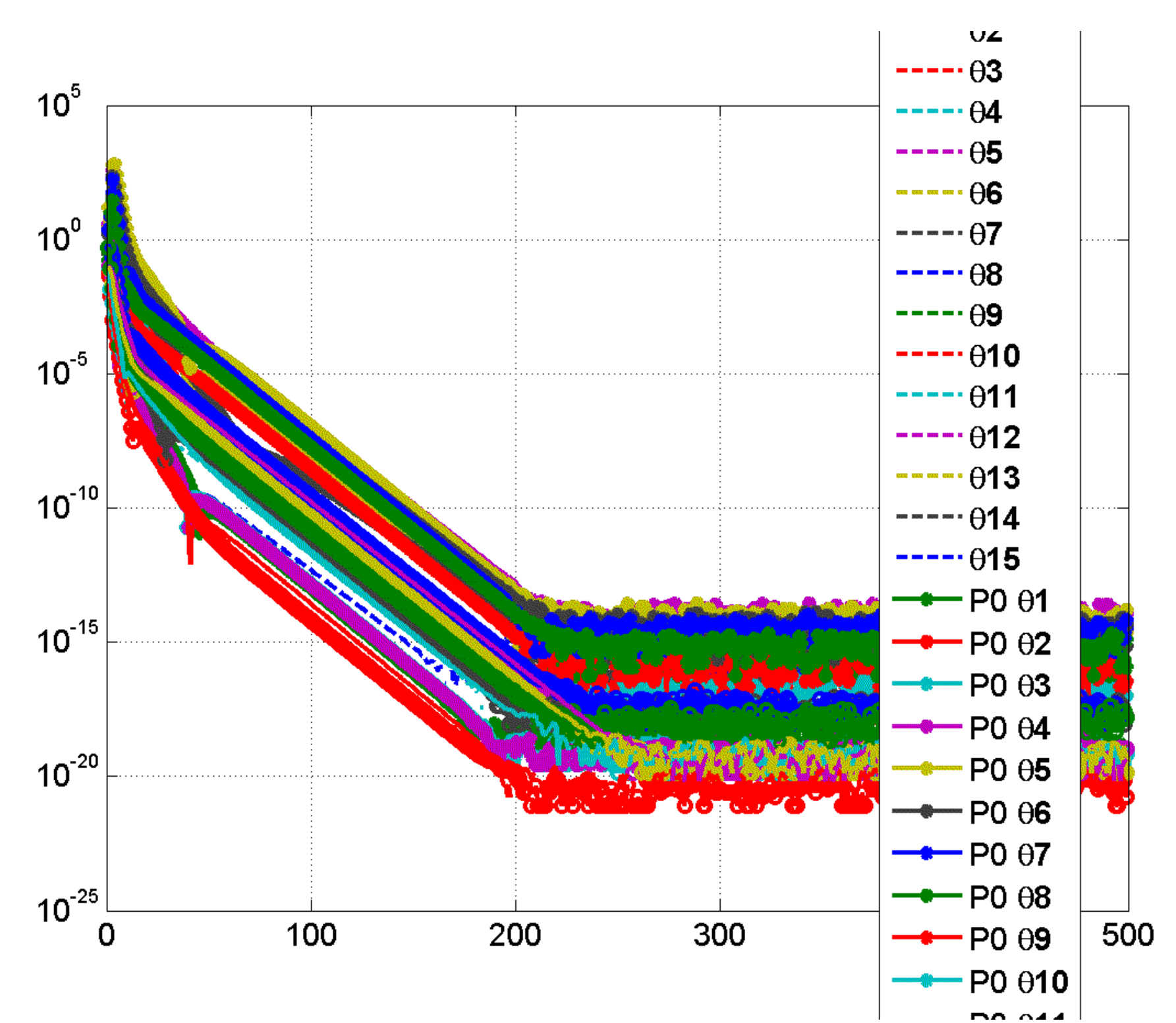}
\caption{The absolute difference between the iterated and final values}
\caption*{with 500 iterations}
\label{lat_err}
\end{figure}

\begin{figure}[h]
\includegraphics[width=6in,height=3.2in]{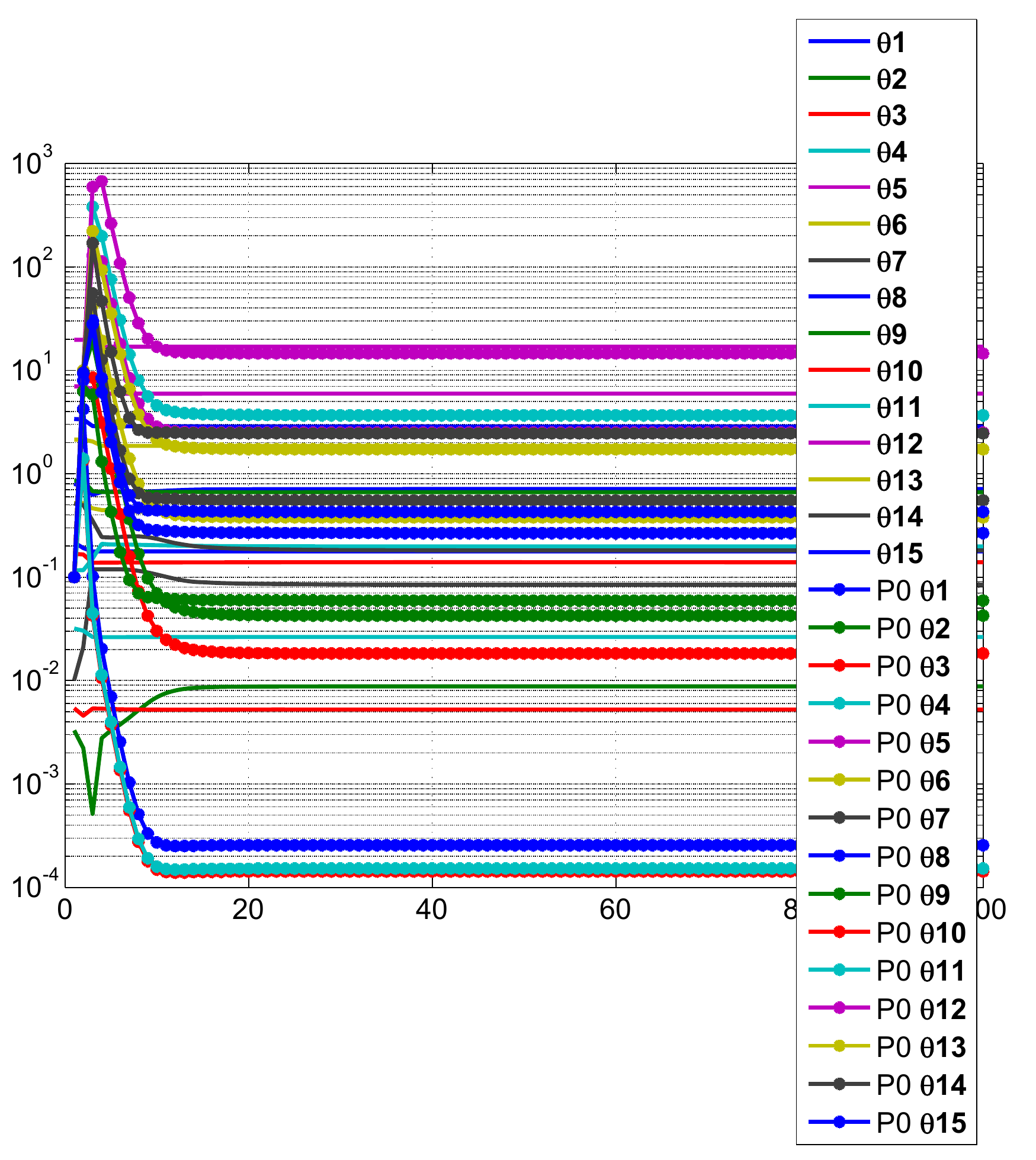}
\caption{Variation of parameter and its initial covariance ($\mathbf{P_0}$) with iterations}
\label{latQ_P0}
\end{figure}

\begin{figure}[h]
\includegraphics[width=6in,height=4in]{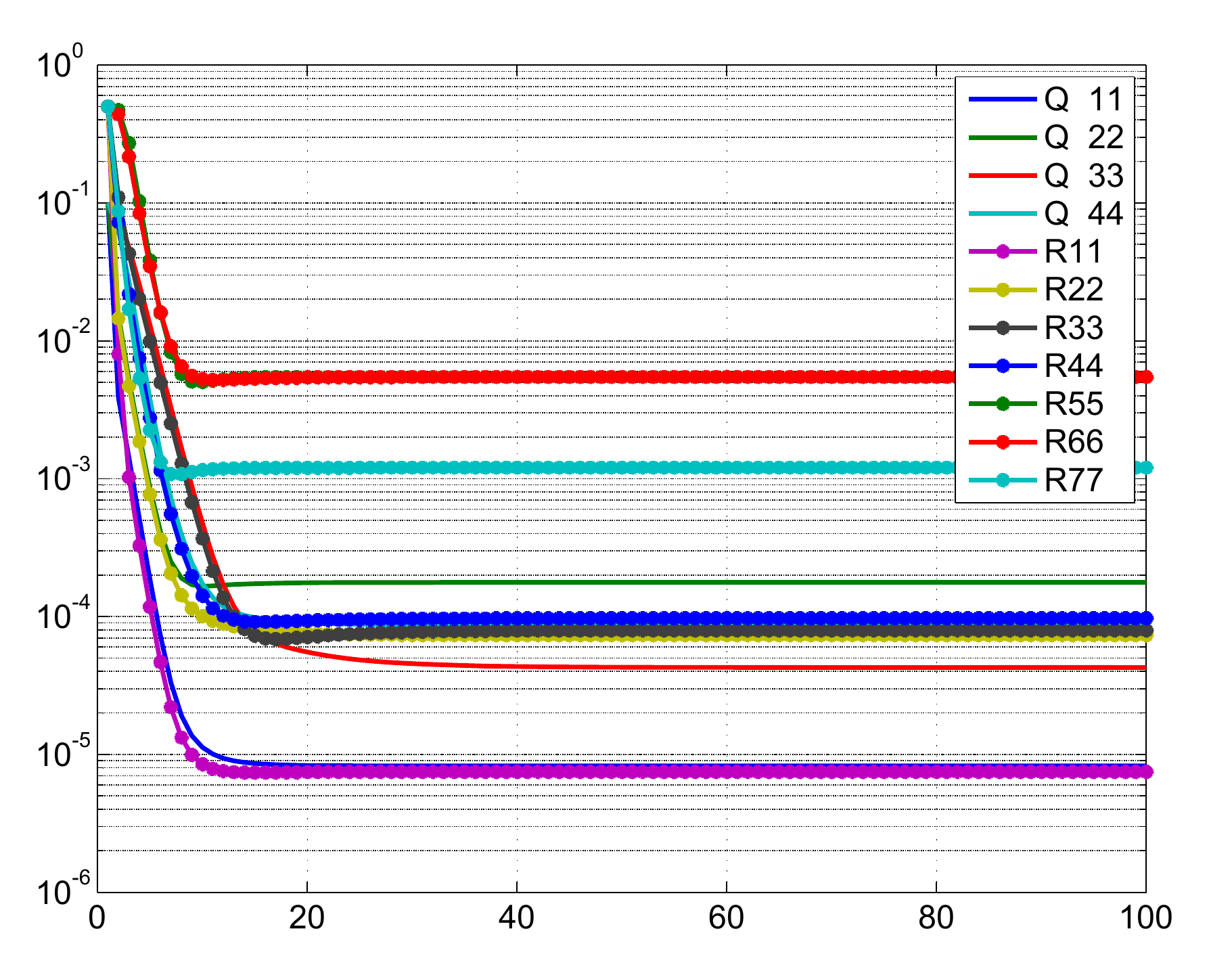}
\caption{Variation of \textbf{Q} and \textbf{R} with iterations}
\label{latQ_R}
\end{figure}

\begin{figure}[h]
\includegraphics[width=6in,height=4in]{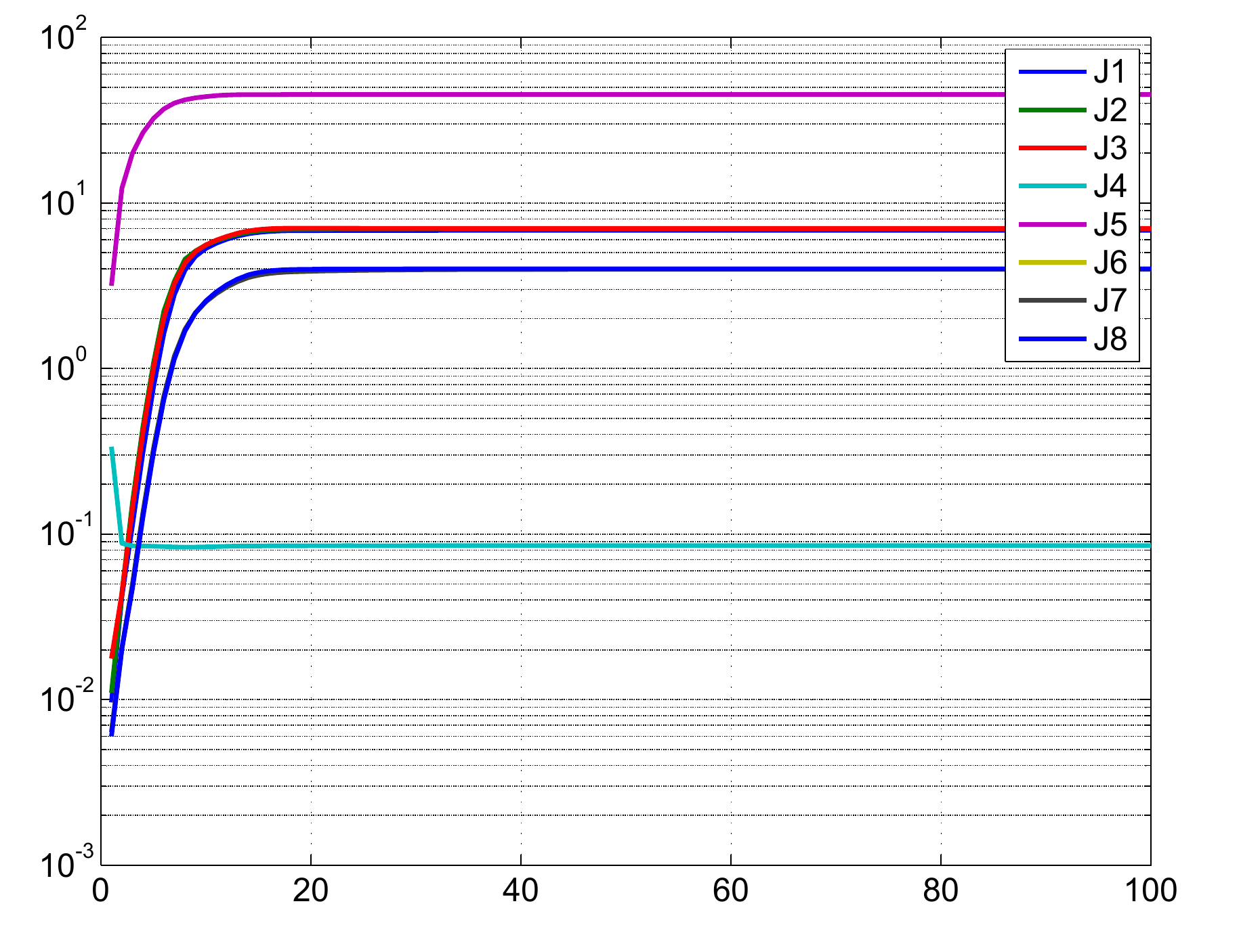}
\caption{Variation of different costs (\textbf{J1-J8}) with iterations}
\label{latQ_J}
\end{figure}

\begin{figure}[h]
\includegraphics[width=6in,height=4in]{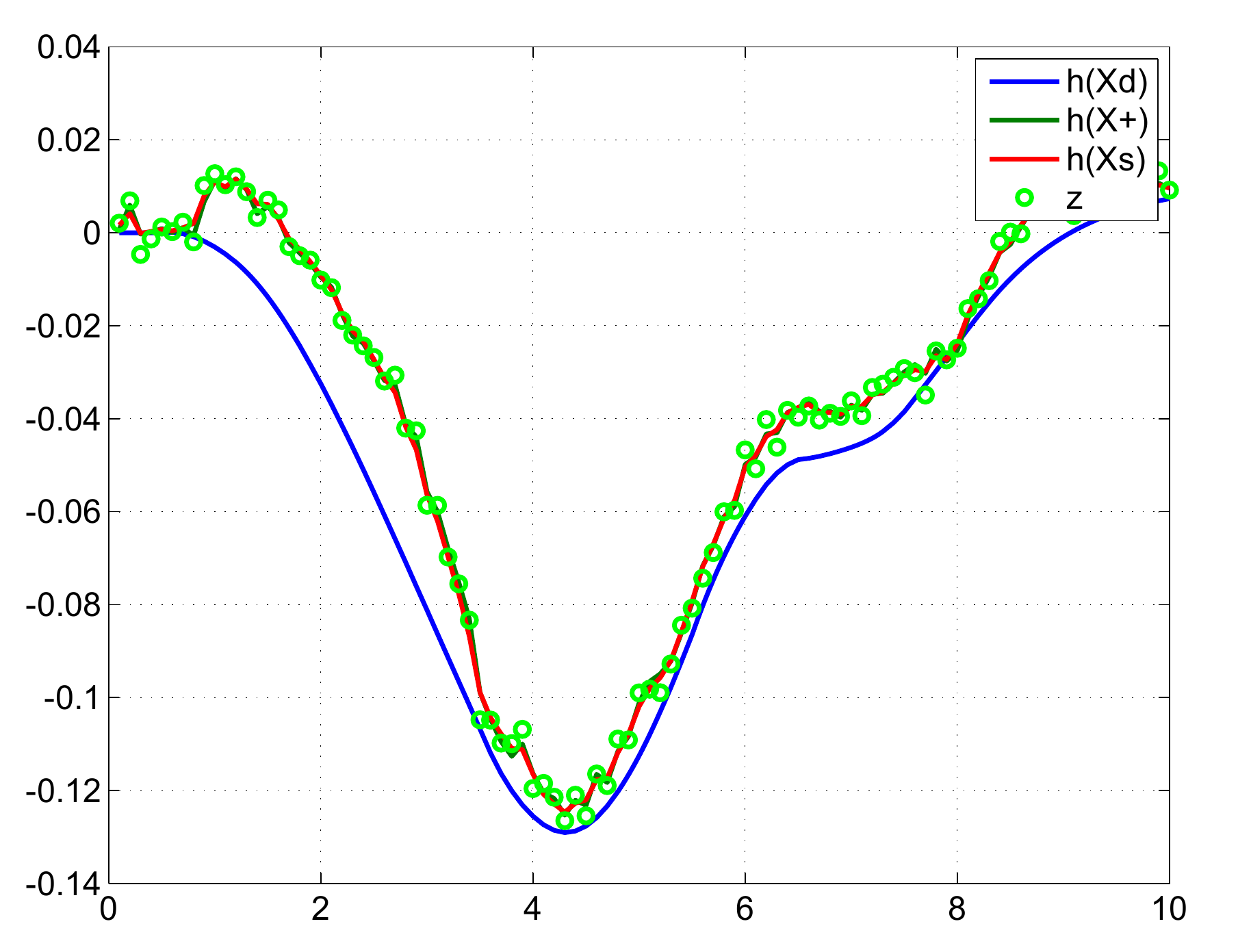}
\caption{Comparison of the predicted dynamics, posterior, smoothed}
\caption*{and the measurement 1 }
\label{latQ_h1}
\end{figure}

\begin{figure}[h]
\includegraphics[width=6in,height=4in]{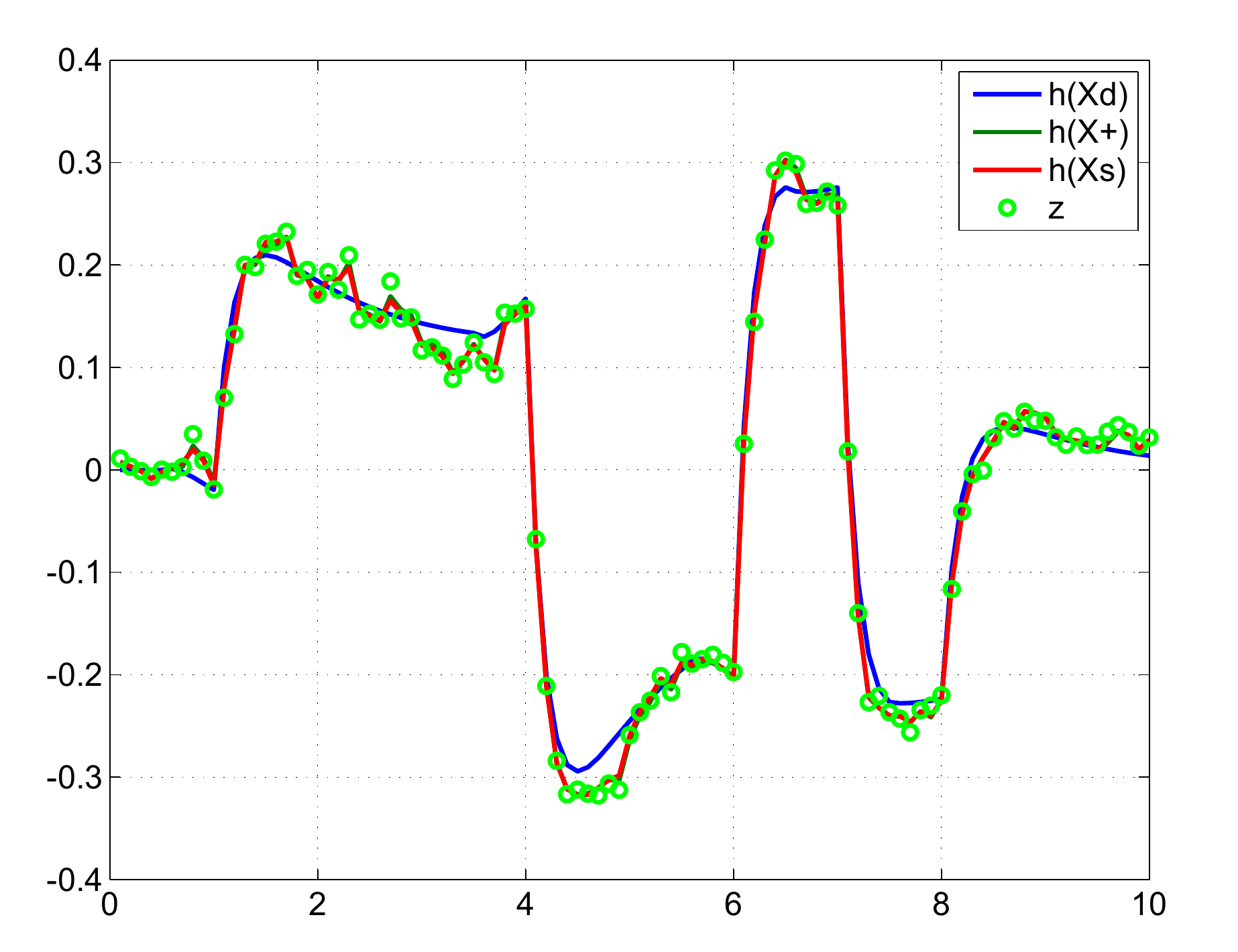}
\caption{Comparison of the predicted dynamics, posterior, smoothed}
\caption*{and the measurement 2}
\label{latQ_h2}
\end{figure}

\begin{figure}[h]
\includegraphics[width=6in,height=4in]{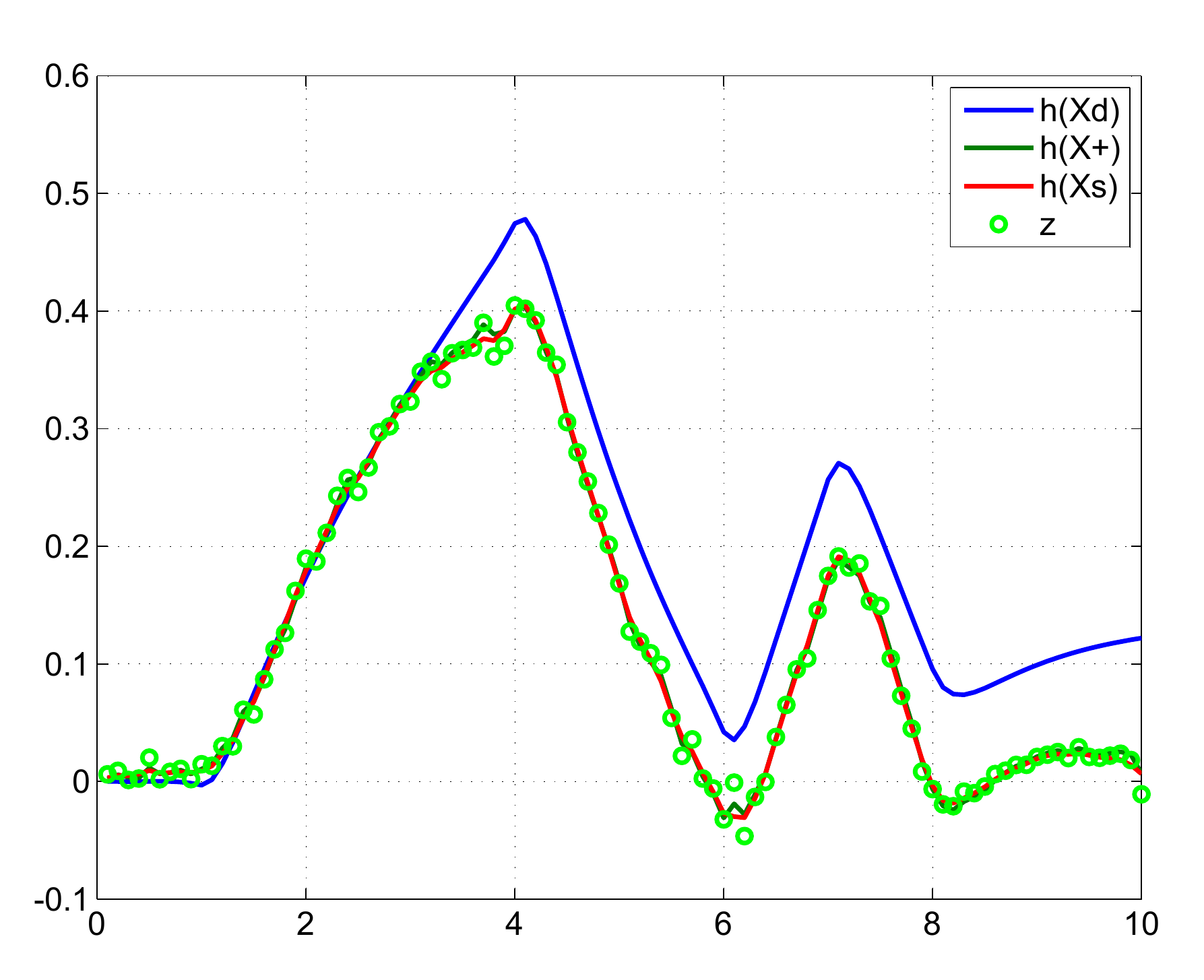}
\caption{Comparison of the predicted dynamics, posterior, smoothed}
\caption*{and the measurement 3 }
\label{laQt_h3}
\end{figure}

\begin{figure}[h]
\includegraphics[width=6in,height=4in]{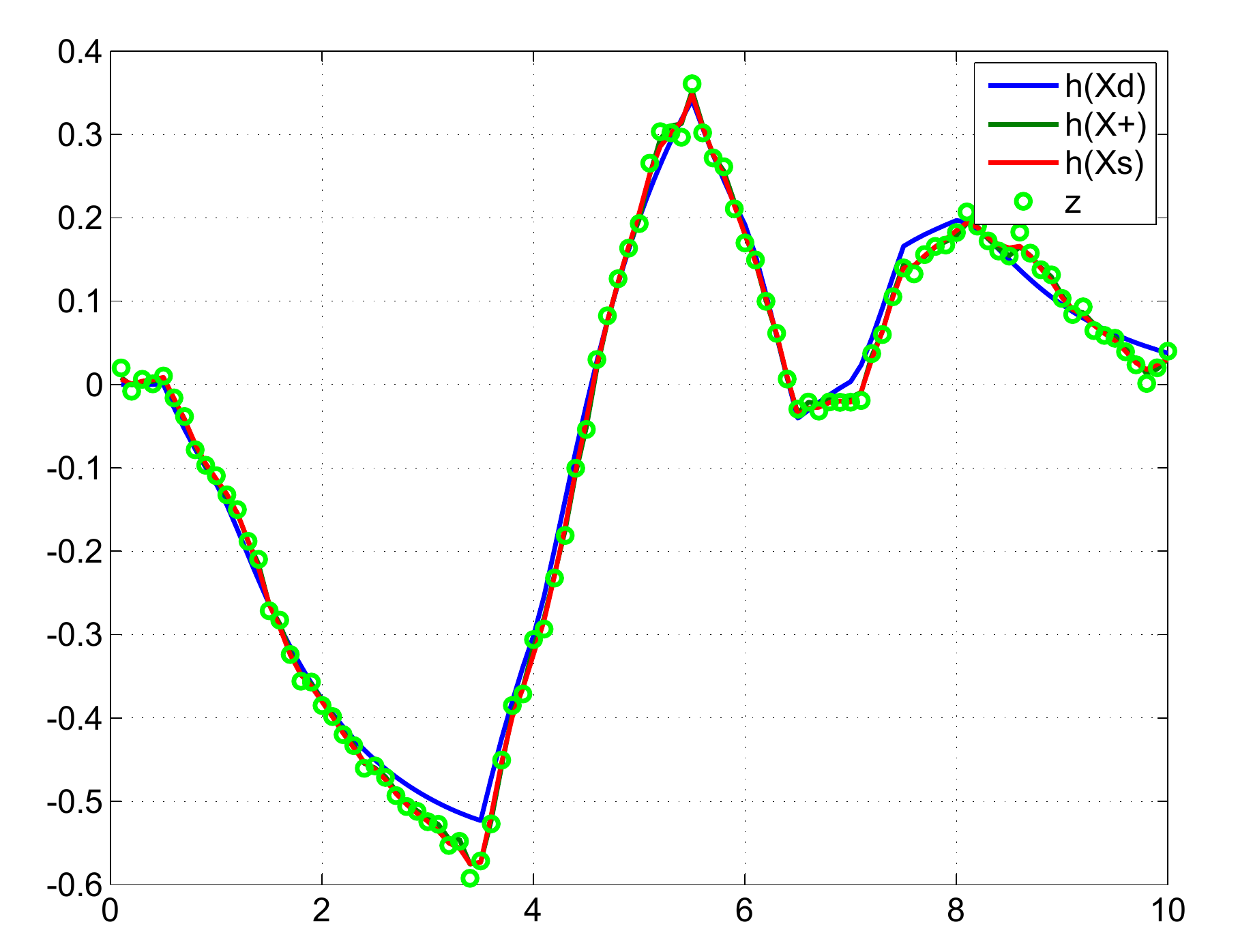}
\caption{Comparison of the predicted dynamics, posterior, smoothed}
\caption*{and the measurement 4}
\label{latQ_h4}
\end{figure}

\begin{figure}[h]
\includegraphics[width=6in,height=4in]{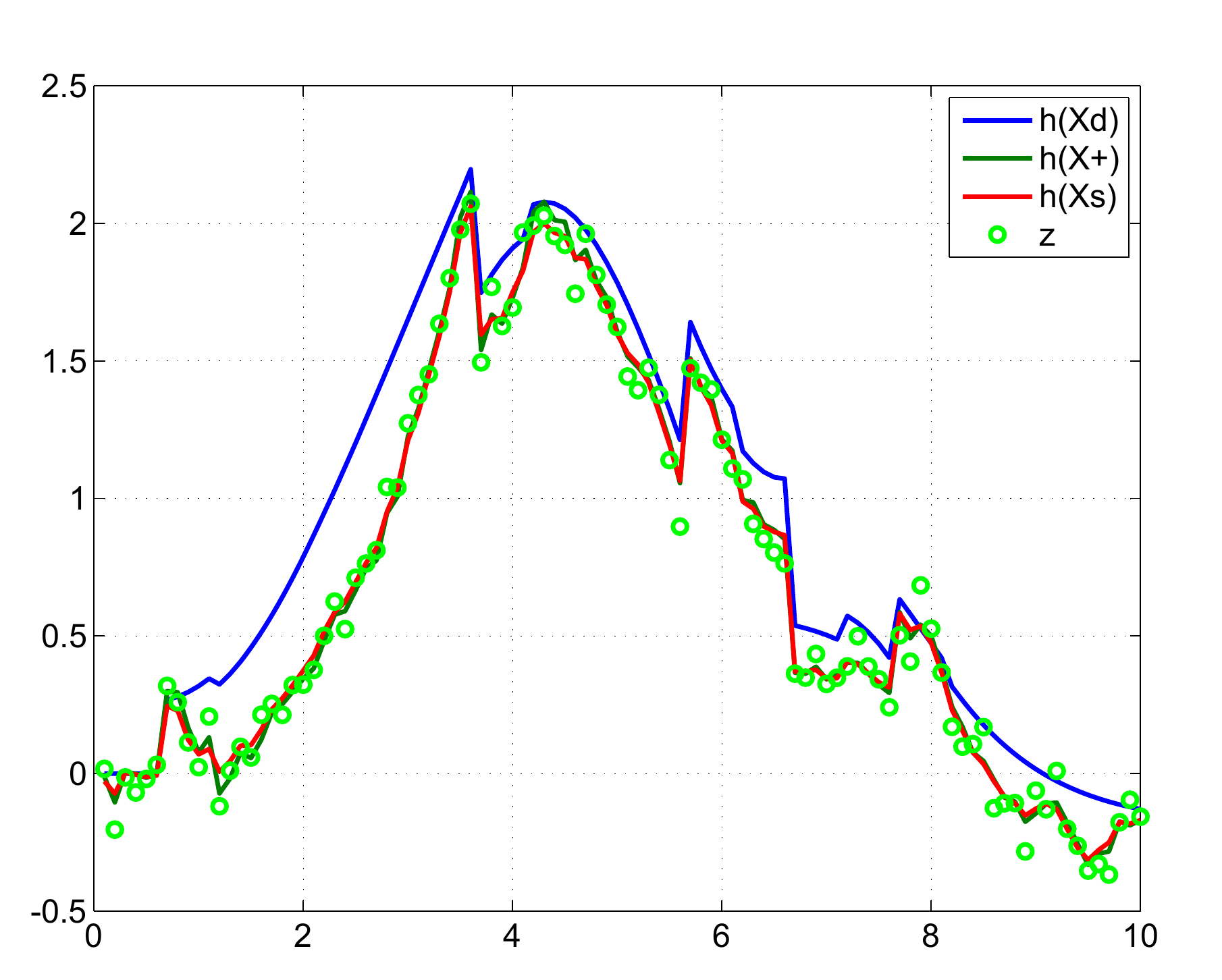}
\caption{Comparison of the predicted dynamics, posterior, smoothed}
\caption*{and the measurement 5}
\label{latQ_h5}
\end{figure}

\begin{figure}[h]
\includegraphics[width=6in,height=4in]{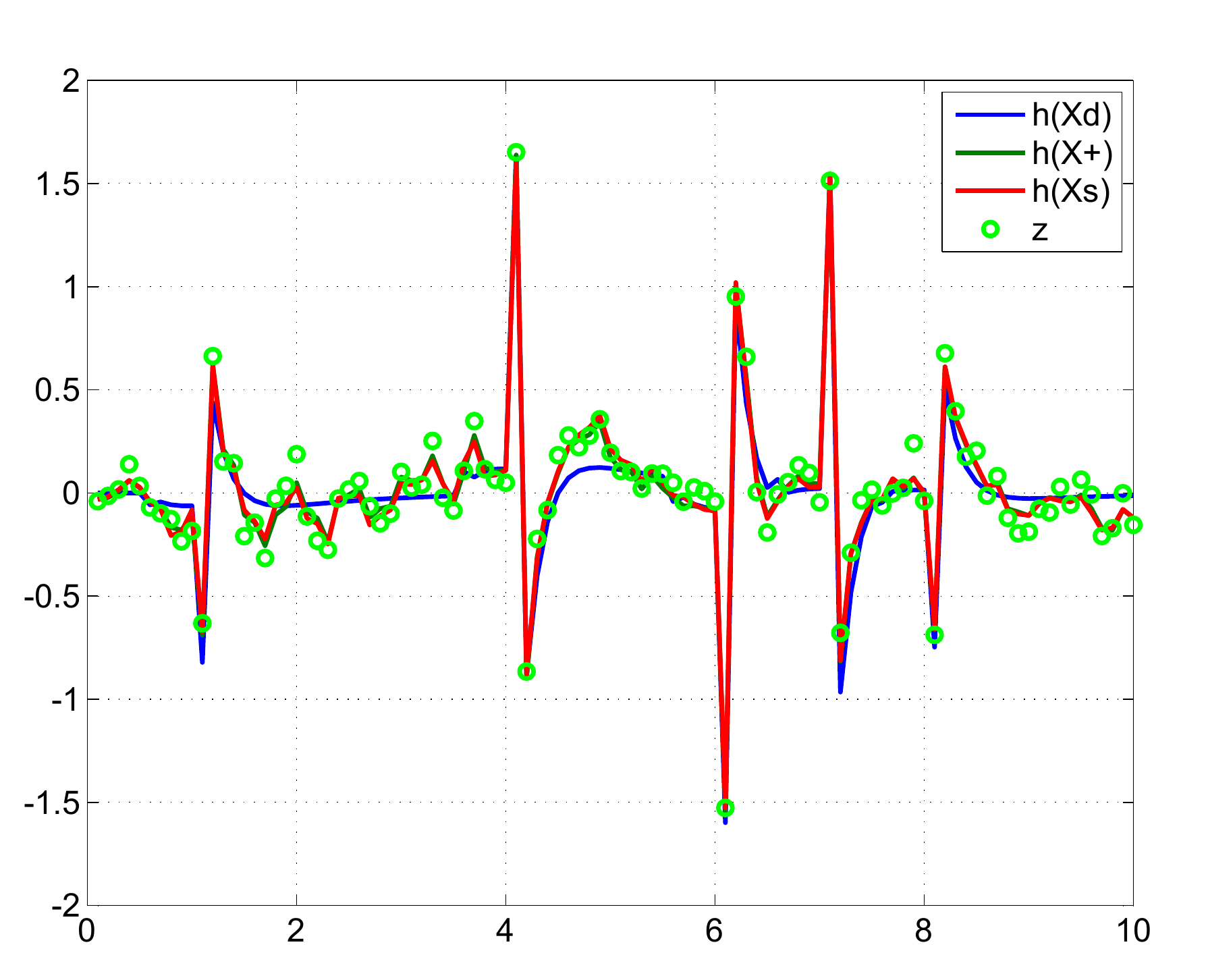}
\caption{Comparison of the predicted dynamics, posterior, smoothed}
\caption*{and the measurement 6}
\label{latQ_h6}
\end{figure}

\begin{figure}[h]
\includegraphics[width=6in,height=4in]{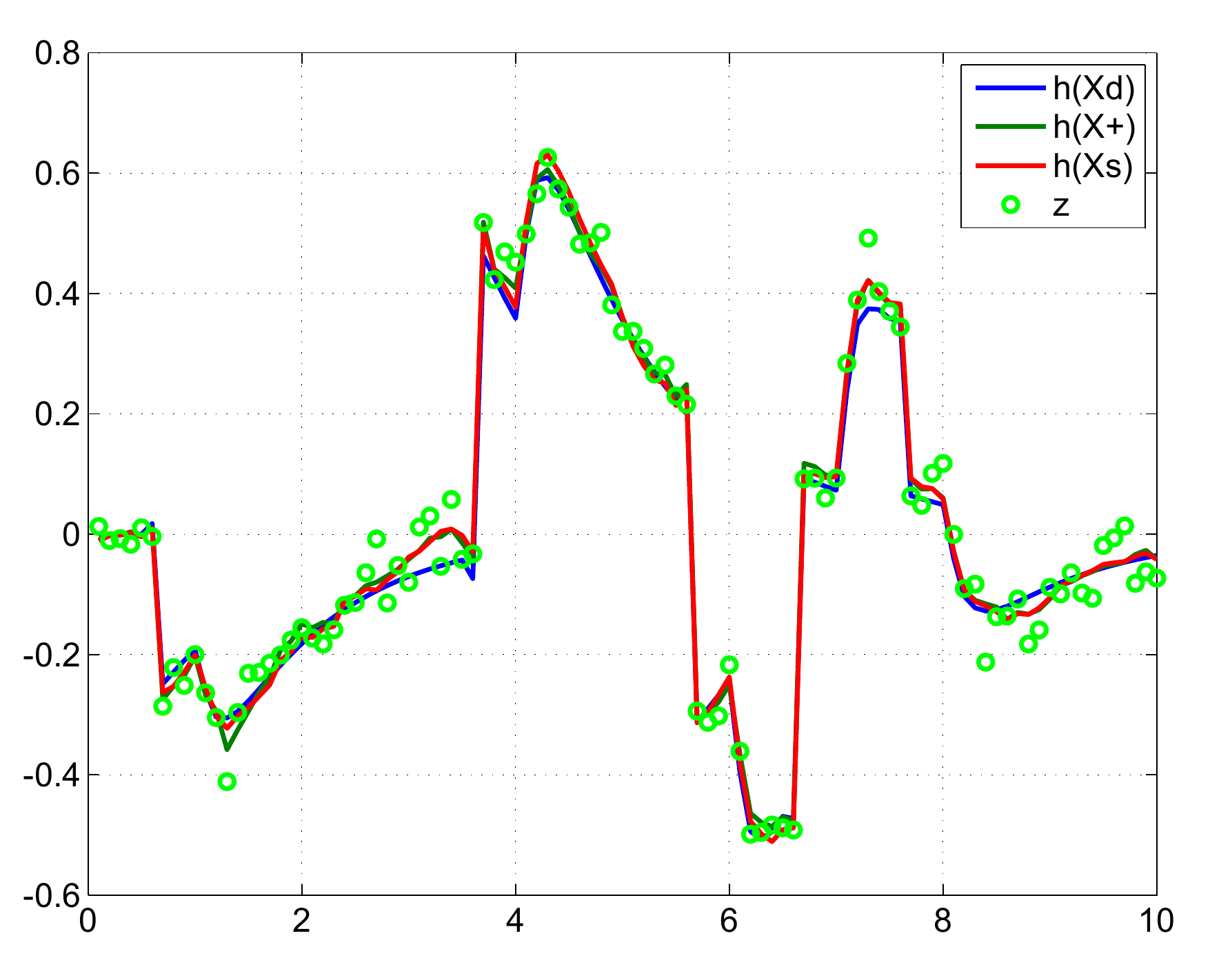}
\caption{Comparison of the predicted dynamics, posterior, smoothed}
\caption*{and the measurement 7}
\label{latQ_h7}
\end{figure}

\begin{figure}[h]
\includegraphics[width=6in,height=4in]{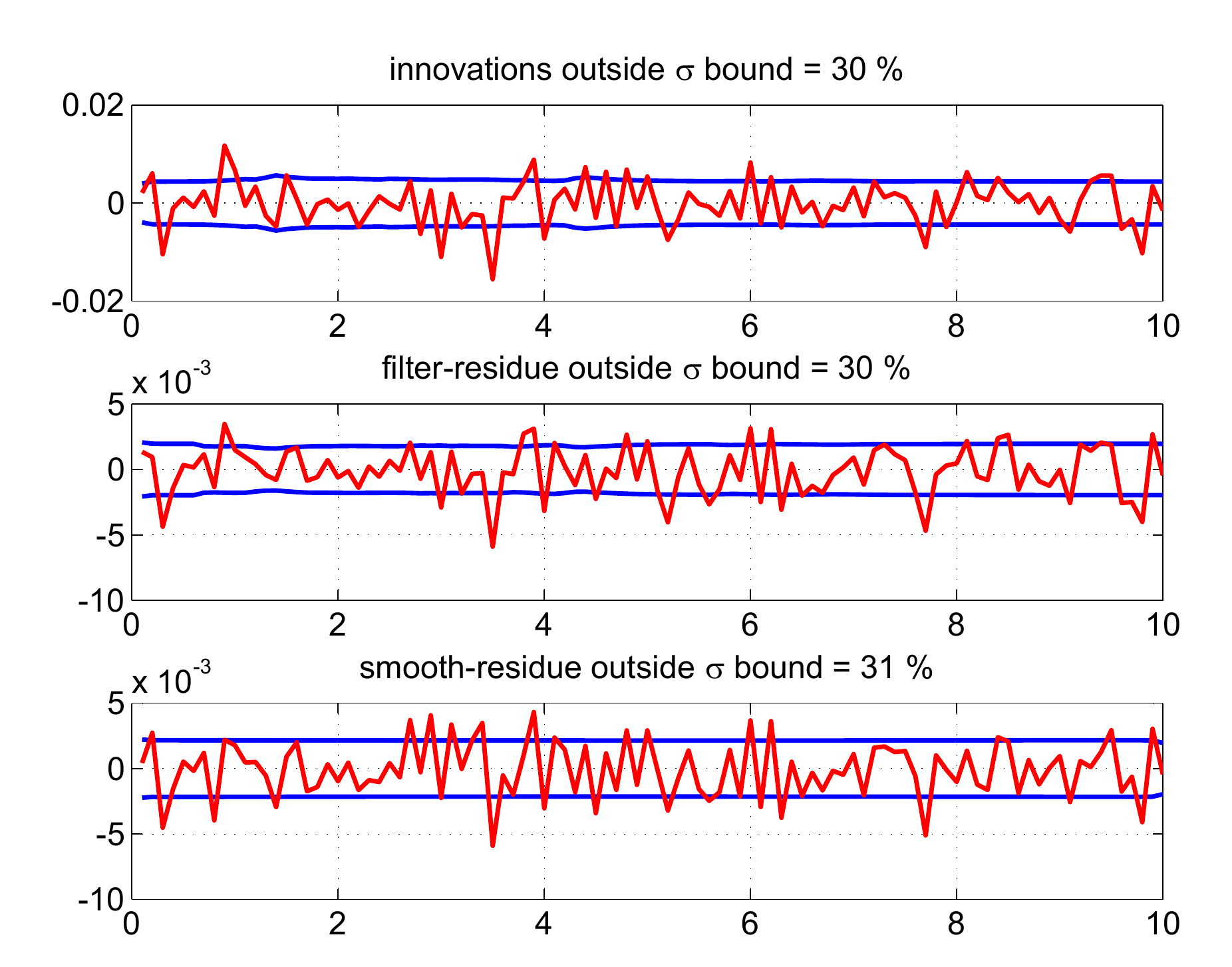}
\caption{The innovations, filtered residue and smoothed residue}
\caption*{corresponding to measurement 1 }
\label{latQ_innov1}
\end{figure}

\begin{figure}[h]
\includegraphics[width=6in,height=4in]{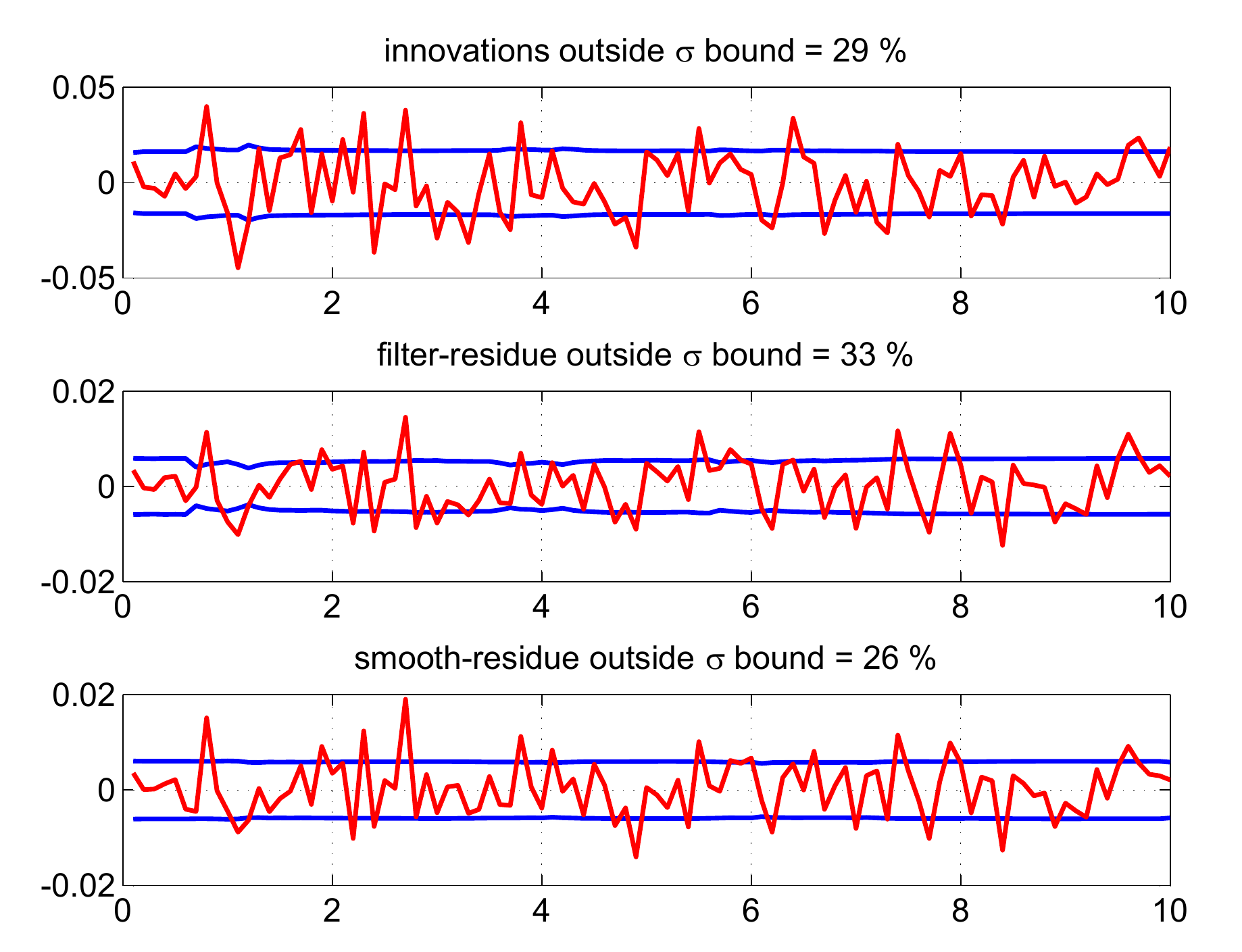}
\caption{The innovations, filtered residue and smoothed residue}
\caption*{corresponding to measurement 2}
\label{latQ_innov2}
\end{figure}

\begin{figure}[h]
\includegraphics[width=6in,height=4in]{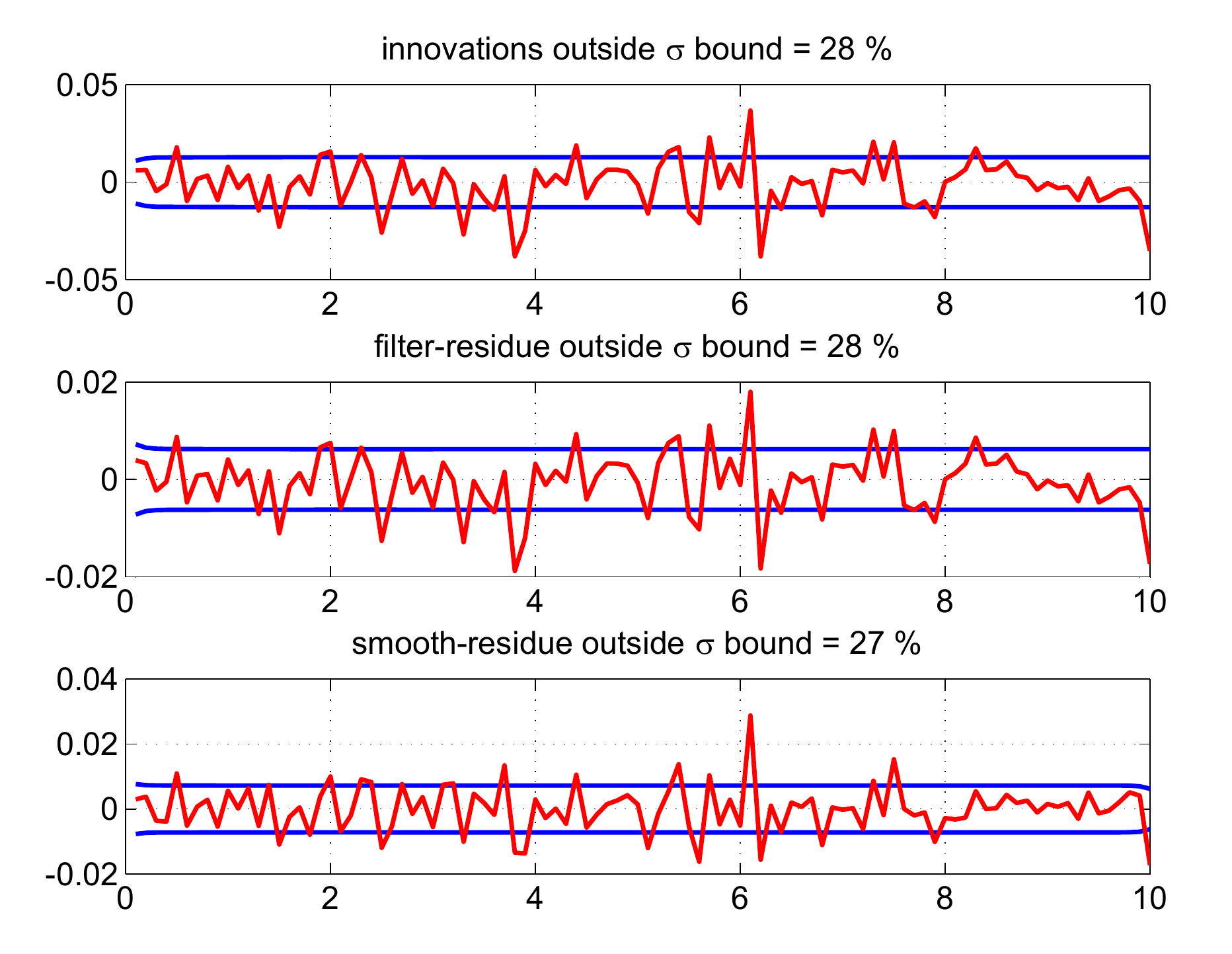}
\caption{The innovations, filtered residue and smoothed residue}
\caption*{corresponding to measurement 3 }
\label{latQ_innov3}
\end{figure}

\begin{figure}[h]
\includegraphics[width=6in,height=4in]{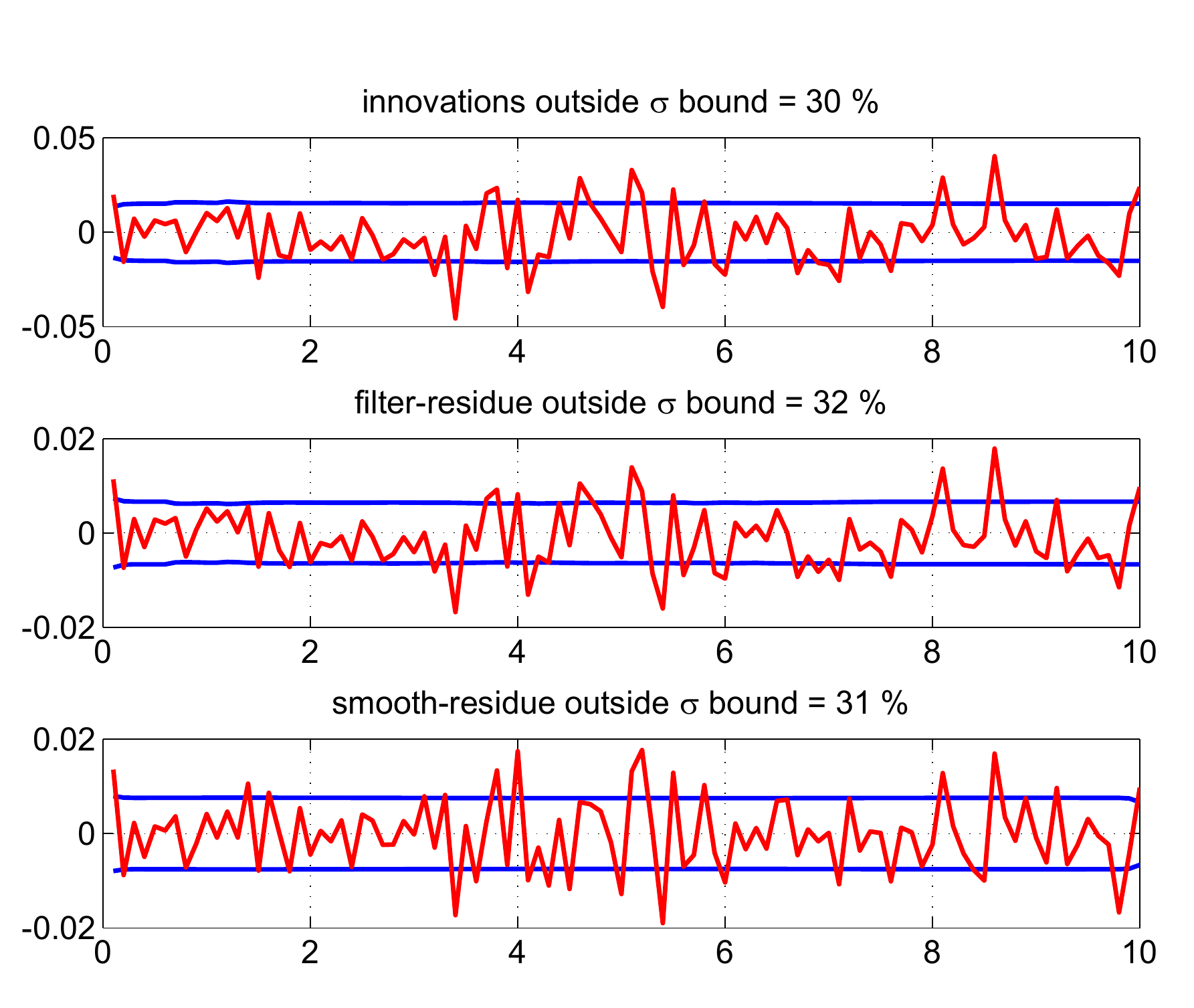}
\caption{The innovations, filtered residue and smoothed residue}
\caption*{corresponding to measurement 4}
\label{latQ_innov4}
\end{figure}

\begin{figure}[h]
\includegraphics[width=6in,height=4in]{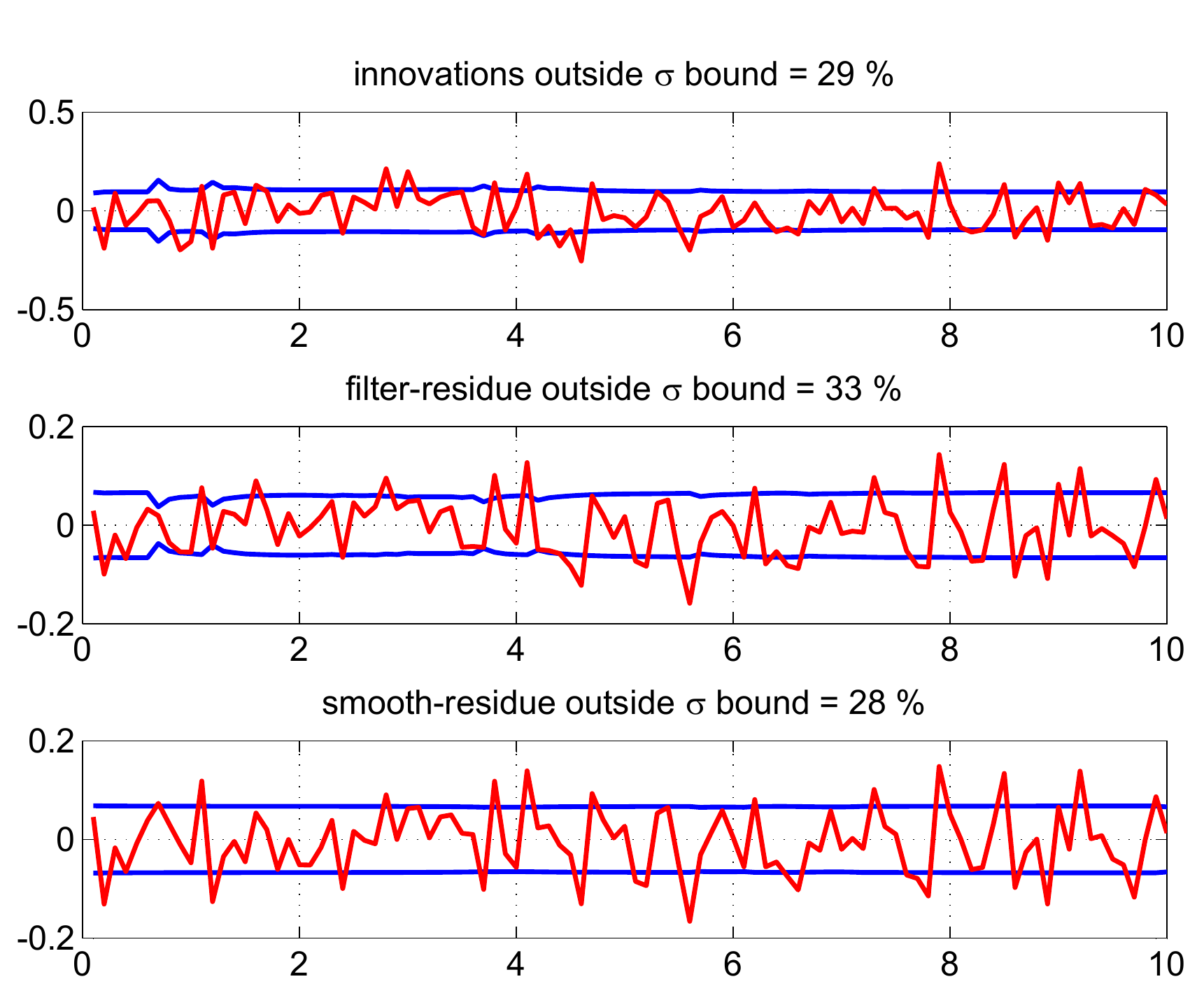}
\caption{The innovations, filtered residue and smoothed residue}
\caption*{corresponding to measurement 5}
\label{latQ_innov5}
\end{figure}

\begin{figure}[h]
\includegraphics[width=6in,height=4in]{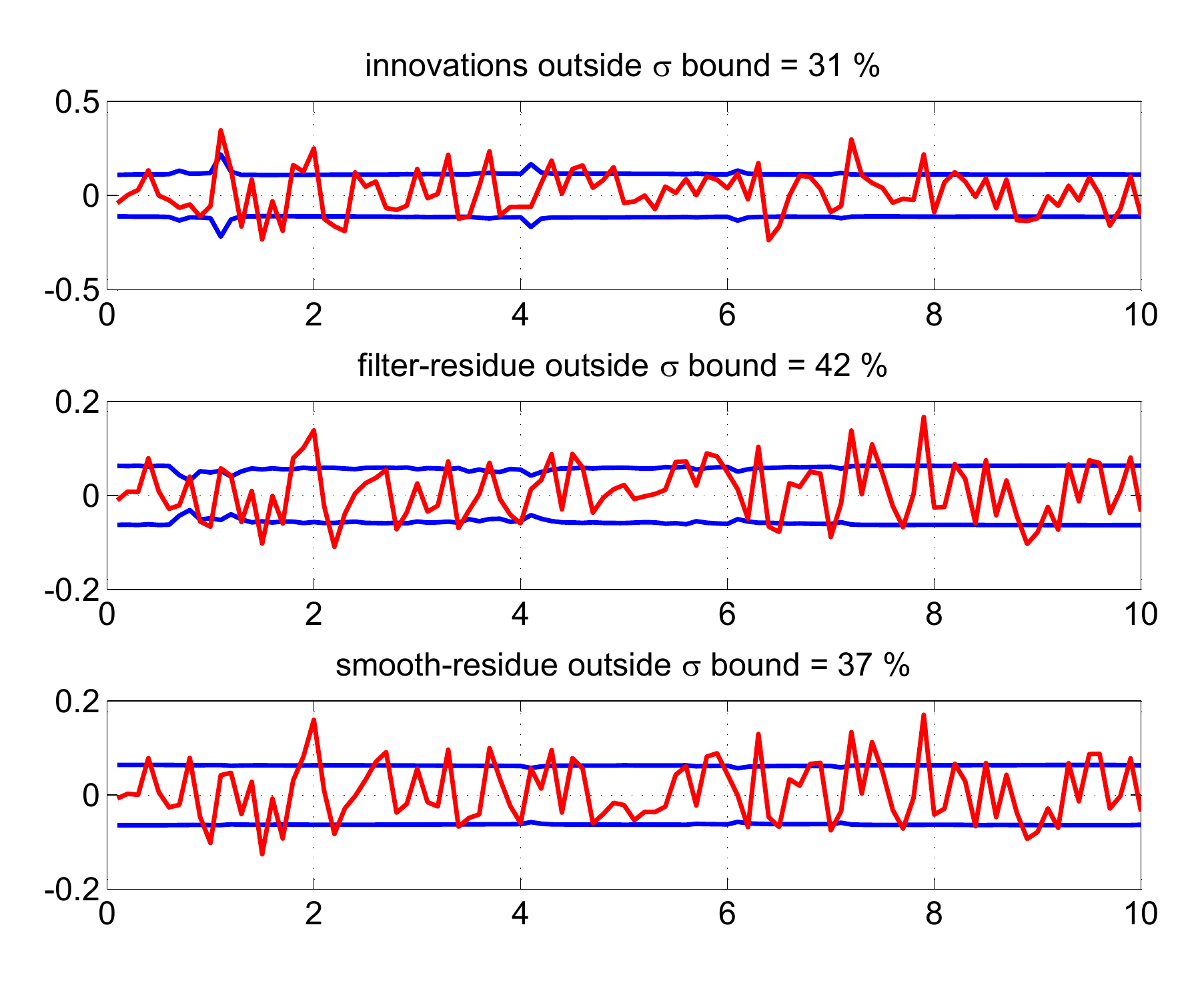}
\caption{The innovations, filtered residue and smoothed residue}
\caption*{corresponding to measurement 6}
\label{latQ_innov6}
\end{figure}

\begin{figure}[h]
\includegraphics[width=6in,height=4in]{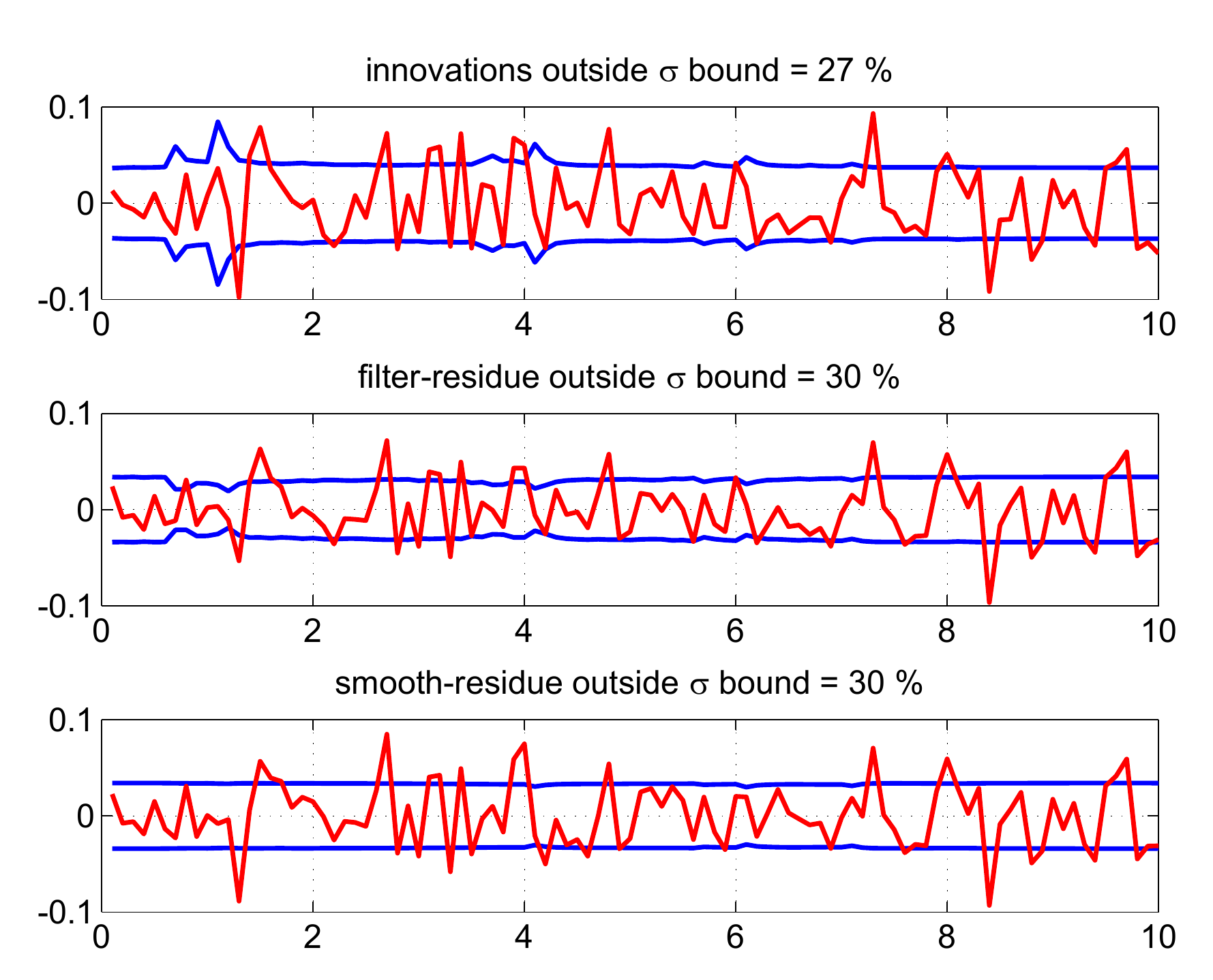}
\caption{The innovations, filtered residue and smoothed residue}
\caption*{corresponding to measurement 7}
\label{latQ_innov7}
\end{figure}

\begin{figure}[h]
\includegraphics[width=6in,height=4in]{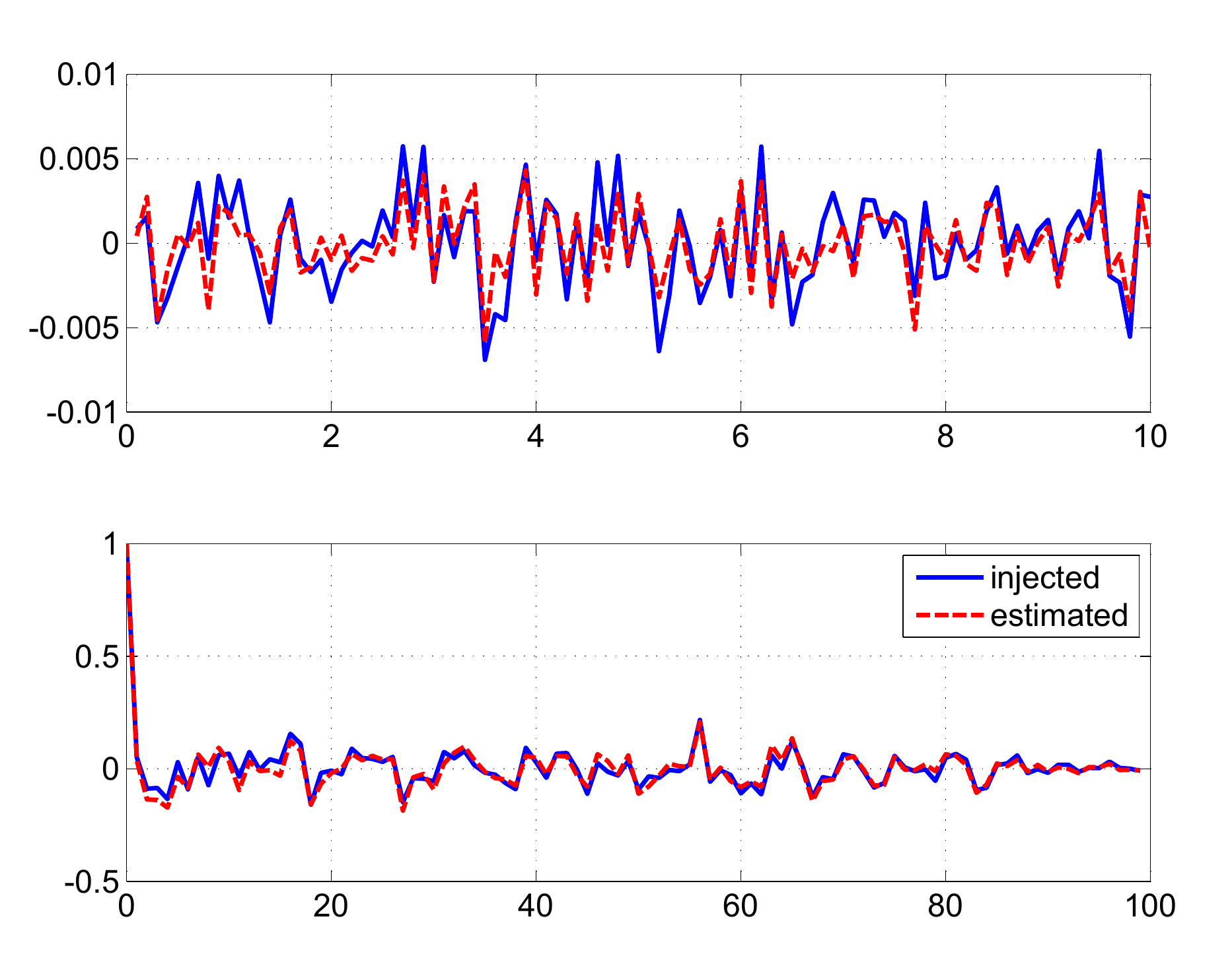}
\caption{Time variation of injected and estimated measurement noise (top) and}
\caption*{their autocorrelation (bottom) for measurement 1}
\label{latQ_mnoise1}
\end{figure}

\begin{figure}[h]
\includegraphics[width=6in,height=4in]{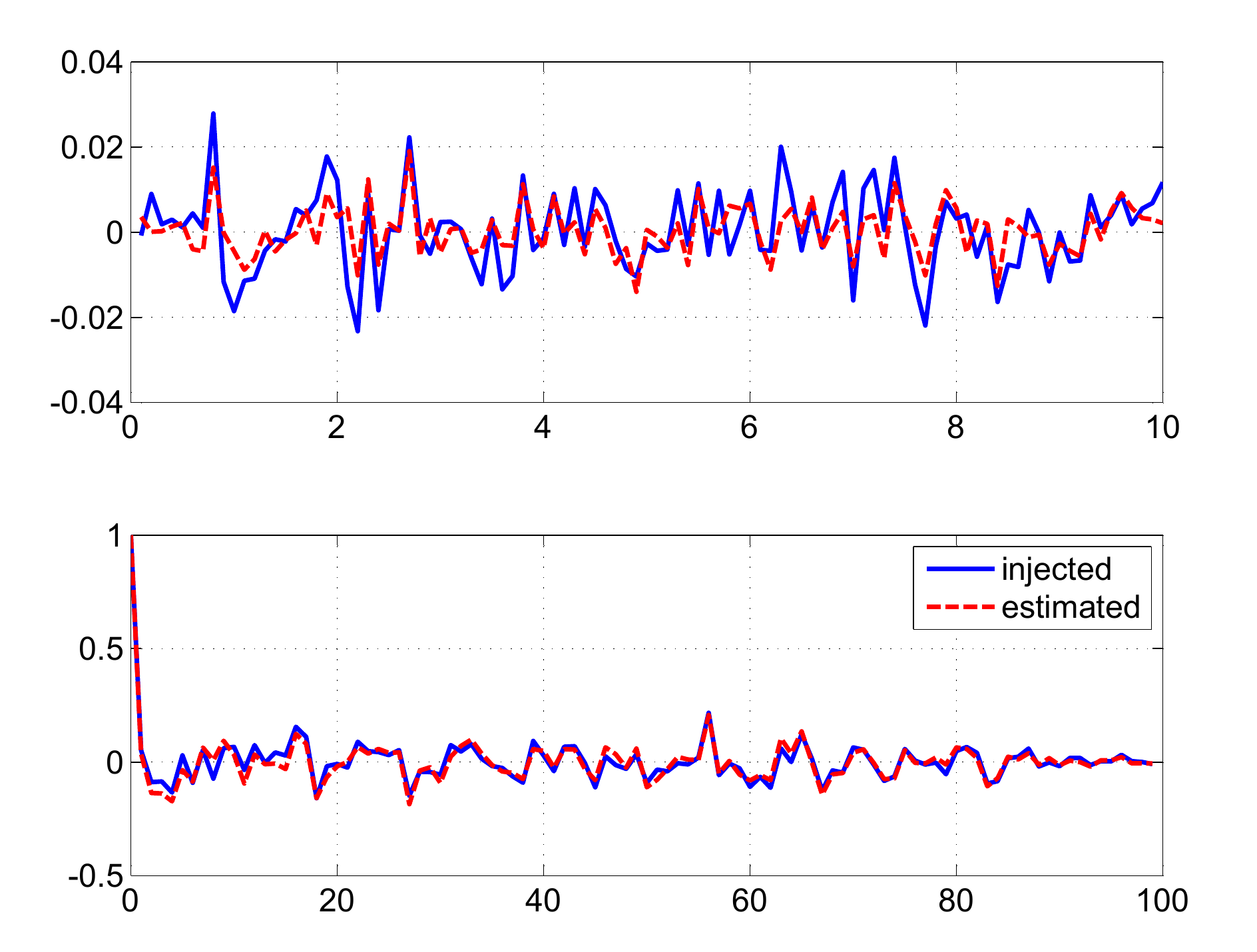}
\caption{Time variation of injected and estimated measurement noise (top) and}
\caption*{their autocorrelation (bottom) for measurement 2}
\label{latQ_mnoise2}
\end{figure}

\begin{figure}[h]
\includegraphics[width=6in,height=4in]{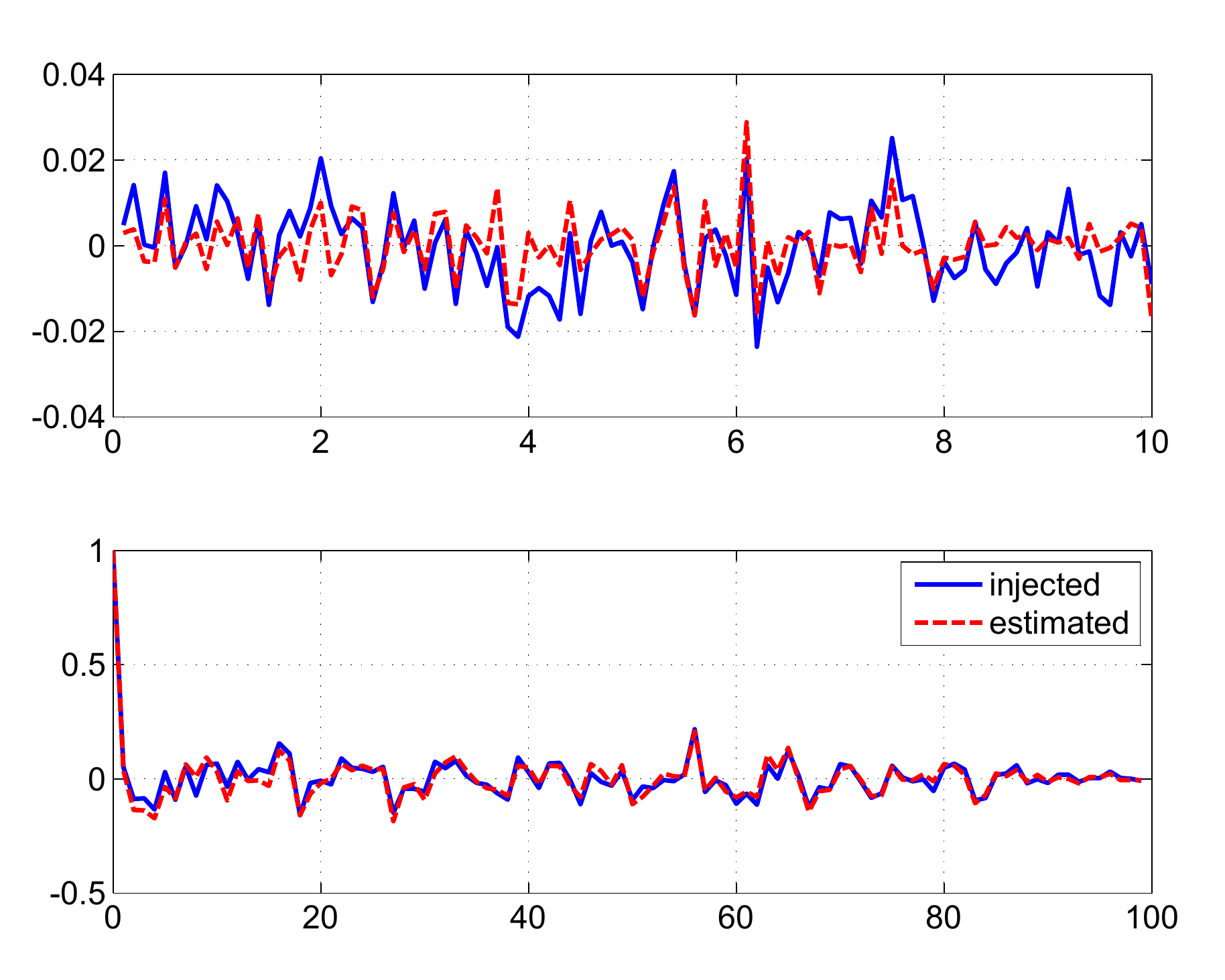}
\caption{Time variation of injected and estimated measurement noise (top) and}
\caption*{their autocorrelation (bottom) for measurement 3}
\label{latQ_mnoise3}
\end{figure}

\begin{figure}[h]
\includegraphics[width=6in,height=4in]{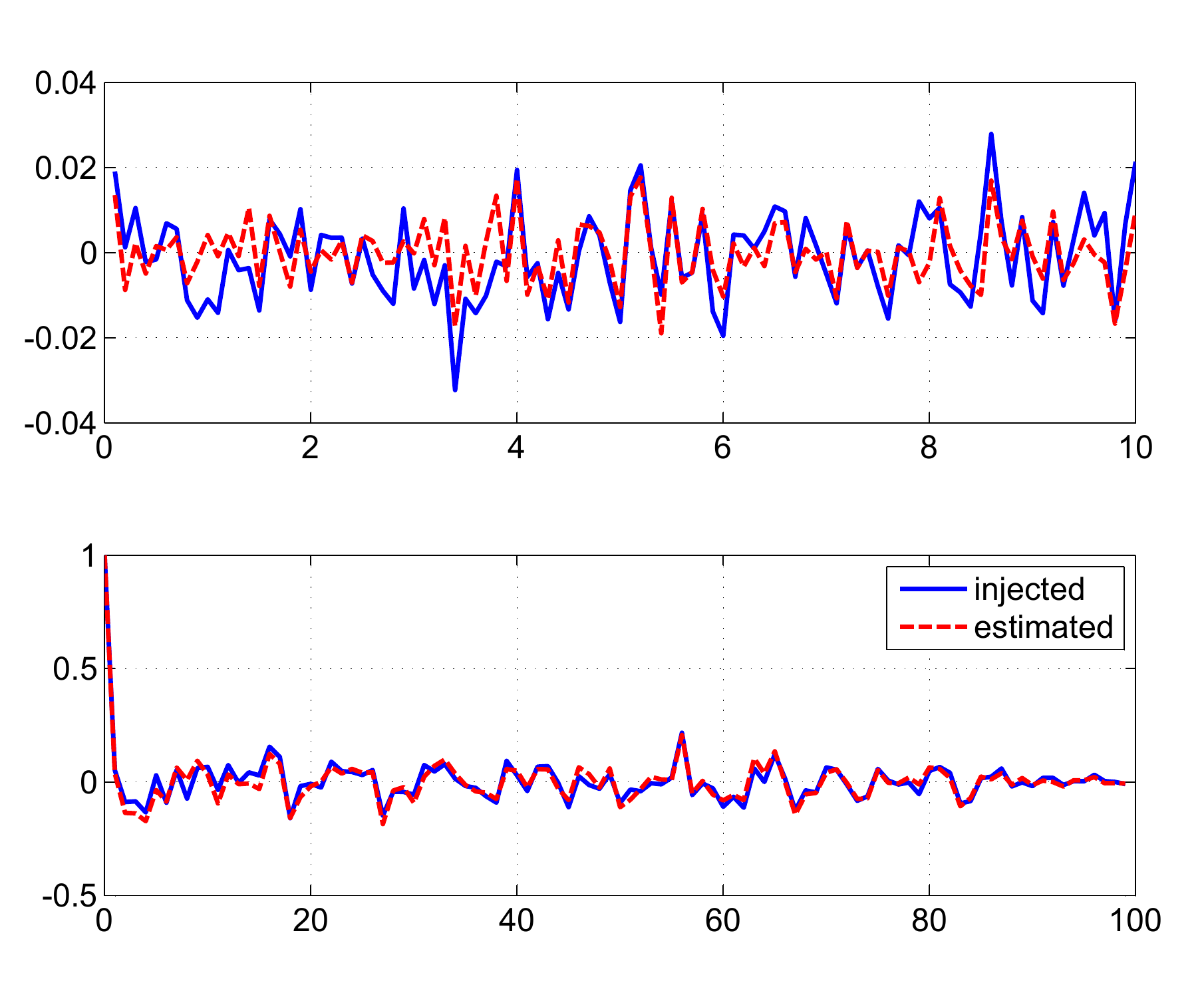}
\caption{Time variation of injected and estimated measurement noise (top) and}
\caption*{their autocorrelation (bottom) for measurement 4}
\label{latQ_mnoise4}
\end{figure}

\begin{figure}[h]
\includegraphics[width=6in,height=4in]{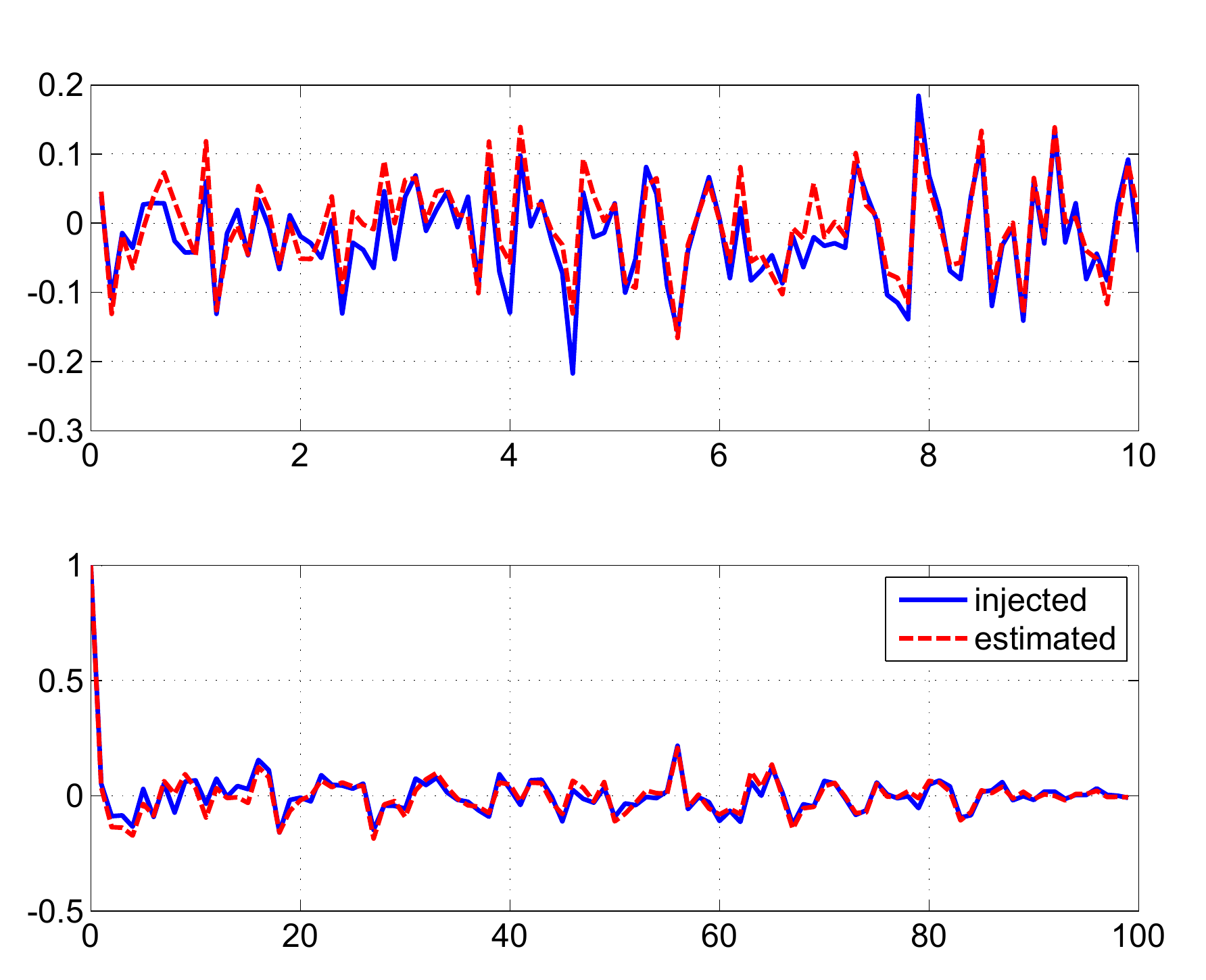}
\caption{Time variation of injected and estimated measurement noise (top) and}
\caption*{their autocorrelation (bottom) for measurement 5}
\label{latQ_mnoise5}
\end{figure}

\begin{figure}[h]
\includegraphics[width=6in,height=4in]{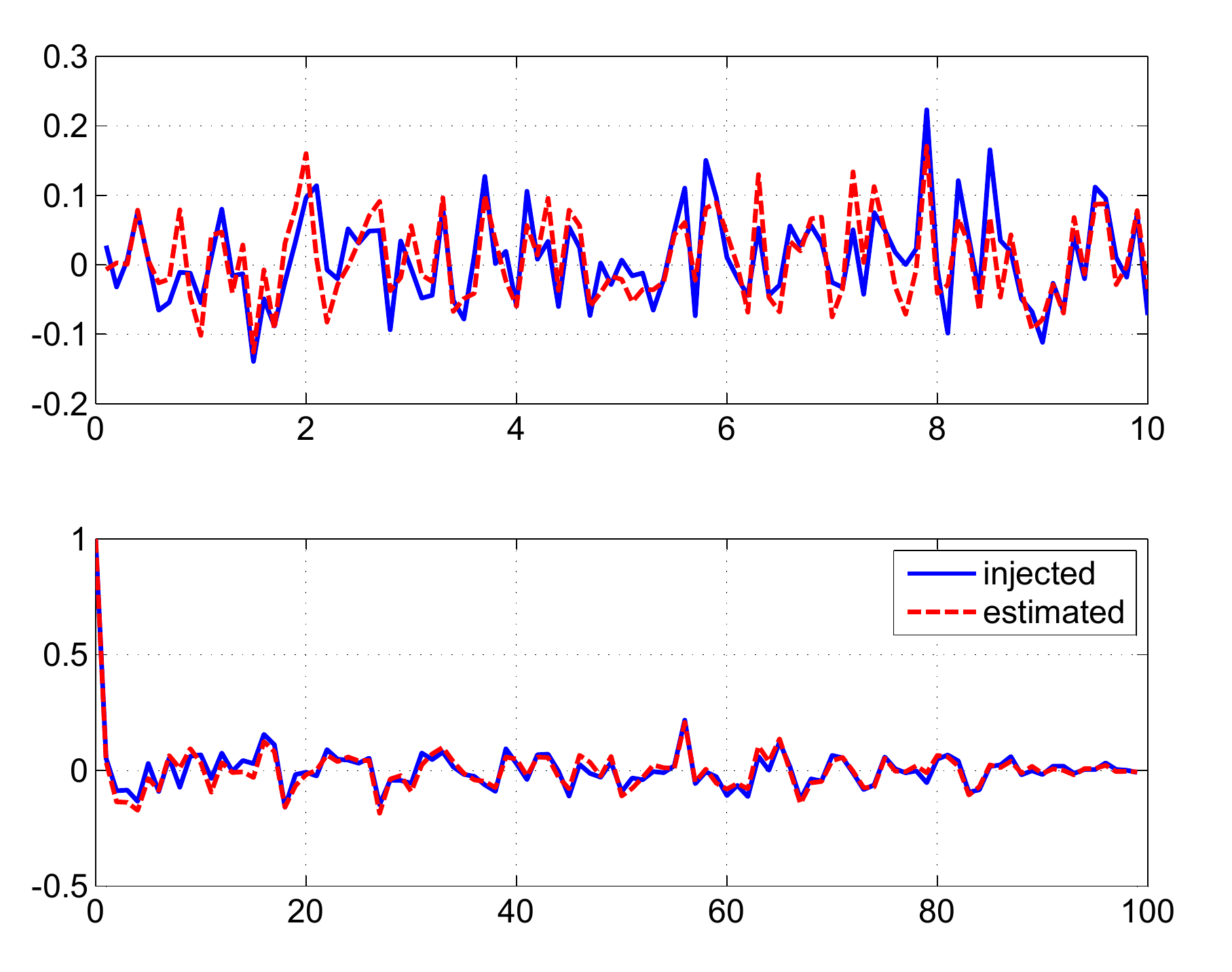}
\caption{Time variation of injected and estimated measurement noise (top) and}
\caption*{their autocorrelation (bottom) for measurement 6}
\label{latQ_mnoise6}
\end{figure}

\begin{figure}[h]
\includegraphics[width=6in,height=4in]{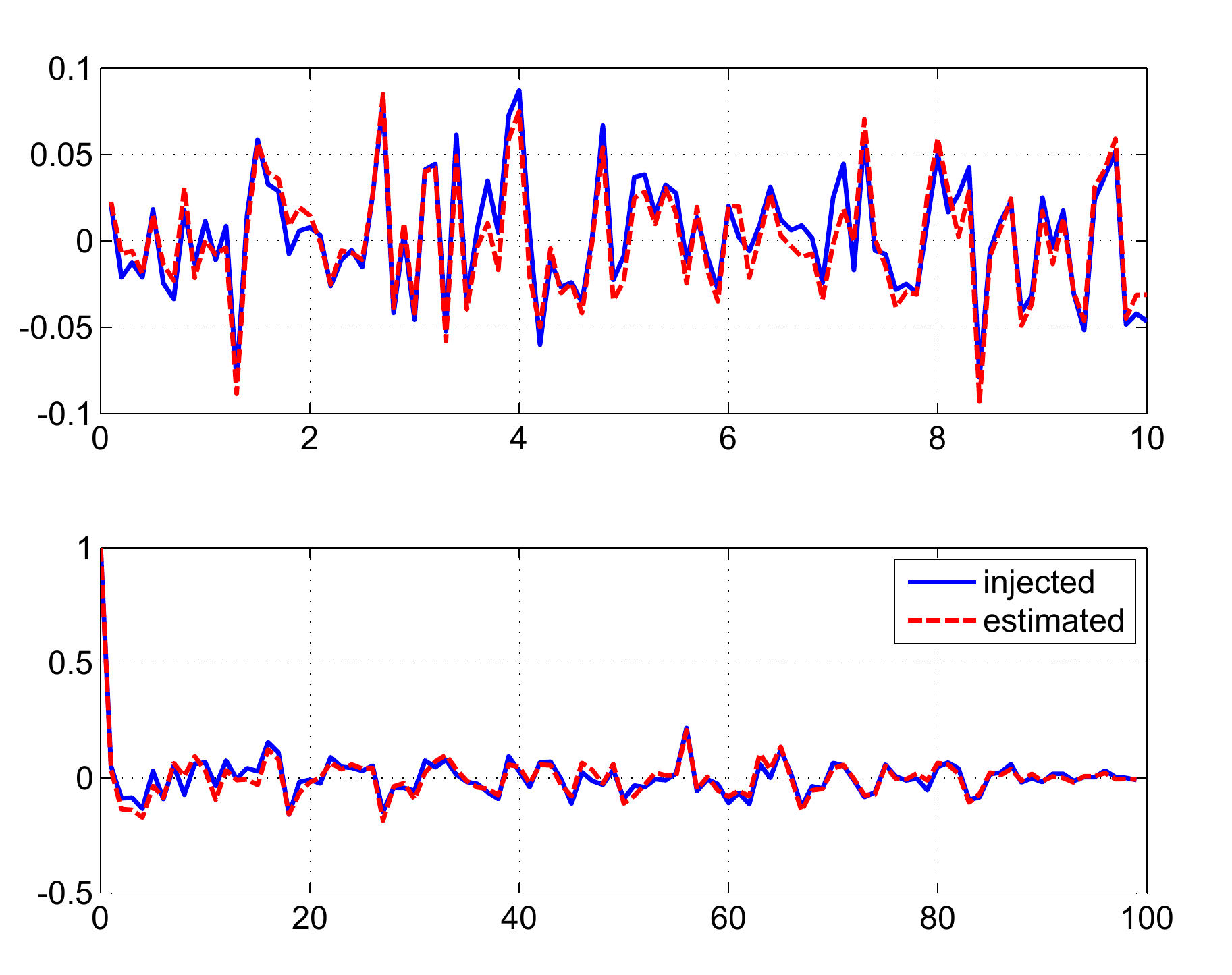}
\caption{Time variation of injected and estimated measurement noise (top) and}
\caption*{their autocorrelation (bottom) for measurement 7}
\label{latQ_mnoise7}
\end{figure}

\begin{figure}[h]
\includegraphics[width=6in,height=4in]{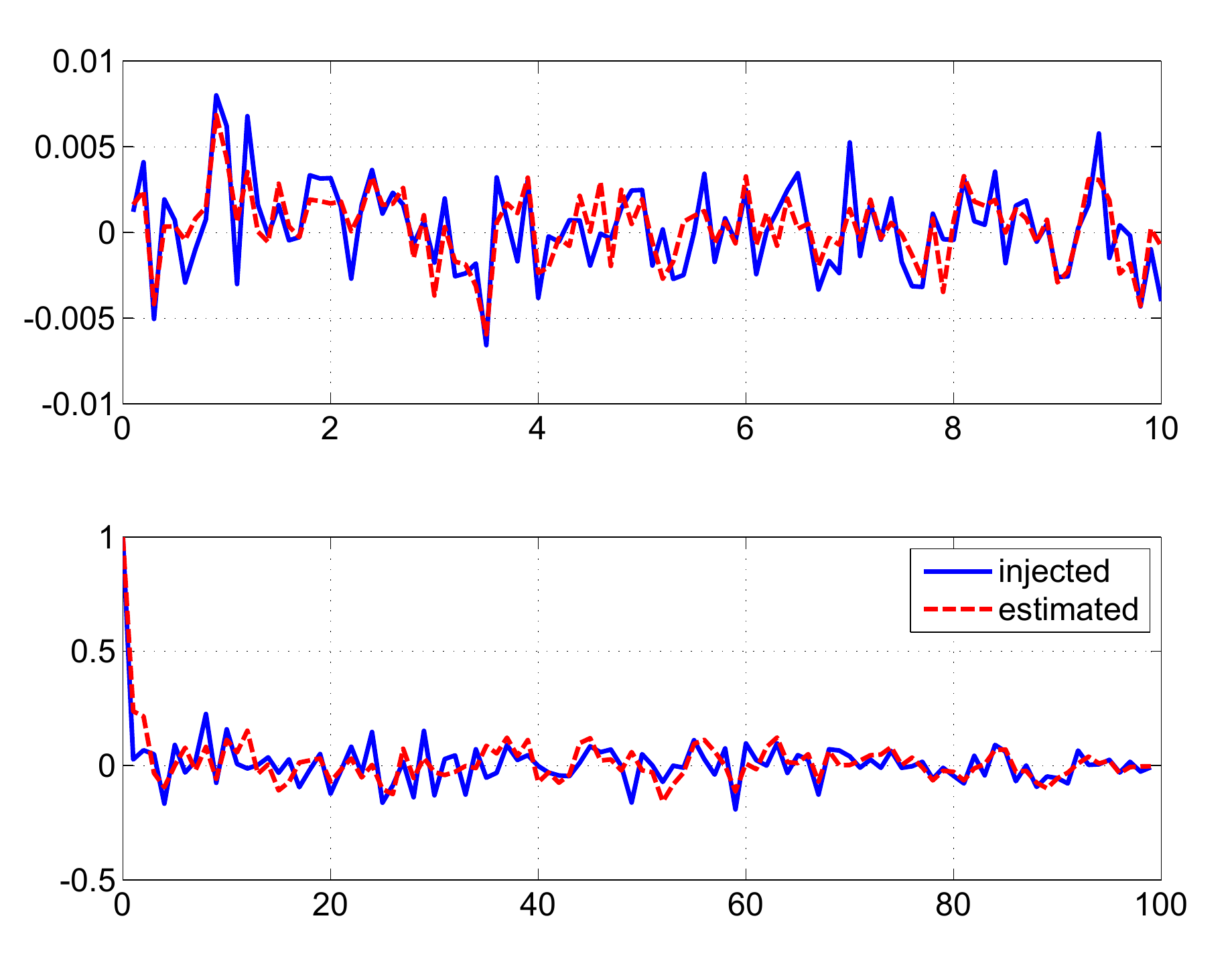}
\caption{Time variation of injected and estimated process noise (top) and}
\caption*{their autocorrelation (bottom) for state 1}
\label{latQ_pnoise1}
\end{figure}

\begin{figure}[h]
\includegraphics[width=6in,height=4in]{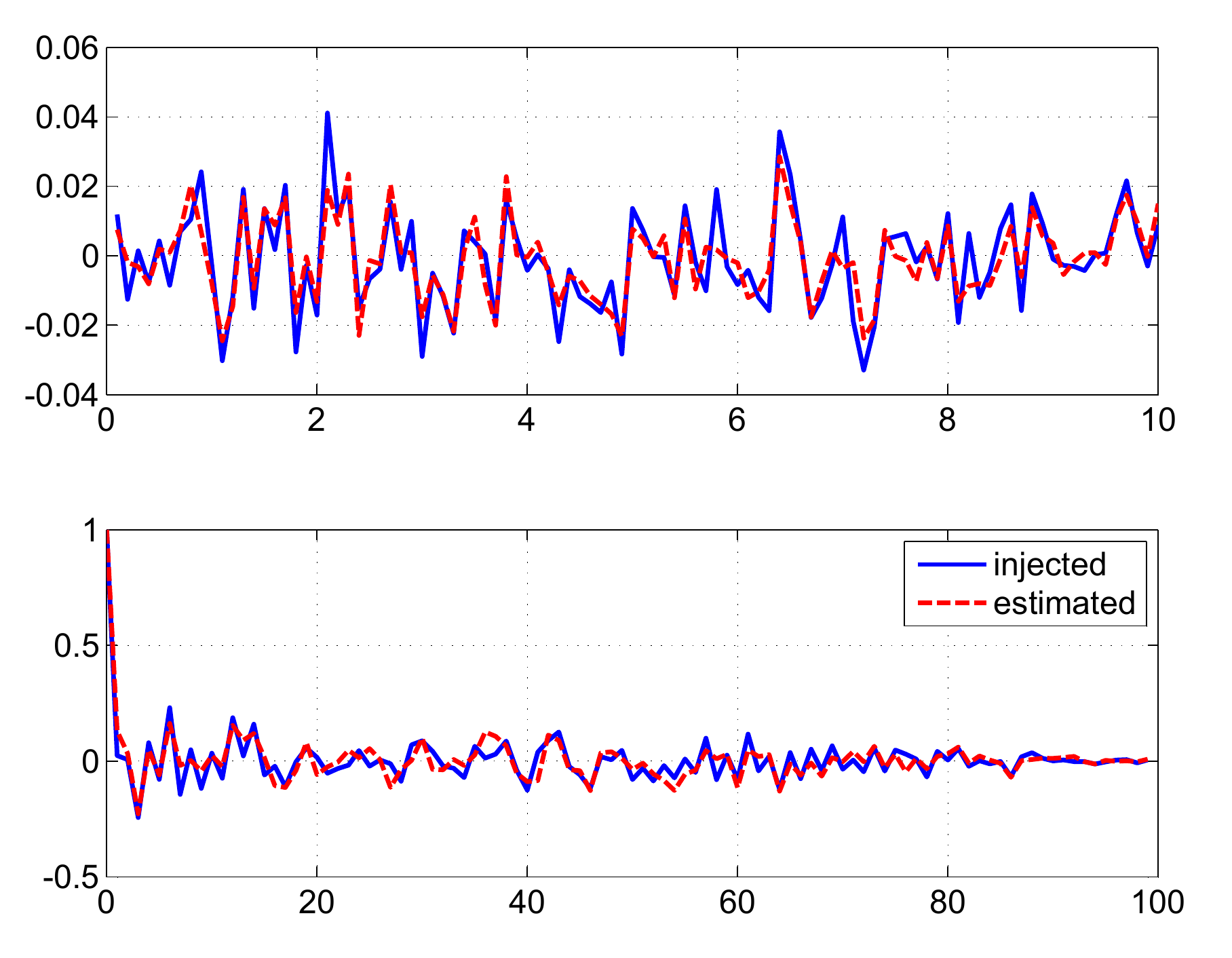}
\caption{Time variation of injected and estimated process noise (top) and}
\caption*{their autocorrelation (bottom) for state 2}
\label{latQ_pnoise2}
\end{figure}

\begin{figure}[h]
\includegraphics[width=6in,height=4in]{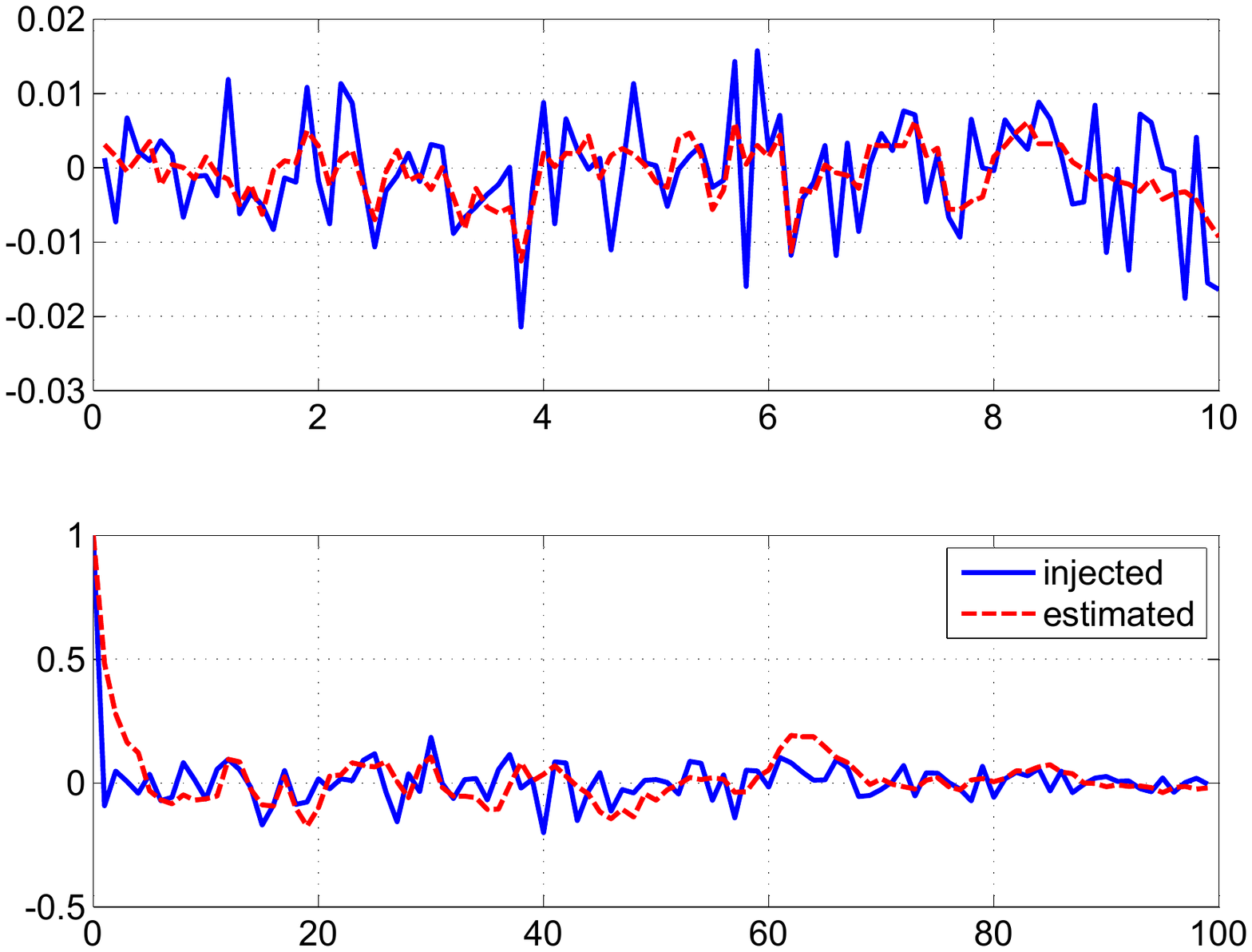}
\caption{Time variation of injected and estimated process noise (top) and}
\caption*{their autocorrelation (bottom) for state 3}
\label{latQ_pnoise3}
\end{figure}

\begin{figure}[h]
\includegraphics[width=6in,height=4in]{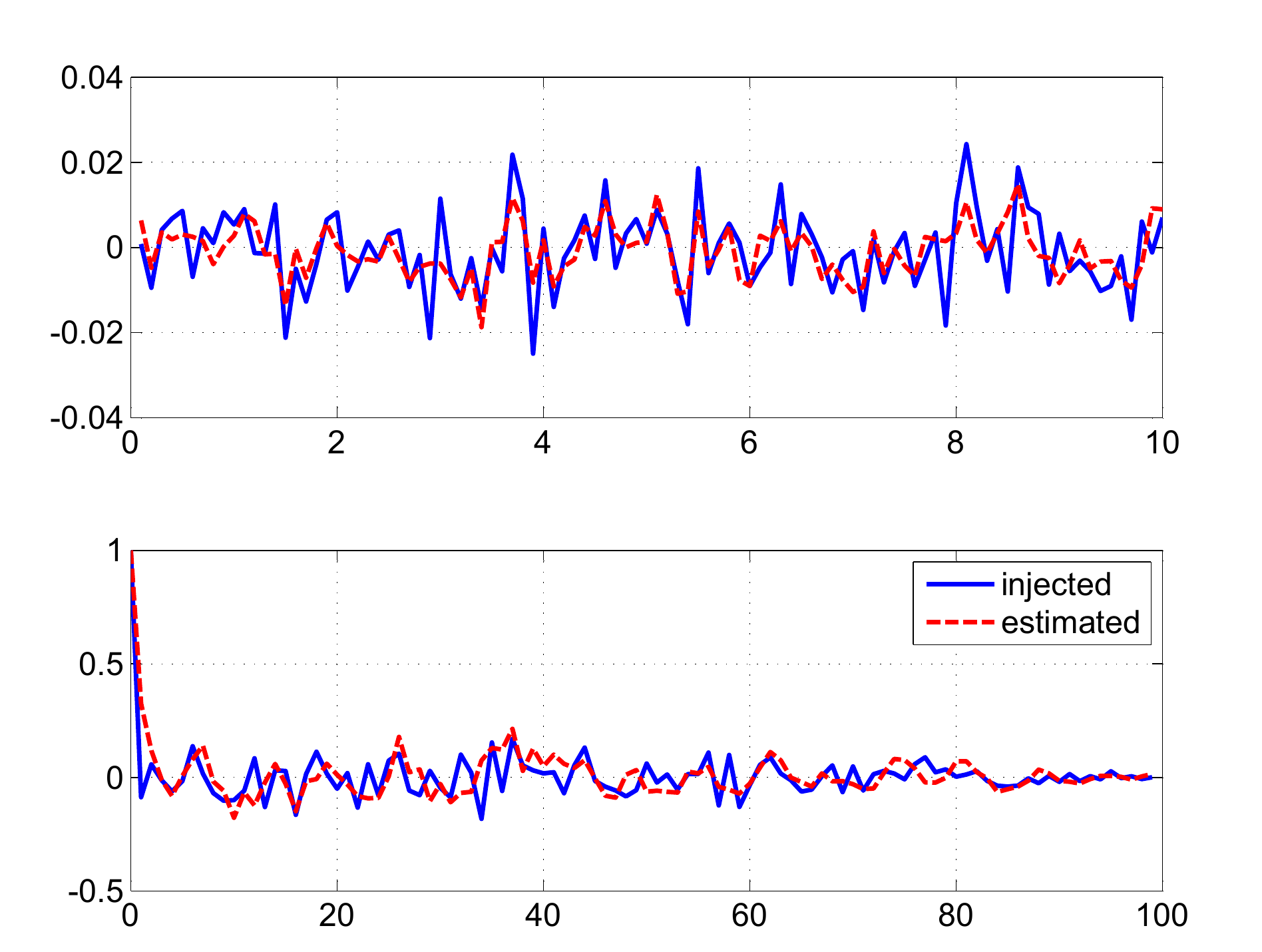}
\caption{Time variation of injected and estimated process noise (top) and}
\caption*{their autocorrelation (bottom) for state 4}
\label{latQ_pnoise4}
\end{figure}

The next step is to process the data with process noise (\textbf{Q} $>$ 0). The Fig. \ref{lat_err} shows the absolute difference between the iterated and final values with iterations which indicates the accuracy level that one needs and it was found that 100 iterations are required. The variation of the estimated initial parameters and their variances through iterations are shown in Fig. \ref{latQ_P0}. The parameter and the uncertainty reach almost their final estimated values in about 5 and 20 iterations respectively. A similar plot in Fig. \ref{latQ_R} shows the variation of the estimated \textbf{R} and \textbf{Q}. The variation of different cost functions through the iterations are shown in Fig. \ref{latQ_J}. The cost functions \textbf{J1-J3} correspond to the number of measurement ($m$=7) and in presence of process noise, \textbf{J6-J8} correspond to the number of states ($n$=4). The \textbf{J4} in absence of process noise corresponds to the trace of \textbf{R}. The \textbf{J5} is the negative log likelihood cost whose absolute value is shown in the plot. The mismatch in the predicted dynamics and the measurements in Fig. \ref{latQ_h1}-\ref{latQ_h7} indicates the presence of \textbf{Q}. The subsequent Fig. \ref{latQ_innov1}-\ref{latQ_innov7} correspond to the earlier Fig. \ref{lat_innov1}-\ref{lat_innov7} of \textbf{Q} = 0 case. The Fig.  \ref{latQ_mnoise1}-\ref{latQ_mnoise7} and Fig. \ref{latQ_pnoise1}-\ref{latQ_pnoise4} shows respectively the injected and estimated measurement and process noise samples during the final iteration.

\begin{landscape}
\begin{table}[h]
\subsection{Lateral Motion of Aircraft System Tables (\textbf{Q} = 0) }
\caption{Sensitivity Study : (\textbf{Q} = 0) : Lateral Aircraft system.\\ No. of iterations=20, No. of simulations=50.}{}
\label{tblat}
\begin{center}
\begin{footnotesize}
\begin{tabular}{| c | c | c | c | c | c | c | c | c | c |}
\hline
Study &
$\Theta$ ratio &
\makecell{CRB ratio\\NR/EKF} &
\makecell{Consistency\\ ratio-EKF}&
\makecell{Spread\\factor\\EKF} &
\makecell{\textbf{R} ratio\\EKF/True} & \makecell{\textbf{R} ratio\\EKF/NR} &
\makecell{$\mu$\ of \\\textbf{J1-J5}} &
\makecell{$\sigma$ of \\\textbf{J1-J5}} &
Remarks
\\ \hline

\makecell{$\mathbf{P_0}$ : Scaled up-[0,0;0,\checkmark]\\\textbf{Q} : EM-[\checkmark,0;0,0] \\ \textbf{R} : EM-diag}
&\multicolumn{9}{c|}{Reference adaptive EKF (\textbf{Q} $>$ 0) gives extremely slow convergence of \textbf{Q} taking more than 100 iterations.} \\ \hline

\multicolumn{10}{|c|}{\makecell{Proposed reference adaptive EKF for zero process noise case, \textbf{Q} is diagonal matrix fixed to $10^{-10}$ for all states \\
Consistency ratio NR=   1.1692,    0.9759,    1.0174,    0.9121,    1.0440,    0.7969,    0.9606, 1.1951,    1.0610   , 1.0618,    0.9218,    1.0657,    0.9702,    1.1985, 1.2637
\\
Spread Factor NR=   1.0150,  153.8666,    1.2514,   60.5416,    1.7035,    2.0545,  199.2658,  8.8451,    2.1629,   32.2073,    5.1724,    1.5438,   22.8668,   36.5847, 1.4793}} \\
\end{tabular}
\end{footnotesize}
\end{center}
\end{table}
\end{landscape}

\begin{landscape}
\begin{table}[h]
\addtocounter{table}{-1}
\caption{Sensitivity Study : (\textbf{Q} = 0) : Lateral Aircraft system.\\ No. of iterations=20, No. of simulations=50.}{}
\begin{center}
\begin{footnotesize}
%
\end{footnotesize}
\end{center}
\end{table}
\end{landscape}

\begin{landscape}
\begin{table}[h]
\addtocounter{table}{-1}
\caption{Sensitivity Study : (\textbf{Q} = 0) : Lateral Aircraft system.\\ No. of iterations=20, No. of simulations=50.}{}
\begin{center}
\begin{footnotesize}
%
\end{footnotesize}
\end{center}
\end{table}
\end{landscape}

\begin{landscape}
\begin{table}[h]
\addtocounter{table}{-1}
\caption{Sensitivity Study : (\textbf{Q} = 0) : Lateral Aircraft system.\\ No. of iterations=20, No. of simulations=50.}{}
\begin{center}
\begin{footnotesize}
%
\end{footnotesize}
\end{center}
\end{table}
\end{landscape}

\begin{landscape}
\begin{table}[h]
\subsection{Lateral Motion of Aircraft System Tables (\textbf{Q} $>$ 0) }
\vspace{14pt}
\caption{Sensitivity Study : (\textbf{Q} $>$ 0) : Lateral Aircraft system.\\ No. of iterations=100, No. of simulations=50.}{}
\label{tblatQ}
\begin{center}
\begin{footnotesize}
%
\end{footnotesize}
\end{center}
\end{table}
\end{landscape}

\begin{landscape}
\begin{table}[h]
\addtocounter{table}{-1}
\caption{Sensitivity Study : (\textbf{Q} $>$ 0) : Lateral Aircraft system.\\ No. of iterations=100, No. of simulations=50.}{}
\begin{center}
\begin{footnotesize}
%
\end{footnotesize}
\end{center}
\end{table}
\end{landscape}

\begin{landscape}
\begin{table}[h]
\addtocounter{table}{-1}
\caption{Sensitivity Study : (\textbf{Q} $>$ 0) : Lateral Aircraft system.\\ No. of iterations=100, No. of simulations=50.}{}
\begin{center}
\begin{footnotesize}
%
\end{footnotesize}
\end{center}
\end{table}
\end{landscape}

\begin{landscape}
\begin{table}[H]
\addtocounter{table}{-1}
\caption{Sensitivity Study : \textbf{Q} $>$ 0 : Lateral Aircraft system.\\ No. of iterations=100, No. of simulations=50.}{}
\begin{center}
\begin{footnotesize}
%
\end{footnotesize}
\end{center}
\end{table}
\end{landscape}

\begin{landscape}
\begin{table}[H]
\addtocounter{table}{-1}
\caption{Sensitivity Study : \textbf{Q} $>$ 0 : Lateral Aircraft system.\\ No. of iterations=100, No. of simulations=50.}{}
\begin{center}
\begin{footnotesize}
%
\end{footnotesize}
\end{center}
\end{table}
\end{landscape}

\clearpage
\subsection{Lateral Motion of Aircraft System Figures (\textbf{Q} = 0) }

\clearpage

\clearpage
\subsection{Lateral Motion of Aircraft System Figures (\textbf{Q} $>$ 0) }

\chapter{Real Flight Test Data Analysis}
\label{ch5}
\par After having studied simulated cases for simple systems and the longitudinal and lateral motion of an airplane it was felt useful to handle more involved real life cases. It turns out that a good choice once again would be the flight test data analysis of an airplane systems involving longitudinal and lateral motion. These generally have many states and measurements and more so a larger number of aerodynamic parameters to be determined. Further the flight tests cannot always be conducted in an ideal situation of the process noise and the measurement noises being white and Gaussian as is generally needed in the Kalman filter. The measurements are not available with respect to the center of gravity, possess scale and bias factors which have to be modelled and accounted for. The coupling between the longitudinal and lateral motion brings in added difficulty making the problem more interesting. At times the noisy measurements from the lateral motion are fed into the longitudinal states and thus are input as process noise. This is another example of introducing subjectivity in estimation theory. However the final results should be meaningful, reasonable, acceptable and useful no matter whatever subjective inputs are introduced into the problem formulation and solution. \par

In the previous Chapter we considered simulated data and established the efficacy of the present reference recursive recipe both for the system parameter estimates and the statistical characteristics of the process and the measurement noises. As in the analysis of simulated data the following filter outputs are studied which provides an insight into the filter performance. The following quantites converges through the iterations

\begin{enumerate}
\item The parameter estimates $\Theta$ and their covariances $P_\Theta$.
\item The noise covariance of  \textbf{Q} and {R}.
\item The state dynamics without measurement and process noises based on the estimated parameter after the filter pass through the data (Xd), the posterior state estimate, the smoothed state and the measurement. 
\item The sample innovation, filtered residue and the smoothed residue along with their bounds which is the square root of the predicted covariances given respectively by ($\textbf{R}+H_kP_{k|k-1}H_k^T$), ($\textbf{R}-H_{k|k}P_{k|k}H_{k|k}^T$) and ($\textbf{R}-H_{k|N}P_{k|N}H_{k|N}^T$) by the filter.
\item The estimated measurement and process noise distribution with time as well as their autocorrelations.
\item The cost functions (\textbf{J1-J8}) in the final iteration as seen in the tables.
\end{enumerate}

 Since the following examples are from aerospace application we have retained their notation as it is and avoided cluttering non aeronautical readers with aeronautical terminology. There are many standard text books such as Roskam \cite{Roskam1973} (1973), or the thesis of Gemson \cite{Gemson1991} (1991) which can be referred to for the notations used here. Only the control inputs and biases are defined in the text. The initial values of $\mathbf{P_0}$, \textbf{Q} and \textbf{R} were chosen as $10^{-1}$, $10^{-1}$ and $2^{-1}$ as used in the simulated case study. All the results have stability over a wide range of the initial values of $\mathbf{P_0}$, \textbf{Q} and \textbf{R}.

\section{Real Flight Test Case-1}
\par
\begin{figure}[h]
\includegraphics[width=6in,height=3.0in]{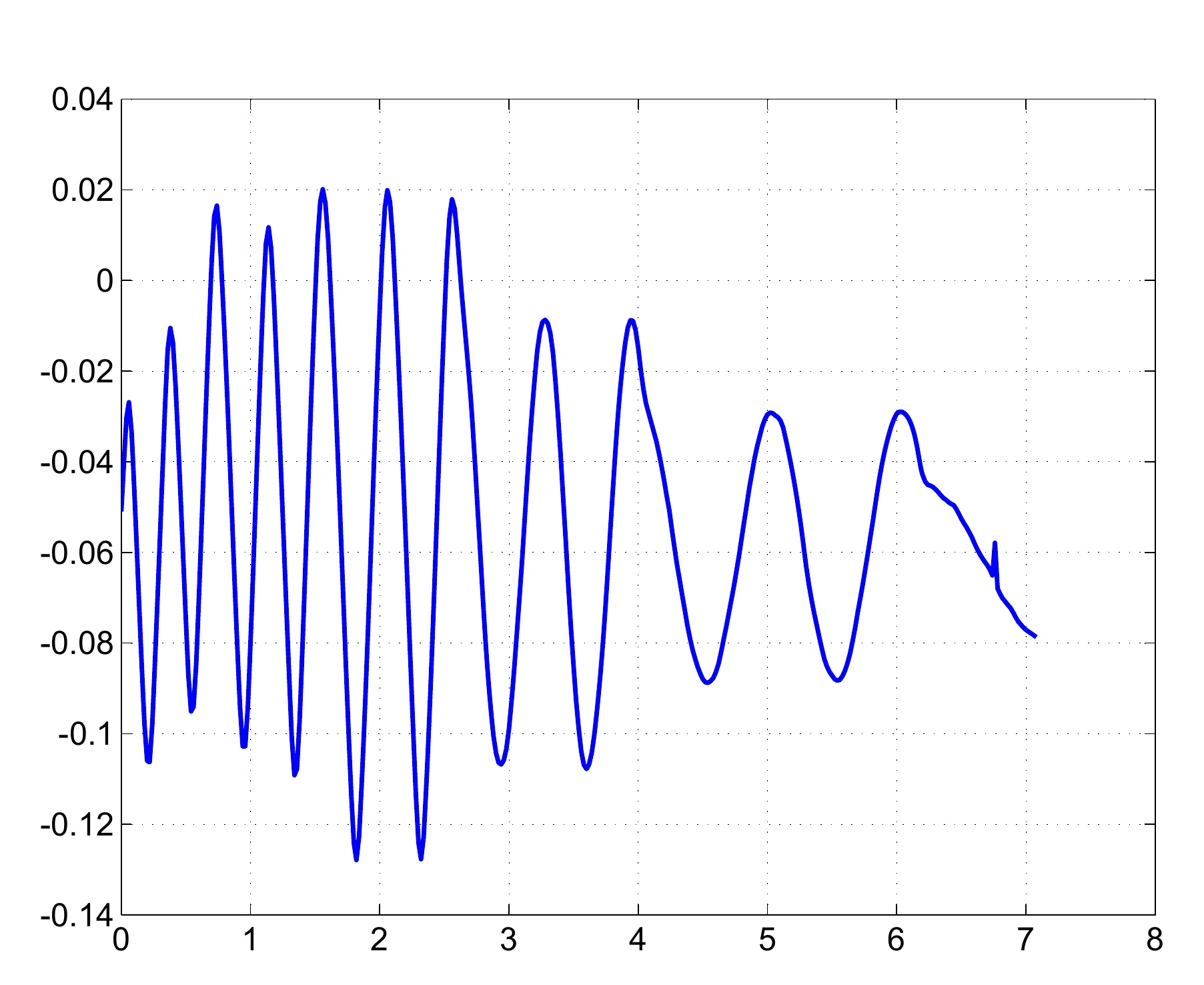}
\caption{Control input ($\delta_e$) versus time (s)}
\label{input1}
\end{figure}

 The data set pertinent to case-1 is the Aircraft B check case in NASA TN D 7831 (Maine \cite{Maine1975} 1975). The short period motion is excited by a elevator control input ($\delta_e$ in radians) as shown in Fig. \ref{input1}. The total number of data points are N = 352 at regular interval of 20 millisecs. The analysis is carried out in FPS units. The data is processed using linear model with dimensional form of stability and control derivatives. The state equations ($n=3$) for the angle of attack ($\alpha$), pitch rate (q) and the pitch angle ($\Theta$) respectively are
\begin{align*}
\dot{\alpha}&=Z_{\alpha}\alpha+q-Z_\Theta\theta+Z_{\delta_e}\delta_e+Z_0\\
\dot{q}&=M_{\alpha}\alpha+M_{q}q+M_{\delta_e}\delta_e+M_0\\
\dot{\Theta}&=Cq+\Theta_0
\end{align*}
The angle of attack ($\alpha$), pitch rate (q), the pitch angle ($\Theta$) and the normal acceleration ($a_n$) are measured (indicated with subscript `m') in units of rad, rad/s, rad and $ft/s^2$  respectively.  The measurement equations ($m$=4) are given by
\begin{align*}
{\alpha_m}&=\alpha \\
{q_m}&=q\\
{\Theta_m}&=\Theta\\
{a_{n_m}}&=-\frac{U_0}{g}(Z_{\alpha}\alpha+Z_{\delta_e}\delta_e+Z_0)+\frac{x_{a_n}}{g}\dot{q}+a_{n_b}
\end{align*}

The unknown parameter set ($p=9$) is $\Theta=(Z_{\alpha},M_\alpha,M_q,Z_{\delta_e},M_{\delta_e},Z_0,M_0,\Theta_0,a_{n_b})^T$. The first three are aerodynamic parameters, the next two are the control derivatives and the last three are the biases in the state and measurements. The initial states are taken as initial measurement and the initial parameter values are $(-0.46,-3.5,-0.47,-0.05,\\-5.74,0.08,0.16,-0.001,1)^T$.

\begin{table}[h]
\begin{center}
\caption*{Other constant values used for case-1}{}
\begin{tabular}{| c | c | c | c | c | }
\hline
g=32.2 & $U_0$=415.2 & $x_{a_n}$=-0.01 & $Z_\Theta$=0.00221  &  C=0.9916 \\ \hline
\end{tabular}
\end{center}
\end{table}

\par
\begin{figure}[h]
\includegraphics[width=6in,height=3.0in]{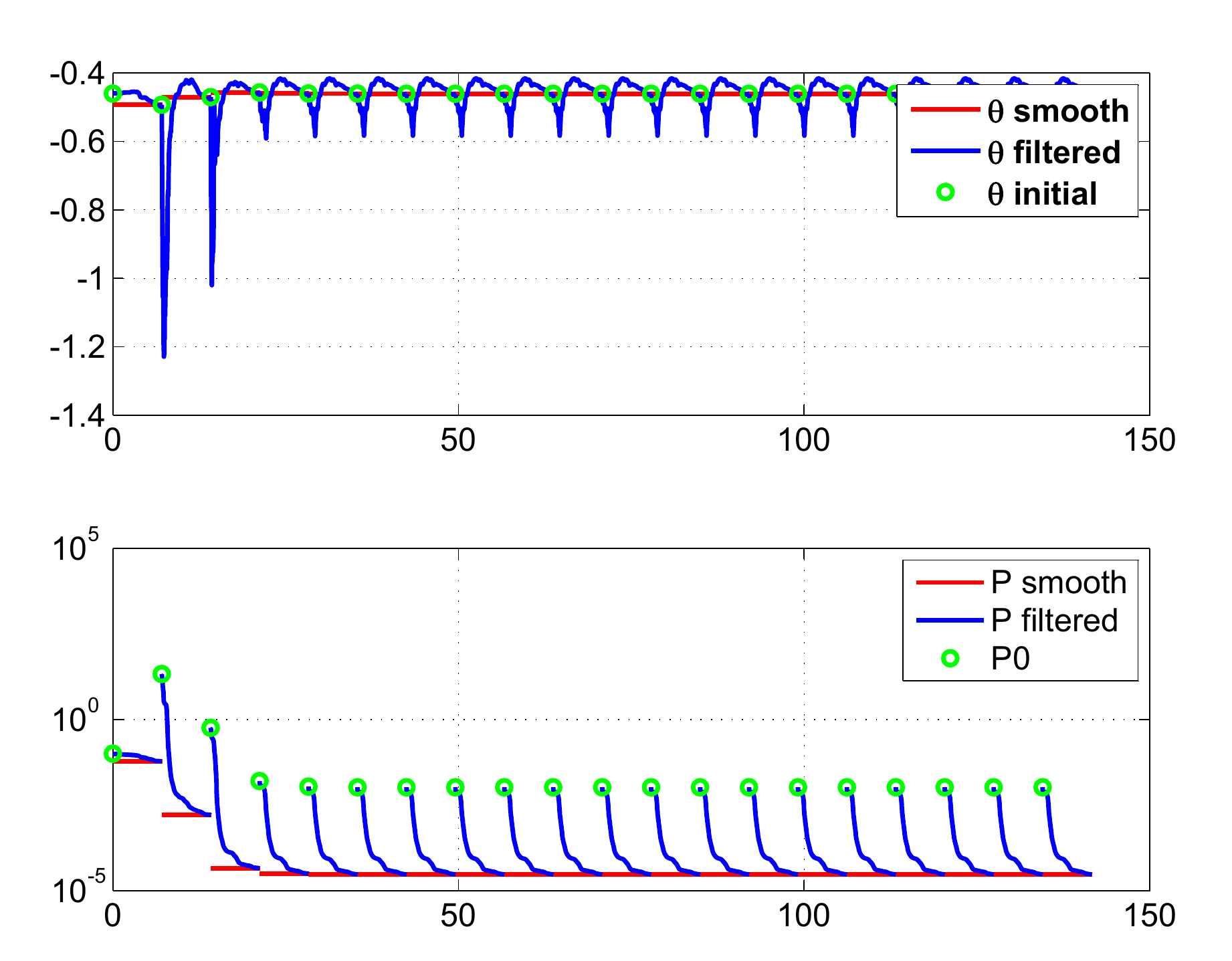}
\caption{The variation of parameter estimate 1 and their filtered }
\caption*{and smoothed covariances through (with the time cumulatively) the iterations}
\label{real1_p1}
\end{figure}

\begin{figure}[h]
\includegraphics[width=6in,height=4in]{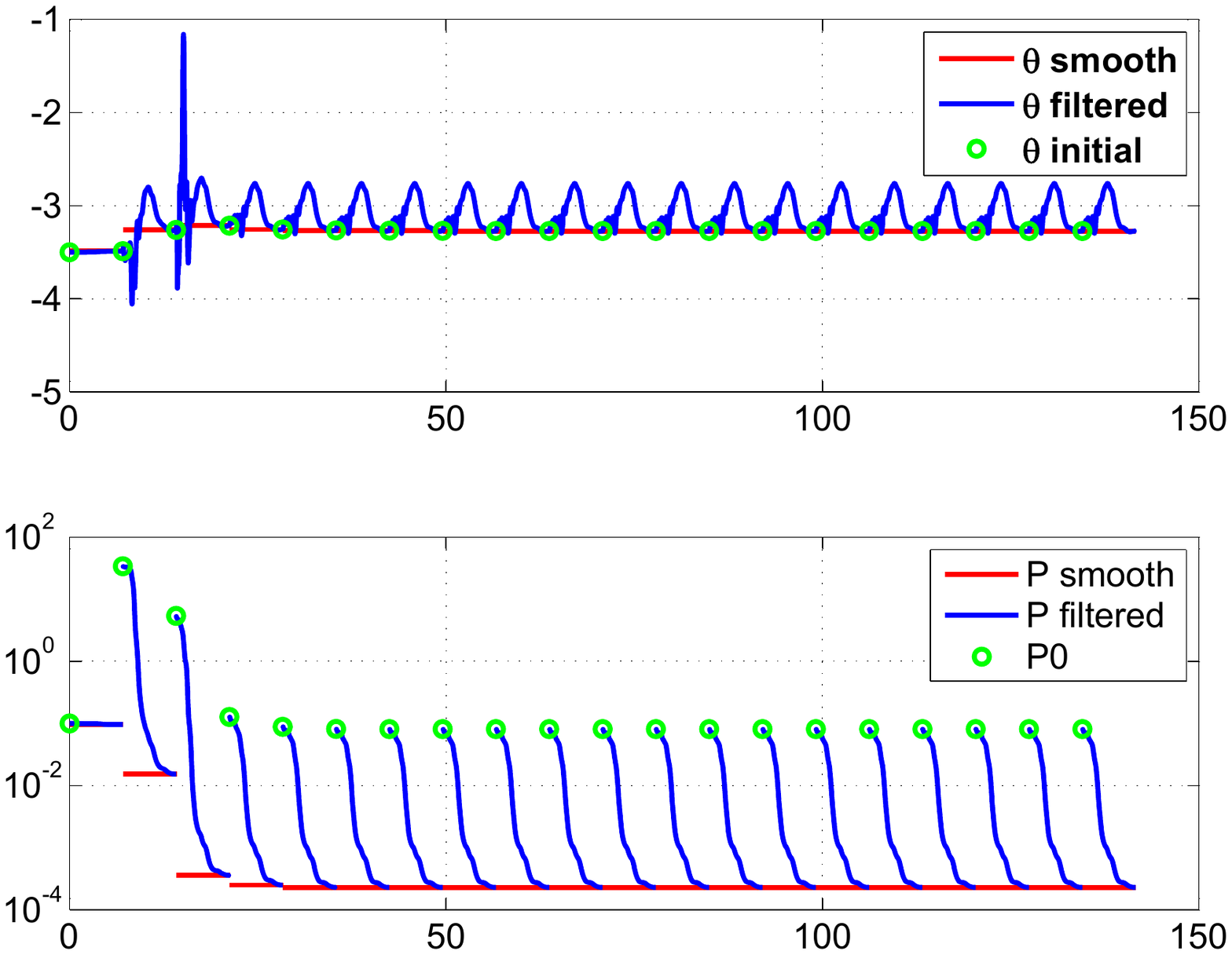}
\caption{The variation of parameter estimate 2 and their filtered }
\caption*{and smoothed covariances through (with the time cumulatively) the iterations}
\label{real1_p2}
\end{figure}

\begin{figure}[h]
\includegraphics[width=6in,height=4in]{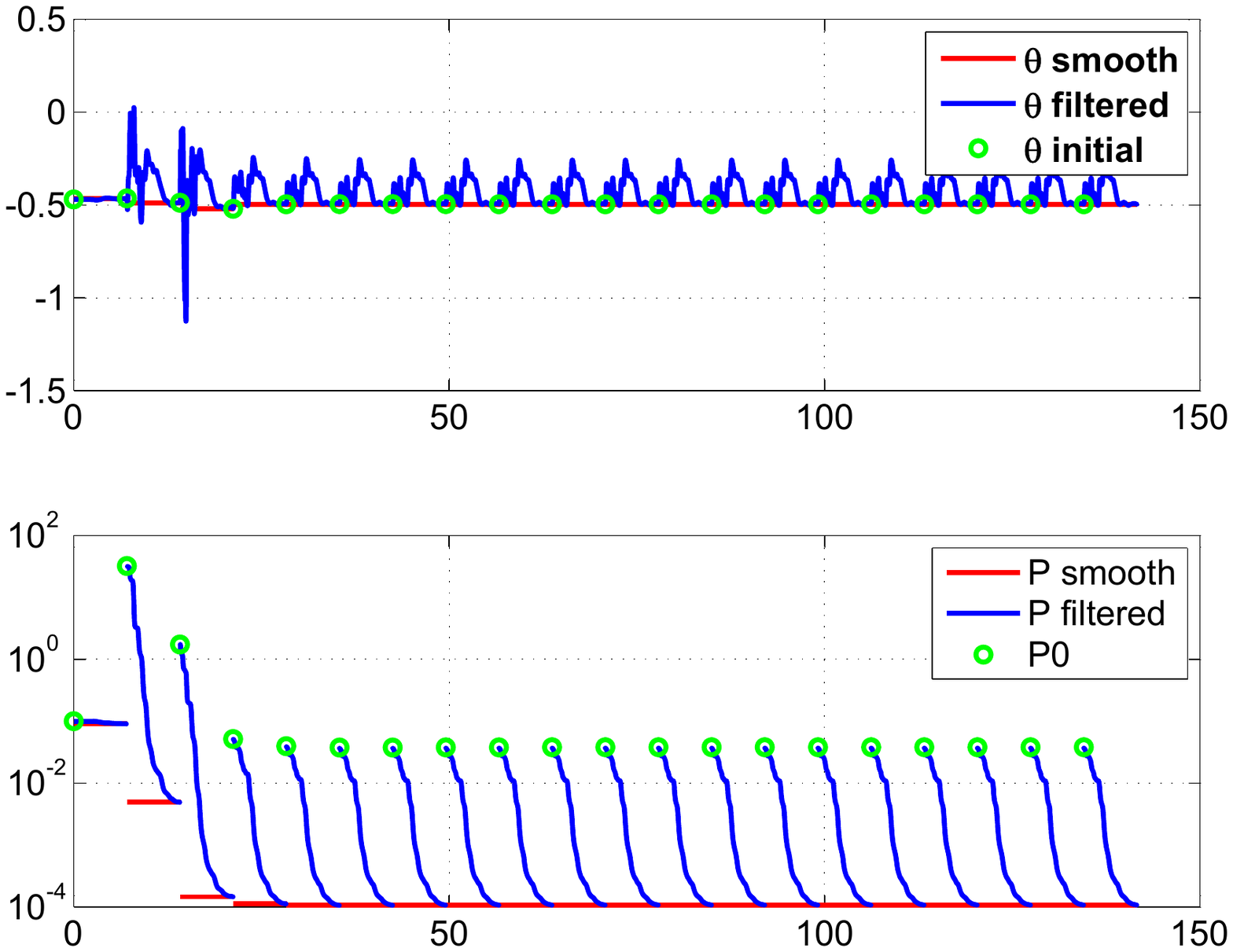}
\caption{The variation of parameter estimate 3 and their filtered }
\caption*{and smoothed covariances through (with the time cumulatively) the iterations}
\label{real1_p3}
\end{figure}

\begin{figure}[h]
\includegraphics[width=6in,height=4in]{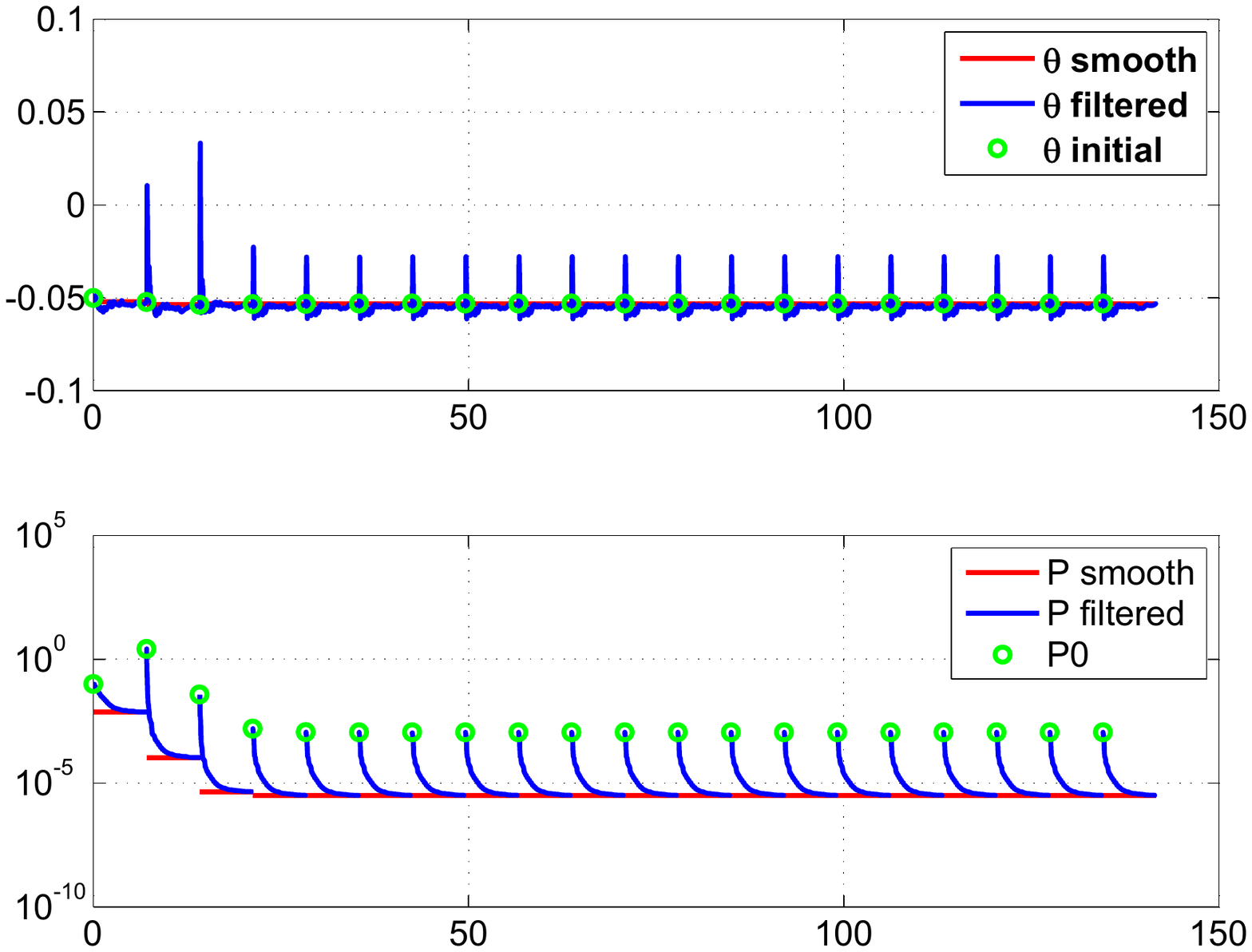}
\caption{The variation of parameter estimate 4 and their filtered }
\caption*{and smoothed covariances through (with the time cumulatively) the iterations}
\label{real1_p4}
\end{figure}

\begin{figure}[h]
\includegraphics[width=6in,height=4in]{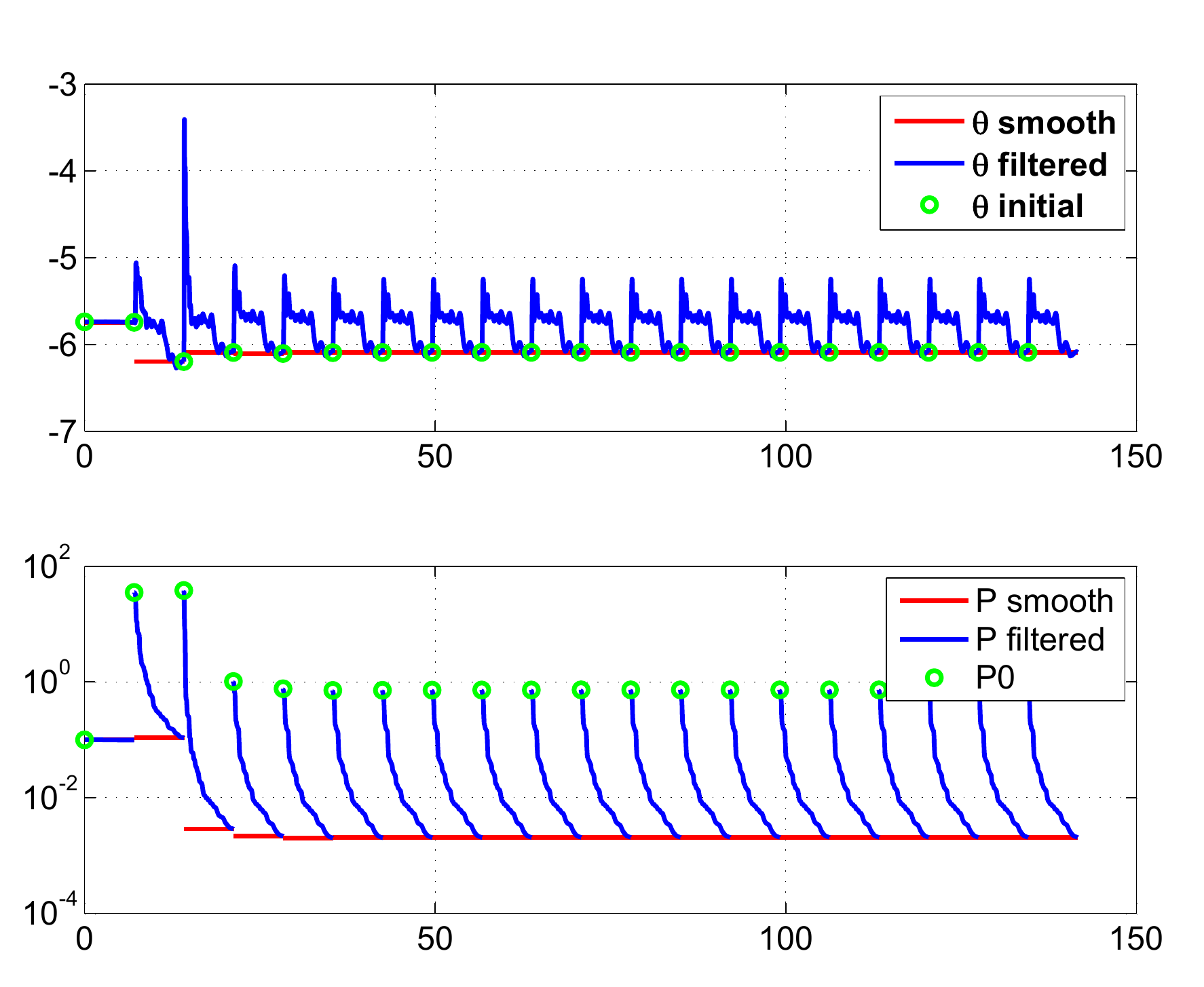}
\caption{The variation of parameter estimate 5 and their filtered }
\caption*{and smoothed covariances through (with the time cumulatively) the iterations}
\label{real1_p5}
\end{figure}

\begin{figure}[h]
\includegraphics[width=6in,height=4in]{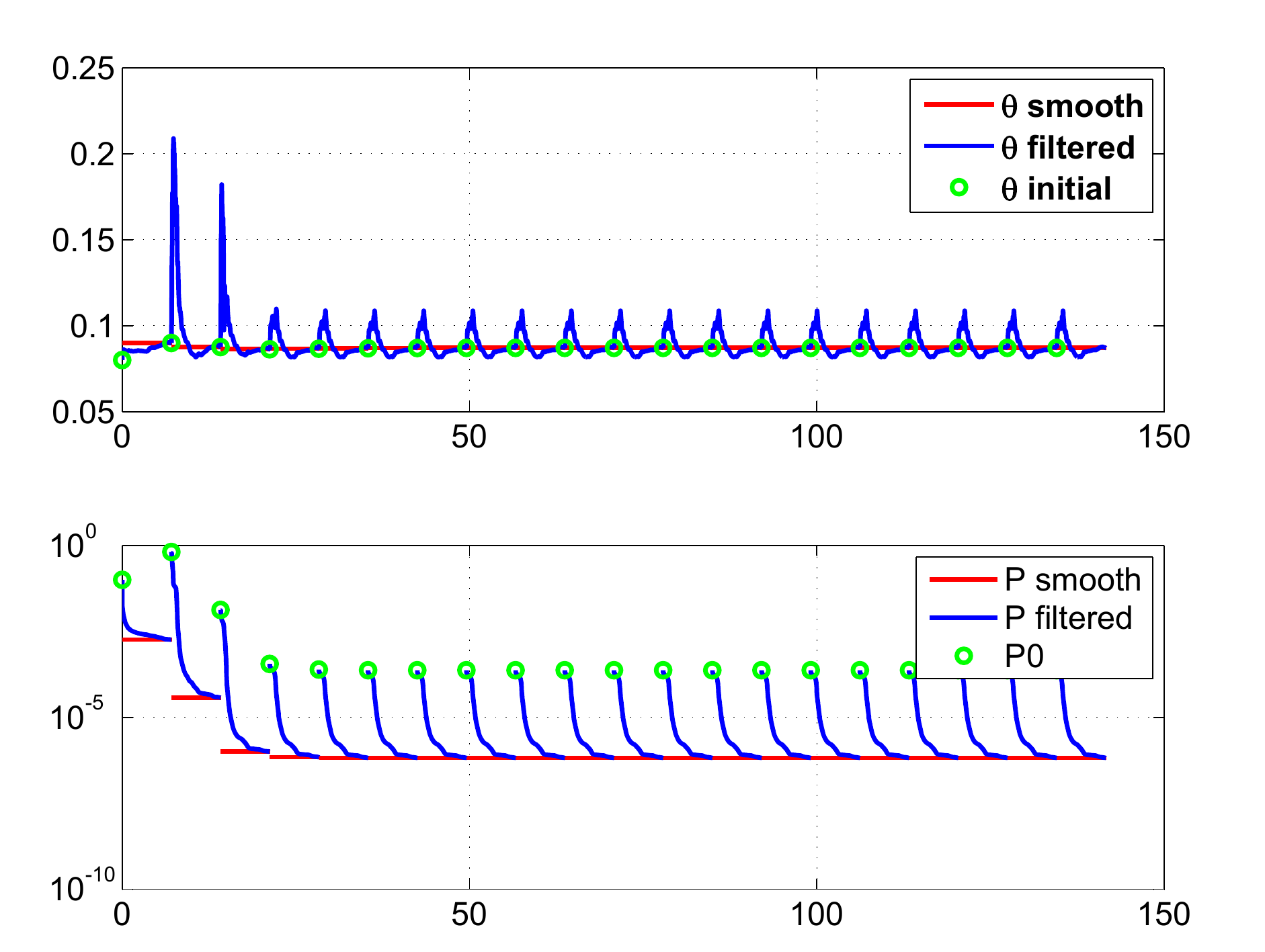}
\caption{The variation of parameter estimate 6 and their filtered }
\caption*{and smoothed covariances through (with the time cumulatively) the iterations}
\label{real1_p6}
\end{figure}

\begin{figure}[h]
\includegraphics[width=6in,height=4in]{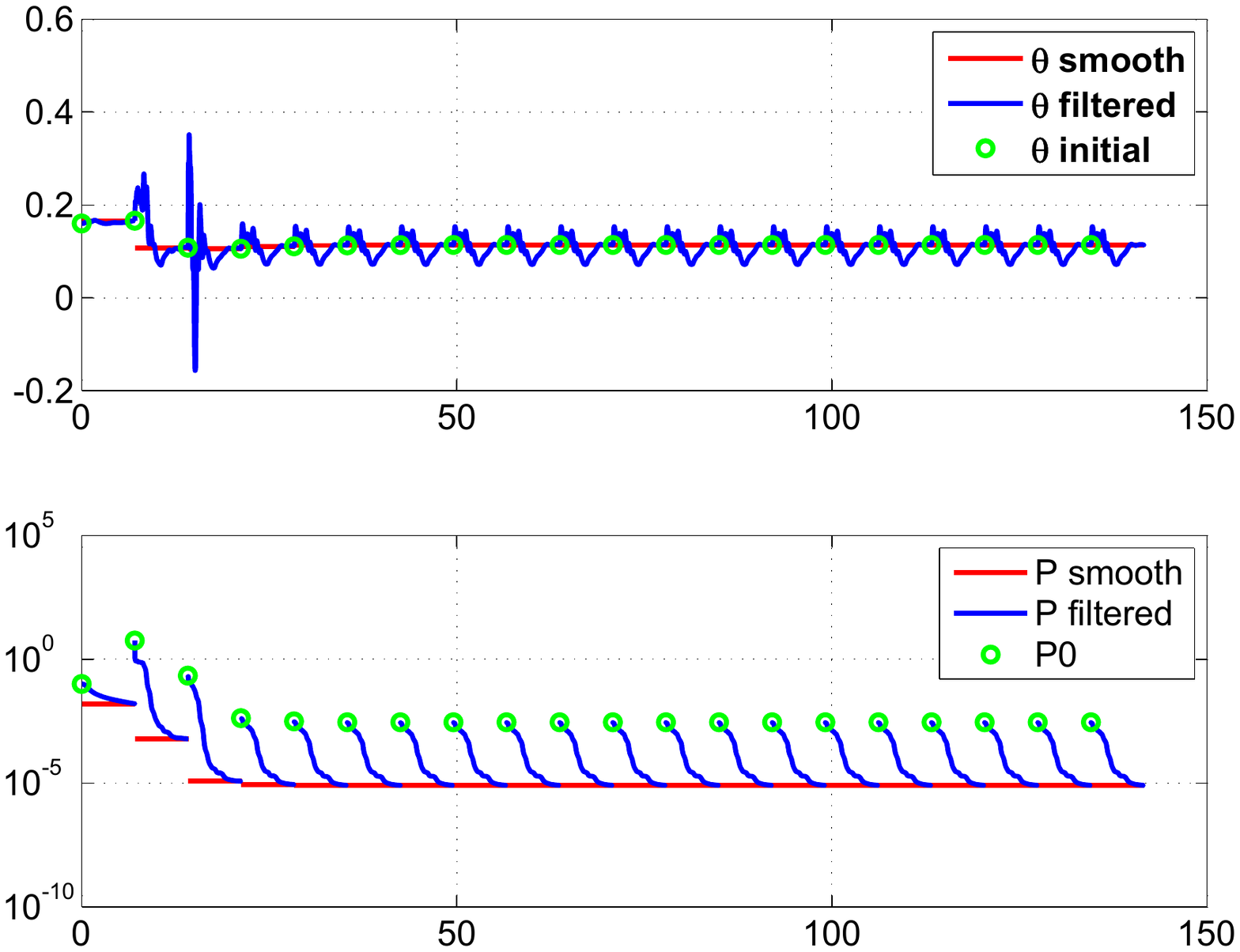}
\caption{The variation of parameter estimate 7 and their filtered }
\caption*{and smoothed covariances through (with the time cumulatively) the iterations}
\label{real1_p7}
\end{figure}

\begin{figure}[h]
\includegraphics[width=6in,height=4in]{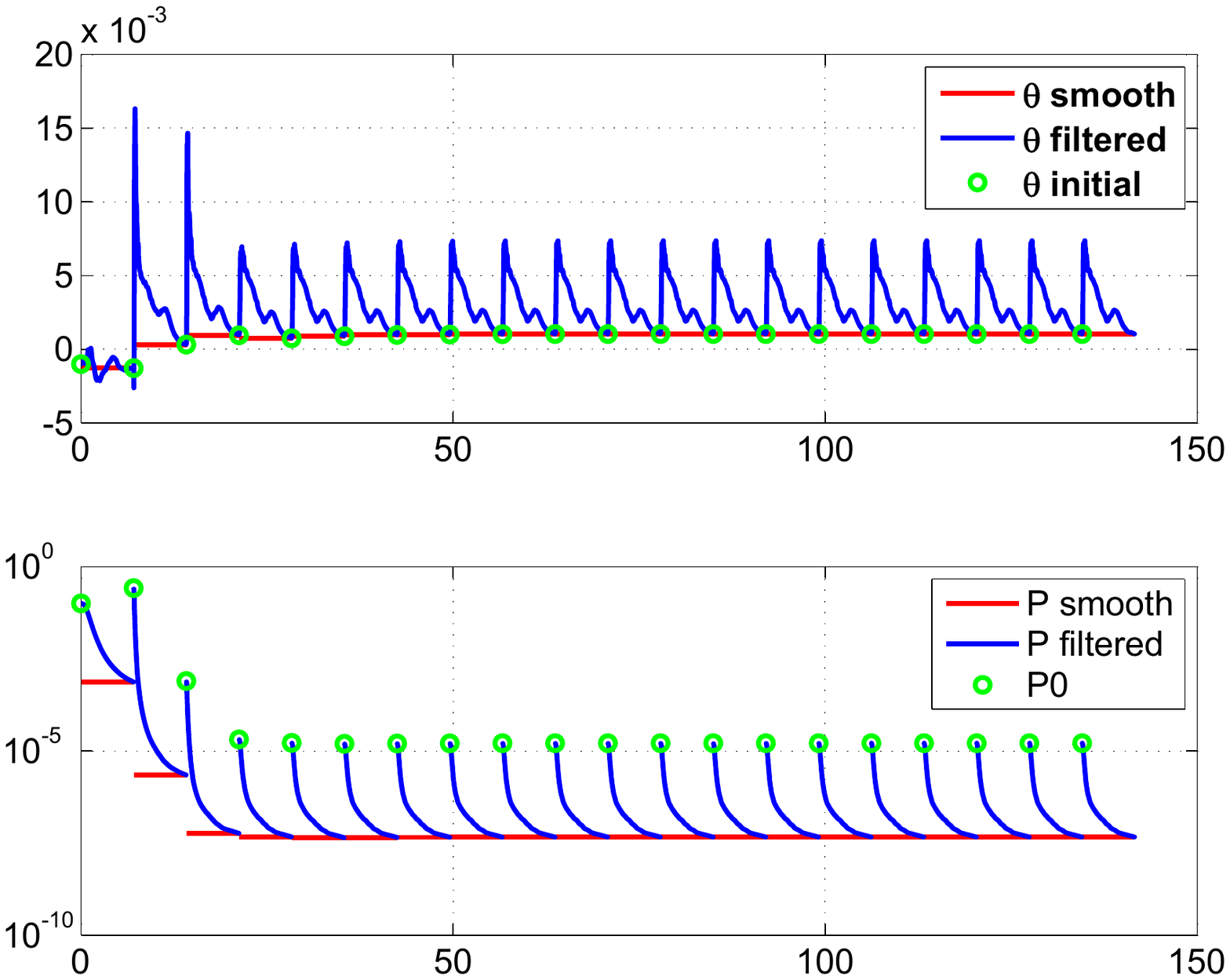}
\caption{The variation of parameter estimate 8 and their filtered }
\caption*{and smoothed covariances through (with the time cumulatively) the iterations}
\label{real1_p8}
\end{figure}

\begin{figure}[h]
\includegraphics[width=6in,height=4in]{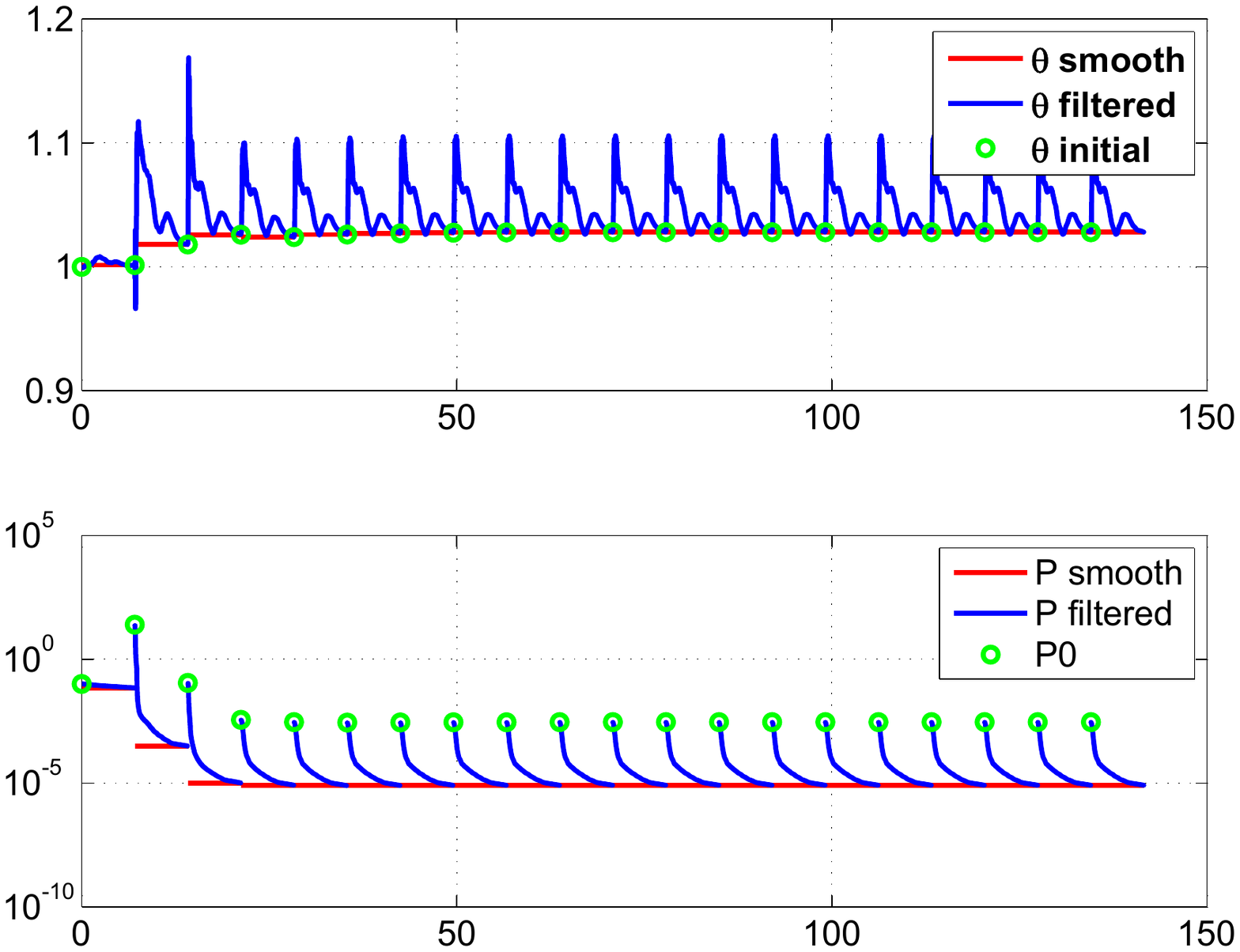}
\caption{The variation of parameter estimate 9 and their filtered }
\caption*{and smoothed covariances through (with the time cumulatively) the iterations}
\label{real1_p9}
\end{figure}

\begin{figure}[h]
\includegraphics[width=6in,height=4in]{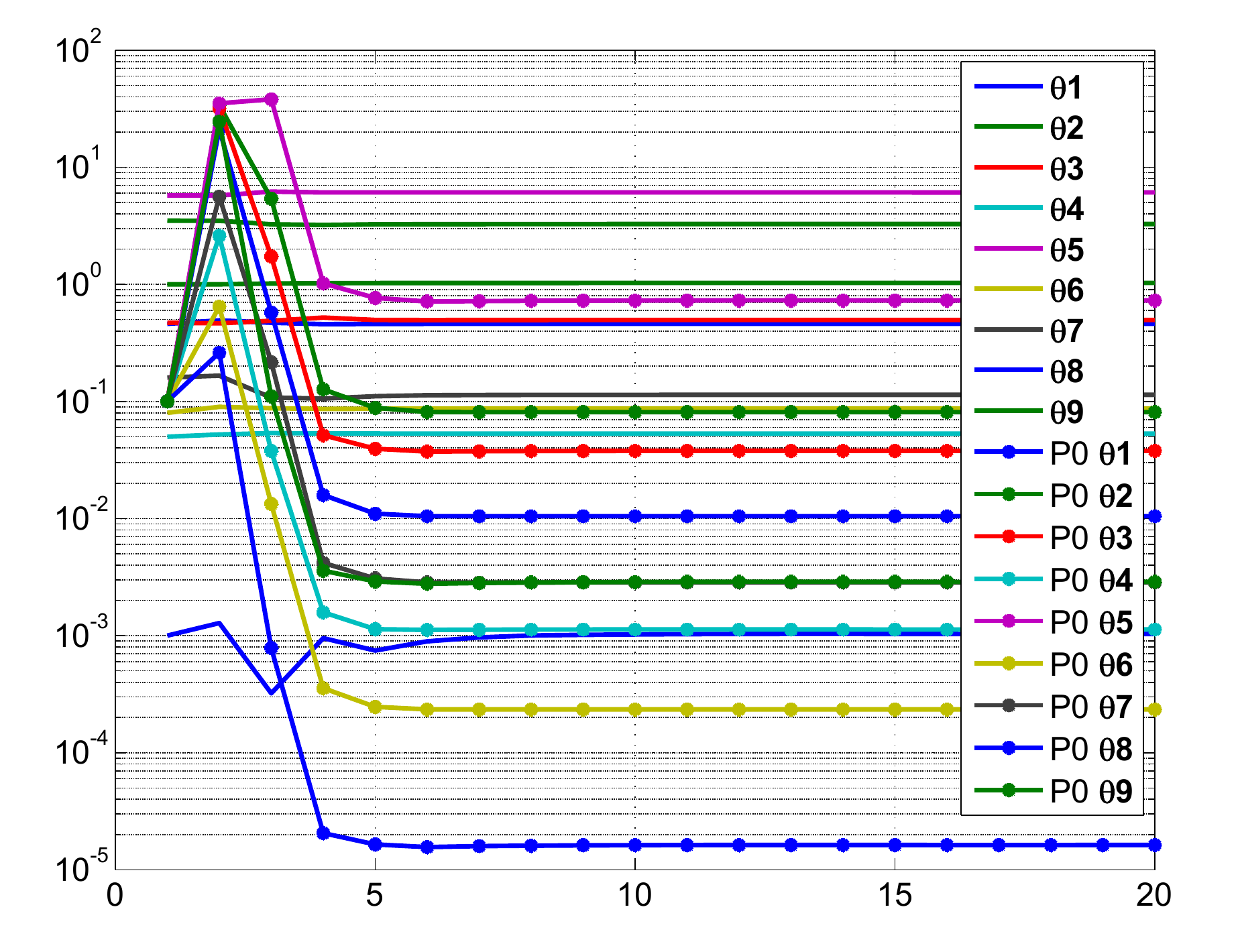}
\caption{Variation of parameter and its initial covariance ($\mathbf{P_0}$) with iterations}
\label{real1_P0}
\end{figure}

\begin{figure}[h]
\includegraphics[width=6in,height=4in]{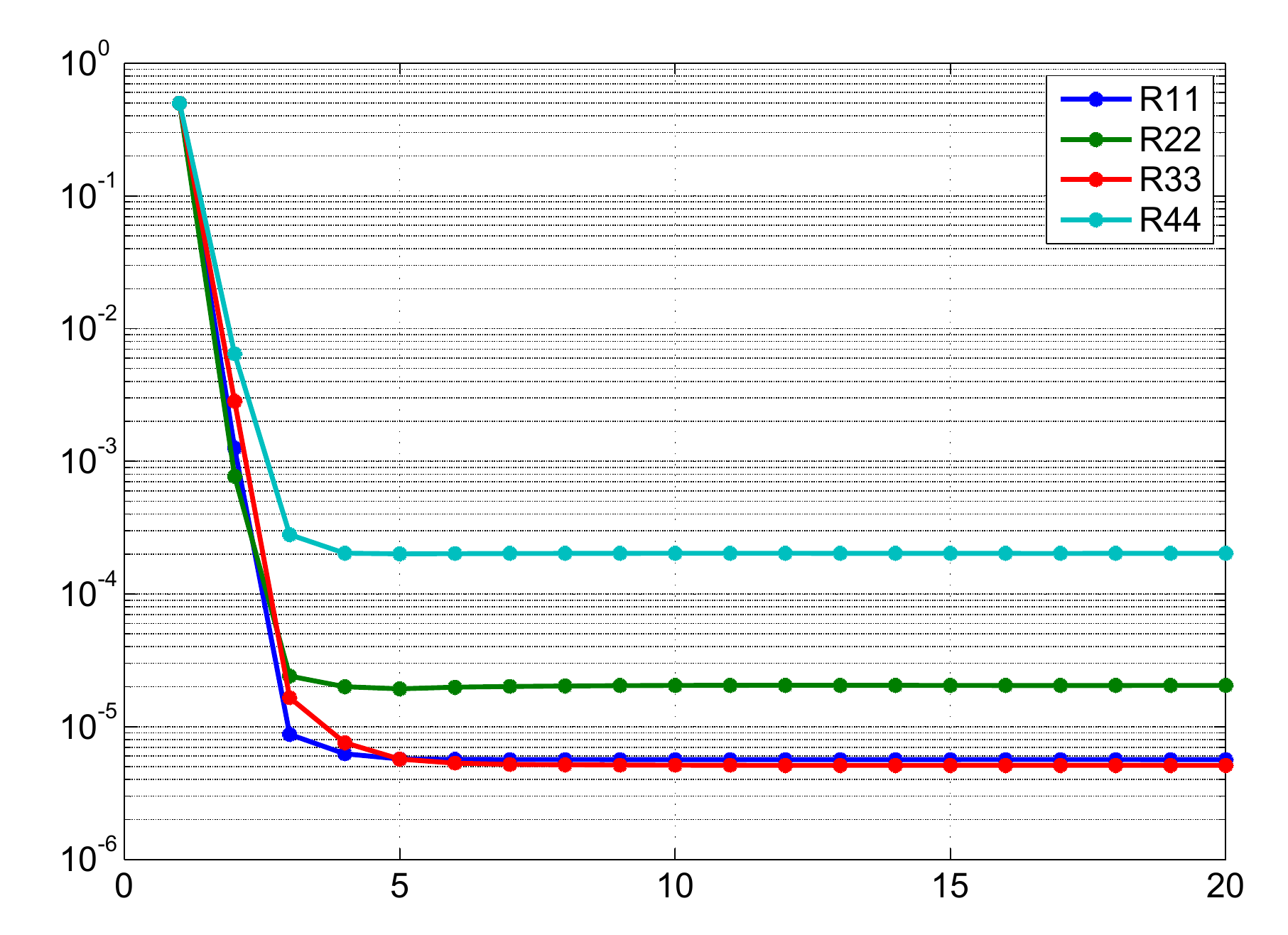}
\caption{Variation of \textbf{R} with iterations}
\label{real1_R}
\end{figure}

\begin{figure}[h]
\includegraphics[width=6in,height=4in]{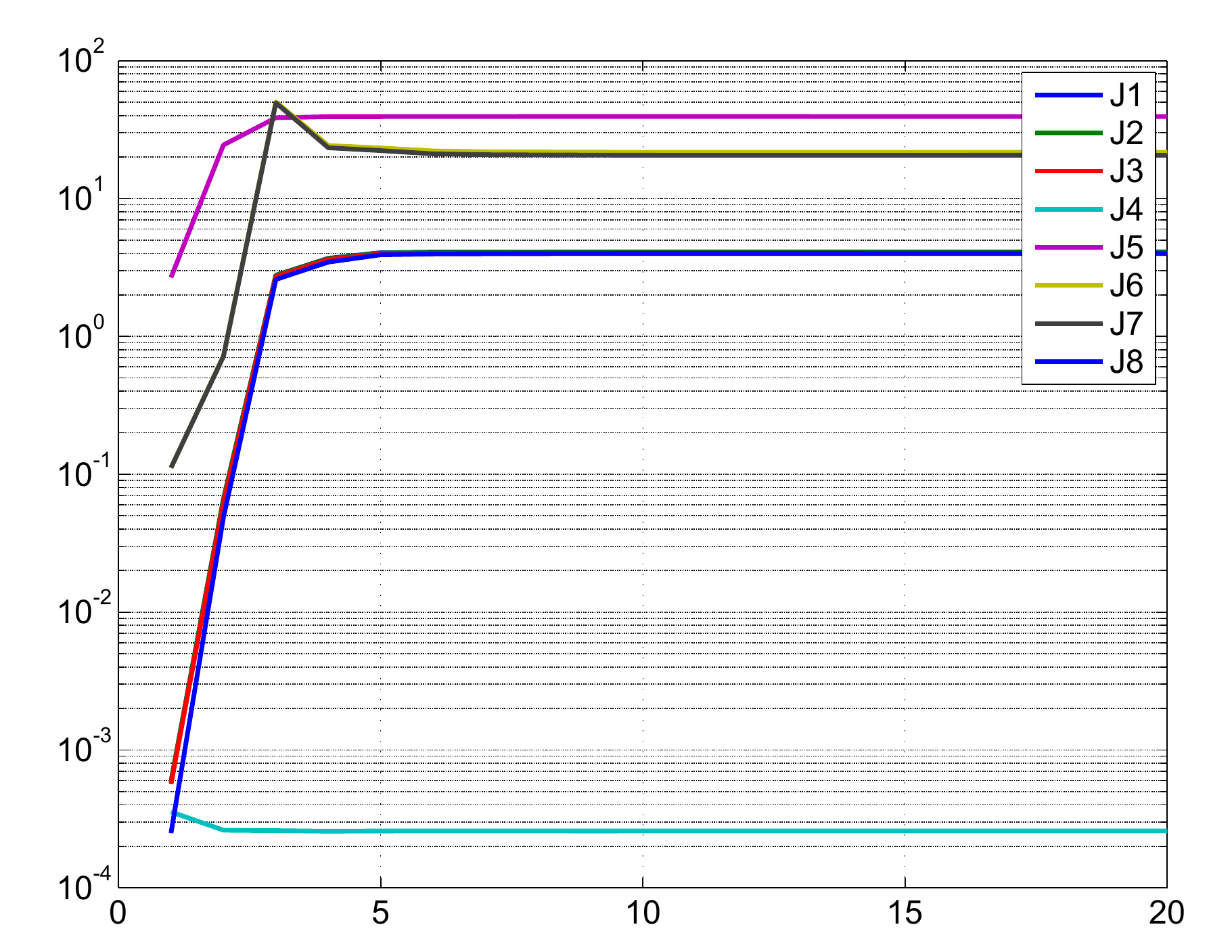}
\caption{Variation of different costs (\textbf{J1-J8}) with iterations}
\label{real1_J}
\end{figure}

\begin{figure}[h]
\includegraphics[width=6in,height=4in]{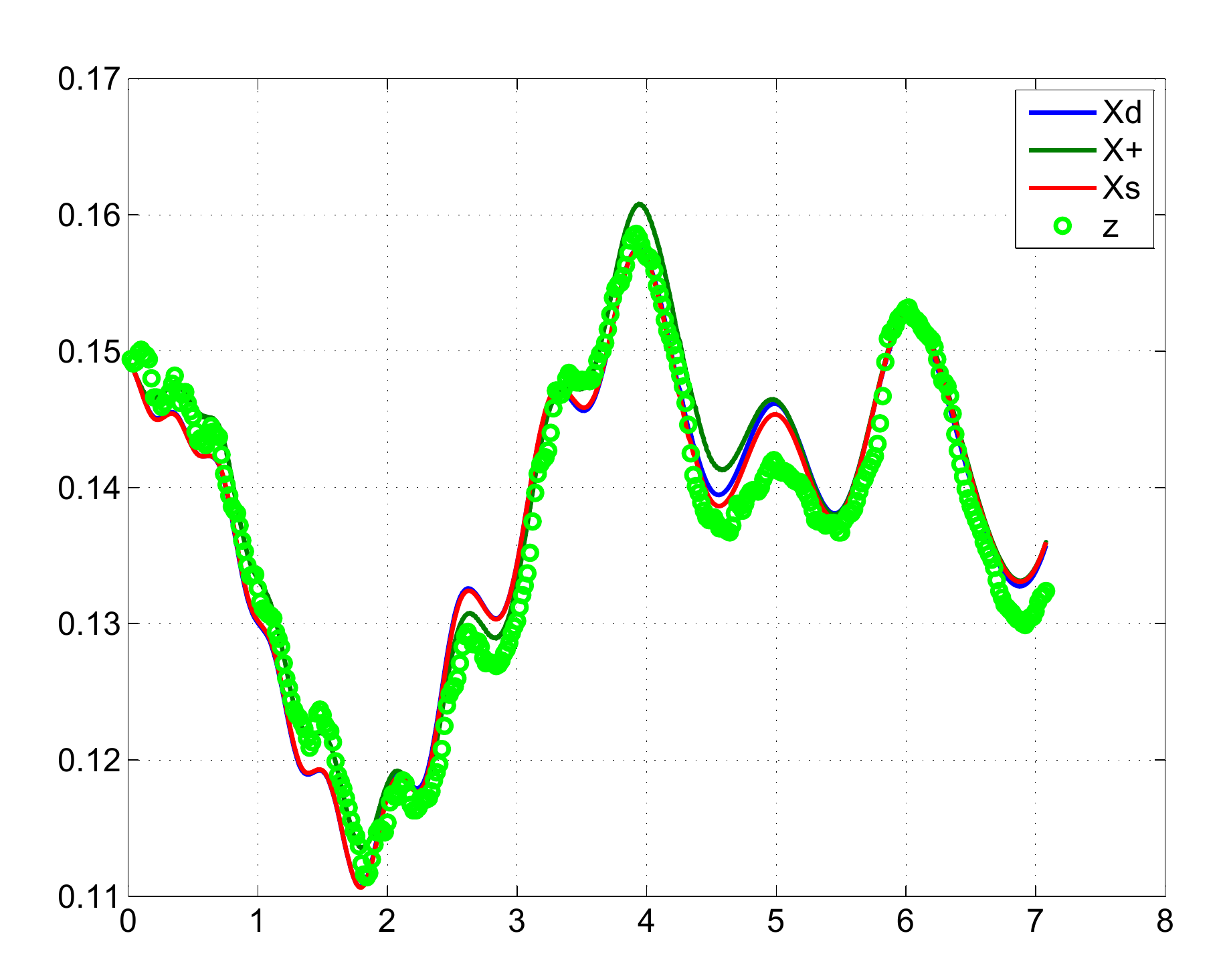}
\caption{Comparison of the predicted dynamics, posterior, smoothed}
\caption*{and the measurement 1}
\label{real1_s1}
\end{figure}

\begin{figure}[h]
\includegraphics[width=6in,height=4in]{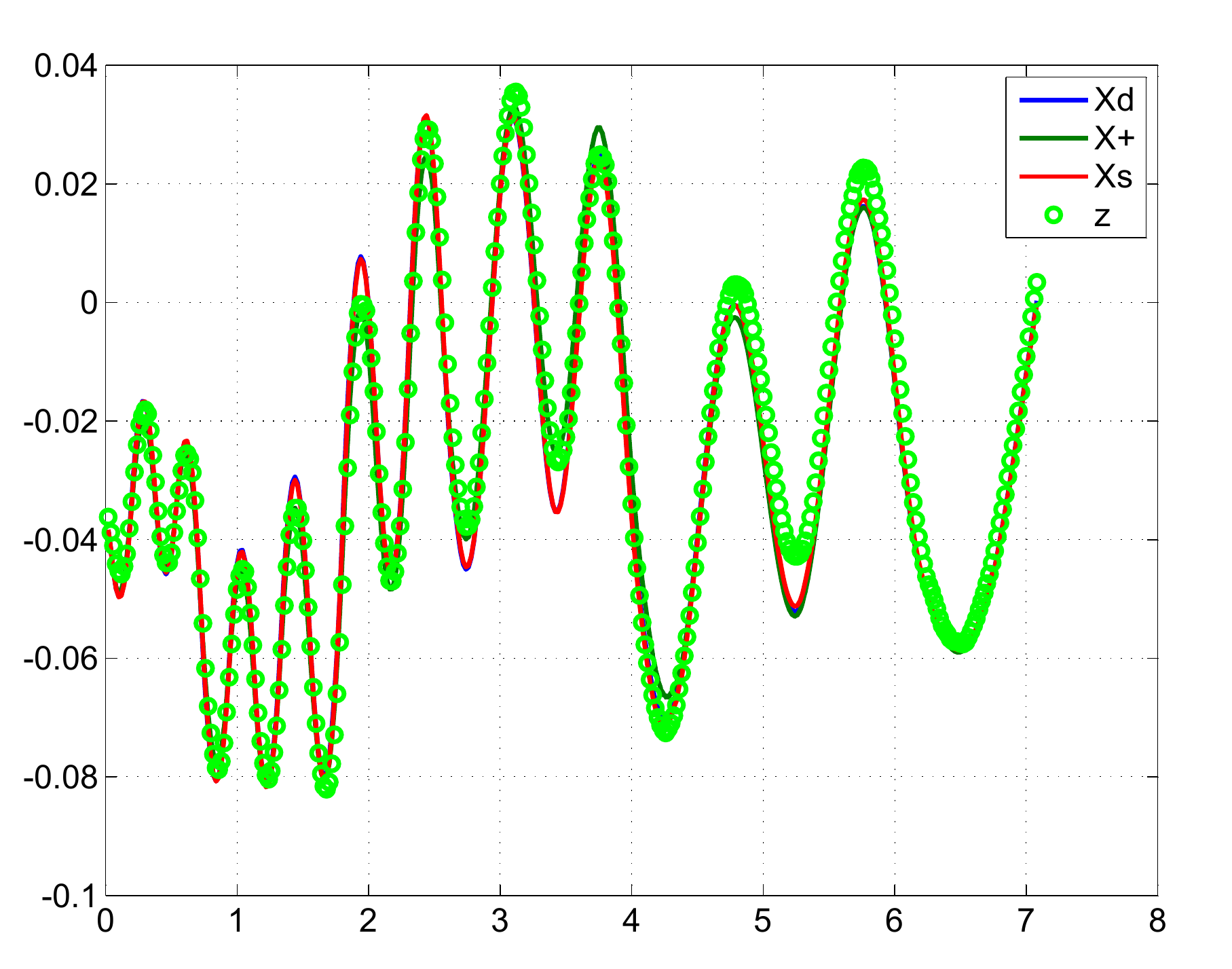}
\caption{Comparison of the predicted dynamics, posterior, smoothed}
\caption*{and the measurement 2}
\label{real1_s2}
\end{figure}

\begin{figure}[h]
\includegraphics[width=6in,height=4in]{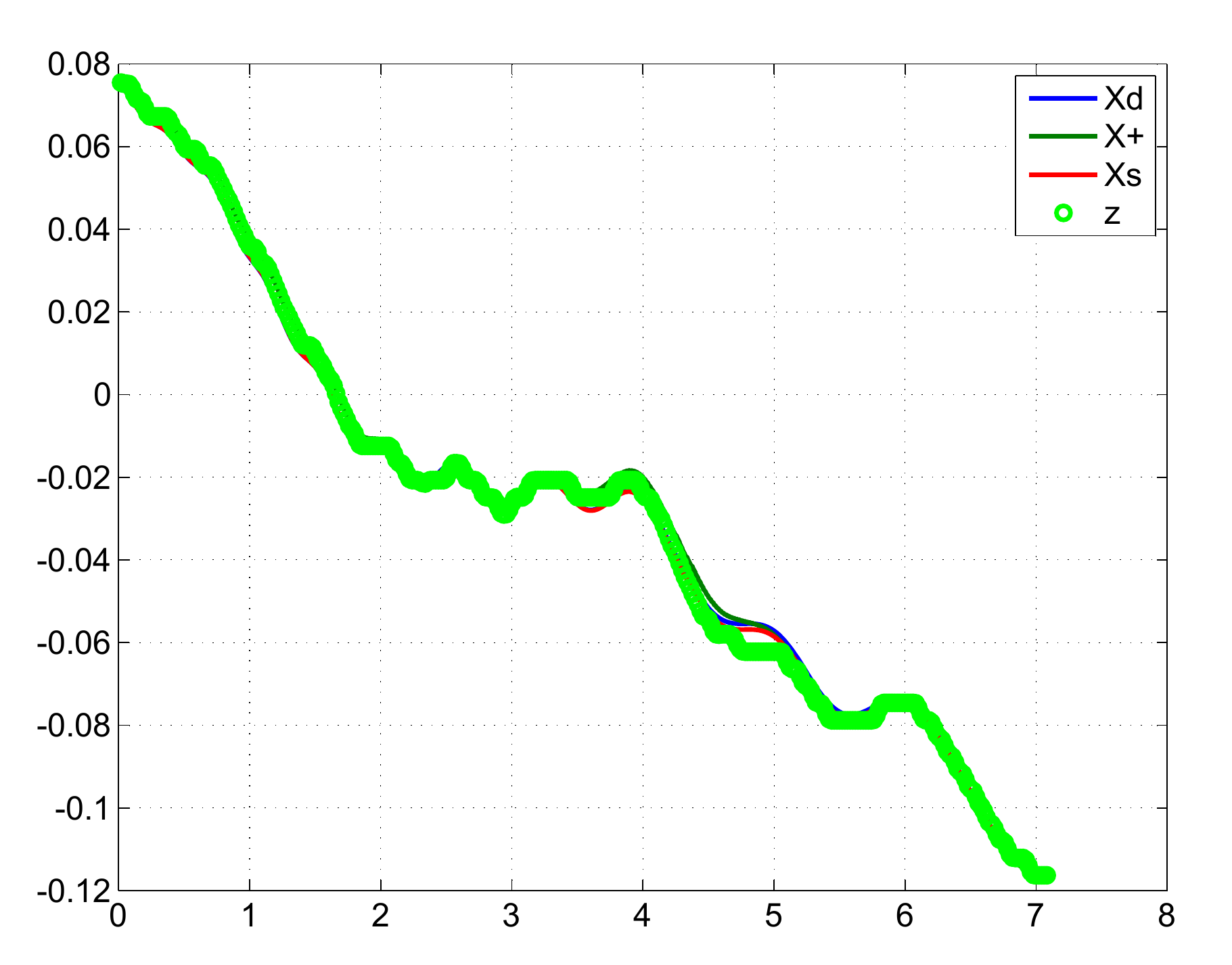}
\caption{Comparison of the predicted dynamics, posterior, smoothed}
\caption*{and the measurement 3}
\label{real1_s3}
\end{figure}

\begin{figure}[h]
\includegraphics[width=6in,height=4in]{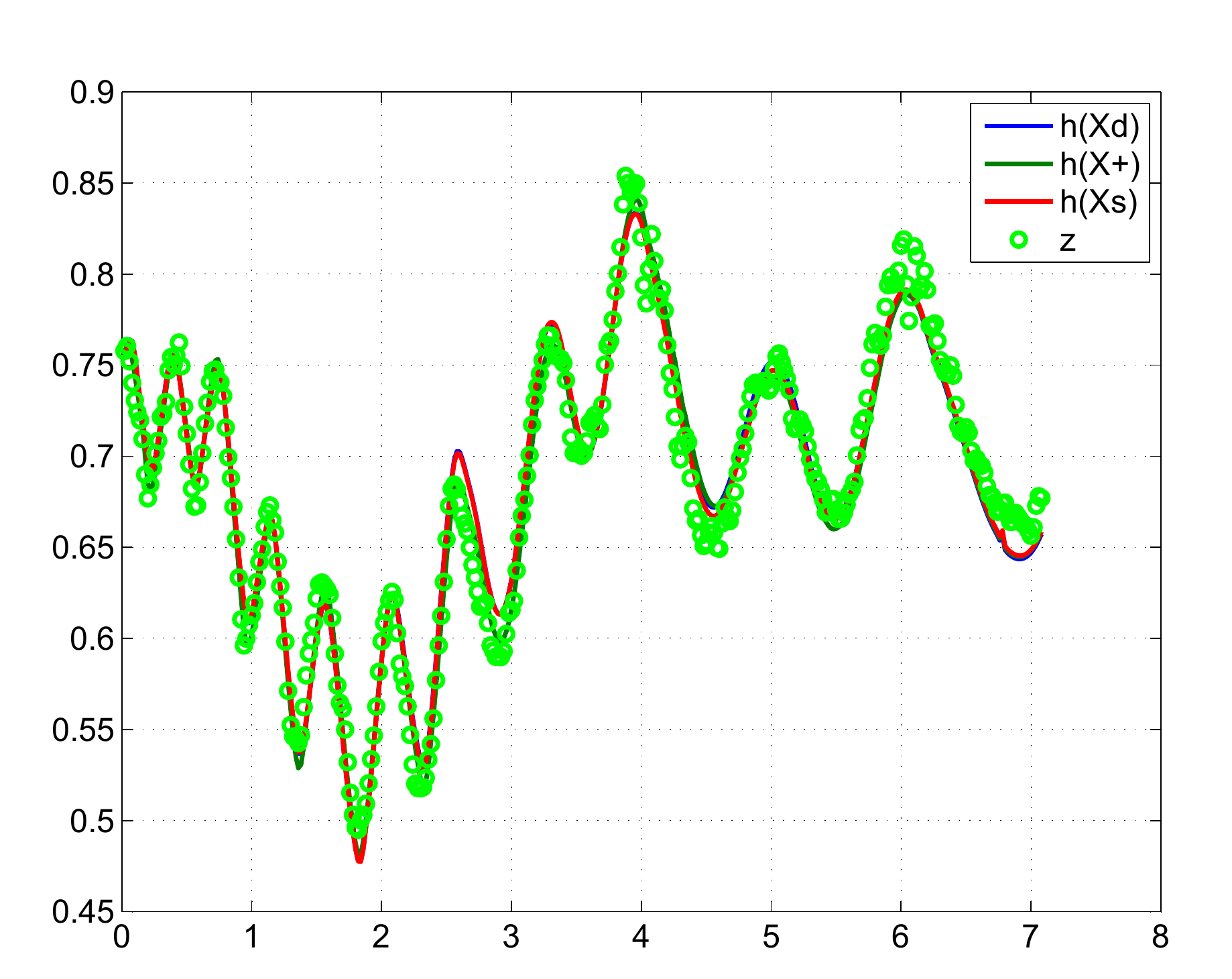}
\caption{Comparison of the predicted dynamics, posterior, smoothed}
\caption*{and the measurement 4}
\label{real1_s4}
\end{figure}

\begin{figure}[h]
\includegraphics[width=6in,height=4in]{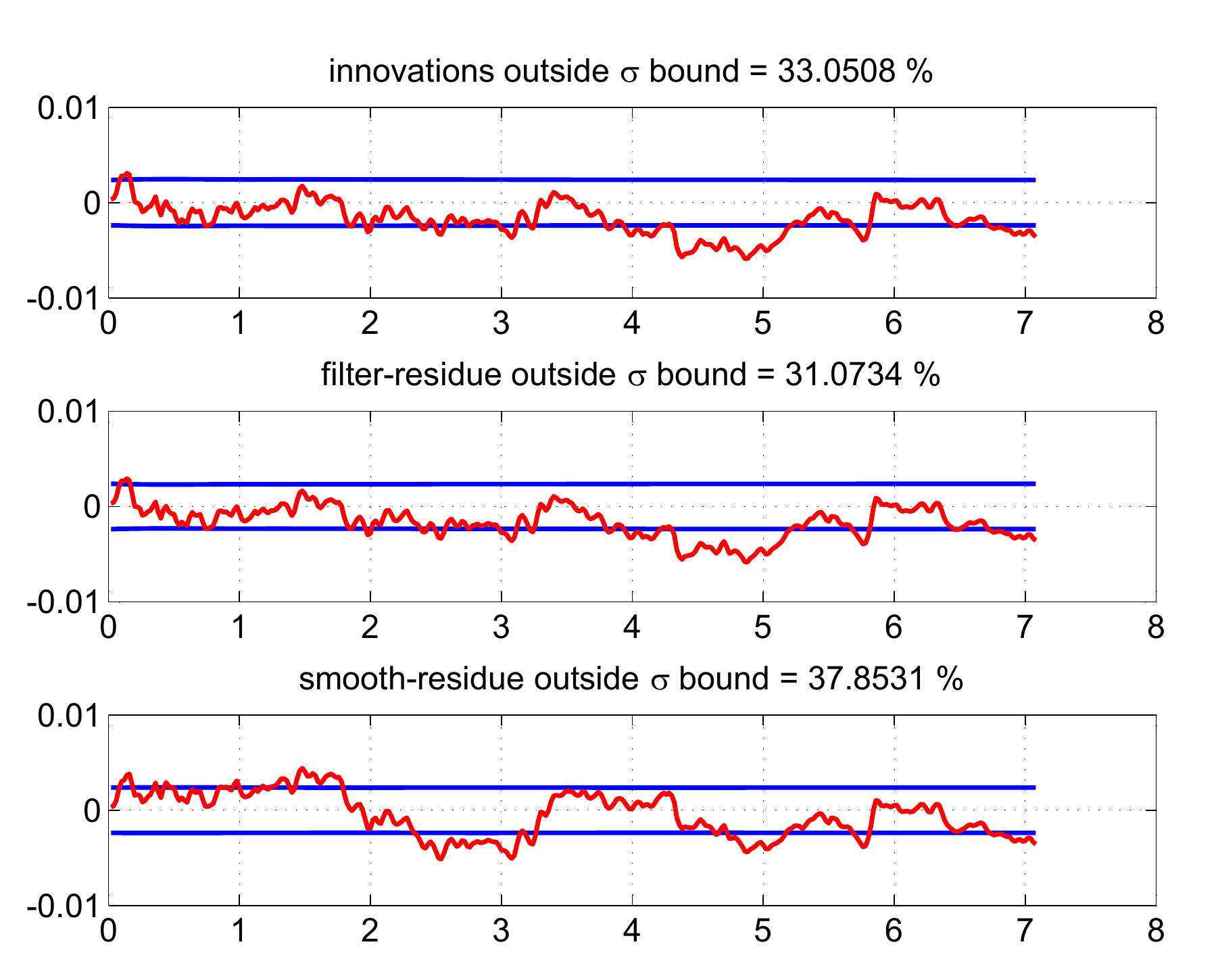}
\caption{The innovations, filtered residue and smoothed residue of measurement 1}
\label{real1_innov1}
\end{figure}

\begin{figure}[h]
\includegraphics[width=6in,height=4in]{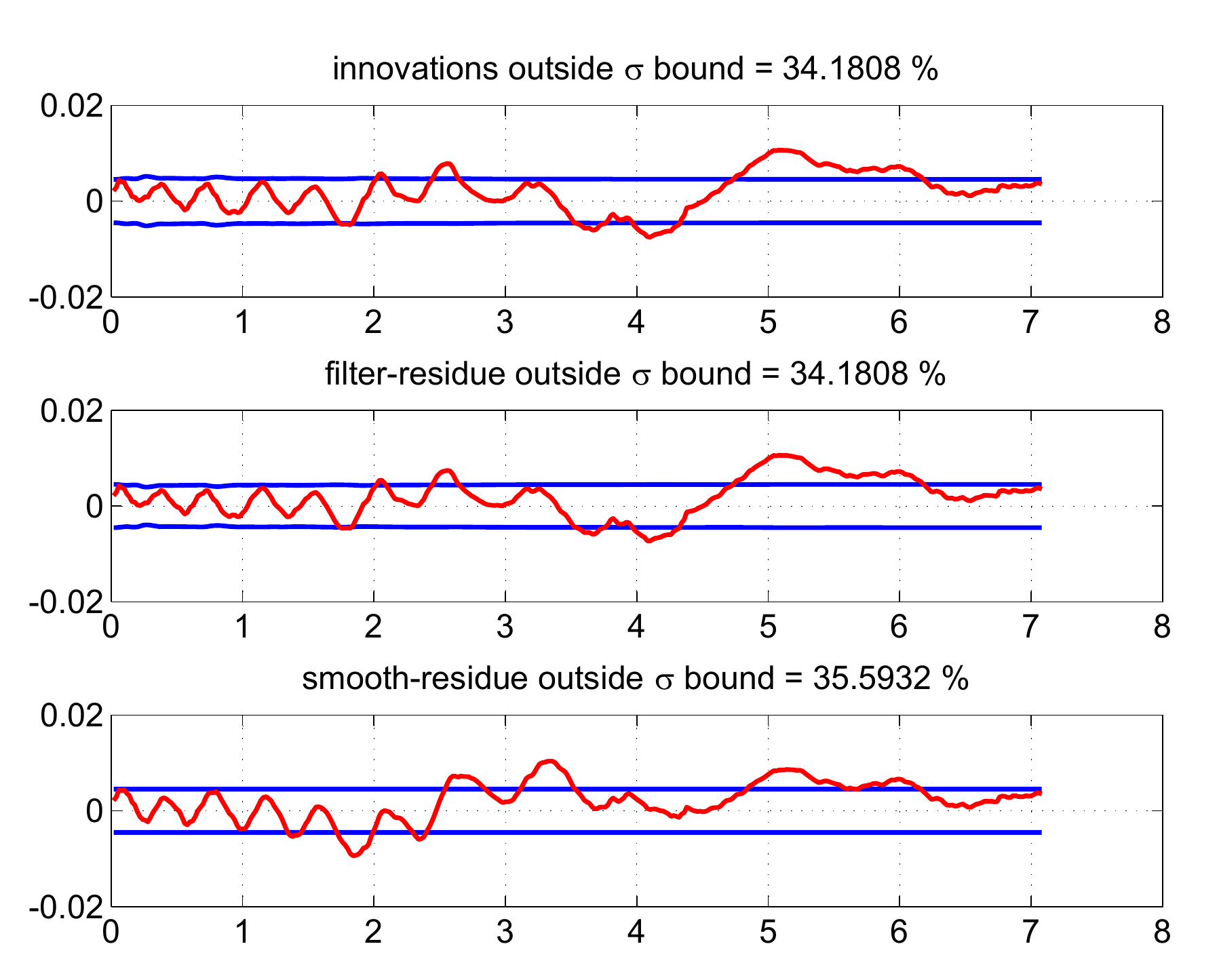}
\caption{The innovations, filtered residue and smoothed residue of measurement 2}
\label{real1_innov2}
\end{figure}

\begin{figure}[h]
\includegraphics[width=6in,height=4in]{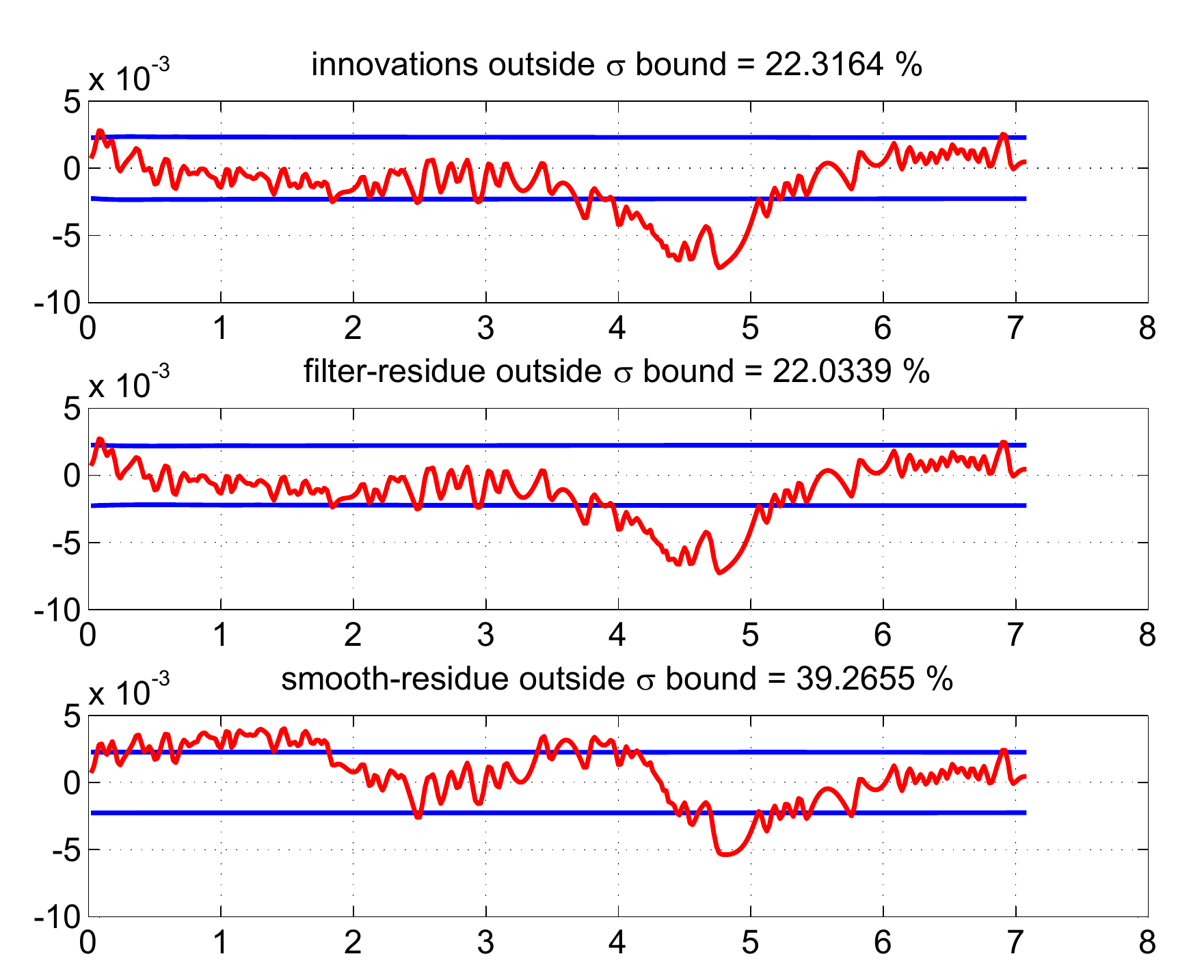}
\caption{The innovations, filtered residue and smoothed residue of measurement 3}
\label{real1_innov3}
\end{figure}

\begin{figure}[h]
\includegraphics[width=6in,height=4in]{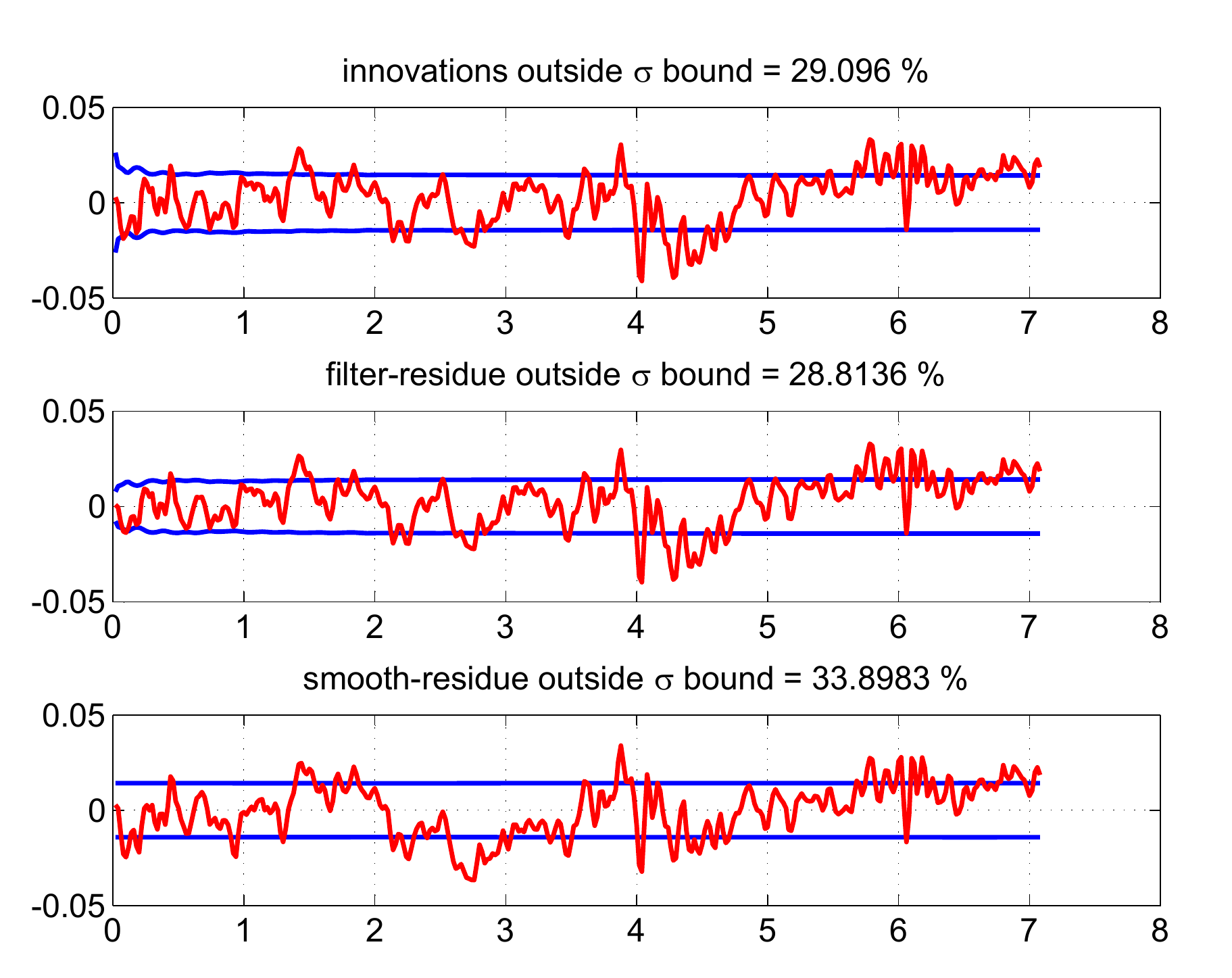}
\caption{The innovations, filtered residue and smoothed residue of measurement 4}
\label{real1_innov4}
\end{figure}

\begin{figure}[h]
\includegraphics[width=6in,height=4in]{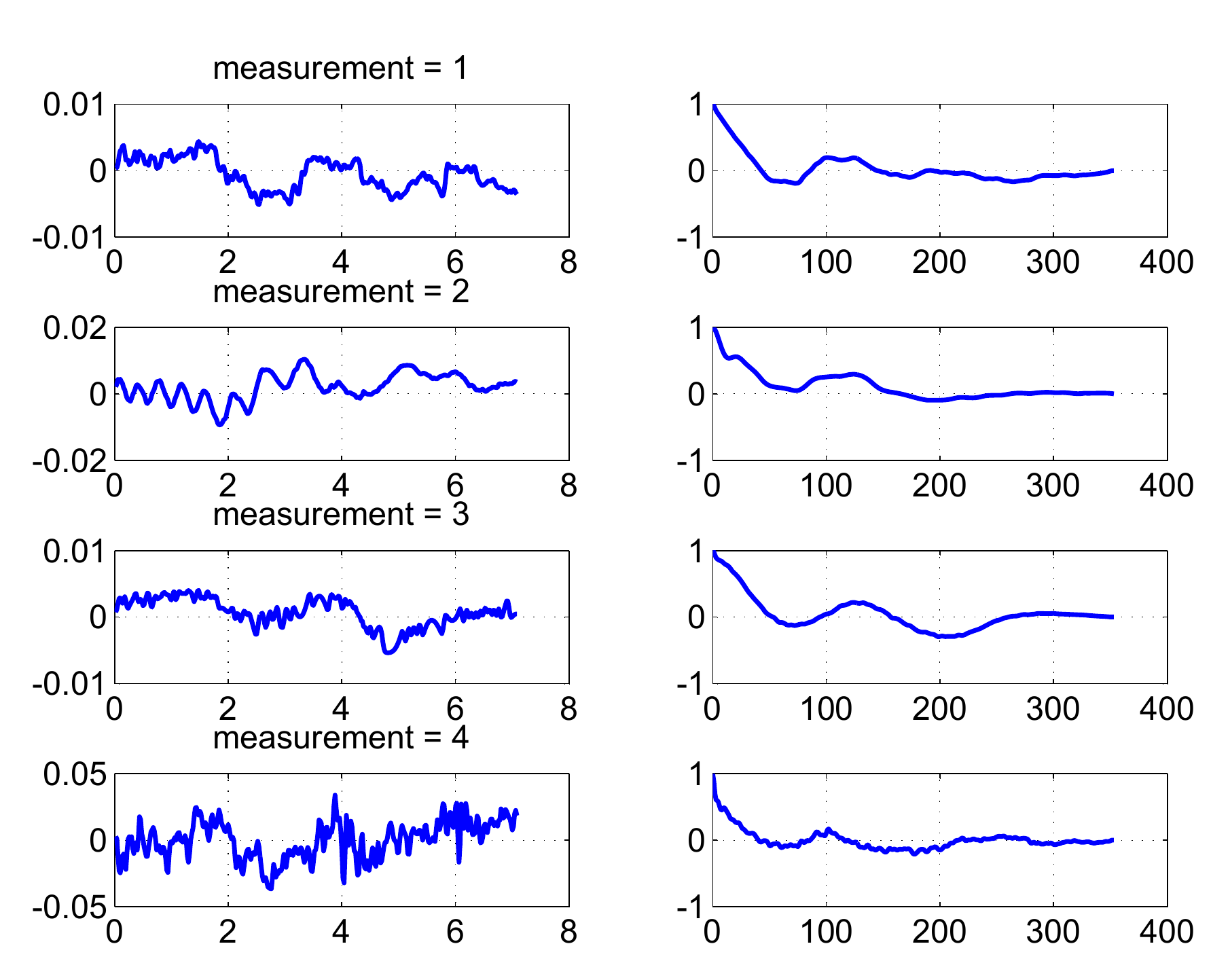}
\caption{Time variation of estimated measurement noise (left) and}
\caption*{their autocorrelation (right)}
\label{real1_mnoise}
\end{figure}

 We first run the filter assuming Q = 0 followed by \textbf{Q} $>$ 0. It was found that about 20 iterations of the data by the filter would suffice. The Fig. \ref{input1} shows the elevator control input. The Fig. \ref{real1_p1}-\ref{real1_mnoise} refer to the \textbf{Q} = 0.  The Fig. \ref{real1_p1} to \ref{real1_p9} show the evolution of the parameter estimates as well as their covariances with iterations. It may be noted that in about 5 to 10 iterations the estimates converge and by about 10 iterations the covariances converge to their final estimates. The Fig. \ref{real1_P0} summarizes the above results. The parameters that control the dynamics strongly converge rapidly with the weaker ones perhaps with more iterations after some fluctuations.  The next Fig. \ref{real1_R} and \ref{real1_J} show respectively the convergence of \textbf{R} and the various cost functions \textbf{J}. The subsequent figures \ref{real1_s1}-\ref{real1_s4} shows (i) the state dynamics based on the estimated parameter after the filter pass through the data, (ii) the estimated state after measurement update, (iii) the smoothed state and (iv) the measurement. It may be seen that the measurements are generally away from all the others quantities. If the filter was tuned properly then all the quantities except the state based on the estimated parameters (without including the process noise) should be close to each other. Thus running the filter assuming zero process noise providing such an above behaviour indicates the compulsive necessity of running the filter by assuming nonzero \textbf{Q} and estimating it as well. The following Fig. \ref{real1_innov1} to \ref{real1_innov4} show the variation of the sample innovation, filtered residue and the smoothed residue along with their $\sigma$ bounds. Since most quantities in the extended Kalman filter follow Gaussian or approximated as a Gaussian distribution the sample values can be expected to be within the square root of the covariances for about two thirds of the time. Though in the above figures this appears to be true the above sample statistics appear to be non white in this real flight test data. The next Fig. \ref{real1_mnoise} shows the estimated measurement noise distribution from different channels to be non white. If the filter works well then the cost functions \textbf{J1-J3} should be close to the number of measurements and \textbf{J6-J8} should be close to the number of states. However the Table-\ref{tbcase1} shows the various cost functions among them \textbf{J1-J3} are close to the number of measurements but the costs \textbf{J6, J7} are far different from the number of states with only \textbf{J8} being close to the number of states. Such a behaviour of the cost functions points to the fact the filter can work better and hence we operate the filter by assuming \textbf{Q} $>$ 0.
\par
\begin{figure}[h]
\includegraphics[width=6in,height=4in]{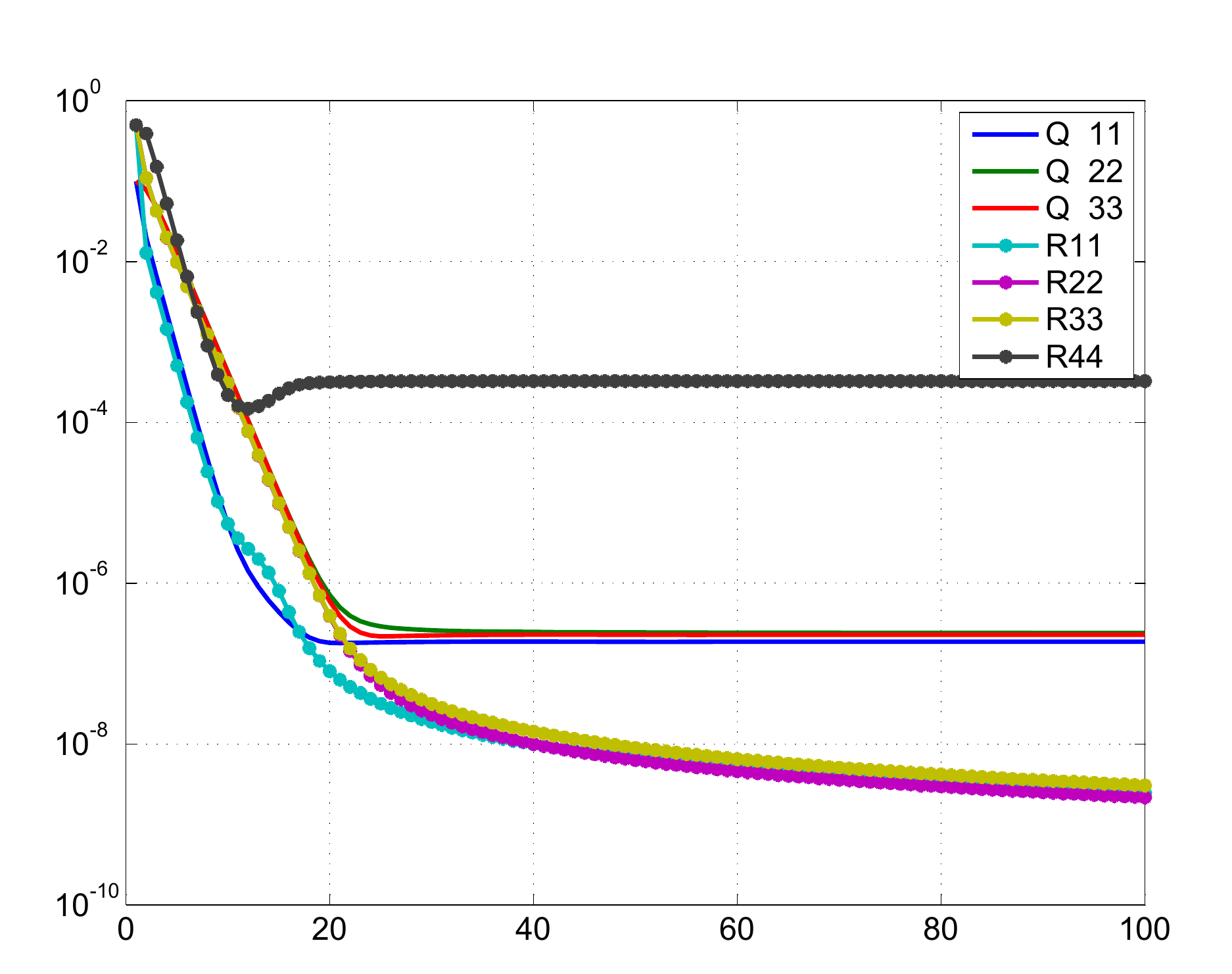}
\caption{Variation of \textbf{R} and \textbf{Q} with iterations}
\label{realQ1_R}
\end{figure}
\begin{figure}[h]
\includegraphics[width=6in,height=4in]{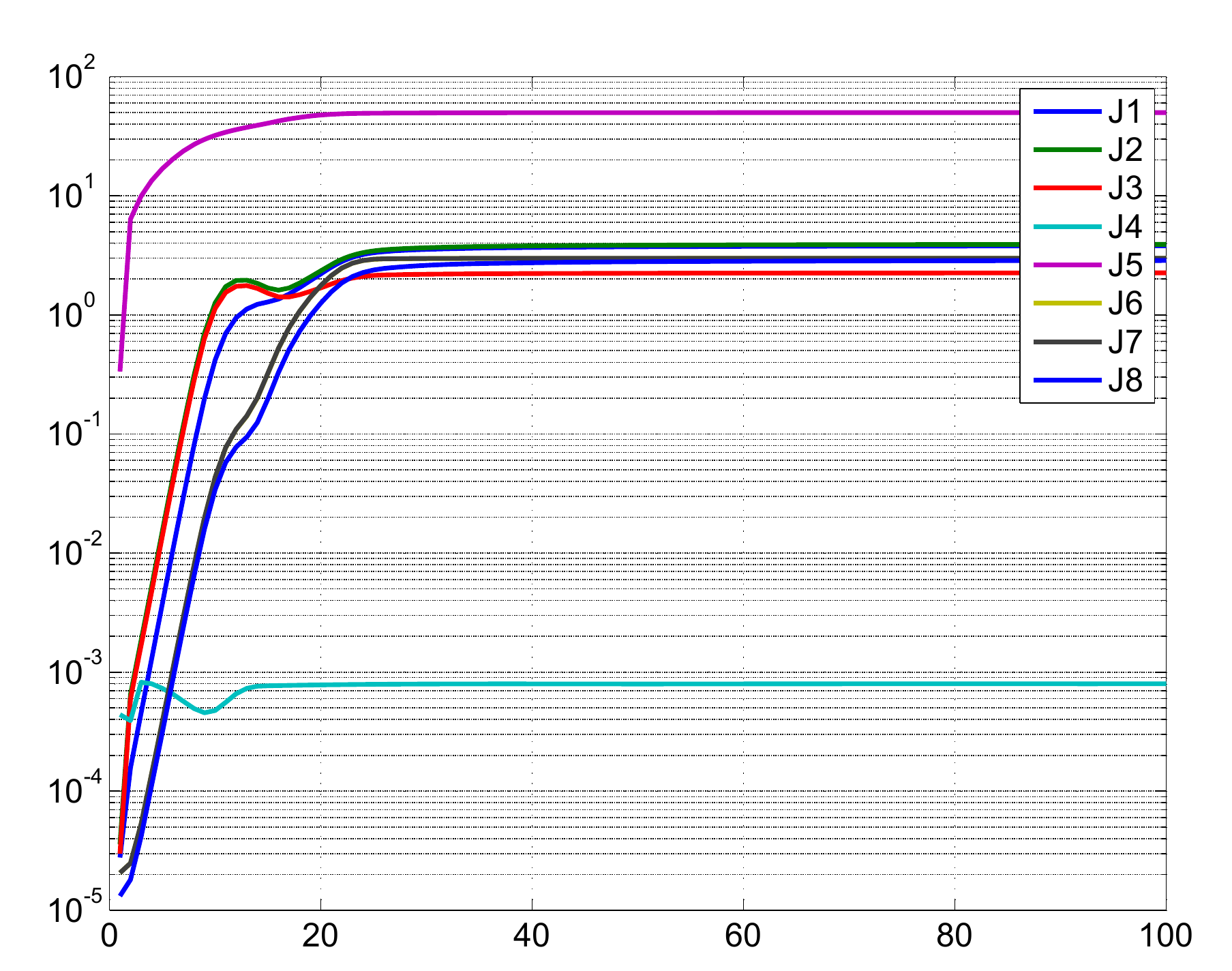}
\caption{Variation of different costs (\textbf{J1-J8}) with iterations}
\label{realQ1_J}
\end{figure}
\begin{figure}[h]
\includegraphics[width=6in,height=4in]{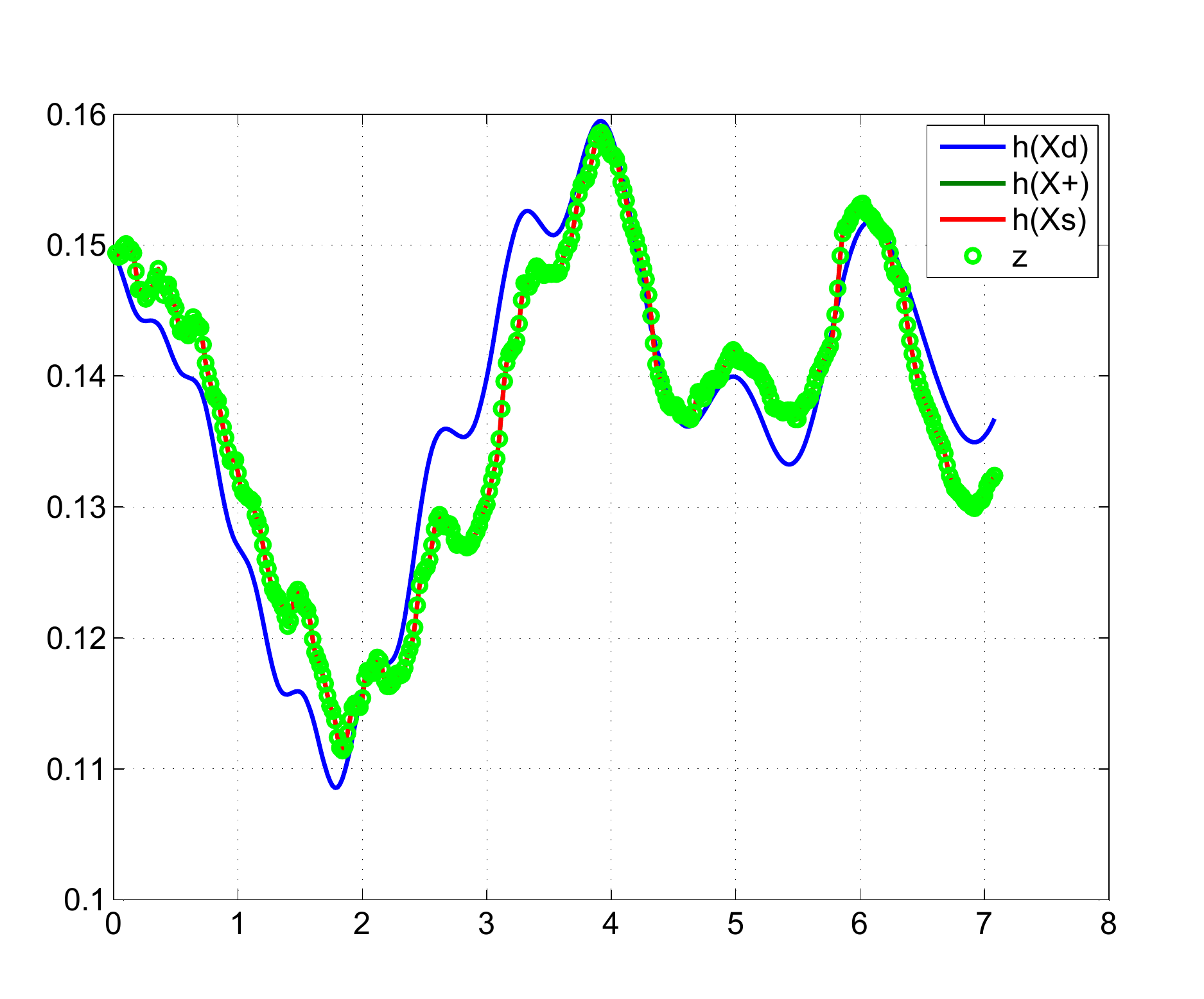}
\caption{Comparison of the predicted dynamics, posterior, smoothed}
\caption*{and the measurement 1}
\label{realQ1_s1}
\end{figure}
\begin{figure}[h]
\includegraphics[width=6in,height=4in]{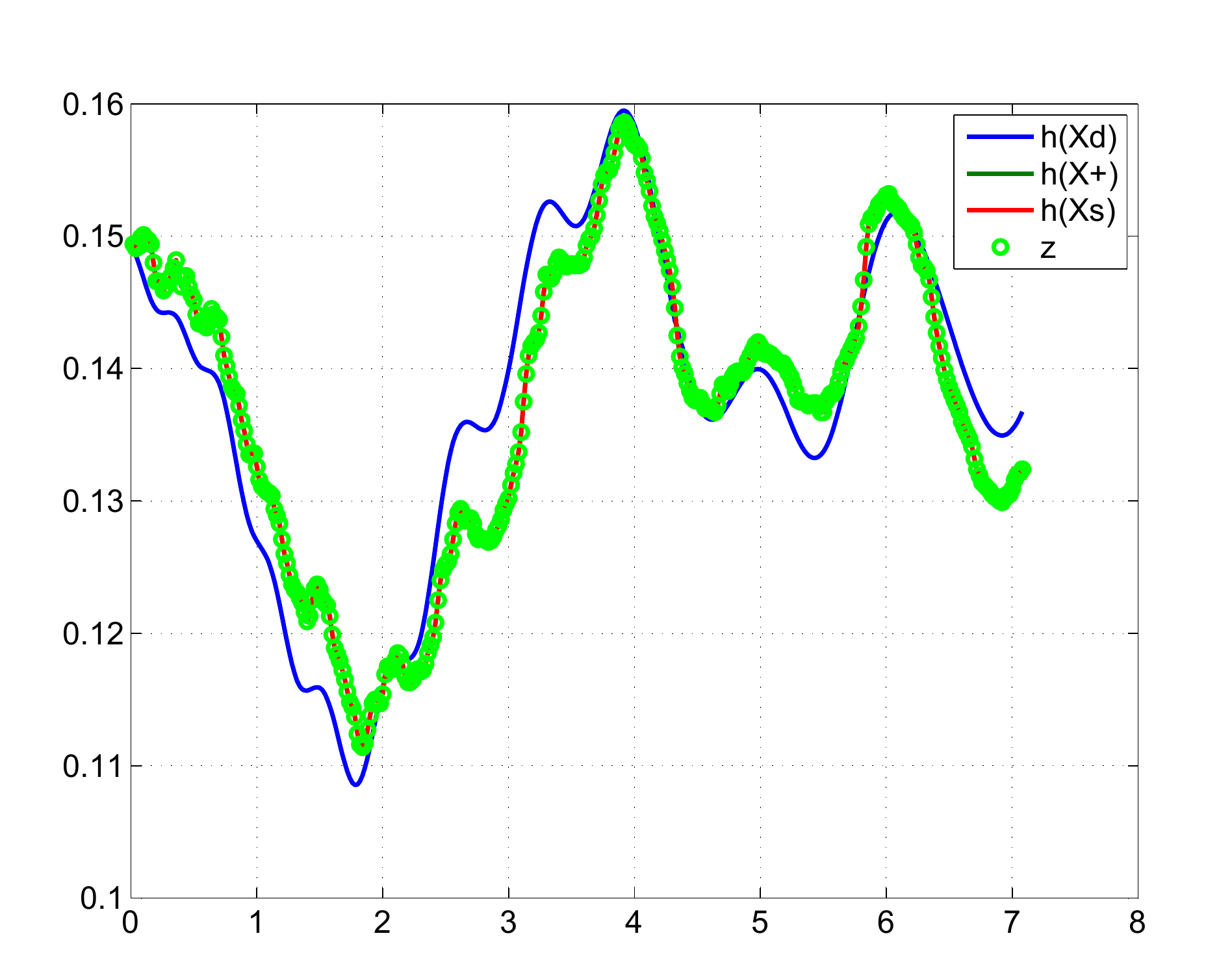}
\caption{Comparison of the predicted dynamics, posterior, smoothed}
\caption*{and the measurement 2}
\label{realQ1_s2}
\end{figure}
\begin{figure}[h]
\includegraphics[width=6in,height=4in]{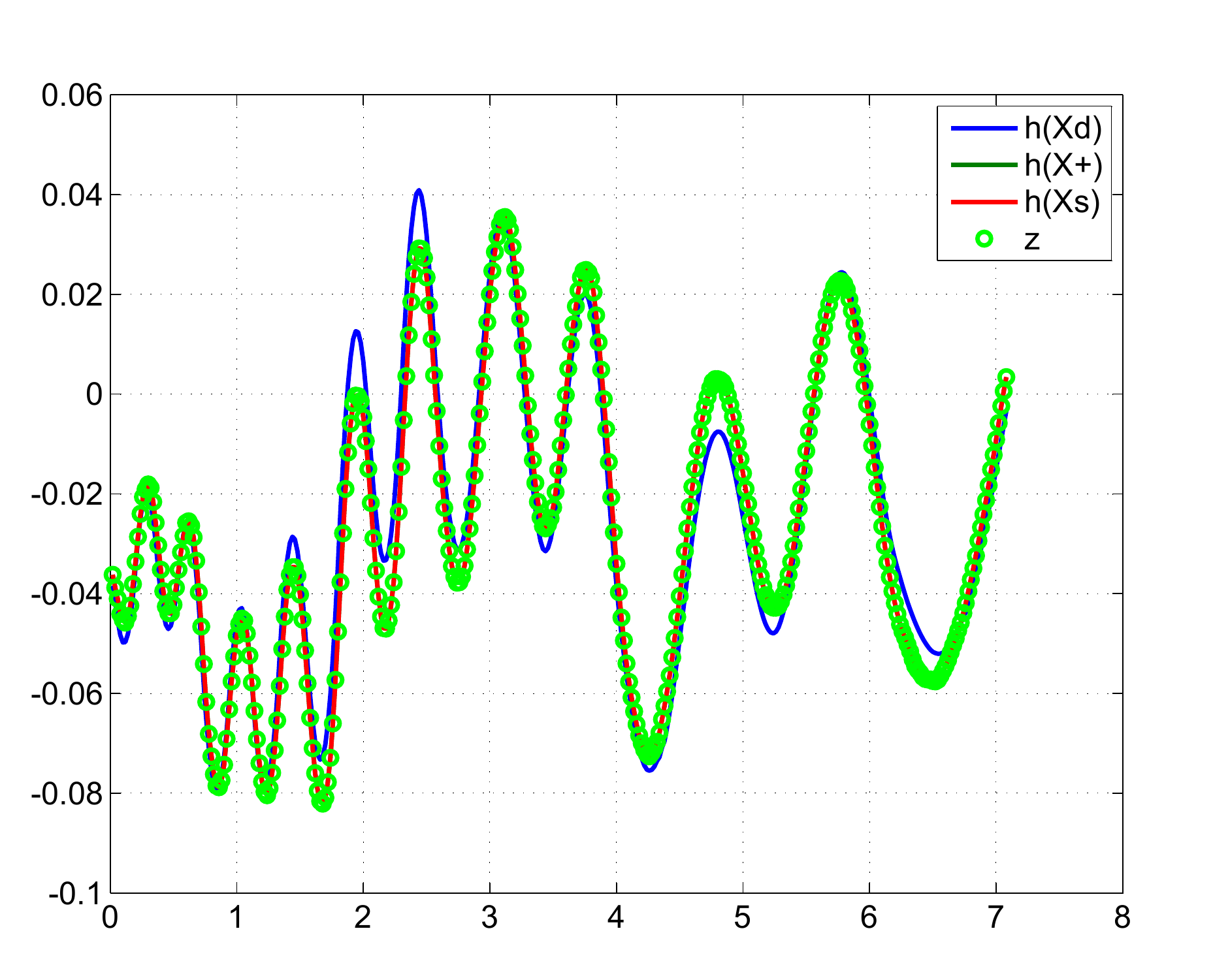}
\caption{Comparison of the predicted dynamics, posterior, smoothed}
\caption*{and the measurement 3}
\label{realQ1_s3}
\end{figure}
\begin{figure}[h]
\includegraphics[width=6in,height=4in]{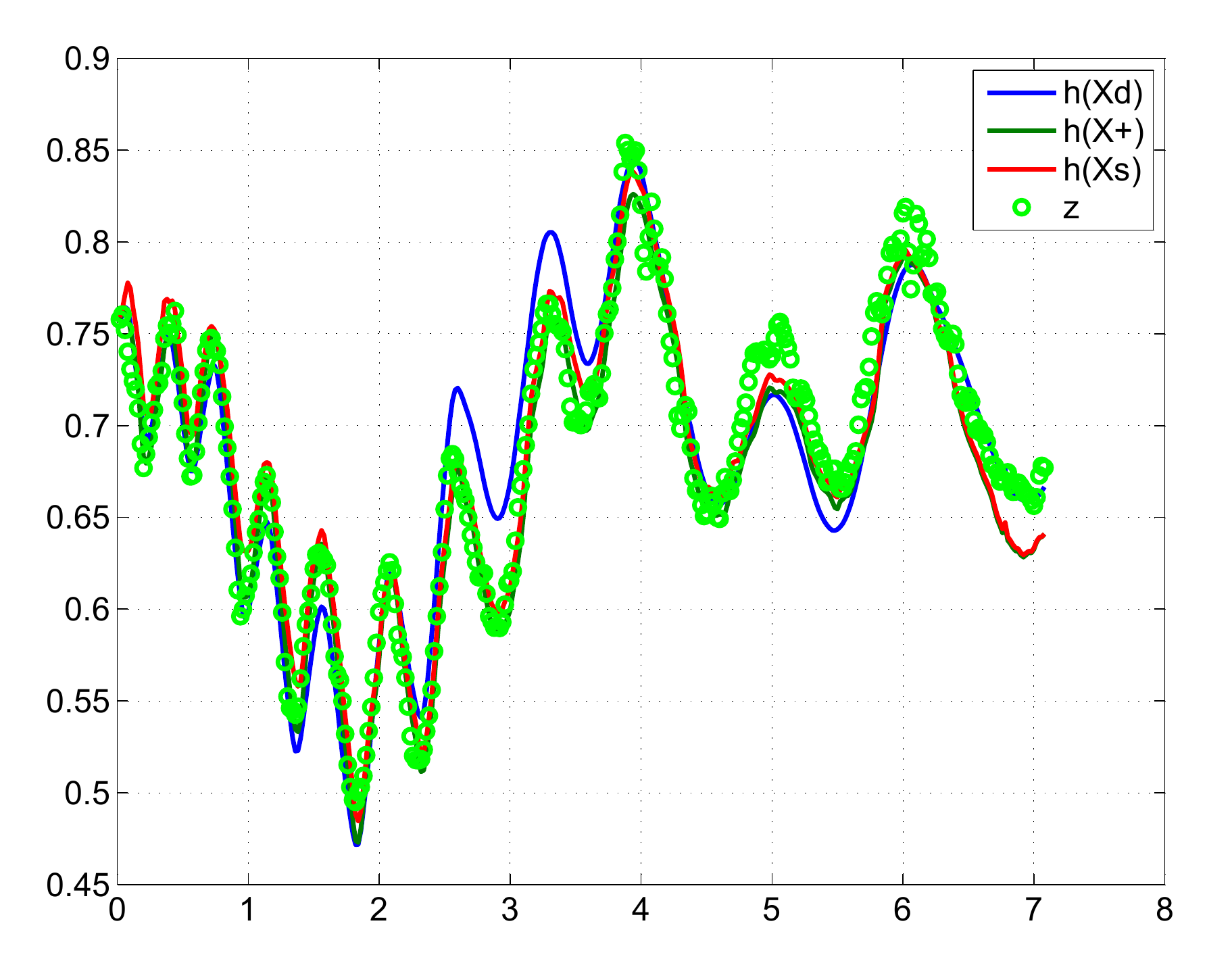}
\caption{Comparison of the predicted dynamics, posterior, smoothed}
\caption*{and the measurement 4}
\label{realQ1_h4}
\end{figure}
\begin{figure}[h]
\includegraphics[width=6in,height=4in]{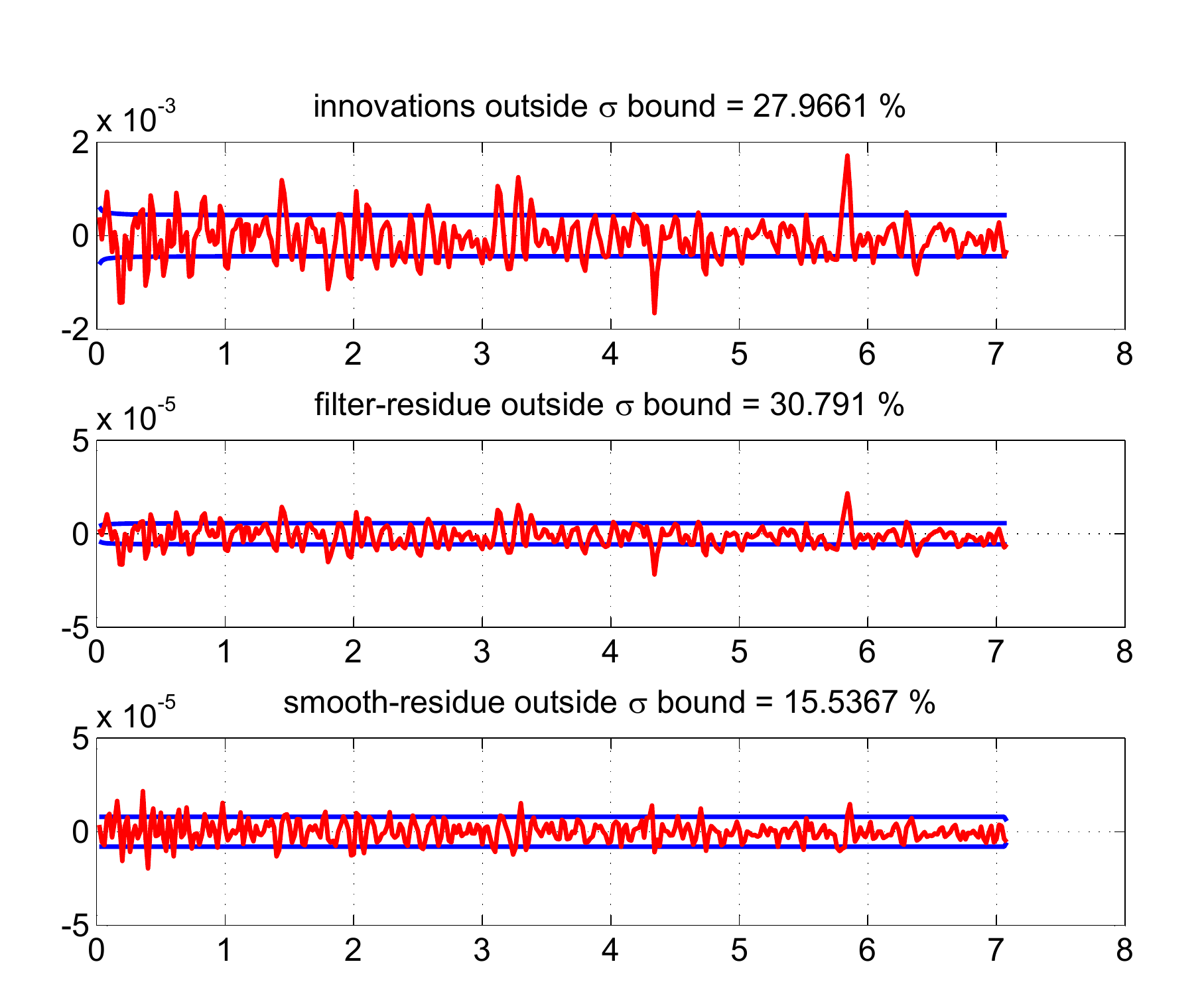}
\caption{The innovations, filtered residue and smoothed residue of measurement 1}
\label{realQ1_innov1}
\end{figure}
\begin{figure}[h]
\includegraphics[width=6in,height=4in]{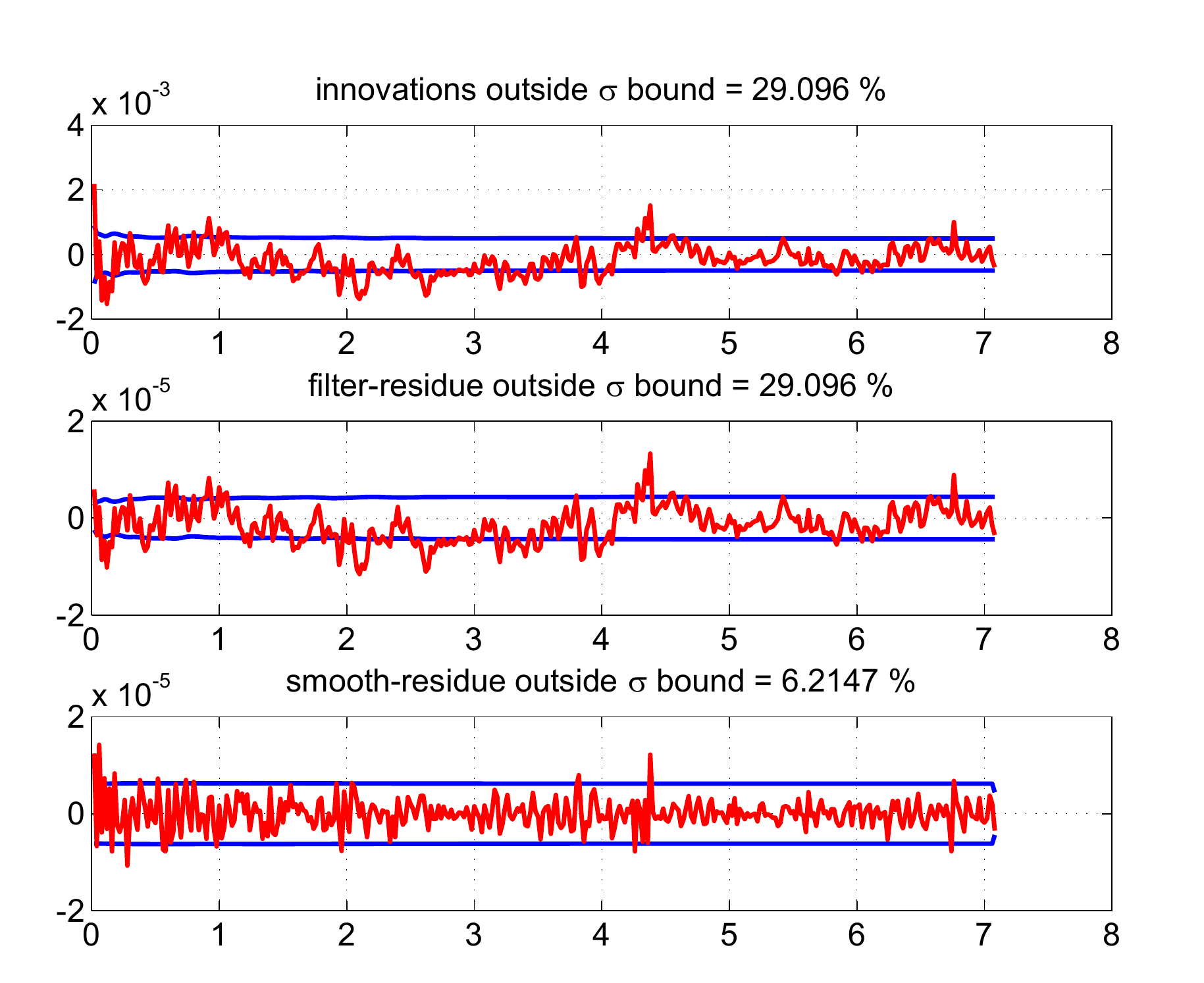}
\caption{The innovations, filtered residue and smoothed residue of measurement 2}
\label{realQ1_innov2}
\end{figure}
\begin{figure}[h]
\includegraphics[width=6in,height=4in]{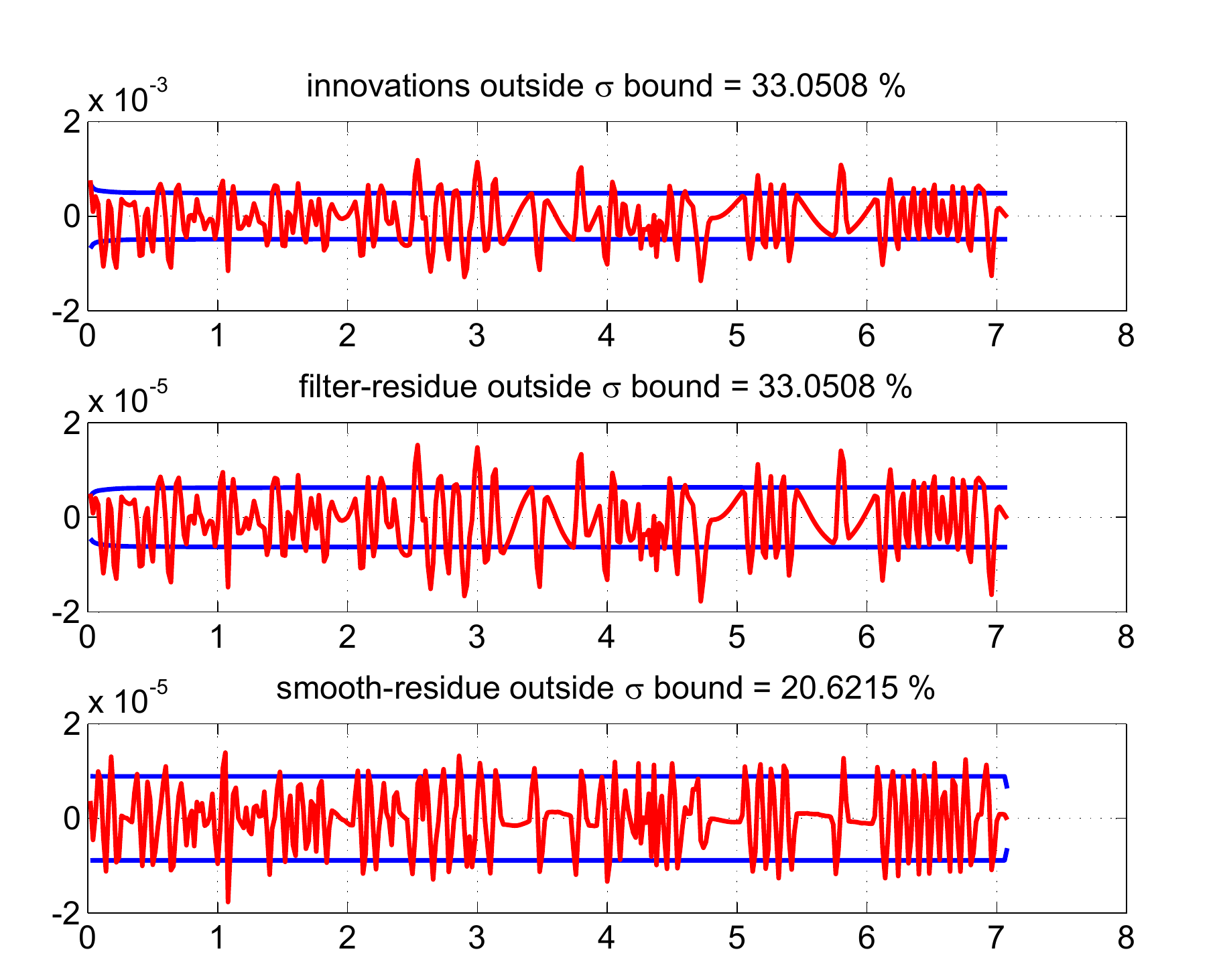}
\caption{The innovations, filtered residue and smoothed residue of measurement 3}
\label{realQ1_innov3}
\end{figure}
\begin{figure}[h]
\includegraphics[width=6in,height=4in]{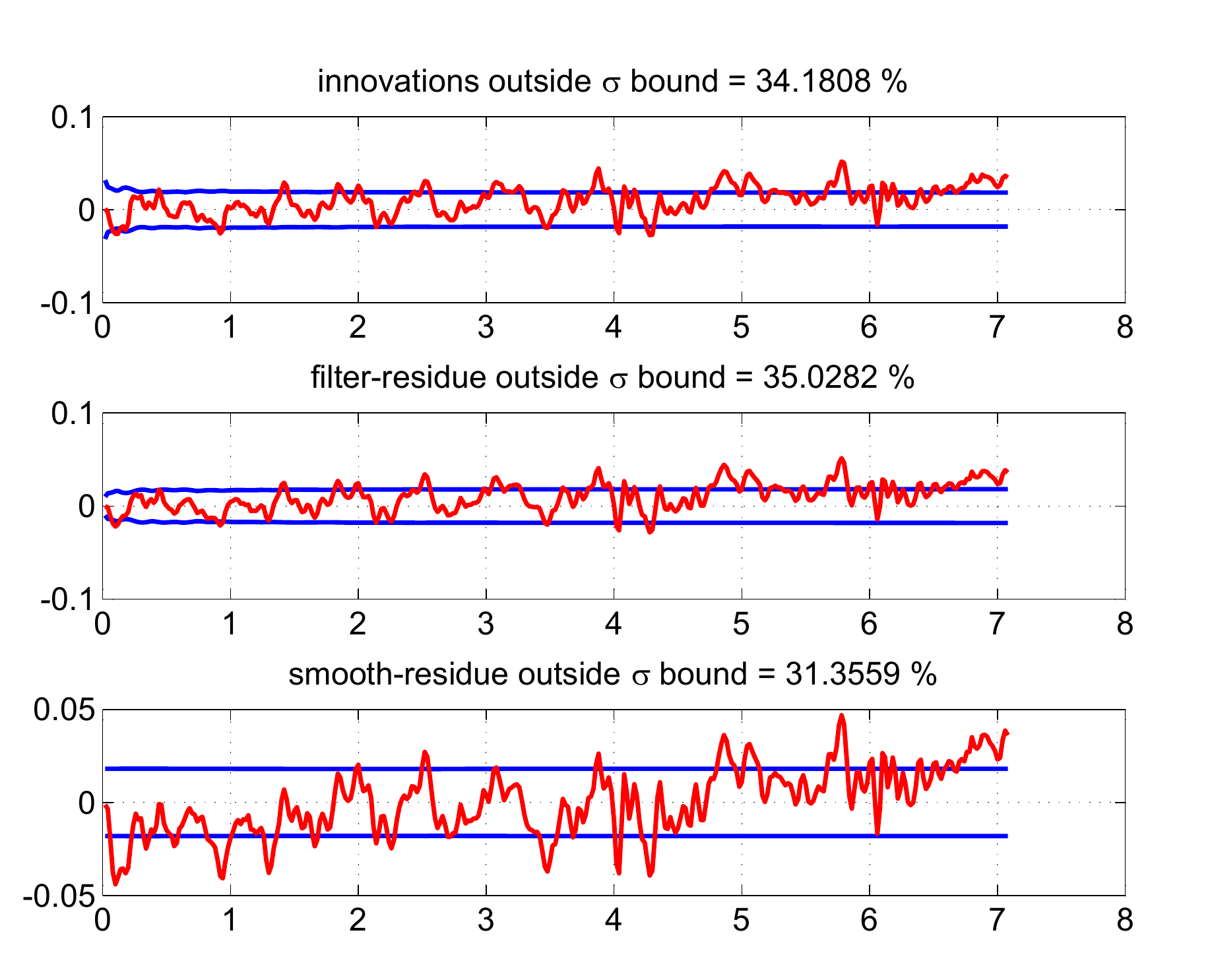}
\caption{The innovations, filtered residue and smoothed residue of measurement 4}
\label{realQ1_innov4}
\end{figure}
\begin{figure}[h]
\includegraphics[width=6in,height=4in]{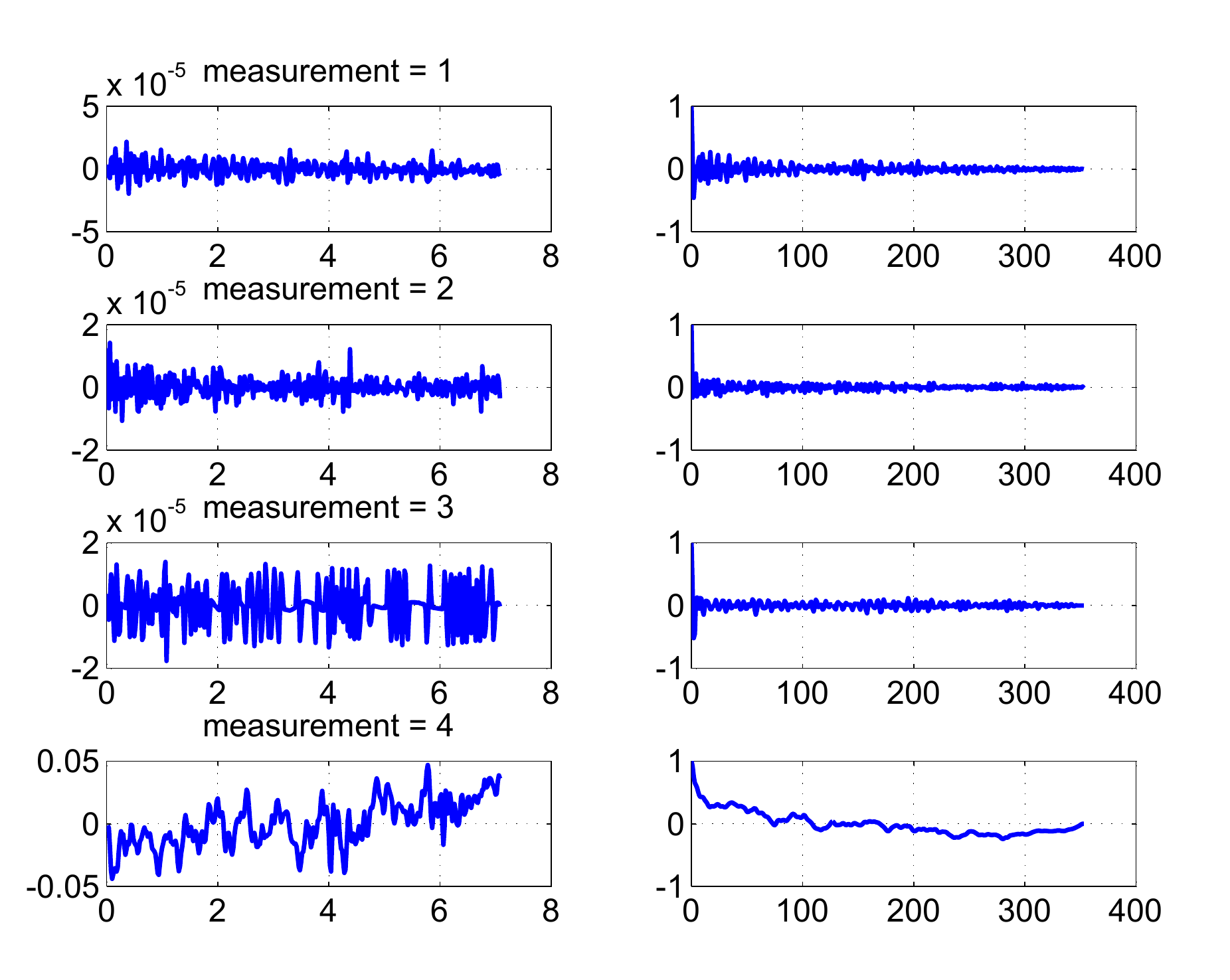}
\caption{Time variation of estimated measurement noise (left) and}
\caption*{their autocorrelation (right)}
\label{realQ1_mnoise}
\end{figure}
\begin{figure}[h]
\includegraphics[width=6in,height=4in]{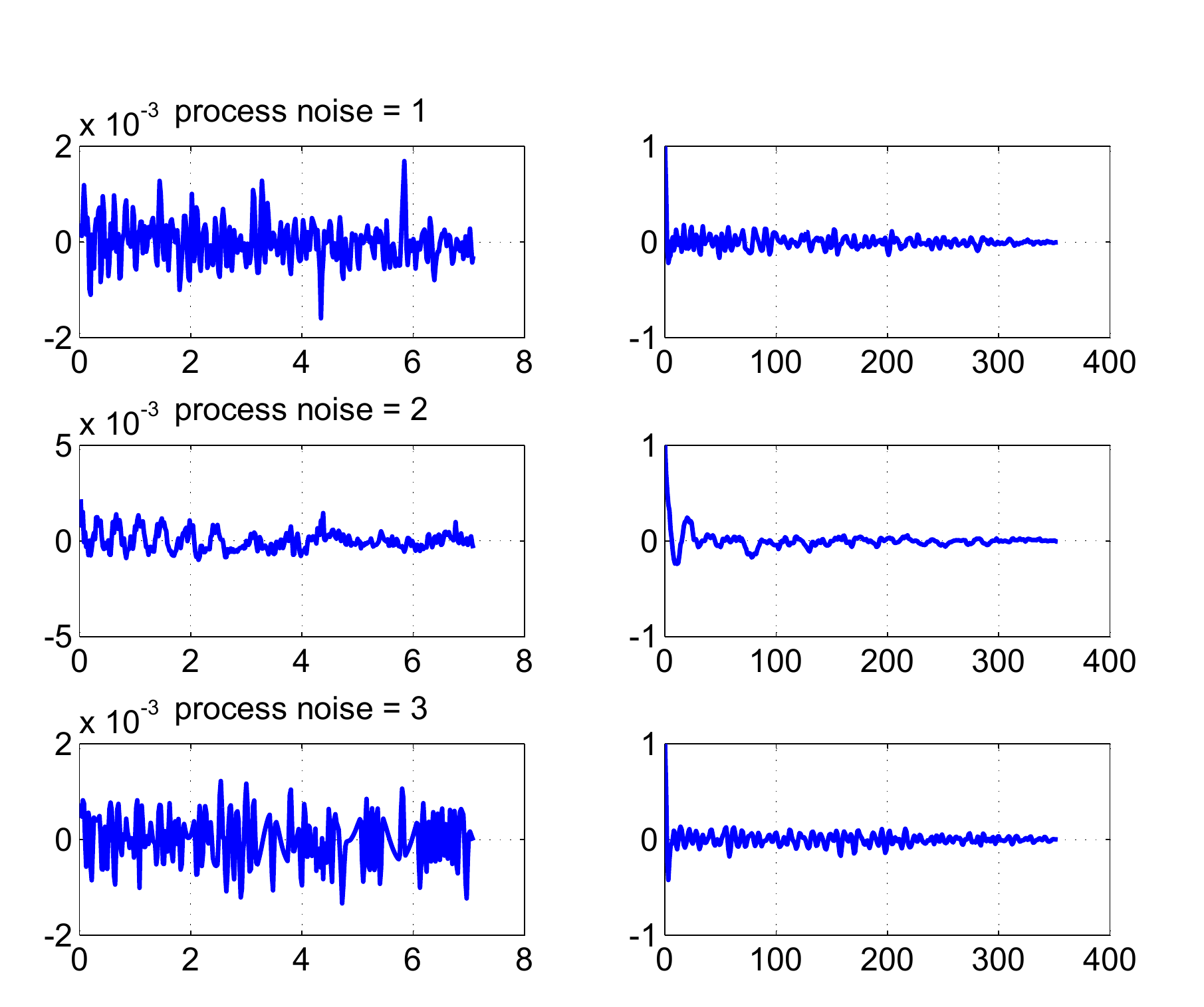}
\caption{Time variation of estimated process noise (left) and}
\caption*{their autocorrelation (right)}
\label{realQ1_pnoise}
\end{figure}

Thus the next step is to process the data by assuming \textbf{Q} $>$ 0 and here 100 iterations was used. This was inferred based on a plot in Fig. \ref{realQ1_err} of the difference between the iterated and final values over 100 iterations. The next Fig. \ref{realQ1_P0} shows the variation of parameter estimates and its initial covariance $\mathbf{P_0}$ with iterations and a similar Fig. \ref{realQ1_R} for \textbf{Q} and \textbf{R} over 100 iterations. The variation of the different costs \textbf{J1-J8} is more revealing and informative. The values of \textbf{J1-J3} are close to the number of measurements ($m$=4) with \textbf{J6-J8} are close to the number of states ($n$=3) as shown in Fig. \ref{realQ1_J} and Table-\ref{tbcase1Q}. This means the measurement and state equations are well balanced. The \textbf{J5} is the negative log likelihood cost function. Hence the filter results with \textbf{Q} $>$ 0 provide more confidence in the parameter $\Theta$ and noise covariance \textbf{Q} and \textbf{R} estimates. The later Fig. \ref{realQ1_s1}-\ref{realQ1_h4} compares (i) the state dynamics based on the estimated parameter after the filter pass through the data, (ii) the state after measurement update, (iii) the smoothed state and (iv) the measurement.  Unlike the case of \textbf{Q} = 0 except the first one as is to be expected the other three are close to each other thus providing more confidence in all the filter results. The Fig. \ref{realQ1_innov1}-\ref{realQ1_innov4} correspond to Fig. \ref{real1_innov1}-\ref{real1_innov4} of the \textbf{Q} = 0 case. Fig. \ref{realQ1_mnoise} shows that the estimated measurements do not appear to have constant statistical characteristics across time and may not be Additive White Gaussian Noise (AWGN). A similar plot for the estimated process noise is shown in Fig. \ref{realQ1_pnoise}.  Another experiment was carried out by generating a typical data set by using the estimated theta and injecting the estimated \textbf{Q} and \textbf{R} as additive white Gaussian noise. This is to determine the effect of non additive, non White and non Gaussian noise distribution in the real data on the CRBs. After each iteration in the reference recipe the $\Theta$, \textbf{Q} and \textbf{R} were reset as from the real data. Similar experiment was also conducted by updating $\Theta$ as well. It turned out that there is not much of a difference in the final CRBs as can be seen from the Table-\ref{tbnew1}.
\par Finally two other filter runs were carried out using the Myers and Tapley \cite{MT1976} (1976) as well as Mohamed and Schwarz \cite{MS1999} (1999) statistics for the estimation of \textbf{Q} and \textbf{R}. The behaviour of the various cost function and in particular \textbf{J6} and \textbf{J7} in Table-\ref{tbcase1QMTMS} shows that the choice of the filter statistics for estimating \textbf{Q} and \textbf{R} in the proposed reference approach is the best possible when compared to other approaches presently considered.

\subsection{Remarks on Case-1 }
\par
\begin{figure}[h]
\includegraphics[width=6in,height=4in]{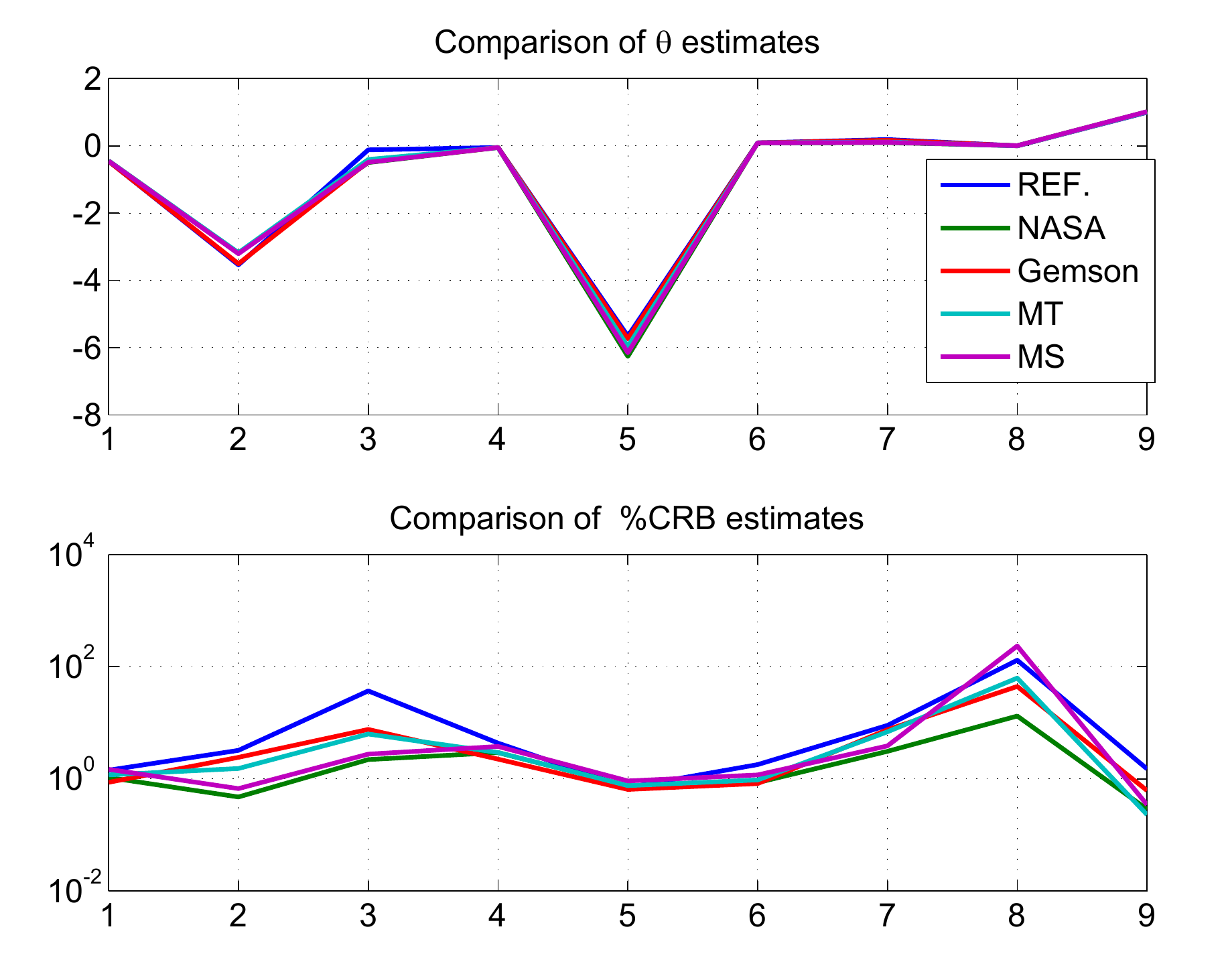}
\caption{Comparison of the parameter estimates and $\%$CRBs by different methods}
\label{comp1}
\end{figure}

 The NASA results have been generated assuming \textbf{Q} = 0 and are comparable with reference procedure for the parameter estimates and their CRBs assuming \textbf{Q} = 0. There are only slight differences in most of the parameter estimates from the reference procedure run with \textbf{Q} = 0 and \textbf{Q} $>$ 0.  However it may be noted that the important parameter $M_q$ representing the pitching moment coefficient with respect to pitch is only one fourth of the value and its CRB is also about 6 times than for \textbf{Q} = 0 case. The weakest parameter in terms of controlling the dynamics parameter $\Theta_0$ has sign changes among the different estimation procedures. It may be noted that its CRB is of the same order as its parameter value which explains the sign change at times. Regarding the noise estimates in the reference case with \textbf{Q} $>$ 0 the \textbf{R} values are lower than in the \textbf{Q} = 0 case perhaps at the cost of the emerging \textbf{Q}. Further the MT and MS methods give quite different estimates for the \textbf{R} values and much lower values for \textbf{Q} than in the reference \textbf{Q} $>$ 0 case. Hence we believe that the present reference procedure provides the best possible parameter estimates and their uncertainties. From the plot of the parameter estimates and their \%CRB in Fig \ref{comp1}, it can be seen that the parameters 1, 2, 3, 4, 5, 6, 7, 9, 10 are strong and the parameter 8 is the weak one.  The \%CRBs as estimated by different methods generally appear to vary widely. However what is interesting is that even the estimate of the strong parameter such as 5 varies widely among the methods. Such a behaviour of the filter estimates shows how important is the tuning of the filter statistics namely $\mathbf{P_0}$, \textbf{Q} and \textbf{R} in parameter estimation and their uncertainties. \par
The correlation co-efficient matrix (C) for i, j = 1, 2,\ldots $p$ is defined  as
\begin{align*}
C_{ij}=\frac{P_{\Theta_{ij}}}{\sqrt{P_{\Theta_{ii}}\times P_{\Theta_{jj}}}}
\end{align*}
where $P_{\Theta}$ is the parameter covariance matrix estimated by the EKF at the last time instant of the final iteration. The rounded off 100$\times$C matrix for case-1 (\textbf{Q} = 0) is given by
\begin{align*}
\begin{bmatrix}
   100  & -79 &   57 &  -91 &   80 &  100 &  -79 &  -73  & -66  \\
   -79  & 100 &  -76 &   72 &  -99 &  -79 &   99 &   73  &  42  \\
    57  & -76 &  100 &  -51 &   73 &   56 &  -78 &  -52  & -30  \\
   -91  &  72 &  -51 &  100 &  -73 &  -91 &   72 &   66  &  61  \\
    80  & -99 &   73 &  -73 &  100 &   81 &  -99 &  -74  & -43  \\
   100  & -79 &   56 &  -91 &   81 &  100 &  -80 &  -74  & -65  \\
   -79  &  99 &  -78 &   72 &  -99 &  -80 &  100 &   73  &  42  \\
   -73  &  73 &  -52 &   66 &  -74 &  -74 &   73 &  100  &   1  \\
   -66  &  42 &  -30 &   61 &  -43 &  -65 &   42 &    1  & 100  \\
\end{bmatrix}
\end{align*}

The rounded off 100$\times$C matrix for case-1 (\textbf{Q} $>$ 0) is given by
\begin{align*}
\begin{bmatrix}
   100   &  0  &   0 &  -84 &    0 &  100 &    0  &   0  & -99 \\
     0  & 100  & -66 &    0 &  -84  &   0 &  100  &   1  &   0  \\
     0  & -66  & 100  &   0 &   53  &   0 &  -67  &  -1  &   0  \\
   -84  &   0  &   0 &  100 &    0  & -85 &    0  &   0  &  85 \\
     0  & -84  &  53 &    0 &  100  &   0 &  -85  &  -1  &   0 \\
   100  &   0  &   0 &  -85 &    0  & 100 &    0  &   0  &-100 \\
     0  & 100  & -67 &    0 &  -85  &   0 &  100  &   1  &   0 \\
     0  &   1  &  -1 &    0  &  -1  &   0 &    1  & 100  &   0 \\
   -99  &   0  &   0 &   85 &    0 & -100 &    0  &   0  & 100 \\
\end{bmatrix}
\end{align*}

Ideally the above matrix is expected to be a diagonal matrix with 100 as its diagonal value. The correlation co-efficient matrix for \textbf{Q} $>$ 0 is more closer to the ideal values. The above matrix comparison indicates the usefullness of the right choice of \textbf{Q} in a real case scenario where one cannot model the system accurately. Hence we will be running only the reference EKF suggested for \textbf{Q} $>$ 0 case for the analysis of all the subsequent real cases.

\clearpage

\begin{landscape}
\begin{table}[h]
\subsection{Case-1 Tables}
\caption{Real flight test data case-1 results using the reference adaptive EKF (\textbf{Q} = 0)\\ No of iterations=20}{}
\label{tbcase1}
\begin{center}
\begin{footnotesize}
\begin{tabular}{|c| c| c| c| c| c|c|c|c|c| }
\hline
Study &
\makecell{$\Theta$\\ (Ref)} &
\makecell{$\Theta$\\ (NASA)} &
\makecell{$\Theta$\\ (Gemson)} &
\makecell{$\sigma_\Theta$ \\(Ref)} &
\makecell{$\sigma_\Theta$\\ (NASA)} &
\makecell{$\sigma_\Theta$\\ (Gemson)} &
\makecell{\textbf{R} \\ $\times10^{-3}$\\ (Ref)}&
\makecell{\textbf{J1-J8} \\(Ref) }&
Remarks
\\ \hline

\multicolumn{10}{|c|}{Case-1, Longitudinal case in dimensional form} \\ \hline

\makecell{$\mathbf{P_0}$ : Scaled up-[0,0;0,\checkmark]\\\textbf{Q} : $10^{-10}$-[\checkmark,0;0,0] \\\textbf{R} : EM-diag} &
\makecell{ -0.4604 \\  -3.2725 \\  -0.4964 \\  -0.0530  \\ -6.0924 \\   0.0871 \\   0.1138 \\
0.0010  \\  1.0280} &
\makecell{ -0.4502 \\  -3.192 \\  -0.5003 \\  -0.05197  \\ -6.264 \\   0.08429 \\   0.09206 \\
0.001517  \\  1.012} &
\makecell{ -0.4616 \\  -3.5093 \\  -0.4737 \\  -0.0534  \\ -5.7407 \\   0.0852 \\   0.1653 \\
-0.0009  \\  1.0038} &
\makecell{ 0.0054  \\  0.0151  \\  0.0103 \\   0.0018  \\  0.0454  \\  0.0008  \\  0.0028 \\ 0.0002  \\  0.0029 } &
\makecell{ 0.004742  \\  0.01515  \\  0.01104 \\   0.001529  \\  0.04719  \\  0.0007219  \\  0.002837 \\ 0.0002  \\  0.0029 } &
\makecell{ 0.0040  \\  0.0845  \\  0.0360 \\   0.0012  \\  0.0372  \\  0.0007  \\  0.0124 \\ 0.0004  \\  0.0062 } &
\makecell{0.0056 \\   0.0205  \\  0.0051  \\  0.2026} &
\makecell{ 4.0886 \\   4.0647  \\  3.9989  \\  0.0003 \\ -39.3360 \\  21.6246 \\  20.5621 \\ 3.9875} &
\makecell{Improper fit as seen in\\ Fig. \ref{real1_s1}-\ref{real1_s4} due \\to the absence of \textbf{Q}}
\\ \hline

\end{tabular}
\end{footnotesize}
\end{center}
\caption{Real flight test data case-1 results using the reference adaptive EKF (\textbf{Q} $>$ 0)\\ No of iterations=100}{}
\label{tbcase1Q}
\begin{center}
\begin{footnotesize}
\begin{tabular}{|c| c| c| c| c| c|c|c|c|c|c| }
\hline
Study &
\makecell{$\Theta$\\ (Ref)} &
\makecell{$\Theta$\\ (NASA)} &
\makecell{$\Theta$\\ (Gemson)} &
\makecell{$\sigma_\Theta$ \\(Ref)} &
\makecell{$\sigma_\Theta$\\ (NASA)} &
\makecell{$\sigma_\Theta$\\ (Gemson)} &
\makecell{\textbf{R} \\ $\times10^{-6}$\\ (Ref)}&
\makecell{\textbf{Q} \\ $\times10^{-6}$\\ (Ref)}&
\makecell{\textbf{J1-J8} \\(Ref) }&
Remarks
\\ \hline

\multicolumn{11}{|c|}{Case-1, Longitudinal case in dimensional form} \\ \hline

\makecell{$\mathbf{P_0}$ : Scaled up-[0,0;0,\checkmark]\\\textbf{Q} : EM-[\checkmark,0;0,0] \\\textbf{R} : EM-diag} &
\makecell{ -0.4571 \\  -3.5397 \\  -0.1227 \\  -0.0507 \\  -5.6628   \\ 0.0839 \\   0.1829\\ -0.0010 \\   0.9941} &

\makecell{ -0.4502 \\  -3.192 \\  -0.5003 \\  -0.05197  \\ -6.264 \\   0.08429 \\   0.09206 \\
0.001517  \\  1.012} &
\makecell{ -0.4616 \\  -3.5093 \\  -0.4737 \\  -0.0534  \\ -5.7407 \\   0.0852 \\   0.1653 \\
-0.0009  \\  1.0038} &

\makecell{  0.0065  \\  0.1141  \\  0.0453 \\   0.0022  \\  0.0396  \\  0.0015  \\  0.0164 \\ 0.0013 \\   0.0149} &
\makecell{ 0.004742  \\  0.01515  \\  0.01104 \\   0.001529  \\  0.04719  \\  0.0007219  \\  0.002837 \\ 0.0002  \\  0.0029 } &
\makecell{ 0.0040  \\  0.0845  \\  0.0360 \\   0.0012  \\  0.0372  \\  0.0007  \\  0.0124 \\ 0.0004  \\  0.0062 } &

\makecell{ 0.0025  \\  0.0022  \\  0.0031 \\ 328.6067} &
\makecell{0.1876 \\   0.2408  \\  0.2293} &
\makecell{3.7982  \\  3.9098  \\  2.2523  \\  0.0008 \\ -49.9638  \\  2.9822  \\  2.9871 \\ 2.8604} &
\makecell{Cost functions \\converge to the\\ expected value}
\\ \hline

\end{tabular}
\end{footnotesize}
\end{center}
\end{table}

\begin{table}[h]
\caption{Case-1 results using simulated Additive White Gaussian Noise}{}
\label{tbnew1}
\begin{center}
\begin{footnotesize}
\begin{tabular}{|c| c| c| c| c|}
\hline
Study &
\makecell{$\sigma_\Theta$ \\(Simulated-without\\ updating $\Theta$)} &
\makecell{$\sigma_\Theta$ \\(Simulated-with\\ updating $\Theta$)} &
\makecell{$\sigma_\Theta$ \\(Ref)} &
Remarks
\\ \hline

\multicolumn{5}{|c|}{\makecell{Case-1 data generated using simulated measurement and process noise (AWGN) \\ of variance \textbf{Q} and \textbf{R} estimated by Reference EKF (\textbf{Q} $>$ 0)} } \\ \hline

\makecell{$\mathbf{P_0}$ : Scaled up-[0,0;0,\checkmark]\\\textbf{Q} : \textbf{Q} (Ref) \\\textbf{R} :  \textbf{R} (Ref)} &

\makecell{ 0.0053 \\   0.0941  \\  0.0416  \\  0.0022  \\  0.0397  \\  0.0014  \\  0.0136 \\ 0.0013   \\ 0.0149} &
\makecell{   0.0053 \\   0.0941  \\  0.0416  \\  0.0022  \\  0.0397  \\  0.0014 \\   0.0136 \\ 0.0013 \\   0.0149} &
\makecell{ 0.0065 \\   0.1141  \\  0.0453  \\  0.0022  \\  0.0396   \\ 0.0015  \\  0.0164  \\ 0.0013   \\ 0.0149 } &

\makecell{No Significant \\ change in $\sigma_\Theta$}
\\ \hline

\end{tabular}
\end{footnotesize}
\end{center}
%
\caption{Real flight test data case-1 results using the MT and MS method. \\ No of iterations=100 }{}
\label{tbcase1QMTMS}
\begin{center}
\begin{footnotesize}
\begin{tabular}{|c| c| c| c| c| c|| c|c|c|c|c|c| }
\hline
Study &
\makecell{$\Theta$\\ (MT)} &
\makecell{$\sigma_\Theta$ \\(MT)} &
\makecell{\textbf{R} (MT)\\ $\times10^{-6}$ }&
\makecell{\textbf{Q} (MT)\\ $\times10^{-6}$}&
\makecell{\textbf{J1-J8} \\(MT) }&

\makecell{$\Theta$\\ (MS)} &
\makecell{$\sigma_\Theta$\\ (MS)} &
\makecell{\textbf{R} (MS) \\ $\times10^{-6}$}&
\makecell{\textbf{Q} (MS)\\ $\times10^{-6}$}&
\makecell{\textbf{J1-J8} \\(MS) }&
Remarks
\\ \hline

\multicolumn{12}{|c|}{Case-1, Longitudinal case in dimensional form} \\ \hline

\makecell{$\mathbf{P_0}$ : Scaled up-[0,0;0,\checkmark]\\\textbf{Q} : MT/MS-[\checkmark,0;0,0] \\\textbf{R} : MT/MS-diag} &

\makecell{ -0.4511 \\  -3.1810 \\  -0.4088 \\  -0.0541 \\  -5.9706 \\   0.0837  \\  0.1088 \\-0.0008 \\   1.0039} &
\makecell{ 0.0053 \\   0.0487 \\   0.0257   \\ 0.0016 \\   0.0450 \\   0.0008 \\   0.0075 \\ 0.0005  \\  0.0023} &
\makecell{8.3  \\ 9.7    \\ 0.5  \\  148.2} &
\makecell{0.00023 \\   0.03652 \\   0.02931} &
\makecell{ 3.2977 \\   3.2896 \\   2.7542 \\   0.0003 \\ -42.8774 \\  21.2035 \\  20.9052 \\ 2.3735} &

\makecell{ -0.4552 \\  -3.2115 \\  -0.4934 \\  -0.0531 \\  -6.1559  \\  0.0853  \\  0.1011 \\ 0.0003 \\   1.0163} &
\makecell{ 0.0067  \\  0.0215 \\   0.0136  \\  0.0020  \\  0.0563  \\  0.0010 \\   0.0039 \\ 0.0007 \\   0.0035} &
\makecell{ 8.2906  \\ 30.0566  \\  0.4620 \\  259.4142} &
\makecell{  0.0001  \\  0.0013  \\  0.0523} &
\makecell{ 2.9032  \\  2.8957 \\   2.5550  \\  0.0003  \\-41.5716 \\  21.5763  \\ 21.3320 \\ 2.3743} &

\makecell{Cost functions are\\ not close to their \\expected values in\\ MT and MS method} \\ \hline

\end{tabular}
\end{footnotesize}
\end{center}
\end{table}
\end{landscape}


\subsection{Case-1 Figures (\textbf{Q} = 0)}

\clearpage
\subsection{Case-1 Figures (\textbf{Q} $>$ 0)}

\begin{figure}[h]
\includegraphics[width=6in,height=3.2in]{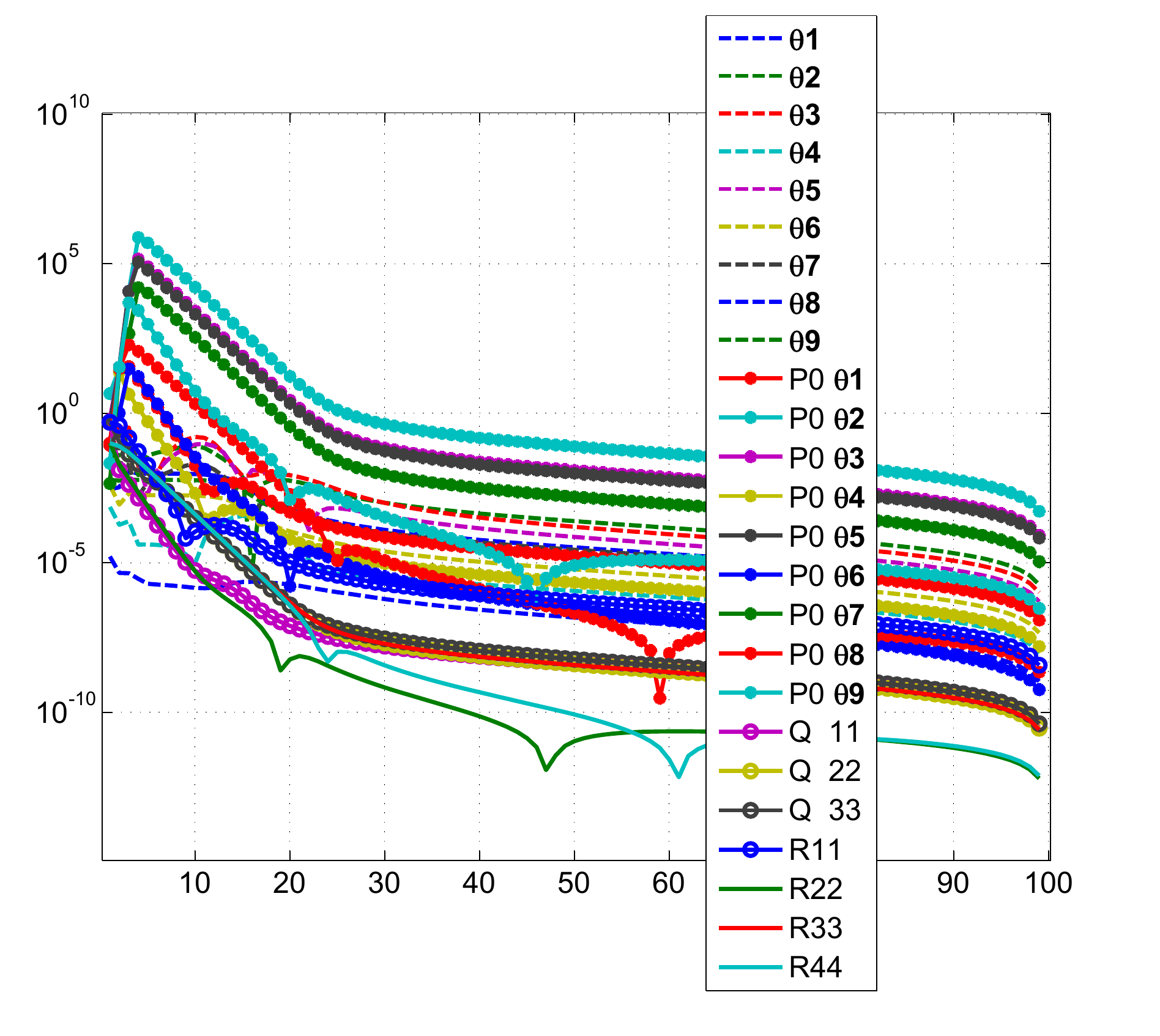}
\caption{The absolute difference between the iterated and final values}
\caption*{with 100 iterations}
\label{realQ1_err}
\end{figure}

\begin{figure}[h]
\includegraphics[width=6in,height=3.2in]{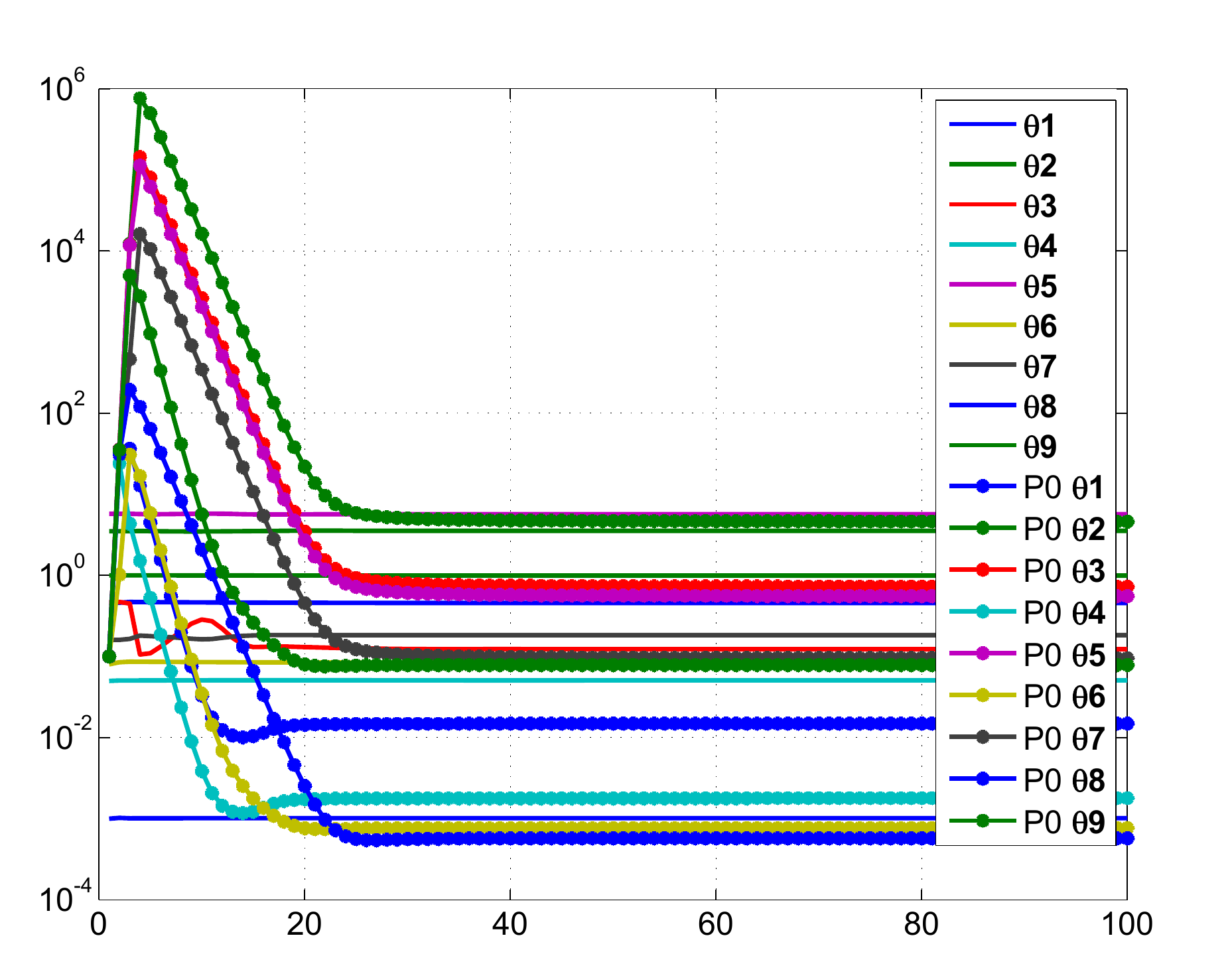}
\caption{Variation of parameter and its initial covariance ($\mathbf{P_0}$) with iterations}
\label{realQ1_P0}
\end{figure}

\clearpage
\section{Real Flight Test Case-2}
\par
\begin{figure}[h]
\includegraphics[width=6in,height=3.2in]{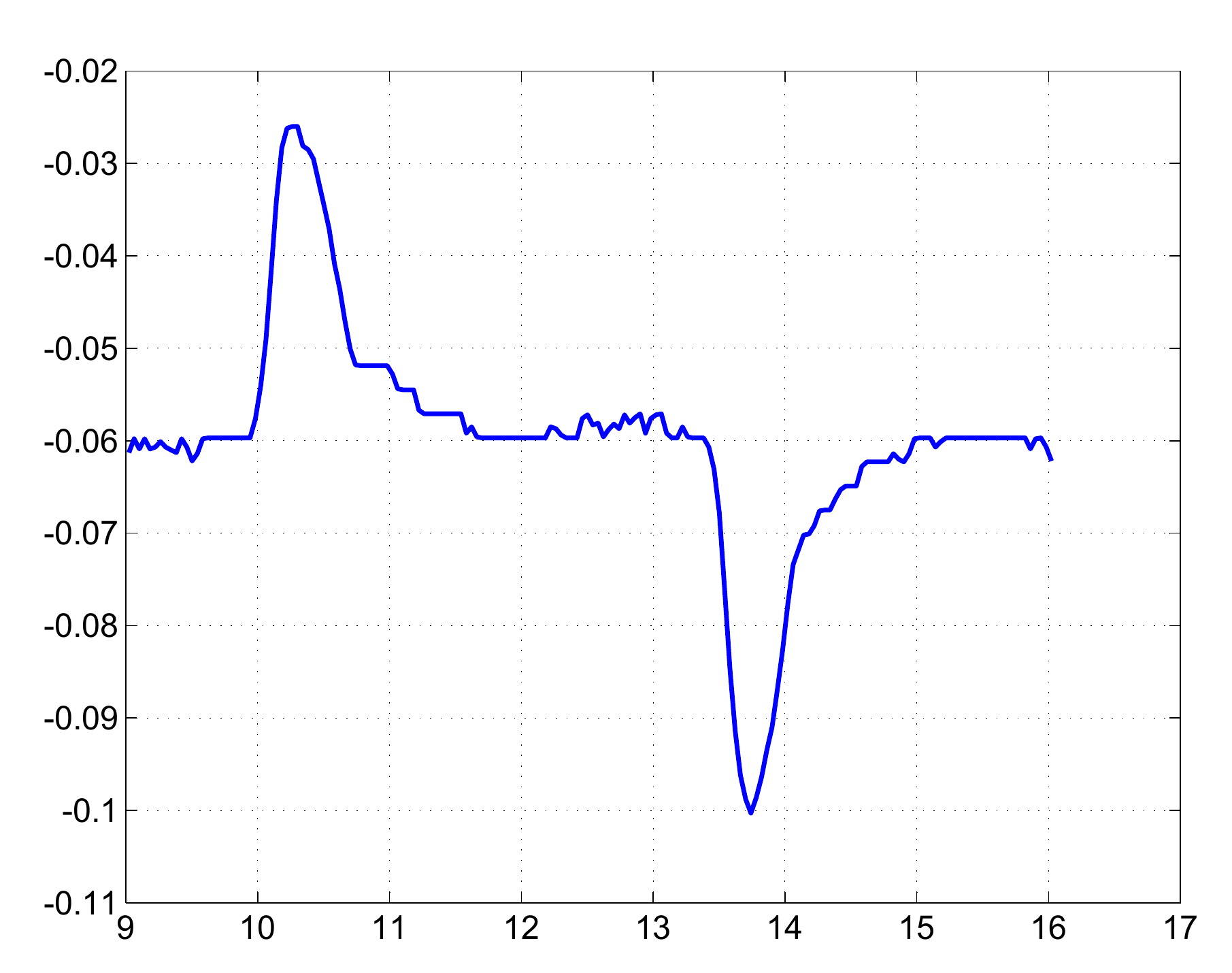}
\caption{Control input ($\delta_e$) versus time (s)}
\label{input2}
\end{figure}

\begin{figure}[h]
\includegraphics[width=6in,height=3.2in]{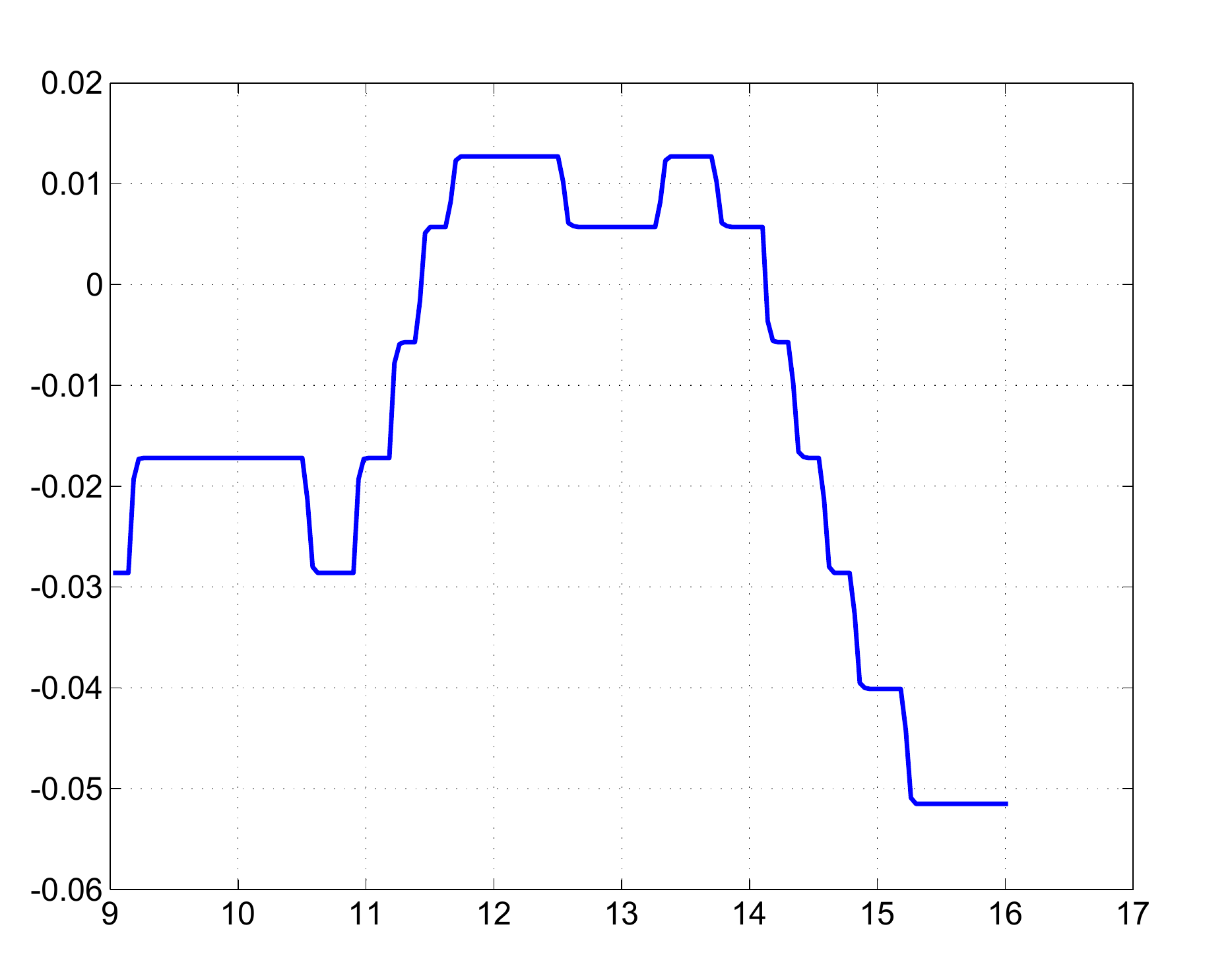}
\caption{Measurement input ($\phi_m$) versus time (s)}
\label{case2_phi}
\end{figure}

\begin{figure}[h]
\includegraphics[width=6in,height=4in]{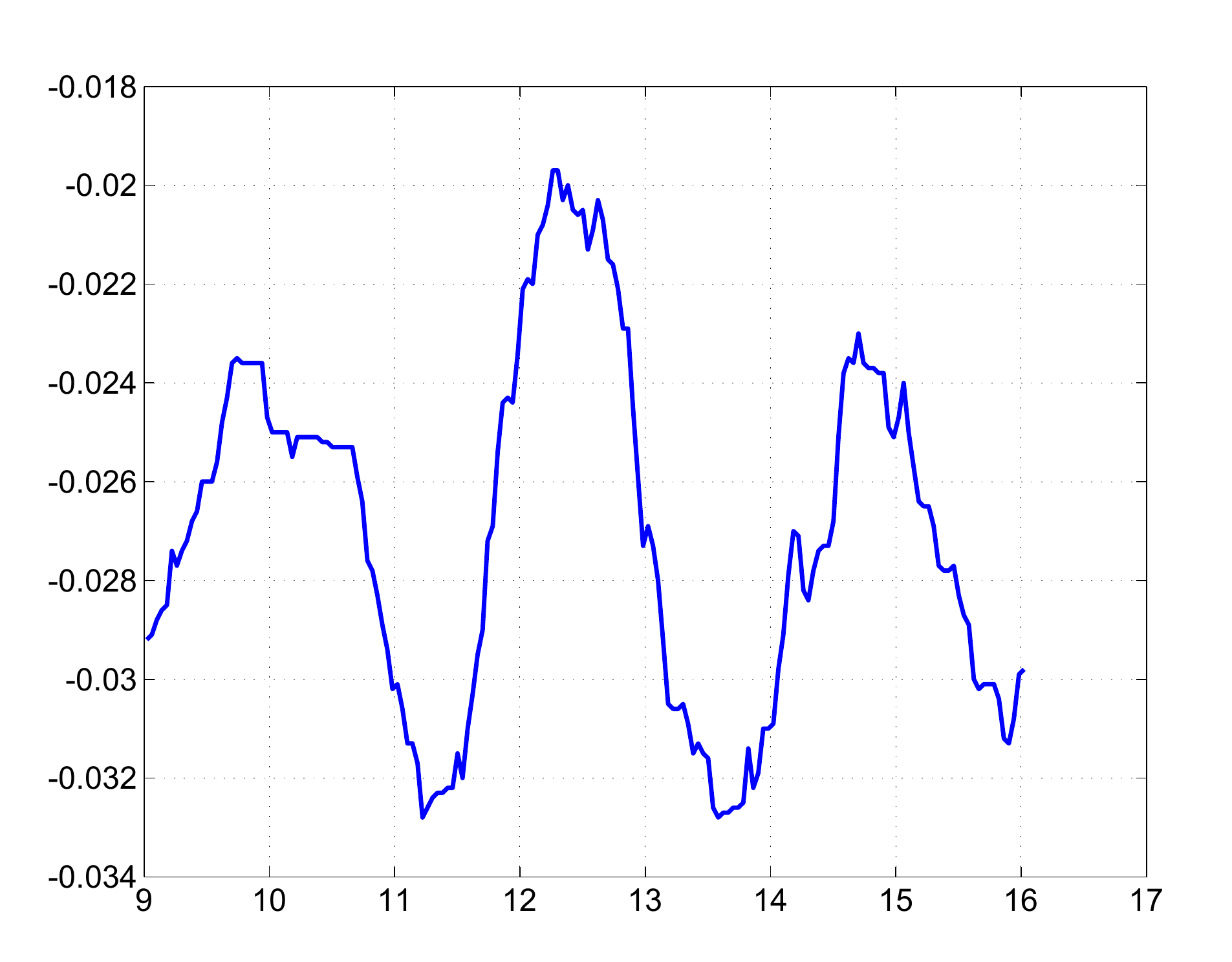}
\caption{Measurement input ($\beta_m$) versus time (s)}
\label{case2_beta}
\end{figure}

\begin{figure}[h]
\includegraphics[width=6in,height=4in]{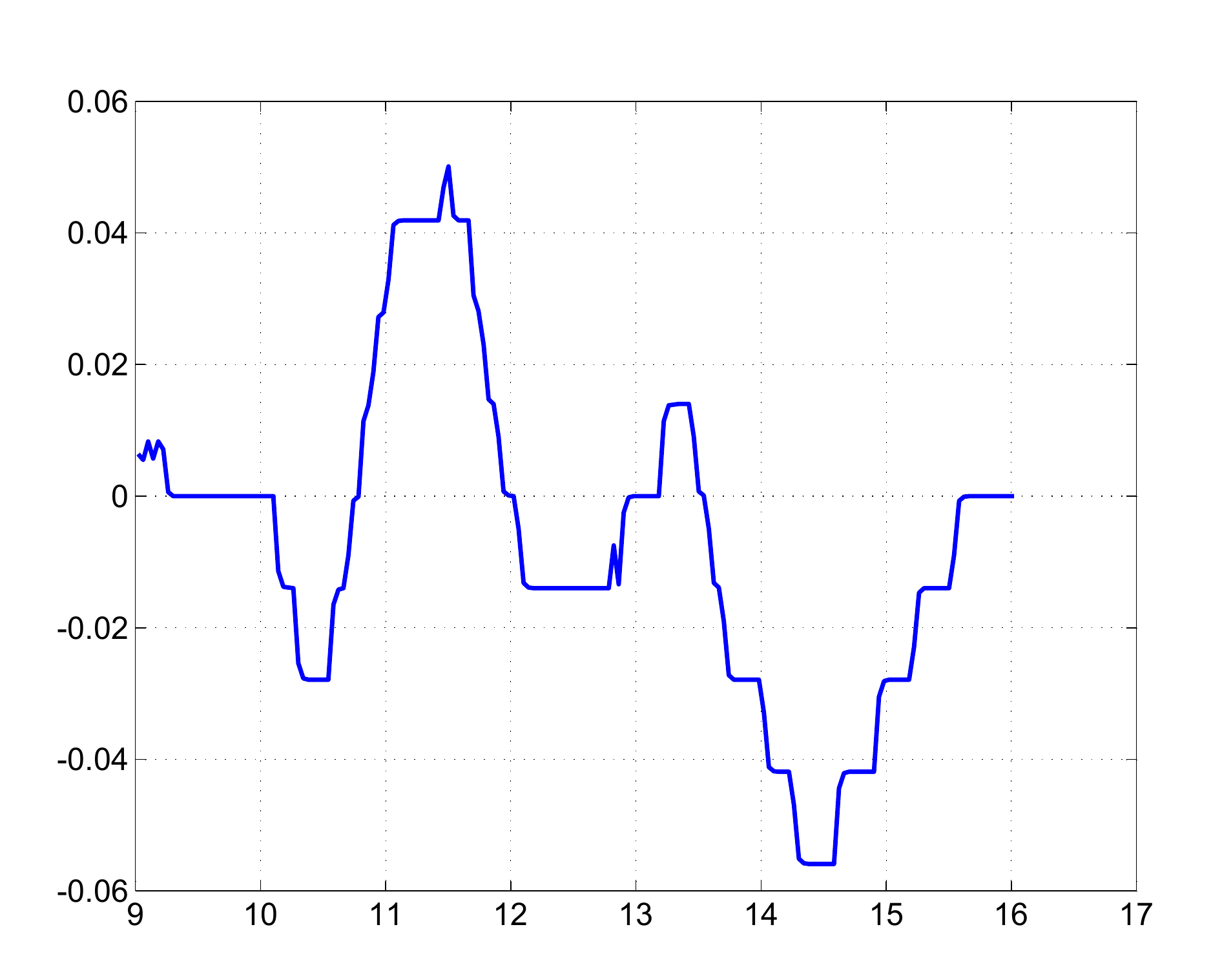}
\caption{Measurement input ($p_m$) versus time (s)}
\label{case2_p}
\end{figure}

\begin{figure}[h]
\includegraphics[width=6in,height=4in]{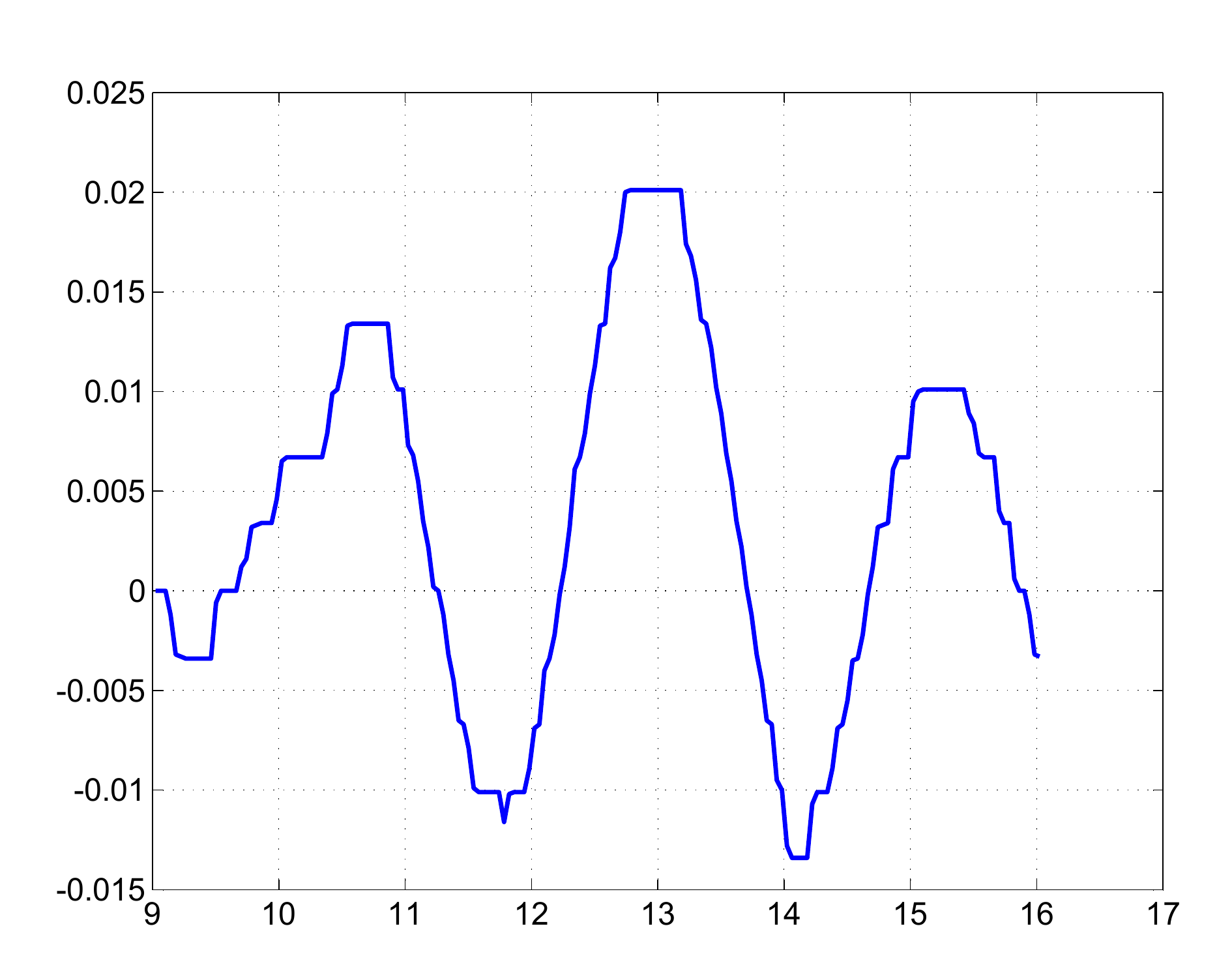}
\caption{Measurement input ($r_m$) versus time (s)}
\label{case2_r}
\end{figure}

\begin{figure}[h]
\includegraphics[width=6in,height=4in]{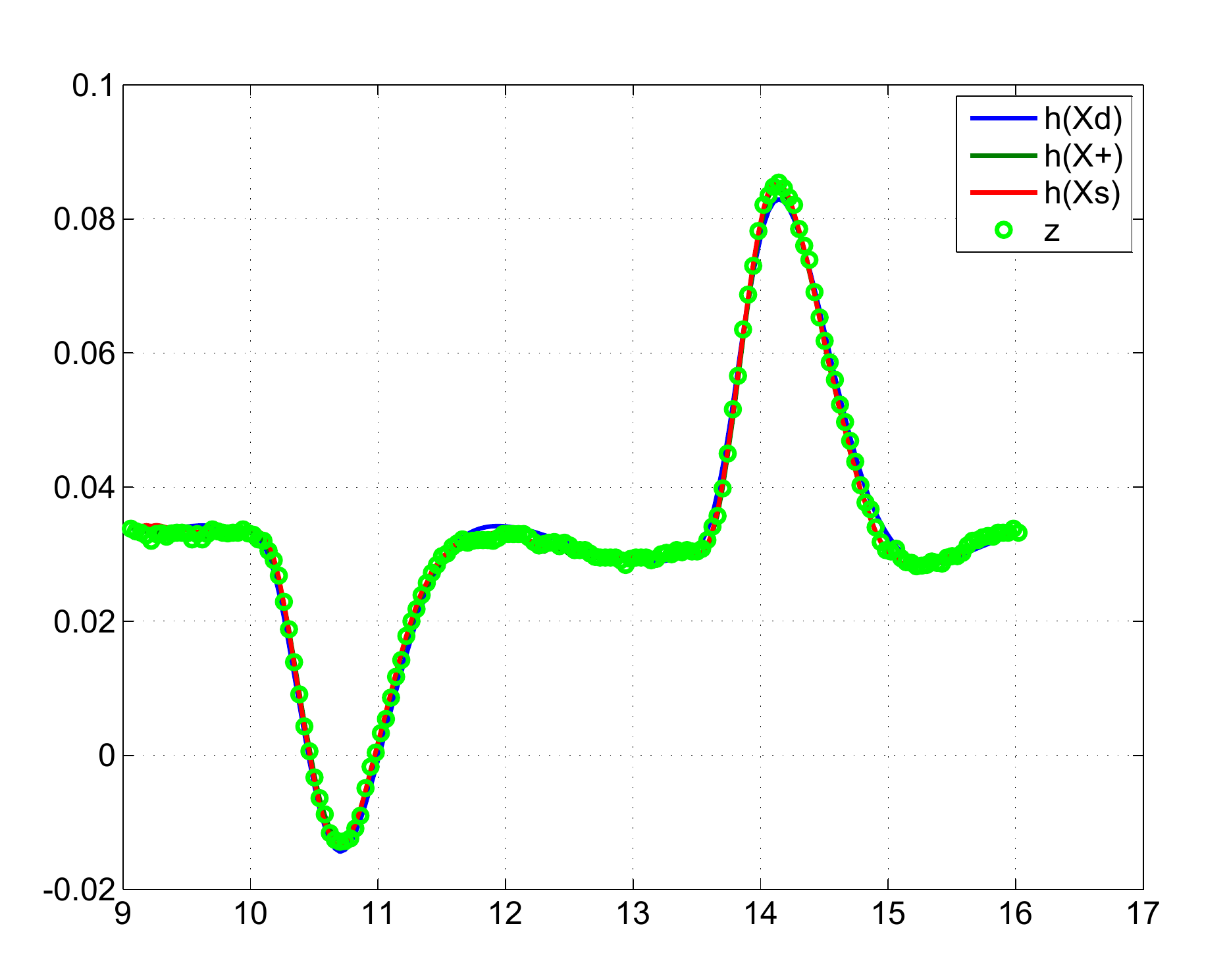}
\caption{Comparison of the predicted dynamics, posterior, smoothed}
\caption*{and the measurement 1}
\label{realQ2b1_s1}
\end{figure}

 The salient features of this aircraft are available in NASA TM-X 56036 (Shafer \cite{Shafer1975} 1975) and in NASA TP 1690  (Maine and Iliff \cite{Maine1981} 1981). The parameters are estimated in dimensionless form.  The data set is obtained is for a short period motion excited by the up and down elevator control input ($\delta_e$ in radians) as shown in Fig. \ref{input2}. Some of the available measurements can also be used as inputs in the state equations which includes roll angle ($\phi_m$), sideslip ($\beta_m$), roll rate ($p_m$), yaw rate ($r_m$) and the angle of attack ($\alpha_m$) are shown in Fig. \ref{case2_phi}, Fig. \ref{case2_beta}, Fig. \ref{case2_p}, Fig. \ref{case2_r} and Fig. \ref{realQ2b1_s1} respectively. The state equations ($n=3$) for the angle of attack ($\alpha$), pitch rate (q) and the pitch angle ($\Theta$) respectively are
\begin{align*}
\dot{\alpha}=&-\frac{\bar qS}{mV}C_L+q+\frac{g}{V} (cos(\phi_m)cos(\alpha_m)cos(\theta)+sin(\alpha_m)sin(\theta))- \\
&\beta_m(p_mcos(\alpha_m)+r_msin(\alpha_m))\\
\dot{q}=&\frac{\bar q S \bar c}{Iyy}(C_{m_\alpha}\alpha+C_{m_q}\frac{\bar c}{2V}q+C_{m_{\delta_e}}\delta_e+C_{m_0})+
\frac{Izz-Ixx}{Iyy}r_mp_m\\
\dot{\theta}=&qcos(\phi_m)-r_msin(\phi_m)+\theta_0
\end{align*}
The angle of attack ($\alpha$), pitch rate (q), the pitch angle ($\theta$), normal acceleration ($a_n$) and the longitudinal acceleration ($a_x$) are measured (indicated with subscript `m') in units of rad, rad/s, rad, $ft/s^2$ and $ft/s^2$ respectively. The measurement equations ($m$=5) are
\begin{align*}
{\alpha_m}&=\alpha-K_\alpha x_\alpha \frac{q}{V} \\
{q_m}&=q\\
{\theta_m}&=\theta\\
{a_{n_m}}&=\frac{\bar q S}{mg}C_N+\frac{x_{a_n}}{g}\dot{q} \\
{a_{x_m}}&=-\frac{\bar q S}{mg}C_A+\frac{z_{a_x}}{g}\dot{q}
\end{align*}
where
\begin{align*}
C_L=&C_Ncos(\alpha)-C_Asin(\alpha)+C_{L_0}\\
C_N=&C_{N_\alpha}\alpha+C_{N_{\delta_e}}\delta_e+C_{N_0}\\
C_A=&C_{A_\alpha}\alpha+C_{A_{\alpha^2}}\alpha^2+C_{A_{\delta_e}}\delta_e+C_{A_0}
\end{align*}


\newpage
The unknown parameter set ($p=13$) is $\Theta=(C_{N_{\alpha}},C_{N_{\delta_e}},C_{L_{0}},C_{m_{\alpha}},C_{m_{q}},C_{m_{\delta_e}},C_{m_{0}},\theta_0,\\C_{N_0},C_{A_{\alpha}},C_{A_{\alpha^2}},C_{A_{\delta_e}},C_{A_0})^T$. The ones with suffix `$\delta_e$' are the control derivatives, the ones with suffix zero are the biases and all others are aerodynamic derivatives. The initial states are taken as initial measurement and the initial parameter values are taken as $(4,0.24,0.17,-0.48,-17,-0.9,-0.05,-0.02,0.175,-0.3,0.03,-0.083,-0.015)^T$.

\begin{table}[h]
\begin{center}
\caption*{Other constant values used for case-2}{}
\begin{tabular}{| c | c | c | c | c | c | }
\hline
$\bar c$=5.58 & S=184 & m=172.667 & Ixx=4142.9  &  Iyy=3922.4 & Izz=7642.5  \\ \hline
g=32.2 & V=403.1 & $\bar q$=83.08 & $K_\alpha x_\alpha$=-0.0279 & $x_{a_n}$=0.101 & $z_{a_x}$=-1.17\\ \hline
\end{tabular}
\end{center}
\end{table}

\par
\begin{figure}[h]
\includegraphics[width=6in,height=4in]{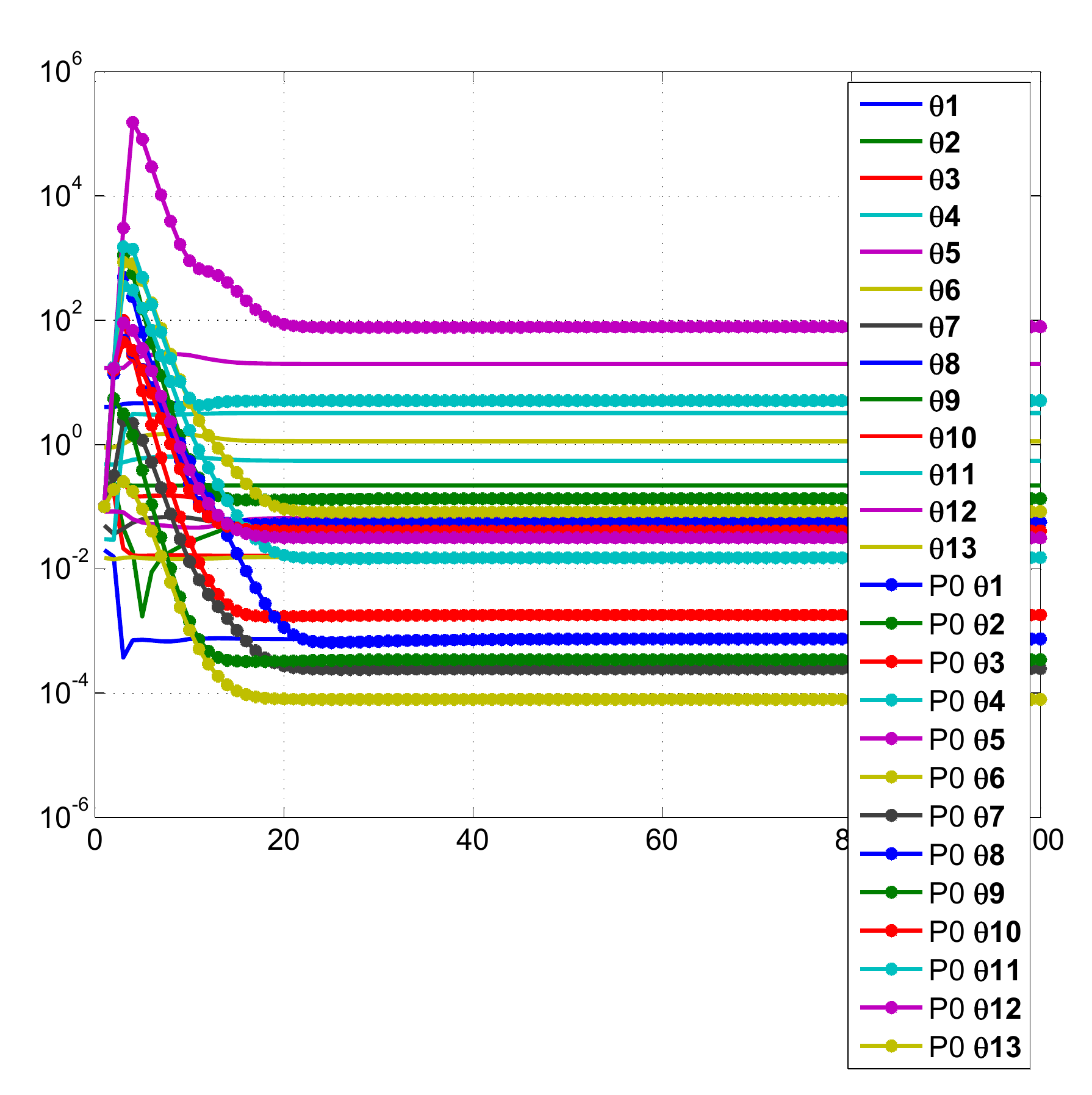}
\caption{Variation of parameter and its initial covariance ($\mathbf{P_0}$) with iterations}
\label{realQ2b1_P0}
\end{figure}

\begin{figure}[h]
\includegraphics[width=6in,height=4in]{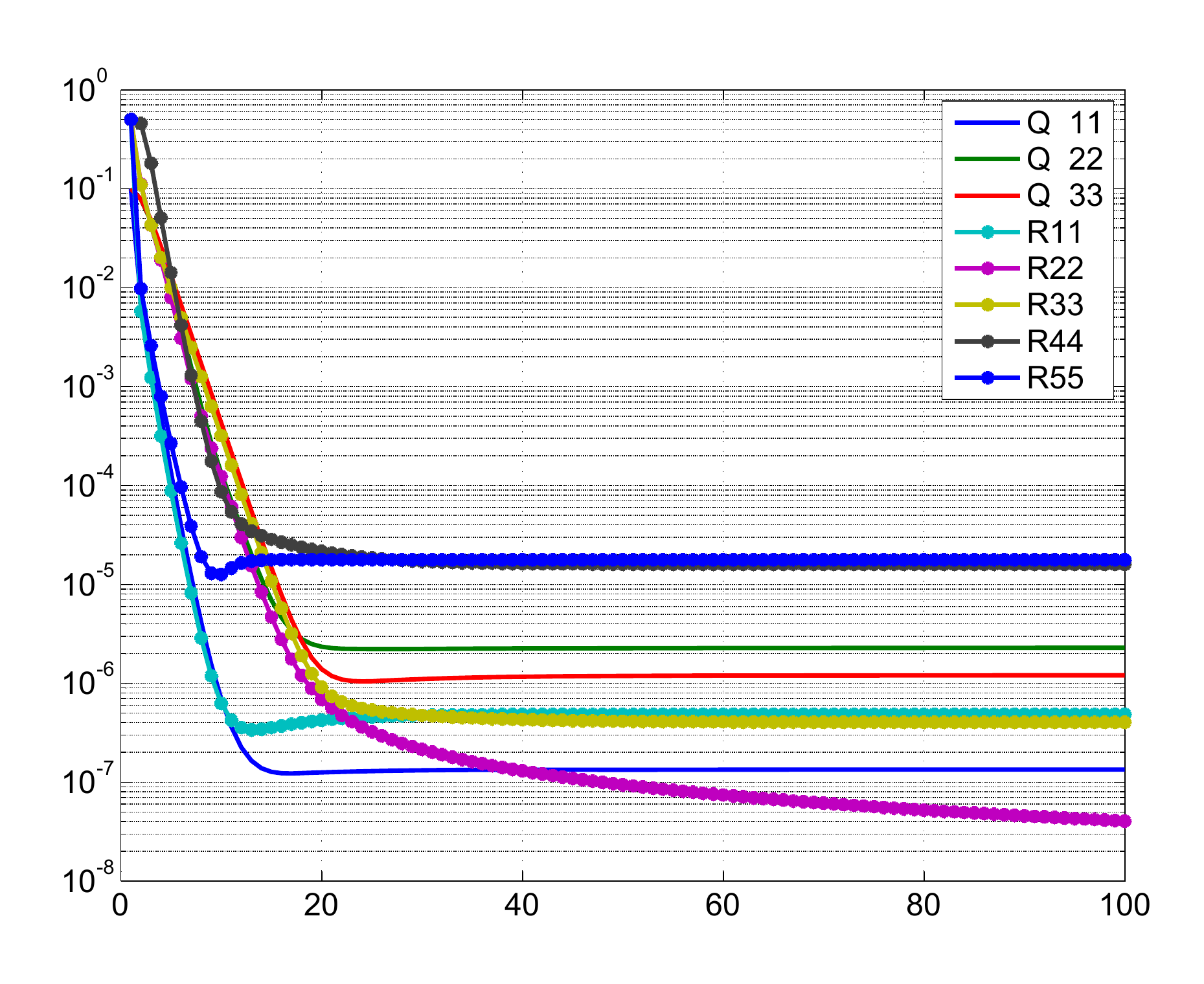}
\caption{Variation of \textbf{Q} and \textbf{R} with iterations }
\label{realQ2b1_R}
\end{figure}

\begin{figure}[h]
\includegraphics[width=6in,height=4in]{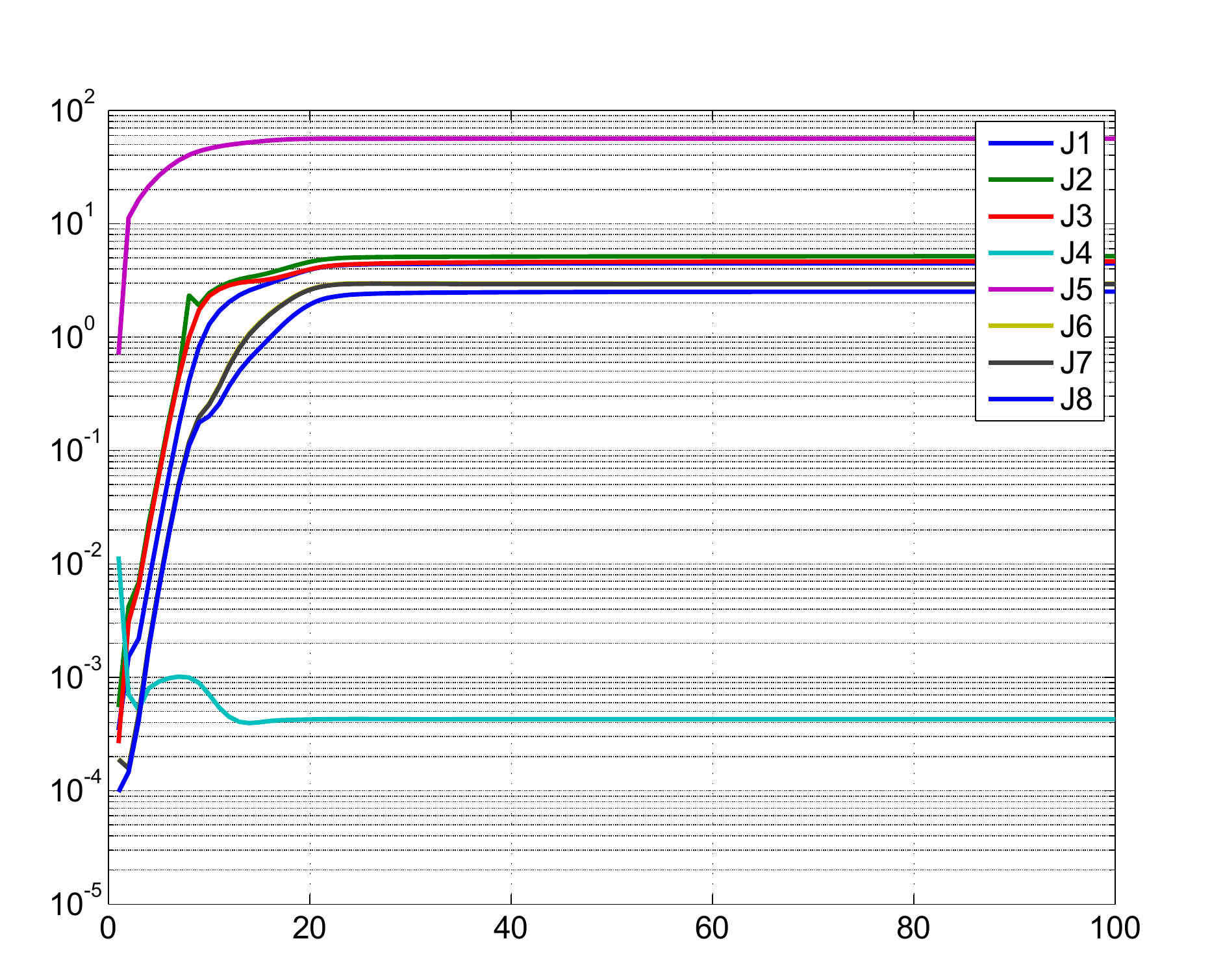}
\caption{Variation of different costs (\textbf{J1-J8}) with iterations}
\label{realQ2b1_J}
\end{figure}

\begin{figure}[h]
\includegraphics[width=6in,height=4in]{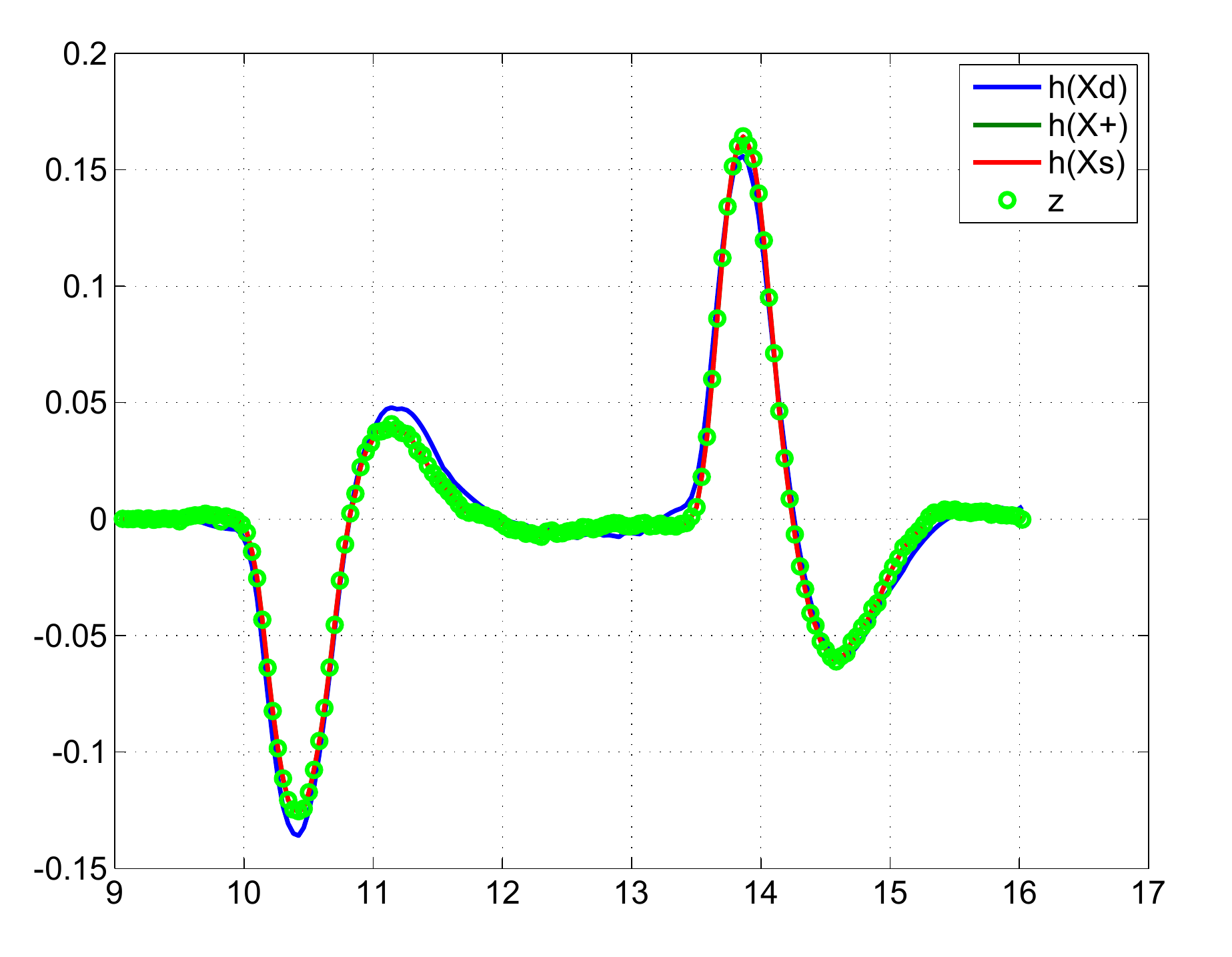}
\caption{Comparison of the predicted dynamics, posterior, smoothed}
\caption*{and the measurement 2}
\label{realQ2b1_s2}
\end{figure}

\begin{figure}[h]
\includegraphics[width=6in,height=4in]{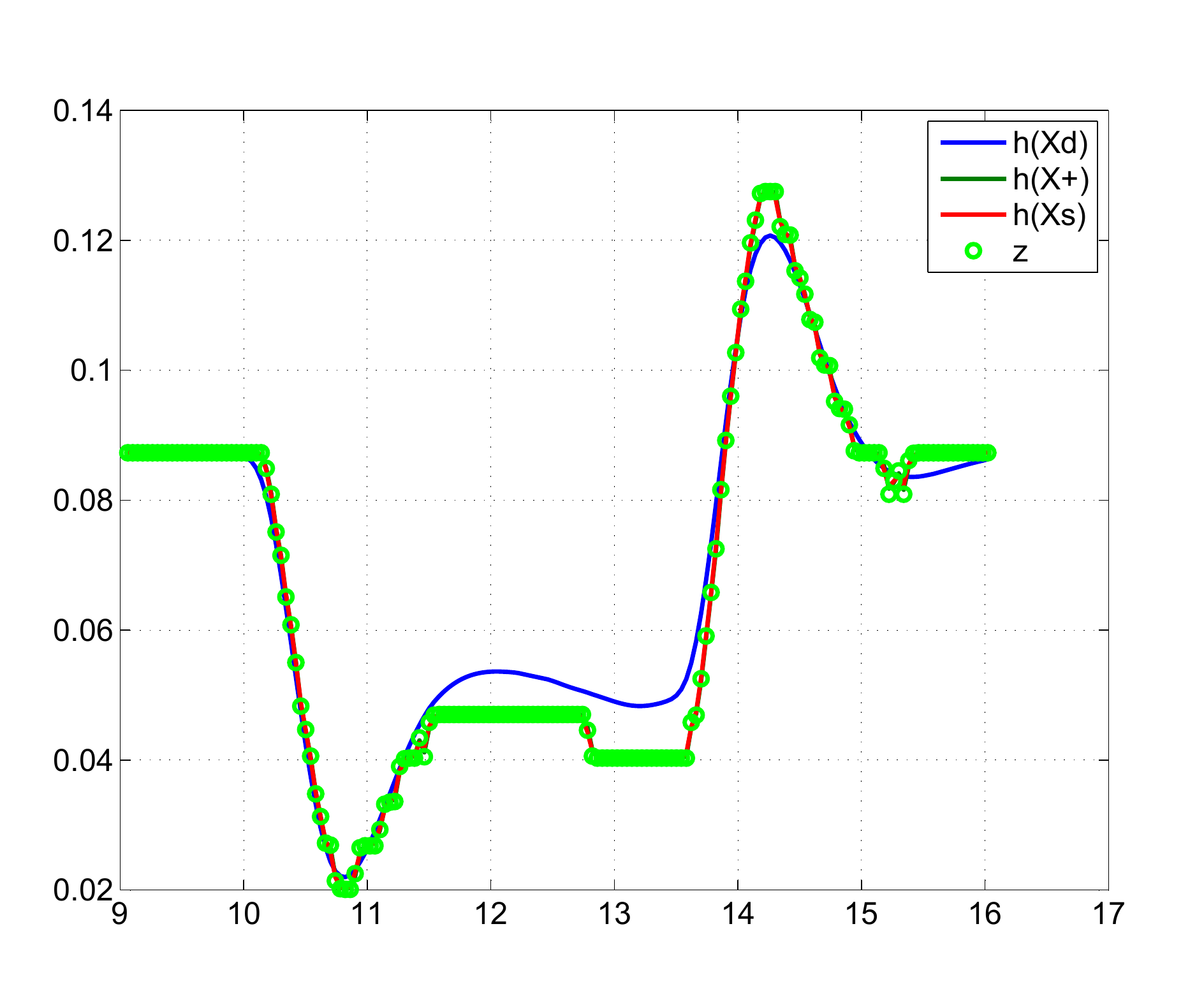}
\caption{Comparison of the predicted dynamics, posterior, smoothed}
\caption*{and the measurement 3}
\label{realQ2b1_s3}
\end{figure}

\begin{figure}[h]
\includegraphics[width=6in,height=4in]{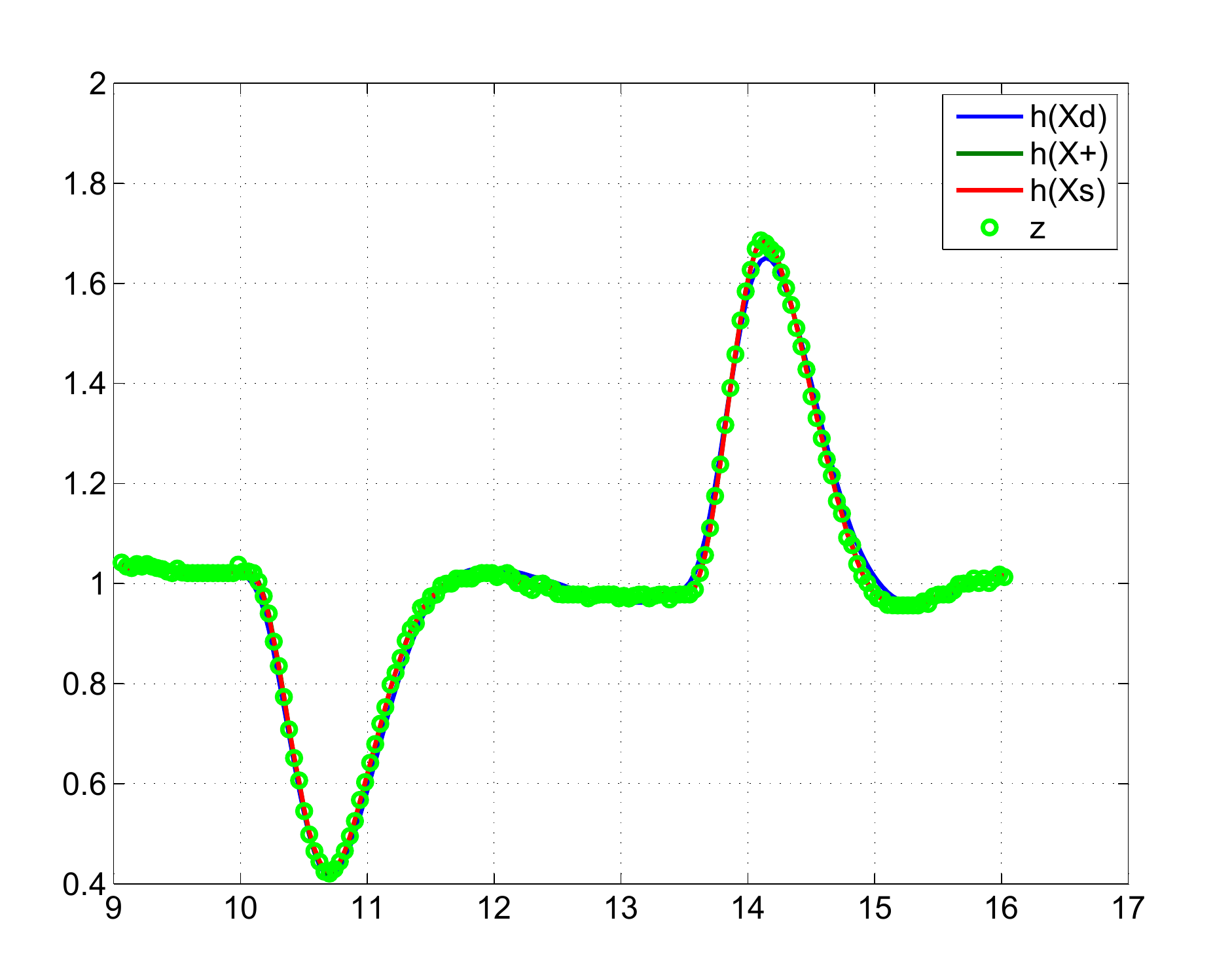}
\caption{Comparison of the predicted dynamics, posterior, smoothed}
\caption*{and the measurement 4}
\label{realQ2b1_h4}
\end{figure}

\begin{figure}[h]
\includegraphics[width=6in,height=4in]{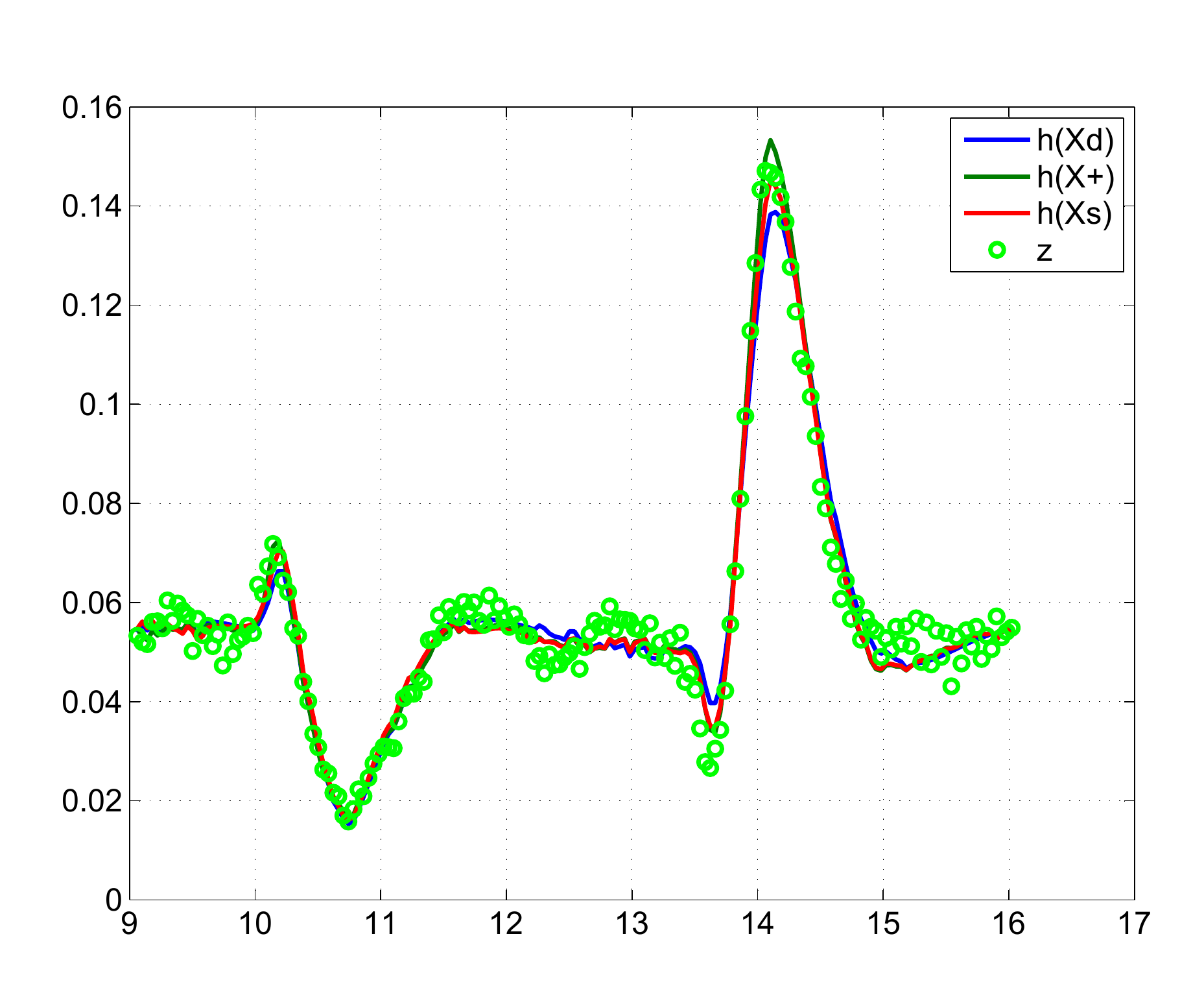}
\caption{Comparison of the predicted dynamics, posterior, smoothed}
\caption*{and the measurement 5}
\label{realQ2b1_h5}
\end{figure}

\begin{figure}[h]
\includegraphics[width=6in,height=4in]{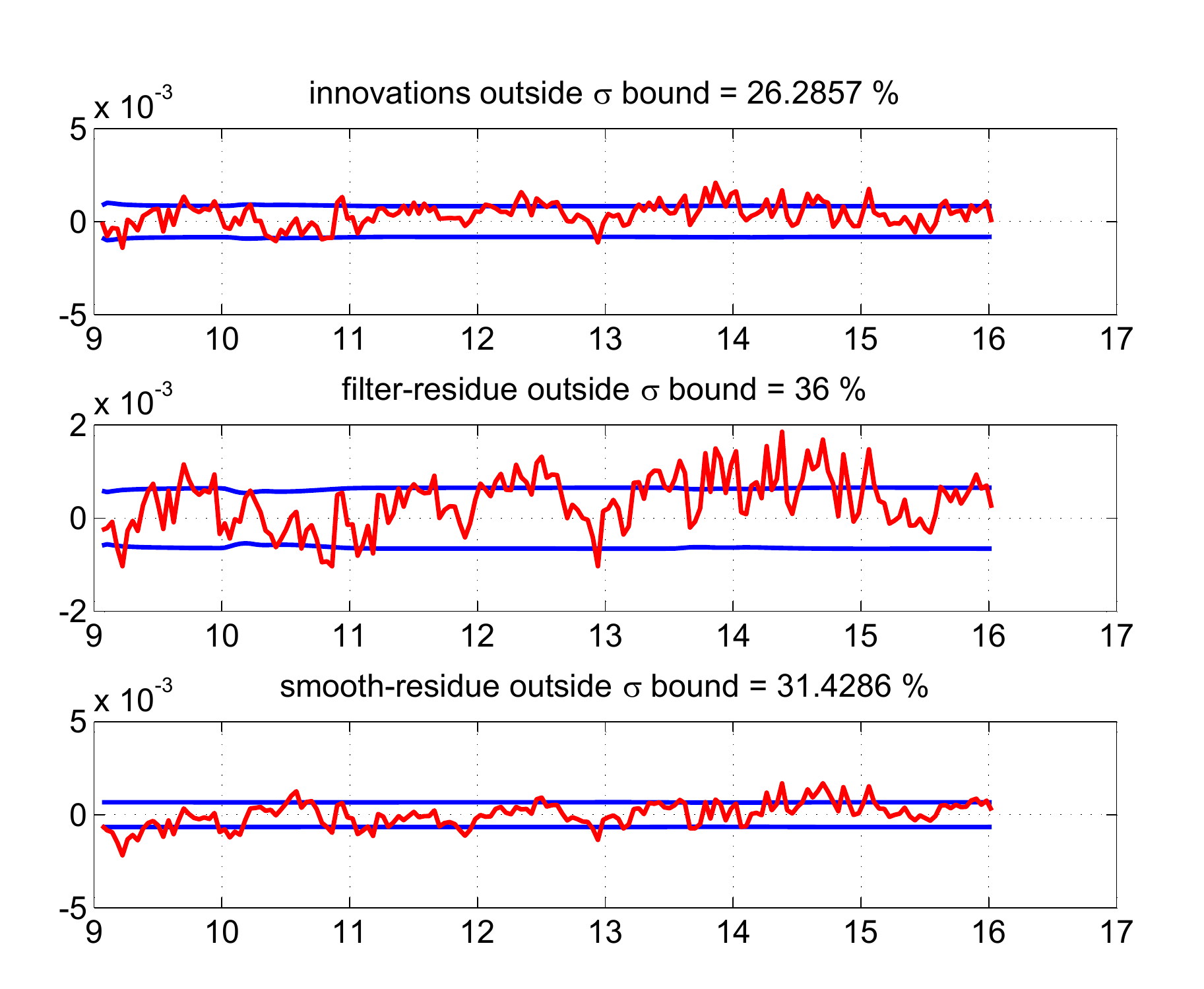}
\caption{The innovations, filtered residue and smoothed residue of measurement 1}
\label{realQ2b1_innov1}
\end{figure}

\begin{figure}[h]
\includegraphics[width=6in,height=4in]{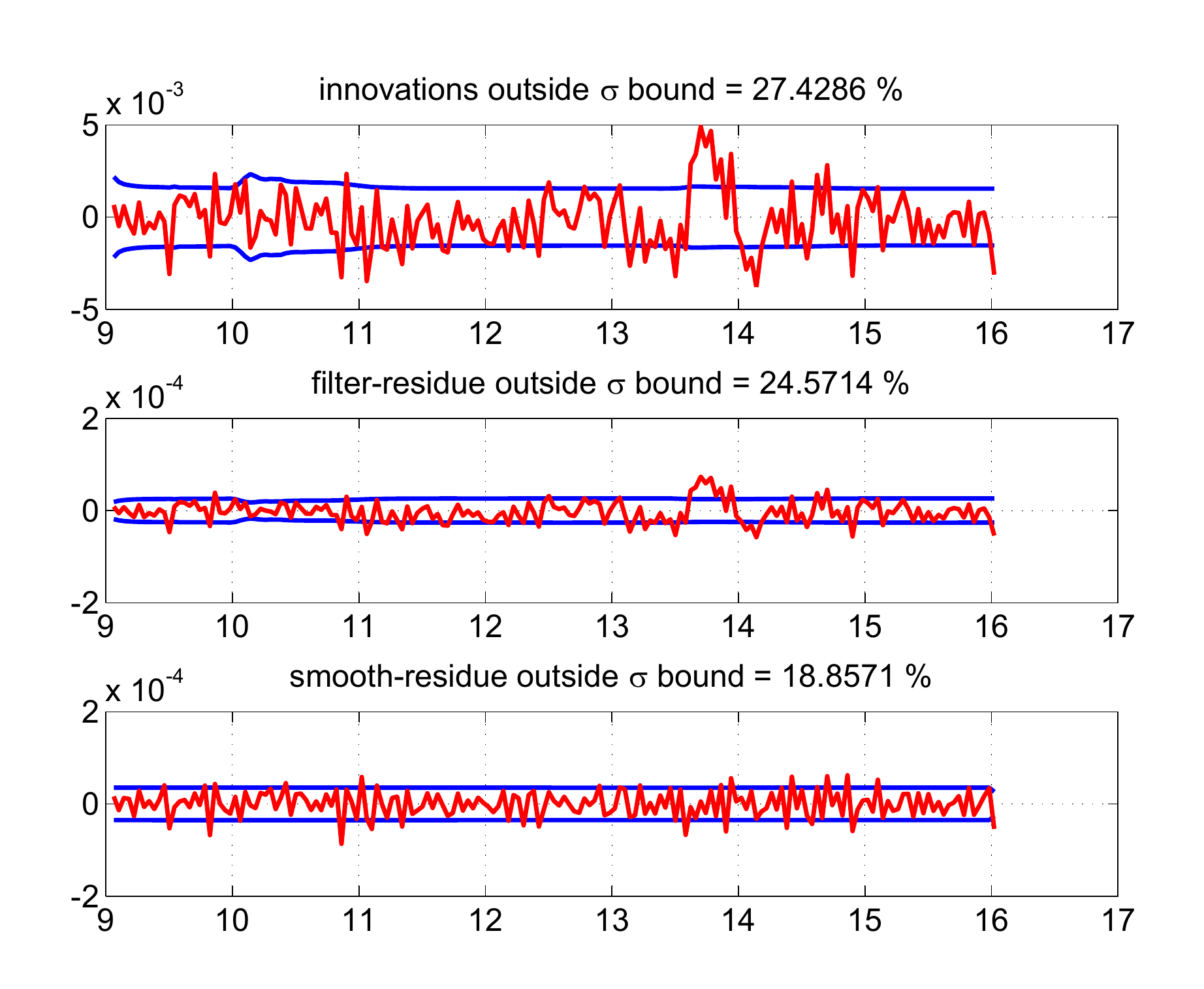}
\caption{The innovations, filtered residue and smoothed residue of measurement 2}
\label{realQ2b1_innov2}
\end{figure}

\begin{figure}[h]
\includegraphics[width=6in,height=4in]{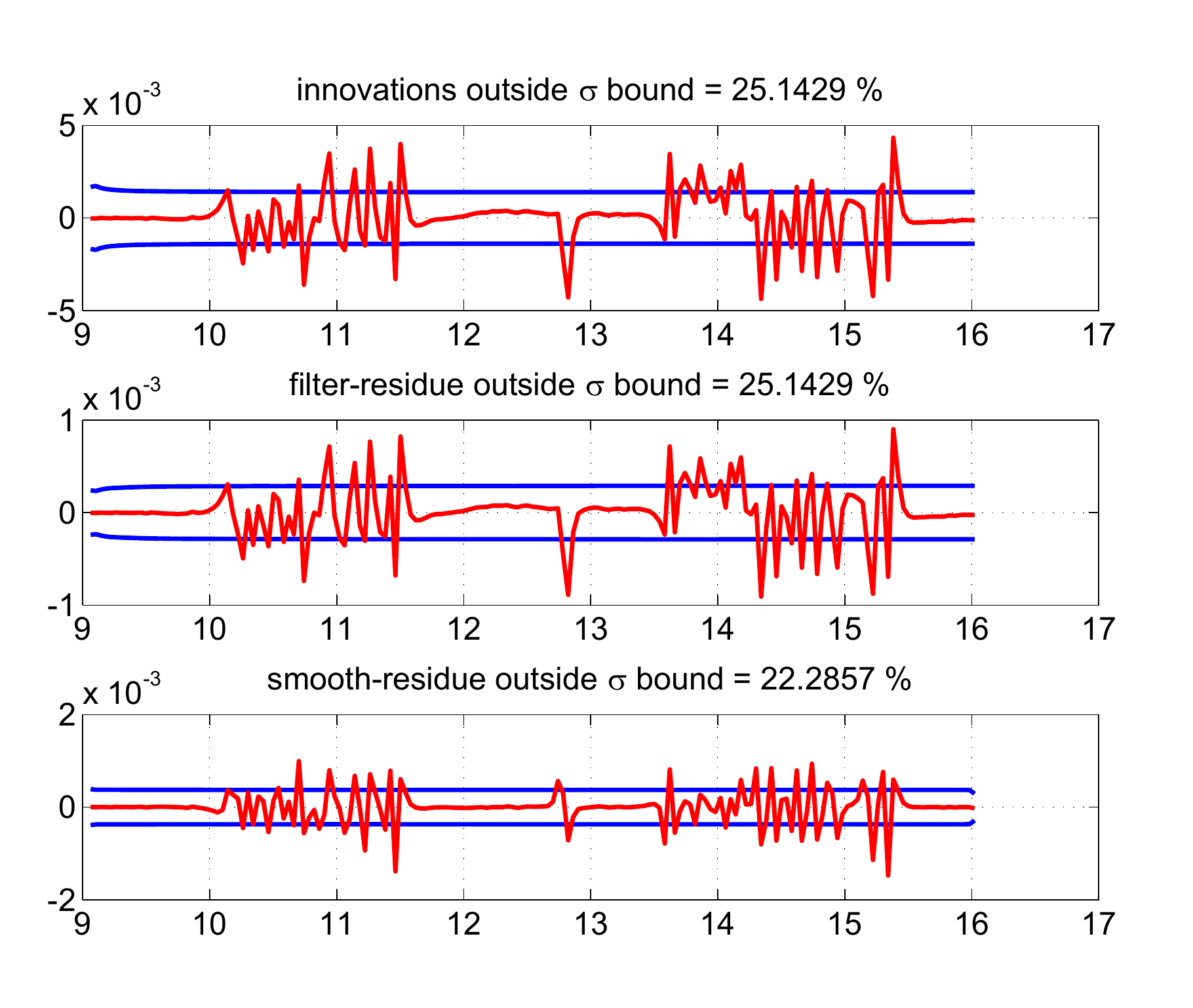}
\caption{The innovations, filtered residue and smoothed residue of measurement 3}
\label{realQ2b1_innov3}
\end{figure}

\begin{figure}[h]
\includegraphics[width=6in,height=4in]{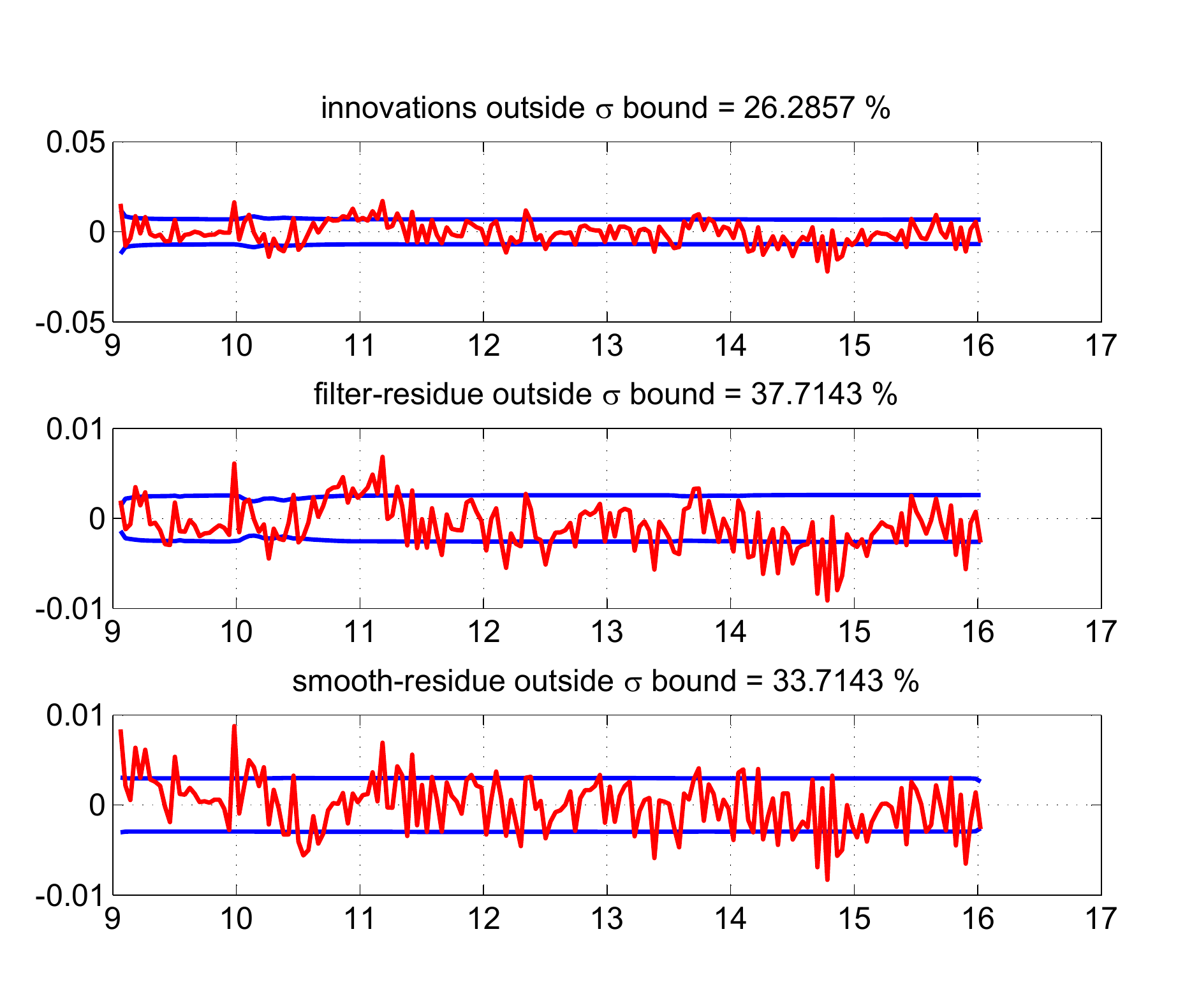}
\caption{The innovations, filtered residue and smoothed residue of measurement 4}
\label{realQ2b1_innov4}
\end{figure}

\begin{figure}[h]
\includegraphics[width=6in,height=4in]{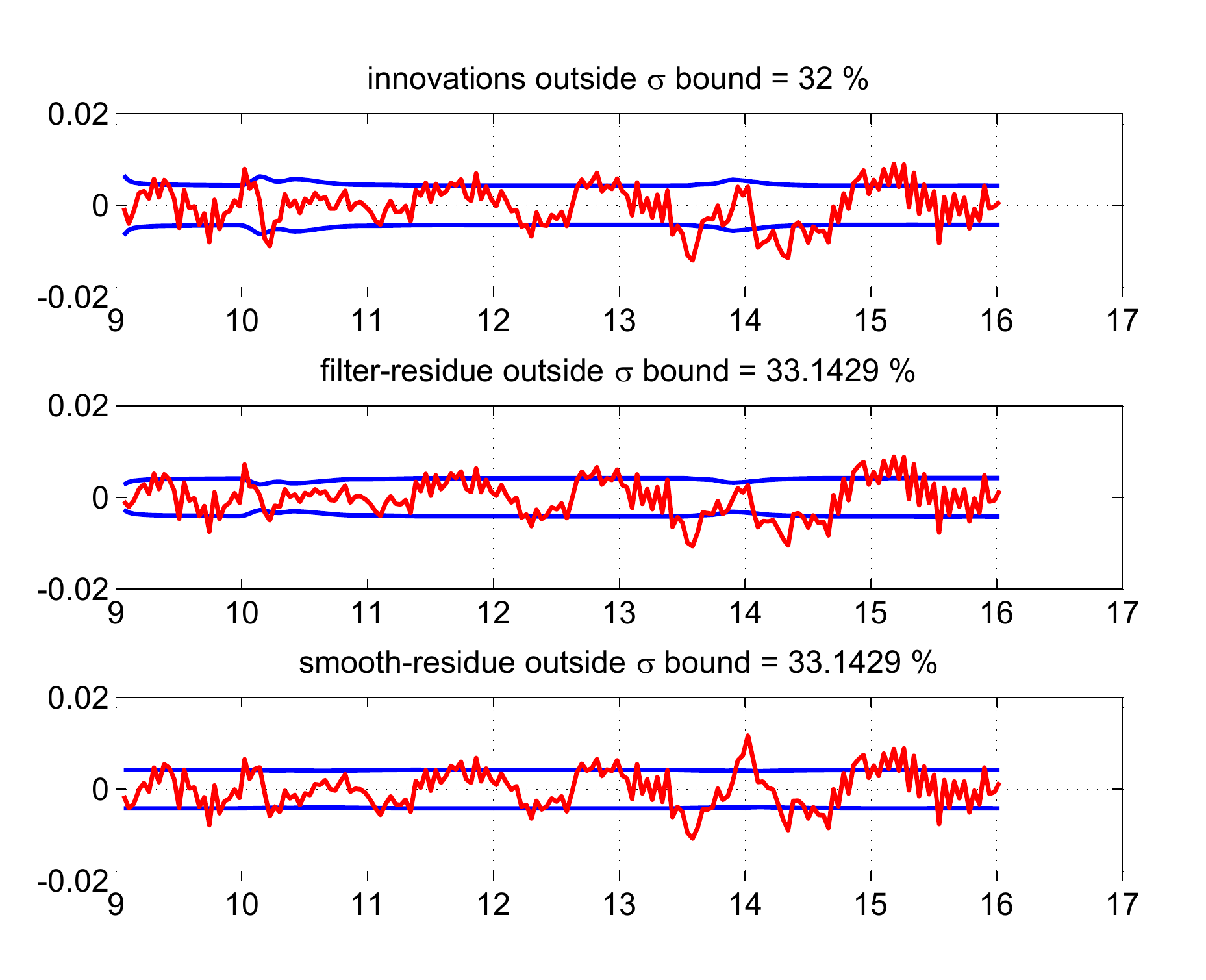}
\caption{The innovations, filtered residue and smoothed residue of measurement 5}
\label{realQ2b1_innov5}
\end{figure}

\begin{figure}[h]
\includegraphics[width=6in,height=4in]{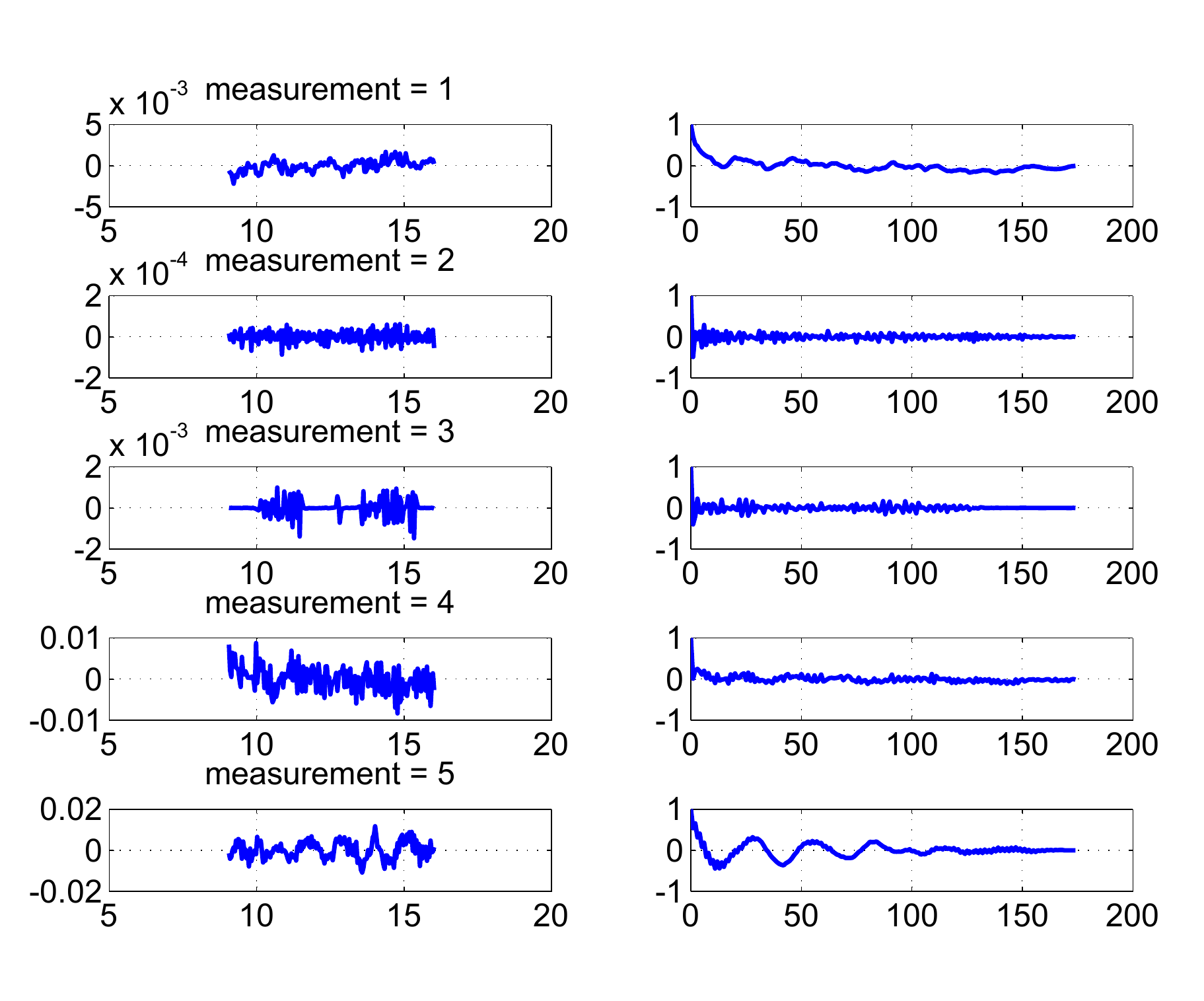}
\caption{Time variation of estimated measurement noise (left) and}
\caption*{their autocorrelation (right)}
\label{realQ2b1_mnoise}
\end{figure}

\begin{figure}[h]
\includegraphics[width=6in,height=4in]{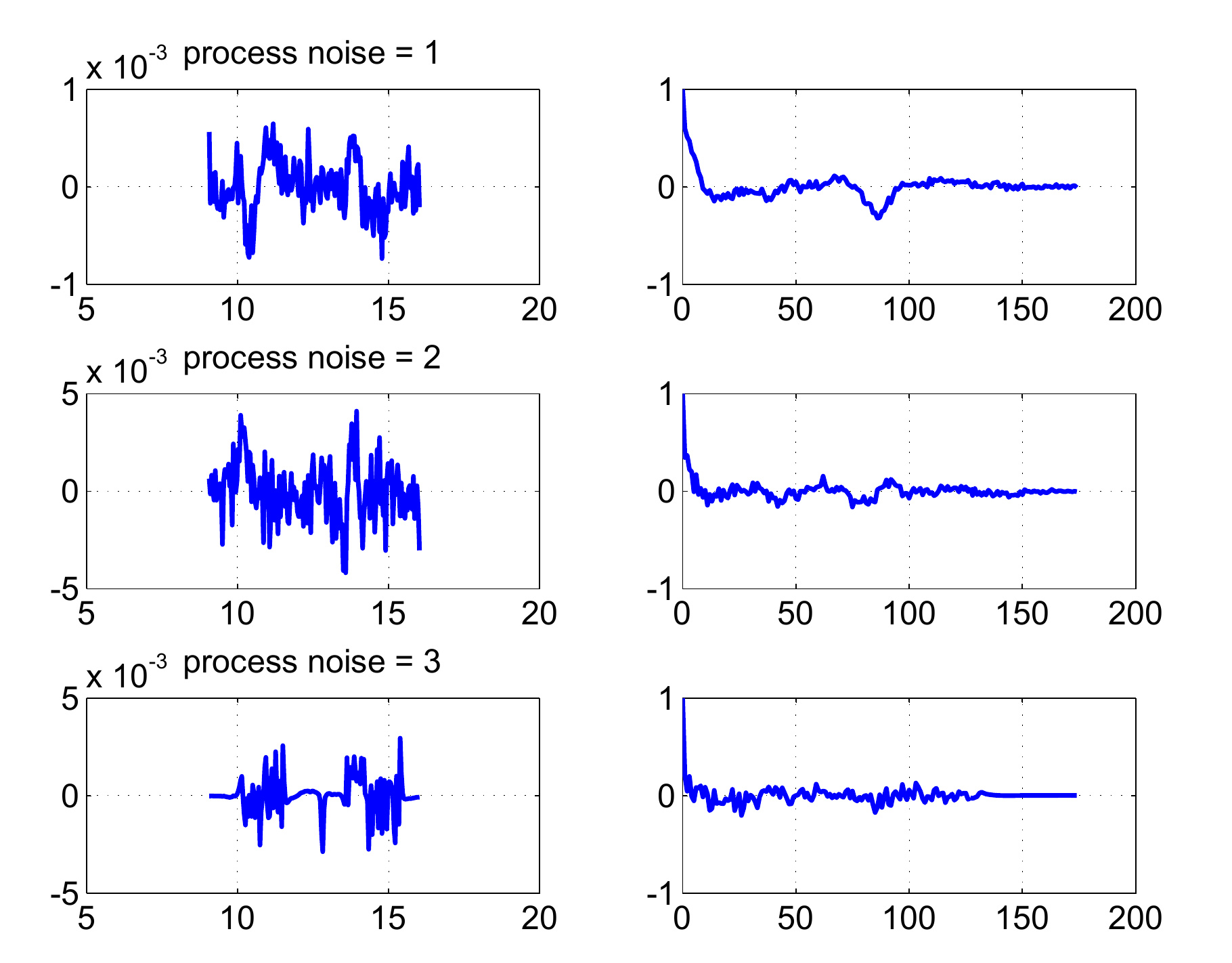}
\caption{Time variation of estimated process noise (left) and}
\caption*{their autocorrelation (right)}
\label{realQ2b1_pnoise}
\end{figure}

 Case-2 real data is run using the reference EKF (\textbf{Q} $>$ 0) with 100 iterations. The Fig. \ref{input2}-\ref{case2_r} are the inputs used in state equations. The Fig. \ref{realQ2b1_P0} shows the variation of parameter estimates and its initial covariance $\mathbf{P_0}$ with iterations and a similar Fig. \ref{realQ2b1_R} for \textbf{Q} and \textbf{R}. The values of \textbf{J1-J3} are close to the number of measurements ($m=5$) with \textbf{J6-J8} are close to the number of states ($n=3$) as shown in Fig. \ref{realQ2b1_J} and Table-\ref{tbcase2Q}. This means the measurement and state equations are well balanced. The \textbf{J5} is the negative log likelihood cost function. The later Fig. \ref{realQ2b1_s1}-\ref{realQ2b1_h5} compares (i) the state dynamics based on the estimated parameter after the filter pass through the data, (ii) the state after measurement update, (iii) the smoothed state and (iv) the measurement.  The Fig. \ref{realQ2b1_innov1}-\ref{realQ2b1_innov5} shows the confidence in the innovations, filtered residue and smoothed residue. The estimated measurement and process noise do not appear to have constant statistical characteristics across time as seen in Fig. \ref{realQ2b1_mnoise} and Fig. \ref{realQ2b1_pnoise}. Another experiment was carried out by generating a typical data set by using the estimated theta and injecting the estimated \textbf{Q} and \textbf{R} as additive white Gaussian noise. This is to determine the effect of non additive, non White and non Gaussian noise distribution in the real data on the CRBs. After each iteration in the reference recipe the $\Theta$, \textbf{Q} and \textbf{R} were reset as from the real data. Similar experiment was also conducted by updating $\Theta$ as well. It turned out that there is not much of a difference in the final CRBs as can be seen from the Table-\ref{tbnew2}.

\par
\begin{figure}[h]
\includegraphics[width=6in,height=4in]{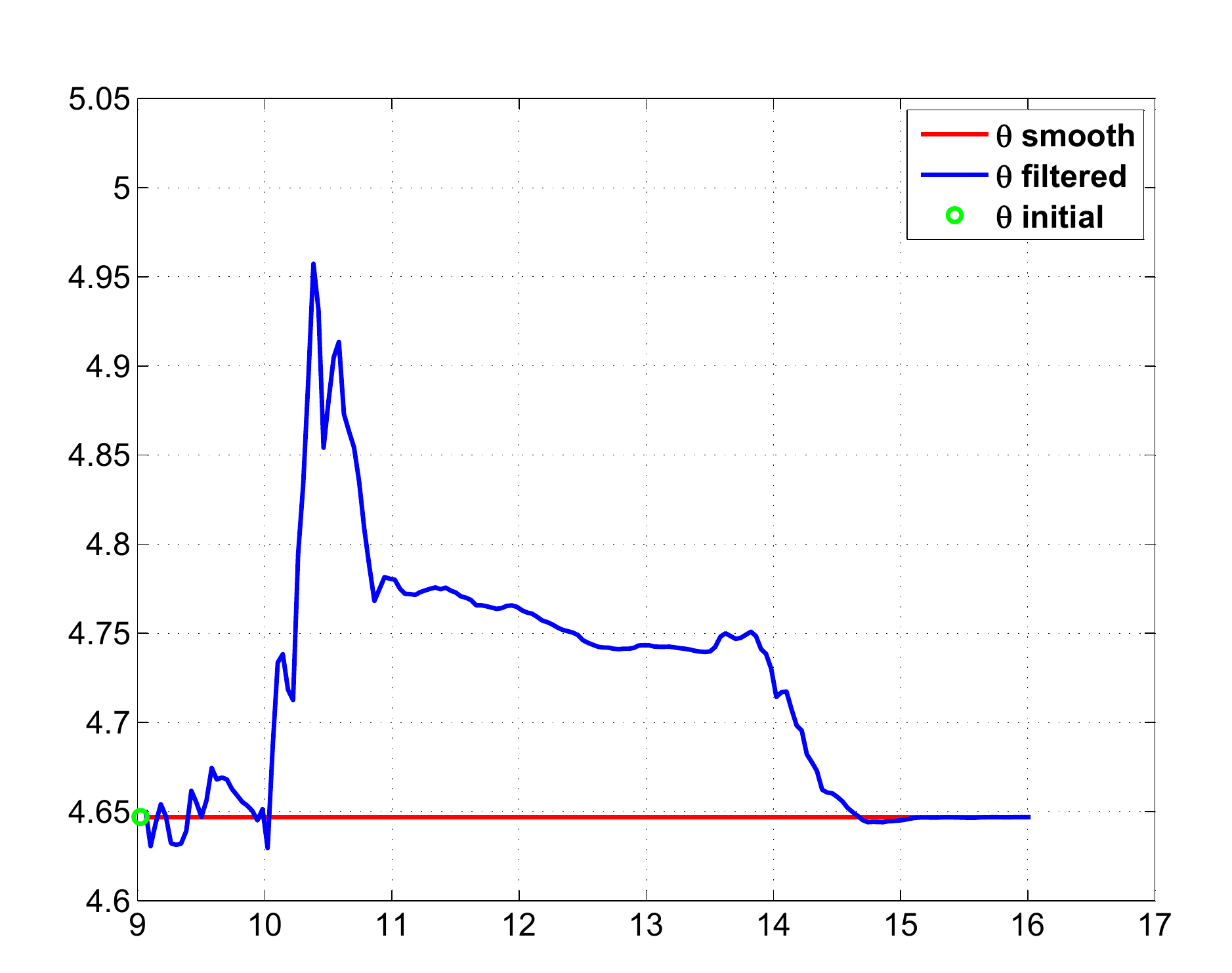}
\caption{Variation of the parameter estimate ($C_{N_\alpha}$) through time instants}
\label{CNalpha}
\end{figure}

 Finally two other filter runs were carried out using the MT and MS statistics for the estimation of \textbf{Q} and \textbf{R}. The behaviour of the various cost function and in particular \textbf{J6} and \textbf{J7} in Table-\ref{tbcase2QMTMS} shows that the choice of the filter statistics for estimating \textbf{Q} and \textbf{R} in the proposed reference approach is the best possible when compared to other approaches presently considered.  Another feature of recursive parameter estimation is that it can vary through time instants and point to two distinct values as reflected in the estimation of $C_{N_{\alpha}}$ in Fig. \ref{CNalpha}. This feature of tracking time varying parameters by the EKF brings in clearly another advantage of sequential processing instead of batch processing of the data by a least squares (LS) procedure. If LS had been used then for the parameter `$C_{N_\alpha}$' only an average value would have been obtained. If one persists in using the LS procedure then to get the varying parameter such a feature should be modelled which would include where to change and one might have to use the data in blocks by splitting them and all such exercises have to be carried out which are not easy. Thus another way of looking at the filter it seems to be doing a better job of using a more appropriate range of data to estimate the parameters. It is doing a good job of time averaging the state estimates in a fluctuating situation created by the measurement noise. However it should be cautioned that the filter has to be tuned very well for such a situation. This may not be an easy task if the system dynamics is extremely fast.

\subsection{Remarks on Case-2}
\par
\begin{figure}[h]
\includegraphics[width=6in,height=4in]{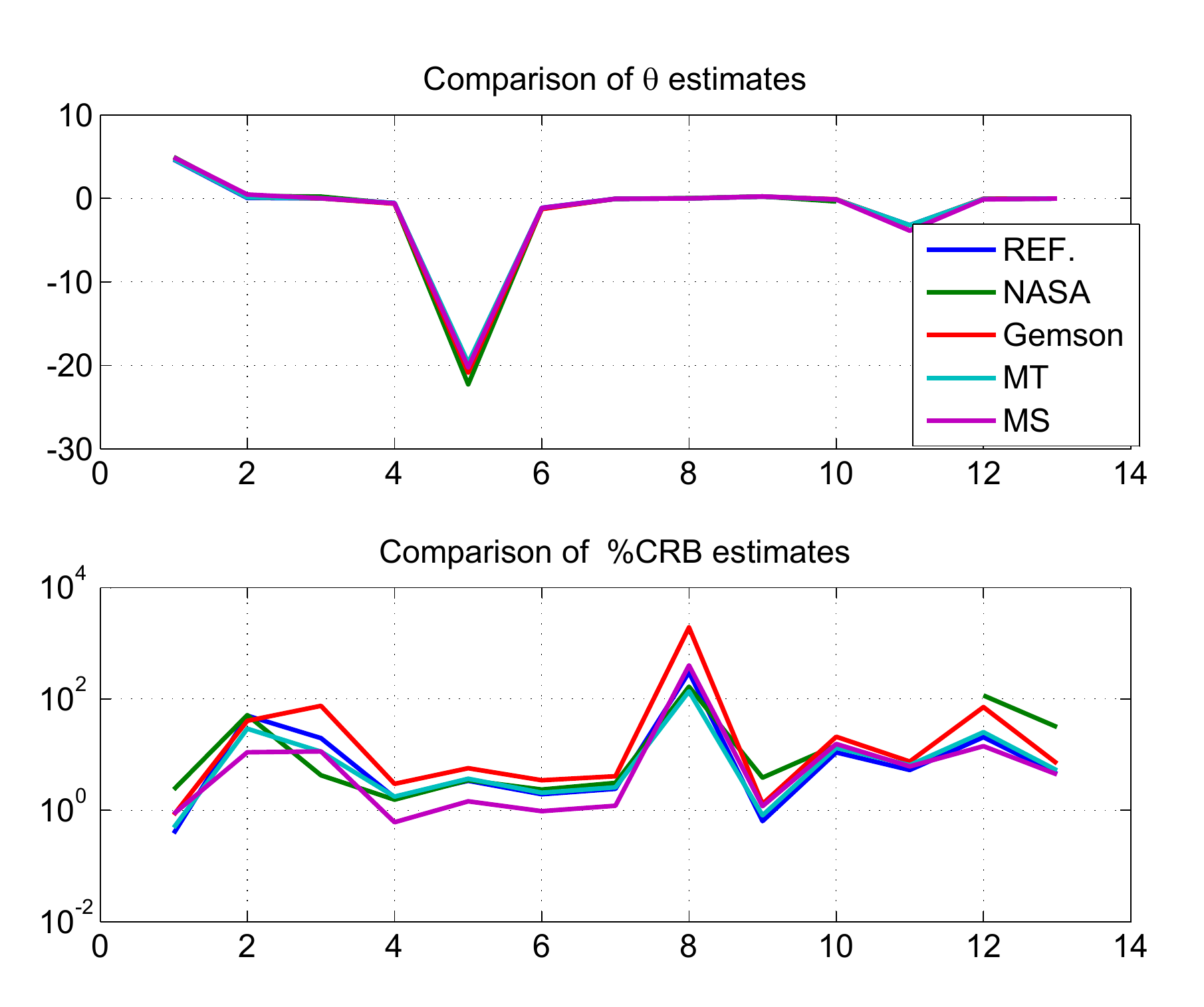}
\caption{Comparison of the parameter estimates and $\%$CRBs by different methods}
\label{comp2}
\end{figure}

 The NASA results have been generated assuming \textbf{Q} = 0 and are comparable with reference procedure for the parameter estimates and their CRBs. Further the MT and MS methods give quite different estimates for the \textbf{Q} and \textbf{R} values than in the reference case. We believe that the reference procedure provides the best possible parameter estimates and their uncertainties. From the plot of the parameter estimates and their \%CRB in Fig \ref{comp2}, it can be seen that the parameters 1-7, 10, 11 are strong and the parameters 8 and 12 are the weak ones.  The CRBs as estimated by different methods generally appear to vary widely. However what is interesting is that even the estimate of the strong parameter such as 5 varies widely among the methods. Such a behaviour of the filter across the parameter estimates shows how important is the tuning of the filter statistics namely $\mathbf{P_0}$, \textbf{Q} and \textbf{R} in parameter estimation and their uncertainties.
\par The rounded off 100$\times$C matrix for case-2 is given by
\begin{footnotesize}
\begin{align*}
\begin{bmatrix}
   100  & -91  &   0  &  26   &   6 &  -23  & 21    &   0  & 86   &  26  &  26   & -22  & 20 \\
   -91  & 100  &   0  &  -23  &  -4 &   25  & -24   &   0  & -98  &  -23 & -20   &  22  & -21 \\
     0  &   0  & 100  &   -2  &   0 &    2  &  -2   &   0  &   0  &   0  &   0   &  0   &  0 \\
    26  & -23  &  -2  &  100  &  18 &  -90  &  86   &   2  &  22  &  32  &  29   & -29  & 27 \\
     6  &  -4  &   0  &   18  & 100 &  -14  &  -2   &   0  &   0  &   6  &   8   &  -5  &  -1 \\
   -23  &  25  &   2  & -90   & -14 &  100  & -98   &  -3  & -24  & -29  & -23   &  32  & -31 \\
    21  & -24  &  -2  &  86   &  -2 &  -98  & 100   &   3  &  24  &  27  &  21   & -31  &  32 \\
     0  &   0  &   0  &   2   &   0 &   -3  &   3   & 100  &   0  &   0  &   0   &   0  &   0 \\
    86  & -98  &   0  &  22   &  0  &  -24  &  24   &   0  & 100  &  21  &  18   &  -21 &   21 \\
    26  & -23  &   0  &  32   &   6 &  -29  &   27  &   0  &  21  & 100  &   91  & -91  &  86 \\
    26  & -20  &   0  &  29   &   8 &  -23  &  21   &   0  &  18  &  91  &  100  & -73  &   67 \\
   -22  &  22  &   0  & -29   &  -5 &   32  & -31   &   0  & -21  &  -91 & -73   & 100  & -98 \\
    20  & -21  &   0  &  27   &  -1 &  -31  &   32  &   0  &  21  &  86  &  67   & -98  &  100
\end{bmatrix}
\end{align*}
\end{footnotesize}

\begin{landscape}
\begin{table}[h]
\subsection{Case-2 Tables}
\vspace{14pt}
\caption{Real flight test data case-2 results using the reference adaptive EKF.\\ No of iterations=100}{}
\label{tbcase2Q}
\begin{center}
\begin{footnotesize}
\begin{tabular}{|c| c| c| c| c| c| c| c|c|c|c|}
\hline
Study &
\makecell{$\Theta$\\ (Ref)} &
\makecell{$\Theta$\\ (NASA)} &
\makecell{$\Theta$\\ (Gemson)} &
\makecell{$\sigma_\Theta$ \\(Ref)} &
\makecell{$\sigma_\Theta$\\ (NASA)} &
\makecell{$\sigma_\Theta$\\ (Gemson)} &
\makecell{\textbf{R} \\ $\times10^{-6}$\\ (Ref)}&
\makecell{\textbf{Q} \\ $\times10^{-6}$\\ (Ref)}&
\makecell{\textbf{J1-J8} \\(Ref) }&
Remarks
\\ \hline

%


\makecell{$\mathbf{P_0}$ : Scaled up-[0,0;0,\checkmark]\\\textbf{Q} : EM-[\checkmark,0;0,0] \\\textbf{R} : EM-diag} &
\makecell{  4.6469  \\  0.0555  \\  0.0162 \\  -0.5468 \\ -19.8027 \\  -1.1229 \\  -0.0495 \\ 0.0007  \\  0.2195  \\ -0.1398 \\  -3.2088 \\  -0.0651  \\ -0.0155} &
\makecell{ 4.9584 \\   0.3023 \\  0.2189 \\  -0.6125  \\ -22.27 \\  -1.2193 \\  -0.0532 \\ 0.0273  \\  0.2254  \\ -0.3639\\ -- \\  -0.07  \\ -0.0131} &
\makecell{ 4.7073 \\   0.1292 \\  -0.0064 \\  -0.63  \\ -20.8623 \\  -1.2763 \\  -0.0561 \\ 0.0007  \\  0.2225  \\ -0.1023 \\  -3.2397  \\ -0.0267 \\-0.0144} &
\makecell{0.0179  \\  0.0277 \\   0.0032  \\  0.0093  \\  0.6692  \\  0.0218 \\   0.0012 \\ 0.0021 \\   0.0014   \\ 0.0153  \\  0.1702 \\   0.0134  \\  0.0007} &
\makecell{ 0.1168 \\   0.1550 \\   0.009344  \\  0.00953 \\   0.7713 \\   0.02881 \\   0.00165 \\ 0.04518  \\  0.008725 \\ 0.05328 \\ -- \\  0.08084 \\   0.004088 } &
\makecell{ 0.039 \\   0.0523 \\   0.0048  \\  0.0188 \\   1.1908 \\   0.0442 \\   0.0023 \\ 0.0135  \\  0.0029   \\ 0.0214 \\  0.2430 \\ 0.0191 \\  0.0010 } &
\makecell{ 0.49  \\  0.04  \\  0.40  \\  15.98  \\  17.70} &
\makecell{ 0.134  \\  2.287  \\  1.204} &
\makecell{4.4752  \\  5.1532  \\  4.6432  \\  0.0004 \\ -56.2206  \\  2.9551 \\   2.9303 \\ 2.5161} &
\makecell{Cost functions converge\\ to the expected values}
\\ \hline

\end{tabular}
\end{footnotesize}
\end{center}
\end{table}

\begin{table}[h]
\caption{Case-2 results using simulated Additive White Gaussian Noise}{}
\label{tbnew2}
\begin{center}
\begin{tabular}{|c| c| c| c| c|}
\hline
Study &
\makecell{$\sigma_\Theta$ \\(Simulated-without\\ updating $\Theta$)} &
\makecell{$\sigma_\Theta$ \\(Simulated-with\\ updating $\Theta$)} &
\makecell{$\sigma_\Theta$ \\(Ref)} &
Remarks
\\ \hline

\multicolumn{5}{|c|}{\makecell{Case-2 data generated using simulated measurement and process noise (AWGN) \\ of variance \textbf{Q} and \textbf{R} estimated by Reference EKF (\textbf{Q} $>$ 0)} } \\ \hline

\makecell{$\mathbf{P_0}$ : Scaled up-[0,0;0,\checkmark]\\\textbf{Q} : \textbf{Q} (Ref) \\\textbf{R} :   \textbf{R} (Ref)} &

\makecell{  0.0178 \\   0.0278 \\   0.0032  \\  0.0095 \\   0.6887 \\   0.0237  \\  0.0013 \\ 0.0021   \\ 0.0014 \\   0.0140  \\  0.1670  \\  0.0136  \\  0.0007} &
\makecell{0.0178  \\  0.0278  \\  0.0032  \\  0.0095  \\  0.6889 \\   0.0238  \\  0.0013 \\ 0.0021  \\  0.0014 \\   0.0140  \\  0.1670 \\   0.0136  \\  0.0007} &
\makecell{0.0179  \\  0.0277 \\   0.0032  \\  0.0093  \\  0.6692  \\  0.0218 \\   0.0012 \\ 0.0021 \\   0.0014   \\ 0.0153  \\  0.1702 \\   0.0134  \\  0.0007} &

\makecell{No Significant \\ change in $\sigma_\Theta$}
\\ \hline

\end{tabular}
\end{center}
\end{table}
\begin{table}[h]
\caption{Real flight test data case-2 results using the MT and MS method.\\ No of iterations=100}{}
\label{tbcase2QMTMS}
\begin{center}
\begin{footnotesize}
\begin{tabular}{|c| c| c| c| c| c|| c|c|c|c|c|c| }
\hline
Study &
\makecell{$\Theta$\\ (MT)} &
\makecell{$\sigma_\Theta$ \\(MT)} &
\makecell{\textbf{R} (MT)\\ $\times10^{-6}$ }&
\makecell{\textbf{Q} (MT)\\ $\times10^{-6}$}&
\makecell{\textbf{J1-J8} \\(MT) }&

\makecell{$\Theta$\\ (MS)} &
\makecell{$\sigma_\Theta$\\ (MS)} &
\makecell{\textbf{R} (MS) \\ $\times10^{-6}$}&
\makecell{\textbf{Q} (MS)\\ $\times10^{-6}$}&
\makecell{\textbf{J1-J8} \\(MS) }&
Remarks
\\ \hline


\makecell{$\mathbf{P_0}$ : Scaled up-[0,0;0,\checkmark]\\\textbf{Q} : MT/MS-[\checkmark,0;0,0] \\\textbf{R} : MT/MS-diag} &

\makecell{  4.6978 \\   0.1225  \\  0.0160 \\  -0.5560 \\ -19.7062  \\ -1.1396 \\  -0.0502 \\ 0.0008 \\   0.2218  \\ -0.1401 \\  -3.2088  \\ -0.0633 \\  -0.0154} &
\makecell{  0.0229  \\  0.0357  \\  0.0018 \\   0.0098 \\   0.7286  \\  0.0236  \\  0.0013 \\ 0.0011  \\  0.0018   \\ 0.0185  \\  0.2070  \\  0.0160 \\   0.0008} &
\makecell{0.4107  \\  0.0312  \\  3.9381 \\  94.5086 \\  26.3511} &
\makecell{0.0393 \\   2.6418  \\  0.3231} &
\makecell{ 4.0090 \\   3.9630  \\  2.9764 \\   0.0004 \\ -54.7596 \\   6.6681 \\   6.4985 \\ 2.4562} &

\makecell{  4.9141 \\   0.4691 \\   0.0184 \\  -0.5885 \\ -20.2395 \\  -1.1503 \\  -0.0497 \\ 0.0003  \\  0.2358  \\ -0.1265 \\  -3.8625  \\ -0.1178 \\  -0.0182} &
\makecell{  0.0422 \\   0.0517  \\  0.0021  \\  0.0036 \\   0.2937  \\  0.0111  \\  0.0006 \\ 0.0012  \\  0.0028   \\ 0.0197 \\   0.2376  \\  0.0167  \\  0.0008} &
\makecell{ 3.2046 \\  37.6770 \\   7.5509 \\ 198.2716 \\  28.9841} &
\makecell{0.0001  \\  0.0015  \\  0.3456} &
\makecell{  3.3893 \\   3.3866  \\  3.2057 \\   0.0002 \\ -49.6223  \\  3.8921   \\ 4.7110 \\ 2.6369} &

\makecell{Cost functions are \\not close to their \\expected values in\\ MT and MS method} \\ \hline

\end{tabular}
\end{footnotesize}
\end{center}
\end{table}
\end{landscape}

\clearpage
\subsection{Case-2 Figures}

\clearpage
\section{Real Flight Test Case-3}
\par
\begin{figure}[h]
\includegraphics[width=6in,height=3.2in]{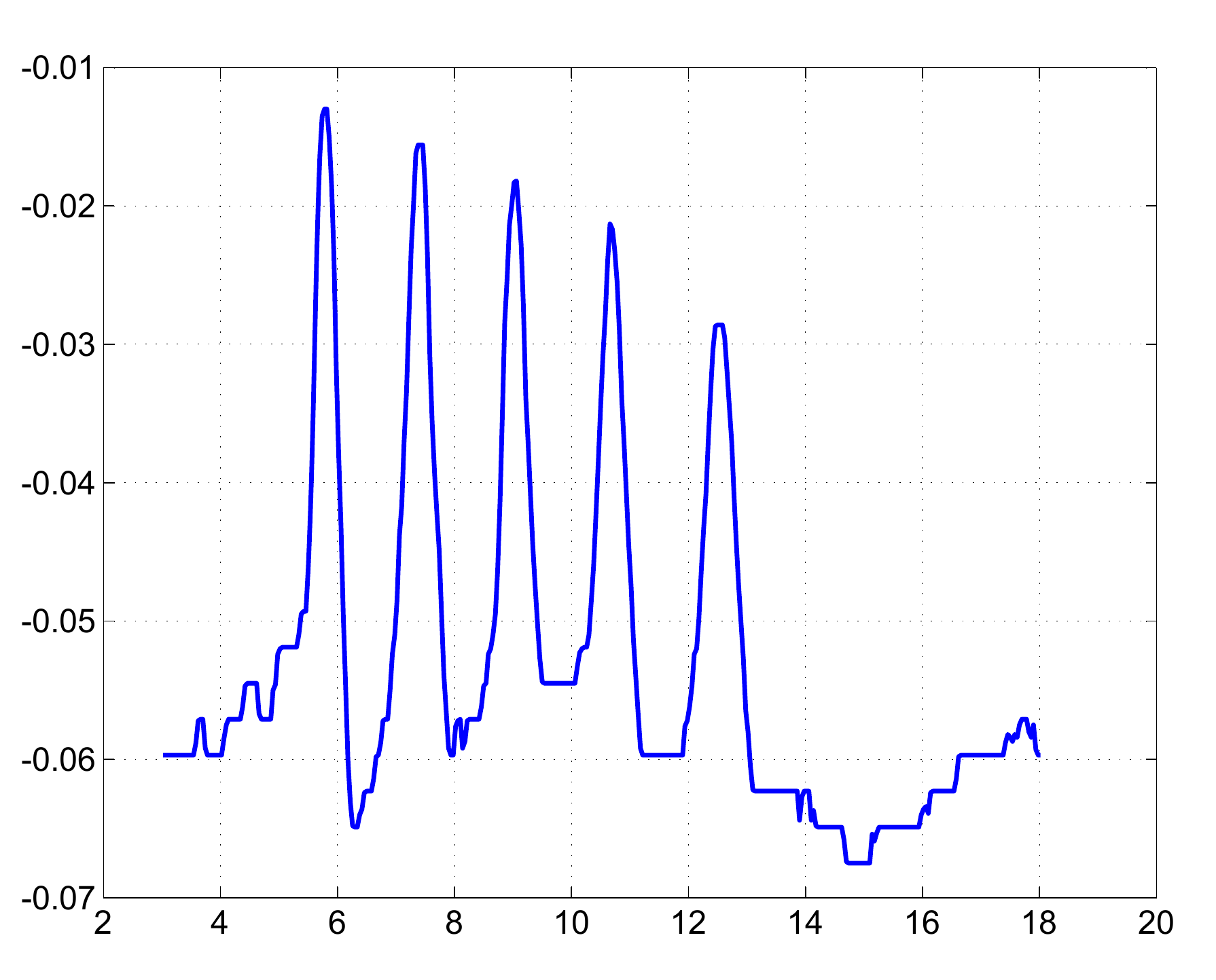}
\caption{Control input ($\delta_e$) versus time (s)}
\label{input3}
\end{figure}

\begin{figure}[h]
\includegraphics[width=6in,height=3.2in]{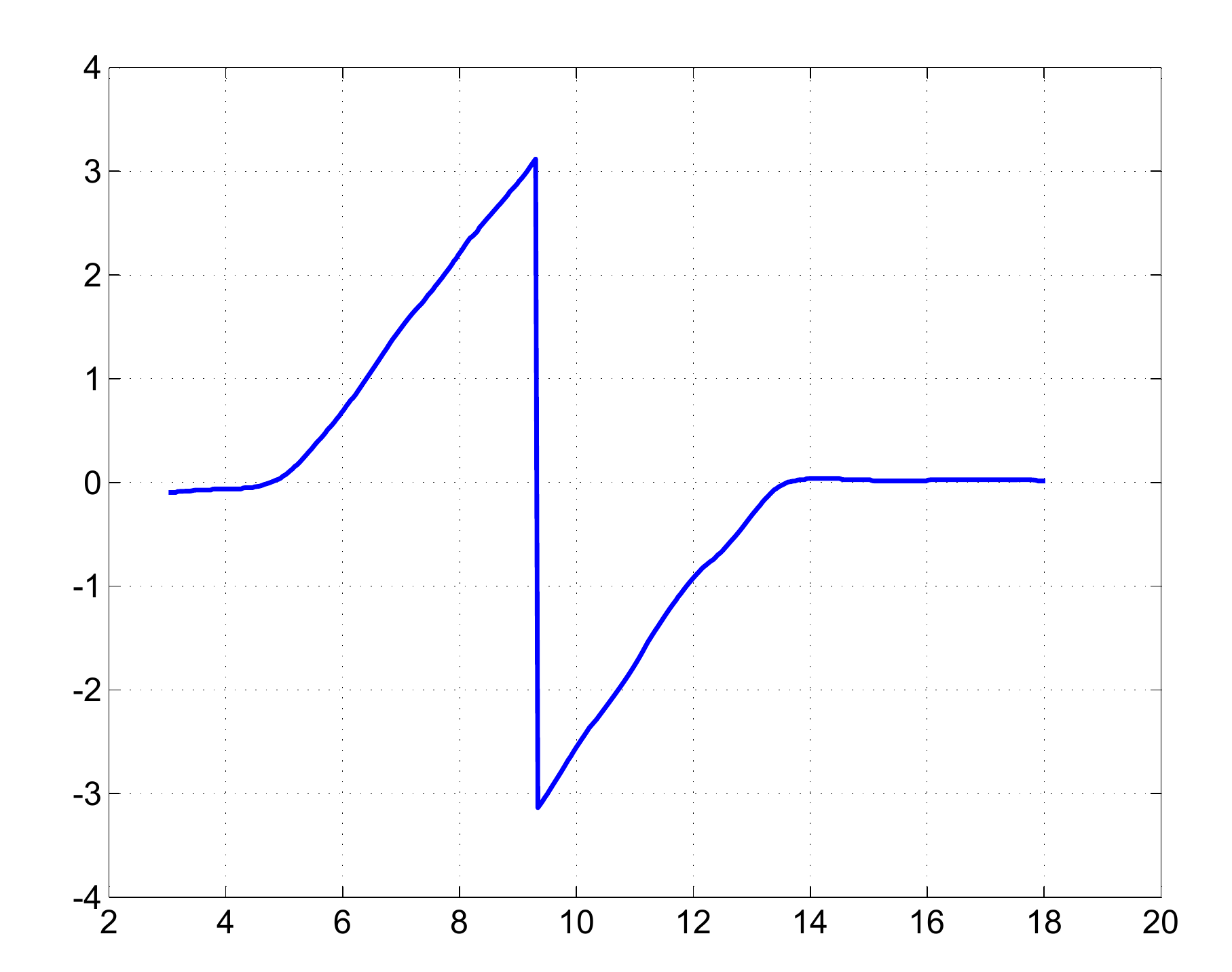}
\caption{Measurement input ($\phi_m$) versus time (s)}
\label{case3_phi}
\end{figure}

\begin{figure}[h]
\includegraphics[width=6in,height=4in]{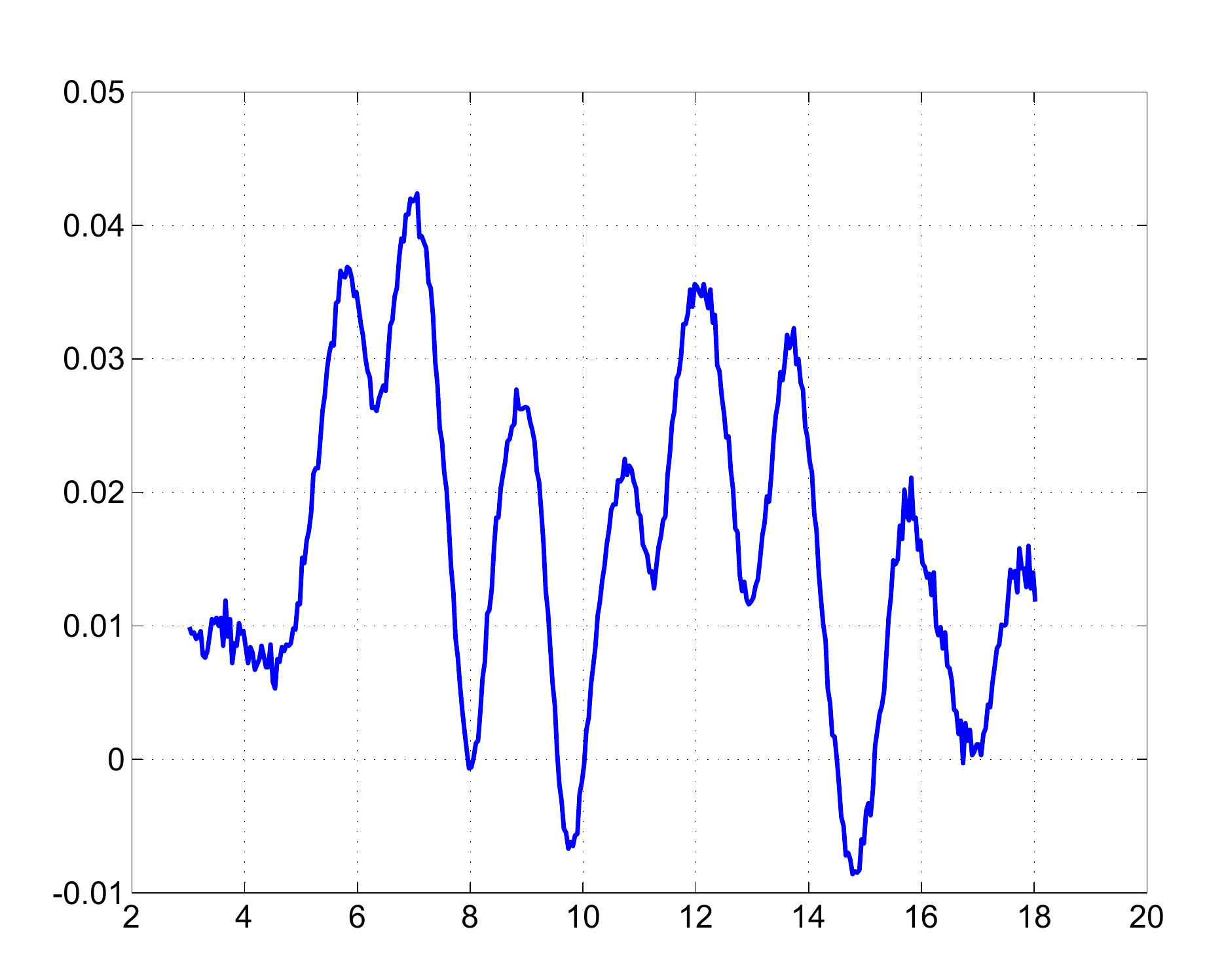}
\caption{Measurement input ($\beta_m$) versus time (s)}
\label{case3_beta}
\end{figure}

\begin{figure}[h]
\includegraphics[width=6in,height=4in]{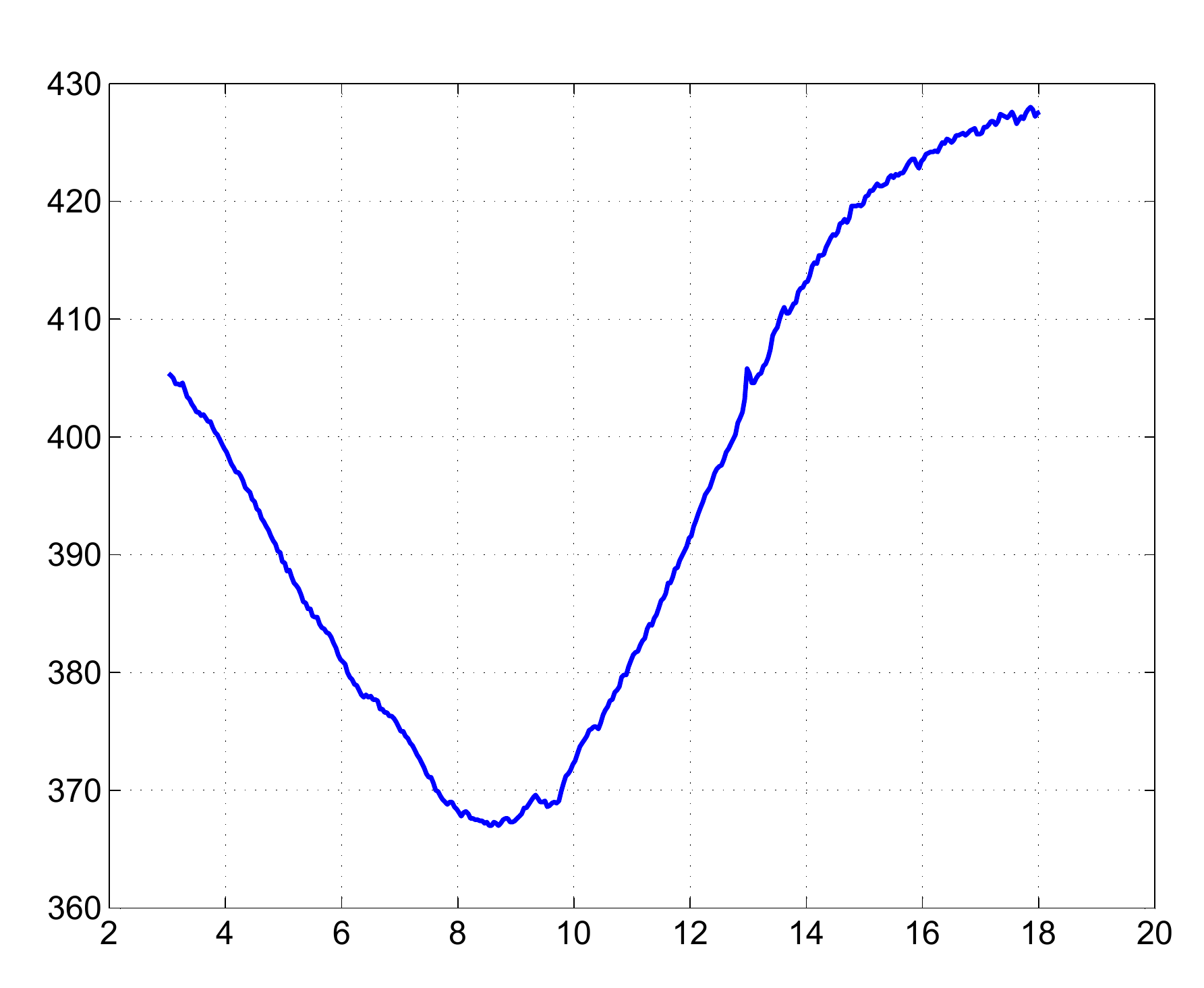}
\caption{Measurement input ($V_m$) versus time (s)}
\label{case3_vel}
\end{figure}

\begin{figure}[h]
\includegraphics[width=6in,height=4in]{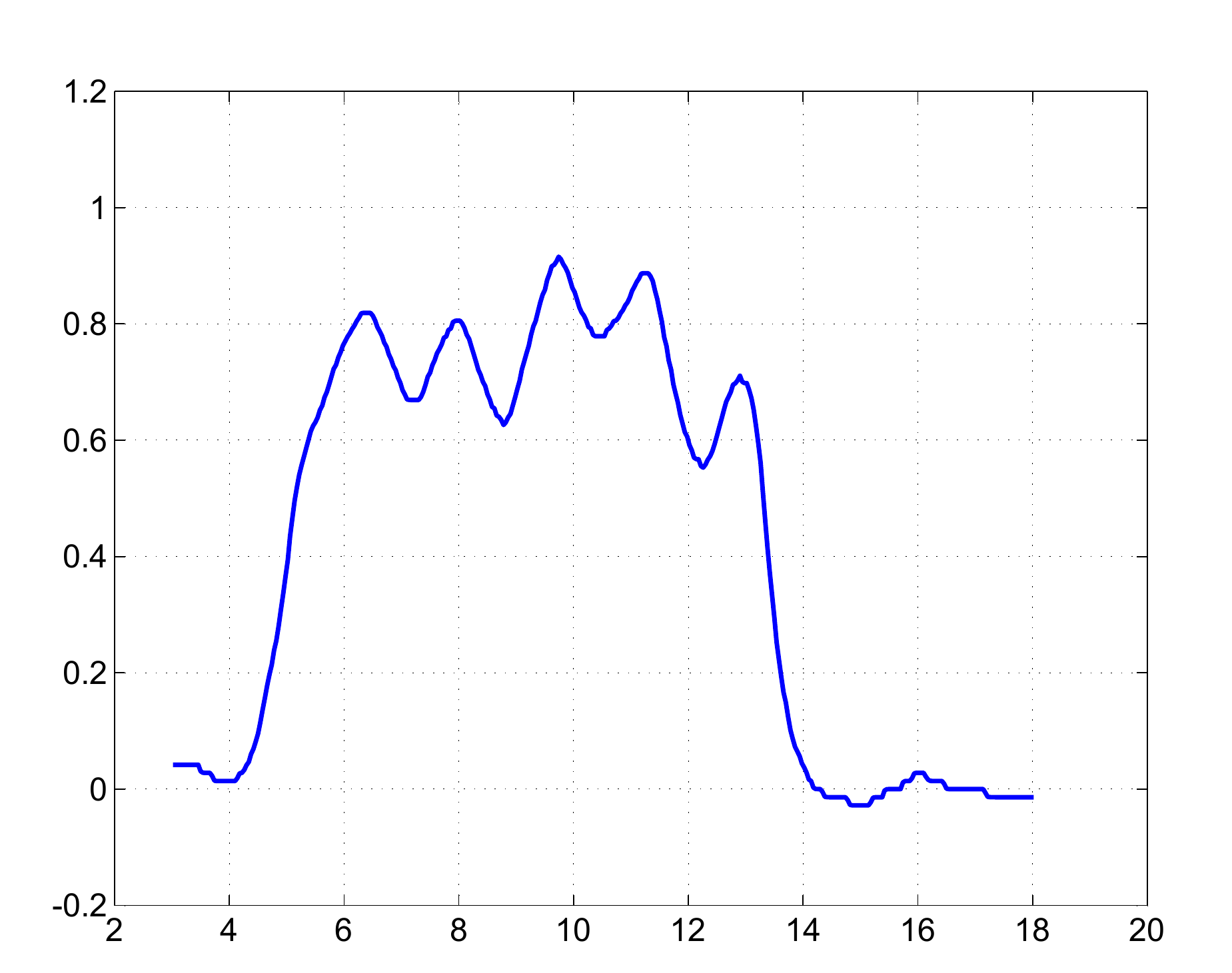}
\caption{Measurement input ($p_m$) versus time (s)}
\label{case3_p}
\end{figure}

\begin{figure}[h]
\includegraphics[width=6in,height=4in]{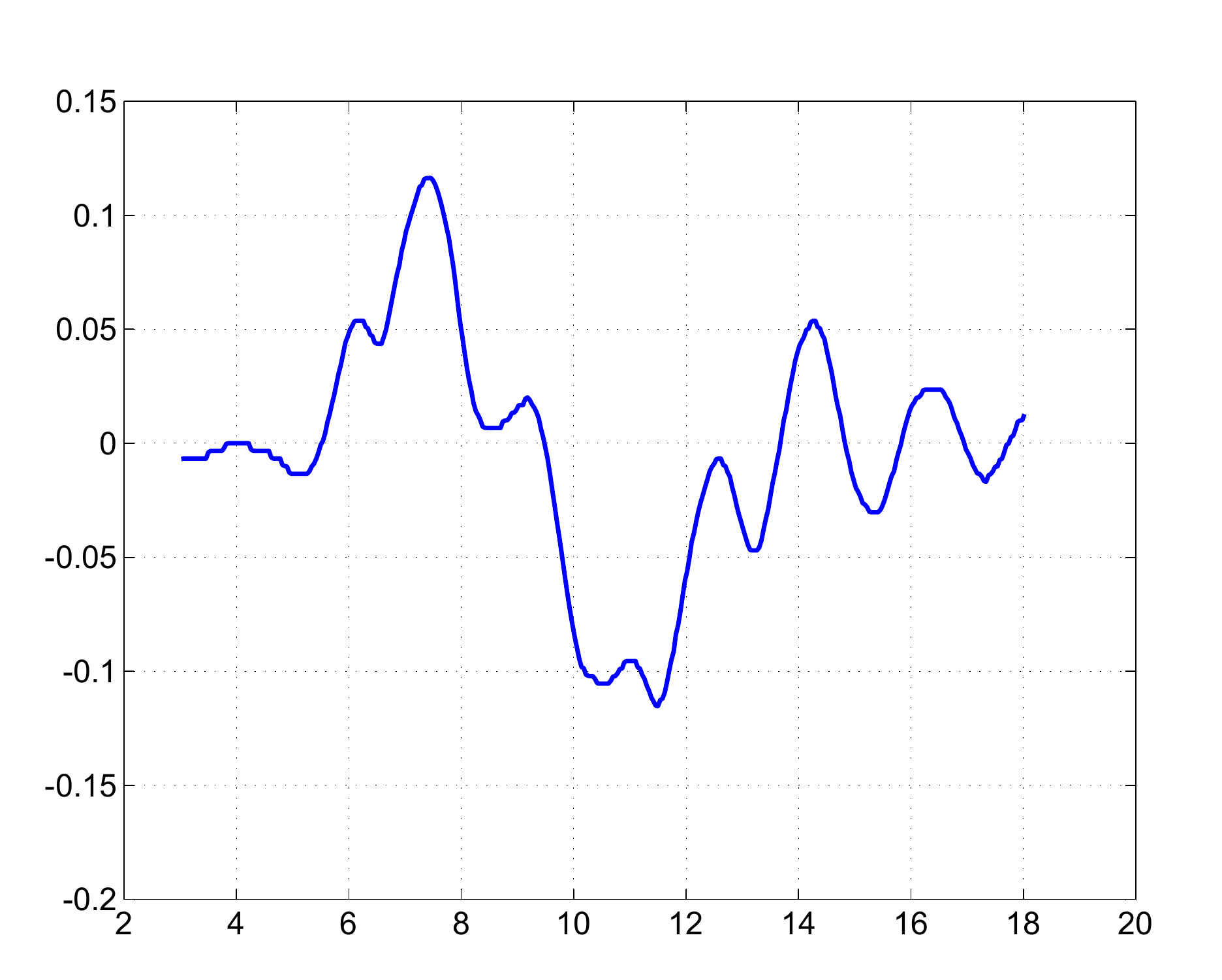}
\caption{Measurement input ($r_m$) versus time (s)}
\label{case3_r}
\end{figure}

\begin{figure}[h]
\includegraphics[width=6in,height=4in]{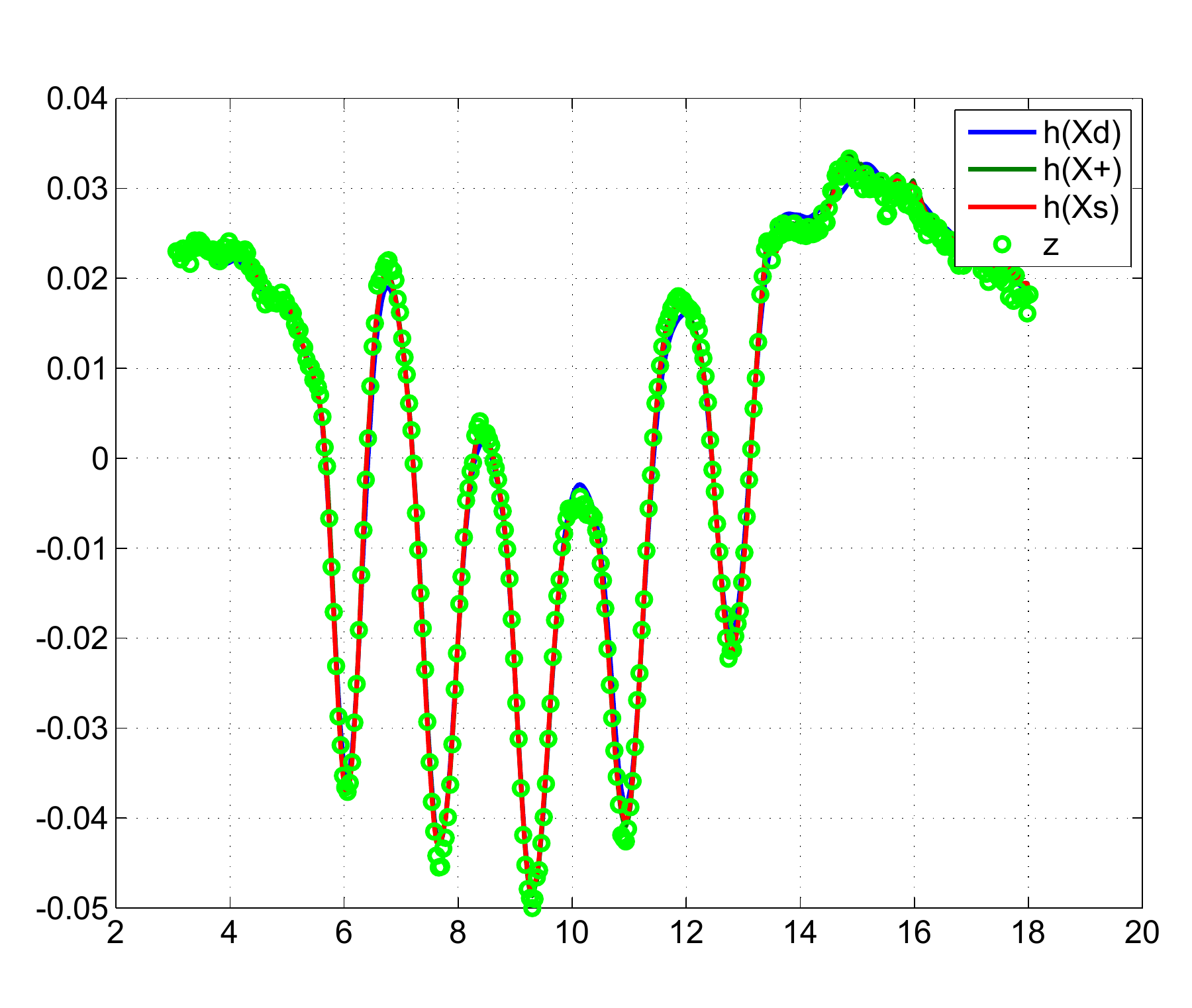}
\caption{Comparison of the predicted dynamics, posterior, smoothed}
\caption*{and the measurement 1}
\label{realQ3_s1}
\end{figure}

 The data set is obtained from NASA TP 1690 (Maine\cite{Maine1981} 1981) by employing a peculiar  manoeuvre  where elevator control input ($\delta_e$ in rad) shown in Fig. \ref{input3} is imparted when the aircraft (T 37 B) is rolling through a full rotation about its x-axis during aileron roll. Some of the available measurements can also be used as inputs in the state equations which includes roll angle ($\phi_m$ in rad), sideslip ($\beta_m$ in rad), velocity ($V_m$ in ft/s), roll rate ($p_m$ in rad/s), yaw rate ($r_m$ in rad/s) and the angle of attack ($\alpha_m$ in rad) are shown in Fig. \ref{case3_beta}, Fig. \ref{case3_vel}, Fig. \ref{case3_p}, Fig. \ref{case3_r}, Fig. \ref{case3_phi} and Fig. \ref{realQ3_s1} respectively. The state equations ($n=3$) for the angle of attack ($\alpha$), pitch rate (q) and the pitch angle ($\theta$) respectively are
\begin{align*}
\dot{\alpha}=&-\frac{\bar qS}{mV_mcos(\beta_m)}(C_{L_\alpha}\alpha+C_{L_{\delta_e}} \delta_e+C_{L_0})+q+\frac{g}{V_mcos(\beta_m)}(cos(\phi_m)cos(\alpha_m)cos(\theta)+\\
&sin(\alpha_m)sin(\theta))-tan(\beta_m)(p_mcos(\alpha_m)+r_msin(\alpha_m))\\
\dot{q}=&\frac{\bar q S \bar c}{Iyy}(C_{m_\alpha}\alpha+C_{m_q}\frac{\bar c}{2V}q+C_{m_{\dot \alpha}}\frac{\bar c}{2V}\dot \alpha+C_{m_{\delta_e}}\delta_e+C_{m_0})+
\frac{Izz-Ixx}{Iyy}r_mp_m\\
\dot{\theta}=&qcos(\phi_m)-r_msin(\phi_m)+\theta_0
\end{align*}
The angle of attack ($\alpha$), pitch rate (q), the pitch angle ($\theta$) and the normal acceleration ($a_n$) are measured (indicated with subscript `m') in units of rad, rad/s, rad and $ft/s^2$ respectively.  The measurement equations ($m$=4) are given by
\begin{align*}
{\alpha_m}&=K_\alpha\alpha-K_\alpha x_\alpha \frac{q}{V} \\
{q_m}&=q\\
{\theta_m}&=\theta\\
{a_{n_m}}&=\frac{\bar q S}{mg}(C_{N_\alpha}\alpha+C_{N_{\delta_e}}\delta_e+C_{N_0})+\frac{x_{a_n}}{g}\dot{q}
\end{align*}

The unknown parameters ($p=10$) is $(C_{L_{\alpha}},C_{L_{\delta_e}},C_{L_{0}},C_{m_{\alpha}},C_{m_{q}},C_{m_{\dot \alpha}},C_{m_{\delta_e}},C_{m_{0}},\theta_0,C_{N_0})^T$ with an approximation $C_{N_\alpha}=C_{L_\alpha}$ and $C_{N_{\delta_e}}=C_{L_{\delta_e}}$. The ones with suffix `$\delta_e$' are the control derivatives, the ones with suffix zero are the biases and all others are aerodynamic derivatives. The initial states are taken as initial measurement and the initial parameter values are taken as $(4,0.15,0.2,-0.5,-11.5,-5,-1.38,-0.06,-0.01,0.2)^T$.

\begin{table}[h]
\begin{center}
\caption*{Other constant values used for case-3}{}
\begin{tabular}{| c | c | c | c | c | c | }
\hline
 S=184 & m=196 & Ixx=6892.7  &  Iyy=3953.2 & Izz=10416.4  \\ \hline
g=32.2 & $\bar c$=5.58 & $K_\alpha x_\alpha$=-0.0279 & $x_{a_n}$=0.101 & $K_\alpha=1$\\ \hline
\end{tabular}
\end{center}
\end{table}

\par
\begin{figure}[h]
\includegraphics[width=6in,height=4in]{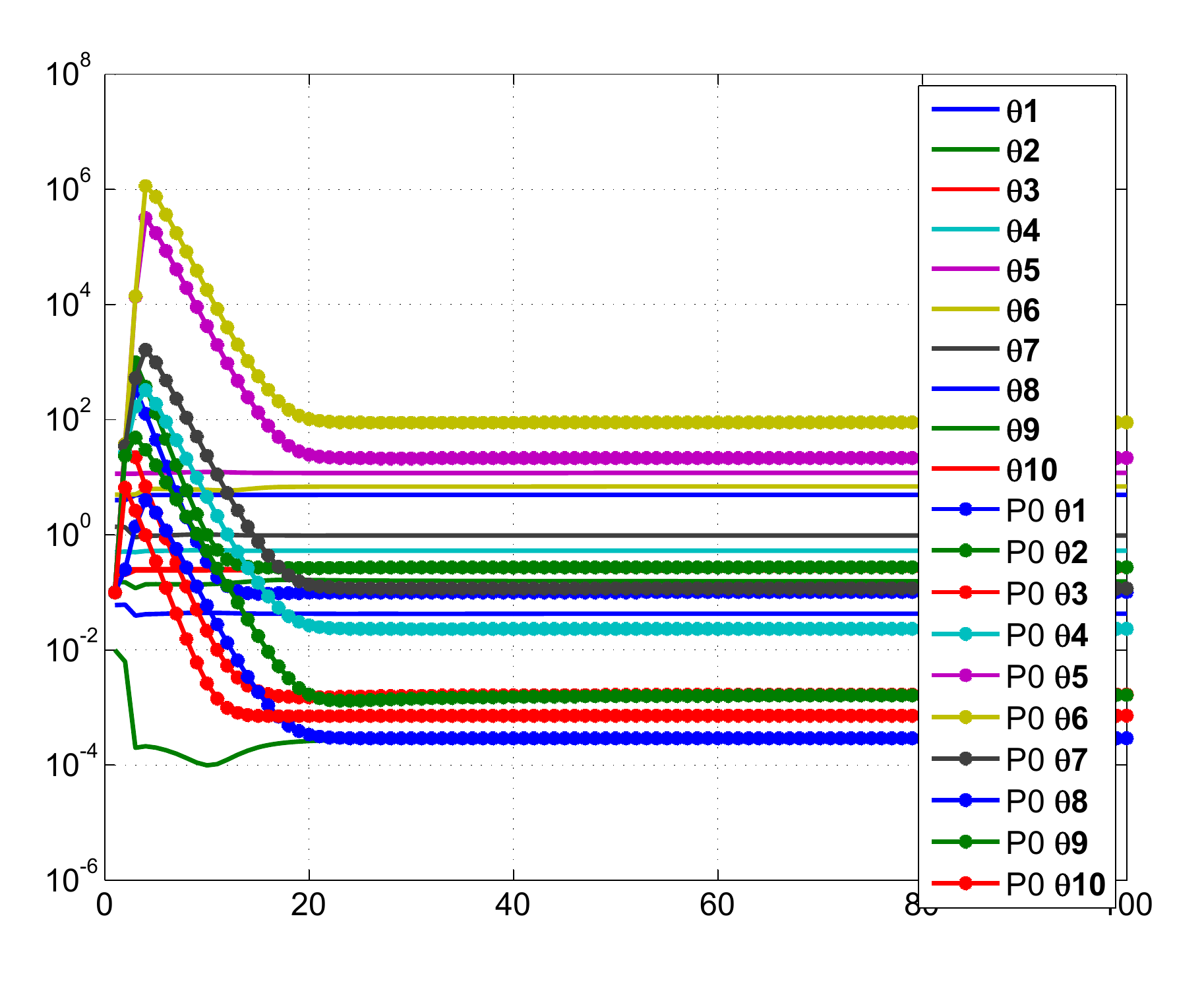}
\caption{Variation of parameter and its initial covariance ($\mathbf{P_0}$) with iterations}
\label{realQ3_P0}
\end{figure}

\begin{figure}[h]
\includegraphics[width=6in,height=4in]{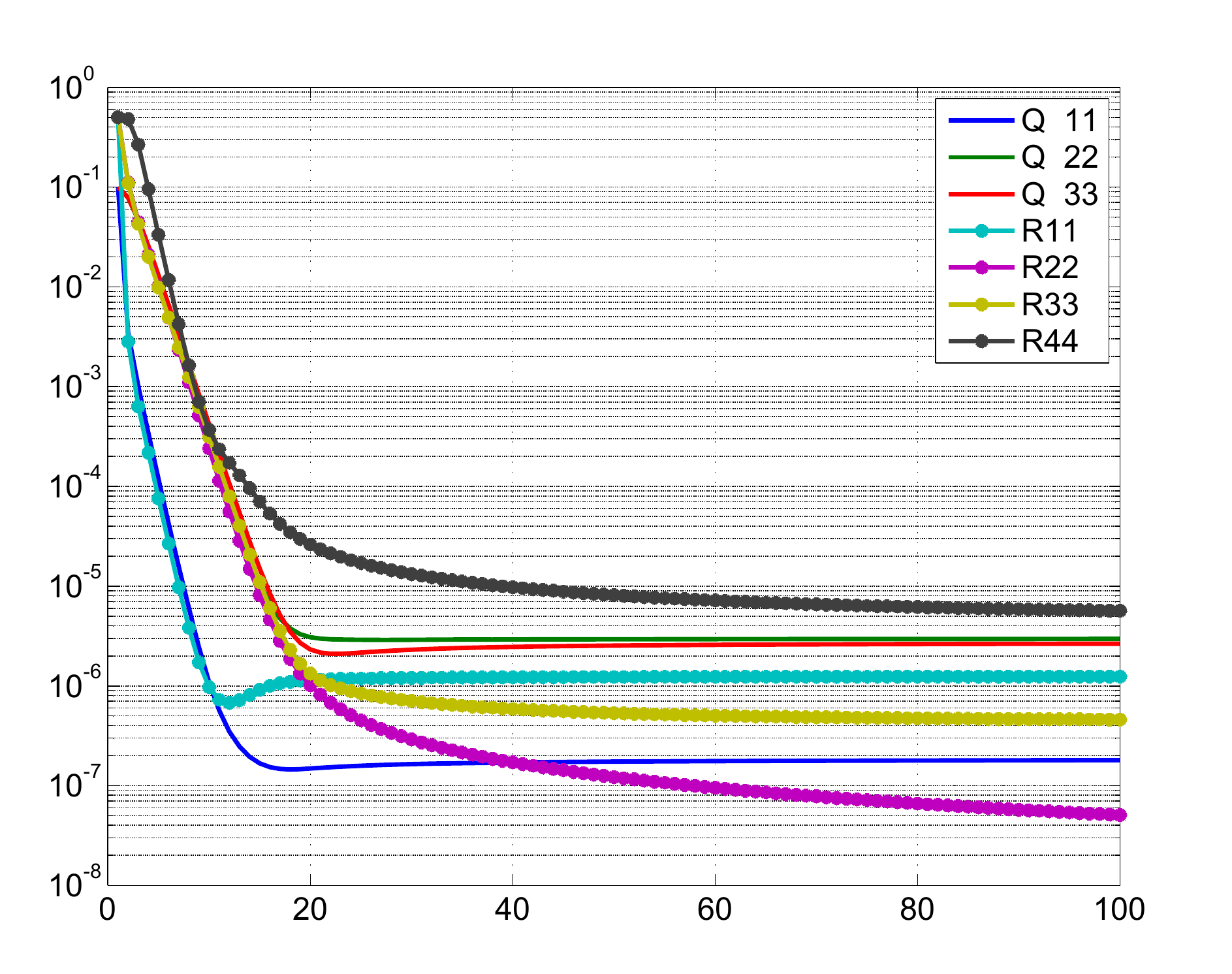}
\caption{Variation of \textbf{R} with iterations of case 3}
\label{realQ3_R}
\end{figure}

\begin{figure}[h]
\includegraphics[width=6in,height=4in]{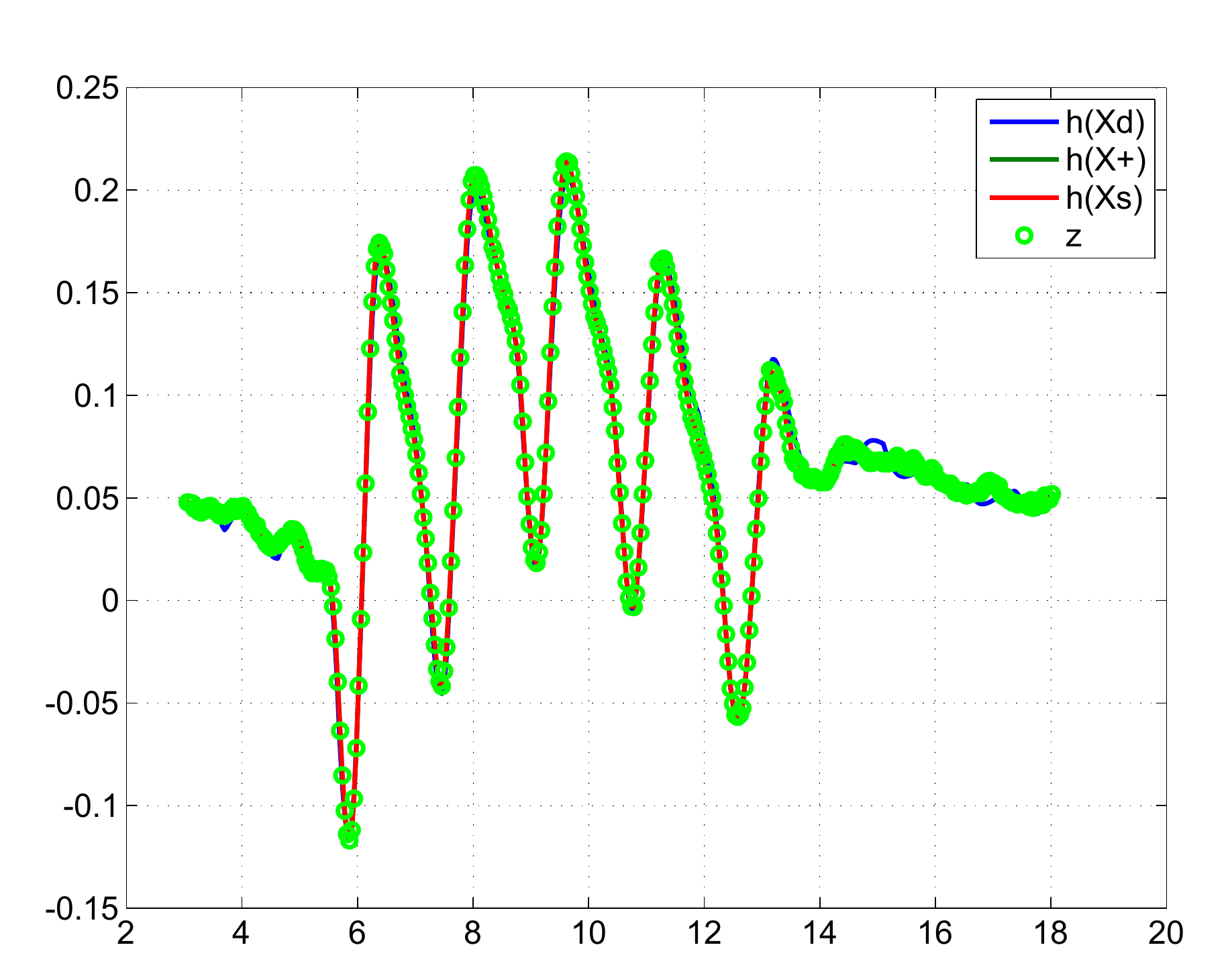}
\caption{Comparison of the predicted dynamics, posterior, smoothed}
\caption*{and the measurement 2}
\label{realQ3_s2}
\end{figure}

\begin{figure}[h]
\includegraphics[width=6in,height=4in]{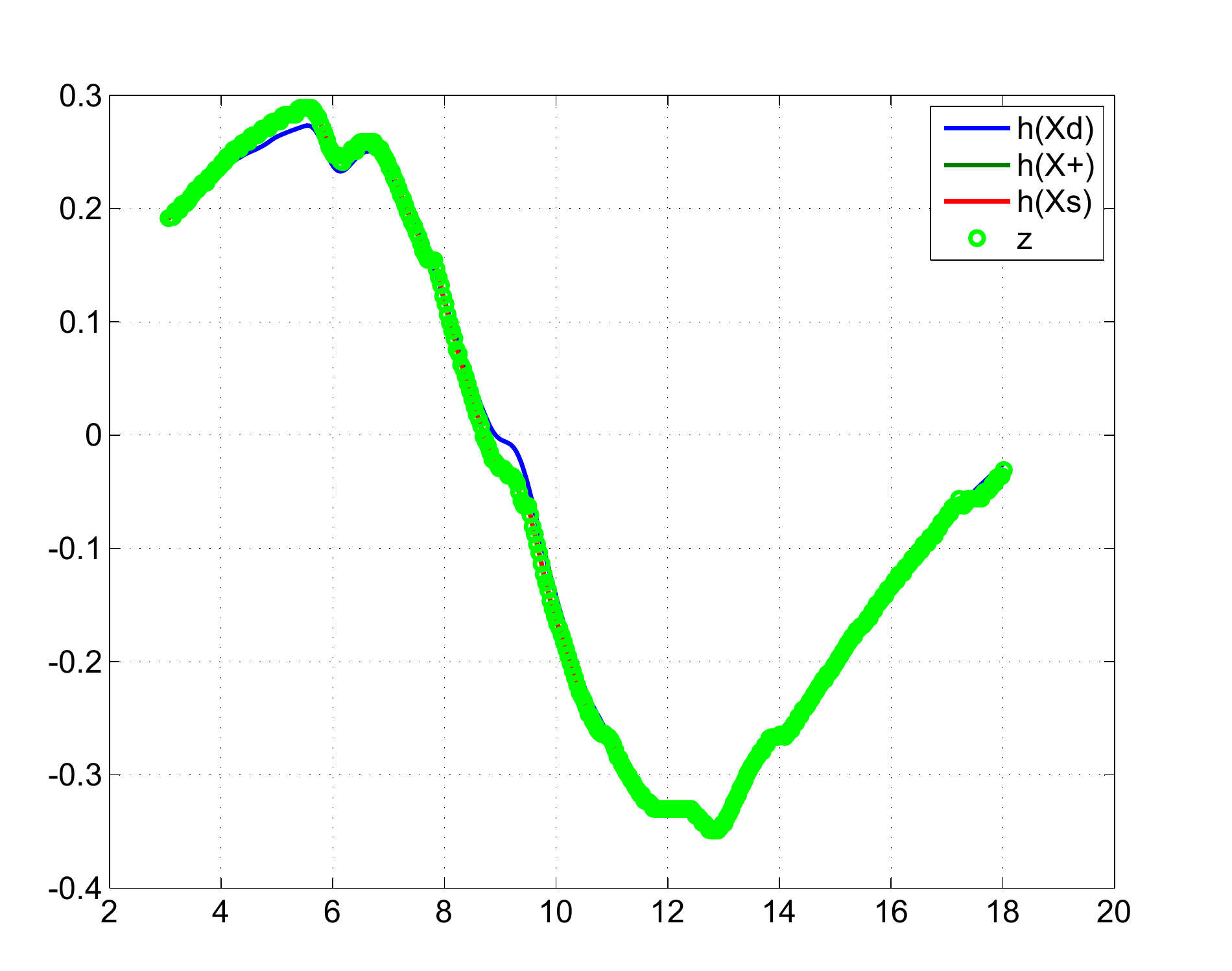}
\caption{Comparison of the predicted dynamics, posterior, smoothed}
\caption*{and the measurement 3}
\label{realQ3_s3}
\end{figure}

\begin{figure}[h]
\includegraphics[width=6in,height=4in]{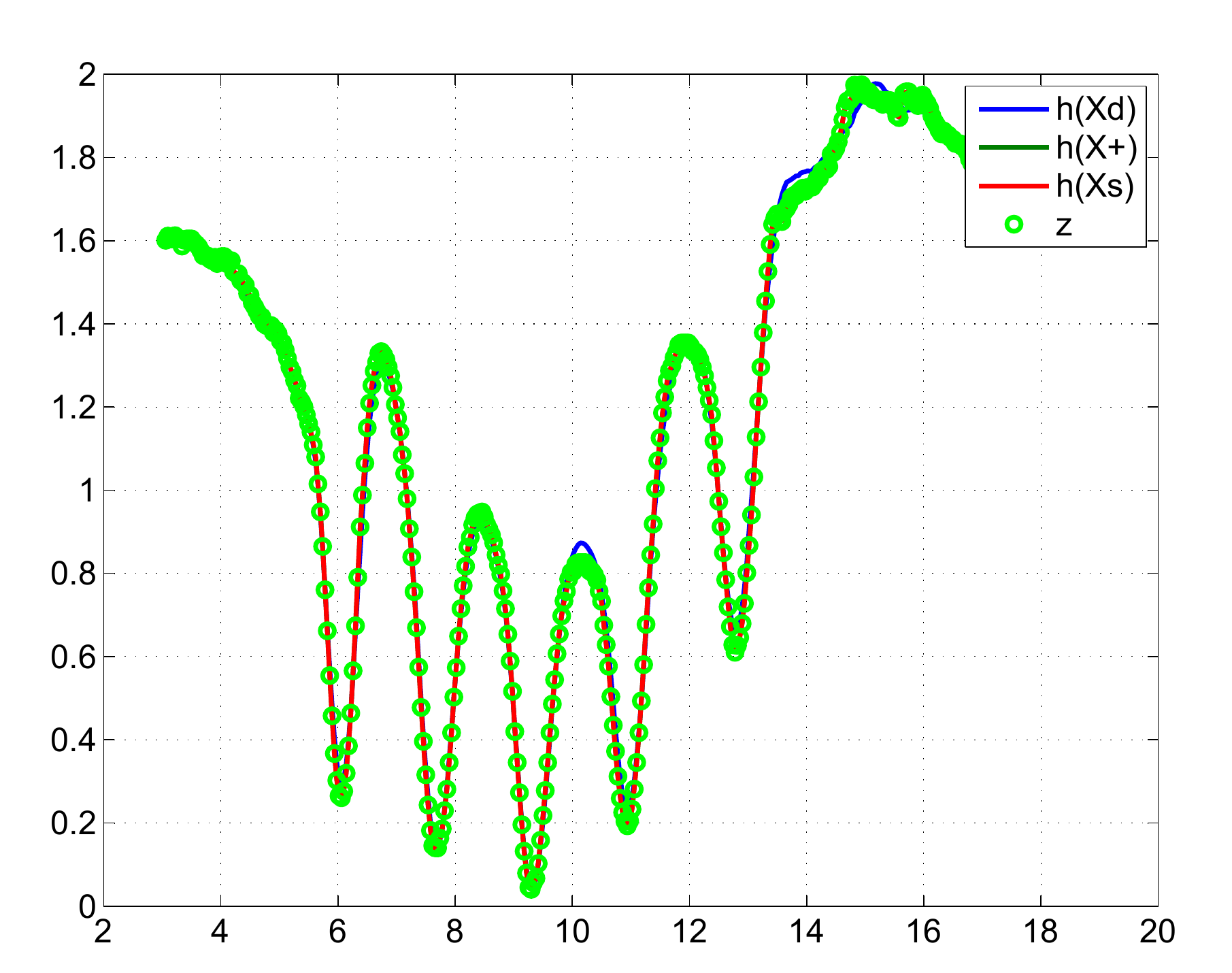}
\caption{Comparison of the predicted dynamics, posterior, smoothed}
\caption*{and the measurement 4}
\label{realQ3_h4}
\end{figure}

\begin{figure}[h]
\includegraphics[width=6in,height=4in]{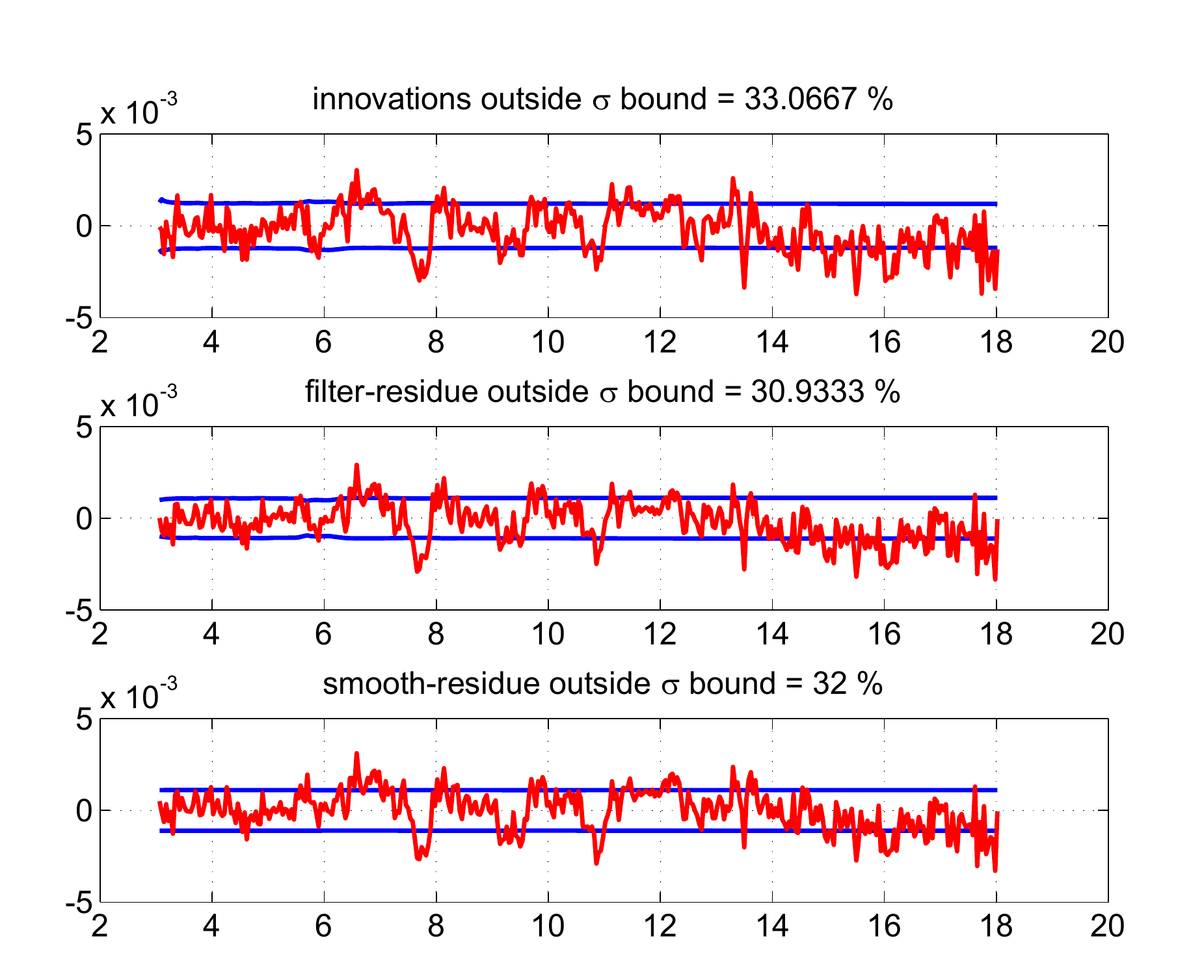}
\caption{The innovations, filtered residue and smoothed residue of measurement 1}
\label{realQ3_innov1}
\end{figure}

\begin{figure}[h]
\includegraphics[width=6in,height=4in]{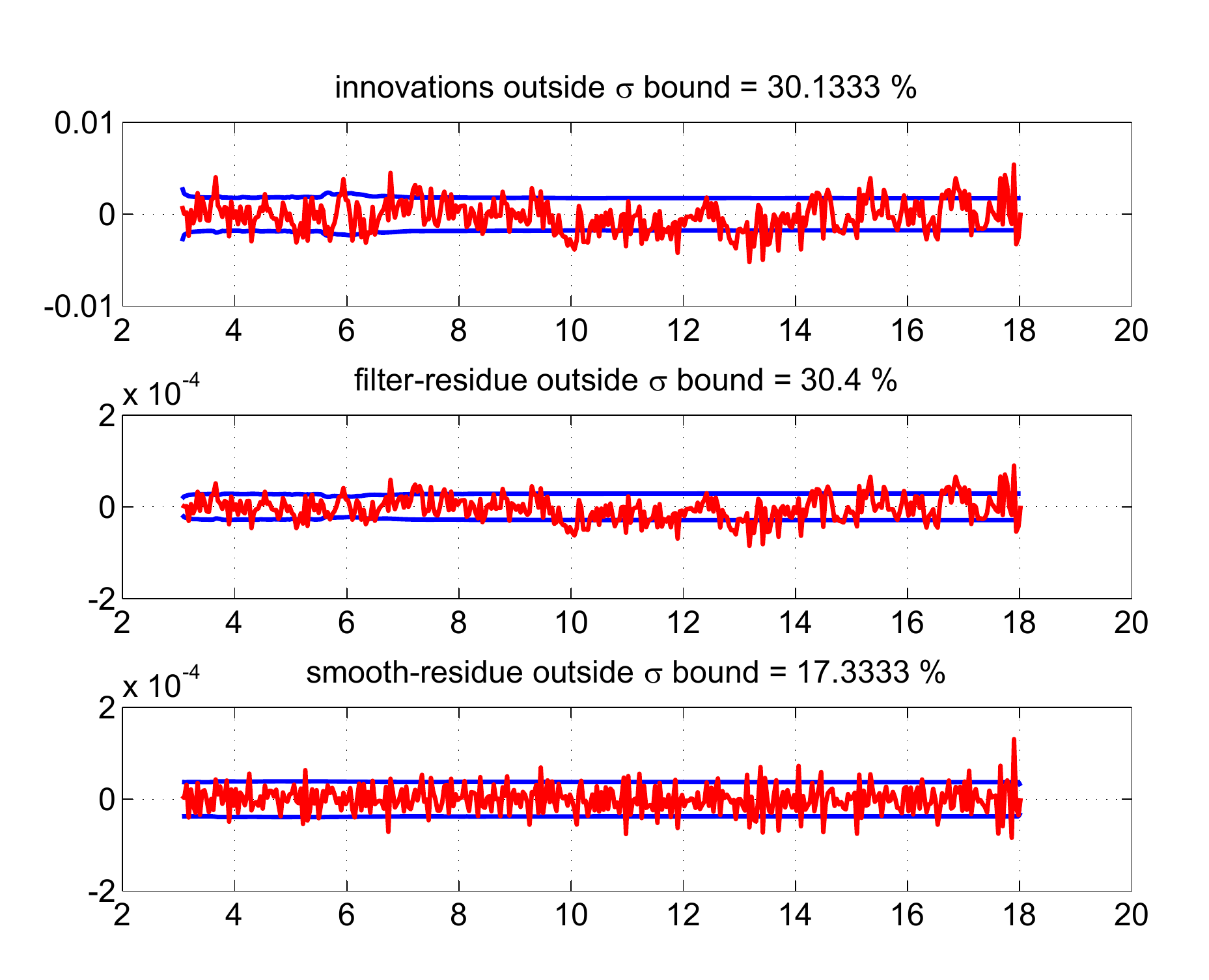}
\caption{The innovations, filtered residue and smoothed residue of measurement 2}
\label{realQ3_innov2}
\end{figure}

\begin{figure}[h]
\includegraphics[width=6in,height=4in]{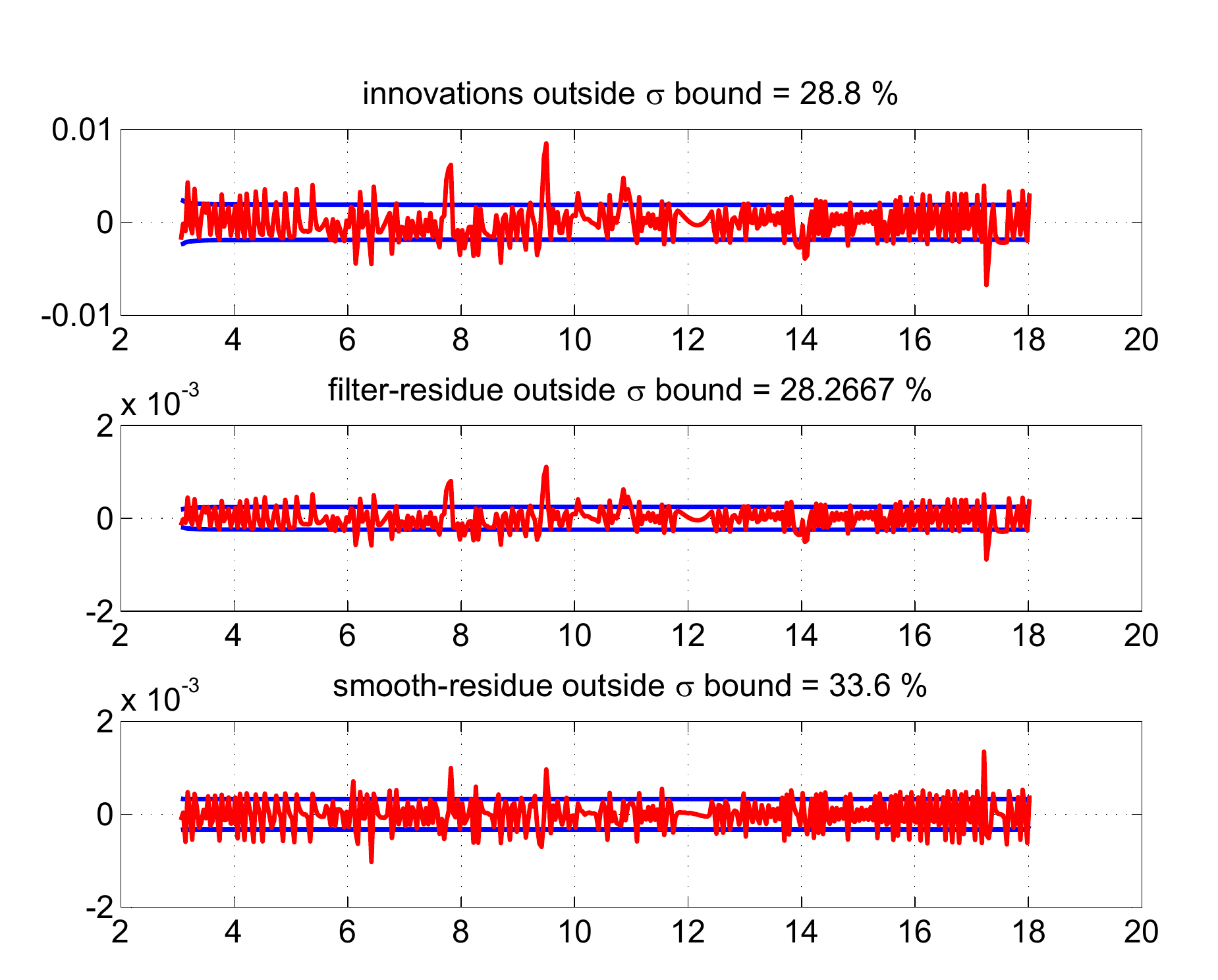}
\caption{The innovations, filtered residue and smoothed residue of measurement 3}
\label{realQ3_innov3}
\end{figure}

\begin{figure}[h]
\includegraphics[width=6in,height=4in]{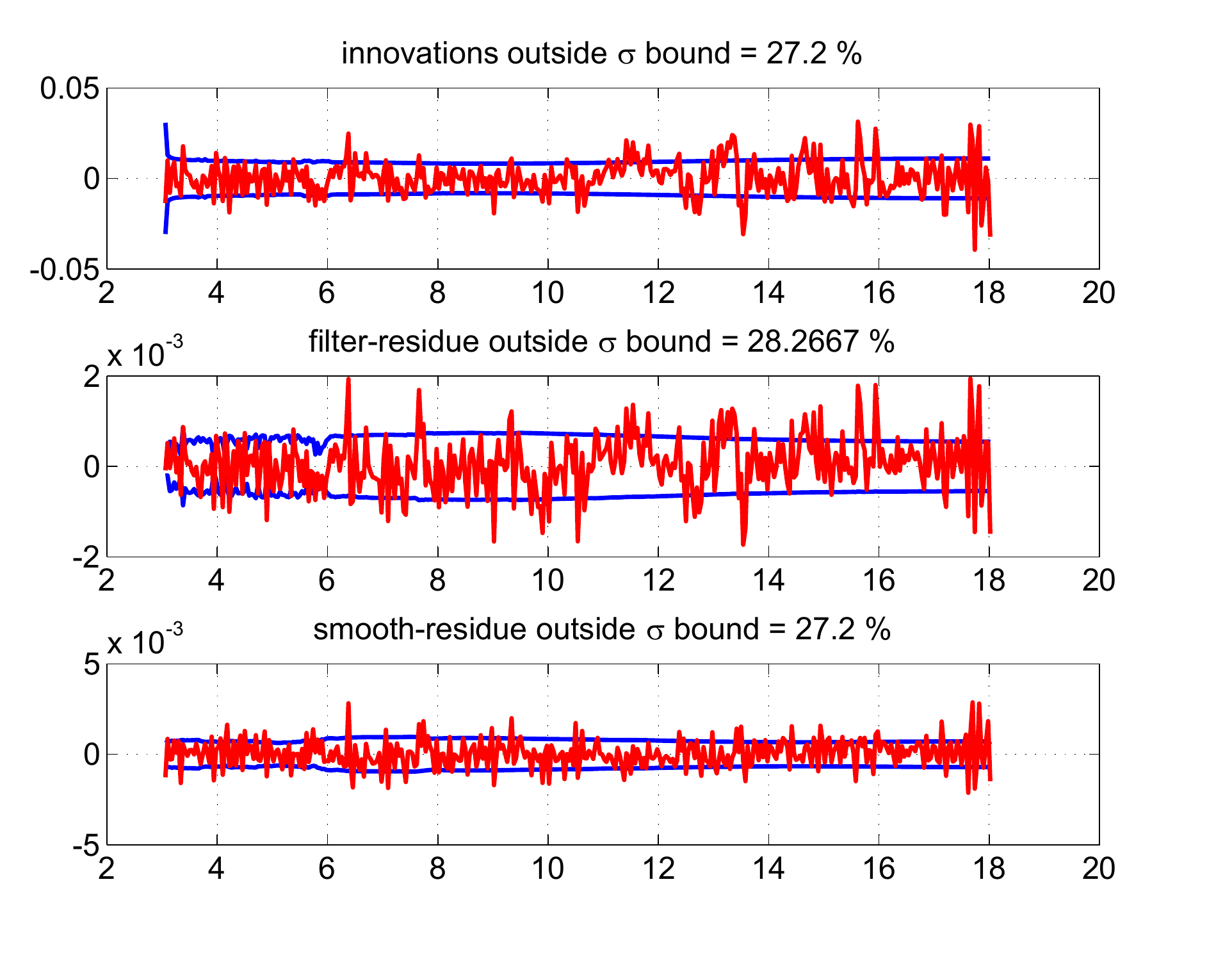}
\caption{The innovations, filtered residue and smoothed residue of measurement 4}
\label{realQ3_innov4}
\end{figure}

\begin{figure}[h]
\includegraphics[width=6in,height=4in]{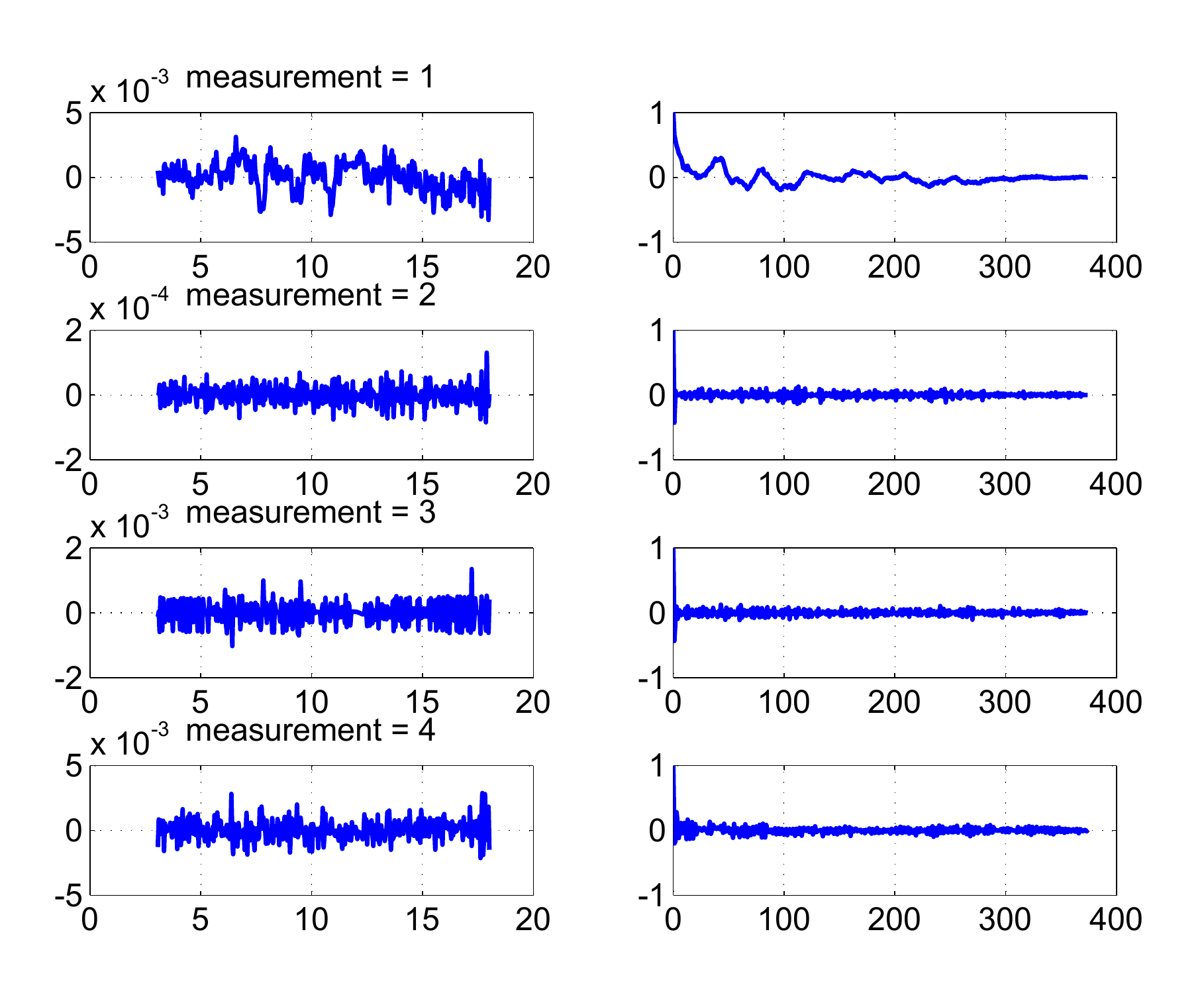}
\caption{Time variation of estimated measurement noise (left) and}
\caption*{their autocorrelation (right)}
\label{realQ3_mnoise}
\end{figure}

\begin{figure}[h]
\includegraphics[width=6in,height=4in]{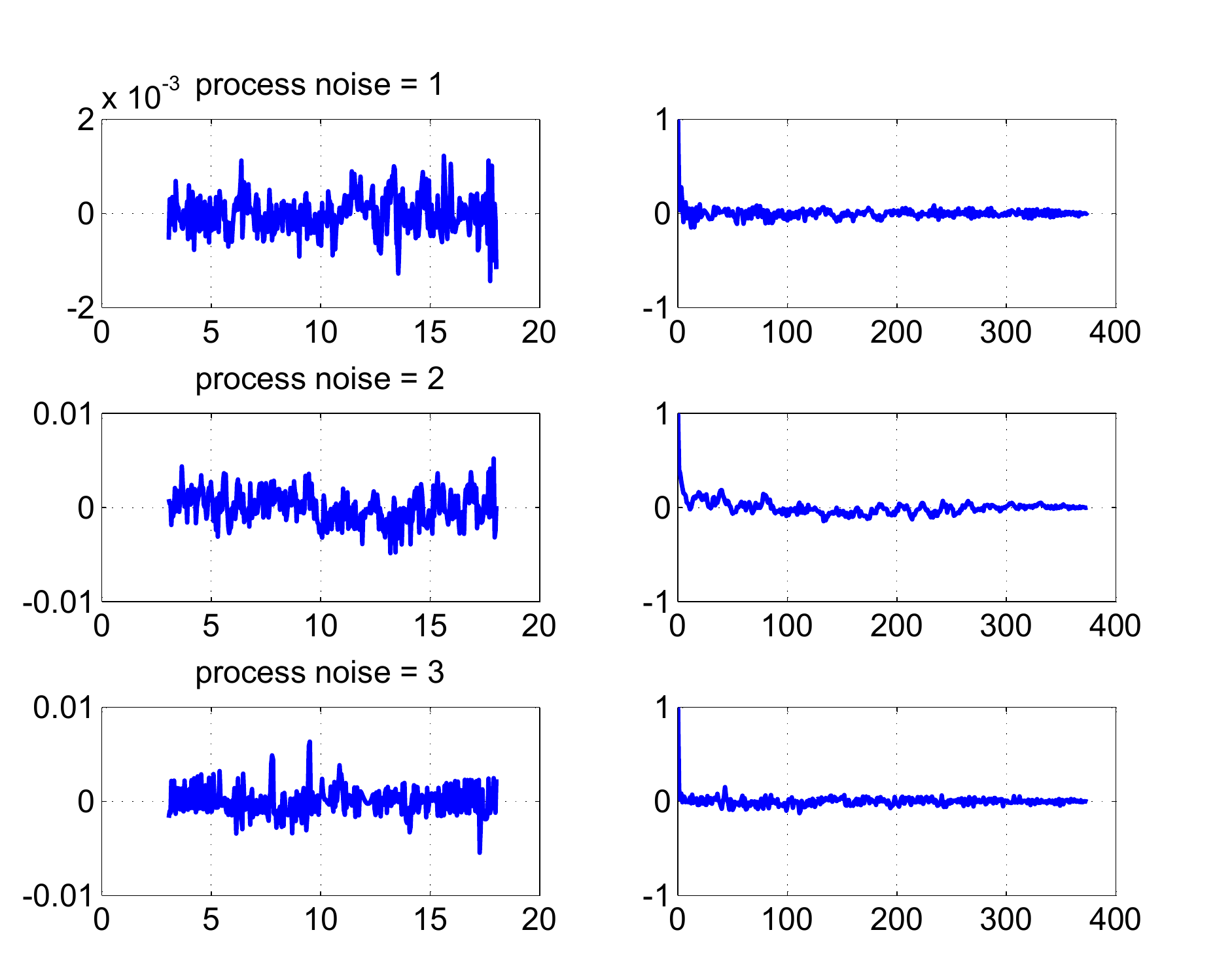}
\caption{Time variation of estimated process noise (left) and}
\caption*{their autocorrelation (right)}
\label{realQ3_pnoise}
\end{figure}

 Case-3 real data is run using the reference EKF (\textbf{Q} $>$ 0) with 100 iterations. The Fig. \ref{input3}-\ref{case3_r} are the inputs used in state equations. The Fig. \ref{realQ3_P0} shows the variation of parameter estimates and its initial covariance $\mathbf{P_0}$ with iterations and a similar Fig. \ref{realQ3_R} for \textbf{Q} and \textbf{R}. The values of \textbf{J1-J3} are close to the number of measurements ($m=4$) with \textbf{J6-J8} are close to the number of states ($n=3$) as shown in Fig. \ref{realQ3_J} and Table-\ref{tbcase3Q}. This means the measurement and state equations are well balanced. The \textbf{J5} is the negative log likelihood cost function. The later Fig. \ref{realQ3_s1}-\ref{realQ3_h4} compares (i) the state dynamics based on the estimated parameter after the filter pass through the data, (ii) the state after measurement update, (iii) the smoothed state and (iv) the measurement.  The Fig. \ref{realQ3_innov1}-\ref{realQ3_innov4} shows the confidence in the innovations, filtered residue and smoothed residue. The estimated measurement and process noise do not appear to have constant statistical characteristics across time as seen in Fig. \ref{realQ3_mnoise} and Fig. \ref{realQ3_pnoise}. Another experiment was carried out by generating a typical data set by using the estimated theta and injecting the estimated \textbf{Q} and \textbf{R} as additive white Gaussian noise. This is to determine the effect of non additive, non White and non Gaussian noise distribution in the real data on the CRBs. After each iteration in the reference recipe the $\Theta$, \textbf{Q} and \textbf{R} were reset as from the real data. Similar experiment was also conducted by updating $\Theta$ as well. It turned out that there is not much of a difference in the final CRBs as can be seen from the Table-\ref{tbnew3}.

\par Finally two other filter runs were carried out using the MT and MS statistics for the estimation of \textbf{Q} and \textbf{R}. The behaviour of the various cost function and in particular \textbf{J6} and \textbf{J7} in Table-\ref{tbcase3QMTMS} shows that the choice of the filter statistics for estimating \textbf{Q} and \textbf{R} in the proposed reference approach is the best possible when compared to other approaches presently considered.

\subsection{Remarks on Case-3}

\par
\begin{figure}[h]
\includegraphics[width=6in,height=4in]{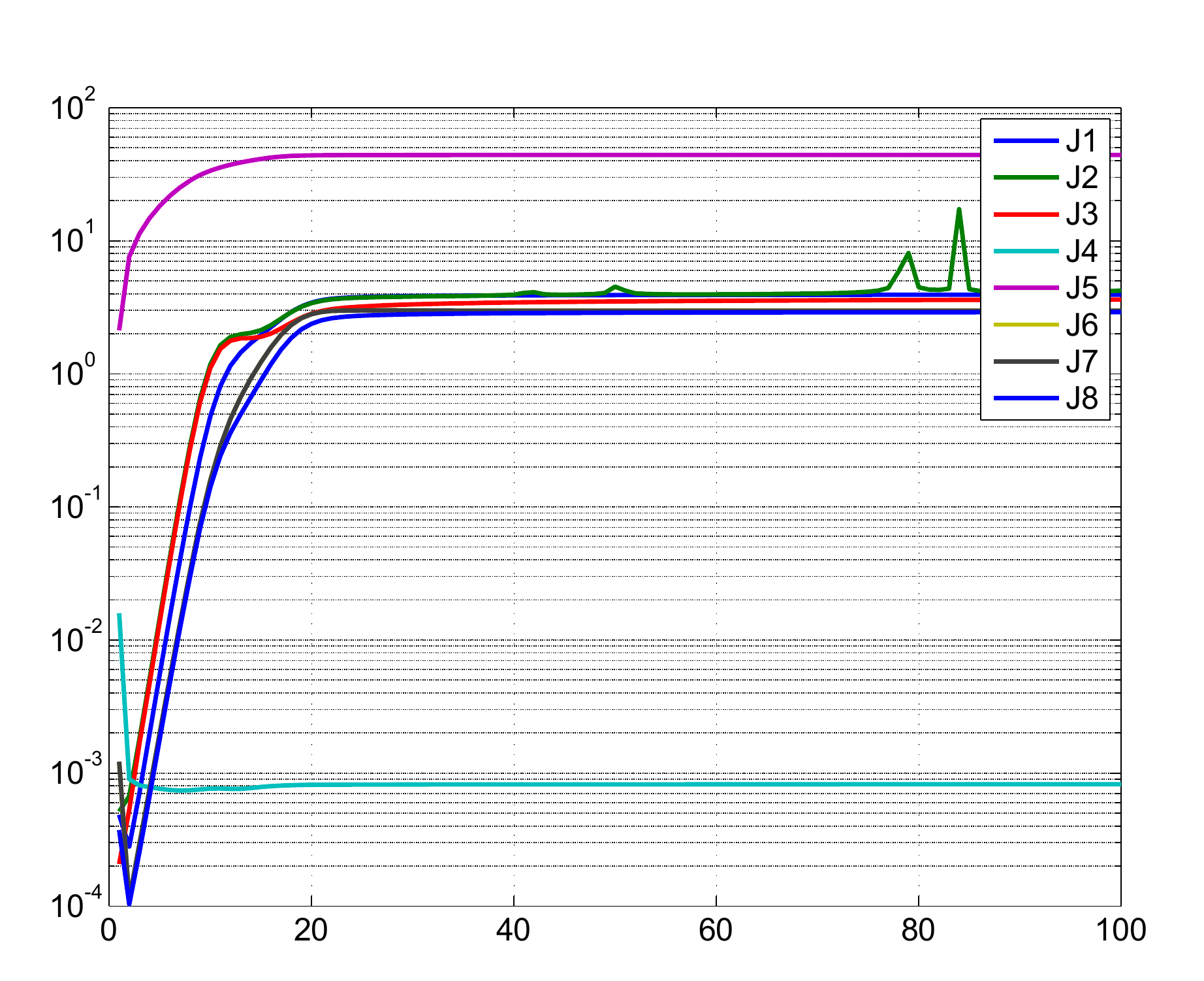}
\caption{Variation of different costs (\textbf{J1-J8}) with iterations}
\label{realQ3_J}
\end{figure}

\begin{figure}[h]
\includegraphics[width=6in,height=4in]{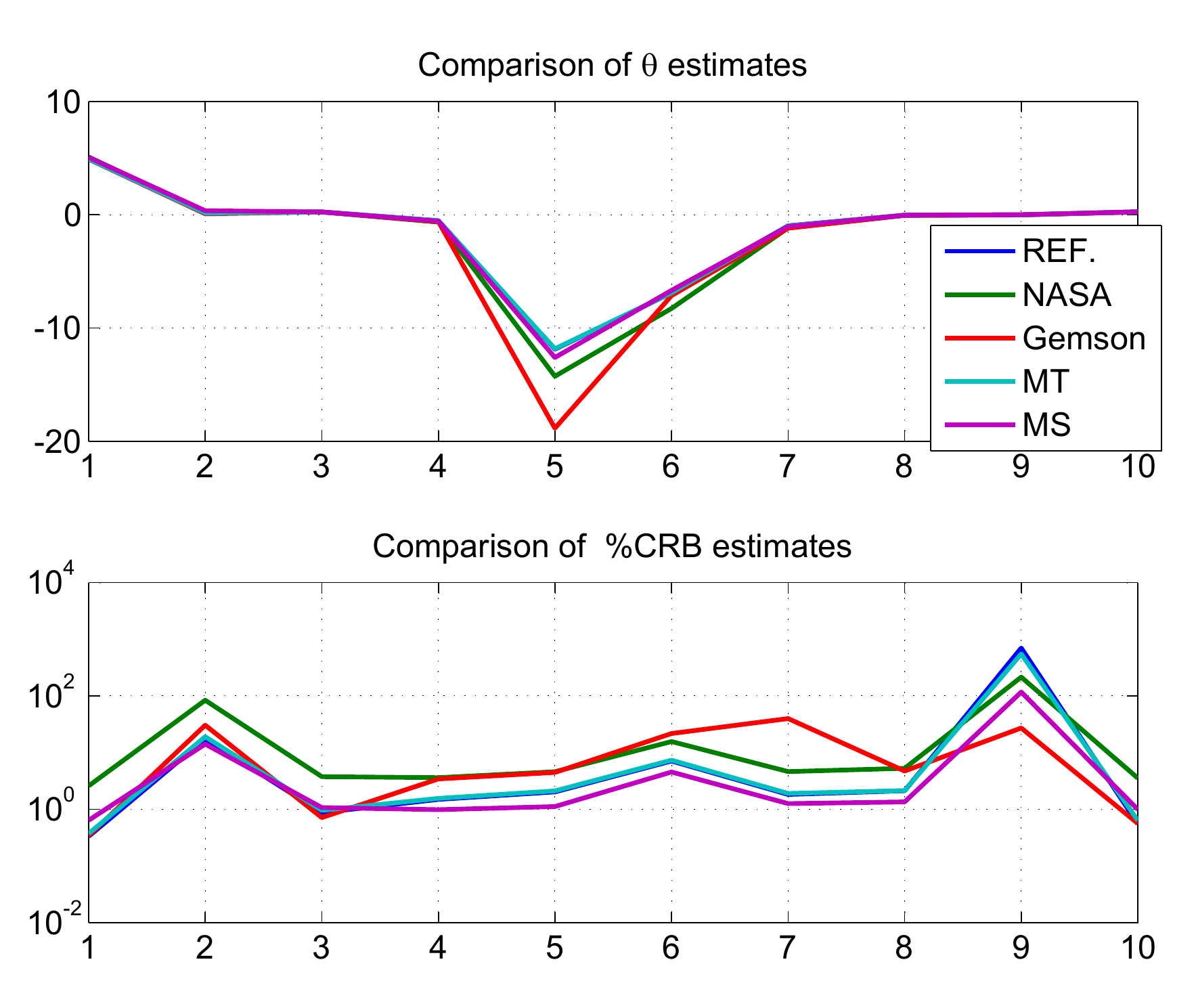}
\caption{Comparison of the parameter estimates and $\%$CRBs by different methods}
\label{comp3}
\end{figure}

 The NASA results have been generated assuming \textbf{Q} = 0 and are comparable with reference procedure for the parameter estimates and their CRBs. Further the MT and MS methods give quite different estimates for the \textbf{Q} and \textbf{R} values than in the reference case. We believe that the reference procedure provides the best possible parameter estimates and their uncertainties. From the plot of the parameter estimates and their \%CRB in Fig \ref{comp3}, it can be seen that the parameters 1, 3, 4, 5, 6, 7, 8, 10 are strong and the parameters 2 and 9 are the weak ones.  The CRBs as estimated by different methods generally appear to vary widely. However what is interesting is that even the estimate of the strong parameter such as 5 varies widely among the methods. Such a behaviour of the filter across the parameter estimates shows how important is the tuning of the filter statistics namely $\mathbf{P_0}$, \textbf{Q} and \textbf{R} in parameter estimation and their uncertainties. It was also observed that for this particular case, the cost \textbf{J2} using filtered residue went negative at some iteration as seen in Fig. \ref{realQ3_J} with a spike whose absolute value was plotted. However the cost \textbf{J3} using smoothed residue that was used for tuning the filter did not show any such peculiarity.
\par The rounded off 100$\times$C matrix for case-3 is given by
\begin{align*}
\begin{bmatrix}
   100  & -44 &    6  &  41  &  10  &   1  & -22 &   16  &   1  &  28 \\
   -44  & 100 &  -14  & -22  & -34  &  -6  &  42 &  -41  &   0  & -95  \\
     6  & -14 &  100  &  14  &  26  &   1  & -32 &   32  &   0  &  -3  \\
    41  & -22 &   14  & 100  &  23  &  -2  & -53 &   41  &   1  &  14 \\
    10  & -34 &   26  &  23  & 100  &  50  & -79 &   73  &   0  &  27 \\
     1  &  -6 &    1  &  -2  &  50  & 100  & -14 &   -1  &   0  &   0 \\
   -22  &  42 &  -32  & -53  & -79  & -14  & 100 &  -97  &  -1  & -35 \\
    16  & -41 &   32  &  41  &  73  &  -1  & -97 &  100  &   1  &  36 \\
     1  &   0  &   0  &   1  &   0  &   0  &  -1 &    1  & 100  &   0 \\
    28  & -95 &   -3  &  14  &  27  &   0  & -35 &   36  &   0  & 100 \\
\end{bmatrix}
\end{align*}

\begin{landscape}
\begin{table}[h]
\subsection{Case-3 Tables}
\caption{Real flight test data case-3 results using the reference adaptive EKF.\\ No of iterations=100}{}
\label{tbcase3Q}
\begin{center}
\begin{footnotesize}
\begin{tabular}{|c| c| c| c| c| c|c|c|c|c|c| }
\hline
Study &
\makecell{$\Theta$\\ (Ref)} &
\makecell{$\Theta$\\ (NASA)} &
\makecell{$\Theta$\\ (Gemson)} &
\makecell{$\sigma_\Theta$ \\(Ref)} &
\makecell{$\sigma_\Theta$\\ (NASA)} &
\makecell{$\sigma_\Theta$\\ (Gemson)} &
\makecell{\textbf{R} \\ $\times10^{-6}$\\ (Ref)}&
\makecell{\textbf{Q} \\ $\times10^{-6}$\\ (Ref)}&
\makecell{\textbf{J1-J8} \\(Ref) }&
Remarks
\\ \hline


\makecell{$\mathbf{P_0}$ : Scaled up-[0,0;0,\checkmark]\\\textbf{Q} : EM-[\checkmark,0;0,0] \\\textbf{R} : EM-diag} &
\makecell{ 4.9235  \\  0.1554  \\  0.2409  \\ -0.5293\\  -11.8596 \\  -6.8959  \\ -0.9731 \\ -0.0425  \\  0.0003   \\ 0.2538} &
\makecell{ 5.1068 \\   0.1909  \\  0.2448  \\ -0.6474 \\ -14.26 \\ -8.27 \\ -1.1614 \\  -0.0505 \\ -0.01177 \\  0.2541 } &
\makecell{ 4.9028 \\   0.0879  \\  0.2529  \\ -0.6174 \\ -18.8339 \\ -7.1290 \\ -1.1841 \\  -0.0507 \\ -0.0037 \\  0.2503} &
\makecell{0.0164 \\    0.0271 \\   0.0021  \\  0.0079  \\  0.2402 \\   0.4891  \\  0.0177 \\ 0.0009 \\   0.0021   \\ 0.0014 } &
\makecell{  0.1322  \\  0.1602 \\   0.009215 \\   0.02339  \\  0.6528 \\ 1.296\\  0.05371  \\  0.002655 \\ 0.02528  \\  0.008935  } &
\makecell{  0.0168  \\  0.0267 \\   0.0018 \\   0.0211  \\  0.8379 \\ 1.544 \\  0.471  \\  0.0024 \\ 0.001  \\  0.0014  } &
\makecell{ 1.241  \\  0.051 \\   0.460 \\   5.668} &
\makecell{ 0.180  \\  2.954  \\  2.646} &
\makecell{ 3.9336   \\ 4.2225  \\  3.6162  \\  0.0008  \\ -44.1347  \\  2.9752 \\   2.9760 \\ 2.9070} &
\makecell{Cost functions converge \\ to the expected values}
\\ \hline

\end{tabular}
\end{footnotesize}
\end{center}
\end{table}

\begin{table}[h]
\caption{Case-3 results using simulated Additive White Gaussian Noise}{}
\label{tbnew3}
\begin{center}
\begin{tabular}{|c| c| c| c| c|}
\hline
Study &
\makecell{$\sigma_\Theta$ \\(Simulated-without\\ updating $\Theta$)} &
\makecell{$\sigma_\Theta$ \\(Simulated-with\\ updating $\Theta$)} &
\makecell{$\sigma_\Theta$ \\(Ref)} &
Remarks
\\ \hline

\multicolumn{5}{|c|}{\makecell{Case-3 data generated using simulated measurement and process noise (AWGN) \\ of variance \textbf{Q} and \textbf{R} estimated by Reference EKF (\textbf{Q} $>$ 0)} } \\ \hline

\makecell{$\mathbf{P_0}$ : Scaled up-[0,0;0,\checkmark]\\\textbf{Q} : \textbf{Q} (Ref) \\\textbf{R} :   \textbf{R} (Ref)} &

\makecell{  0.0173  \\  0.0277  \\  0.0022 \\   0.0082  \\  0.2441\\    0.5068  \\  0.0182 \\0.0009  \\  0.0021  \\  0.0014} &
\makecell{ 0.0172  \\  0.0276 \\   0.0022 \\   0.0082  \\  0.2445 \\   0.5086  \\  0.0182 \\ 0.0009 \\   0.0021  \\  0.0014} &
\makecell{0.0164 \\    0.0271 \\   0.0021  \\  0.0079  \\  0.2402 \\   0.4891  \\  0.0177 \\ 0.0009 \\   0.0021   \\ 0.0014} &

\makecell{No Significant \\ change in $\sigma_\Theta$}
\\ \hline

\end{tabular}
\end{center}
\end{table}

\begin{table}[h]
\caption{Real flight test data case-3 results using the MT and MS method. \\ No of iterations=100 }{}
\label{tbcase3QMTMS}
\begin{center}
\begin{footnotesize}
\begin{tabular}{|c| c| c| c| c| c|| c|c|c|c|c|c| }
\hline
Study &
\makecell{$\Theta$\\ (MT)} &
\makecell{$\sigma_\Theta$ \\(MT)} &
\makecell{\textbf{R} (MT)\\ $\times10^{-6}$ }&
\makecell{\textbf{Q} (MT)\\ $\times10^{-6}$}&
\makecell{\textbf{J1-J8} \\(MT) }&

\makecell{$\Theta$\\ (MS)} &
\makecell{$\sigma_\Theta$\\ (MS)} &
\makecell{\textbf{R} (MS) \\ $\times10^{-6}$}&
\makecell{\textbf{Q} (MS)\\ $\times10^{-6}$}&
\makecell{\textbf{J1-J8} \\(MS) }&
Remarks
\\ \hline


\makecell{$\mathbf{P_0}$ : Scaled up-[0,0;0,\checkmark]\\\textbf{Q} : MT/MS-[\checkmark,0;0,0] \\\textbf{R} : MT/MS-diag} &

\makecell{   4.9260 \\   0.1587 \\   0.2408 \\  -0.5285 \\ -11.8255 \\  -6.8798  \\ -0.9711 \\ -0.0424 \\   0.0002   \\ 0.2540} &
\makecell{ 0.0184 \\   0.0302 \\   0.0023 \\   0.0082  \\  0.2483  \\  0.5062 \\   0.0184 \\ 0.0009 \\   0.0011   \\ 0.0016} &
\makecell{ 1.6135   \\ 0.2395  \\  2.3155  \\  2.9290} &
\makecell{0.2025 \\   3.1532  \\  0.6666} &
\makecell{ 3.7662  \\  4.5191  \\  3.8384  \\  0.0008 \\ -43.7340  \\  4.2266  \\  4.2284 \\ 2.9489} &

\makecell{  5.0620  \\  0.3594  \\  0.2517 \\  -0.5590 \\ -12.5965 \\  -6.6713 \\  -1.0247 \\ -0.0447  \\ -0.0006   \\ 0.2635} &
\makecell{0.0323  \\  0.0508  \\  0.0027  \\  0.0055  \\  0.1400 \\   0.3021  \\  0.0129 \\ 0.0006 \\   0.0007   \\ 0.0026} &
\makecell{ 3.1599 \\  37.2424 \\   9.3413 \\ 841.5496} &
\makecell{  0.00005  \\  0.0003  \\  0.2386} &
\makecell{  3.1621   \\ 3.1507  \\  2.5900 \\   0.0007 \\ -38.0517  \\  8.4768  \\  8.4655 \\ 3.0215} &

\makecell{Cost functions are \\not close to their \\expected values in\\ MT and MS method} \\ \hline

\end{tabular}
\end{footnotesize}
\end{center}
\end{table}
\end{landscape}

\clearpage
\subsection{Case-3 Figures}

\clearpage
\section{Real Flight Test Case-4}
\par
\begin{figure}[h]
\includegraphics[width=6in,height=3.2in]{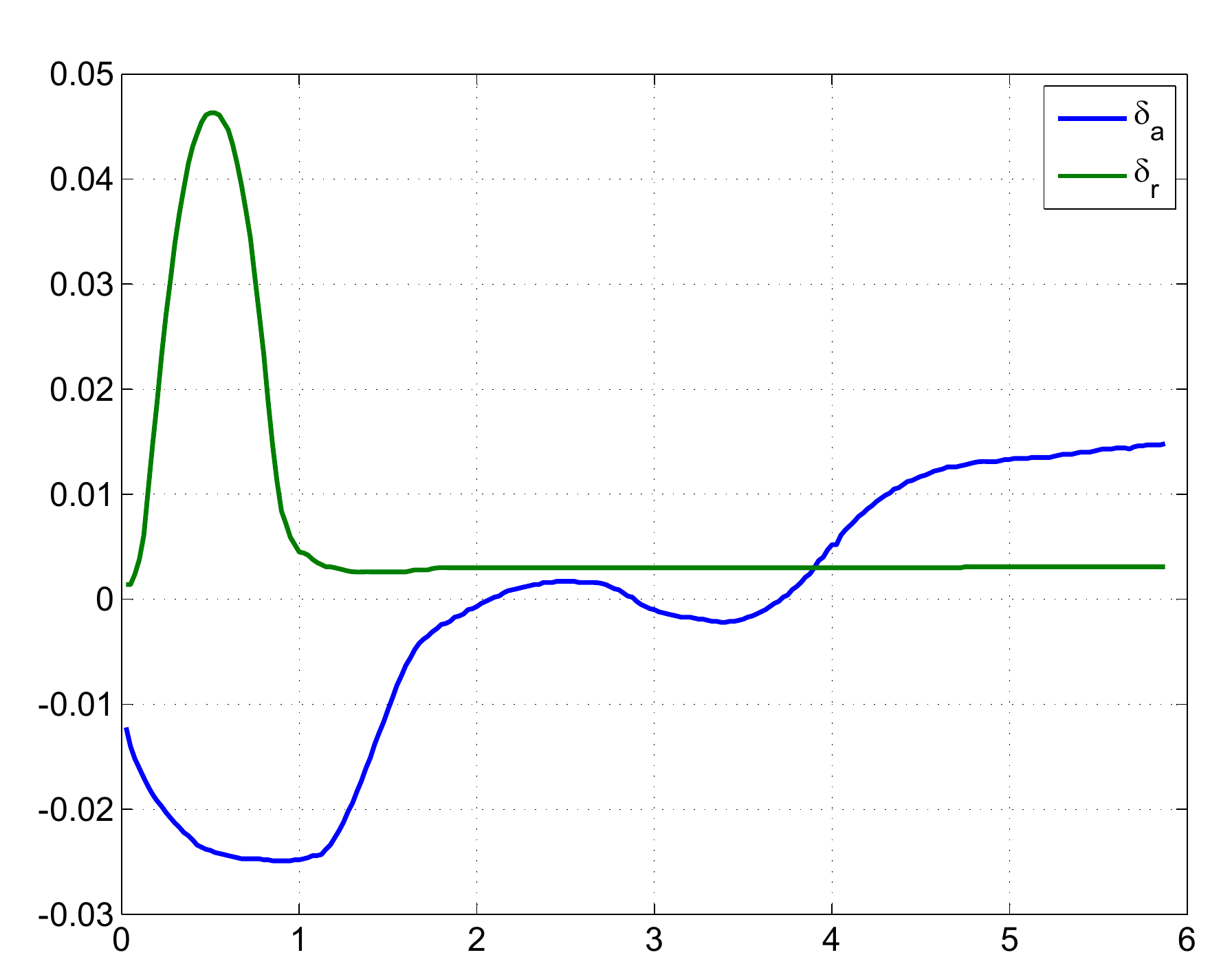}
\caption{Control input versus time (s)}
\label{input4}
\end{figure}

 The data set is obtained from NASA TN D-7831 (Maine \cite{Maine1975} 1975) which pertains to a high speed (4665.0 ft/s) flight in lateral motion excited by a rudder ($\delta_r$ in rad) and aileron ($\delta_a$ in rad) pulse as shown in Fig. \ref{input4}. The state equations ($n=4$) for the sideslip ($\beta$), roll rate (p), roll angle ($\phi$) and yaw rate (r) respectively are
\begin{align*}
\dot{\beta}&=Y_{\beta}\beta+Y_pp+Y_\phi\phi-r+Y_{\delta_a}\delta_a +Y_{\delta_r}\delta_r+Y_0\\
\dot{p}&=L_{\beta}\beta+L_pp+L_rr+L_{\delta_a}\delta_a +L_{\delta_r}\delta_r+L_0\\
\dot\phi&=p+\phi_0\\
\dot{r}&=N_{\beta}\beta+N_pp+N_rr+N_{\delta_a}\delta_a +N_{\delta_r}\delta_r+N_0
\end{align*}
The angle of sideslip ($\beta$), roll rate (p), roll angle ($\phi$), yaw rate (r), lateral acceleration ($a_y$), roll acceleration ($\dot p$) and yaw acceleration ($\dot r$) are measured (indicated with subscript `m') in units of rad, rad/s, rad, rad/s, $ft/s^2$, $rad/s^2$ and $rad/s^2$ respectively. The measurement equations ($m$=7) are given by
\begin{align*}
{\beta_m}&=\beta \\
{p_m}&=p\\
{\phi_m}&=\phi\\
{r_m}&=r\\
{a_{y_m}}&=\frac{U_0}{g}(Y_{\beta}\beta+Y_{\delta_a}\delta_a +Y_{\delta_r}\delta_r+Y_0)+a_{y_b} \\
\dot{p_m}&=L_{\beta}\beta+L_pp+L_rr+L_{\delta_a}\delta_a +L_{\delta_r}\delta_r+p_b\\
\dot{r_m}&=N_{\beta}\beta+N_pp+N_rr+N_{\delta_a}\delta_a +N_{\delta_r}\delta_r+r_b
\end{align*}

The unknown parameter set ($p=21$) is $\theta=(Y_\beta,Y_p,Y_{\delta_a},Y_{\delta_r},Y_0,L_\beta,L_p,L_r,L_{\delta_a},L_{\delta_r},L_0,\\ \phi_0,N_\beta,N_p,N_r,N_{\delta_a},N_{\delta_r},N_0,a_{y_b},p_b,r_b)^T$. The ones with suffix `$\delta_a$' or `$\delta_r$' are the control derivatives, the ones with suffix zero or b are the biases and all others are aerodynamic derivatives. The initial states are taken as initial measurement and the initial parameter values are taken as $(-0.04,0.01,0.0001,0.01,0.0001,-16.8,-0.24,0.4,12,20,0.0001,\\0.0001,1.5,-0.003,-0.04,0.35,-2.5,0.0001,0.2,0.01,0.001)^T$.

\begin{table}[h]
\begin{center}
\caption*{Other constant values used for case-4}{}
\begin{tabular}{| c | c | c | c | c | c | }
\hline
$Y_\phi=0.0069$ & $U_0=4665$ & g=32.2  \\ \hline
\end{tabular}
\end{center}
\end{table}

\newpage
\par
\begin{figure}[h]
\includegraphics[width=6in,height=3.2in]{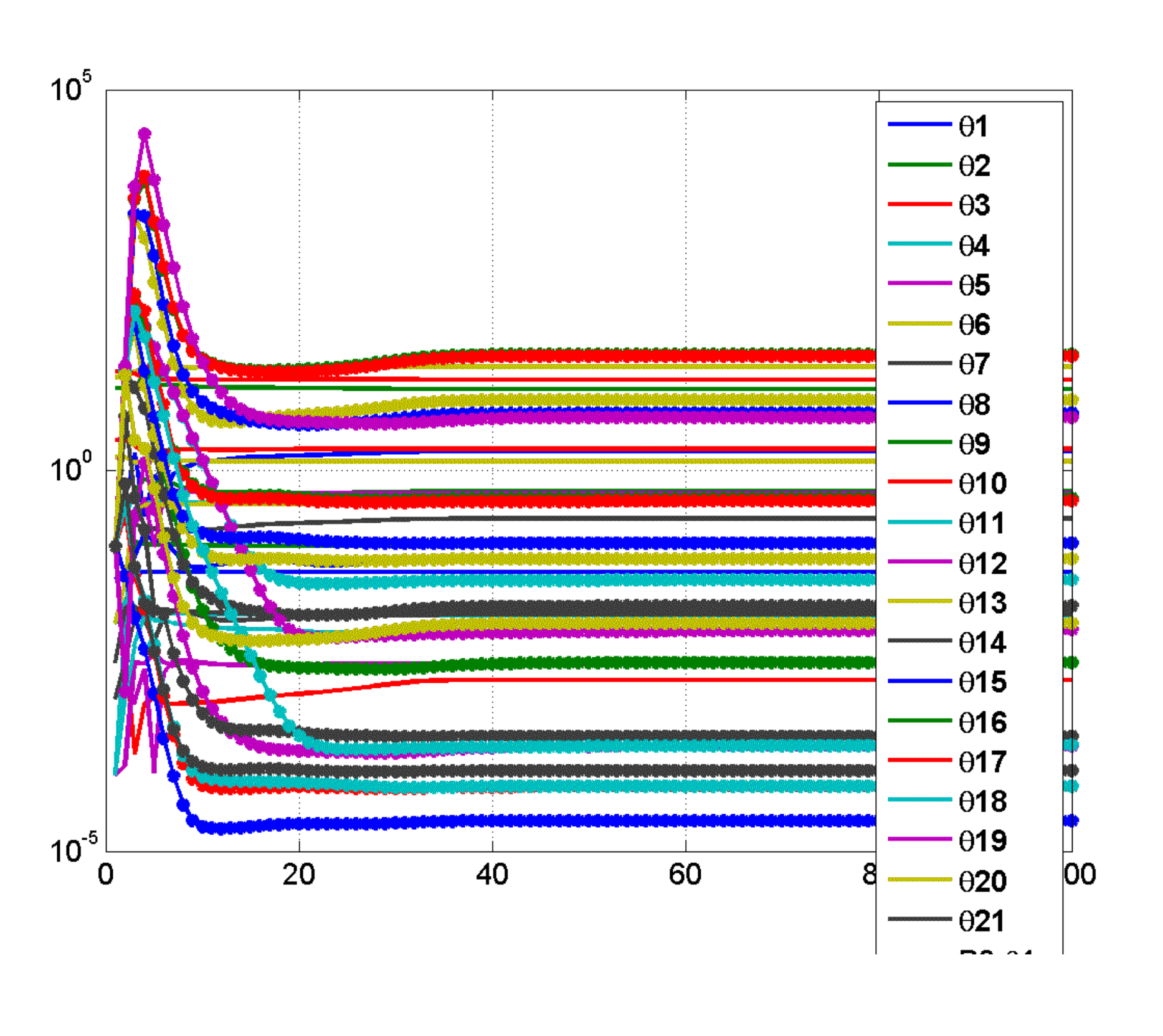}
\caption{Variation of parameter and its initial covariance ($\mathbf{P_0}$) with iterations}
\label{realQ4_P0}
\end{figure}

\begin{figure}[h]
\includegraphics[width=6in,height=4in]{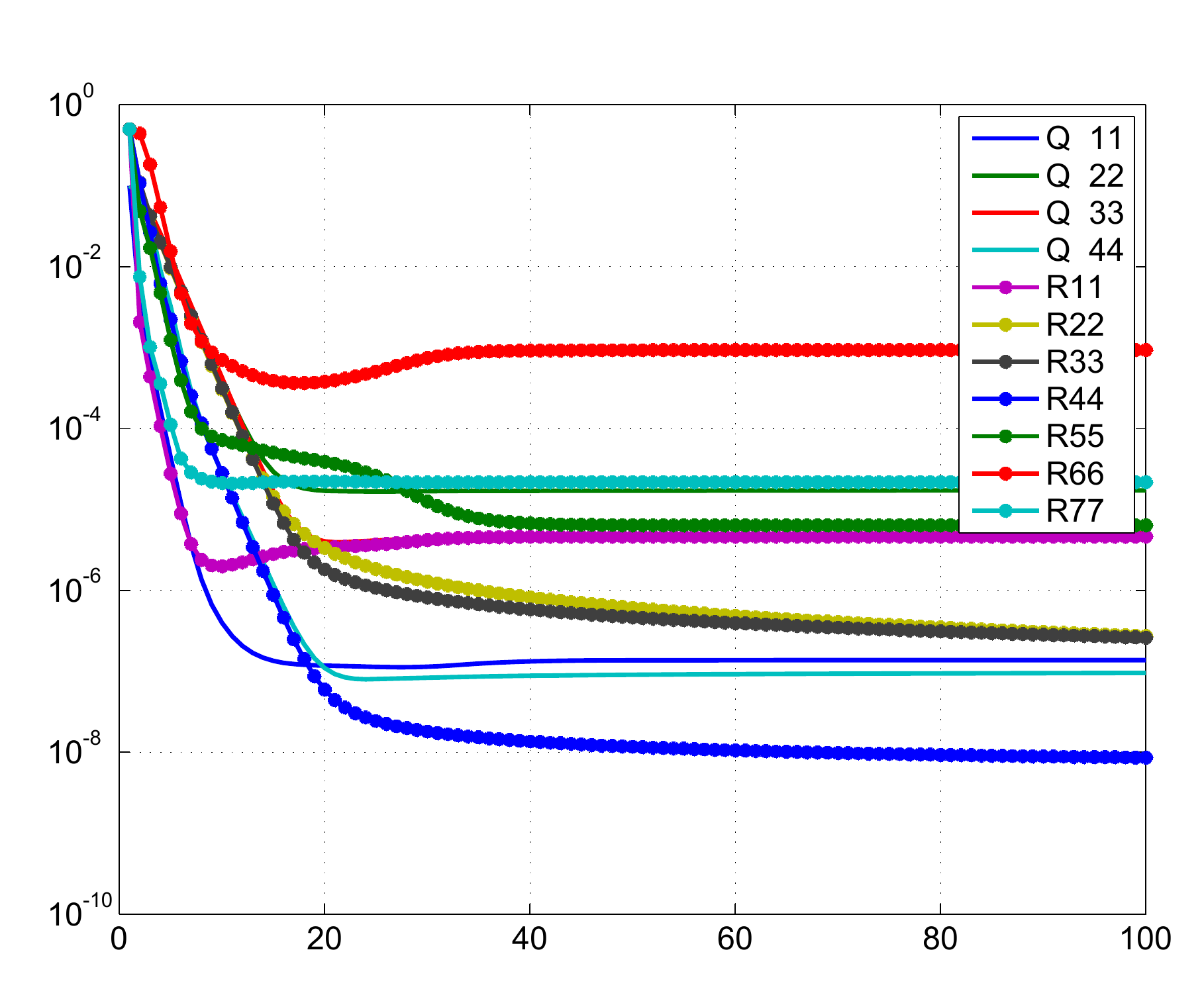}
\caption{Variation of \textbf{Q} and \textbf{R} with iterations}
\label{realQ4_R}
\end{figure}

\begin{figure}[h]
\includegraphics[width=6in,height=4in]{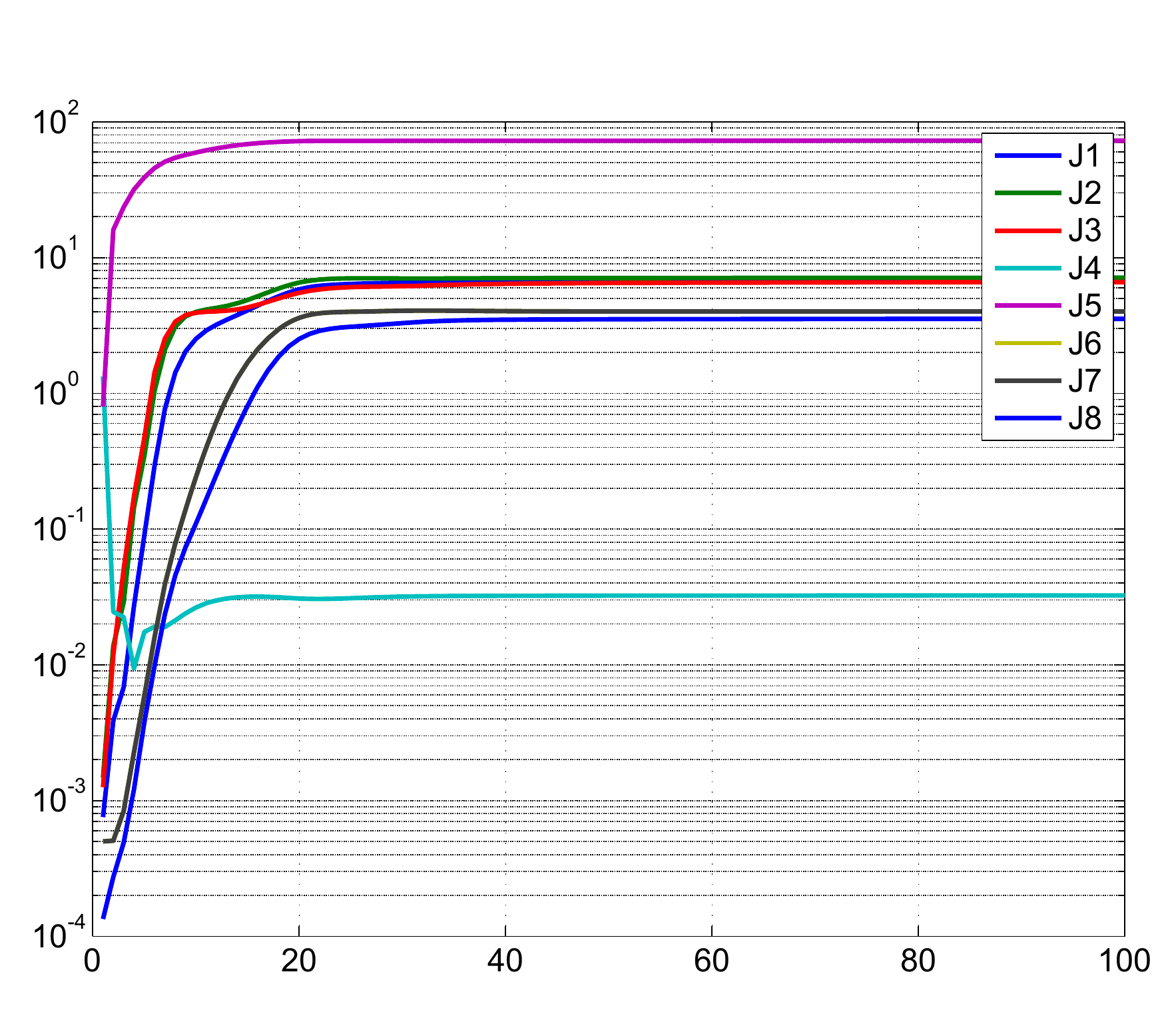}
\caption{Variation of different costs (\textbf{J1-J8}) with iterations}
\label{realQ4_J}
\end{figure}

\begin{figure}[h]
\includegraphics[width=6in,height=4in]{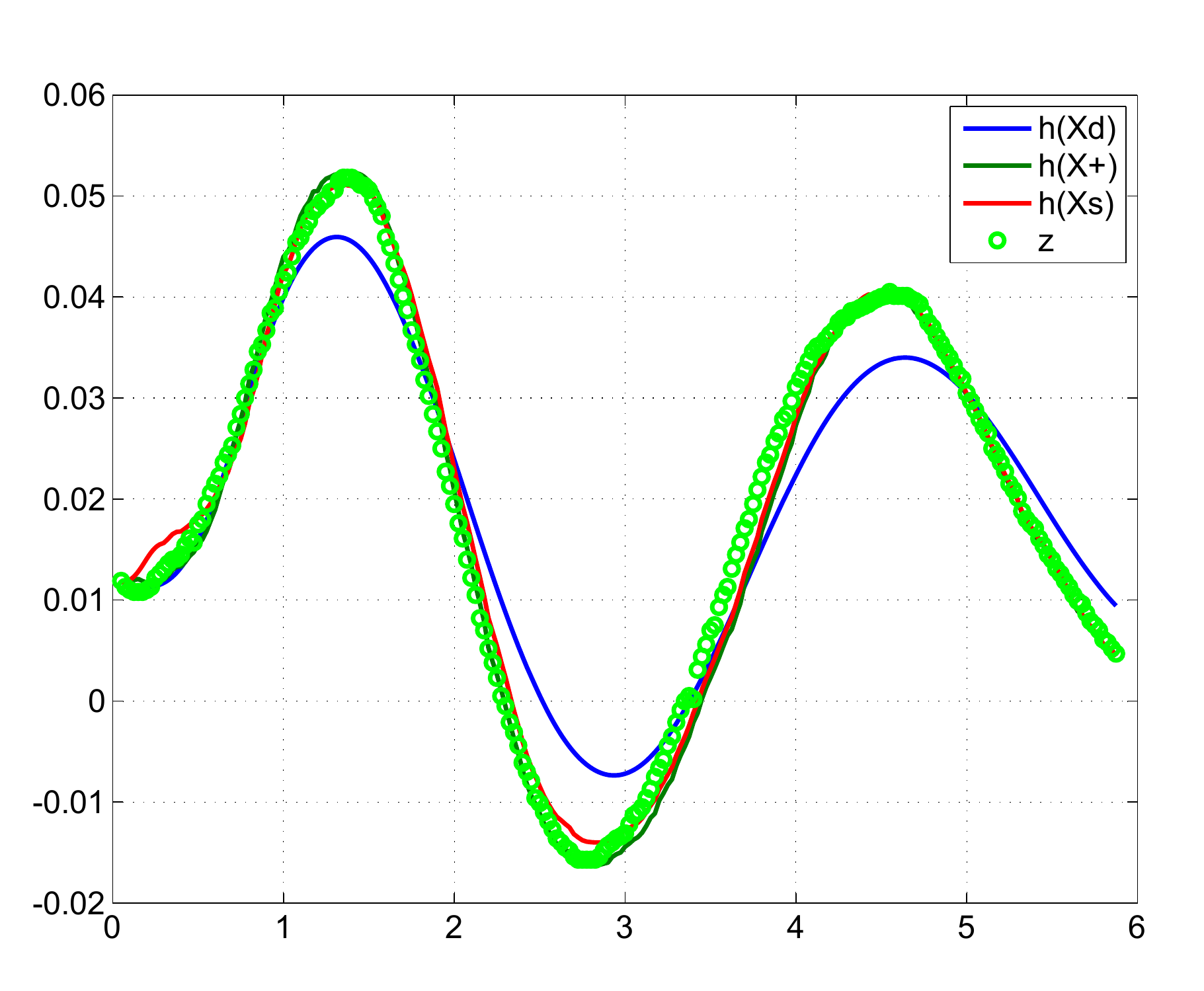}
\caption{Comparison of the predicted dynamics, posterior, smoothed}
\caption*{and the measurement 1}
\label{realQ4_s1}
\end{figure}

\begin{figure}[h]
\includegraphics[width=6in,height=4in]{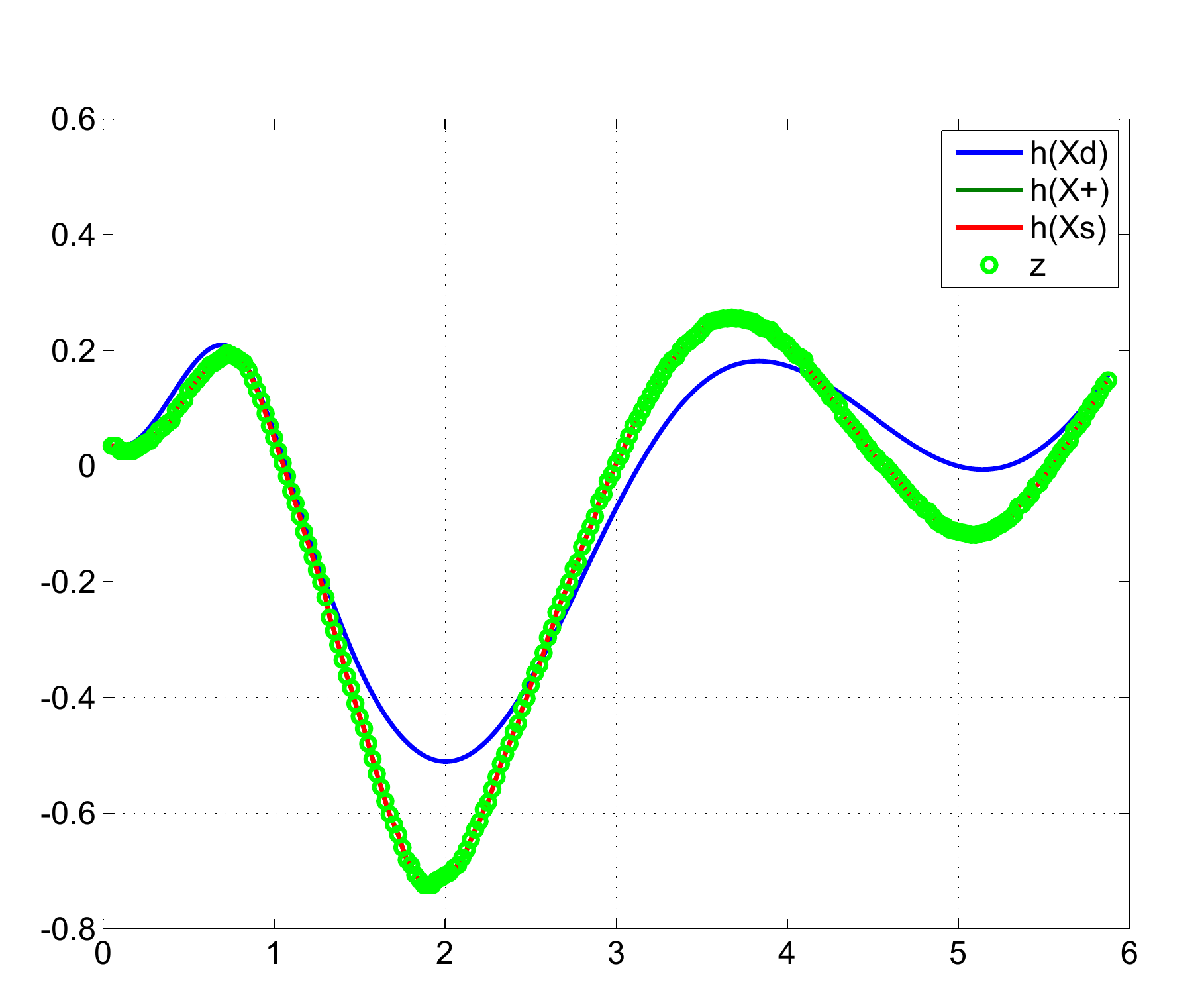}
\caption{Comparison of the predicted dynamics, posterior, smoothed}
\caption*{and the measurement 2}
\label{realQ4_s2}
\end{figure}

\begin{figure}[h]
\includegraphics[width=6in,height=4in]{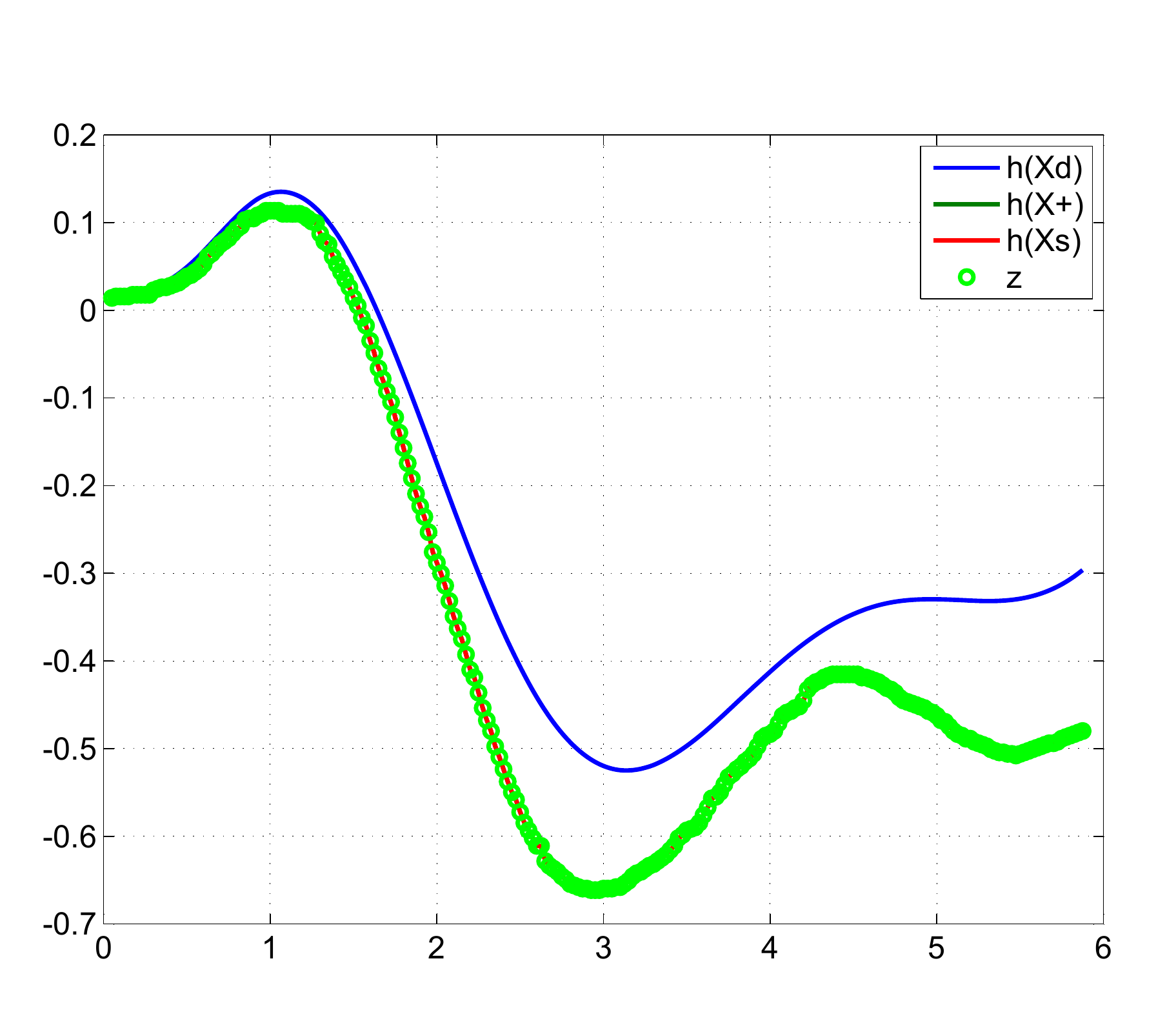}
\caption{Comparison of the predicted dynamics, posterior, smoothed}
\caption*{and the measurement 3}
\label{realQ4_s3}
\end{figure}

\begin{figure}[h]
\includegraphics[width=6in,height=4in]{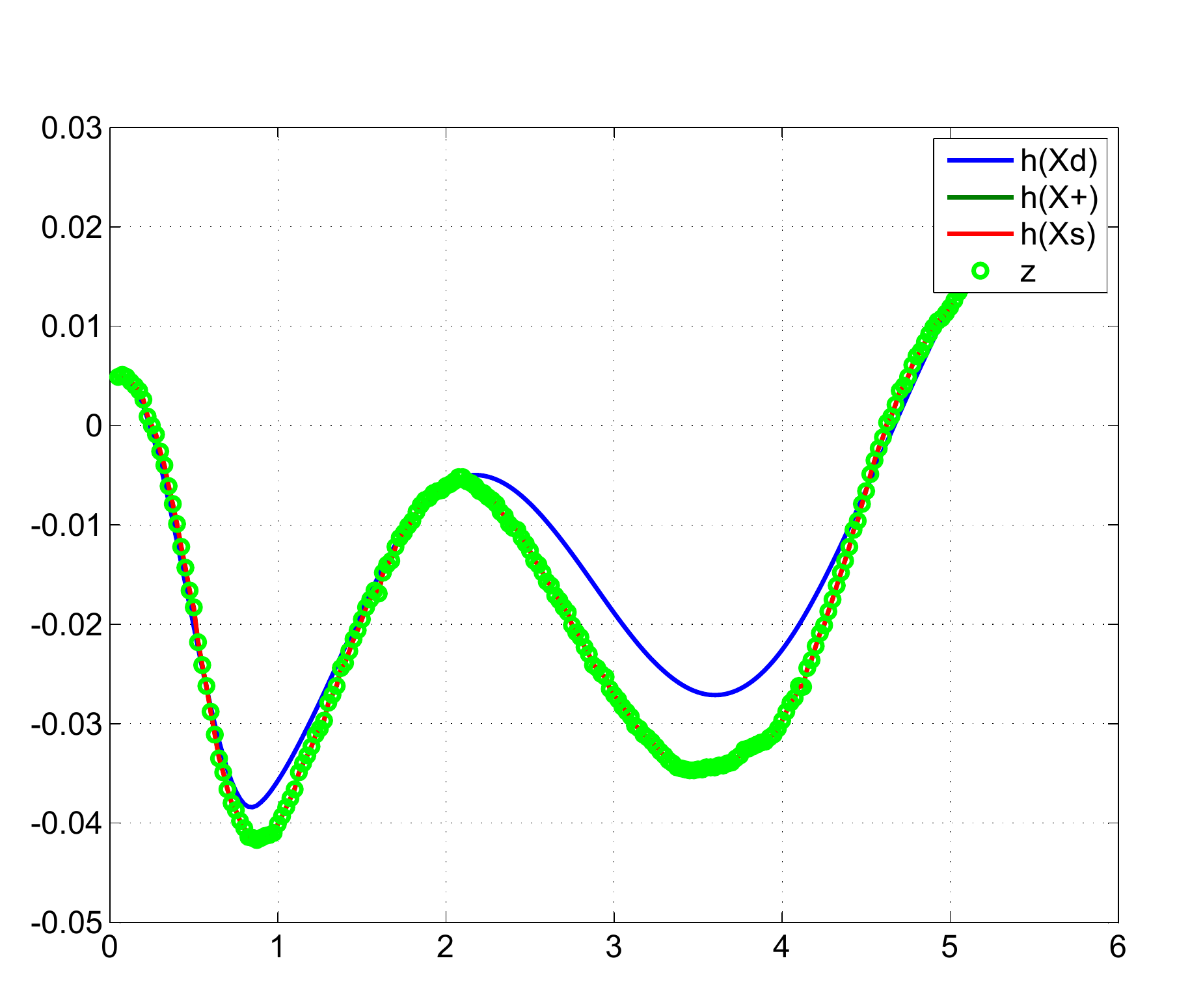}
\caption{Comparison of the predicted dynamics, posterior, smoothed}
\caption*{and the measurement 4}
\label{realQ4_s4}
\end{figure}

\begin{figure}[h]
\includegraphics[width=6in,height=4in]{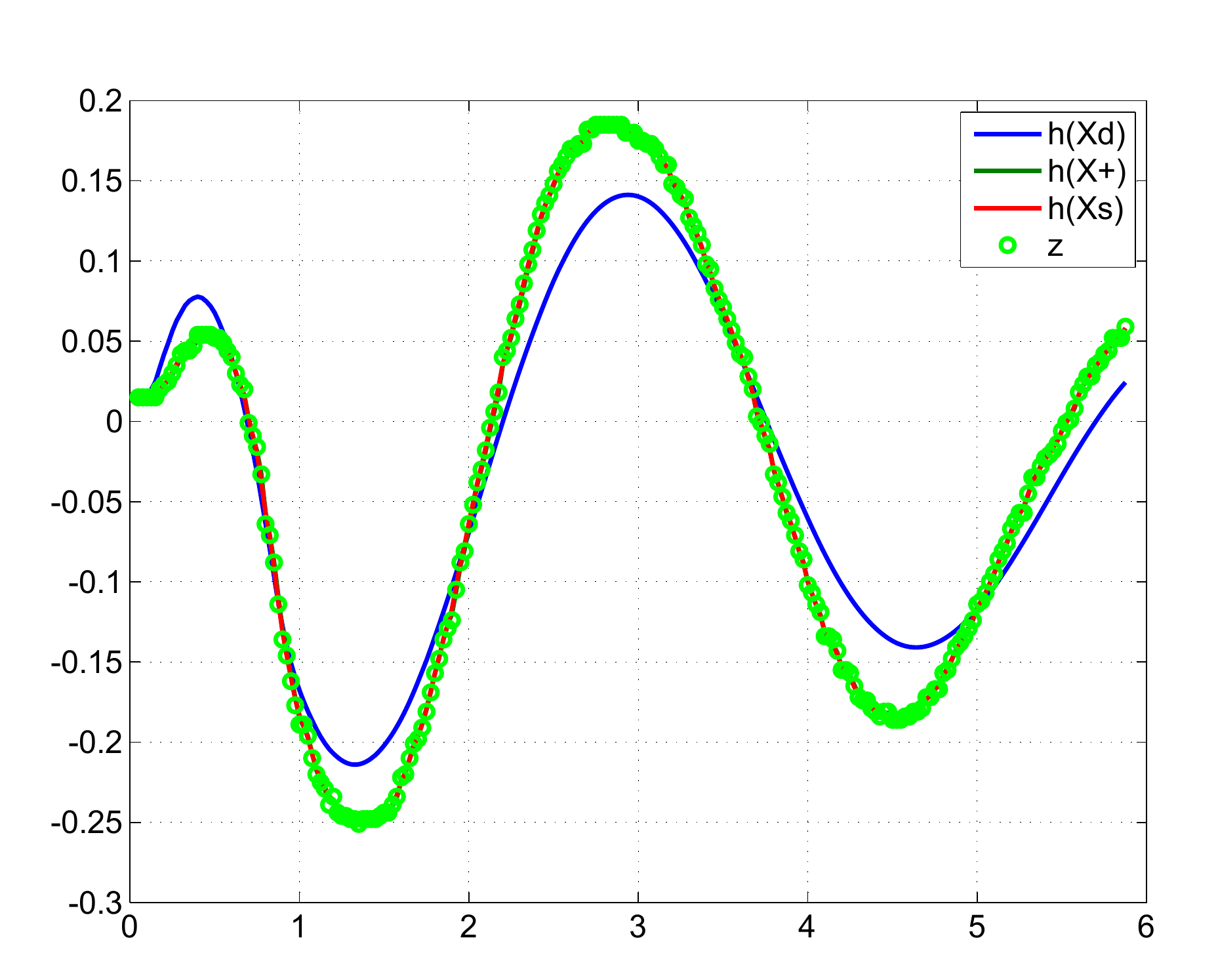}
\caption{Comparison of the predicted dynamics, posterior, smoothed}
\caption*{and the measurement 5}
\label{realQ4_h5}
\end{figure}

\begin{figure}[h]
\includegraphics[width=6in,height=4in]{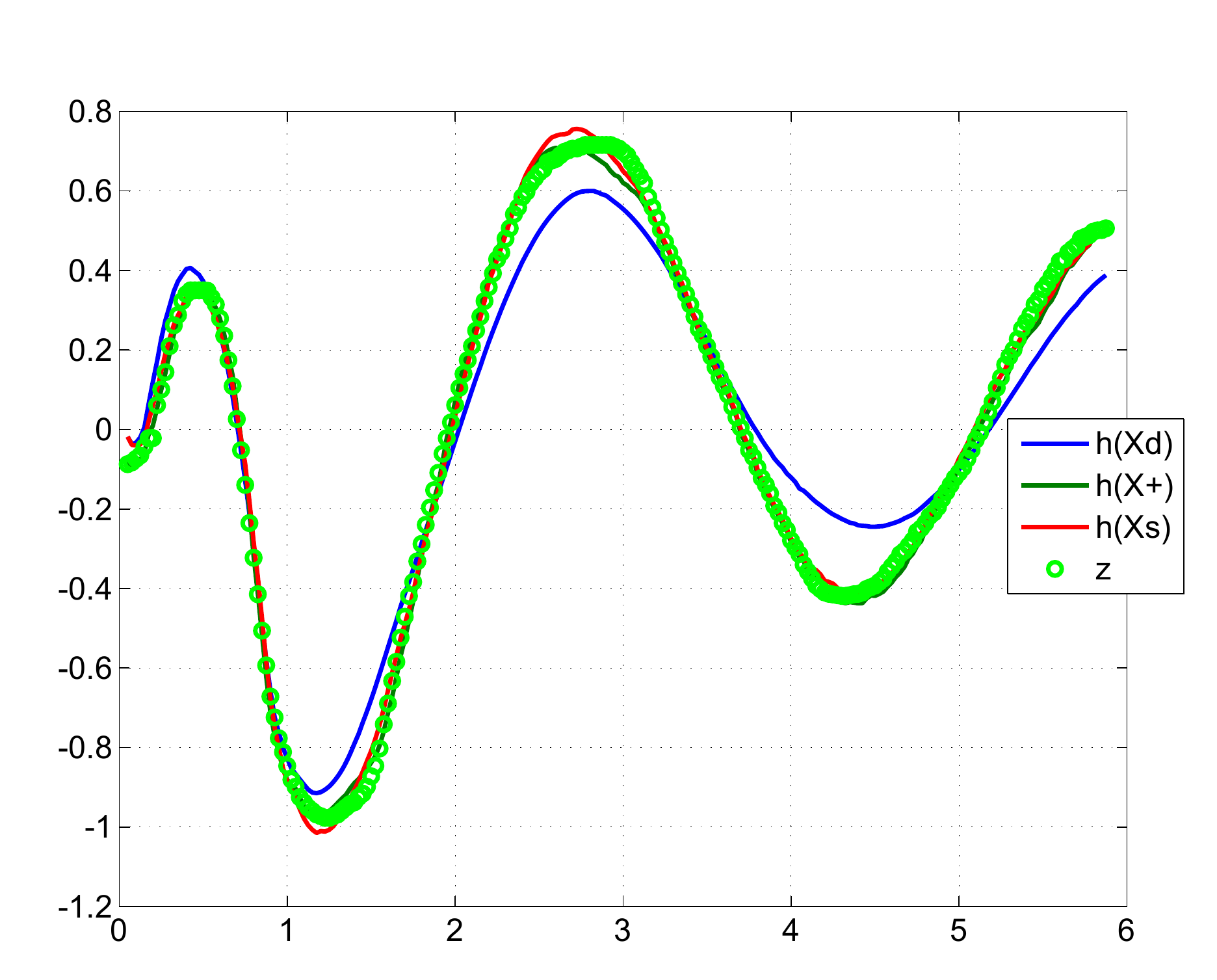}
\caption{Comparison of the predicted dynamics, posterior, smoothed}
\caption*{and the measurement 6}
\label{realQ4_h6}
\end{figure}

\begin{figure}[h]
\includegraphics[width=6in,height=4in]{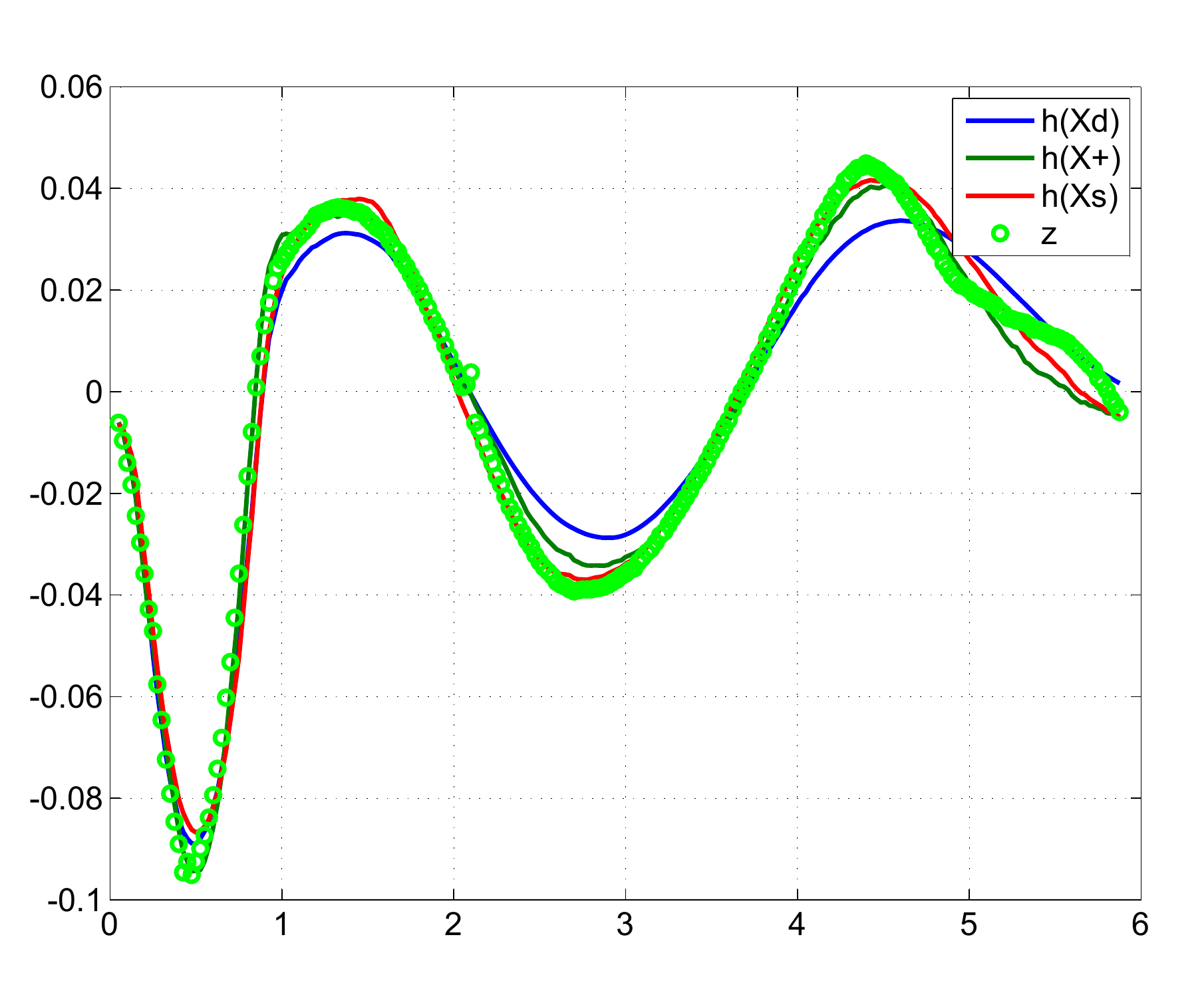}
\caption{Comparison of the predicted dynamics, posterior, smoothed}
\caption*{and the measurement 7}
\label{realQ4_h7}
\end{figure}

\begin{figure}[h]
\includegraphics[width=6in,height=4in]{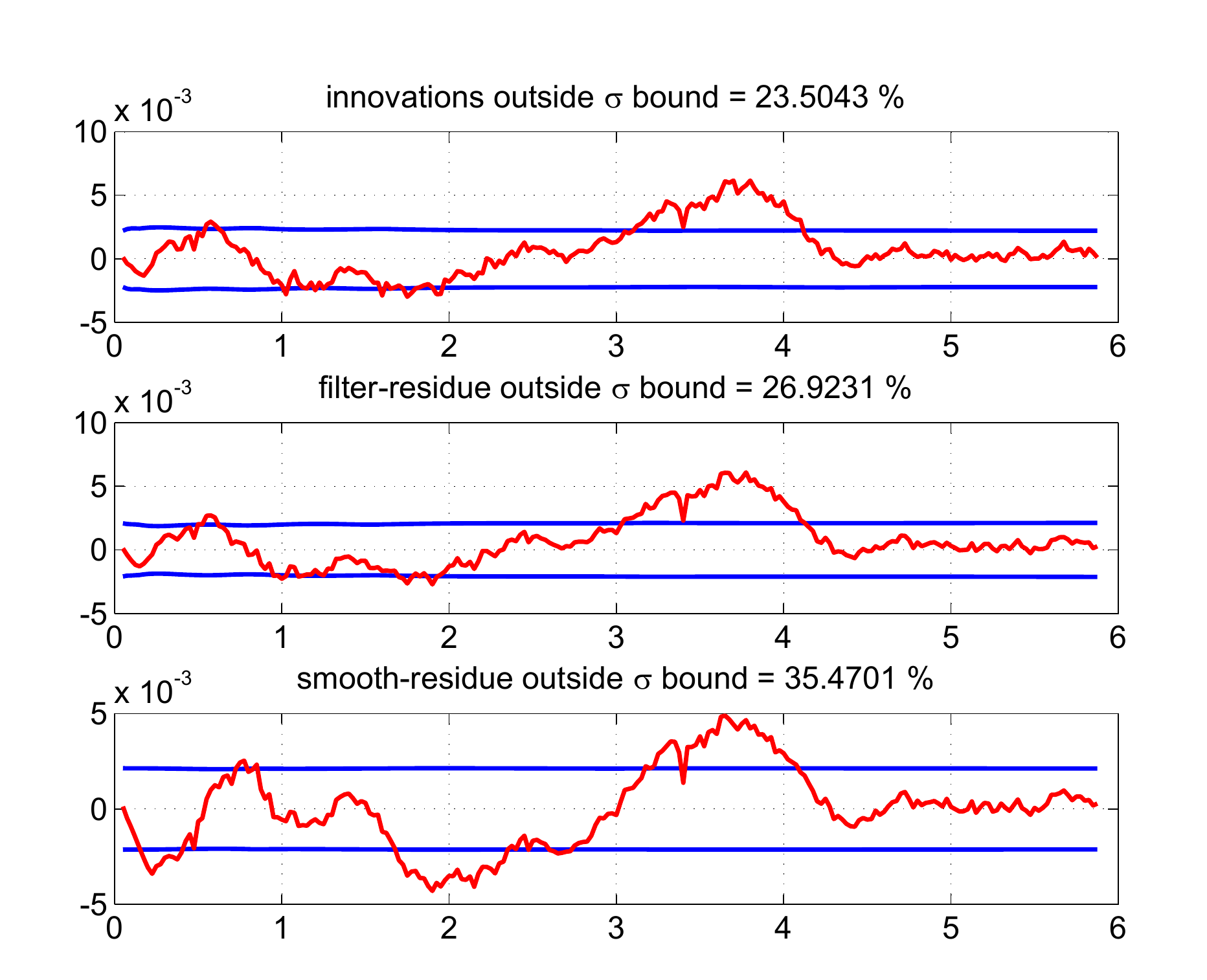}
\caption{The innovations, filtered residue and smoothed residue of measurement 1}
\label{realQ4_innov1}
\end{figure}

\begin{figure}[h]
\includegraphics[width=6in,height=4in]{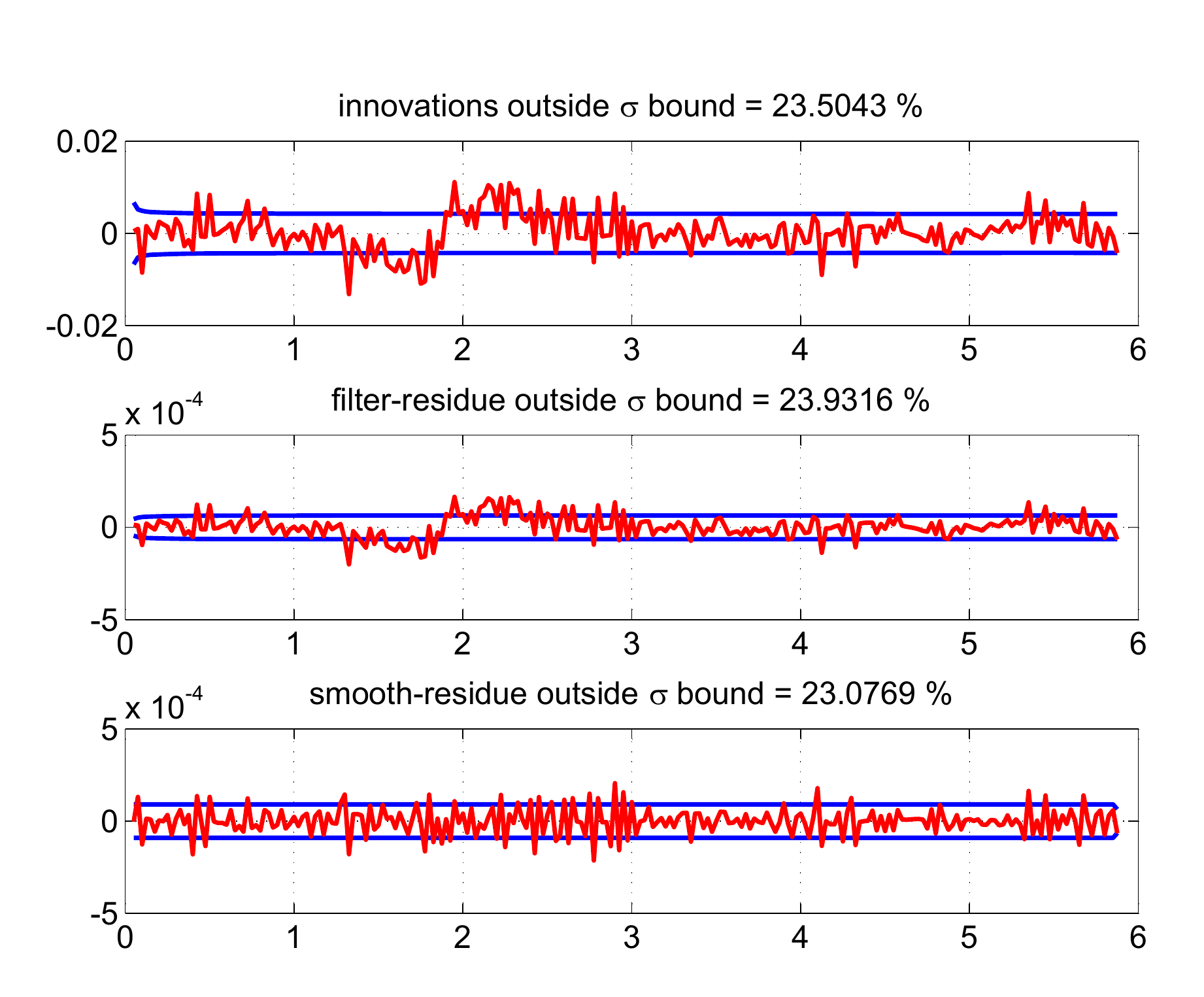}
\caption{The innovations, filtered residue and smoothed residue of measurement 2}
\label{realQ4_innov2}
\end{figure}

\begin{figure}[h]
\includegraphics[width=6in,height=4in]{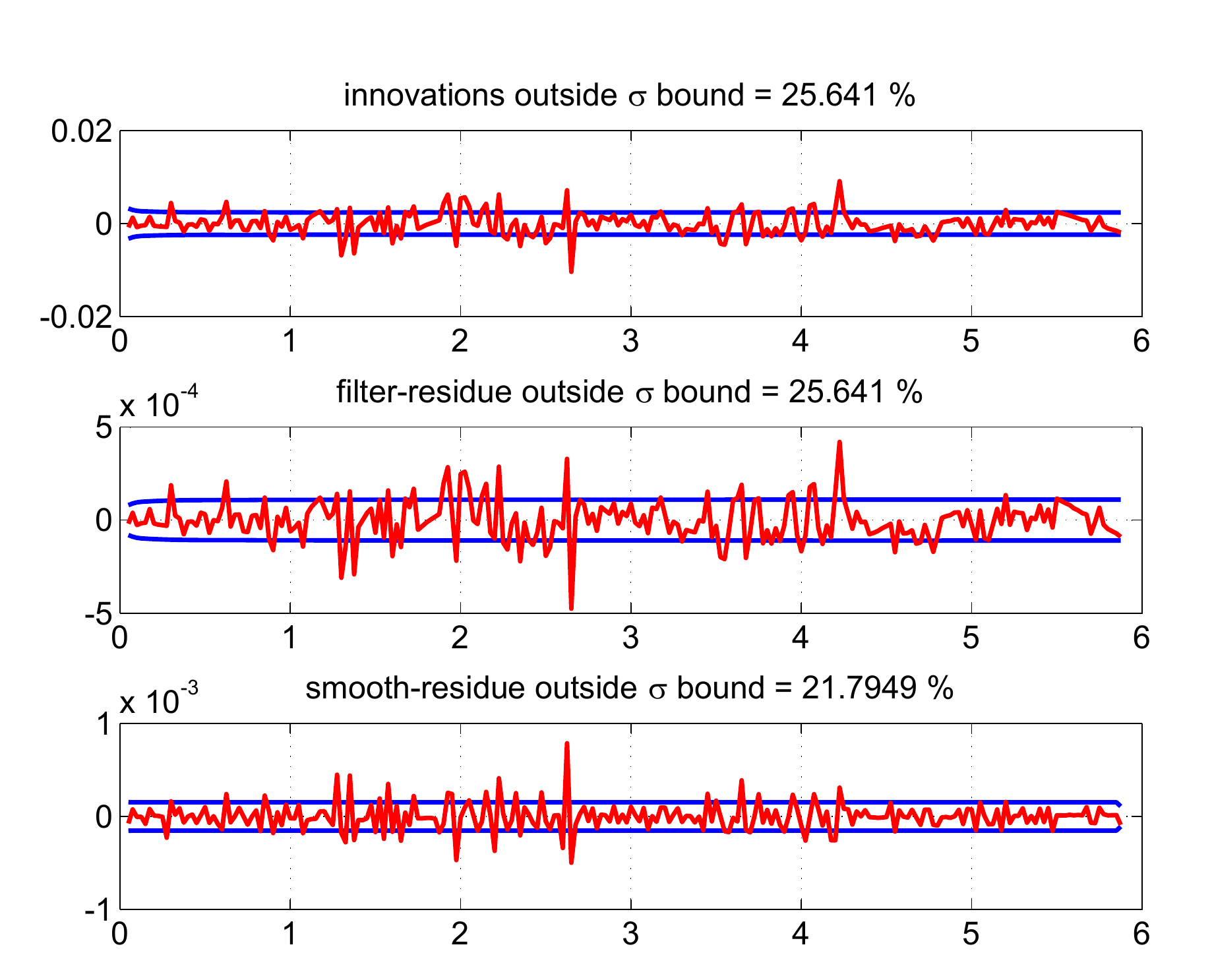}
\caption{The innovations, filtered residue and smoothed residue of measurement 3}
\label{realQ4_innov3}
\end{figure}

\begin{figure}[h]
\includegraphics[width=6in,height=4in]{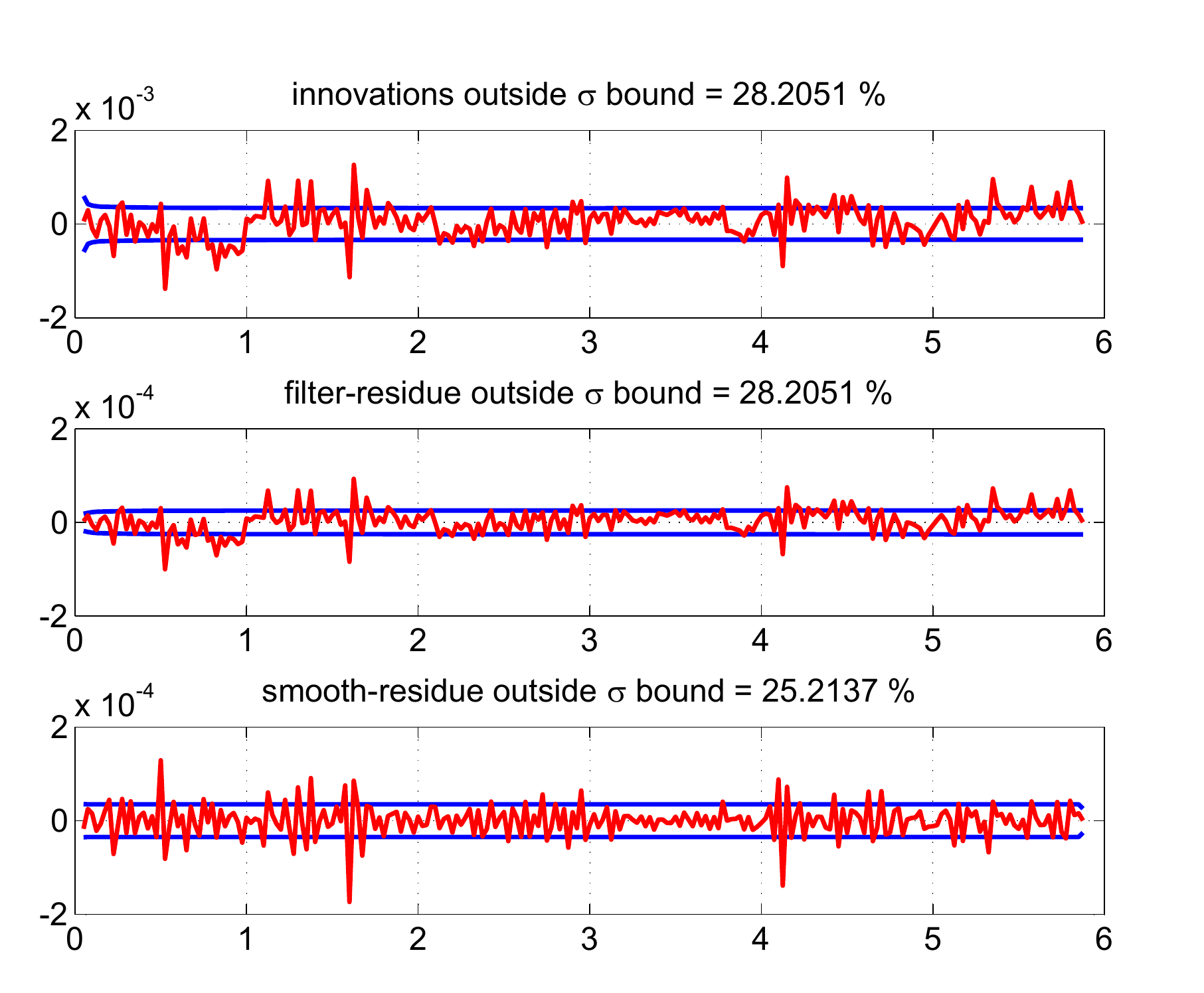}
\caption{The innovations, filtered residue and smoothed residue of measurement 4}
\label{realQ4_innov4}
\end{figure}

\begin{figure}[h]
\includegraphics[width=6in,height=4in]{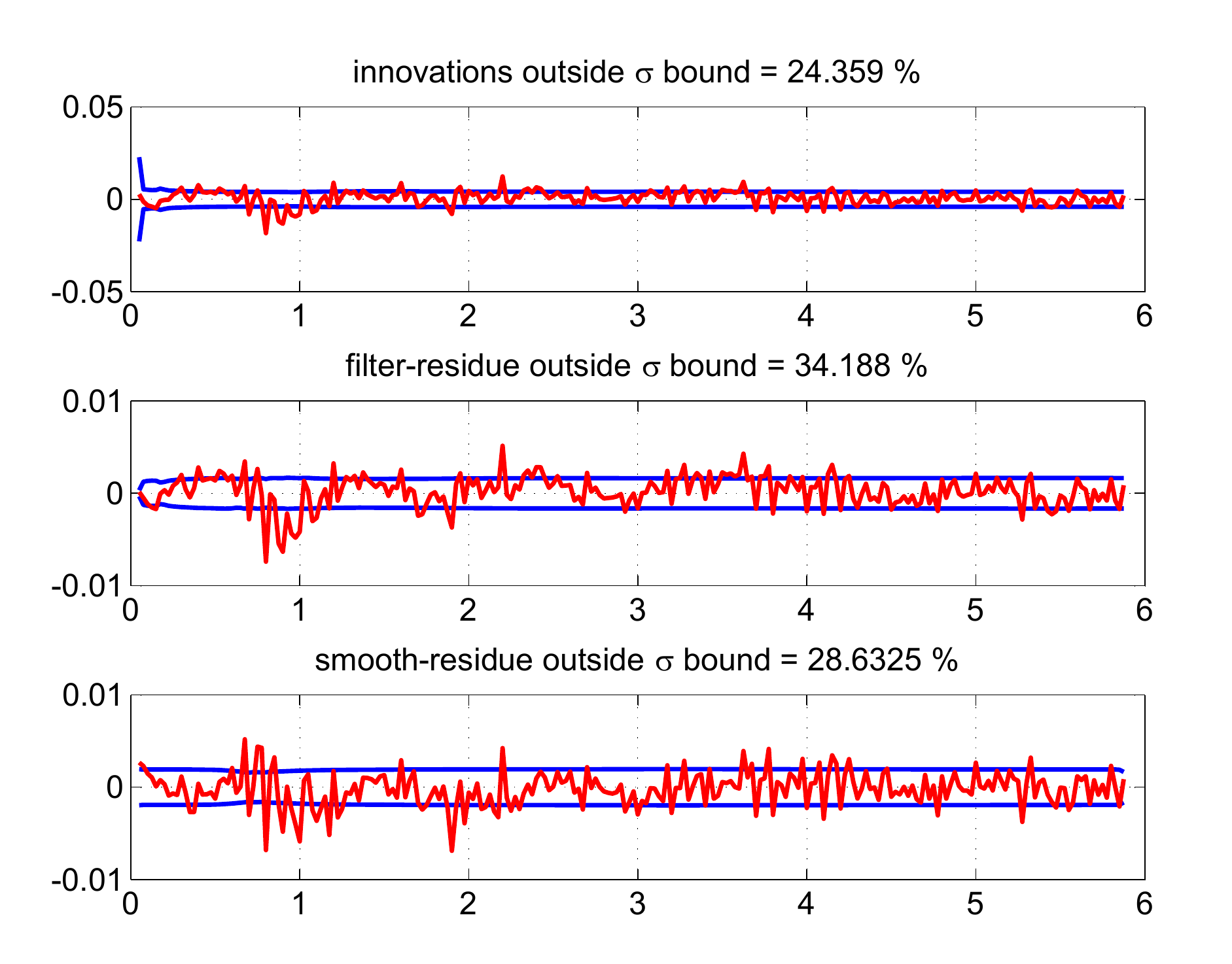}
\caption{The innovations, filtered residue and smoothed residue of measurement 5}
\label{realQ4_innov5}
\end{figure}

\begin{figure}[h]
\includegraphics[width=6in,height=4in]{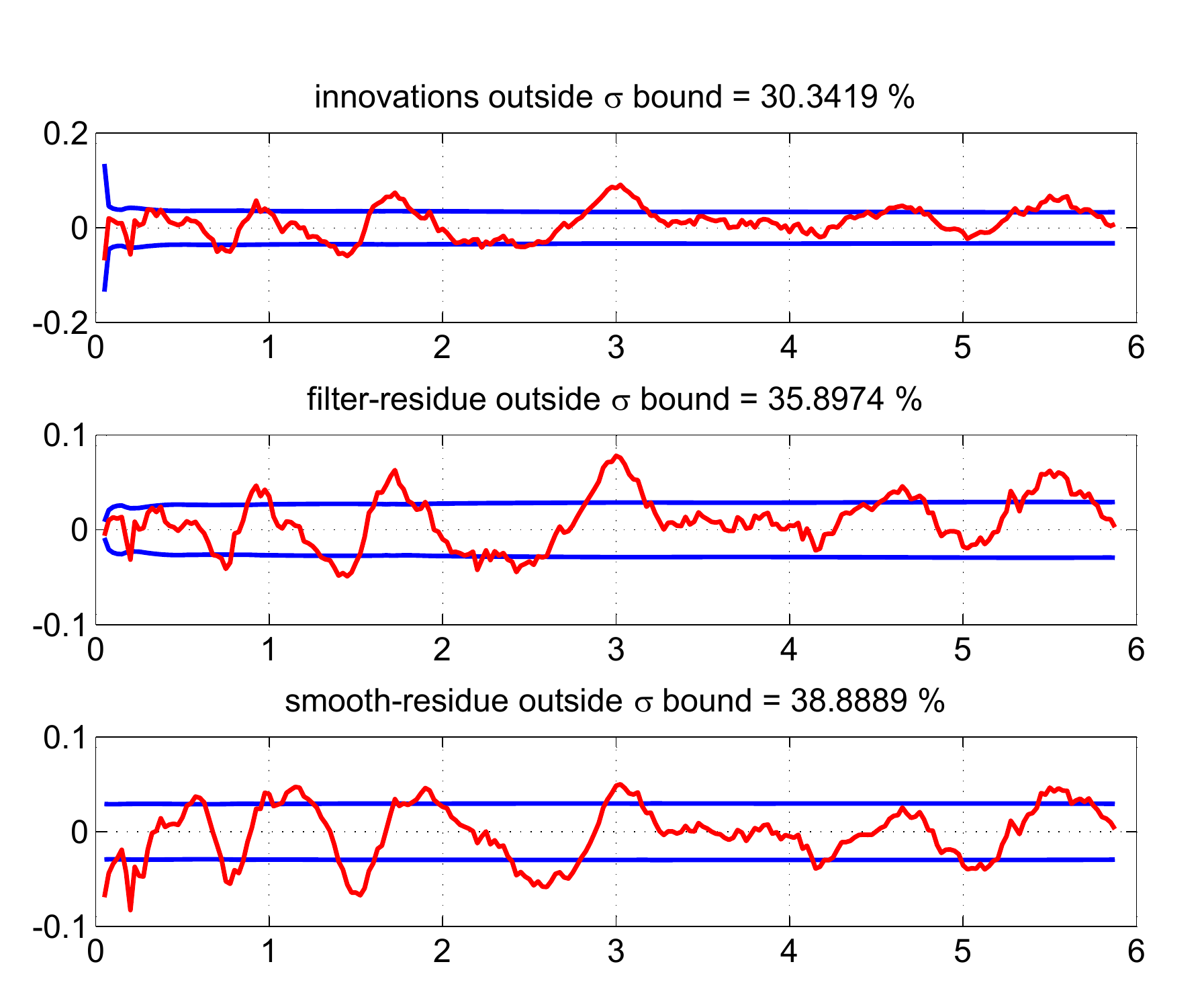}
\caption{The innovations, filtered residue and smoothed residue of measurement 6}
\label{realQ4_innov6}
\end{figure}

\begin{figure}[h]
\includegraphics[width=6in,height=4in]{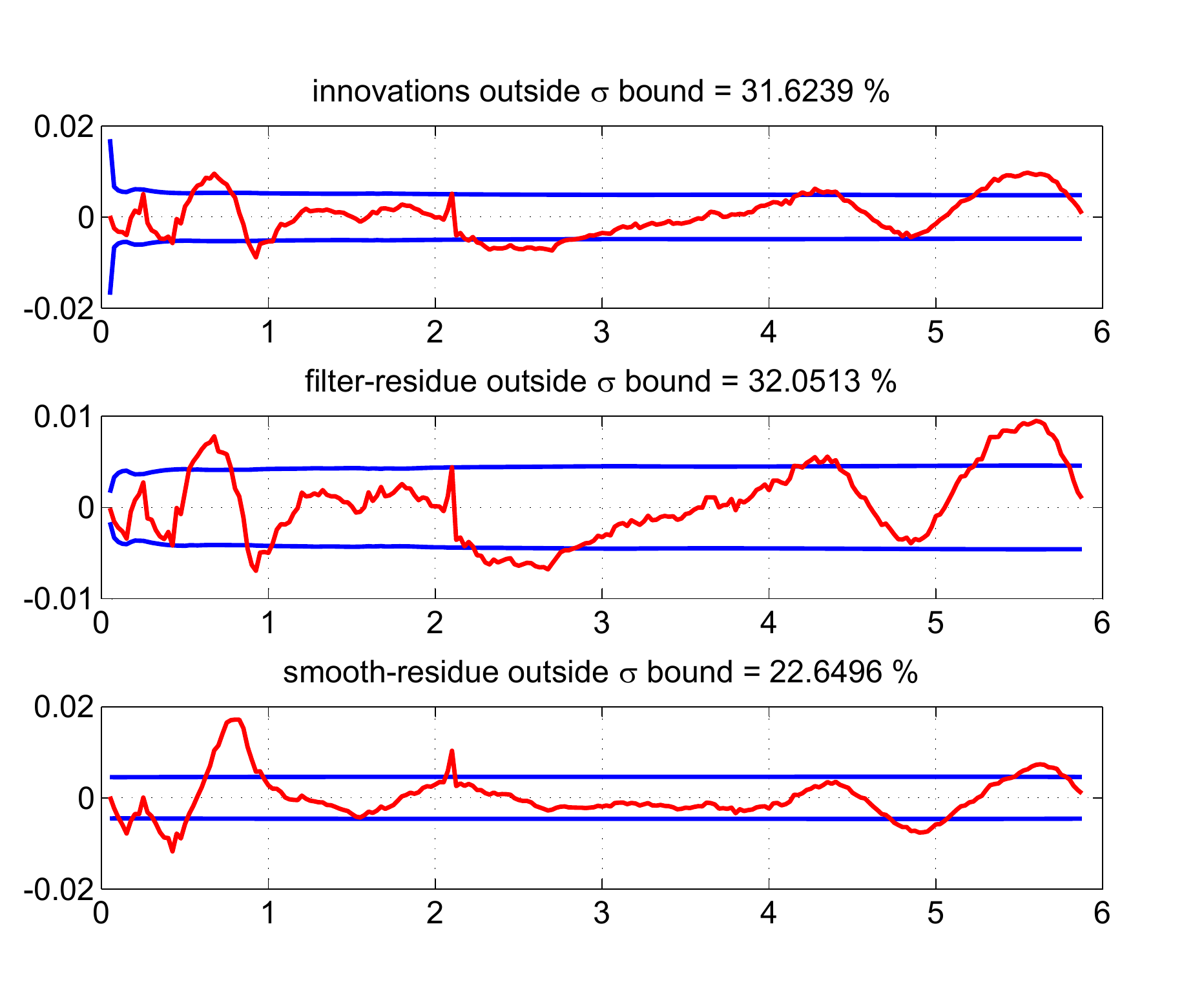}
\caption{The innovations, filtered residue and smoothed residue of measurement 7}
\label{realQ4_innov7}
\end{figure}

\begin{figure}[h]
\includegraphics[width=6in,height=4in]{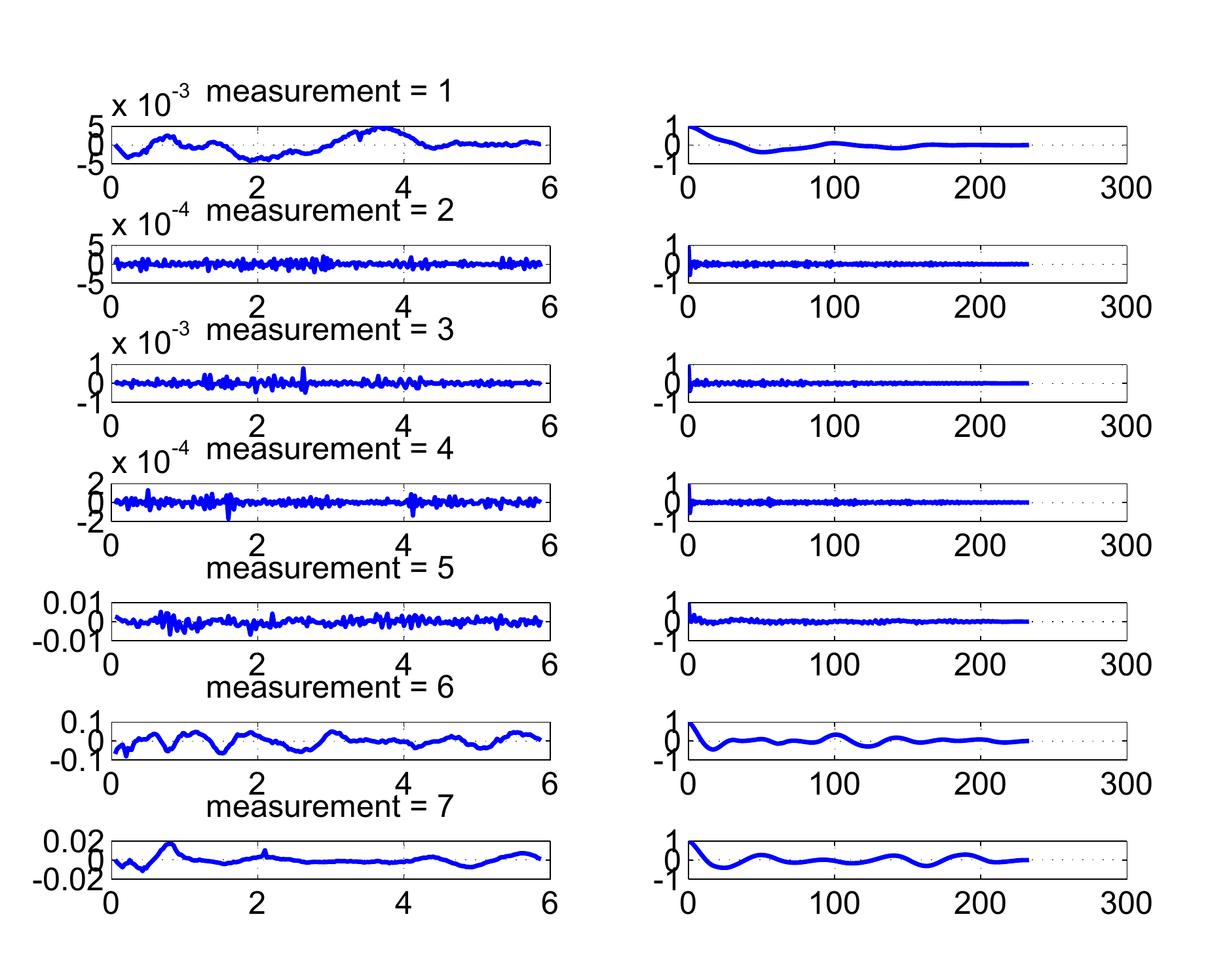}
\caption{Time variation of estimated measurement noise (left) and}
\caption*{their autocorrelation (right)}
\label{realQ4_mnoise}
\end{figure}

\begin{figure}[h]
\includegraphics[width=6in,height=4in]{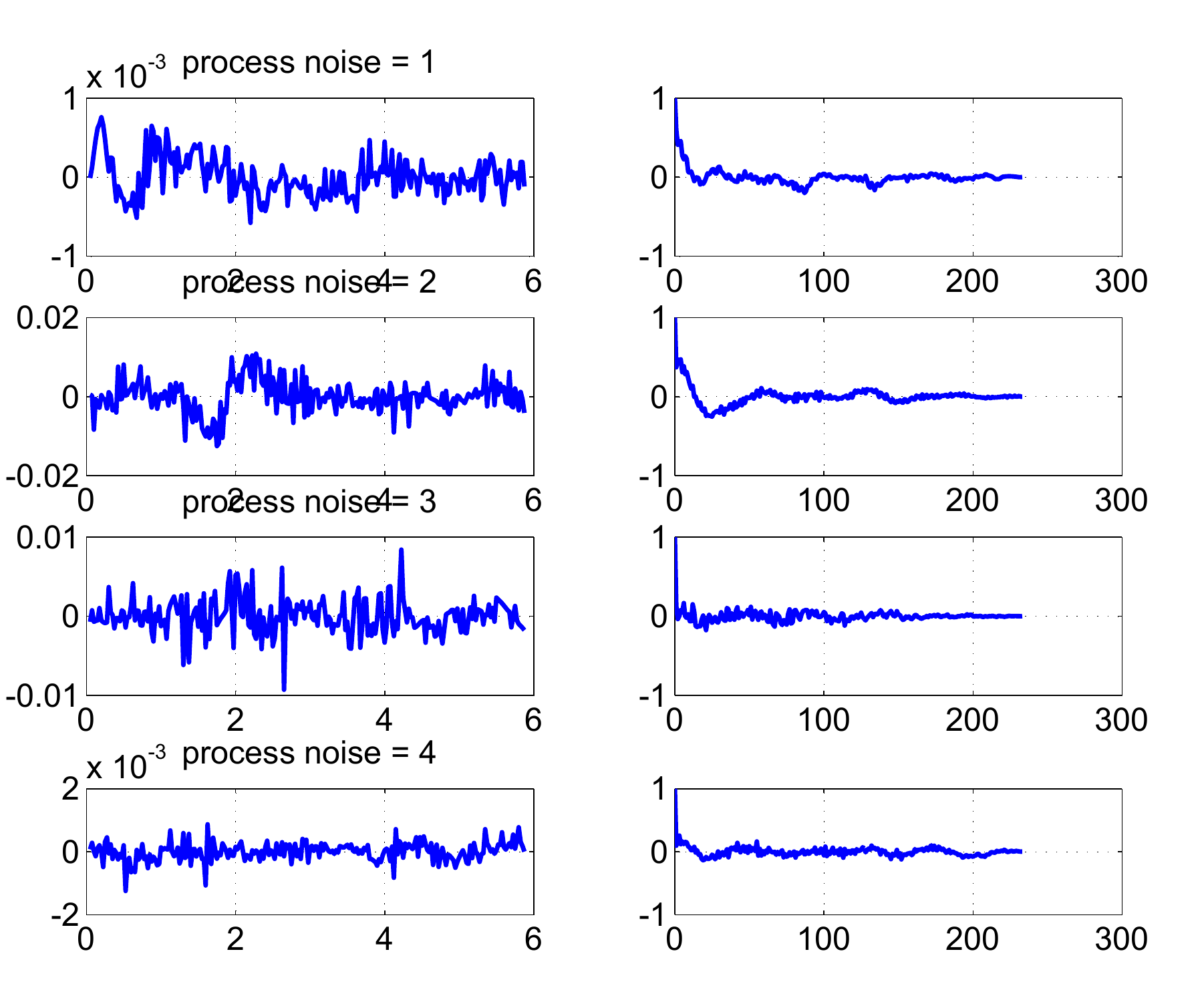}
\caption{Time variation of estimated process noise (left) and}
\caption*{their autocorrelation (right)}
\label{realQ4_pnoise}
\end{figure}

 Case-4 real data is run using the reference EKF (\textbf{Q} $>$ 0) with 100 iterations. The Fig. \ref{input4} is the control input. The Fig. \ref{realQ4_P0} shows the variation of parameter estimates and its initial covariance $\mathbf{P_0}$ with iterations and a similar Fig. \ref{realQ4_R} for \textbf{Q} and \textbf{R}. The values of \textbf{J1-J3} are close to the number of measurements ($m=7$) with \textbf{J6-J8} are close to the number of states ($n=4$) as shown in Fig. \ref{realQ4_J} and Table-\ref{tbcase4Q}. This means the measurement and state equations are well balanced. The \textbf{J5} is the negative log likelihood cost function. The later Fig. \ref{realQ4_s1}-\ref{realQ4_h7} compares (i) the state dynamics based on the estimated parameter after the filter pass through the data, (ii) the state after measurement update, (iii) the smoothed state and (iv) the measurement.  The Fig. \ref{realQ4_innov1}-\ref{realQ4_innov7} shows the confidence in the innovations, filtered residue and smoothed residue. The estimated measurement and process noise do not appear to have constant statistical characteristics across time as seen in Fig. \ref{realQ4_mnoise} and Fig. \ref{realQ4_pnoise}. Another experiment was carried out by generating a typical data set by using the estimated theta and injecting the estimated \textbf{Q} and \textbf{R} as additive white Gaussian noise. This is to determine the effect of non additive, non White and non Gaussian noise distribution in the real data on the CRBs. After each iteration in the reference recipe the $\Theta$, \textbf{Q} and \textbf{R} were reset as from the real data. Similar experiment was also conducted by updating $\Theta$ as well. It turned out that there is not much of a difference in the final CRBs as can be seen from the Table-\ref{tbnew4}.

\par Finally two other filter runs were carried out using the MT and MS statistics for the estimation of \textbf{Q} and \textbf{R}. The behaviour of the various cost function and in particular \textbf{J6} and \textbf{J7} in Table-\ref{tbcase4QMTMS} shows that the choice of the filter statistics for estimating \textbf{Q} and \textbf{R} in the proposed reference approach is the best possible when compared to other approaches presently considered.

\newpage
\subsection{Remarks on Case-4}
\begin{figure}[h]
\includegraphics[width=6in,height=4in]{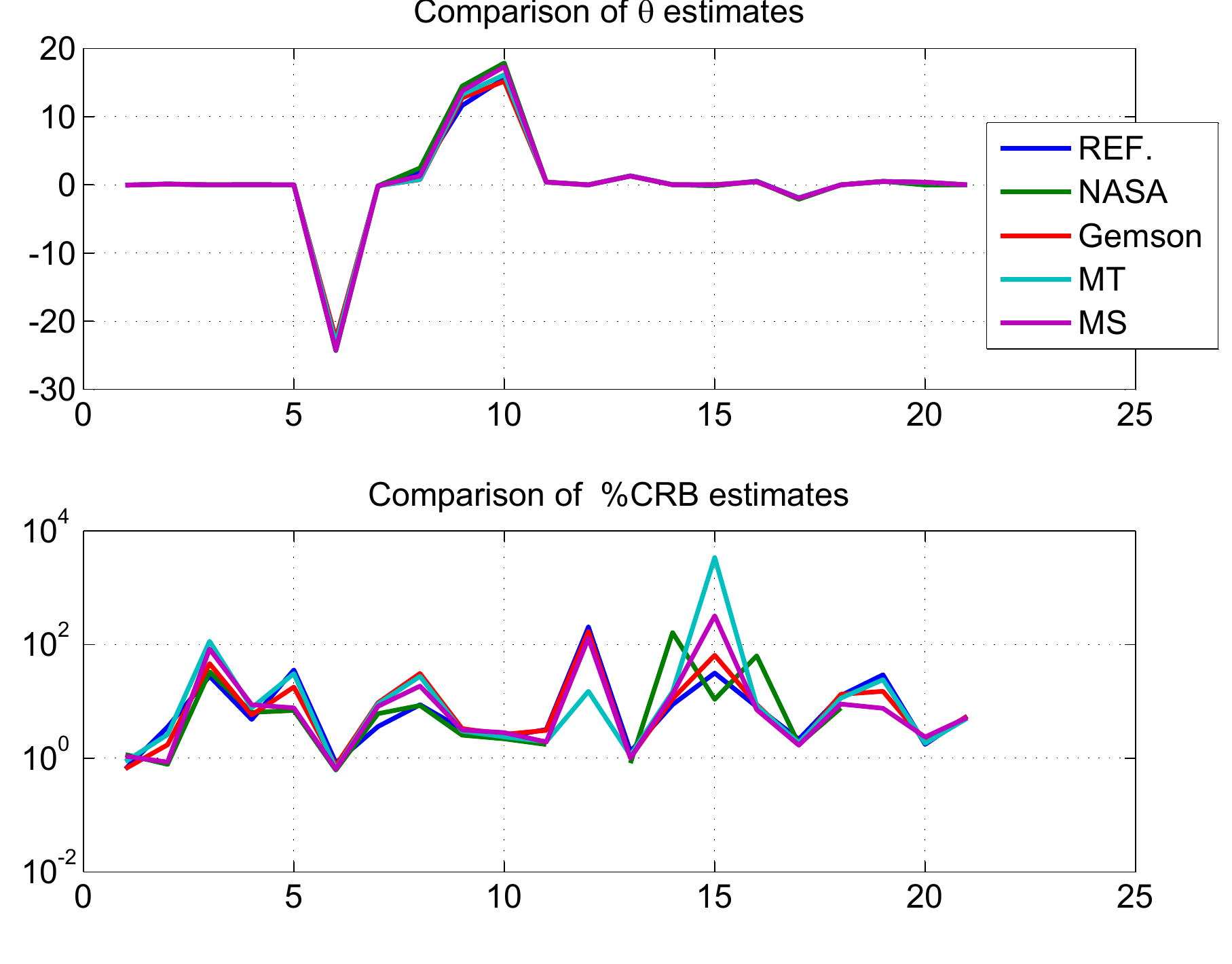}
\caption{Comparison of the parameter estimates and $\%$CRBs by different methods}
\label{comp4}
\end{figure}

The NASA results have been generated assuming \textbf{Q} = 0 and are comparable with reference procedure for the parameter estimates and their CRBs. Further the MT and MS methods give quite different estimates for the \textbf{Q} and \textbf{R} values than in the reference case. We believe that the reference procedure provides the best possible parameter estimates and their uncertainties. From the plot of the parameter estimates and their \%CRB in Fig \ref{comp4}, it can be seen that the parameters 1, 2, 4, 6, 9-11, 13, 17, 20 and 21 are strong and the parameters 3, 8, 12, 14-16 and 19 are the weak ones and others can be considered as modestly controlling the dynamics of the system.  The CRBs as estimated by different methods generally appear to vary widely. However what is interesting is that even the estimate of the strong parameters such as 6, 9 and 10 varies widely among the methods. Such a behaviour of the filter across the parameter estimates shows how important is the tuning of the filter statistics namely $\mathbf{P_0}$, \textbf{Q} and \textbf{R} in parameter estimation and their uncertainties.
\begin{landscape}
The rounded off 100$\times$C  matrix for case-4 is given by
\begin{footnotesize}
\begin{align*}
\begin{bmatrix}
   100 &   -9  & -24  &  36   & 69  & -77   & 21  &  31 &   17 &  -29   & -9  & 0  &  29   & -7  & -12  &  -7  &12    & 7   & 69   & -53   & 19 \\
    -9 & 100   &   3  &   1 &   -1  &  -1  &   0  &   0 &    0  &   0 &    0   & 0  &   0  &   0  &   0 &    0  &  0  &   0   &  0    & 0   &  0 \\
   -24  &  3  & 100  & -58  &  -16  &  16  &   0 &  -44 &  -75  &  43  &   2 & 0 &   -7  &  -1  &  18  &  30   & -18 &   -1  & -16    & 9  &  -4 \\
    36 &  1  &  -58  &  100  &   52   & -27  &  -5  &  31 &   42  & -73 &   -7 & 0  &  11  &   3  & -12 &  -17  & 29  &   5  &  52  & -39   &  15 \\
    69 &  -1  &  -16  &  52  &  100  & -53   &  23  &  44   & 10  & -42 &  -14 & 0  &  20  &  -7   & -16   & -4   & 16   & 10   & 100  & -77   & 27 \\
   -77  &  -1  &  16  &  -27 &  -53 &  100 &  -27  & -40  & -23  &  37   & 12  &  0   &  3   & -1   & -1  &  -1   & 1   &  1  & -53  &  68   &  2 \\
    21 &  0  &  0  & -5  &  23   & -27  &  100   & -3   &  0   &  7   & -5  & 0  &  -1   &  3   &  0  &   0    & 0 & 0  &  23  & -27   & -1 \\
    31 &  0 & -44 & 31  &  44  & -40  &  -3  & 100  &  58  & -42  & -10 & 0  &  -1 &    0  &   3   &  2
   & -1  &  -1   & 44  & -55  &  -2 \\
    17 &  0 & -75 & 42  &  10  & -23  &   0  &  58  & 100  & -57   & -2  & 0   & -1   &  0   &  2   &  3
    & -2   &  0  &  10  & -13   &  0 \\
   -29 &  0 &  43 & -73  & -42   & 37   &  7  & -42  & -57 &  100   &  9 & 0    & 1    & 0    & -1  &  -2
   & 3  &   1 &  -42   & 53  &   2 \\
    -9 &  0 & 2   & -7  & -14   & 12   & -5  & -10   & -2   &  9  & 100  & 2    & 0   &  0   &  0   &  0
    &0   &  1   &-13   & -1   &  0 \\
    0  &  0 & 0  & 0   &  0  &   0  &   0   &  0  &   0  &   0   &  2  & 100   &  0   &  0   &  0    & 0
    & 0   &  0   &  0    & 0   &  0 \\
    29 &  0 & -7 & 11  &  20   &  3  &  -1  &  -1  &  -1   &  1    & 0  &  0  & 100 &  -27  & -40  & -22
    & 37  &  24 &   20    & 2  &  65 \\
    -7 &  0 & -1 &  3  &  -7  &  -1  &   3  &   0   &  0  &   0   &  0 &  0  & -27  & 100  &  -3   & -1
    &  7  & -10   & -7   & -1  & -26  \\
   -12 &  0 & 18  & -12 &  -16   & -1   &  0   &  3  &   2 &   -1  &   0  & 0  & -40  &  -3  & 100  &  58
   &  -42  & -20   &-16   & -2  & -52   \\
    -7 &  0 & 30 & -17  &  -4  &  -1  &   0  &   2 &    3 &   -2  &   0 &  0  & -22&    -1  &  58  & 100
    &  -57   & -5   & -4    & 0 &  -13  \\
    12 &  0 & -18 &   29  &  16   &  1  &   0 &   -1  &  -2  &   3  &   0  &  0  &  37  &   7  & -42 &  -57
    &  100   & 19   & 16    & 2  &  50 \\
     7 & 0  & -1  &   5  &  10  &   1  &   0   & -1  &   0  &   1  &   1 & 0  &  24  & -10 &  -20  &  -5
     &  19  & 100 &   10    & 1   &  0  \\
    69 & 0  & -16 &    52 &  100  & -53  &  23  &  44  &  10  & -42 &  -13 & 0  &  20  &  -7 &  -16  &  -4   & 16  &  10  & 100  & -77  &  27 \\
   -53 & 0 &  9 &  -39  & -77  &  68  & -27  & -55  & -13  &  53   & -1  & 0   &  2 &   -1  &  -2 &    0
   &  2 &    1 &  -77  & 100   &  3 \\
    19 & 0 & -4 &   15  &  27  &   2  &  -1   & -2 &    0  &   2   &  0  & 0   & 65  & -26&   -52  & -13
    &  50   &  0   & 27    & 3  & 100
\end{bmatrix}
\end{align*}
\end{footnotesize}
\end{landscape}

\begin{landscape}
\begin{table}[h]
\subsection{Case-4 Tables}
\caption{Real flight test data case-4 results using the reference adaptive EKF\\ No of iterations=100}{}
\label{tbcase4Q}
\begin{center}
\begin{footnotesize}
\begin{tabular}{|c| c| c| c| c| c|c|c|c|c|c| }
\hline
Study &
\makecell{$\Theta$\\ (Ref)} &
\makecell{$\Theta$\\ (NASA)} &
\makecell{$\Theta$\\ (Gemson)} &
\makecell{$\sigma_\Theta$ \\(Ref)} &
\makecell{$\sigma_\Theta$\\ (NASA)} &
\makecell{$\sigma_\Theta$\\ (Gemson)} &
\makecell{\textbf{R} \\ $\times10^{-6}$\\ (Ref)}&
\makecell{\textbf{Q} \\ $\times10^{-6}$\\ (Ref)}&
\makecell{\textbf{J1-J8} \\(Ref) }&
Remarks
\\ \hline


\makecell{$\mathbf{P_0}$ : Scaled up-[0,0;0,\checkmark]\\\textbf{Q} : EM-[\checkmark,0;0,0] \\\textbf{R} : EM-diag} &
\makecell{ -0.0465 \\   0.1017  \\ -0.0018  \\  0.0125 \\  -0.0028 \\ -22.8656 \\  -0.2335 \\ 1.7855 \\  11.6289  \\ 15.6179 \\   0.3920 \\  -0.0029  \\  1.3150 \\   0.0137 \\ -0.0698  \\  0.5326  \\ -1.9137  \\ -0.0079  \\  0.4982   \\ 0.3703 \\  -0.0128} &
\makecell{  -0.04670 \\   0.1026  \\ 0.002753 \\   0.01594 \\  -0.003035 \\ -24.32 \\  -0.1505 \\ 2.4640 \\  14.47  \\ 17.87 \\   0.4092 \\  -0.008423  \\  1.2900 \\   0.0004483 \\ -0.1514  \\  0.5062  \\ -2.125 \\  -0.007553  \\  0.5179   \\ -0.02581 \\  -0.004472} &
\makecell{  -0.04590 \\   0.10430  \\ -0.0013 \\   0.0121 \\  -0.0028 \\ -22.5872 \\  -0.1346 \\ 0.7565 \\  12.7465  \\ 15.1983 \\   0.3859 \\  -0.0019  \\  1.2708 \\   0.0090 \\ 0.0299  \\  0.4043  \\ -1.9494 \\  -0.006  \\  0.4908   \\ 0.3617 \\  -0.0110} &
\makecell{  0.0003 \\   0.0036  \\  0.0005  \\  0.0006  \\  0.0010  \\  0.1902  \\  0.0084 \\ 0.1554  \\  0.3795   \\ 0.3697 \\   0.0125  \\  0.0059  \\  0.0172 \\   0.0012 \\ 0.0219  \\  0.0428 \\   0.0412 \\   0.0010  \\  0.1466   \\ 0.0065  \\  0.0007} &
\makecell{  0.0005401  \\  0.0008  \\  0.0009128 \\    0.001019  \\  0.0002101  \\  0.1500  \\  0.009149 \\ 0.2088  \\  0.3688   \\ 0.3906  \\  0.007108  \\  --  \\  0.01056  \\  0.0007281 \\ 0.01637  \\  0.3197  \\  0.03726  \\  0.0005759  \\  --   \\ --  \\  --} &
\makecell{  0.0003  \\  0.0018  \\  0.0006 \\    0.0007  \\  0.0005  \\  0.1672  \\  0.0127 \\ 0.2344  \\  0.4221   \\ 0.4004  \\  0.0119  \\  0.0032  \\  0.0131  \\  0.0010 \\ 0.0194  \\  0.0355  \\  0.0343  \\  0.0008  \\  0.0739   \\ 0.0072  \\  0.0006} &
\makecell{4.6243  \\  0.2728  \\  0.2611  \\  0.0086  \\  6.3789 \\ 930.0747 \\  21.7747} &
\makecell{ 0.14    \\17.23 \\   5.12 \\   0.10} &
\makecell{6.6805  \\  7.0957  \\  6.6025  \\  0.0324 \\ -72.6142  \\  4.0024  \\  4.0024 \\ 3.5376} &
\makecell{Cost functions converge\\ to the expected values.}
\\ \hline

\end{tabular}
\end{footnotesize}
\end{center}
\end{table}

\begin{table}[h]
\caption{Case-4 results using simulated Additive White Gaussian Noise}{}
\label{tbnew4}
\begin{center}
\begin{tabular}{|c| c| c| c| c|}
\hline
Study &
\makecell{$\sigma_\Theta$ \\(Simulated-without\\ updating $\Theta$)} &
\makecell{$\sigma_\Theta$ \\(Simulated-with\\ updating $\Theta$)} &
\makecell{$\sigma_\Theta$ \\(Ref)} &
Remarks
\\ \hline

\multicolumn{5}{|c|}{\makecell{Case-4 data generated using simulated measurement and process noise (AWGN) \\ of variance $  \textbf{Q}$ and \textbf{R} estimated by Reference EKF (\textbf{Q} $>$ 0)} } \\ \hline

\makecell{$\mathbf{P_0}$ : Scaled up-[0,0;0,\checkmark]\\\textbf{Q} : \textbf{Q} (Ref) \\\textbf{R} :   \textbf{R} (Ref)} &

\makecell{ 0.0004 \\   0.0043  \\  0.0006  \\  0.0006  \\  0.0011  \\  0.2225  \\  0.0108 \\ 0.1642 \\   0.3668 \\   0.3775 \\   0.0130  \\  0.0059  \\  0.0204  \\  0.0015 \\ 0.0232  \\  0.0395  \\  0.0419   \\ 0.0011  \\  0.1550  \\  0.0074  \\  0.0008} &
\makecell{  0.0004 \\   0.0043  \\  0.0006 \\   0.0006  \\  0.0011  \\  0.2370  \\  0.0108 \\ 0.1644   \\ 0.3724 \\   0.3813  \\  0.0132 \\   0.0059  \\  0.0214  \\  0.0015 \\ 0.0232  \\  0.0398  \\  0.0420   \\ 0.0011  \\  0.1550 \\   0.0078  \\  0.0009 } &
\makecell{  0.0003 \\   0.0036  \\  0.0005  \\  0.0006  \\  0.0010  \\  0.1902  \\  0.0084 \\ 0.1554  \\  0.3795   \\ 0.3697 \\   0.0125  \\  0.0059  \\  0.0172 \\   0.0012 \\ 0.0219  \\  0.0428 \\   0.0412 \\   0.0010  \\  0.1466   \\ 0.0065  \\  0.0007} &

\makecell{No Significant \\ change in $\sigma_\Theta$}
\\ \hline

\end{tabular}
\end{center}
\end{table}

\begin{table}[h]
\caption{Real flight test data case-4 results using the MT and MS method. \\ No of iterations=100 }{}
\label{tbcase4QMTMS}
\begin{center}
\begin{footnotesize}
\begin{tabular}{|c| c| c| c| c| c|| c|c|c|c|c|c| }
\hline
Study &
\makecell{$\Theta$\\ (MT)} &
\makecell{$\sigma_\Theta$ \\(MT)} &
\makecell{\textbf{R} (MT)\\ $\times10^{-6}$ }&
\makecell{\textbf{Q} (MT)\\ $\times10^{-6}$}&
\makecell{\textbf{J1-J8} \\(MT) }&

\makecell{$\Theta$\\ (MS)} &
\makecell{$\sigma_\Theta$\\ (MS)} &
\makecell{\textbf{R} (MS) \\ $\times10^{-6}$}&
\makecell{\textbf{Q} (MS)\\ $\times10^{-6}$}&
\makecell{\textbf{J1-J8} \\(MS) }&
Remarks
\\ \hline


\makecell{$\mathbf{P_0}$ : Scaled up-[0,0;0,\checkmark]\\\textbf{Q} : MT/MS-[\checkmark,0;0,0] \\\textbf{R} : MT/MSMS-diag} &

\makecell{  -0.0460 \\   0.1040  \\ -0.0007  \\  0.0130 \\  -0.0026 \\ -23.2819\\   -0.1308 \\ 0.7872 \\  13.2291  \\ 16.1444  \\  0.3926 \\  -0.0020   \\ 1.2867 \\   0.0074 \\ 0.0006  \\  0.4469  \\ -1.9271  \\ -0.0070  \\  0.4652   \\ 0.3710 \\  -0.0119} &
\makecell{  0.0004 \\   0.0027 \\   0.0008  \\  0.0010  \\  0.0008 \\   0.1526 \\   0.0118 \\ 0.2170 \\   0.3960   \\ 0.3826 \\   0.0077  \\  0.0003  \\  0.0144  \\  0.0011 \\ 0.0204  \\  0.0374 \\   0.0360 \\   0.0008 \\   0.1118   \\ 0.0068 \\   0.0006} &
\makecell{0.00002  \\  0.00310 \\   0.09153 \\   0.00008  \\  0.35034 \\  2.85722 \\   0.02140} &
\makecell{0.0802 \\   3.4875 \\   0.0048 \\   0.0495} &
\makecell{  7.8309  \\  7.9831  \\  5.5817  \\  0.0187  \\-68.6021 \\  36.8377 \\  36.8379 \\ 5.1064} &

\makecell{  -0.0461  \\  0.1051 \\  -0.0012  \\  0.0126 \\  -0.0026 \\ -24.2547 \\  -0.1537 \\ 1.2883 \\  13.7752  \\ 17.3261 \\   0.4087  \\ -0.0031  \\  1.2826  \\  0.0074 \\ -0.0058  \\  0.4704  \\ -1.9647 \\  -0.0067  \\  0.4598   \\ 0.3864  \\ -0.0115} &
\makecell{ 0.0005  \\  0.0009  \\  0.0010 \\   0.0011 \\   0.0002 \\   0.1558  \\  0.0126 \\ 0.2398 \\   0.4337   \\ 0.4870 \\   0.0079  \\  0.0040 \\   0.0129  \\  0.0010 \\ 0.0186   \\ 0.0335  \\  0.0333  \\  0.0006 \\   0.0348   \\ 0.0091 \\   0.0006} &
\makecell{  0.0005 \\   0.2160  \\  0.0026 \\   0.0005 \\   0.4559 \\   7.0779  \\  0.0212} &
\makecell{ 0.0056  \\  0.8155 \\   2.1818 \\   0.0194} &
\makecell{  7.1837  \\  7.2237 \\   5.0069  \\  0.0092 \\ -64.7950 \\  58.5572 \\  58.5577 \\ 4.2938} &

\makecell{Cost functions are \\not close to their \\expected values in\\ MT and MS method} \\ \hline

\end{tabular}
\end{footnotesize}
\end{center}
\end{table}
\end{landscape}

\clearpage
\subsection{Case-4 Figures}

\clearpage
\section{Real Flight Test Case-5}
\par
\begin{figure}[h]
\includegraphics[width=6in,height=3.2in]{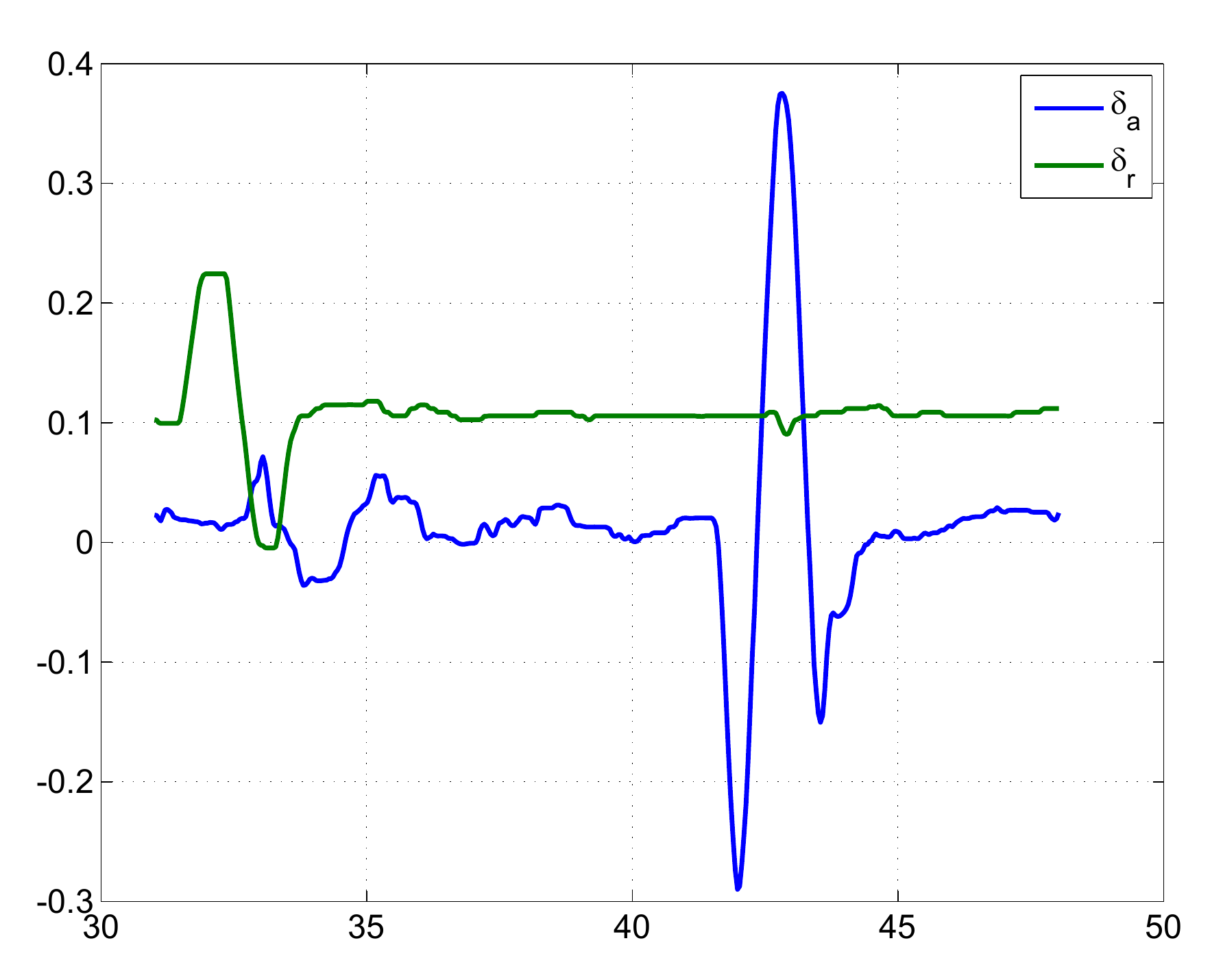}
\caption{Control input ($\delta_a,\delta_r$) versus time (s)}
\label{input5}
\end{figure}

\begin{figure}[h]
\includegraphics[width=6in,height=3.2in]{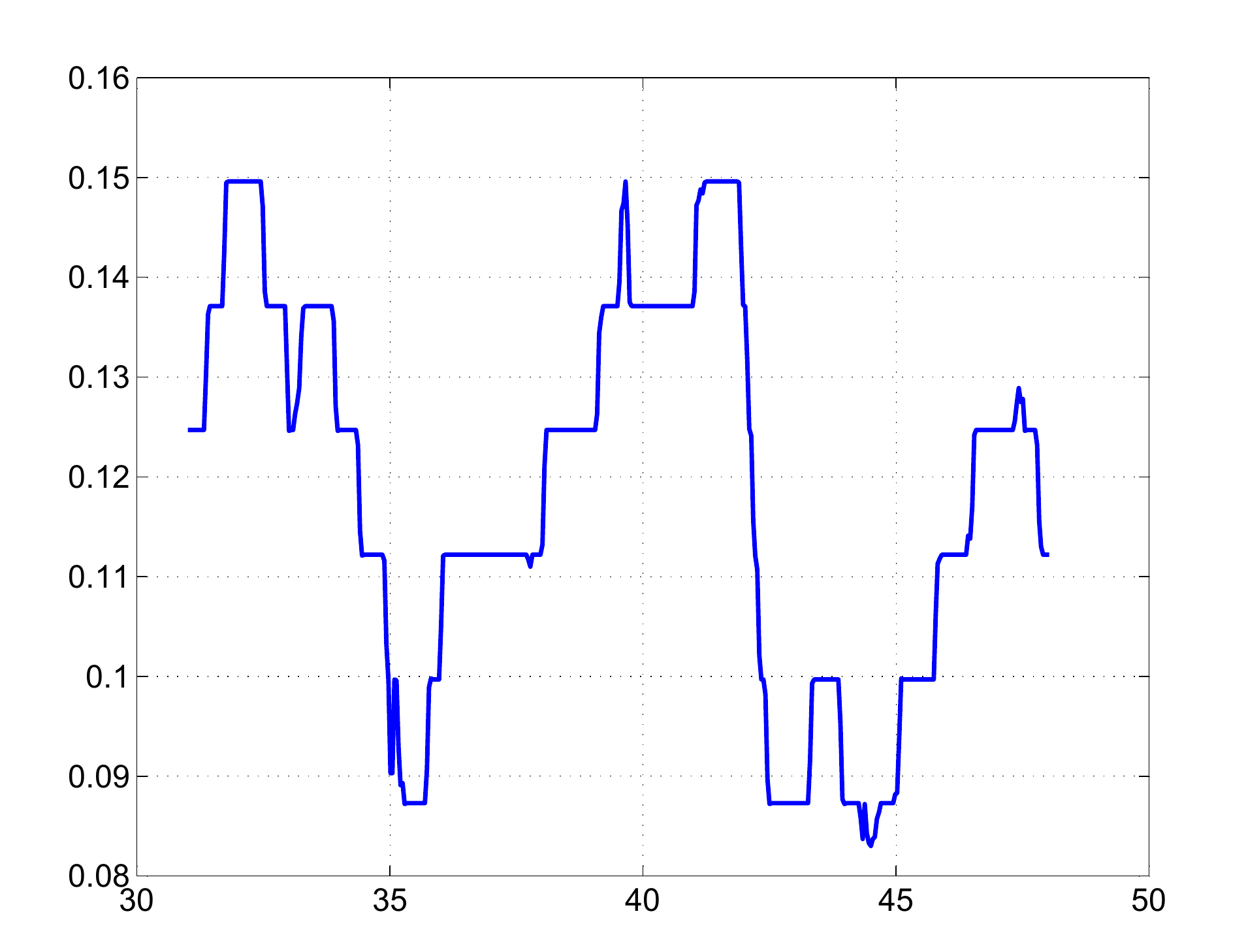}
\caption{Measurement input versus ($\Theta_m$) time (s)}
\label{case5_theta}
\end{figure}

\begin{figure}[h]
\includegraphics[width=6in,height=4in]{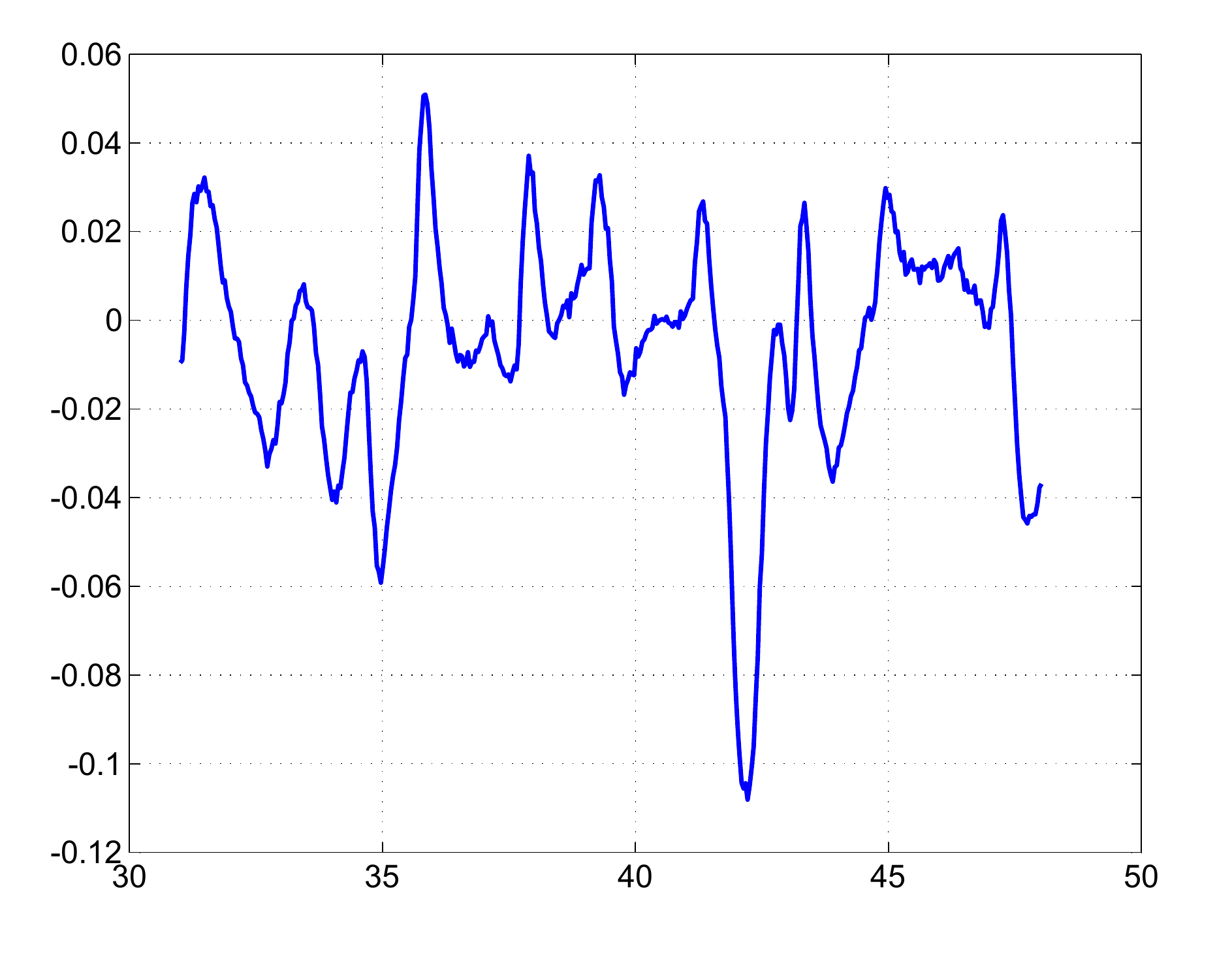}
\caption{Measurement input versus ($q_m$) time (s)}
\label{case5_q}
\end{figure}

\begin{figure}[h]
\includegraphics[width=6in,height=4in]{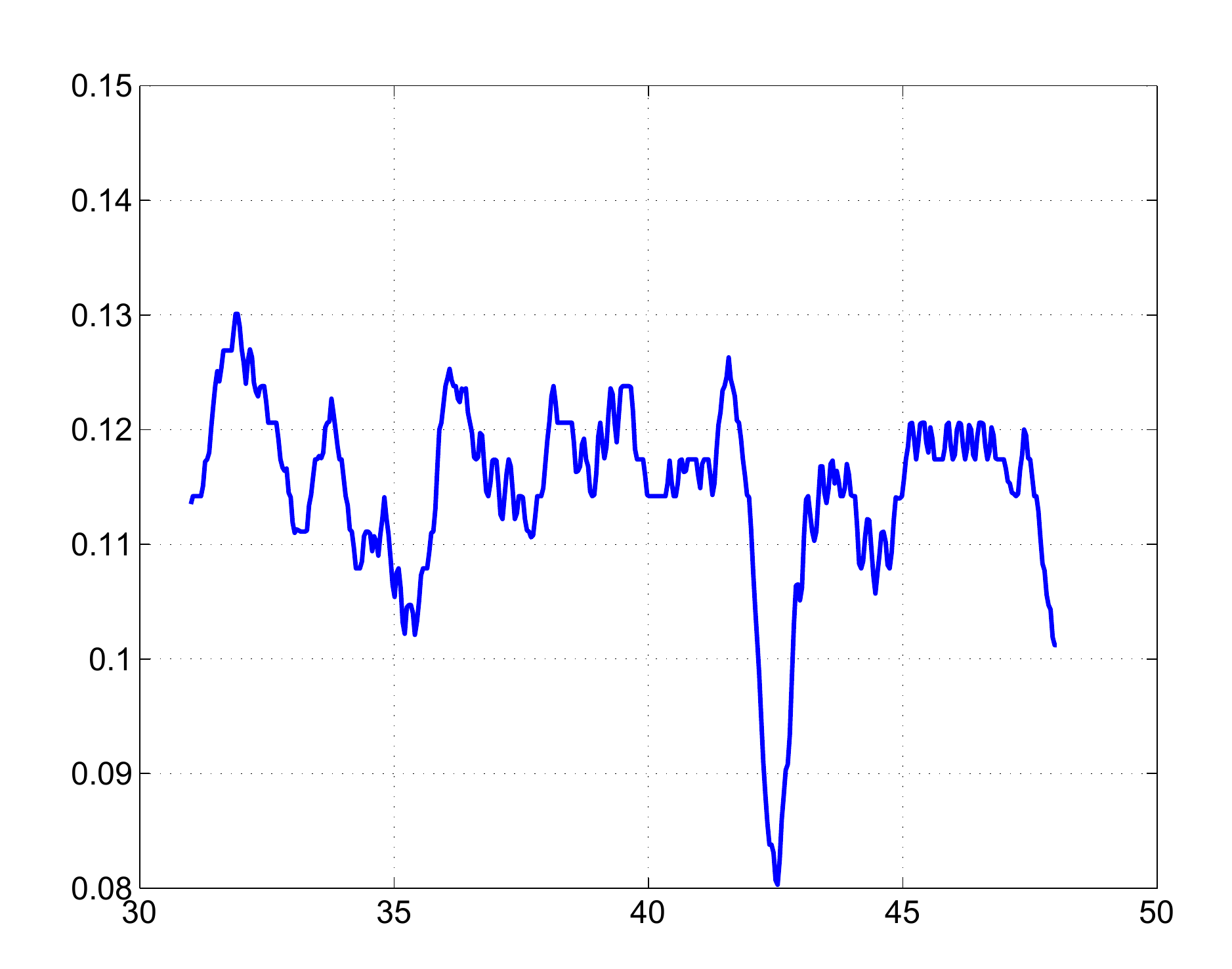}
\caption{Measurement input versus ($\alpha_m$) time (s)}
\label{case5_alpha}
\end{figure}

 The data set is obtained from NASA TP 1690 which describes the lateral motion of a oblique wing aircraft with zero wing skew excited by the control input ($\delta_a$ and $\delta_r$ in rad) as shown in Fig. \ref{input5}. Some of the available measurements can also be used as inputs in the state equations which includes pitch angle ($\theta_m$ in rad), pitch rate ($q_m$ in rad/s) and the angle of attack ($\alpha_m$ in rad) are shown in Fig. \ref{case5_theta}, Fig. \ref{case5_q} and Fig. \ref{case5_alpha} respectively. The state equations ($n$=4) for the angle of sideslip ($\beta$), roll rate (p), roll angle ($\phi$) and yaw rate (r) are
\begin{align*}
&\dot{\beta}=-\frac{\bar qS}{mV}(C_{Y_\beta}\beta+C_{Y_p}\frac{\bar c}{2V}p+C_{Y_r}\frac{\bar c}{2V}r+C_{Y_{\delta_a}} \delta_a+\beta_0)+\frac{g}{V}sin(\phi)cos(\theta_m)+psin(\alpha_m)-rcos(\alpha_m) \\
&\dot{p}-\dot{r}\frac{Izx}{Ixx}=\frac{\bar q S b}{Ixx}(C_{L_\beta}\beta+C_{L_p}\frac{\bar c}{2V}p+C_{L_r}\frac{\bar c}{2V}r+C_{L_{\delta_a}} \delta_a+C_{L_0})+\frac{Iyy-Izz}{Ixx}rq_m+\frac{Izx}{Ixx}pq_m\\
&\dot{\phi}=p+q\text{ }tan(\theta_m)sin(\phi)+r\text{ }tan(\theta_m)cos(\phi)+\phi_0\\
&\dot{r}-\dot{p}\frac{Izx}{Izz}=\frac{\bar q S b}{Izz}(C_{N_\beta}\beta+C_{N_p}\frac{\bar c}{2V}p+C_{N_r}\frac{\bar c}{2V}r+C_{N_{\delta_a}} \delta_a+C_{N_0})+\frac{Ixx-Iyy}{Izz}pq_m-\frac{Izx}{Izz}rq_m
\end{align*}
The angle of sideslip ($\beta$), roll rate (p), roll angle ($\phi$), yaw rate ($\dot r$) and lateral acceleration ($a_y$) are measured (indicated with subscript `m') in units of are rad, rad/s, rad, rad/s and $ft/s^2$ respectively. The measurement equations ($m$=5) are given by
\begin{align*}
{\beta_m}&=\beta-K_\beta z_\beta \frac{p}{V}+K_\beta x_\beta \frac{r}{V} \\
{p_m}&=p\\
{\phi_m}&=\phi\\
r_m&=r\\
{a_{y_m}}&=\frac{\bar q S}{mg}(C_{Y_\beta}\beta+C_{Y_{\delta_a}}\delta_a+C_{Y_{\delta_r}}\delta_r+C_{Y_0})-\frac{z_{a_y}}{g}\dot{p}+\frac{x_{a_y}}{g}\dot{r}
\end{align*}

The unknown parameter set ($p=20$) is $\Theta=(C_{Y_{\beta}},C_{Y_{\delta_r}},\beta_0,C_{L_{\beta}},C_{L_{p}},C_{L_r},C_{L_{\delta_a}},C_{L_{\delta_r}},C_{L_{0}},\\ \phi_0,C_{N_{\beta}},C_{N_{p}},C_{N_r},C_{N_{\delta_a}},C_{N_{\delta_r}},C_{N_0},C_{Y_0},C_{Y_p},C_{Y_r},C_{Y_{\delta_a}})^T$. The ones with suffix `$\delta_a$' and `$\delta_r$' are the control derivatives, the ones with suffix zero are the biases and all others are aerodynamic derivatives. The initial states are taken as initial measurement  and the initial parameter values are taken as $(-0.5,0.1,-0.01,0.01,-0.35,0.01,0.06,0.01,-0.002,\\0.002,0.07,-0.055,-0.05,0.003,-0.04,0.0068,-0.025,0.5,-1,0.005)^T$.

\begin{table}[h]
\begin{center}
\caption*{Other constant values used for case-5}{}
\begin{tabular}{| c | c | c | c | c | c | c | }
\hline
$\bar q=865.3$ & S=9.3 & m=387.7 & Ixx=314  &  Iyy=488 & Izz=698 & Izx=69   \\ \hline
V=39.41 & g=9.81 & b=6.81 & $K_\beta z_\beta$=0.305 & $K_\beta x_\beta$=2.73 & $z_{a_y}$=-0.098 & $x_{a_y}$=0.651 \\ \hline
\end{tabular}
\end{center}
\end{table}

\newpage
\par
\begin{figure}[h]
\includegraphics[width=6in,height=4in]{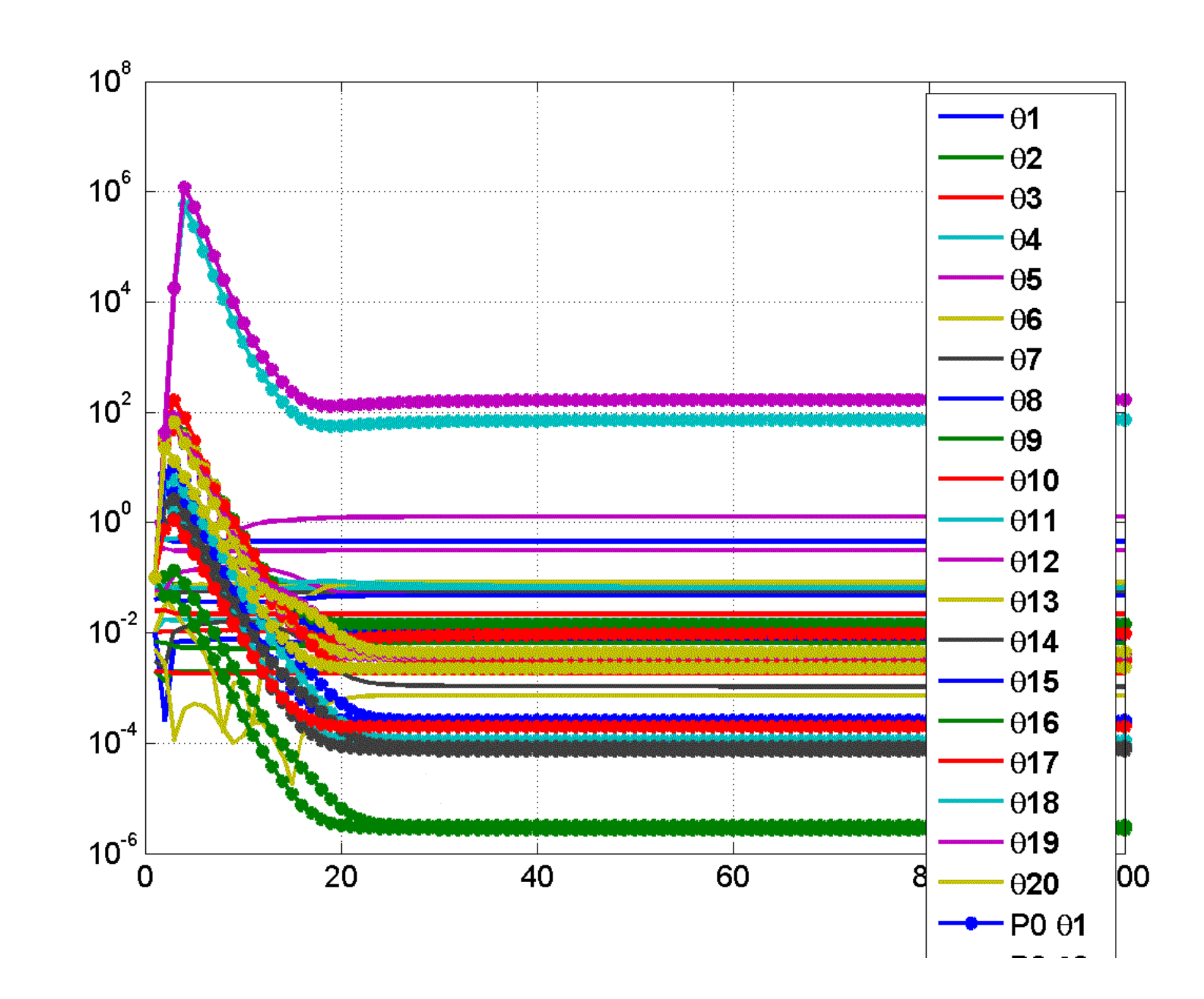}
\caption{Variation of parameter and its initial covariance ($\mathbf{P_0}$) with iterations}
\label{realQ5_P0}
\end{figure}

\begin{figure}[h]
\includegraphics[width=6in,height=4in]{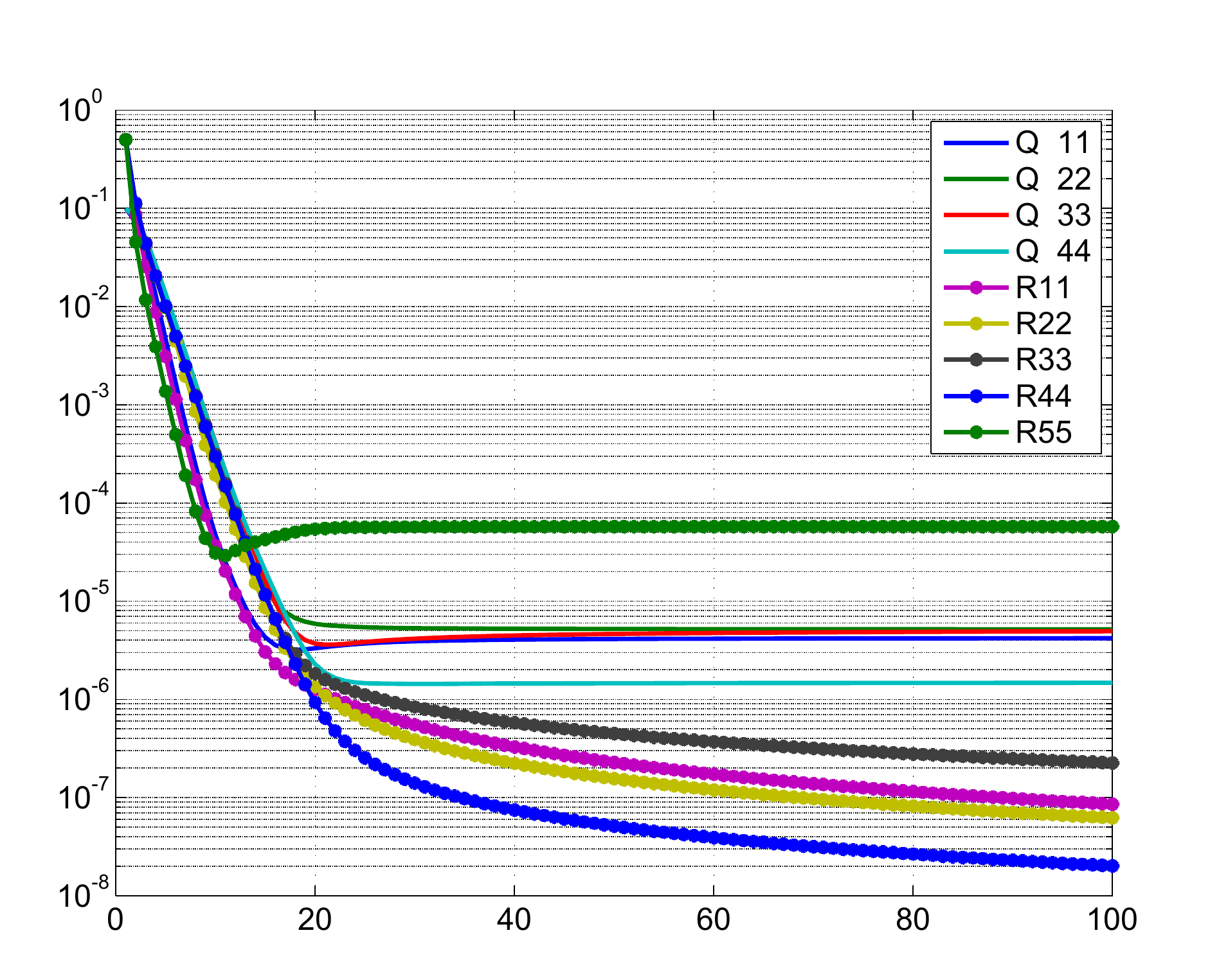}
\caption{Variation of \textbf{R} with iterations}
\label{realQ5_R}
\end{figure}

\begin{figure}[h]
\includegraphics[width=6in,height=4in]{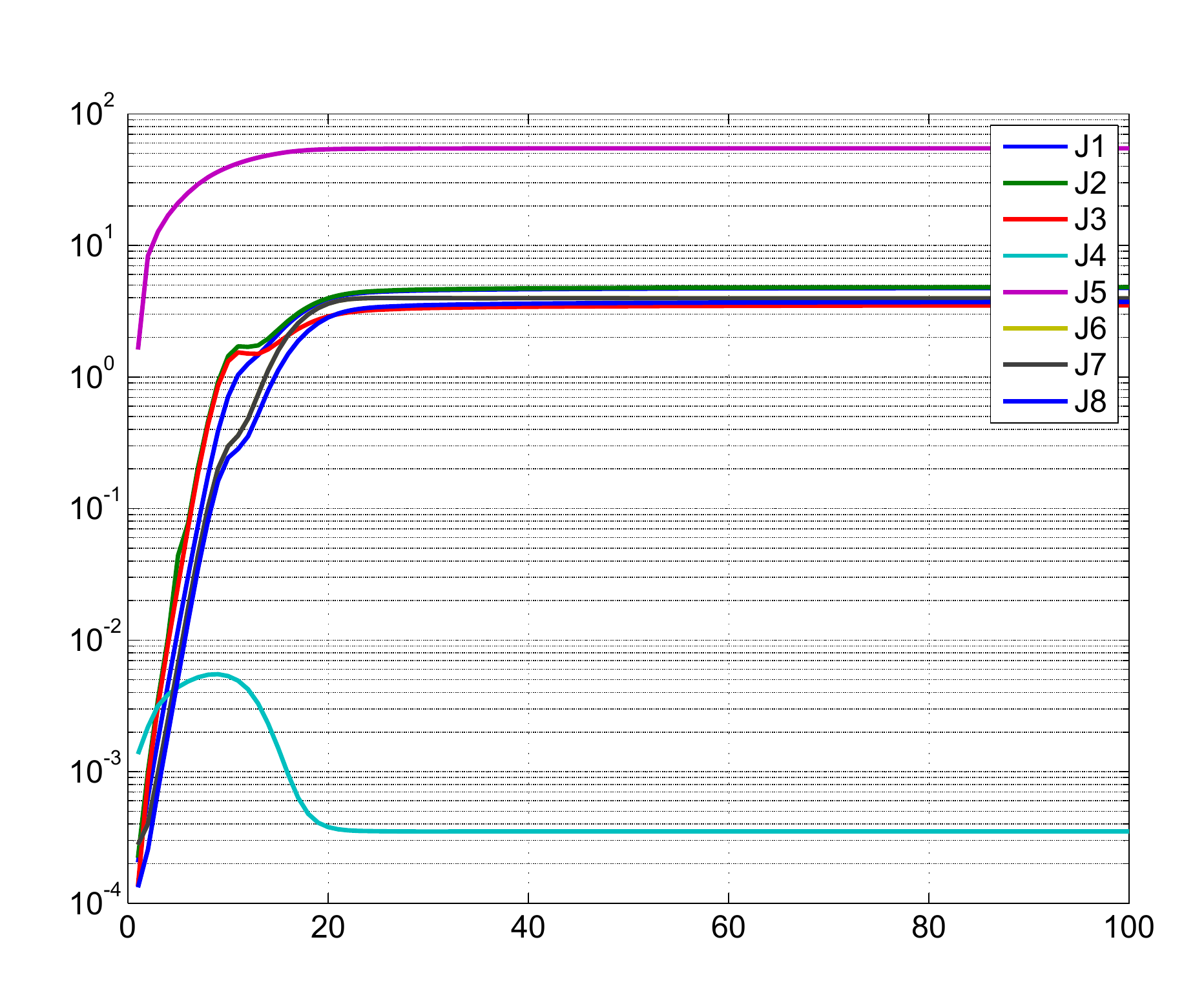}
\caption{Variation of different costs (\textbf{J1-J8}) with iterations}
\label{realQ5_J}
\end{figure}

\begin{figure}[h]
\includegraphics[width=6in,height=4in]{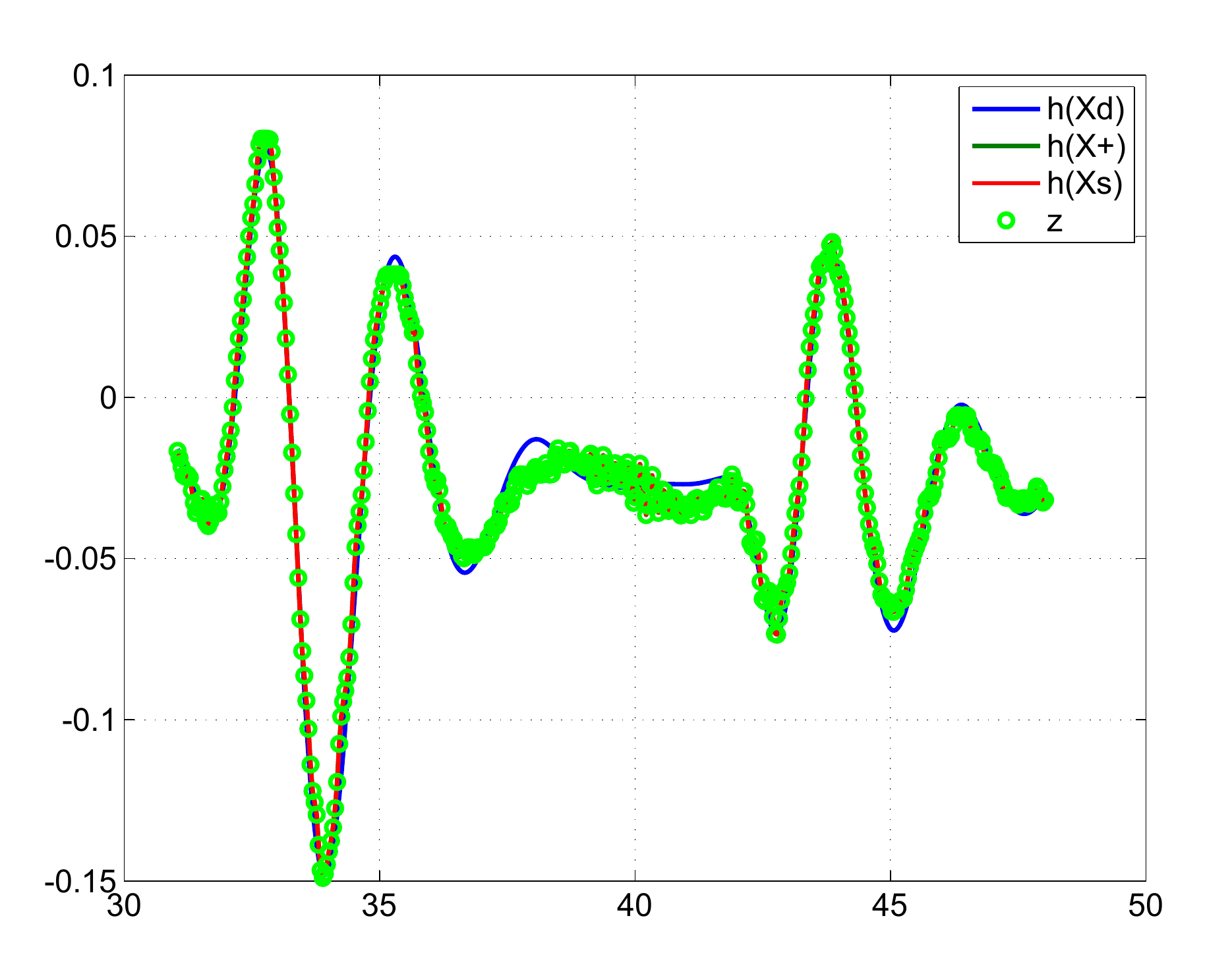}
\caption{Comparison of the predicted dynamics, posterior, smoothed}
\caption*{and the measurement 1}
\label{realQ5_s1}
\end{figure}

\begin{figure}[h]
\includegraphics[width=6in,height=4in]{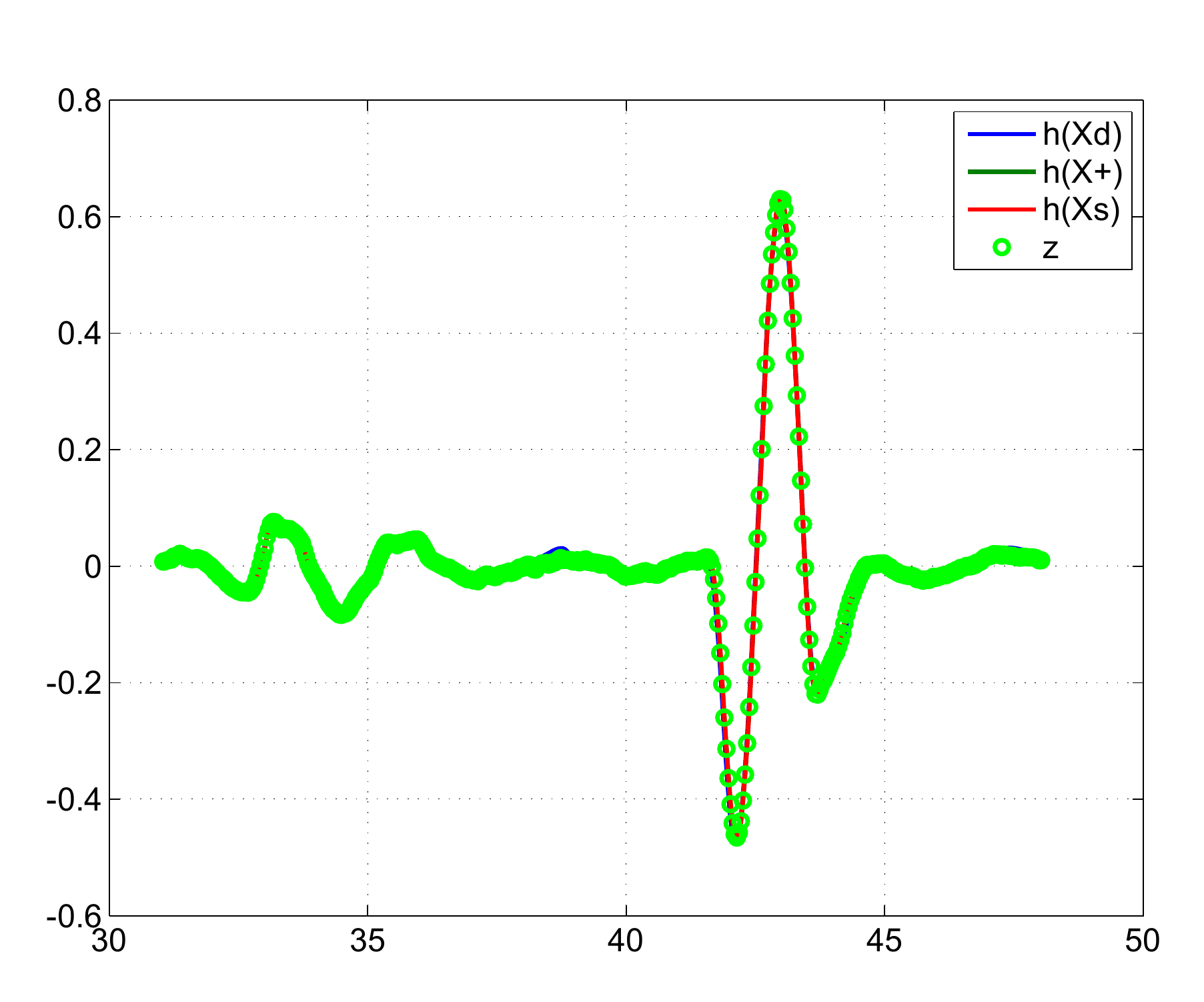}
\caption{Comparison of the predicted dynamics, posterior, smoothed}
\caption*{and the measurement 2}
\label{realQ5_s2}
\end{figure}

\begin{figure}[h]
\includegraphics[width=6in,height=4in]{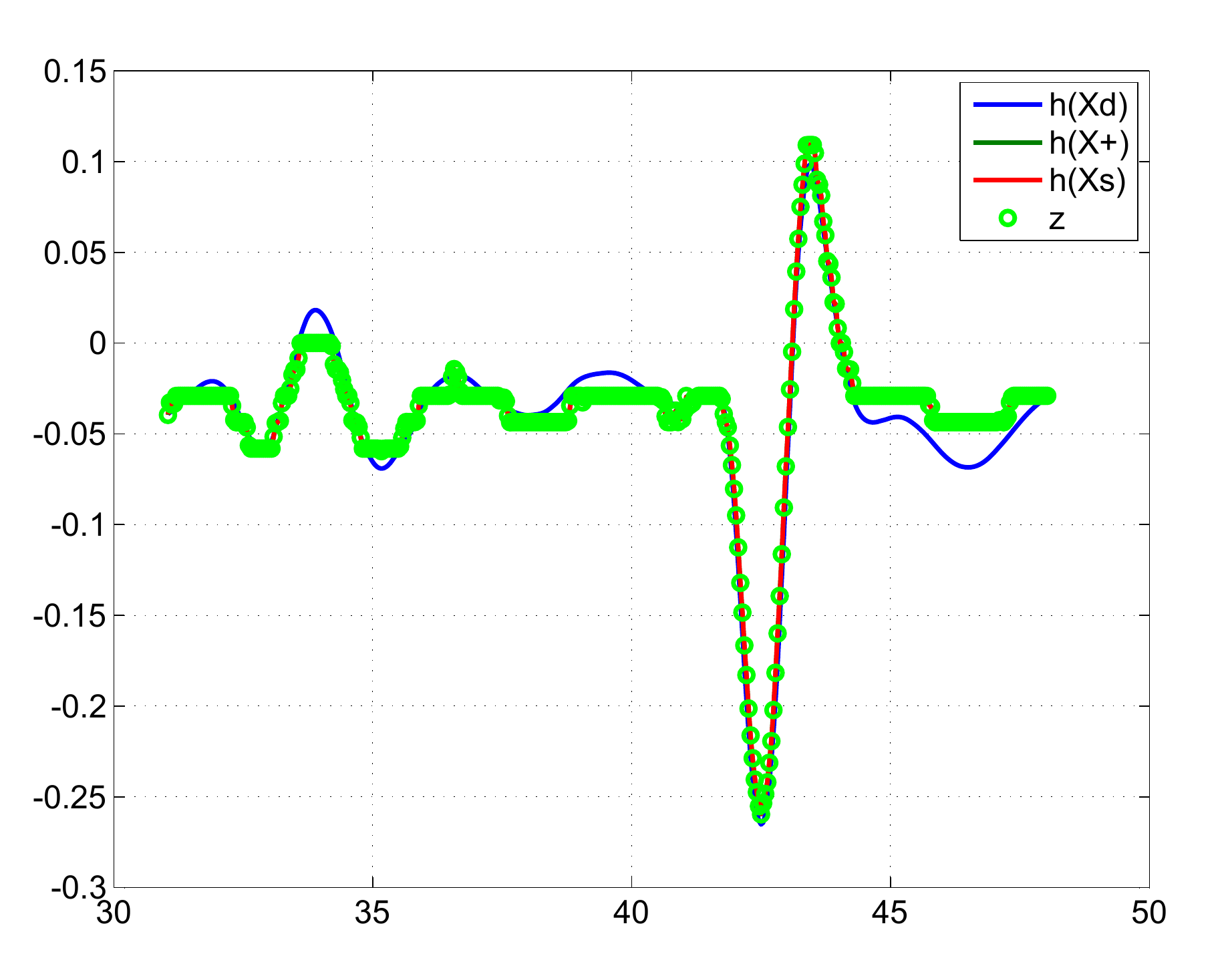}
\caption{Comparison of the predicted dynamics, posterior, smoothed}
\caption*{and the measurement 3}
\label{realQ5_s3}
\end{figure}

\begin{figure}[h]
\includegraphics[width=6in,height=4in]{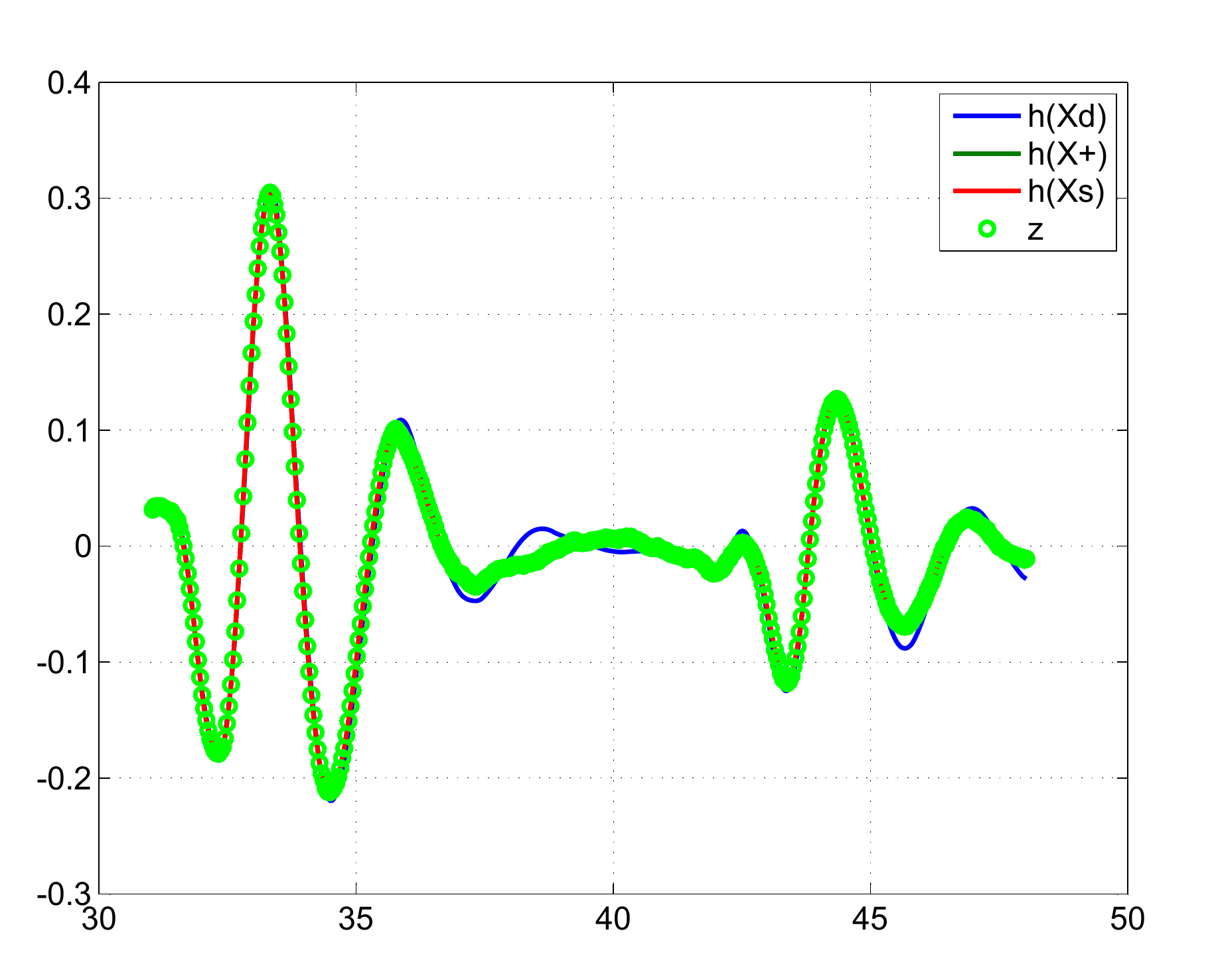}
\caption{Comparison of the predicted dynamics, posterior, smoothed}
\caption*{and the measurement 4}
\label{realQ5_s4}
\end{figure}

\begin{figure}[h]
\includegraphics[width=6in,height=4in]{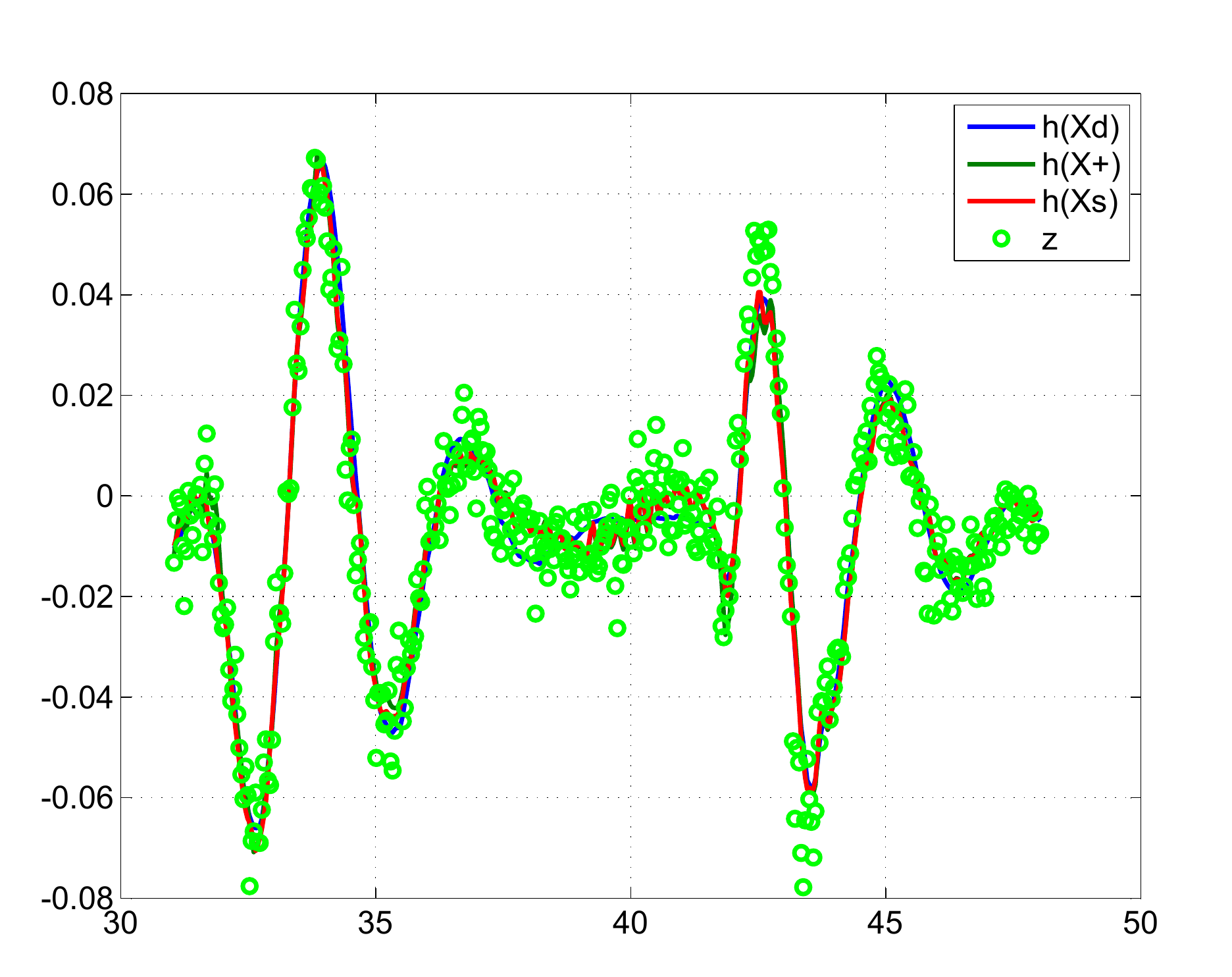}
\caption{Comparison of the predicted dynamics, posterior, smoothed}
\caption*{and the measurement 5}
\label{realQ5_h5}
\end{figure}

\begin{figure}[h]
\includegraphics[width=6in,height=4in]{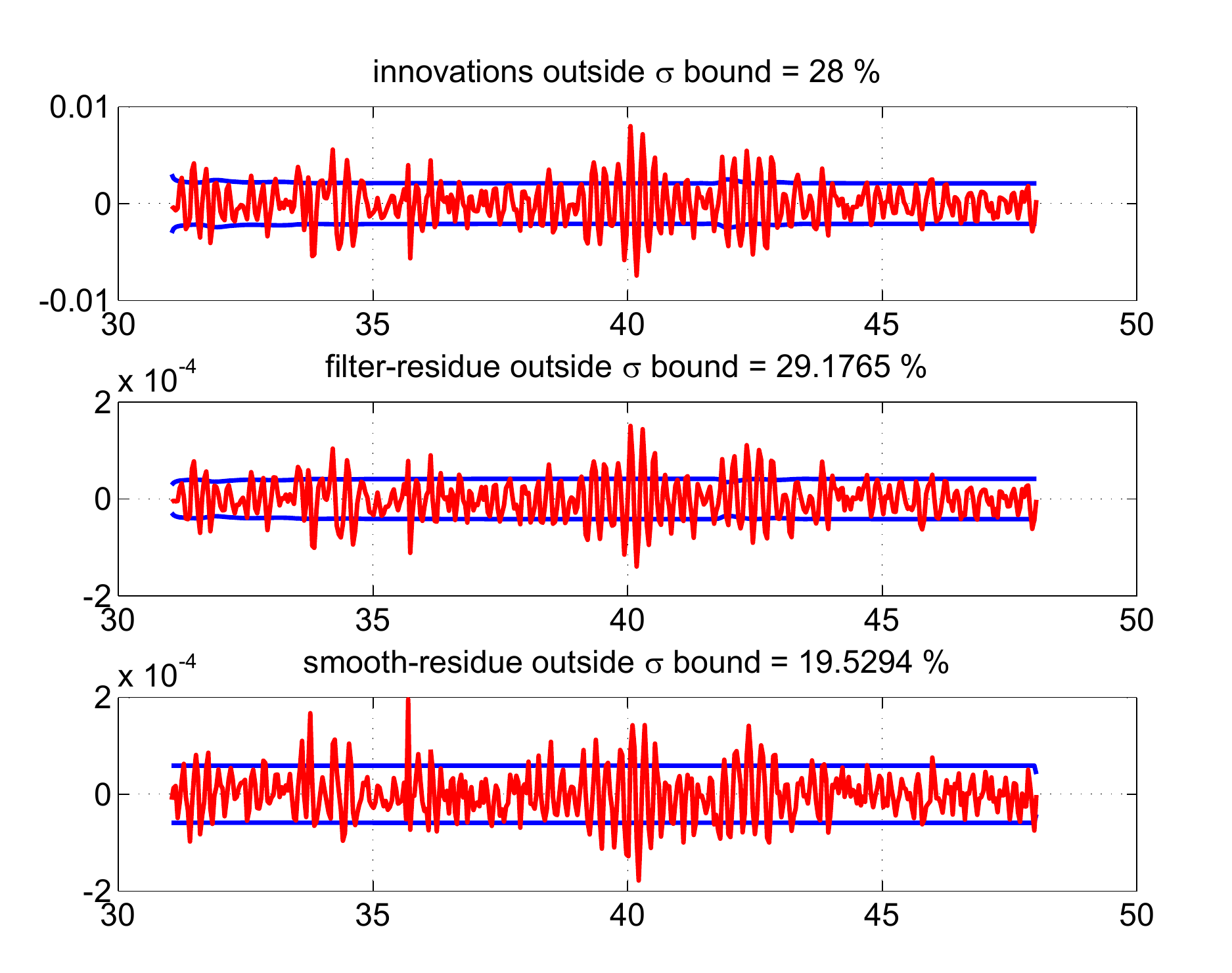}
\caption{The innovations, filtered residue and smoothed residue of measurement 1}
\label{realQ5_innov1}
\end{figure}

\begin{figure}[h]
\includegraphics[width=6in,height=4in]{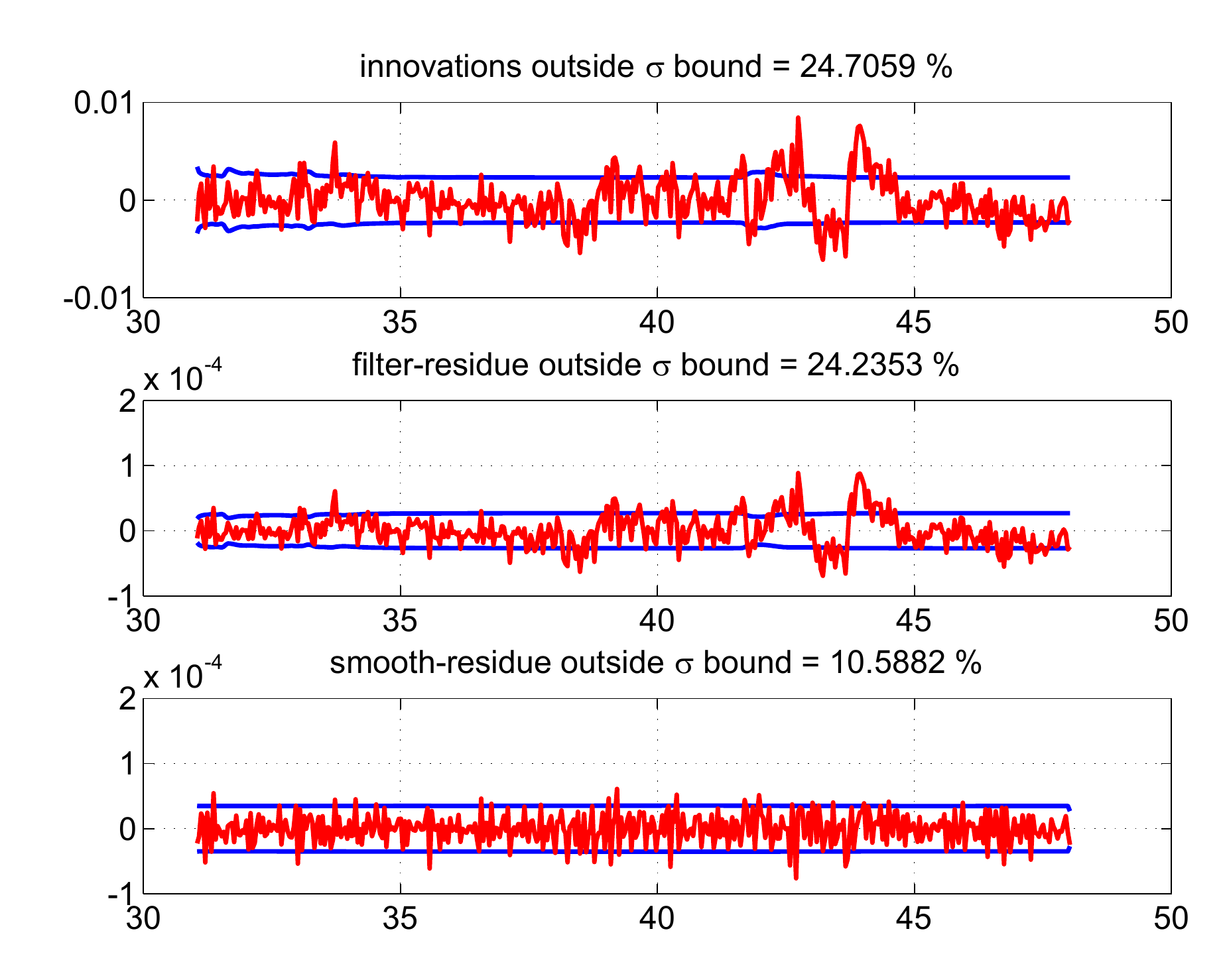}
\caption{The innovations, filtered residue and smoothed residue of measurement 2}
\label{realQ5_innov2}
\end{figure}

\begin{figure}[h]
\includegraphics[width=6in,height=4in]{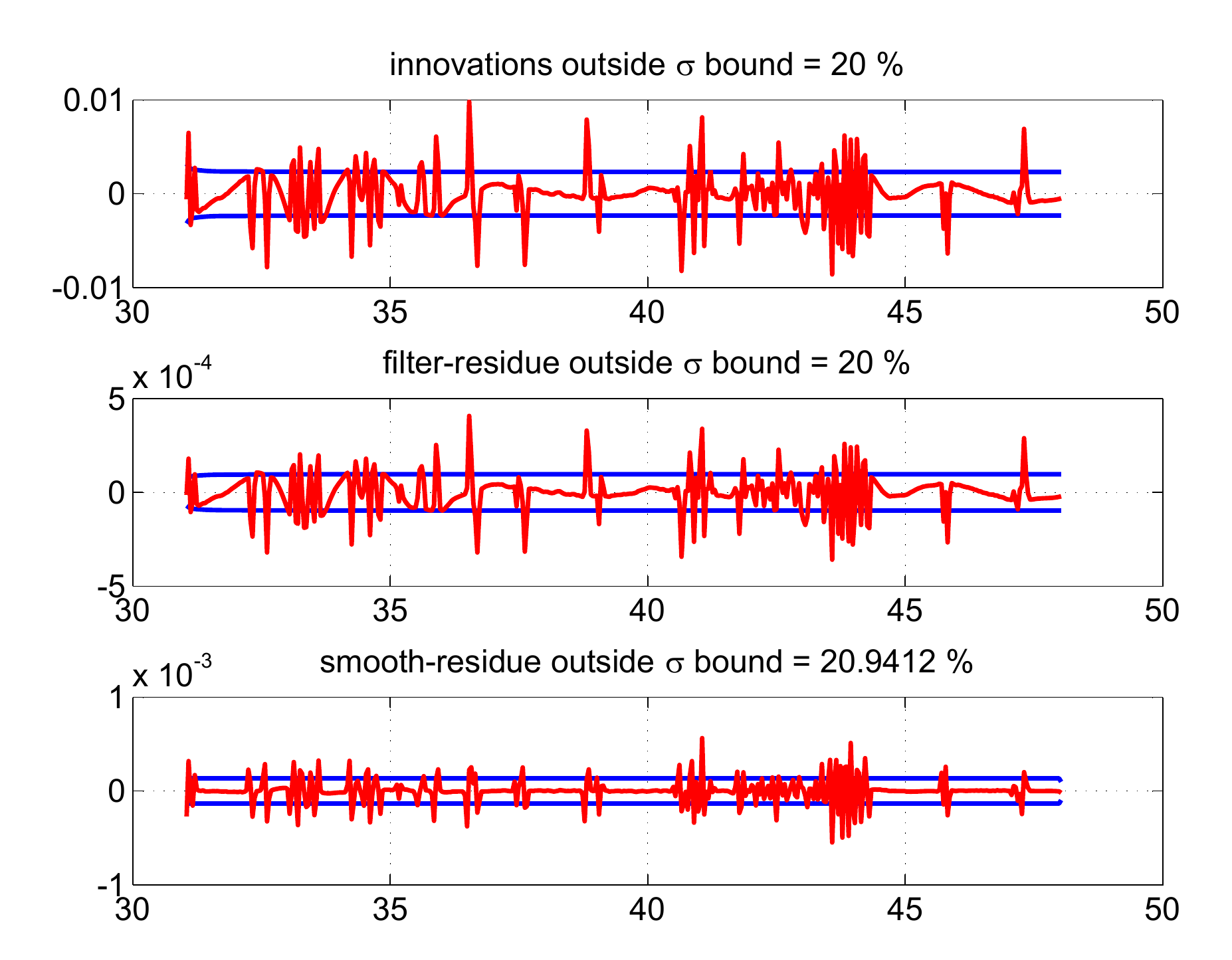}
\caption{The innovations, filtered residue and smoothed residue of measurement 3}
\label{realQ5_innov3}
\end{figure}

\begin{figure}[h]
\includegraphics[width=6in,height=4in]{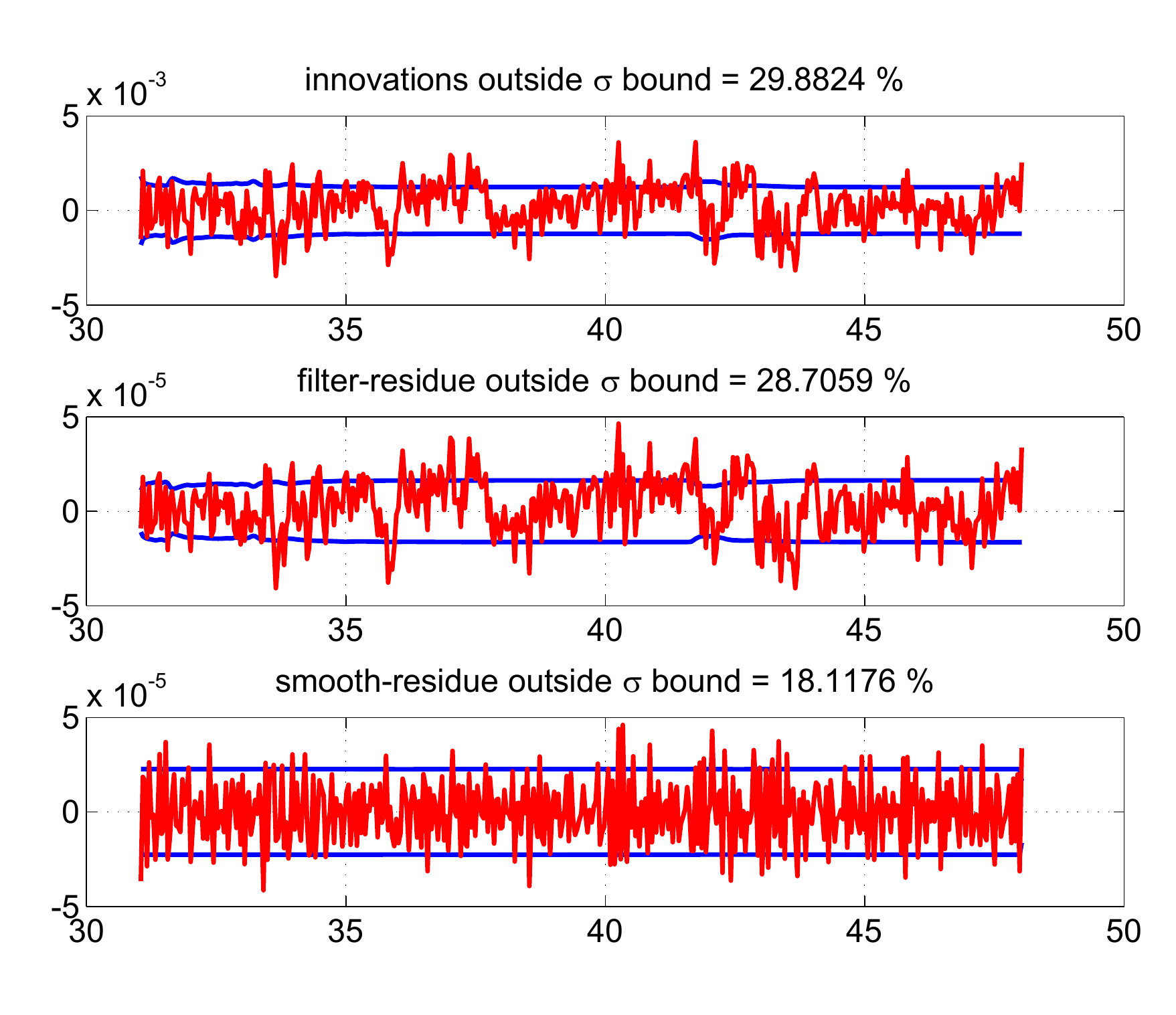}
\caption{The innovations, filtered residue and smoothed residue of measurement 4}
\label{realQ5_innov4}
\end{figure}

\begin{figure}[h]
\includegraphics[width=6in,height=4in]{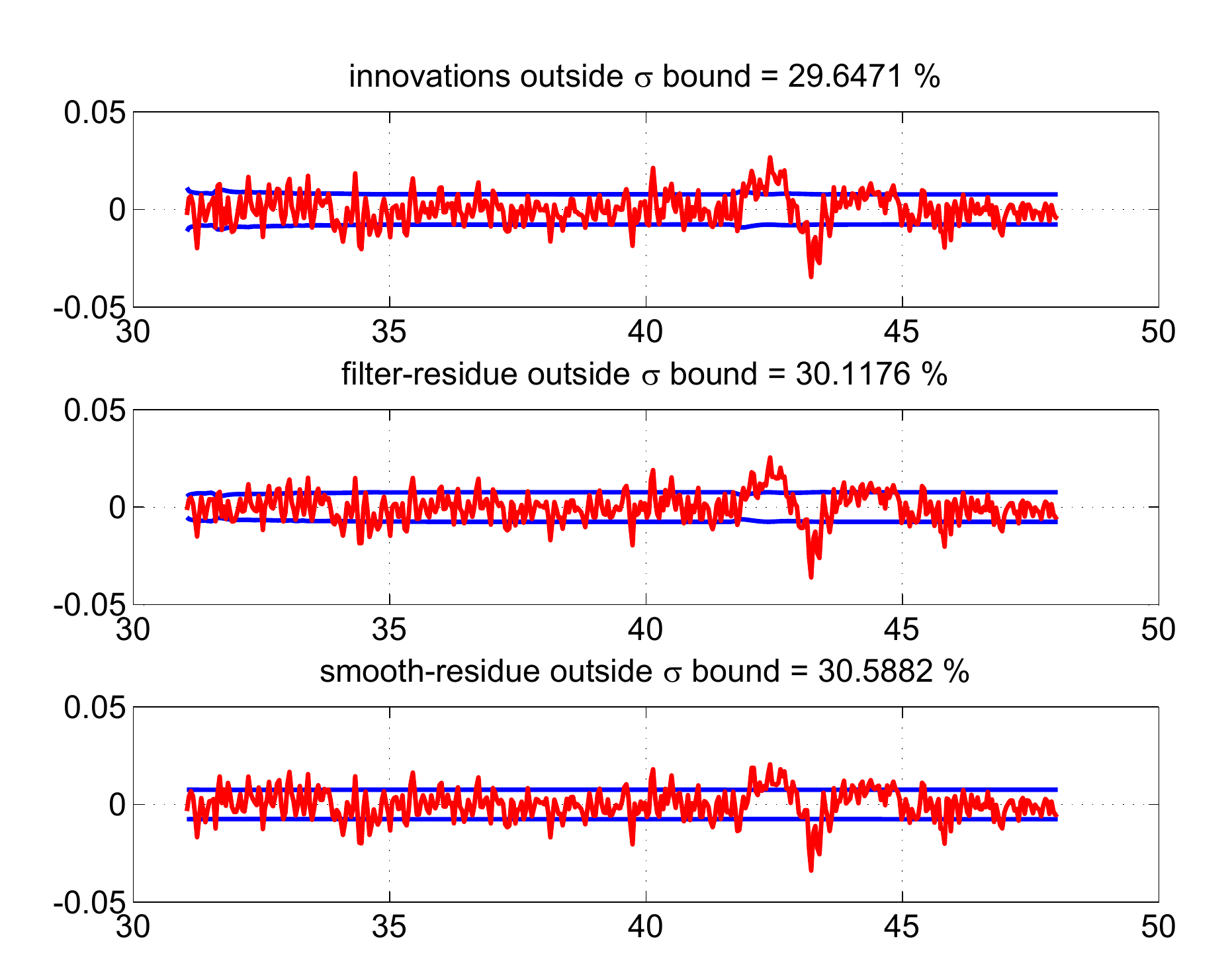}
\caption{The innovations, filtered residue and smoothed residue of measurement 5}
\label{realQ5_innov5}
\end{figure}

\begin{figure}[h]
\includegraphics[width=6in,height=4in]{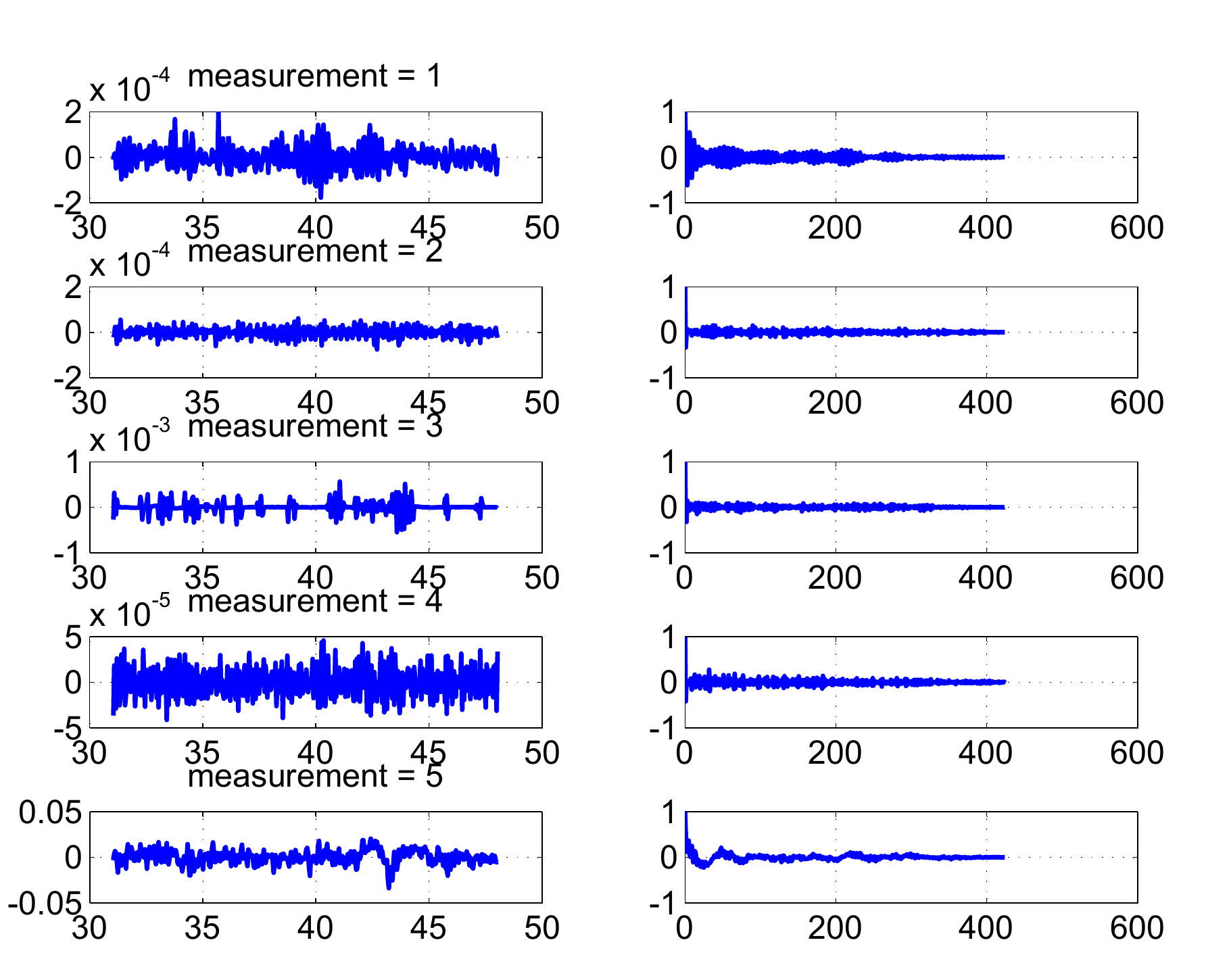}
\caption{Time variation of estimated measurement noise (left) and}
\caption*{their autocorrelation (right)}
\label{realQ5_mnoise}
\end{figure}

\begin{figure}[h]
\includegraphics[width=6in,height=4in]{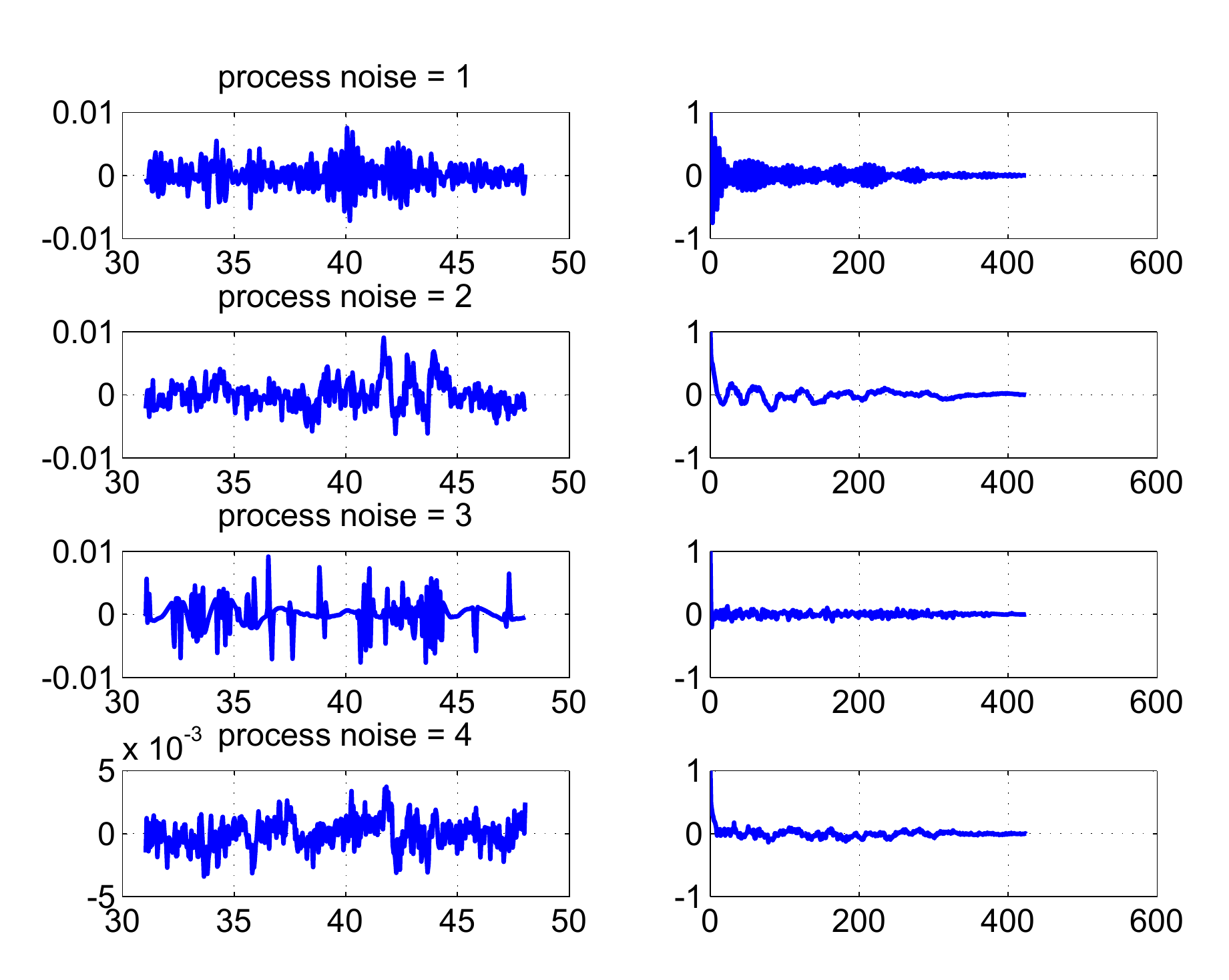}
\caption{Time variation of estimated process noise (left) and}
\caption*{their autocorrelation (right)}
\label{realQ5_pnoise}
\end{figure}

 Case-5 real data is run using the reference EKF (\textbf{Q} $>$ 0) with 100 iterations. Fig. \ref{input5}-\ref{case5_alpha} are the inputs used in state equations. The Fig. \ref{realQ5_P0} shows the variation of parameter estimates and its initial covariance $\mathbf{P_0}$ with iterations and a similar Fig. \ref{realQ5_R} for \textbf{Q} and \textbf{R}. The values of \textbf{J1-J3} are close to the number of measurements ($m=5$) with \textbf{J6-J8} are close to the number of states ($n=4$) as shown in Fig. \ref{realQ5_J} and Table-\ref{tbcase5Q}. The \textbf{J5} is the negative log likelihood cost function. The later Fig. \ref{realQ5_s1}-\ref{realQ5_h5} compares (i) the state dynamics based on the estimated parameter after the filter pass through the data, (ii) the state after measurement update, (iii) the smoothed state and (iv) the measurement.  The Fig. \ref{realQ5_innov1}-\ref{realQ5_innov5} shows the confidence in the innovations, filtered residue and smoothed residue. The estimated measurement and process noise do not appear to have constant statistical characteristics across time as seen in Fig. \ref{realQ5_mnoise} and Fig. \ref{realQ5_pnoise}. Another experiment was carried out by generating a typical data set by using the estimated theta and injecting the estimated \textbf{Q} and \textbf{R} as additive white Gaussian noise. This is to determine the effect of non additive, non White and non Gaussian noise distribution in the real data on the CRBs. After each iteration in the reference recipe the $\Theta$, \textbf{Q} and \textbf{R} were reset as from the real data. Similar experiment was also conducted by updating $\Theta$ as well. It turned out that there is not much of a difference in the final CRBs as can be seen from the Table-\ref{tbnew5}.

\par Finally two other filter runs were carried out using the MT and MS statistics for the estimation of \textbf{Q} and \textbf{R}. The behaviour of the various cost function and in particular \textbf{J6} and \textbf{J7} in Table-\ref{tbcase5QMTMS} shows that the choice of the filter statistics for estimating \textbf{Q} and \textbf{R} in the proposed reference approach is the best possible when compared to other approaches presently considered.

\newpage
\subsection{Remarks on Case-5}
\begin{figure}[h]
\includegraphics[width=6in,height=4in]{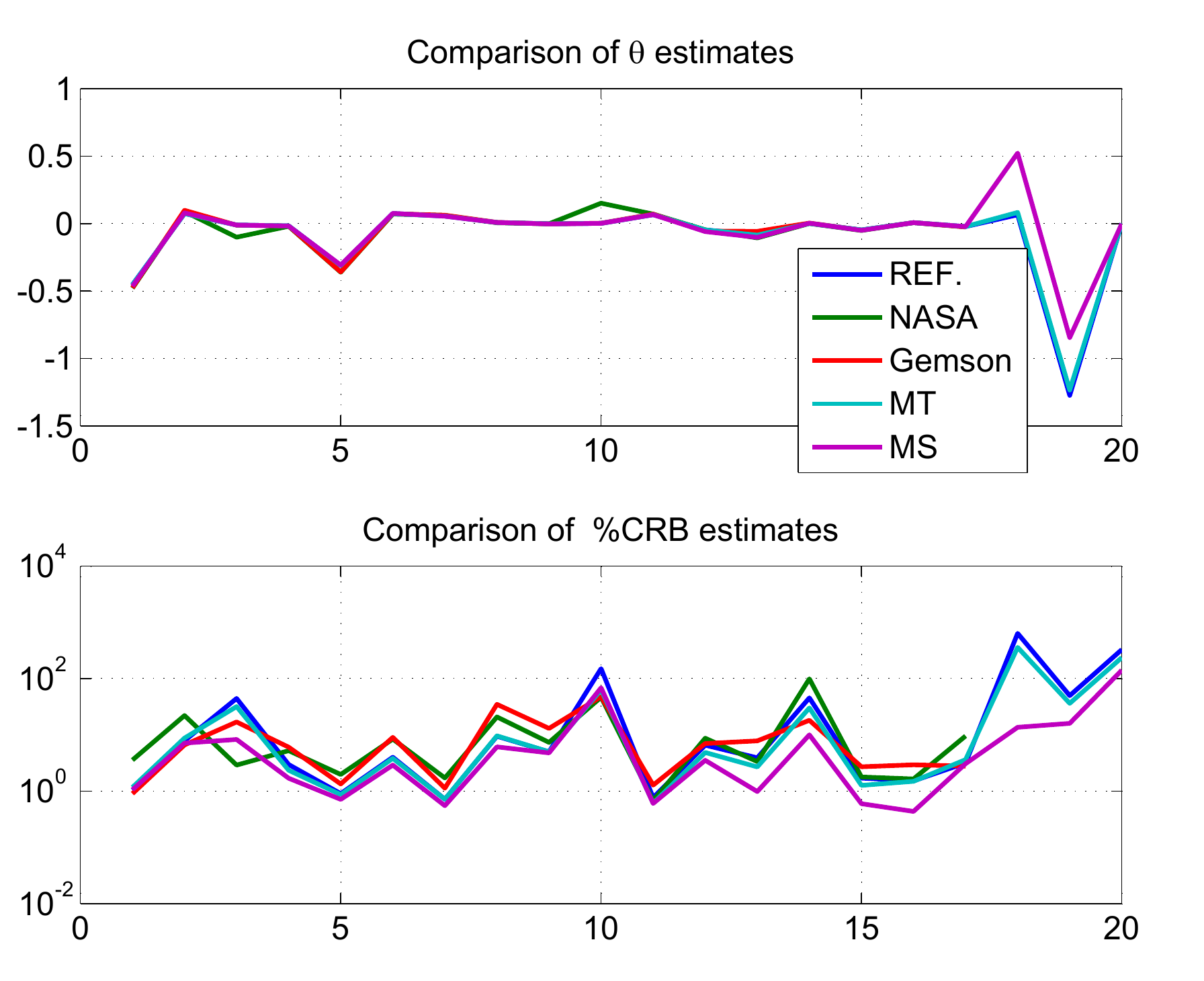}
\caption{Comparison of the parameter estimates and $\%$CRBs by different methods}
\label{comp5}
\end{figure}

The NASA results have been generated assuming \textbf{Q} = 0 and are comparable with reference procedure for the parameter estimates and their CRBs. Further the MT and MS methods give quite different estimates for the \textbf{Q} and \textbf{R} values than in the reference case. We believe that the reference procedure provides the best possible parameter estimates and their uncertainties. From the plot of the parameter estimates and their \%CRB in Fig \ref{comp5}, it can be seen that the parameters 1, 2, 4-9, 11-13, 15, 16 and 17 are strong and the parameters 3, 10, 14, 18, 19 and 20 are the weak ones and others can be considered as modestly controlling the dynamics of the system.  The CRBs as estimated by different methods generally appear to vary widely. However what is interesting is that even the estimate of the strong parameter such as 5 varies widely among the methods. Such a behaviour of the filter across the parameter estimates shows how important is the tuning of the filter statistics namely $\mathbf{P_0}$, \textbf{Q} and \textbf{R} in parameter estimation and their uncertainties.
\begin{landscape}
 The rounded off 100$\times$C  matrix for case-5 is given by
\begin{footnotesize}
\begin{align*}
\begin{bmatrix}
   100 &  -48  &  -2  &  13  &  -2  &  -1 &   -2  &  -6  &  -7 &   0  &  27 & -3   & -2  &  -4 &  -13  & -14 &  -53  &   0  &   0 &  -14 \\
   -48 & 100   &  4   & -6   &  0   & -3  &   2   & 13   & 13  &   0  & -13 & -1 &   -6 &    4  &  27  &  26 &   96  &   0  &  -1 &   16 \\
    -2 &   4  & 100   &  0   &  0   &  0  &   0   &  0   &  0  &   1  &  -8 &0 &   -1  &   3  &  14  &  14 &    0  &  -1  &  -5 &    1 \\
    13 & -6   &  0  & 100   &-12    &-9   &-15    &-50   &-53  &  -1  &  30 & -4 &   -3  &  -5  & -15  & -16 &   -7  &   0  &   0 &   -2 \\
    -2 &   0  &   0 &  -12  & 100   &  5  &  83    &-4   & -1  &   0  &  -4 & 31 &    2  &  26  &  -1  &   0 &    0  &   0  &   0 &   12 \\
    -1 &  -3  &   0 &   -9  &   5   &100  &   2   &-22   & -5  &   0  &  -3 & 2 &   31  &   1  &  -7  &  -1 &   -1  &   0  &   1 &    0 \\
    -2 &   2  &   0 &  -15  &  83   &  2  & 100   & 16   & 18  &   0  &  -4 &  26 &    1  &  31  &   5  &   6 &    2  &   0  &   0 &   13 \\
    -6 &  13  &   0 &  -50  &  -4   &-22  &  16   &100   & 96  &   2  & -15 & -1 &   -7  &   5  &  30  &  29 &   13  &   0  &   0 &    2 \\
    -7 &  13  &   0 &  -53  &  -1   & -5  &  18   & 96   &100  &   2  & -16 &  0 &   -1  &   6  &  29  &  30 &   13  &   0  &   0 &    2 \\
     0 &   0  &   1 &   -1  &   0   &  0  &   0   &  2    & 2  & 100  &   0 & 0 &    0  &   0  &   0  &   0 &    0  &   0  &   0 &    0 \\
    27 & -13  &  -8 &   30  &  -4   & -3  &  -4   &-15   &-16  &   0  & 100 &  -12 &   -9  & -15  & -49  & -53 &  -14  &  -2   & -1 &   -4 \\
    -3 &  -1  &   0 &   -4  &  31   &  2  &  26   & -1    & 0  &   0  & -12 & 100 &    5  &  83  &  -3  &  -1 &    0  &  15   &  1 &   24 \\
    -2 &  -6  &  -1 &   -3  &   2   & 31  &   1   & -7    &-1  &   0   & -9 &  5 &  100  &   2  & -22  &  -5 &   -1  &   1   & 14 &    0 \\
    -4 &   4  &   3 &   -5  &  26   &  1  &  31   &  5    & 6  &   0  & -15 &  83 &    2  & 100  &  16  &  18 &    5  &  12   &  0 &   27 \\
   -13 &  27  &  14 &  -15  &  -1   & -7  &   5   & 30   & 29  &   0  & -49 &  -3 &  -22  &  16  & 100  &  96 &   26  &  -1   & -3 &    4 \\
   -14 &  26  &  14 &  -16  &   0   & -1  &   6   & 29   & 30  &   0  & -53 &  -1 &   -5  &  18  &  96  & 100 &   27  &   0   & -1 &    5 \\
   -53 &  96  &   0 &   -7  &   0   & -1  &   2   & 13   & 13  &   0  & -14 &  0  &  -1   &  5  &  26  &  27 &  100  &   0   &  0 &   18 \\
     0 &   0  &  -1 &    0  &   0   &  0  &   0   &  0   &  0  &   0  &  -2 & 15  &   1 &   12  &  -1  &   0 &    0  & 100   &  5 &    3 \\
     0 &  -1  &  -5 &    0  &   0   &  1  &   0   &  0   &  0  &   0  &  -1 & 1  &  14  &   0  &  -3  &  -1 &    0  &   5   &100 &    0 \\
   -14 &  16  &   1 &   -2  &  12   &  0  &  13   &  2   &  2  &    0 &   -4  &  24  &   0  &  27  &   4  &   5 &   18  &   3   &  0 &  100
\end{bmatrix}
\end{align*}
\end{footnotesize}
\end{landscape}

\begin{landscape}
\begin{table}[h]
\subsection{Case-5 Tables}
\caption{Real flight test data case-5 results using the reference adaptive EKF\\ No of iterations=100}{}
\label{tbcase5Q}
\begin{center}
\begin{footnotesize}
\begin{tabular}{|c| c| c| c| c| c|c|c|c|c|c| }
\hline
Study &
\makecell{$\Theta$\\ (Ref)} &
\makecell{$\Theta$\\ (NASA)} &
\makecell{$\Theta$\\ (Gemson)} &
\makecell{$\sigma_\Theta$ \\(Ref)} &
\makecell{$\sigma_\Theta$\\ (NASA)} &
\makecell{$\sigma_\Theta$\\ (Gemson)} &
\makecell{\textbf{R} \\ $\times10^{-6}$\\ (Ref)}&
\makecell{\textbf{Q} \\ $\times10^{-6}$\\ (Ref)}&
\makecell{\textbf{J1-J8} \\(Ref) }&
Remarks
\\ \hline


\makecell{$\mathbf{P_0}$ : Scaled up-[0,0;0,\checkmark]\\\textbf{Q} : EM-[\checkmark,0;0,0] \\\textbf{R} : EM-diag} &
\makecell{ -0.4529  \\  0.0739 \\  -0.0108 \\  -0.0168 \\  -0.3108 \\   0.0749  \\  0.0558 \\ 0.0074  \\ -0.0020   \\ 0.0018 \\   0.0655  \\ -0.0474 \\  -0.0818  \\  0.0011 \\ -0.0470  \\  0.0066 \\  -0.0221  \\  0.0650  \\ -1.2731   \\ 0.0007} &
\makecell{ -0.4792 \\   0.0887 \\  -0.10116  \\ -0.0205 \\  -0.36 \\  0.0697  \\  0.0612 \\ 0.006 \\  -0.002   \\ 0.1506 \\   0.0705  \\ -0.046 \\  -0.1062 \\   0.0006 \\-0.0513  \\  0.0072  \\ -0.0242  \\  --  \\ --  \\
 --} &
\makecell{ -0.4761 \\   0.0981 \\  -0.0124  \\ -0.0182 \\  -0.3585  \\  0.0731  \\  0.0622 \\ 0.0089 \\  -0.0023   \\ 0.0023 \\   0.0703  \\ -0.0557 \\  -0.0576 \\   0.0033 \\-0.048  \\  0.0068  \\ -0.0251  \\ --  \\ --  \\
--} &
\makecell{    0.0047 \\   0.0058  \\  0.0048 \\   0.0005  \\  0.0028 \\   0.0030  \\  0.0004 \\ 0.0007 \\   0.0001   \\ 0.0027  \\  0.0005  \\  0.0031  \\  0.0032  \\  0.0005 \\ 0.0008 \\   0.0001  \\  0.0007  \\  0.4146  \\  0.6334   \\ 0.0023 } &
\makecell{ 0.01711  \\  0.01955  \\  0.00294  \\  0.00107 \\   0.00713 \\   0.005884  \\  0.001050 \\ 0.001252 \\   0.0001467  \\ 0.07034 \\  0.000478  \\  0.004006 \\   0.003562 \\   0.0005924 \\ 0.0009139  \\  0.0001181 \\   0.002307 \\ --  \\  --   \\ -- } &
\makecell{ 0.0043  \\  0.0065  \\  0.0021  \\  0.0011 \\   0.0048 \\   0.0066  \\  0.0007 \\ 0.0031 \\   0.0003   \\ 0.0011  \\  0.0009  \\  0.0039 \\   0.0045 \\   0.0006 \\ 0.0013  \\  0.0002 \\   0.0007  \\  --  \\  --   \\ -- } &
\makecell{ 0.09  \\  0.06   \\ 0.23  \\  0.02 \\   57.68} &
\makecell{4.191  \\  5.171 \\   4.943  \\  1.472} &
\makecell{ 4.7460  \\  4.8152  \\  3.5033 \\   0.0004 \\ -54.7239  \\  3.9627  \\  3.9620 \\ 3.7135} &
\makecell{Cost functions converge\\ to the expected values}
\\ \hline

\end{tabular}
\end{footnotesize}
\end{center}
\end{table}

\begin{table}[h]
\caption{Case-5 results using simulated Additive White Gaussian Noise}{}
\label{tbnew5}
\begin{center}
\begin{tabular}{|c| c| c| c| c|}
\hline
Study &
\makecell{$\sigma_\Theta$ \\(Simulated-without\\ updating $\Theta$)} &
\makecell{$\sigma_\Theta$ \\(Simulated-with\\ updating $\Theta$)} &
\makecell{$\sigma_\Theta$ \\(Ref)} &
Remarks
\\ \hline

\multicolumn{5}{|c|}{\makecell{Case-5 data generated using simulated measurement and process noise (AWGN) \\ of variance \textbf{Q} and \textbf{R} estimated by Reference EKF (\textbf{Q} $>$ 0)} } \\ \hline

\makecell{$\mathbf{P_0}$ : Scaled up-[0,0;0,\checkmark]\\\textbf{Q} : \textbf{Q} (Ref) \\\textbf{R} :  \textbf{R} (Ref)} &

\makecell{  0.0052 \\   0.0059  \\  0.0048 \\   0.0005 \\   0.0030 \\   0.0031  \\  0.0004 \\ 0.0007   \\ 0.0001 \\   0.0027  \\  0.0006  \\  0.0032  \\  0.0034  \\  0.0005 \\0.0007 \\   0.0001  \\  0.0007  \\  0.4092  \\  0.6996  \\  0.0023} &
\makecell{  0.0052  \\  0.0059 \\   0.0048 \\   0.0005   \\ 0.0030 \\   0.0031  \\  0.0004 \\ 0.0007   \\ 0.0001 \\   0.0027 \\   0.0006  \\  0.0032  \\  0.0034  \\  0.0005 \\ 0.0007 \\   0.0001  \\  0.0007   \\ 0.4092  \\  0.6996 \\   0.0023} &
\makecell{    0.0047 \\   0.0058  \\  0.0048 \\   0.0005  \\  0.0028 \\   0.0030  \\  0.0004 \\ 0.0007 \\   0.0001   \\ 0.0027  \\  0.0005  \\  0.0031  \\  0.0032  \\  0.0005 \\ 0.0008 \\   0.0001  \\  0.0007  \\  0.4146  \\  0.6334   \\ 0.0023 } &

\makecell{No Significant \\ change in $\sigma_\Theta$}
\\ \hline

\end{tabular}
\end{center}
\end{table}

\begin{table}[h]
\caption{Real flight test data case-5 results using the MT and MS method. \\ No of iterations=100 }{}
\label{tbcase5QMTMS}
\begin{center}
\begin{footnotesize}
\begin{tabular}{|c| c| c| c| c| c|| c|c|c|c|c|c| }
\hline
Study &
\makecell{$\Theta$\\ (MT)} &
\makecell{$\sigma_\Theta$ \\(MT)} &
\makecell{\textbf{R} (MT)\\ $\times10^{-6}$ }&
\makecell{\textbf{Q} (MT)\\ $\times10^{-6}$}&
\makecell{\textbf{J1-J8} \\(MT) }&

\makecell{$\Theta$\\ (MS)} &
\makecell{$\sigma_\Theta$\\ (MS)} &
\makecell{\textbf{R} (MS) \\ $\times10^{-6}$}&
\makecell{\textbf{Q} (MS)\\ $\times10^{-6}$}&
\makecell{\textbf{J1-J8} \\(MS) }&
Remarks
\\ \hline


\makecell{$\mathbf{P_0}$ : Scaled up-[0,0;0,\checkmark]\\\textbf{Q} : MT/MS-[\checkmark,0;0,0] \\\textbf{R} : MT/MS-diag} &

\makecell{  -0.4541 \\   0.0741 \\  -0.0107  \\ -0.0170  \\ -0.3112 \\   0.0733  \\  0.0557\\ 0.0073 \\  -0.0020   \\ 0.0019 \\   0.0657  \\ -0.0473 \\  -0.0854 \\   0.0010 \\ -0.0476 \\   0.0067  \\ -0.0221 \\   0.0821 \\  -1.2336  \\  0.0011} &
\makecell{  0.0053 \\   0.0065 \\   0.0034  \\  0.0004 \\   0.0027  \\  0.0028 \\   0.0004 \\0.0007  \\  0.0001   \\ 0.0012\\    0.0004 \\   0.0023  \\  0.0023  \\  0.0003 \\ 0.0006 \\   0.0001  \\  0.0008  \\  0.2941  \\  0.4470   \\ 0.0026} &
\makecell{ 0.0028  \\  0.0189  \\  0.0039  \\  0.0041 \\   0.0736} &
\makecell{2.0481  \\  3.7876  \\  1.0057  \\  0.5502} &
\makecell{ 4.3450 \\   4.3888 \\   3.1039 \\   0.0003 \\ -51.3490  \\  9.3006  \\  9.3005 \\ 3.5105} &

\makecell{  -0.4642 \\   0.0797  \\ -0.0109 \\  -0.0177  \\ -0.3080  \\  0.0757 \\   0.0546 \\ 0.0082 \\  -0.0021   \\ 0.0019  \\  0.0662 \\  -0.0596 \\  -0.1021 \\   0.0030 \\ -0.0502  \\  0.0069  \\ -0.0228   \\ 0.5223 \\  -0.8452  \\ -0.0017} &
\makecell{0.0049  \\  0.0057 \\   0.0009 \\   0.0003 \\   0.0022 \\   0.0022 \\   0.0003 \\ 0.0005 \\   0.0001\\    0.0013 \\   0.0004  \\  0.0021  \\  0.0010 \\   0.0003 \\ 0.0003 \\   0.00003 \\   0.0007  \\  0.0710  \\  0.1355   \\ 0.0024} &
\makecell{  0.0130  \\  0.0881 \\   0.0045  \\  0.0364  \\  0.0601} &
\makecell{  0.0005 \\   0.0007  \\  1.0975  \\  0.0016} &
\makecell{  4.8127 \\   4.8200  \\  4.5173 \\   0.0003 \\ -47.2441 \\   7.5931  \\  7.5896 \\ 3.9681} &

\makecell{Cost functions are \\not close to their \\expected values in\\ MT and MS method} \\ \hline

\end{tabular}
\end{footnotesize}
\end{center}
\end{table}
\end{landscape}

\clearpage
\subsection{Case-5 Figures}

\chapter{Conclusions and Suggestions for Further Work}
\label{ch6}
One can now summarize the present study as follows,
\begin{enumerate}
\item A comparative study of the three different adaptive estimation technique is carried out, namely extended Expectation Maximisation (EM) suggested by Bavdekar et al. \cite{Bavdekar2011} (2011), Maximum Likelihood suggested by Mohamed and Schwarz \cite{MS1999} (1999), Covariance matching suggested by Myers and Tapley \cite{MT1976} (1976) and Gemson\cite{Gemson1991} (1991). A reference adaptive EKF procedure is proposed which is stable for a large range of initial guess values for the tuning parameters but sensitive to the measurement data.
\item In order to tune the filter and estimate \textbf{Q, R} simultaneously, proper choice of their estimation `statistics' is necessary.
\item The connection between the Newton Raphson (NR) results and the Kalman filter has been established for \textbf{Q} = 0 case. The filter would give very close results to the NR optimization technique when the filter statistics ($\mathbf{P_0}$ and \textbf{R}) are tuned properly.
\item The confidence in the filter results is best based on cost functions (\textbf{J1-J8}) and are summarised in Table-\ref{tbconcl} for \textbf{Q} $>$ 0 case.
\item A correct choice of $\mathbf{P_0}$ is necessary to achieve the proper CRB which can be estimated either by a simple heuristic Scaling up method, Inverse of the information matrix (IIM) or the smoothed $\mathbf{P_0}$ being scaled up.
\item A constant signal model with zero process noise is used for the unknown constant parameters rather than a random walk model. The augmented parameters learns recursively through different time and iterations with the right choice of $\mathbf{P_0}$.
\item The reference adaptive EKF achieves statistical equilibrium after few iterations and the mean value of the estimated unknown parameters and the noise statistics over many simulations tend to a stable value.
\item The reference adaptive EKF is robust to the change in initial conditions over a wide range. A thorough sensitivity study is conducted on simple and complex systems and the results are provided.
\item Real flight test data from NASA open source is processed using the proposed reference adaptive EKF and the results are compared with the existing techniques.
\item The estimated measurement and process noise may not be Additive White Gaussian Noise (AWGN) in a real case scenario which needs to be studied further.
\item It is useful to rework out the case studies, and real flight test data available in many of the text books, and references, and compare the results therein with the present method.
\item The estimated noise statistics refer to the complete data processed off line. In many applications when the system dynamics is very fast a rapid adjustment of \textbf{Q} becomes necessary. Based on some of the ideas in the present work it should be possible to tune \textbf{Q} rapidly on line.
\item In most variants of the Kalman filter implementation such as Unscented, Particle, and Ensemble filters generally researchers do not appear to be concerned about tuning the Kalman filter. The exploration of the tuning the statistics in such implementation in other fields of application would be interesting.
\item One should routinely tune off line using simulated data and then apply it for on line real and real time data analysis.
\item The present work has concentrated on tuning the filter statistics. The constant gain Kalman filter approach offers a much faster implementation for many real time applications to solve many involved problems using the generalised cost functions suggested in the present work.
\end{enumerate}

\clearpage
\begin{table}[h]
\begin{center}
\caption{Cost convergence of different methods.}
\caption*{C-Converge, U/D-Under Perform/Diverge}{}
\label{tbconcl}
\begin{footnotesize}
\begin{tabular}{|c| c| c| c| c| c|c| }
\hline
Adaptive EKF &
Constant  &
Ramp &
SMD &
Longitudinal &
Lateral &
Remarks
\\[10pt] \hline

Reference & C & C & C & C & C & \makecell{Cost functions\\ converge to the\\ expected value} \\[10pt] \hline
MT & U/D & U/D &U/D & U/D & U/D & \makecell{Converge only if \\\textbf{R} is known} \\[10pt] \hline
MS & U/D & U/D & U/D & U/D & U/D & \makecell{Converge only if \\\textbf{R} is known}  \\[10pt] \hline
\makecell{Gemson} & C & C & C & C & C & \makecell{Initial \textbf{R} should be close\\ to true value} \\[10pt] \hline

\end{tabular}
\end{footnotesize}
\end{center}
\end{table}

Finally a few suggested term paper topics are
\begin{itemize}
\item The Least Squares (LS) history and Gauss procedure.
\item Bayesian and Frequentist approaches.
\item Prior probabilities.
\item Rework the earlier test case studies in books and references.
\item Try to automate the evaluation of the partial derivatives of functions needed in EKF.
\item Use of simple Euler integration for propagating the differential equations and see how the estimates and CRBs are affected.
\item Tuning the other variants, combinations, and implementations of the Kalman filter.
\item Try to fine tune the scaled up full $\mathbf{P_0}$ of the parameters for statistical equilibrium.
\item Attempt if a combination of some arbitrary combination $\mathbf{P_0}$ and \textbf{Q} can lead to the same CRB even in the zero process noise case?
\end{itemize}

\begin{appendices}

\titleformat{\chapter}[display]
{\bfseries\Large}
{Appendix \ \thechapter\Large}{5pt}{}[\vspace{-1cm}\rule{\textwidth}{2pt}]

\afterpage{\null\newpage}
\chapter{MMLE and Cramer Rao Bound}
\par The following section explain the parameter estimation technique using modified Newton Raphson (NR) optimisation for zero process noise (\textbf{Q} = 0) case. It also gives a bound on the estimated parameters which is a check for efficiency of any estimator.
\section{Newton Raphson Optimisation and CRB}
\label{NR}
\par Method of maximum likelihood estimation (MMLE) is one of the fundamental method in estimation theory (ET) developed by Fisher \cite{Fisher1922} (1922). The likelihood function is given by
\begin{equation*}
L(\Theta|Z)=p(Z|\Theta)
\end{equation*}
where p(.) denotes the joint probability density function. The multivariate Gaussian analysis gives us
\begin{align*}
L(\Theta|Z)=&\frac{(2\pi)^{-N/2}}{|\textbf{R}|^N}exp\left[-\frac{1}{2}\sum_{k=1}^{N}{\left(Z_k-h(x_{k},\Theta)\right)^T \textbf{R}^{-1}(Z_k-h(x_{k},\Theta))}\right]
\end{align*}
Maximizing the log likelihood function leads to the minimisation of cost function \textbf{J} given by
\begin{equation*}
\mathbf{J}(\Theta)=\sum_{k=1}^{N}{(Z_k-h(x_{k},\Theta))^TR^{-1}(Z_k-h(x_{k},\Theta))}
\end{equation*}
The above function is minimized for $\Theta$ using the well known modified Newton Raphson iterative procedure,
\begin{equation}
  \Theta^{i+1}=  \Theta^i-(D^2J^i)^{-1}DJ^i
\end{equation}
where `i' is the iteration number. The first and second order gradients of \textbf{J} are respectively,
\begin{align*}
DJ^i&=\sum_{k=1}^{N} {\frac{\partial h(  x_{k},  \Theta^i)}{\partial \Theta}}^TR^{-1}(Z_k-h(  x_{k},  \Theta^i))\\
D^2J^i&\approx \sum_{k=1}^{N}{\frac{\partial h(  x_{k},  \Theta^i)}{\partial \Theta}}^TR^{-1}{\frac{\partial h(  x_{k},  \Theta^i)}{\partial \Theta}}
\end{align*}
where the partial derivative (${\partial h}/{\partial \Theta}$) of size $m \times p$ can be found using finite forward differences. The gradients are evaluated using the true initial state and the initial parameter, $\Theta^1$ is chosen within $\pm 5\%$ error. If the measurement noise \textbf{R} is unknown then it can be estimated as
\begin{align}
\label{R0}  {\textbf{R}}^{i+1}=\frac{1}{N}\sum_{k=1}^{N}{(Z_k-h(  x_k,  \Theta^i))(Z_k-h(  x_k,  \Theta^i))^T}
\end{align}
The Cramer Rao Bound or the uncertainty in the parameter estimate, $\Theta^{i_{max}}$ is given by
\begin{align}
CRB=(D^2J^{i_{max}})^{-1}
\end{align}
where `$i_{max}$' is the maximum iteration number and `diag' is the MATLAB\textsuperscript{\textregistered} notation for the diagonal values of a square matrix. The iterations are stopped when there is no significant change (say $\gamma=10^{-5}$) in the cost function \textbf{J},
\begin{align*}
\frac{|\mathbf{J}^{i}-\mathbf{J}^{i-1}|}{|\mathbf{J}^i|}<\gamma
\end{align*}
As per Cramer Rao Bound criterion, for any estimator $P_{\Theta_{jj}}\geq CRB_{jj}$ for $j^{th}$ parameter. If the estimator gives the CRB ratio $\frac{CRB_{jj}}{P_{\Theta_{jj}}}\approx$1 then it is said to be an efficient estimator.

\newpage
\section{Posterior Cramer Rao Bound (PCRB)}
\label{PCRB}
\par A recursive equation for the lower bound on the discrete state with process noise was given by Tichavsky \cite{Tichavsky1998} (1998) based on Van Trees version of Cramer Rao inequality which is termed as Posterior Cramer Rao Bound (PCRB) and is given by
\begin{align}
PCRB_{k}=D_k^{22}-D_k^{21}(PCRB_{k-1}+D_k^{11})^{-1}D_k^{12}
\end{align}
where
\begin{align*}
D_k^{11}=&E[f'^T(x_{k-1},\Theta)\textbf{Q}^{-1}f'(x_{k-1},\Theta)]\\
D_k^{22}=&\textbf{Q}^{-1}+E[h'^T(x_k,\Theta)\textbf{R}^{-1}h'(x_k,\Theta)]\\
D_k^{12}=&[D_k^{21}]^T=E[f'^T(x_{k-1},\Theta)]\textbf{Q}^{-1}
\end{align*}
where $f'(x_{k-1},\Theta)=\left[\frac{\partial{f}}{\partial{x}}\right]_{x=x_{k-1}}$ $\&$ $h'(x_{k},\theta)=\left[\frac{\partial{h}}{\partial{x}}\right]_{x=x_{k}}$ are the Jacobians evaluated at true state and parameter values. The expectation operator used in calculating $D_k$'s can be replaced by the average values taken over many simulations (say 50). The diagonal values of the PCRB evaluated at the last time instant ($PCRB_N$) is used as reference check for the state covariance ($P_x$) obtained from $P_{N|N}=[P_x, \checkmark;\checkmark,\checkmark]$ for \textbf{Q} $>$ 0 case.


\chapter{Simple Modeling and Testing of the Estimates}
\label{QM}

\par Consider a second order polynomial with about 100 points added at each point by a white Gaussian noise. Now, instead of choosing a second order polynomial let us say one chooses a lower order 1, appropriate order 2, higher orders 5 and 25. Then using a cost function \textbf{J} based on a minimum mean square error it is possible to estimate all the unknown parameters in the appropriate order polynomial.

\par
\begin{figure}[t]
\includegraphics[width=6in,height=4in]{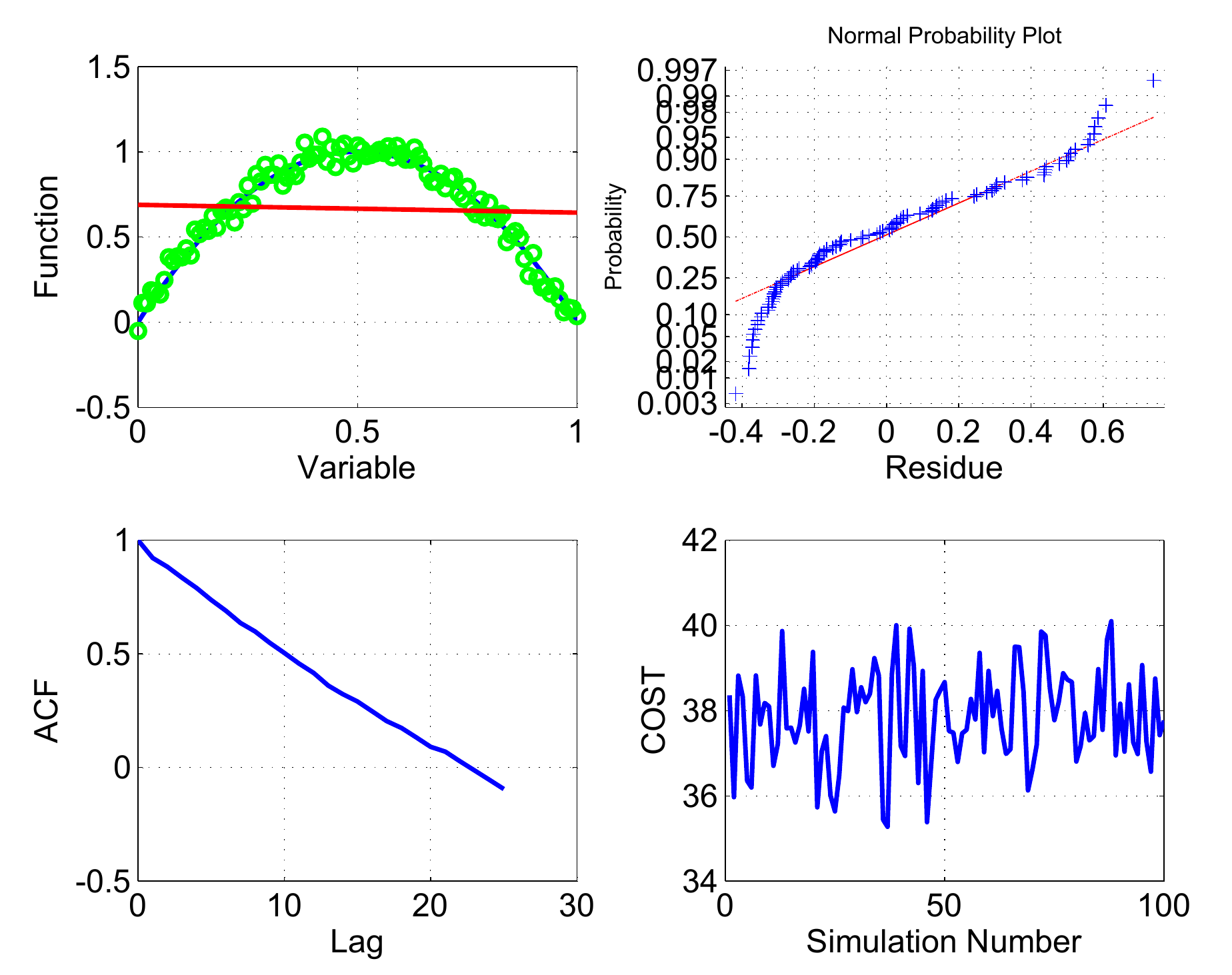}
\caption{Second order Polynomial :  Fit Order  =  1}
\label{ft1}
\end{figure}

\begin{figure}[b]
\includegraphics[width=6in,height=4in]{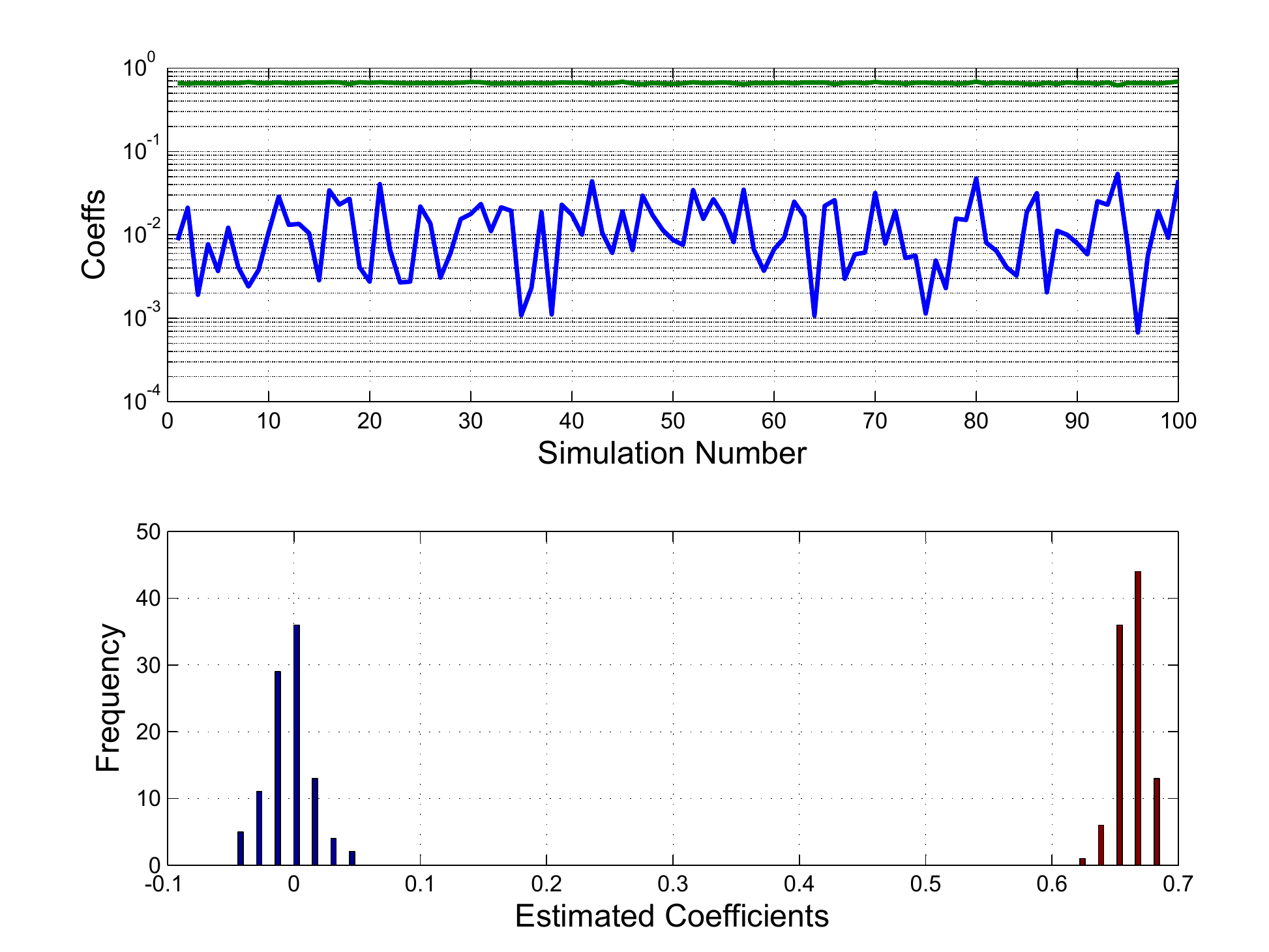}
\caption{Second order Polynomial :  Fit Order  =  1}
\caption*{Coefficient estimates and their Histogram over 100 ensembles}
\label{ft1_hist}
\end{figure}

\begin{figure}[h]
\includegraphics[width=6in,height=4in]{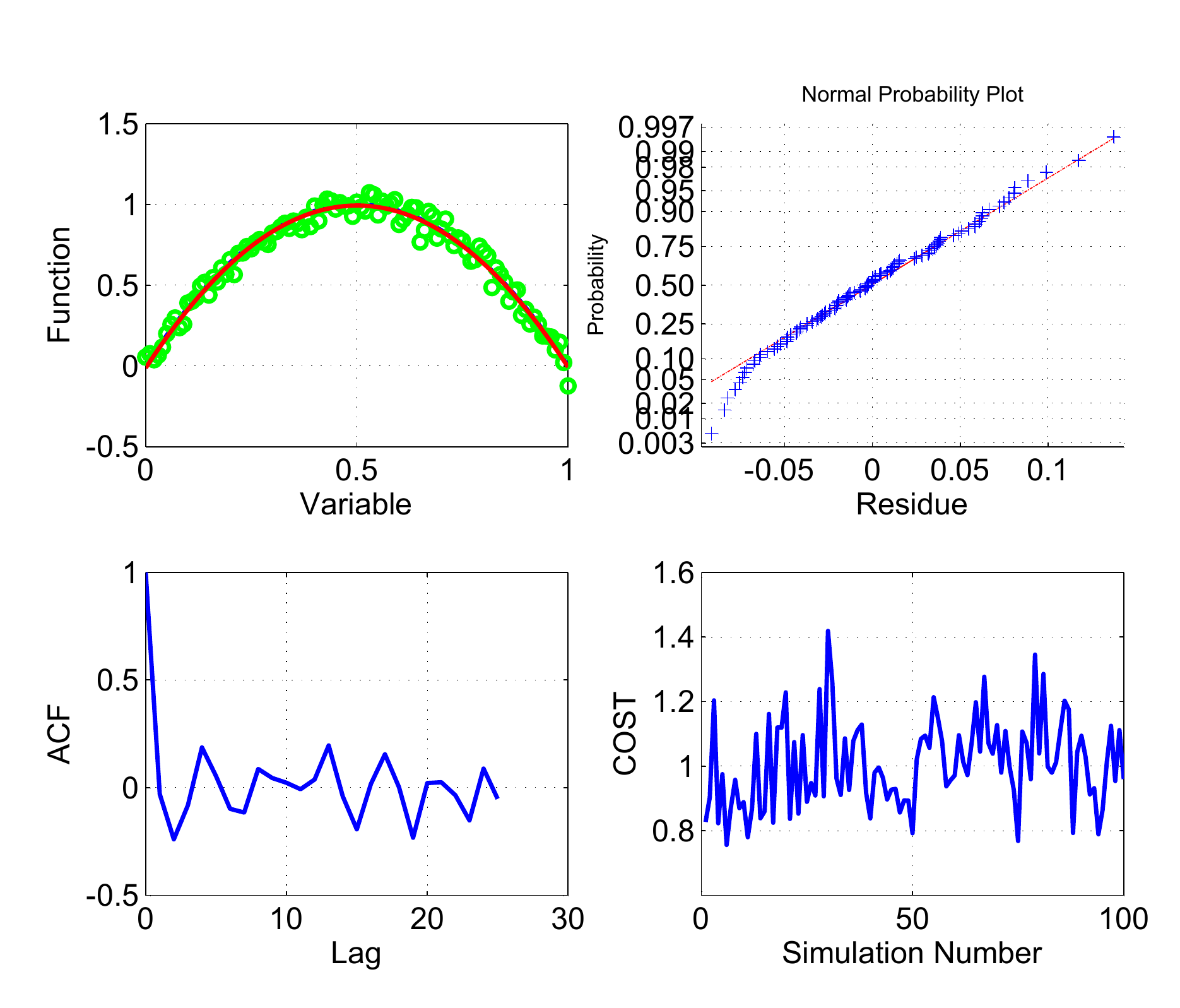}
\caption{Second order Polynomial :  Fit Order  =  2}
\label{ft2}
\end{figure}

\begin{figure}[h]
\includegraphics[width=6in,height=4in]{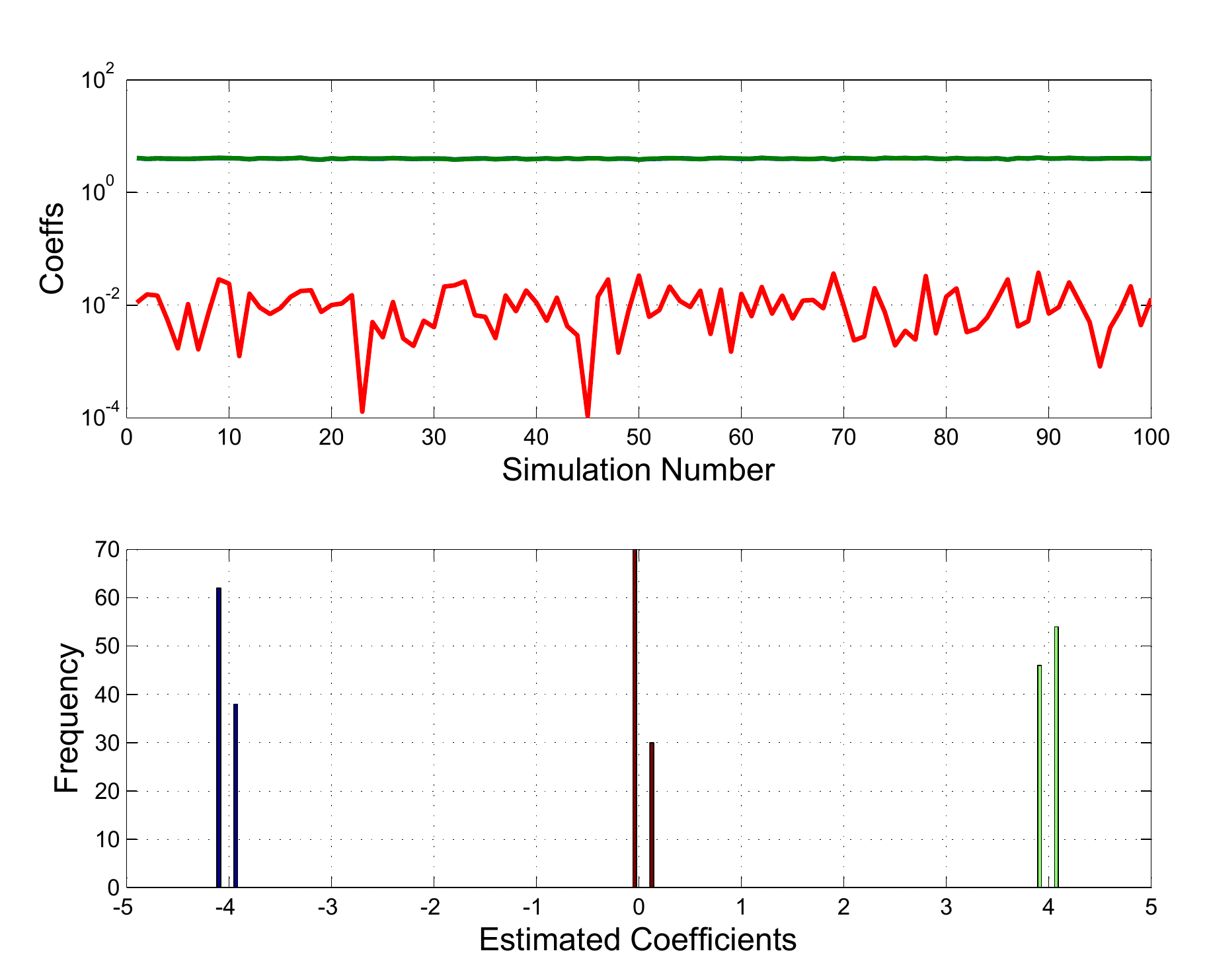}
\caption{Second order Polynomial :  Fit Order  =  2}
\caption*{Coefficient estimates and their Histogram over 100 ensembles}
\label{ft2_hist}
\end{figure}

\begin{figure}[h]
\includegraphics[width=6in,height=4in]{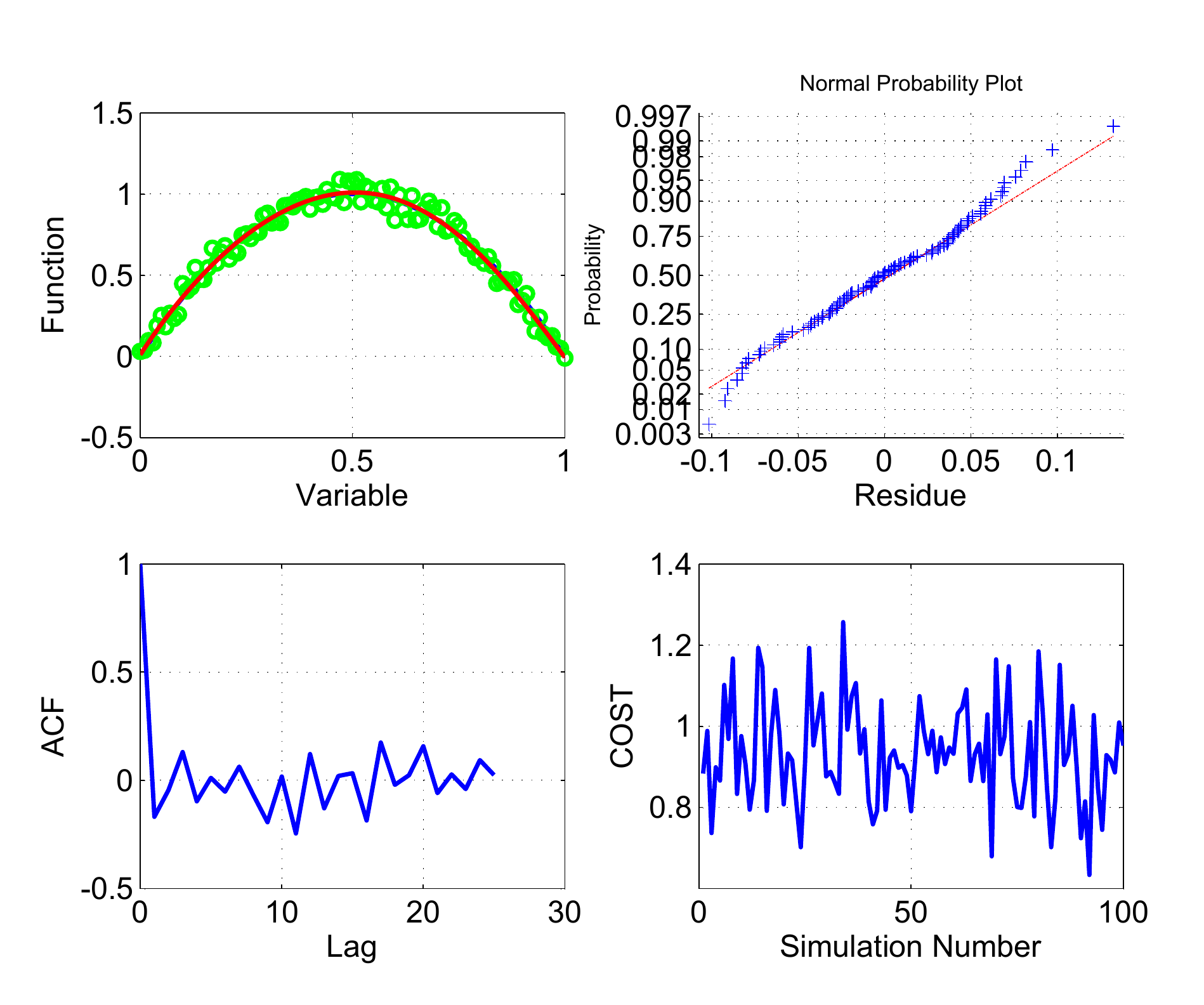}
\caption{Second order Polynomial :  Fit Order  =  5}
\label{ft5}
\end{figure}

\begin{figure}[h]
\includegraphics[width=6in,height=4in]{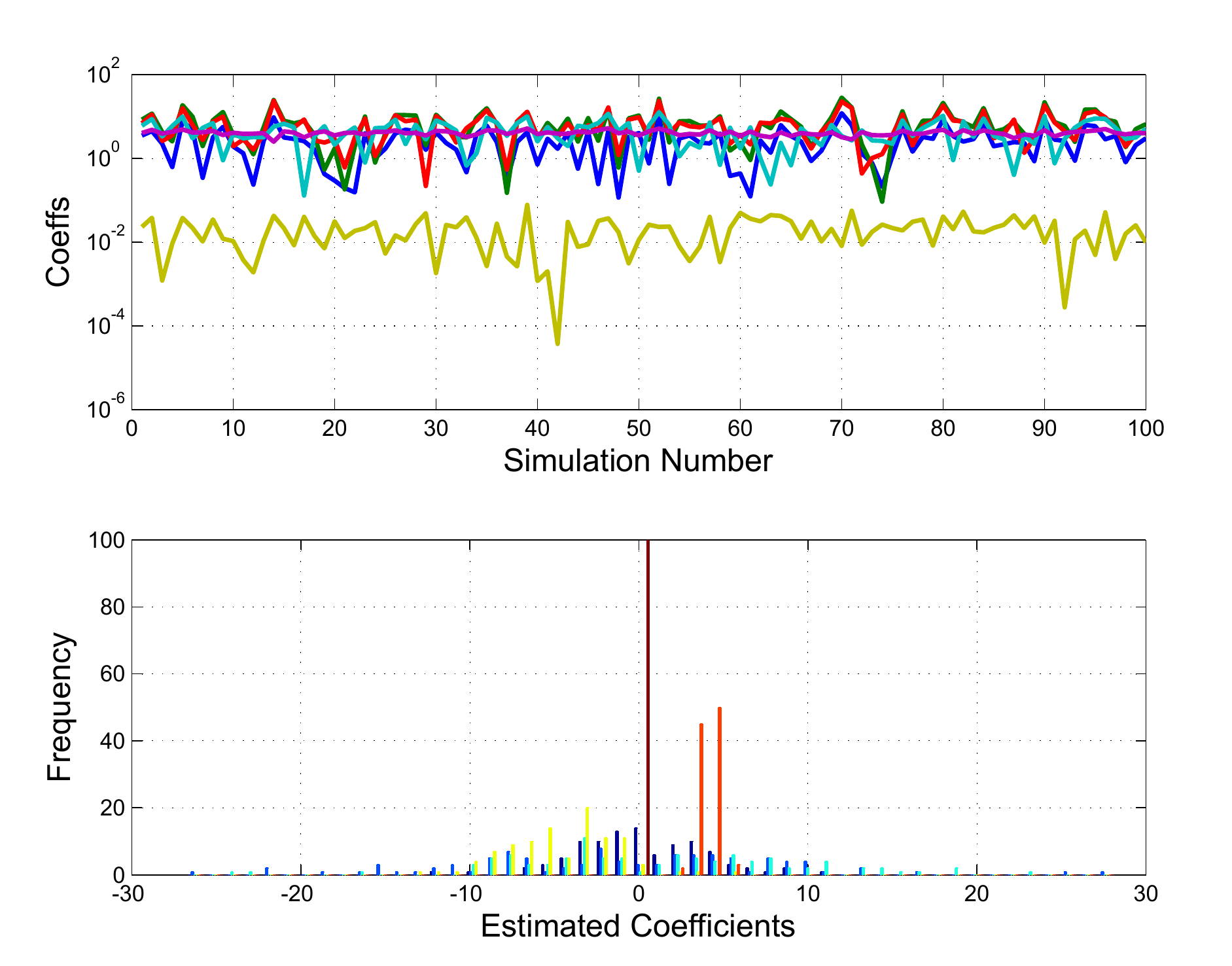}
\caption{Second order Polynomial :  Fit Order  =  5}
\caption*{Coefficient estimates and their Histogram over 100 ensembles}
\label{ft5_hist}
\end{figure}

\begin{figure}[h]
\includegraphics[width=6in,height=4in]{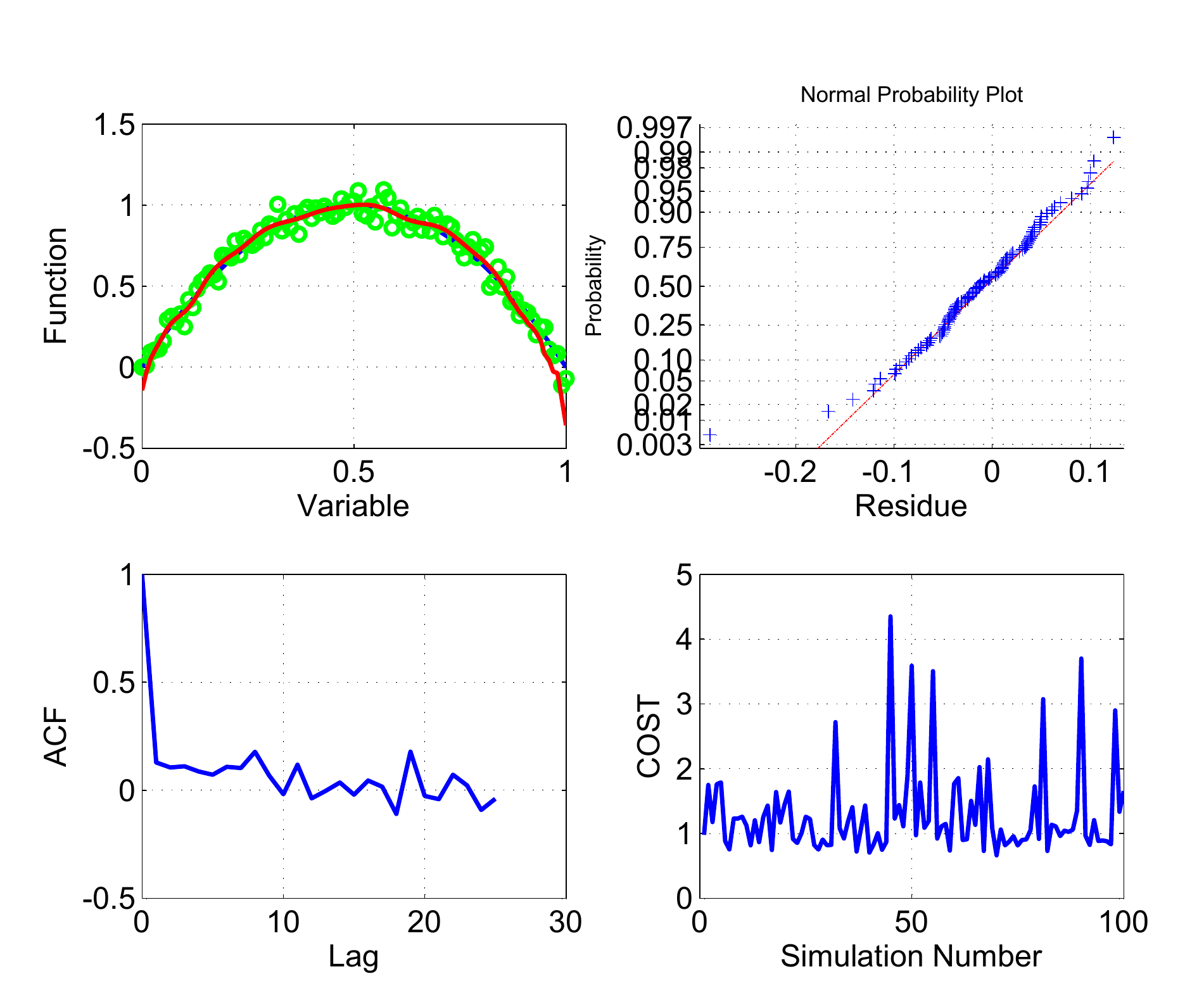}
\caption{Second order Polynomial :  Fit Order  =  25}
\label{ft25}
\end{figure}

\begin{figure}[h]
\includegraphics[width=6in,height=4in]{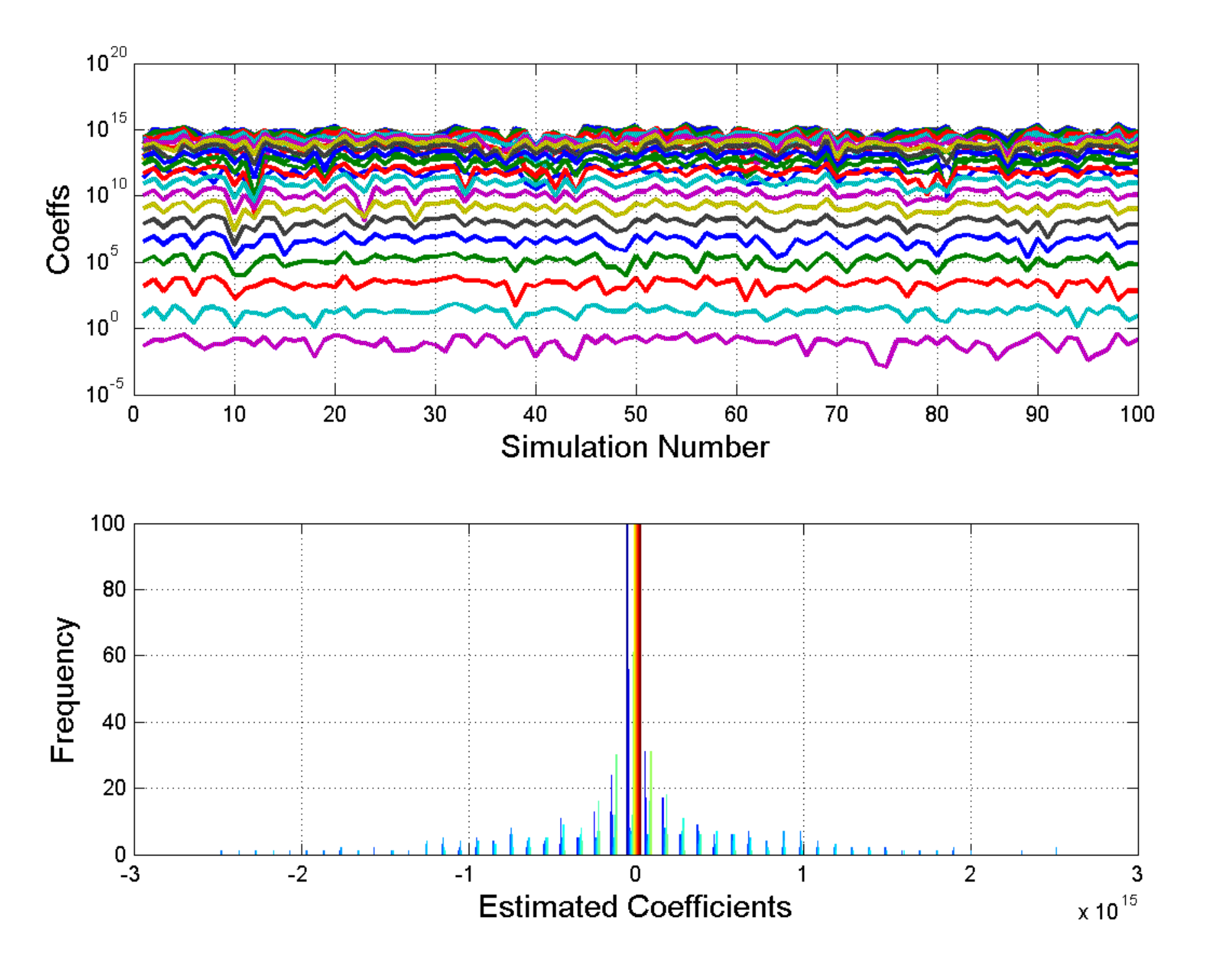}
\caption{Second order Polynomial :  Fit Order  =  25}
\caption*{Coefficient estimates and their Histogram over 100 ensembles}
\label{ft25_hist}
\end{figure}

 The Fig. \ref{ft1}-\ref{ft25_hist} shows a very simple example of a second order polynomial using a MATLAB\textsuperscript{\textregistered} least squares program (shown below) illustrating the Output Error Method. It has no fancy numerical solution of differential equations and the use of sophisticated optimization techniques! If this is studied and further experiments are conducted then you have understood almost everything except for the CRB about the OEM.

\par We consider the following namely (i) A noisy second order polynomial with the fit based on the estimated parameters and the measurements, (ii) a typical normal probability distribution of the difference in the previous items, (iii) a typical autocorrelation of the above difference, and then based on 100 similar data sets the (iv) the normalized cost, (v) the distribution of the coefficients, and (vi) the histogram of the estimated fit coefficients. The first order fit fails the autocorrelation function (ACF) test, and the expected cost behaviour. The second order fit passes ACF test and all the other quantities follow their expected behaviour. The fifth order fit gives no clue about its behaviour except some small wiggles appear in the fit. The twenty fifth order fit is also similar to the fifth but the wiggles are more pronounced in its fit with the second order polynomial. The correlation coefficient matrix (C) of the estimated parameters for 100 simulations for different order fit is shown below. There is a very high positive and negative correlation among the coefficients for higher order fits. The reasonable correlation coefficient among the fit parameters show the appropriateness of the fit of order 2 and the over fitting of the higher order ones. Ideally for the correct order the correlation coefficient matrix should be an identity matrix. It is thus clear just one or two test quantities are not enough but more will have to be formed to avoid deceptive and reach decisive conclusions.

The rounded off 100$\times$C  matrix for order = 1 is given by
\begin{align*}
\begin{bmatrix}
 100  & -83 \\
 -83 &  100
\end{bmatrix}
\end{align*}

The rounded off 100$\times$C  matrix for order = 2 is given by
\begin{align*}
\begin{bmatrix}
  100  & -96  &  72 \\
   -96  & 100 &  -86 \\
    72  & -86  & 100
\end{bmatrix}
\end{align*}

The rounded off 100$\times$C  matrix for order = 5 is given by
\begin{align*}
\begin{bmatrix}
   100 & -100   & 98  & -95  &  89  & -63 \\
  -100 &  100 & -100  &  97  & -91  &  66 \\
    98  &-100  & 100 &  -99 &   94  & -70 \\
   -95  &  97  & -99  & 100  & -98  &  76 \\
    89  & -91  &  94 &  -98 &  100  & -86 \\
   -63  &  66   &-70   & 76  & -86  & 100
\end{bmatrix}
\end{align*}

\begin{landscape}
 The rounded off 100$\times$C  matrix for order = 25 is given by
\begin{scriptsize}
\begin{align*}
\begin{bmatrix}
100	 &	-100	 &	100	 &	-99	 &	99	 &	-97	 &	93	 &	-87	 &	74	 &	-54	 &	29	 &	-6	 &	-11	 &	21	 &	-27	 &	30	 &	-30	 &	29	 &	-26	 &	22	 &	-17	 &	12	 &	-7	 &	7	 &	-30	 &	69	\\
-100	 &	100	 &	-100	 &	100	 &	-99	 &	97	 &	-94	 &	88	 &	-76	 &	56	 &	-31	 &	8	 &	9	 &	-20	 &	26	 &	-29	 &	29	 &	-28	 &	25	 &	-21	 &	16	 &	-11	 &	7	 &	-6	 &	30	 &	-70	\\
100	 &	-100	 &	100	 &	-100	 &	99	 &	-98	 &	95	 &	-89	 &	78	 &	-58	 &	34	 &	-11	 &	-6	 &	17	 &	-24	 &	27	 &	-27	 &	26	 &	-23	 &	19	 &	-14	 &	9	 &	-5	 &	5	 &	-29	 &	72	\\
-99	 &	100	 &	-100	 &	100	 &	-100	 &	99	 &	-97	 &	91	 &	-80	 &	62	 &	-38	 &	15	 &	2	 &	-14	 &	20	 &	-23	 &	24	 &	-23	 &	20	 &	-17	 &	12	 &	-7	 &	3	 &	-3	 &	28	 &	-75	\\
99	 &	-99	 &	99	 &	-100	 &	100	 &	-100	 &	98	 &	-94	 &	84	 &	-66	 &	43	 &	-20	 &	3	 &	8	 &	-15	 &	19	 &	-20	 &	19	 &	-16	 &	13	 &	-8	 &	4	 &	0	 &	0	 &	-27	 &	79	\\
-97	 &	97	 &	-98	 &	99	 &	-100	 &	100	 &	-99	 &	96	 &	-88	 &	72	 &	-50	 &	28	 &	-11	 &	-1	 &	8	 &	-12	 &	13	 &	-12	 &	10	 &	-7	 &	3	 &	2	 &	-5	 &	4	 &	24	 &	-83	\\
93	 &	-94	 &	95	 &	-97	 &	98	 &	-99	 &	100	 &	-99	 &	93	 &	-79	 &	59	 &	-39	 &	22	 &	-10	 &	2	 &	2	 &	-3	 &	3	 &	-1	 &	-2	 &	5	 &	-9	 &	13	 &	-11	 &	-20	 &	88	\\
-87	 &	88	 &	-89	 &	91	 &	-94	 &	96	 &	-99	 &	100	 &	-98	 &	88	 &	-72	 &	53	 &	-37	 &	25	 &	-18	 &	13	 &	-11	 &	11	 &	-12	 &	14	 &	-18	 &	21	 &	-23	 &	21	 &	13	 &	-93	\\
74	 &	-76	 &	78	 &	-80	 &	84	 &	-88	 &	93	 &	-98	 &	100	 &	-96	 &	85	 &	-70	 &	56	 &	-45	 &	38	 &	-33	 &	31	 &	-30	 &	31	 &	-32	 &	34	 &	-36	 &	38	 &	-34	 &	-2	 &	96	\\
-54	 &	56	 &	-58	 &	62	 &	-66	 &	72	 &	-79	 &	88	 &	-96	 &	100	 &	-96	 &	86	 &	-76	 &	67	 &	-61	 &	56	 &	-54	 &	52	 &	-52	 &	52	 &	-53	 &	54	 &	-54	 &	49	 &	-13	 &	-93	\\
29	 &	-31	 &	34	 &	-38	 &	43	 &	-50	 &	59	 &	-72	 &	85	 &	-96	 &	100	 &	-97	 &	91	 &	-85	 &	80	 &	-76	 &	74	 &	-72	 &	71	 &	-70	 &	70	 &	-69	 &	68	 &	-62	 &	28	 &	83	\\
-6	 &	8	 &	-11	 &	15	 &	-20	 &	28	 &	-39	 &	53	 &	-70	 &	86	 &	-97	 &	100	 &	-98	 &	95	 &	-92	 &	89	 &	-87	 &	85	 &	-84	 &	82	 &	-81	 &	79	 &	-76	 &	70	 &	-40	 &	-69	\\
-11	 &	9	 &	-6	 &	2	 &	3	 &	-11	 &	22	 &	-37	 &	56	 &	-76	 &	91	 &	-98	 &	100	 &	-99	 &	97	 &	-96	 &	94	 &	-93	 &	91	 &	-89	 &	87	 &	-84	 &	81	 &	-75	 &	49	 &	56	\\
21	 &	-20	 &	17	 &	-14	 &	8	 &	-1	 &	-10	 &	25	 &	-45	 &	67	 &	-85	 &	95	 &	-99	 &	100	 &	-100	 &	99	 &	-97	 &	96	 &	-95	 &	93	 &	-90	 &	87	 &	-83	 &	77	 &	-55	 &	-45	\\
-27	 &	26	 &	-24	 &	20	 &	-15	 &	8	 &	2	 &	-18	 &	38	 &	-61	 &	80	 &	-92	 &	97	 &	-100	 &	100	 &	-100	 &	99	 &	-98	 &	97	 &	-95	 &	92	 &	-89	 &	85	 &	-79	 &	59	 &	38	\\
30	 &	-29	 &	27	 &	-23	 &	19	 &	-12	 &	2	 &	13	 &	-33	 &	56	 &	-76	 &	89	 &	-96	 &	99	 &	-100	 &	100	 &	-100	 &	99	 &	-98	 &	96	 &	-94	 &	91	 &	-87	 &	82	 &	-63	 &	-34	\\
-30	 &	29	 &	-27	 &	24	 &	-20	 &	13	 &	-3	 &	-11	 &	31	 &	-54	 &	74	 &	-87	 &	94	 &	-97	 &	99	 &	-100	 &	100	 &	-100	 &	99	 &	-98	 &	96	 &	-93	 &	89	 &	-84	 &	66	 &	32	\\
29	 &	-28	 &	26	 &	-23	 &	19	 &	-12	 &	3	 &	11	 &	-30	 &	52	 &	-72	 &	85	 &	-93	 &	96	 &	-98	 &	99	 &	-100	 &	100	 &	-100	 &	99	 &	-97	 &	95	 &	-92	 &	87	 &	-69	 &	-31	\\
-26	 &	25	 &	-23	 &	20	 &	-16	 &	10	 &	-1	 &	-12	 &	31	 &	-52	 &	71	 &	-84	 &	91	 &	-95	 &	97	 &	-98	 &	99	 &	-100	 &	100	 &	-100	 &	99	 &	-97	 &	94	 &	-89	 &	72	 &	31	\\
22	 &	-21	 &	19	 &	-17	 &	13	 &	-7	 &	-2	 &	14	 &	-32	 &	52	 &	-70	 &	82	 &	-89	 &	93	 &	-95	 &	96	 &	-98	 &	99	 &	-100	 &	100	 &	-100	 &	98	 &	-96	 &	92	 &	-75	 &	-32	\\
-17	 &	16	 &	-14	 &	12	 &	-8	 &	3	 &	5	 &	-18	 &	34	 &	-53	 &	70	 &	-81	 &	87	 &	-90	 &	92	 &	-94	 &	96	 &	-97	 &	99	 &	-100	 &	100	 &	-100	 &	98	 &	-95	 &	78	 &	34	\\
12	 &	-11	 &	9	 &	-7	 &	4	 &	2	 &	-9	 &	21	 &	-36	 &	54	 &	-69	 &	79	 &	-84	 &	87	 &	-89	 &	91	 &	-93	 &	95	 &	-97	 &	98	 &	-100	 &	100	 &	-99	 &	97	 &	-81	 &	-35	\\
-7	 &	7	 &	-5	 &	3	 &	0	 &	-5	 &	13	 &	-23	 &	38	 &	-54	 &	68	 &	-76	 &	81	 &	-83	 &	85	 &	-87	 &	89	 &	-92	 &	94	 &	-96	 &	98	 &	-99	 &	100	 &	-99	 &	84	 &	35	\\
7	 &	-6	 &	5	 &	-3	 &	0	 &	4	 &	-11	 &	21	 &	-34	 &	49	 &	-62	 &	70	 &	-75	 &	77	 &	-79	 &	82	 &	-84	 &	87	 &	-89	 &	92	 &	-95	 &	97	 &	-99	 &	100	 &	-90	 &	-30	\\
-30	 &	30	 &	-29	 &	28	 &	-27	 &	24	 &	-20	 &	13	 &	-2	 &	-13	 &	28	 &	-40	 &	49	 &	-55	 &	59	 &	-63	 &	66	 &	-69	 &	72	 &	-75	 &	78	 &	-81	 &	84	 &	-90	 &	100	 &	-10	\\
69	 &	-70	 &	72	 &	-75	 &	79	 &	-83	 &	88	 &	-93	 &	96	 &	-93	 &	83	 &	-69	 &	56	 &	-45	 &	38	 &	-34	 &	32	 &	-31	 &	31	 &	-32	 &	34	 &	-35	 &	35	 &	-30	 &	-10	 &	100				
\end{bmatrix}
\end{align*}
\end{scriptsize}
\end{landscape}

\includepdf[pages={1},scale=.9,offset=20 0,pagecommand=\section*{Source Code for Polynomial Fit :}]{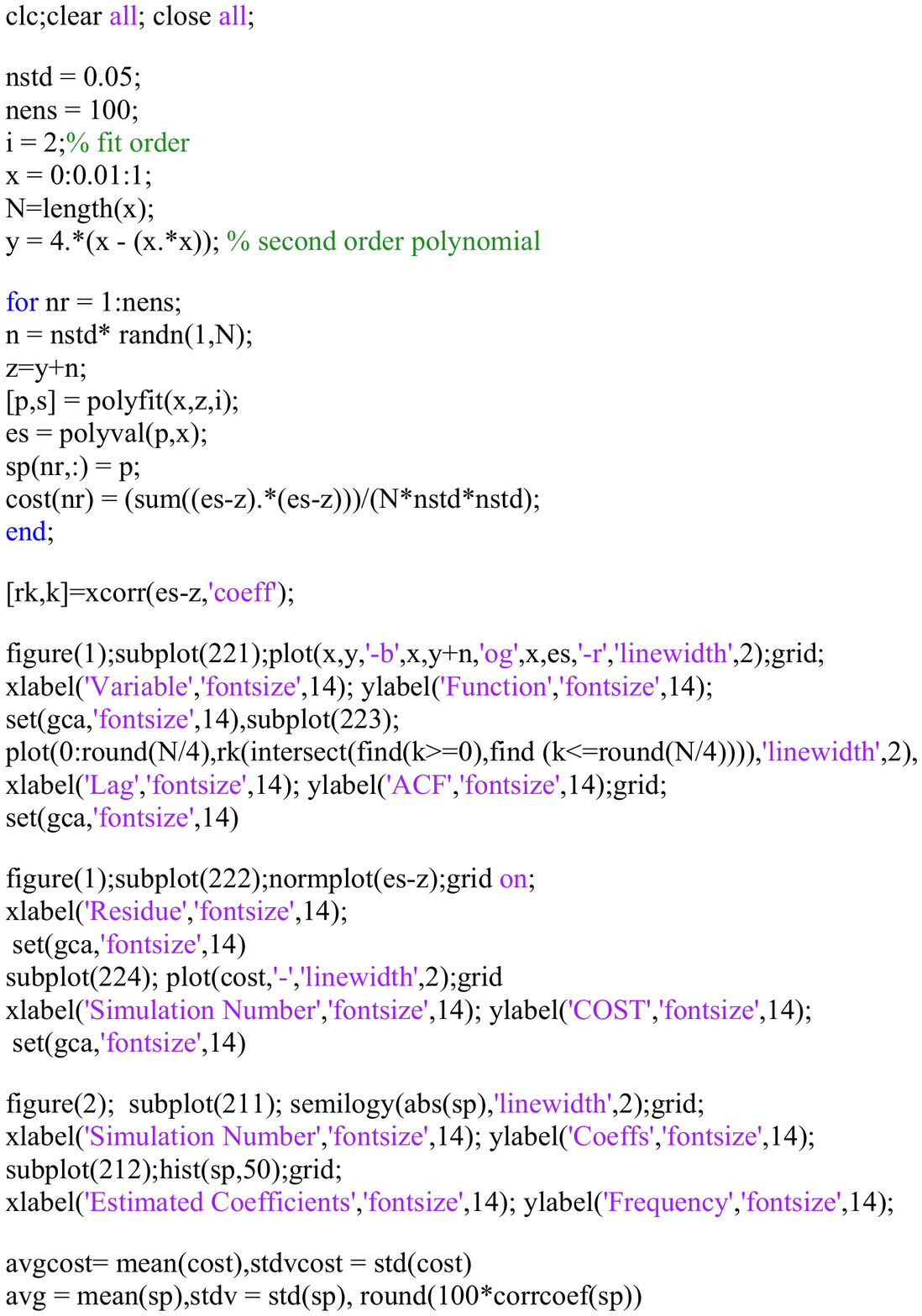}

\chapter{Filtering of a Noisy Constant Signal with and without $\mathbf{P_0}$ Scaling}
\label{X0P0}

Consider a data of N = 100 data points obtained from a constant signal of value zero added with noise. This data is processed by known value of the measurement noise variance (\textbf{R}=0.25). The standard filter steps are used and the filter iterates over the data many times. The value of initial $\mathbf{X_0}$ and $\mathbf{P_0}$ is varied over a large range of respectively (-10:10:10) and (-10:5:10). The number of iterations was set at 10. The filter uses the estimated parameter value of X at the end of each pass for the next pass through the data. The final covariance at the end of each pass can be scaled up by the number of data points to be used as the initial covariance for the next pass or not. This makes the difference in the final estimates of the parameter value X and its uncertainty P (= CRB).

\begin{figure}[h]
\includegraphics[width=6in,height=3.2in]{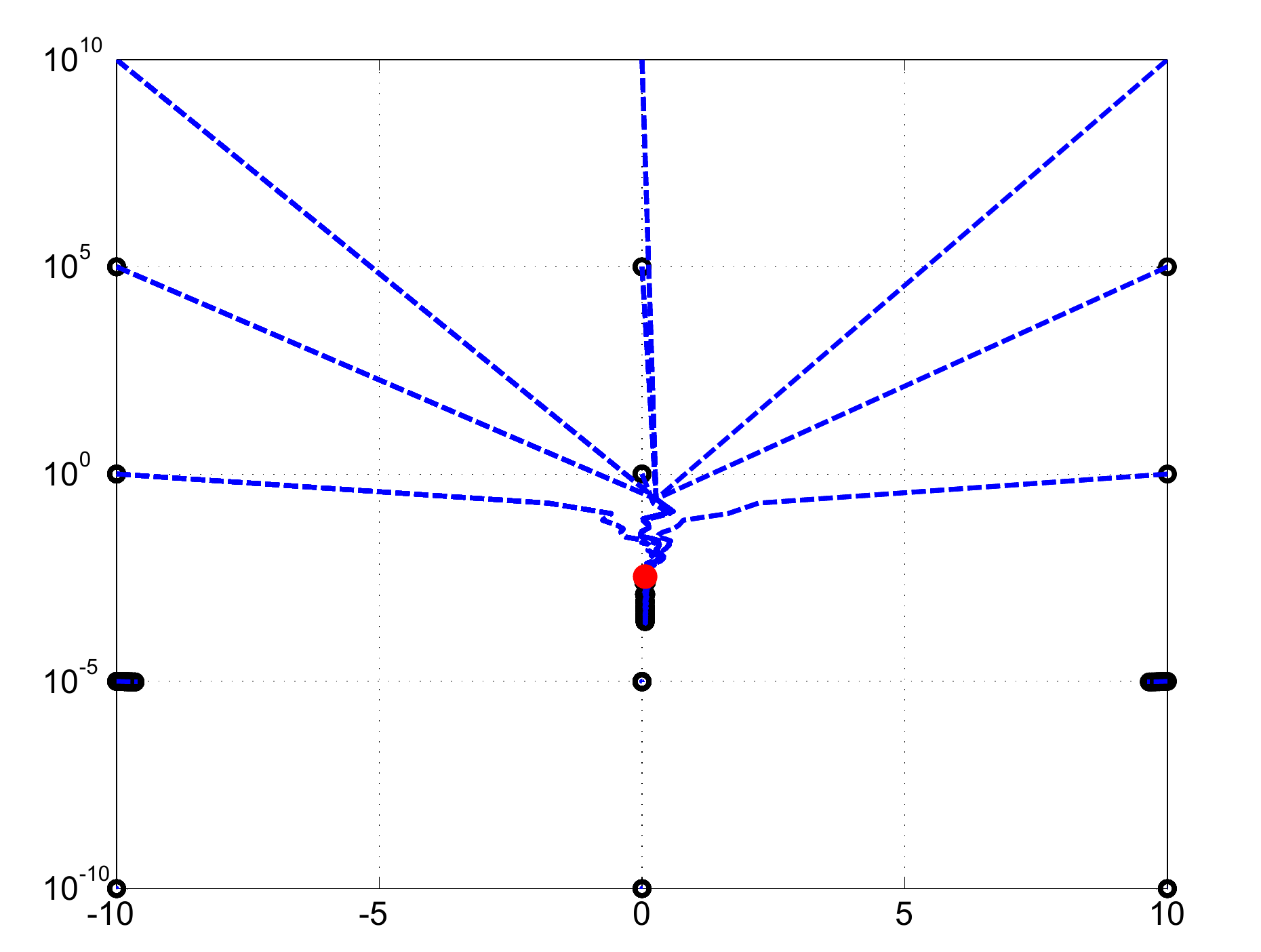}
\caption{Plot of P versus X without Scaling}
\label{X0P0_1}
\end{figure}

The Fig. \ref{X0P0_1} shows if there is no scaling then starting from very low values of $\mathbf{P_0}$ the parameter is not updated much even if the initial X value is close to the true value of zero. For higher and very high initial $\mathbf{P_0}$ the parameter estimates get updated and reaches close to the true value indicated by the red circled point (showing the mean and variance/N of the data using the simple formulae in statistics) during the first few iterations. Subsequently though the parameter estimate X does not change but the P value keeps decreasing continuously which is not acceptable.

\begin{figure}[h]
\includegraphics[width=6in,height=3.2in]{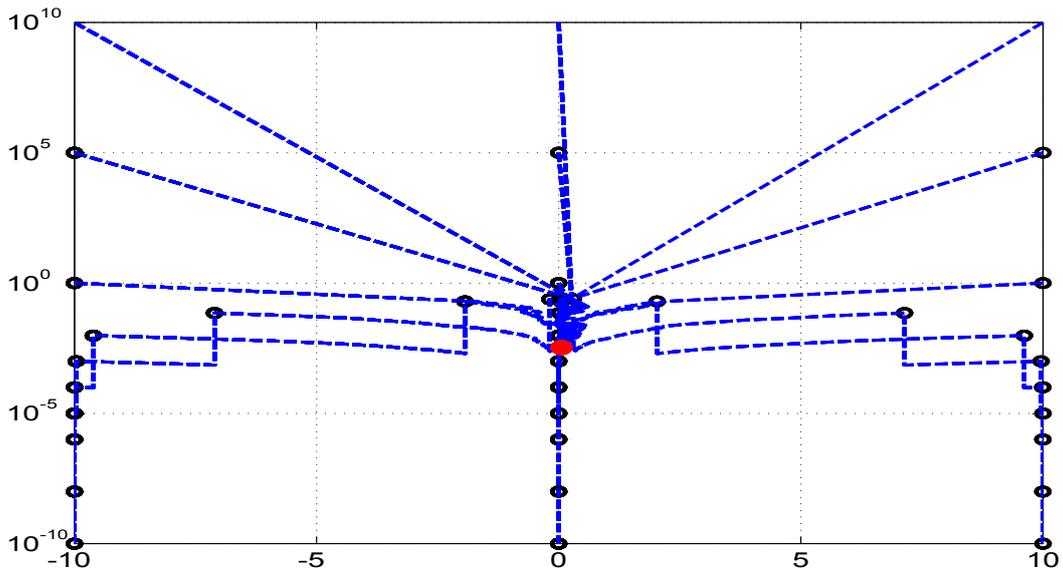}
\caption{Plot of P versus X with Scaling}
\label{X0P0_2}
\end{figure}

However if there is scaling then starting from even very low to very high values of P0 the initial parameter $\mathbf{X_0}$ gets updated as shown in Fig. \ref{X0P0_2}. Then commencing from any initial ($\mathbf{X_0}$, $\mathbf{P_0}$) the filter after some iterations reaches a statistically steady state trajectory (X, P) between the initial and the final states as it processes each one of the data point. This feature can be seen by expanding the figure around this region. The above simple experiment shows the importance $\mathbf{P_0}$ to be  scaled after every filter pass. This feature is taken over even for more involved problems.

\section*{Effect of a \textbf{Q} pulse during filter pass in lieu of $\mathbf{P_0}$ = 0 }

\begin{figure}[h]
\includegraphics[width=6in,height=3.2in]{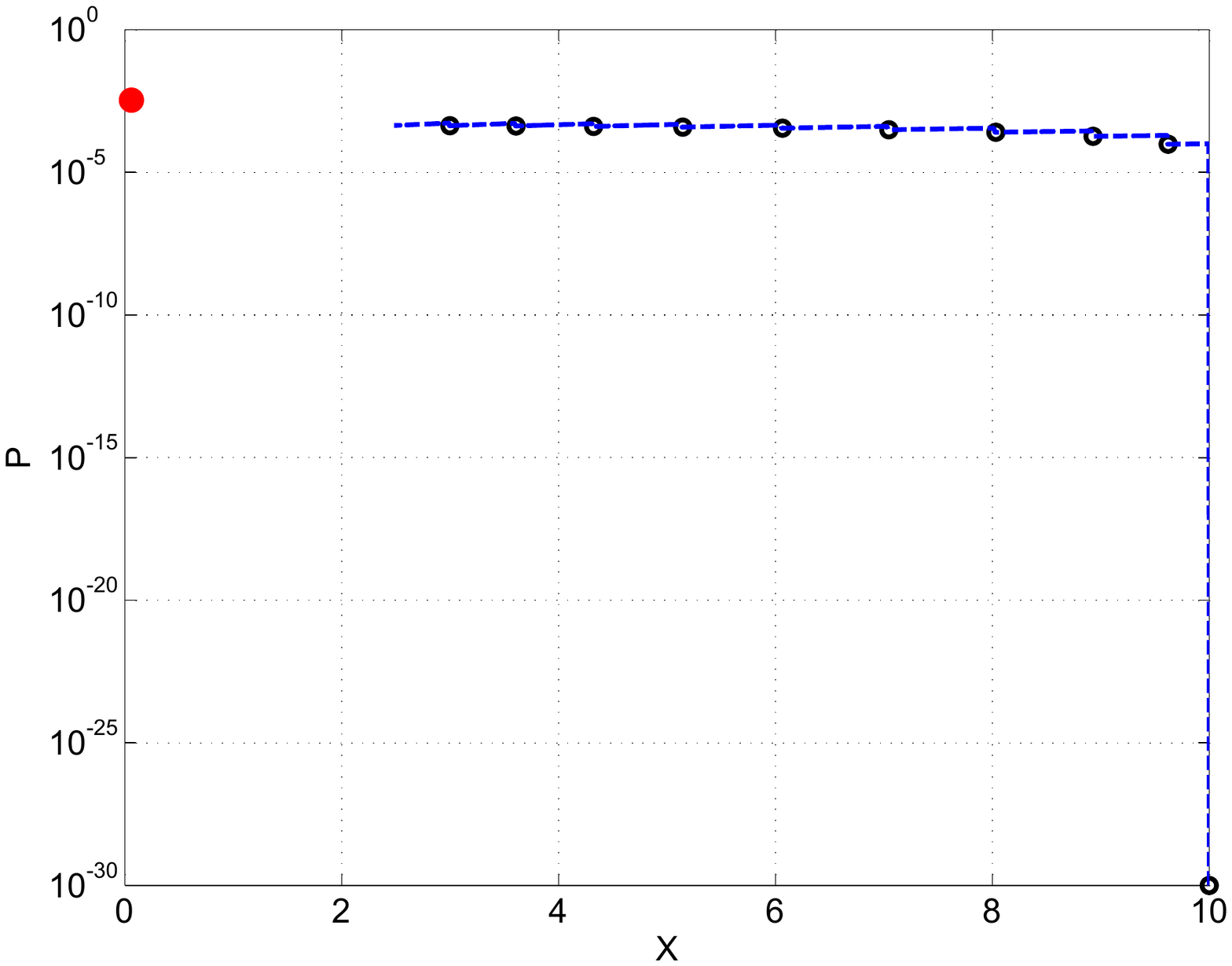}
\caption{Plot of P versus X with impulse \textbf{Q}= $10^{-4}$}
\label{X0P0_3}
\end{figure}

\begin{figure}[h]
\includegraphics[width=6in,height=3.2in]{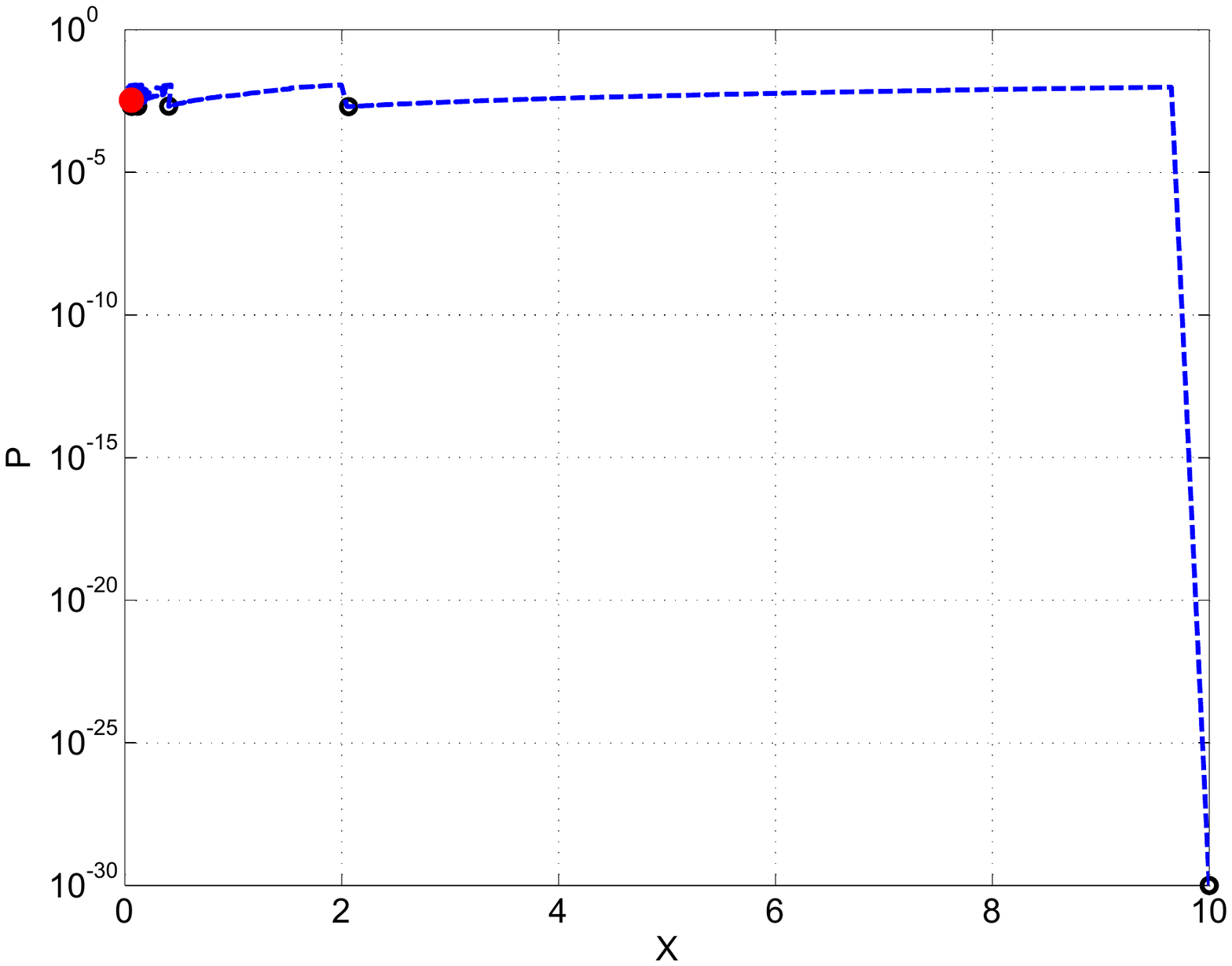}
\caption{Plot of P versus X with impulse \textbf{Q}= $10^{-2}$}
\label{X0P0_4}
\end{figure}

This experiment demonstrates that even if $\mathbf{P_0}$ = 0 it can be substituted by injecting \textbf{Q} during the filter passes. The experiment was carried out from an initial ($\mathbf{X_0}$, $\mathbf{P_0}$) anywhere in the range say (-10:10, $10^{-30}:10^{15}$). Starting from (-10, $10^{-30})$ a very small \textbf{Q} = $10^{-4}$ was injected as a single impulse before processing the second data point in every pass and nowhere else. The filter estimates in Fig. \ref{X0P0_3} shows no convergence towards the true value even after about 10 iterations but with 100 iterations the estimate was close but the covariance was lower than the true value. When \textbf{Q} was increased to $10^{-2}$ then in about 10 passes the filter reached an estimate close to the true value as shown in Fig. \ref{X0P0_4}. The reason being when a large \textbf{Q} is injected at any point the filter ignores the state estimate but believes the then measurement. Thus the filter estimate drops from a far off value to be within the measurement error band beyond which it improves the estimate by assimilating the subsequent measurement data.

An analogy from human life can be given though may not be quite perfect. The initial condition ($\mathbf{X_0}$, $\mathbf{P_0}$) represents the random conditions at birth. The sample measurements Z based on \textbf{R} are  varied individual experiences in life. Based on the experience the goal is to realize the true state represented by the parameter value zero or sunya. The filter iterations represent rebirths. The injection of \textbf{Q} continuously or a few times denotes opportunities to learn. The state improves in each birth. If one is far away like $\mathbf{X_0}$ (= -10) and confident that he knows the ultimate truth (little $\mathbf{P_0}$ = $10^{-30}$) with very low \textbf{Q} ( = $10^{-4}$) it would take far too many births to attain realization. The injection of a small \textbf{Q} = $10^{-2}$ helps to reach the goal in fewer births. However if a very large \textbf{Q} is injected in one's life at some point then the final goal is reached in that birth itself. This is what happened to Buddha after which he meditated for many years and obtained realization. But how and from where does this large \textbf{Q} come about? It is God's Grace for one who has extraordinary compassion and humility. One can also read the analogy between the `State Estimation' and `The meaning of Life' in Dan Simon \cite{Dan2006} (2006).

\includepdf[pages={1},scale=.9,offset=20 0,pagecommand=\section*{Source code showing the importance of $\mathbf{P_0}$ Scaling :}]{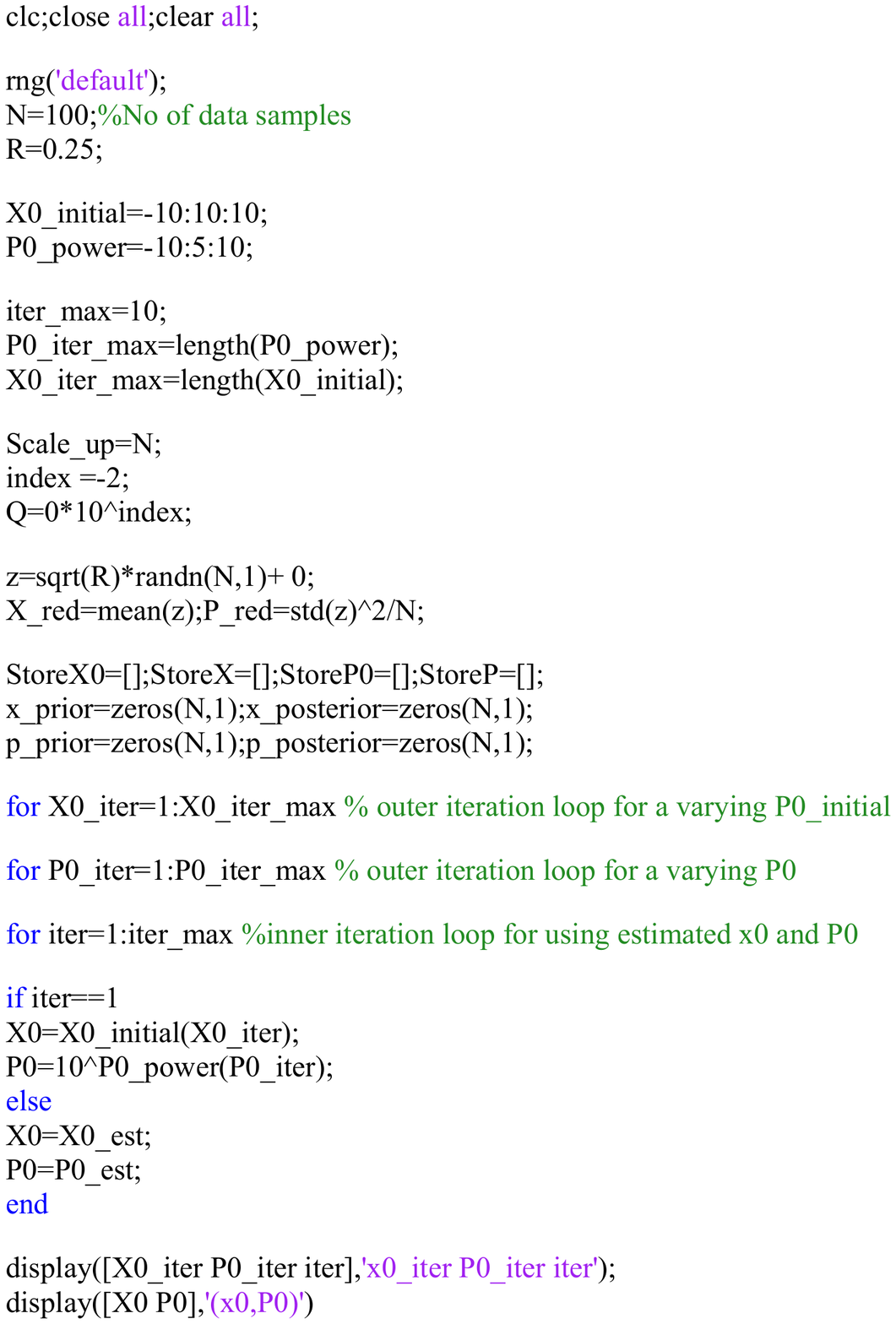}
\includepdf[pages={2-last},scale=.9,offset=20 0,pagecommand={}]{Prog_X0P0.pdf}

\chapter{Source Code used for the Simulated and Real Data Cases}

\par In this section, the MATLAB\textsuperscript{\textregistered} source code (written in version 2012b) used in generating the results are printed for reference. There are two programs one corresponding to the simulated system case (\ref{sd}) and the other corresponding to the real data case (\ref{rd}). All the input options are provided in the first few lines of the program for the user to conduct sensitivity studies.

The soft copy of the source code and the data files are embedded below. One can right click on the below icons and select the option `Save Embedded File to Disk' to save it in a desired folder.

\begin{center}
\attachfile[icon=PushPin,description=Simulated System Program]{EKF_REF_NEW.m}
\attachfile[icon=PushPin,description=Real Data Program]{EKF_REAL_NEW.m}
\attachfile[icon=Paperclip,description=case 1 data]{case1.dat}
\attachfile[icon=Paperclip,description=case 2 data]{case2.dat}
\attachfile[icon=Paperclip,description=case 3 data]{case3.dat}
\attachfile[icon=Paperclip,description=case 4 data]{case4.dat}
\attachfile[icon=Paperclip,description=case 5 data]{case5.dat}
\end{center}

\includepdf[pages={1},scale=.9,offset=20 0,pagecommand=\section{Simulated System Program :}\label{sd}]{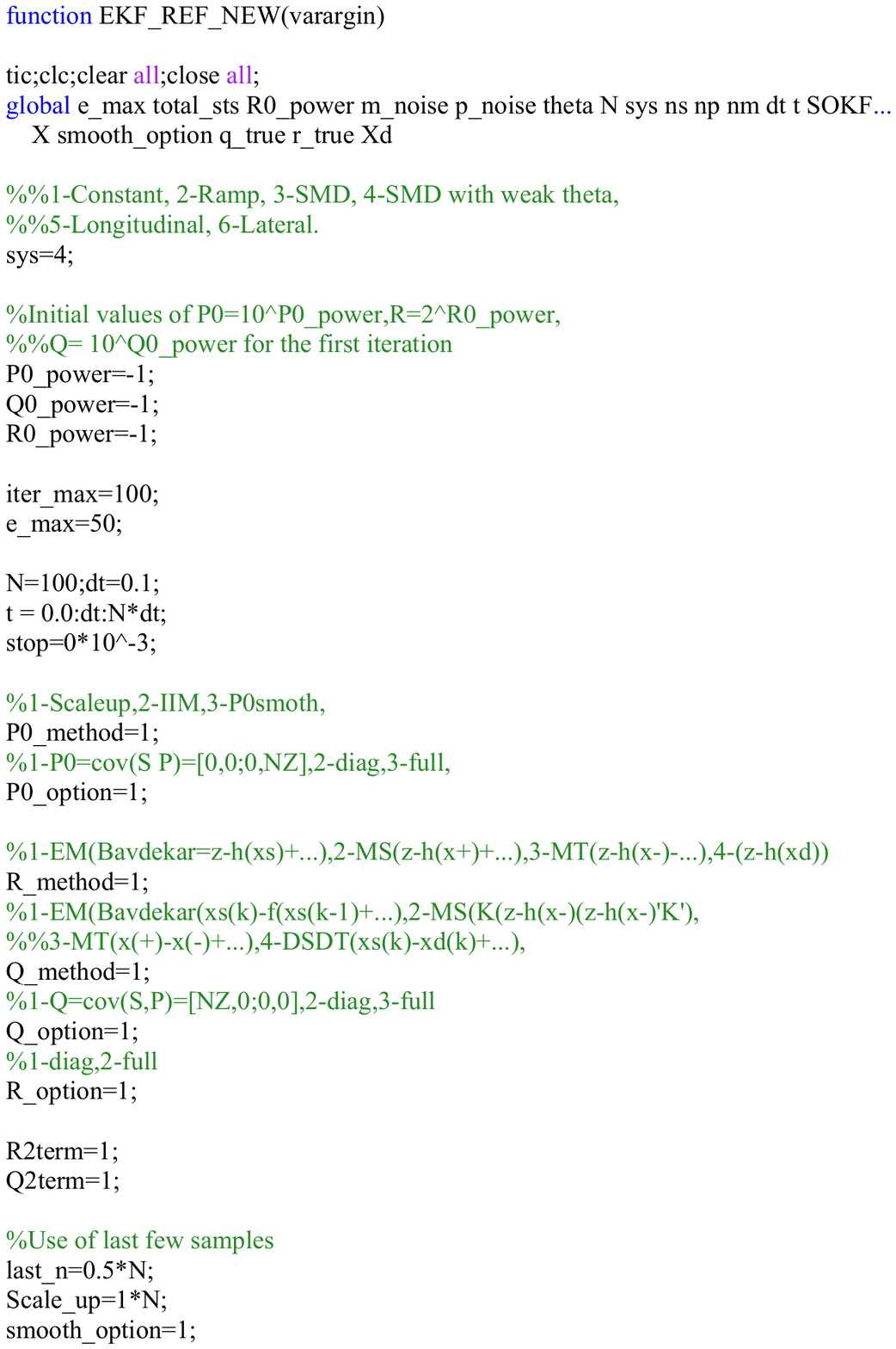}
\includepdf[pages={2-22},scale=.9,offset=20 0,pagecommand={}]{Prog1.pdf}
\includepdf[pages={23},scale=.9,offset=20 0,pagecommand=\section{Real Data Program :}\label{rd}]{Prog1.pdf}
\includepdf[pages={24-},scale=.9,offset=20 0,pagecommand={}]{Prog1.pdf}

\end{appendices}

\newpage
\renewcommand{\bibname}{References}






\end{spacing}
\end{document}